\documentclass{article}
\usepackage[margin=.75in]{geometry}
\usepackage{amsthm,amsmath,amssymb}
\usepackage{graphicx}
\usepackage[hidelinks,breaklinks]{hyperref}
\newtheorem{theorem}{Theorem}[]

\newtheorem{remark1}[theorem]{Remark}
\newenvironment{remark}{\begin{remark1} \rm}{\end{remark1}}

\DeclareMathOperator{\Bernoulli}{Bernoulli}
\DeclareMathOperator{\E}{\mathop{}\mathbb{E}}

\title{Cumulative deviation of a subpopulation from the full population}
\author{Mark Tygert\\{\normalsize Facebook Artificial Intelligence Research}\\
{\normalsize 1 Facebook Way, Menlo Park, CA 94025}\\
{\normalsize Main e-mail address:\ \ {\tt mark@tygert.com}}\\
{\normalsize \today}}
\date{\vspace{-.25in}}

\begin{document}

\maketitle

\begin{abstract}
Assessing equity in treatment of a subpopulation often involves assigning
numerical ``scores'' to all individuals in the full population such that
similar individuals get similar scores; matching via propensity scores
or appropriate covariates is common, for example. Given such scores,
individuals with similar scores may or may not attain similar outcomes
independent of the individuals' memberships in the subpopulation.
The traditional graphical methods for visualizing inequities are known as
``reliability diagrams'' or ``calibrations plots,'' which bin the scores
into a partition of all possible values, and for each bin plot
both the average outcomes for only individuals in the subpopulation
as well as the average outcomes for all individuals; comparing the graph
for the subpopulation with that for the full population gives some sense
of how the averages for the subpopulation deviate from the averages
for the full population. Unfortunately, real data sets contain
only finitely many observations, limiting the usable resolution of the bins,
and so the conventional methods can obscure important variations
due to the binning. Fortunately, plotting cumulative deviation
of the subpopulation from the full population as proposed in this paper
sidesteps the problematic coarse binning.
The cumulative plots encode subpopulation deviation directly
as the slopes of secant lines for the graphs.
Slope is easy to perceive even when the constant offsets of the secant lines
are irrelevant. The cumulative approach avoids binning that smooths over
deviations of the subpopulation from the full population.
Such cumulative aggregation furnishes both high-resolution graphical methods
and simple scalar summary statistics (analogous to those
of Kuiper and of Kolmogorov and Smirnov used
in statistical significance testing for comparing probability distributions).

\medskip

\end{abstract}

\section{Introduction}
\label{intro}

Analysis of whether a subpopulation has had equitable treatment
often assesses whether similar individuals attain similar outcomes
irrespective of the individuals' memberships in the subpopulation.
To effect the comparison, each individual gets a real-valued ``score''
used to match that individual with other similar individuals
(such matching could use propensity scores or a suitable covariate,
for instance).
Each individual also gets a real-valued outcome of the treatment.
A formal assessment could then consider how much the average outcome
for individuals with a given score that belong to the subpopulation
deviates from the average outcome for all individuals from the full population
that have that same score.

When there are only finitely many observations,
questions of statistical significance arise;
for example, if the scores are predicted probabilities
and the outcomes are drawn from independent (but not necessarily
identically distributed) Bernoulli distributions with parameters
given by the predicted probabilities, then the average outcome
for individuals with a given score fluctuates across different random samples
defining the subpopulation being analyzed.
In such scenarios, the average outcome for the observed subpopulation
would be expected to deviate stochastically from the average
for the full population.
Furthermore, each individual in the sampled subpopulation may very well
have a different score from all the others, requiring some aggregation
of scores in order to average away the statistical noise.
The conventional approach is to partition the scores into some number of bins
and calculate averages separately for every bin, trading off resolution
in the scores for increased confidence in the statistical estimates.
The present paper proposes an alternative based on cumulative statistics
that avoids the necessarily arbitrary binning
(binning that typically follows heuristics discussed shortly);
the cumulative approach yields graphical methods
as well as scalar summary statistics that are similar
to the Kolmogorov-Smirnov and Kuiper metrics familiar
from statistical significance testing for comparing probability distributions.

More concretely, subpopulations commonly considered include those associated
with protected classes (such as those defined by race, color, religion, gender,
national origin, age, disability, veteran status, or genetic information)
and those associated with biomedicine (such as diseased, infected, treated,
or recovered).
The present paper discusses generic methodology applicable to all cases,
illustrating the methods via interesting subpopulations that are unlikely
to court controversy via their consideration; this paper avoids
giving examples based on subpopulations defined by sensitive classes
of direct interest to hot-button issues.
The illustrative examples presented below are deliberately anodyne
(but hopefully still sufficiently engaging) to avoid distracting
from the focus on the statistical methodology being proposed.

Mathematically speaking,
we consider $m$ real-valued observations $R_1$,~$R_2$, \dots, $R_m$
of the outcomes of independent trials with corresponding real-valued scores
$S_1$,~$S_2$, \dots, $S_m$ (where the scores very well may determine
the probability distributions of the trials; for example,
if $0 \le S_i \le 1$, then $R_i$ could be the outcome of a Bernoulli trial
whose probability of success is $S_i$). We view $R_1$,~$R_2$, \dots, $R_m$
(but not $S_1$,~$S_2$, \dots, $S_m$) as random. Without loss of generality,
we order the scores (preserving the pairing of $R_i$ with $S_i$ for every $i$)
such that $S_1 \le S_2 \le \dots \le S_m$, ordering any ties at random,
perturbed so that $S_1 < S_2 < \dots < S_m$.
We consider a subset of indices corresponding to members
of a subpopulation of interest, say $i_1$, $i_2$, \dots, $i_n$, with $n < m$;
without loss of generality, we order the indices
such that $1 \le i_1 < i_2 < \dots < i_n \le m$.
Each observation $R_i$ and score $S_i$ may also come
with a positive weight $W_i$; however,
we focus first on the simpler, more common case of uniform weighting,
in which $W_1 = W_2 = \dots = W_m$, and generalize to arbitrary weights later
(in Subsection~\ref{weighted} below).

The classical methods require choosing some partitions of the real line
into $\ell$ disjoint intervals with endpoints $B_1$, $B_2$, \dots, $B_{\ell-1}$
and another (possibly the same) $\ell$ disjoint intervals with endpoints
$\tilde{B}_1$, $\tilde{B}_2$, \dots, $\tilde{B}_{\ell-1}$
such that $B_1 < B_2 < \dots < B_{\ell-1}$
and $\tilde{B}_1 < \tilde{B}_2 < \dots < \tilde{B}_{\ell-1}$.
We can then form the averages for the subpopulation
\begin{equation}
\label{subY}
Y_k = \frac{\sum_{j : B_{k-1} < S_{i_j} \le B_k} R_{i_j}}
           {\#\{j : B_{k-1} < S_{i_j} \le B_k\}}
\end{equation}
and for the full population
\begin{equation}
\label{fullY}
\tilde{Y}_k = \frac{\sum_{i : \tilde{B}_{k-1} < S_i \le \tilde{B}_k} R_i}
                   {\#\{i : \tilde{B}_{k-1} < S_i \le \tilde{B}_k\}}
\end{equation}
for $k = 1$, $2$, \dots, $\ell$,
under the convention that $B_0 = \tilde{B}_0 = -\infty$
and $B_{\ell} = \tilde{B}_{\ell} = \infty$.
We also calculate the average scores in the bins for the subpopulation
\begin{equation}
\label{subX}
X_k = \frac{\sum_{j : B_{k-1} < S_{i_j} \le B_k} S_{i_j}}
           {\#\{j : B_{k-1} < S_{i_j} \le B_k\}}
\end{equation}
and for the full population
\begin{equation}
\label{fullX}
\tilde{X}_k = \frac{\sum_{i : \tilde{B}_{k-1} < S_i \le \tilde{B}_k} S_i}
                   {\#\{i : \tilde{B}_{k-1} < S_i \le \tilde{B}_k\}}
\end{equation}
for $k = 1$, $2$, \dots, $\ell$,
under the same convention that $B_0 = \tilde{B}_0 = -\infty$
and $B_{\ell} = \tilde{B}_{\ell} = \infty$.
A graphical method for assessing the deviation of the subpopulation
from the full population is then to scatterplot the pairs
$(X_1, Y_1)$, $(X_2, Y_2)$, \dots, $(X_{\ell}, Y_{\ell})$ in black
and the pairs $(\tilde{X}_1, \tilde{Y}_1)$, $(\tilde{X}_2, \tilde{Y}_2)$,
\dots, $(\tilde{X}_{\ell}, \tilde{Y}_{\ell})$ in gray.
Comparing the black plotted points (possibly connected with black lines)
to the gray plotted points (possibly connected with gray lines)
then indicates how much the subpopulation deviates from the full population.
Especially when assessing the calibration or reliability
of probabilistic predictions, this graphical method is known
as a ``reliability diagram'' or ``calibration plot,''
as reviewed, for example, by~\cite{corbett-davies-pierson-feller-goel-huq}
and~\cite{crowson-atkinson-therneau}.

A full review of the literature is available in Subsection~\ref{aintro}
of Appendix~\ref{calibration}.

There are at least two common choices of the bins
whose endpoints are $B_1$, $B_2$, \dots, $B_{\ell-1}$
(and similarly for the bins whose endpoints are
$\tilde{B}_1$, $\tilde{B}_2$, \dots, $\tilde{B}_{\ell-1}$).
The first is to make $B_1$, $B_2$, \dots, $B_{\ell-1}$ be equispaced.
The second is to select $B_1$, $B_2$, \dots, $B_{\ell-1}$ such that
the number of scores from the subpopulation that fall in the $k$th bin,
that is, $\#\{j : B_{k-1} < S_{i_j} \le B_k\}$, is the same for all $k$
(aside from the rightmost bin, that for $k = \ell$,
if $n$ is not perfectly divisible by $\ell$).
In both cases, the number $\ell$ of bins is a parameter that we can vary
to trade-off higher-confidence estimates for finer resolution
in detecting deviation as a function of score (and vice versa).
Unfortunately, no choice can fully offset how the difference between
the subpopulation and the full population is typically the primary interest,
whereas the standard plot bins the subpopulation and the full population
separately (potentially smoothing away information due to the discretization).
Plotting the difference directly would solve this particular problem.
Even then, however, no choice of bins can be optimal
for all possible distributions of scores or for all possible distributions
of deviations between the subpopulation and the full population.
Binning will always discretize the distributions,
smoothing away potentially important information.

Fortunately, binning so coarsely is unnecessary in the methods proposed below.
The methods discussed below employ exactly one bin per score $S_{i_j}$,
$j = 1$,~$2$, \dots, $n$, thus viewing the subpopulation on its own terms,
dictated by the observations from the subpopulation.
Employing only one bin per score would be nonsensical in the classical plots,
as then the classical plots would trade-off all statistical confidence
for the finest resolution possible.
The cumulative methods sacrifice no resolution that is available
in the observed data for the subpopulation.

For a simple illustrative example,
Figure~\ref{5000} displays both the conventional reliability diagrams
as well as the cumulative plot proposed below.
A detailed description is available in Subsection~\ref{synthetic} below.
The lowermost two rows of Figure~\ref{5000} are the classical diagrams,
with $m =$ 50,000 and $n =$ 5,000;
there are $\ell = 10$ bins for each diagram in the second row
and $\ell = 50$ for each diagram in the third row.
In the lowermost two rows, the bins are equispaced along the scores
in the leftmost plots, whereas each bin contains the same number of scores
from $S_{i_1}$,~$S_{i_2}$, \dots, $S_{i_n}$
(or from $S_1$,~$S_2$, \dots, $S_m$) in the rightmost plots.
The diagram of Figure~\ref{5000e} plots the ordered pairs
$(S_1, P_1)$,~$(S_2, P_2)$, \dots, $(S_m, P_m)$ in gray and
$(S_{i_1}, P_{i_1})$,~$(S_{i_2}, P_{i_2})$, \dots, $(S_{i_n}, P_{i_n})$
in black, where $P_1$,~$P_2$, \dots, $P_m$ are the expected values
of $R_1$,~$R_2$, \dots, $R_m$, respectively;
for this example, $R_1$,~$R_2$, \dots, $R_m$ are drawn independently
from Bernoulli distributions with probabilities of success
$P_1$,~$P_2$, \dots, $P_m$.
Thus, Figure~\ref{5000e} depicts the ``ground-truth'' expectations
that the lowermost two rows of classical plots in Figure~\ref{5000}
are trying to characterize.
The gray points correspond to the full population,
while the solid black points correspond to the subpopulation.

The topmost row of Figure~\ref{5000} displays both
the cumulative plot introduced below as well as its ideal noiseless
``ground-truth'' constructed using the expected values
$P_1$,~$P_2$, \dots, $P_m$ of the random observations
$R_1$,~$R_2$, \dots, $R_m$.
Leaving elucidation of the cumulative plots and their construction
to Section~\ref{methods} below,
we just point out here that the deviation of the subpopulation
from the averages over the full population across an interval
is equal to the slope of the secant line for the graph over that interval,
aside from expected stochastic fluctuations
detailed in Subsection~\ref{significance} below.
Steep slopes correspond to substantial deviations across the ranges of scores
where the slopes are steep, with the deviation over an interval exactly equal
to the expected value of the slope of the secant line for the graph
over that interval. The cumulative plot closely resembles
its ideal ground-truth in Figure~\ref{5000},
and the Kolmogorov-Smirnov and Kuiper metrics conveniently summarize
the statistical significance of the overall deviation across all scores,
in accord with Subsections~\ref{scalarstats} and~\ref{significance} below.

The structure of the remainder of the present paper is as follows:
Section~\ref{methods} details the statistical methodology that
Section~\ref{results} illustrates
via numerical examples.\footnote{Permissively licensed open-source software
implementing (in Python modules) all these methods --- software that also
reproduces all figures and statistics reported below --- is available at
\url{https://github.com/facebookresearch/fbcdgraph}}
Section~\ref{results} also proposes avenues for further development.
Section~\ref{conclusion} then briefly summarizes the methods
and numerical results. Appendix~\ref{calibration} describes methods
for calibrating probabilistic predictions that are analogous
to the cumulative methodology introduced in Section~\ref{methods}
for assessing deviation of a subpopulation from the full population.
Appendix~\ref{calibration} also contains a literature review,
in Subsection~\ref{aintro}; please consult Subsection~\ref{aintro}
for a discussion of related work.
Appendix~\ref{caution} warns about potentially tempting overinterpretations
of the various plots.
Table~\ref{notation} gives a glossary of the notation used throughout
except in the appendices, while Table~\ref{anotation} gives a glossary
of the notation used in the appendices.

\newlength{\vertsep}
\setlength{\vertsep}{.1in}
\newlength{\imsize}
\setlength{\imsize}{.465\textwidth}

\begin{figure}
\begin{centering}

\parbox{\imsize}{\includegraphics[width=\imsize]
                {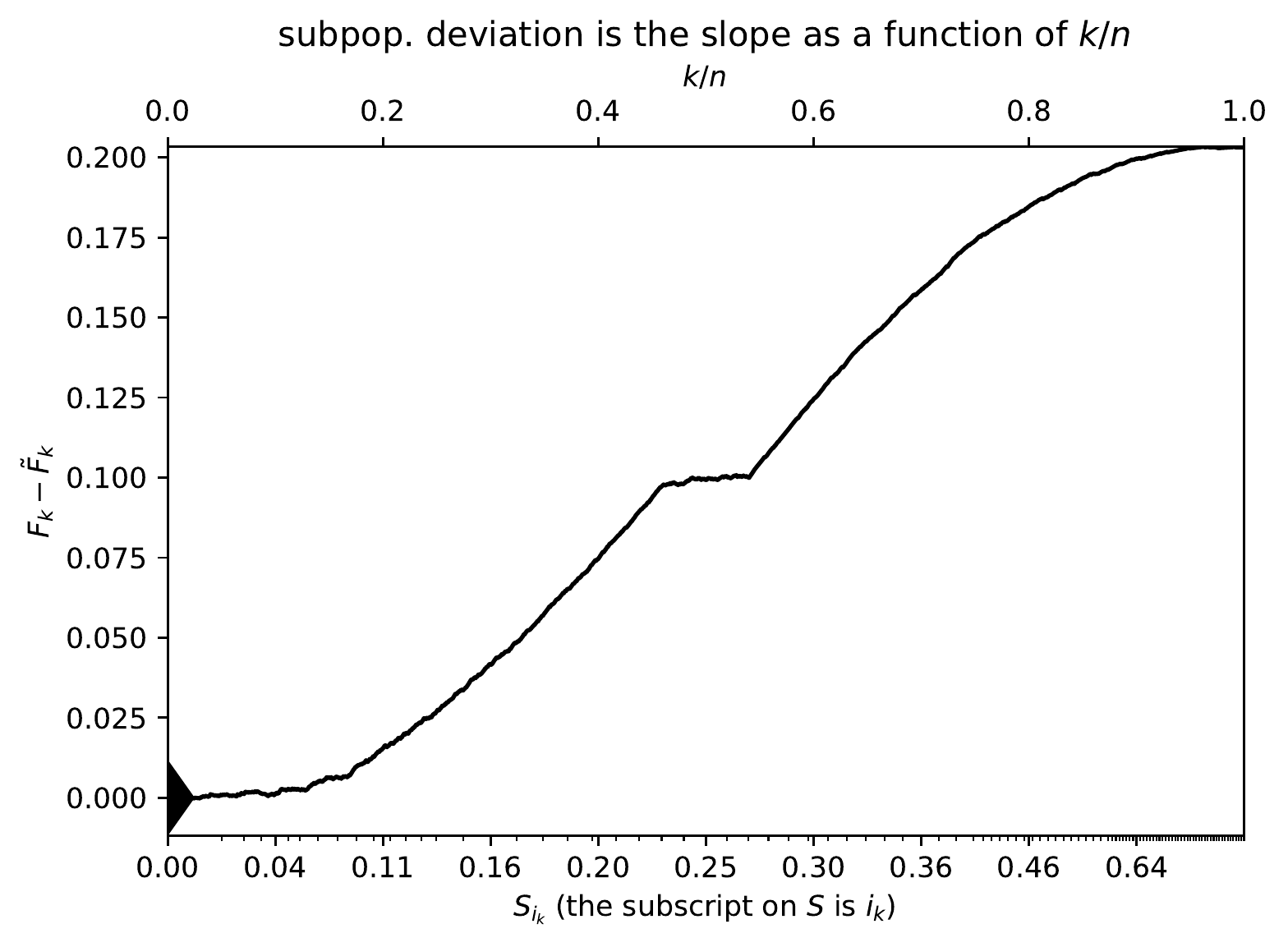}}
\quad\quad
\parbox{\imsize}{\includegraphics[width=\imsize]
                {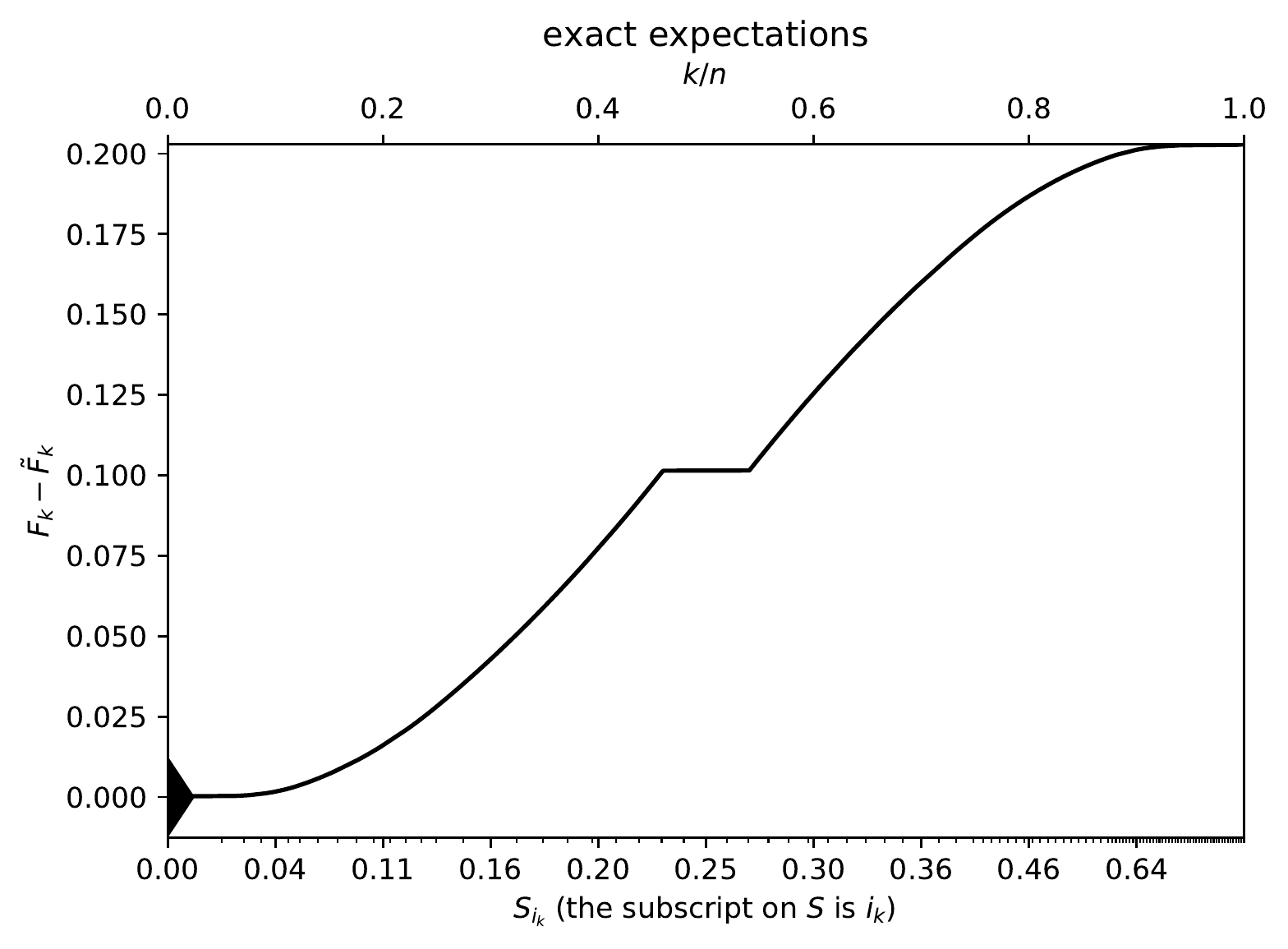}}

\vspace{\vertsep}

\parbox{\imsize}{\includegraphics[width=\imsize]
                {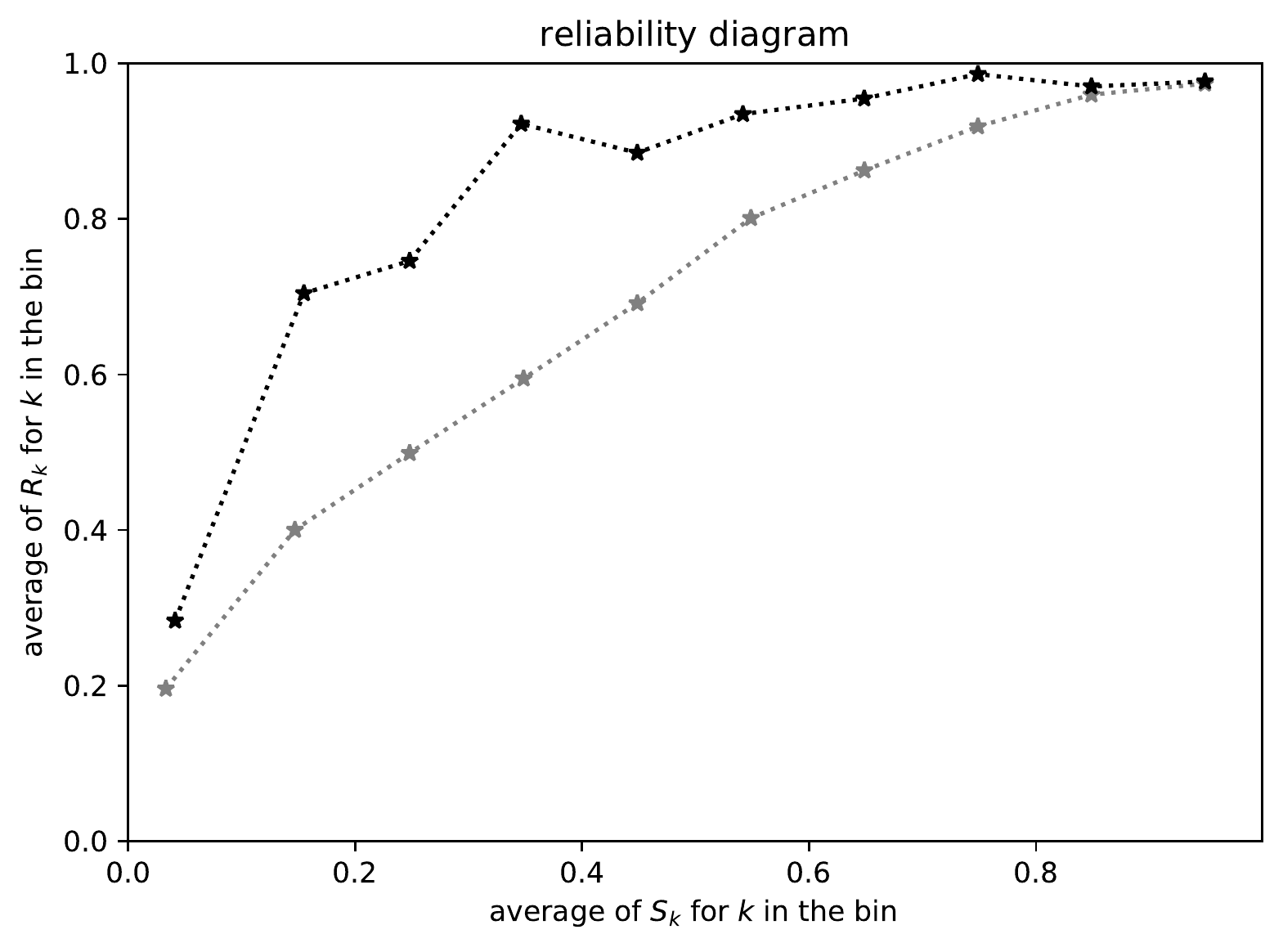}}
\quad\quad
\parbox{\imsize}{\includegraphics[width=\imsize]
                {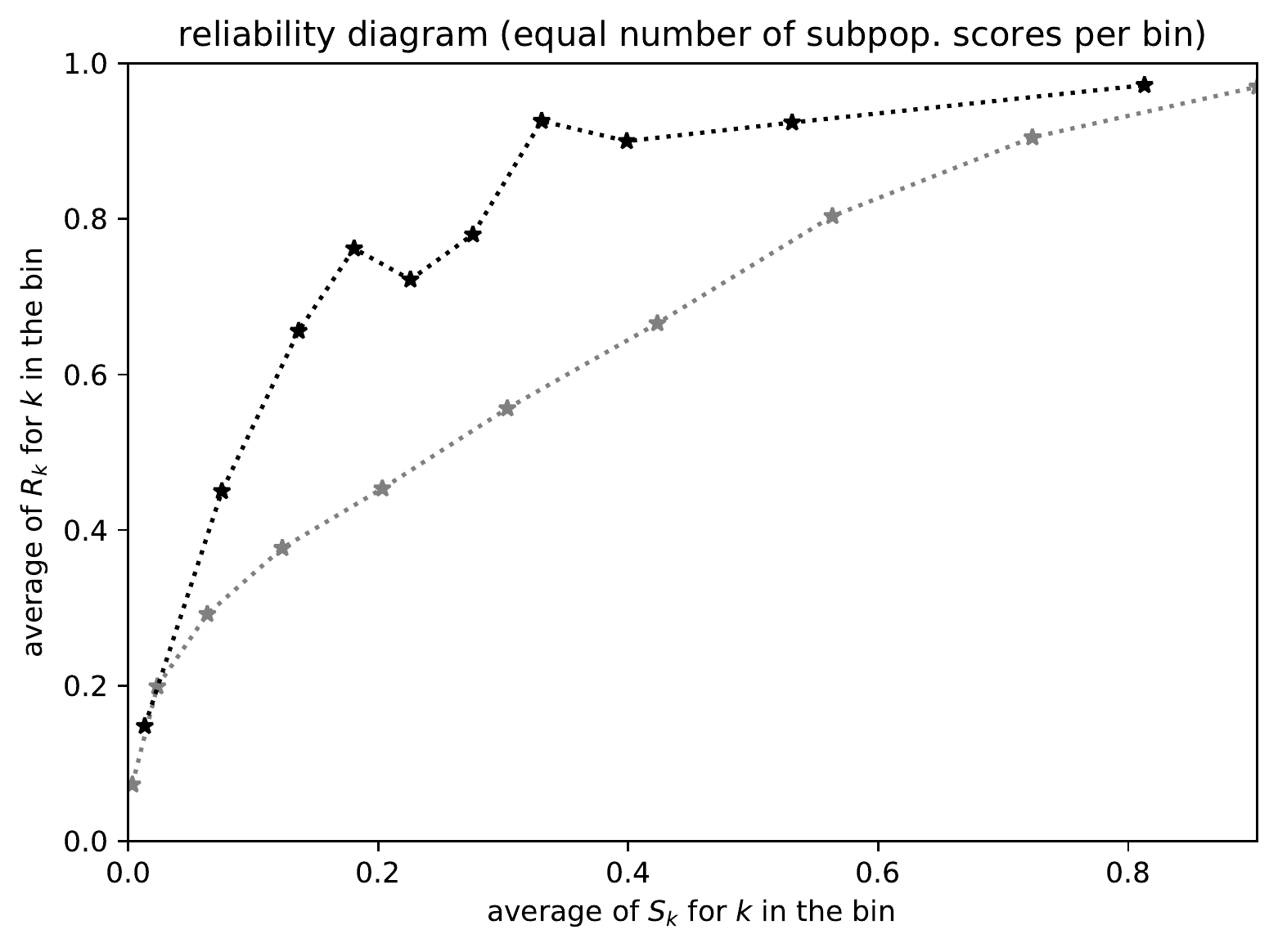}}

\vspace{\vertsep}

\parbox{\imsize}{\includegraphics[width=\imsize]
                {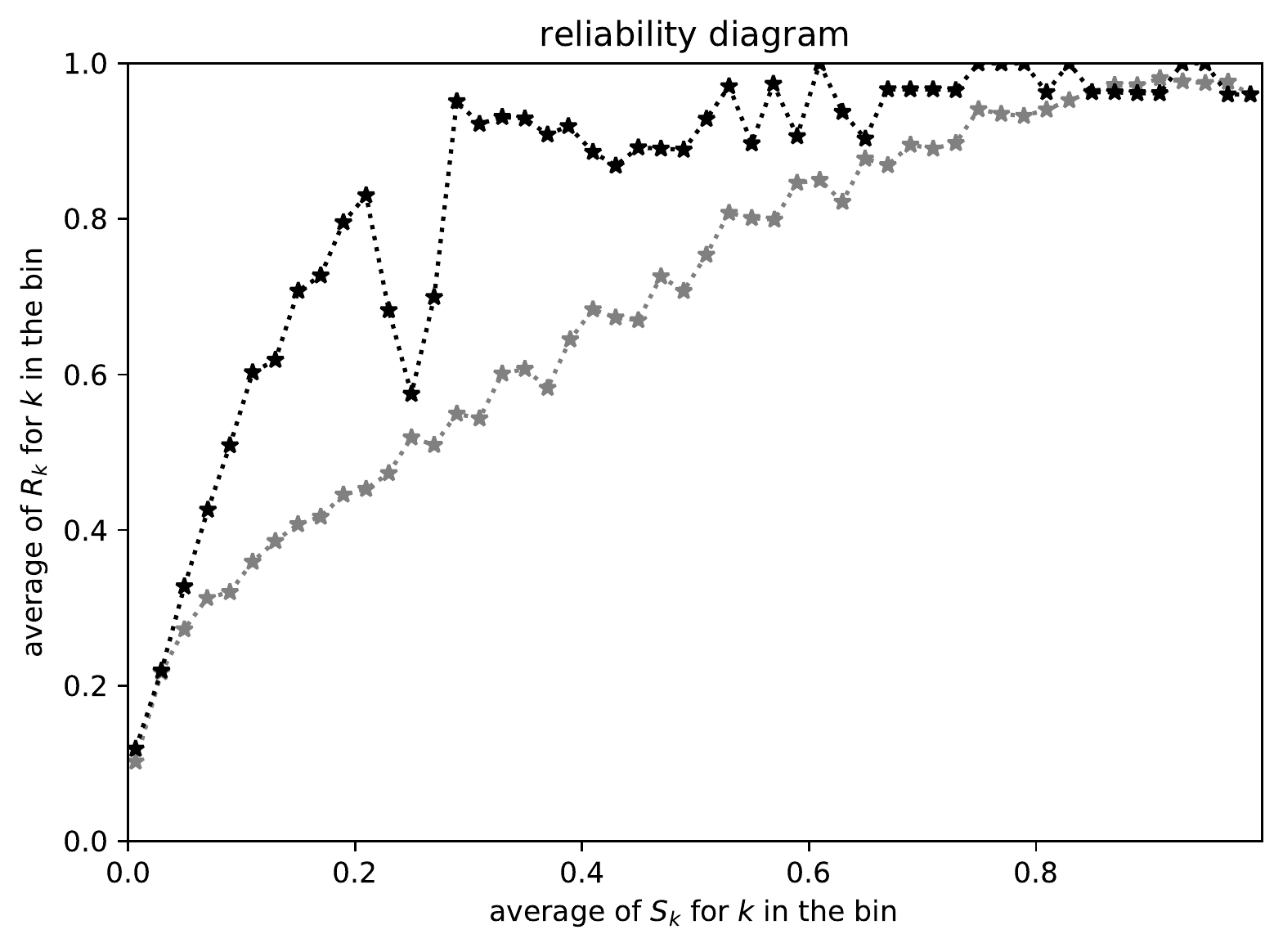}}
\quad\quad
\parbox{\imsize}{\includegraphics[width=\imsize]
                {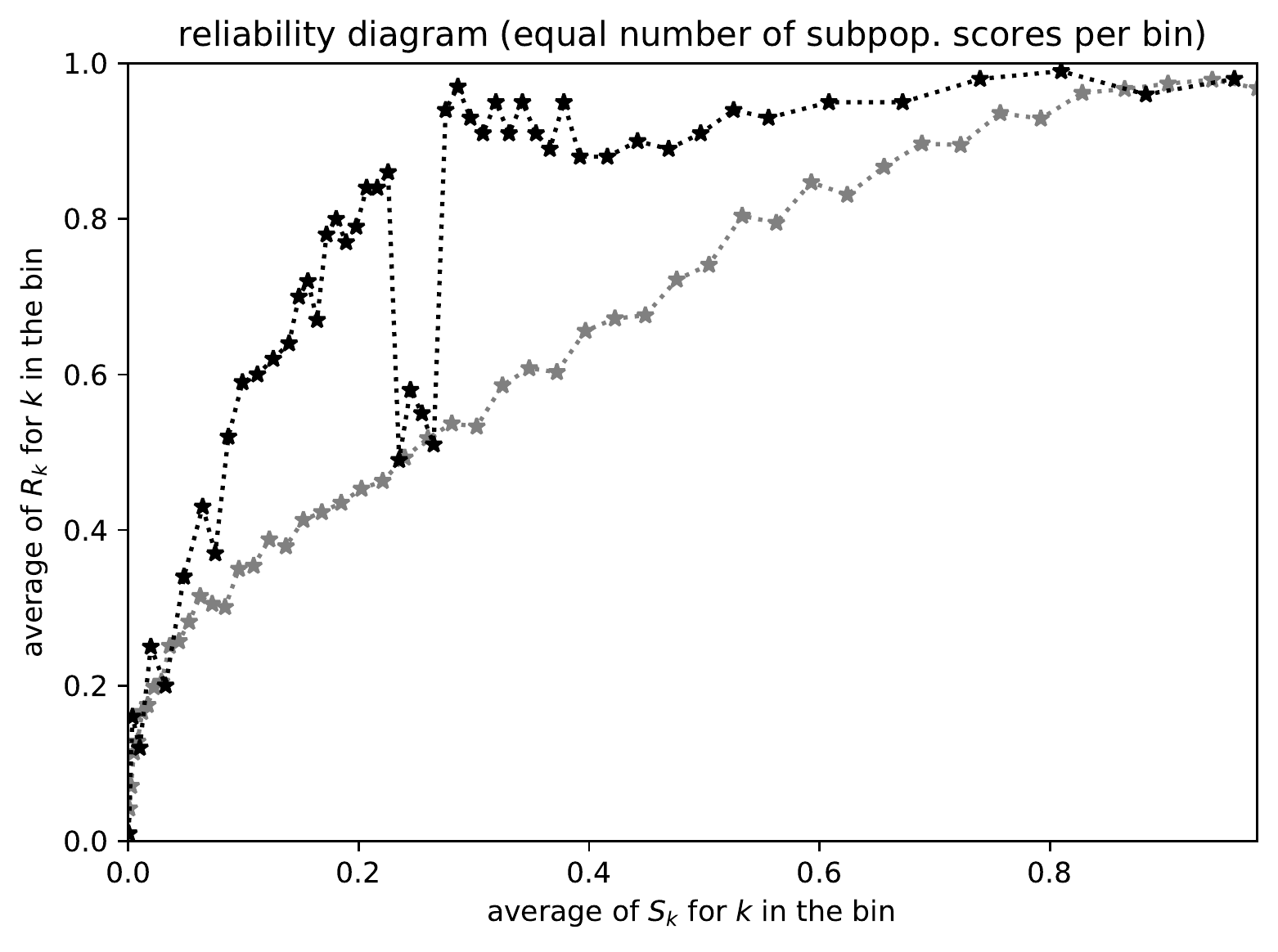}}

\end{centering}
\caption{$n =$ 5,000;
         Kuiper's statistic is $0.2037 / \sigma = 34.34$,
         Kolmogorov's and Smirnov's is $0.2033 / \sigma = 34.27$.
Figure~\ref{5000e} displays the ground-truth reliability diagram.
The reliability diagram with 50 bins
that each contain the same number of scores from the subpopulation
is able to detect the notch around scores of 0.25;
however, the oscillation of the bin frequencies
for the subpopulation complicates disentangling real variations
from statistical noise. The reliability diagrams that each have only 10 bins
exhibit fewer random oscillations, but smear out the notch.
In the reliability diagrams, the averages for the subpopulation are black,
while the averages for the full population are gray.
In the top row, the plot of cumulative deviation resolves the notch nicely
while displaying minimal random fluctuations across the full range of scores.
The scalar summary statistics of Kuiper and of Kolmogorov and Smirnov
very successfully detect the statistically significant deviation
of the subpopulation from the full population.
}
\label{5000}
\end{figure}

\begin{figure}
\begin{centering}

\parbox{\imsize}{\includegraphics[width=\imsize]
                {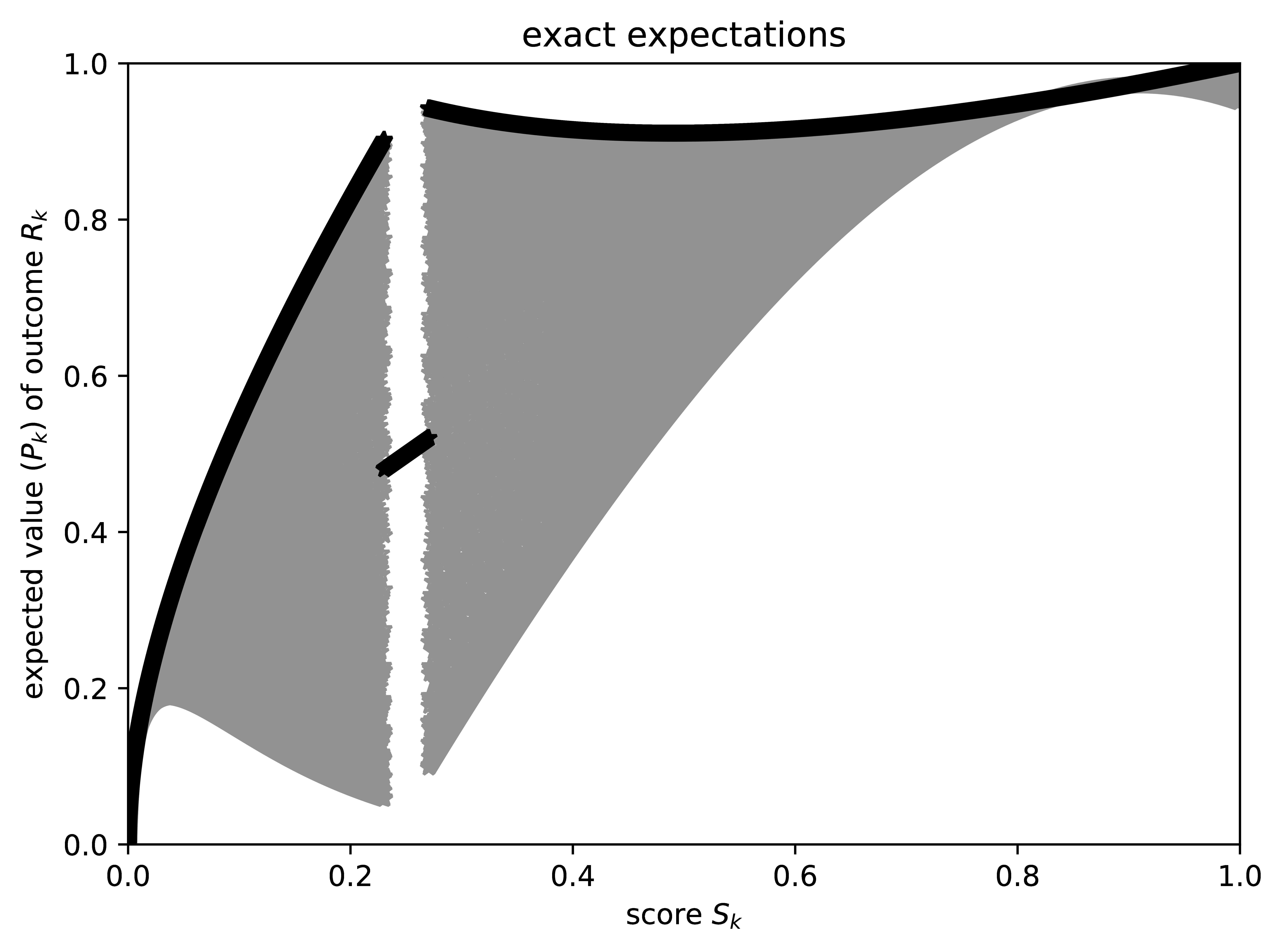}}

\end{centering}
\caption{Ground-truth reliability diagram for Figure~\ref{5000}}
\label{5000e}
\end{figure}

\section{Methods}
\label{methods}

This section formulates the cumulative statistics mathematically,
with Subsection~\ref{high-level} proposing a workflow for large-scale analysis.
Subsection~\ref{graphical} details the graphical methods.
Subsection~\ref{scalarstats} details the scalar summary metrics.
Subsection~\ref{significance} discusses statistical significance
for both the graphical methods and the summary statistics.
Subsection~\ref{weighted} presents a generalization
of these statistical methodologies to the case of weighted samples
(beyond just equally or uniformly weighted).

\begin{table}
\caption{Notational conventions for all but the appendices
(Table~\ref{anotation} summarizes the notation used in the appendices.
The symbols in the tables are in alphabetical order.)}
\label{notation}
\begin{center}
\begin{tabular}{llll}
\hline
& & equation for the & equation for the \\
symbol & meaning & unweighted case & case with weights \\\hline
$A_k$ & abscissa for the cumulative graph in the case with weights &
(Not applicable) & (\ref{abscissae}) \\
$D$ & Kuiper statistic & (\ref{Kuiper}) & (\ref{Kuiper}) \\
$\Delta_k$ & expected slope of $F_j - \tilde{F}_j$ from $j = k-1$ to $j = k$ &
(\ref{delta}) & (\ref{deltaw}) \\
$F_k$ & cumulative response for the subpopulation &
(\ref{empirical}) & (\ref{empiricalw}) \\
$\tilde{F}_k$ & cumulative average response for the full population &
(\ref{full}) & (\ref{fullw}) \\
$G$ & Kolmogorov-Smirnov statistic & (\ref{Kolmogorov-Smirnov}) &
(\ref{Kolmogorov-Smirnov}) \\
$i_k$ & index of an individual from the subpopulation &
(\ref{subY}) & (\ref{subYw}) \\
$P_k$ & actual prob.\ of success for a Bernoulli trial in synthetic data &
(Subsection~\ref{synthetic}) & (Subsection~\ref{synthetic}) \\
$R_k$ & response --- (random) dependent variable, outcome,
or result & (\ref{subY}) & (\ref{subYw}) \\
$\tilde{R}_k$ & response for the full population averaged to the subpopulation
& (\ref{average}) & (\ref{averagew}) \\
$S_k$ & score --- (non-random) independent variable &
(\ref{subX}) & (\ref{subXw}) \\
$\sigma$ & scale of random fluctuations over the full range of scores &
(\ref{stddev}) or~(\ref{empiricalstddev}) &
(\ref{stddevw}) or~(\ref{empiricalstddevw}) \\
$V_{i_k}$ & estimate of variance in responses for a narrow bin around $S_{i_k}$
& (\ref{varest}) & (\ref{varestw}) \\
$W_k$ & weight & (Not applicable) & (\ref{abscissae}) \\
$X_k$ & abscissa of the subpopulation for reliability diagrams & (\ref{subX}) &
(\ref{subXw}) \\
$\tilde{X}_k$ & abscissa of the full population for reliability diagrams &
(\ref{fullX}) & (\ref{fullXw}) \\
$Y_k$ & ordinate of the subpopulation for reliability diagrams & (\ref{subY}) &
(\ref{subYw}) \\
$\tilde{Y}_k$ & ordinate of the full population for reliability diagrams &
(\ref{fullY}) & (\ref{fullYw}) \\
\hline
\end{tabular}
\end{center}
\end{table}

\subsection{High-level strategy}
\label{high-level}

This subsection suggests a hybrid method for large-scale data analysis.
When there are many data sets and subpopulations to assess,
a two-step approach may be the most practical:
\begin{enumerate}
\item A screening stage assigning a single scalar summary statistic
to each pair of data set and subpopulation (where the size of the statistic
measures the deviation of the subpopulation from the full population).
\item A detailed drill-down for each pair of data set and subpopulation
whose scalar summary statistic is large, drilling down into how the deviation
of the subpopulation from the full population varies as a function of score.
\end{enumerate}

The drill-down relies on graphically displaying the deviation
of the subpopulation from the full population as a function of score;
the scalar statistic for the first stage simply summarizes
the overall deviation across all scores,
as either the maximum absolute deviation of the graph
or the size of the range of deviations in the graphical display.
Thus, for each pair of data set and subpopulation,
both stages leverage a graph; the first stage collapses the graph
into a single scalar summary statistic.
The following subsection constructs the graph.

\subsection{Graphical method}
\label{graphical}

This subsection details the construction of cumulative plots.

Two sequences will define the cumulative plot
based on the notation set in the introduction.
The cumulative sequence for the subpopulation is
\begin{equation}
\label{empirical}
F_k = \frac{1}{n} \sum_{j=1}^k R_{i_j}
\end{equation}
for $k = 1$,~$2$, \dots, $n$.

Facilitating comparison of the subpopulation with the full population,
the average result for the full population in a bin around $S_{i_k}$ is
\begin{equation}
\label{average}
\tilde{R}_{i_k} = \frac{\sum_{i : B_{k-1} < S_i \le B_k} R_i}
                       {\#\{i : B_{k-1} < S_i \le B_k\}}
\end{equation}
for $k = 1$,~$2$, \dots, $n$, where the thresholds for the bins are
\begin{equation}
\label{B}
B_k = \frac{S_{i_k} + S_{i_{k+1}}}{2}
\end{equation}
for $k = 0$,~$1$,~$2$, \dots, $n$, under the convention that
$S_{i_0} = -\infty$ and $S_{i_{n+1}} = \infty$
(so $B_0 = -\infty$ and $B_n = \infty$).

The cumulative sequence for the full population at the subpopulation's subset
of scores is
\begin{equation}
\label{full}
\tilde{F}_k = \frac{1}{n} \sum_{j=1}^k \tilde{R}_{i_j}
\end{equation}
for $k = 1$,~$2$, \dots, $n$,
where $\tilde{R}_{i_j}$ is defined in~(\ref{average}).

Although the accumulation from lower scores might at first glance appear
to overwhelm the contributions from higher scores,
a plot of $F_k-\tilde{F}_k$ as a function of $k$ will display deviation
of the subpopulation from the full population
for any score solely in slopes that deviate significantly from 0;
any problems accumulated from earlier, lower scores pertain
only to the constant offset from 0, not to the slope deviating from 0.
In fact, the increment in the expected difference $F_j-\tilde{F}_j$
from $j = k-1$ to $j = k$ is
\begin{equation}
\E[ (F_k-\tilde{F}_k) - (F_{k-1}-\tilde{F}_{k-1}) ]
= \frac{\E[ R_{i_k} ] - \E[ \tilde{R}_{i_k} ]}{n};
\end{equation}
thus, on a plot with the values for $k$ spaced $1/n$ apart,
the slope from $j = k-1$ to $j = k$ is
\begin{equation}
\label{delta}
\Delta_k = \E[ R_{i_k} ] - \E[ \tilde{R}_{i_k} ],
\end{equation}
where the expected value of the average result $\tilde{R}_{i_k}$
in the bin around $S_{i_k}$ is equal to the average of the expected results,
that is,
\begin{equation}
\label{linearity}
\E[ \tilde{R}_{i_k} ] = \frac{\sum_{i : B_{k-1} < S_i \le B_k} \E[ R_i ]}
                             {\#\{i : B_{k-1} < S_i \le B_k\}}
\end{equation}
for $k = 1$,~$2$, \dots, $n$,
and $\tilde{R}_{i_k}$ is defined in~(\ref{average}).
The subpopulation deviates from the full population for the scores near
$S_{i_k}$ when $\Delta_k$ is significantly nonzero, that is, when the slope
of the plot of $F_k-\tilde{F}_k$ deviates significantly from horizontal
over a significantly long range.

To emphasize: {\it the deviation of the subpopulation from the full population
over a contiguous range of $S_{i_k}$
is the slope of the secant line for the plot of $F_k-\tilde{F}_k$
as a function of $\frac{k}{n}$ over that range,
aside from expected random fluctuations}.

Figure~\ref{5000} presents a simple illustrative example,
and many examples analyzing data from a popular data set in computer vision,
``ImageNet,'' are available in Subsection~\ref{imagenetex}.
The leftmost plot in the topmost row of Figure~\ref{5000}
graphs $F_k-\tilde{F}_k$ versus $k/n$; the rightmost plot in the top row
is the ideal noiseless ``ground-truth'' constructed using the precise
expected values of the random observations $R_1$,~$R_2$, \dots, $R_m$
(as detailed in Subsection~\ref{synthetic},
the exact expectations are available for constructing the ground-truth
in the synthetic example corresponding to Figure~\ref{5000}).
Steep slopes correspond to substantial deviations across the ranges of scores
where the slopes are steep, with the deviation over an interval exactly equal
to the expected value of the slope of the secant line for the graph
over that interval. The cumulative plot nicely matches its ideal ground-truth
in Figure~\ref{5000}.

The following subsection discusses metrics summarizing
how much the graph deviates from 0 (needless to say,
if the slopes of the secant lines are all nearly 0,
then the whole graph cannot deviate much from 0).
Subsection~\ref{significance} then leverages these metrics in a discussion
of the expected stochastic fluctuations mentioned above.

\subsection{Scalar summary statistics}
\label{scalarstats}

This subsection details the construction of scalar statistics
which provide broad-brush summaries of the plots introduced
in the previous subsection.

Two standard metrics for the overall deviation of the subpopulation
from the full population over the full range of scores
that account for expected random fluctuations are that due
to Kolmogorov and Smirnov, the maximum absolute deviation
\begin{equation}
\label{Kolmogorov-Smirnov}
G = \max_{1 \le k \le n} |F_k-\tilde{F}_k|,
\end{equation}
and that due to Kuiper, the size of the range of the deviations
\begin{equation}
\label{Kuiper}
D = \max_{0 \le k \le n} (F_k-\tilde{F}_k)
  - \min_{0 \le k \le n} (F_k-\tilde{F}_k),
\end{equation}
where $F_0 = 0 = \tilde{F}_0$;
the following remark explains the reason for including $F_0$ and $\tilde{F}_0$
(a reason that often makes $D$ modestly preferable to $G$).
Under appropriate statistical models,
$G$ and $D$ can form the basis for tests of statistical significance,
the context in which they originally appeared;
see, for example, Section~14.3.4 of~\cite{press-teukolsky-vetterling-flannery}.
For assessing statistical significance (rather than overall effect size),
$G$ and $D$ should be rescaled larger by a factor proportional to $\sqrt{n}$;
further discussion of the rescaling is available in the next subsection.
The captions of the figures report the values of these summary statistics
for all examples.

\begin{remark}
\label{zero}
The statistic $D$ from~(\ref{Kuiper}) has the same statistical power
across all indices.
Indeed, shifting the index in the definitions of $F_k$ and $\tilde{F}_k$
where the summation starts has no effect on the value of $D$,
for the following reasons:
Recall that an integral whose lower limit of integration
is greater than the upper limit is simply the negative of the integral
with the lower and upper limits of integration interchanged.
Thus, the natural generalization when starting from arbitrary values of $j$
the summations defining $F_k$ in~(\ref{empirical}) and $\tilde{F}_k$
in~(\ref{full}) is
\begin{equation}
\label{circular}
F_k^{(\ell)}
= \frac{1}{n} \sum_{j\,:\,\ell < j \le k} R_{i_j}
- \frac{1}{n} \sum_{j\,:\,k < j \le \ell} R_{i_j}
\end{equation}
and
\begin{equation}
\label{circularfull}
\tilde{F}_k^{(\ell)}
= \frac{1}{n} \sum_{j\,:\,\ell < j \le k} \tilde{R}_{i_j}
- \frac{1}{n} \sum_{j\,:\,k < j \le \ell} \tilde{R}_{i_j}
\end{equation}
for $k = 0$, $1$, $2$, \dots, $n$; $\ell = 0$, $1$, $2$, \dots, $n$,
with the case $\ell = 0$ reducing to~(\ref{empirical}) and~(\ref{full}).
Of course, the first summation in the right-hand side of~(\ref{circular})
vanishes when $k \le \ell$ and the second summation
vanishes when $k \ge \ell$; similarly,
the first summation in the right-hand side of~(\ref{circularfull})
vanishes when $k \le \ell$ and the second summation
vanishes when $k \ge \ell$.
Consideration of each case, for $k < \ell$, for $k = \ell$, and for $k > \ell$,
yields that
\begin{equation}
\label{shift}
F_k^{(\ell)} = F_k^{(0)} - F_{\ell}^{(0)}
\end{equation}
and
\begin{equation}
\label{shiftfull}
\tilde{F}_k^{(\ell)} = \tilde{F}_k^{(0)} - \tilde{F}_{\ell}^{(0)}
\end{equation}
for $k = 0$, $1$, $2$, \dots, $n$; $\ell = 0$, $1$, $2$, \dots, $n$.
The definition
\begin{equation}
D^{(\ell)}
= \max_{0 \le k \le n} (F_k^{(\ell)}-\tilde{F}_k^{(\ell)})
- \min_{0 \le k \le n} (F_k^{(\ell)}-\tilde{F}_k^{(\ell)}),
\end{equation}
when combined with~(\ref{shift}) and~(\ref{shiftfull}),
then yields that
\begin{equation}
D^{(\ell)} = D^{(0)} = D
\end{equation}
for $\ell = 0$, $1$, $2$, \dots, $n$, where $D$ is from~(\ref{Kuiper}).
This shows that the statistic $D$ has the same statistical power
for any index, as the statistic is invariant to shifts
in where the summation for the cumulative differences starts.
\end{remark}

\subsection{Significance of stochastic fluctuations}
\label{significance}

This subsection discusses statistical significance
both for the graphical methods of Subsection~\ref{graphical}
and for the summary statistics of Subsection~\ref{scalarstats}.

The plot of $F_k - \tilde{F}_k$ as a function of $k/n$ automatically
includes some ``confidence bands'' courtesy
of the discrepancy $F_k - \tilde{F}_k$ fluctuating randomly
as the index $k$ increments --- at the very least, the ``thickness''
of the plot coming from the random fluctuations gives a sense
of ``error bars.'' To give a rough indication of the size of the fluctuations
of the maximum deviation expected under the hypothesis that
the subpopulation does not deviate from the full population
in the actual underlying distributions, the plots should include
a triangle centered at the origin whose height above the origin
is proportional to $1/\sqrt{n}$.
Such a triangle can be a proxy for the classic confidence bands
around an empirical cumulative distribution function
introduced by Kolmogorov and Smirnov, as reviewed by~\cite{doksum}.
Indeed, a driftless, purely random walk deviates from zero
by roughly $\sqrt{n}$ after $n$ steps, so a random walk scaled by $1/n$
deviates from zero by roughly $1/\sqrt{n}$.
Identification of deviation between the subpopulation and the full population
is reliable when focusing on long ranges (as a function of $k/n$)
of steep slopes for $F_k - \tilde{F}_k$;
the triangle gives a sense of the length scale for variations
that arise solely due to randomness even in the absence
of any actual underlying deviation between the subpopulation
and the full population.
A simple illustrative example is available in Figure~\ref{5000},
and many examples analyzing data from a popular data set in computer vision,
``ImageNet,'' are available in Subsection~\ref{imagenetex}.

In cases for which either $R_i = 0$ or $R_i = 1$
for each $i = 1$, $2$, \dots, $m$,
and for which the scores are nothing but the probabilities of success,
that is, $S_i$ is the probability that $R_i = 1$
(where $i = 1$, $2$, \dots, $m$),
the tip-to-tip height of the triangle centered at the origin should be
$4/n$ times the standard deviation of the sum of independent Bernoulli variates
with success probabilities $S_{i_1}$, $S_{i_2}$, \dots, $S_{i_n}$, that is,
$4 \sqrt{\sum_{k=1}^n S_{i_k} (1-S_{i_k})} / n$.
This height will be representative to within a factor of $\sqrt{2}$ or so
provided that the subpopulation is a minority of the full population ---
see Remark~\ref{proofs} and Subsubsection~\ref{asignificance}
of Appendix~\ref{calibration} below.
Needless to say, similar remarks pertain whenever the variance of $R_i$
is a known function of $S_i$ for each $i = 1$, $2$, \dots, $m$.

In cases for which either $R_i = 0$ or $R_i = 1$
for each $i = 1$, $2$, \dots, $m$,
and for which there are many scores from the full population in the bin
for each score from the subpopulation, that is, 
$\#\{i : B_{k-1} < S_i \le B_k\}$ is large for every $k = 1$,~$2$, \dots, $n$,
the average of the outcomes for each bin will be a good approximation
to the average of the underlying probabilities of success for that bin,
that is,
\begin{equation}
\label{approx}
\frac{\sum_{i : B_{k-1} < S_i \le B_k} R_i}{\#\{i : B_{k-1} < S_i \le B_k\}}
\approx
\frac{\sum_{i : B_{k-1} < S_i \le B_k} \E[ R_i ]}
     {\#\{i : B_{k-1} < S_i \le B_k\}}
= \E[ \tilde{R}_{i_k} ]
\end{equation}
for $k = 1$, $2$, \dots, $n$, where $B_k$ is from~(\ref{B}),
$\E[ \tilde{R}_{i_k} ]$ is from~(\ref{linearity}),
and $\E[ R_i ]$ is the probability that the outcome is a success,
that is, the probability that $R_i = 1$.
In such cases, the tip-to-tip height of the triangle at the origin should be
$4/n$ times the standard deviation of the sum of independent Bernoulli variates
with success probabilities $\E[ \tilde{R}_{i_1} ]$, $\E[ \tilde{R}_{i_2} ]$,
\dots, $\E[ \tilde{R}_{i_n} ]$, that is,
$4 \sqrt{\sum_{k=1}^n \E[ \tilde{R}_{i_k} ] \, (1-\E[ \tilde{R}_{i_k} ])} / n$
--- and we may use~(\ref{approx}) to approximate this height as four times
\begin{equation}
\label{stddev}
\sigma = \frac{1}{n} \sqrt{\sum_{k=1}^n \tilde{R}_{i_k} (1-\tilde{R}_{i_k})},
\end{equation}
where $\tilde{R}_{i_k}$ is the left-hand side of~(\ref{approx}),
as seen from~(\ref{average}).
The triangles in the figures all have this height
when each $R_i$ is either 0 or 1,
since the numerical results reported in the following section
pertain to the case in which there are quite a few scores
from the full population in the bin for each score from the subpopulation.
When $R_i$ can take on more than two possible values,
the figures instead use the empirical variance
of the members from the full population for each bin from $B_{k-1}$ to $B_k$,
that is, we replace~(\ref{stddev}) with
\begin{equation}
\label{empiricalstddev}
\sigma = \frac{1}{n} \sqrt{\sum_{k=1}^n V_{i_k}},
\end{equation}
where the empirical variance is
\begin{equation}
\label{varest}
V_{i_k} = \frac{\sum_{i : B_{k-1} < S_i \le B_k} (R_i - \tilde{R}_{i_k})^2}
               {\#\{i : B_{k-1} < S_i \le B_k\}}
\end{equation}
for $k = 1$,~$2$, \dots, $n$,
with $\tilde{R}_{i_k}$ defined in~(\ref{average});
$\tilde{R}_{i_k}$ is also the left-hand side of~(\ref{approx}).

[Rigorous justification of~(\ref{approx}) is straightforward:
the expected value of the left-hand side of~(\ref{approx})
is the right-hand side of~(\ref{approx}), and
$0 \le \E[ R_i ] \le 1$ implies that $\E[ R_i ] \, (1-\E[ R_i ]) \le 1/4$, so
the standard deviation of the left-hand side of~(\ref{approx}) is
\begin{equation}
\frac{\sqrt{\sum_{i : B_{k-1} < S_i \le B_k} \E[ R_i ] \, (1-\E[ R_i ])}}
     {\#\{i : B_{k-1} < S_i \le B_k\}}
\le \frac{1}{2 \sqrt{\#\{i : B_{k-1} < S_i \le B_k\}}},
\end{equation}
which converges to 0 as $\#\{i : B_{k-1} < S_i \le B_k\}$ increases.]

\begin{remark}
\label{interpretation}
Interpreting the scalar summary statistics $G$ and $D$
from~(\ref{Kolmogorov-Smirnov}) and~(\ref{Kuiper}) is straightforward
in these latter cases, using $\sigma$ defined in~(\ref{stddev})
or~(\ref{empiricalstddev}).
Indeed, under the null hypothesis that the subpopulation has no deviation
from the average values of the full population at the corresponding scores,
the expected value of $G/\sigma$ is less than or equal to the expected value
of the maximum (over a subset of the unit interval $[0, 1]$)
of the absolute value of the standard Brownian motion over $[0, 1]$,
in the limit $n \to \infty$ and $\#\{i : B_{k-1} < S_i \le B_k\} \to \infty$
for all $k = 1$, $2$, \dots, $n$.
As reviewed below in Remark~\ref{proofs},
the expected value of the maximum of the absolute value
of the standard Brownian motion over the unit interval $[0, 1]$
is $\sqrt{\pi/2} \approx 1.25$; and the discussion by~\cite{masoliver}
immediately following Formula~44
of the associated arXiv publication\footnote{A freely available preprint
of~\cite{masoliver} is available at \url{https://arxiv.org/pdf/1401.4939.pdf}}
shows that the probability distribution of the maximum of the absolute value
of the standard Brownian motion over $[0, 1]$ is sub-Gaussian,
decaying past its mean $\sqrt{\pi/2} \approx 1.25$.
So, values of $G/\sigma$ much greater than 1.25 imply that
the subpopulation's deviation from the full population is significant,
while values of $G/\sigma$ close to 0 imply that
$G$ did not detect any statistically significant deviation.
Similar remarks pertain to $D$, since $G \le D \le 2G$.
\end{remark}

\begin{remark}
\label{zooming}
Zooming in on the origin of the plot can reveal relative deviations
(or small absolute deviations)
that may be of interest beyond just the absolute deviations;
Figure~\ref{zoom} displays such zooming, while setting the height
of the triangle at the origin based on only the scores and deviations appearing
in the restricted domain of the plot, rather than on the full domain depicted
in the other figures.
\end{remark}

\subsection{Weighted sampling}
\label{weighted}

This subsection generalizes the methods of the preceding subsections
to the case of weighted samples.

Specifically, some data sets include a weight for how much each observation
should contribute to the data analysis. In the setting
of Subsection~\ref{graphical} above,
such a data set would supplement the results $R_1$,~$R_2$, \dots, $R_m$
and scores $S_1$,~$S_2$, \dots, $S_m$ with positive weights
$W_1$,~$W_2$, \dots, $W_m$.
Subsection~\ref{graphical} will correspond to the special case that
$W_1 = W_2 = \dots = W_m$ (admittedly, the ``special'' case is the standard
in practice).

With weights, the cumulative sequence for the subpopulation,
replacing~(\ref{empirical}), becomes
\begin{equation}
\label{empiricalw}
F_k = \frac{\sum_{j=1}^k W_{i_j} R_{i_j}}{\sum_{j=1}^n W_{i_j}}
\end{equation}
for $k = 1$,~$2$, \dots, $n$.

The average result for the full population in a bin around $S_{i_k}$,
replacing~(\ref{average}), becomes
\begin{equation}
\label{averagew}
\tilde{R}_{i_k} = \frac{\sum_{i : B_{k-1} < S_i \le B_k} W_i R_i}
                       {\sum_{i : B_{k-1} < S_i \le B_k} W_i}
\end{equation}
for $k = 1$,~$2$, \dots, $n$, where (\ref{B})
defines $B_0$,~$B_1$, \dots, $B_n$, the thresholds for the bins.

The cumulative sequence for the full population at the subpopulation's subset
of scores, replacing~(\ref{full}), becomes
\begin{equation}
\label{fullw}
\tilde{F}_k
= \frac{\sum_{j=1}^k W_{i_j} \tilde{R}_{i_j}}{\sum_{j=1}^n W_{i_j}}
\end{equation}
for $k = 1$,~$2$, \dots, $n$,
where $\tilde{R}_{i_j}$ is defined in~(\ref{averagew}).

The cumulative sequence of weights is
\begin{equation}
\label{abscissae}
A_k = \frac{\sum_{j=1}^k W_{i_j}}{\sum_{j=1}^n W_{i_j}}
\end{equation}
for $k = 1$,~$2$, \dots, $n$.

In a plot of the weighted cumulative differences
$F_1-\tilde{F}_1$,~$F_2-\tilde{F}_2$, \dots, $F_n-\tilde{F}_n$
from~(\ref{empiricalw}) and~(\ref{fullw})
versus the cumulative weights
$A_1$,~$A_2$, \dots, $A_n$ from~(\ref{abscissae}), that is, in a plot where 
$F_1-\tilde{F}_1$,~$F_2-\tilde{F}_2$, \dots, $F_n-\tilde{F}_n$
are the ordinates (vertical coordinates), and $A_1$,~$A_2$, \dots, $A_n$
are the corresponding abscissae (horizontal coordinates),
the expected value of the slope from $A_{k-1}$ to $A_k$ is
\begin{equation}
\label{deltaw}
\Delta_k
= \frac{\E[(F_k-\tilde{F}_k) - (F_{k-1}-\tilde{F}_{k-1})]}{A_k-A_{k-1}}
= \frac{W_{i_k}\!\E[R_{i_k}] - W_{i_k}\!\E[\tilde{R}_{i_k}]}{W_{i_k}}
= \E[R_{i_k}] - \E[\tilde{R}_{i_k}],
\end{equation}
where the expected value of the weighted average result $\tilde{R}_{i_k}$
in the bin around $S_{i_k}$ is equal to the weighted average
of the expected results, that is,
\begin{equation}
\label{linearityw}
\E[ \tilde{R}_{i_k} ]
= \frac{\sum_{i : B_{k-1} < S_i \le B_k} W_i \E[ R_i ]}
       {\sum_{i : B_{k-1} < S_i \le B_k} W_i}
\end{equation}
for $k = 1$,~$2$, \dots, $n$,
and $\tilde{R}_{i_k}$ is defined in~(\ref{averagew}).
Thus, $\Delta_k$ defined in~(\ref{deltaw})
is equal to the expected value of the deviation of the subpopulation
from the full population for the scores near $S_{i_k}$,
with $W_{i_k}$ canceling in the rightmost identity of~(\ref{deltaw}). Hence,
the subpopulation deviates from the full population for the scores near
$S_{i_k}$ when $\Delta_k$ is significantly nonzero, that is, when the slope
of the plot of $F_k-\tilde{F}_k$ versus $A_k$ deviates significantly
from horizontal over a significantly long range.

To emphasize: {\it the deviation of the subpopulation from the full population
over a contiguous range of $S_{i_k}$
is the slope of the secant line for the plot of $F_k-\tilde{F}_k$
as a function of $A_k$ over that range,
aside from expected random fluctuations}.

A simple illustrative example is available in Figure~\ref{2500w}
of Subsection~\ref{synthetic}, and many examples analyzing data
from the U.S.\ Census Bureau are available in Subsection~\ref{census}.

The slope of line segments connecting the points
in the plot of $F_k-\tilde{F}_k$ versus $A_k$
is constant between successive values of $k$,
and those successive values are spaced further apart on the horizontal axis
when the weight $W_{i_k}$ is larger.
A plotted line that is straight for a wide horizontal range is therefore
indicative of a large weight.
Moreover, setting the (major) ticks on the upper horizontal axis
at the positions corresponding to equispaced values for $k$ visually depicts
the distribution of weights; including equispaced minor ticks
on the same upper horizontal axis provides a comparison to the case
of uniform weighting.

In cases for which either $R_i = 0$ or $R_i = 1$
for each $i = 1$, $2$, \dots, $m$,
and for which there are many scores from the full population in the bin
for each score from the subpopulation, that is,
$\#\{i : B_{k-1} < S_i \le B_k\}$ is large for every $k = 1$,~$2$, \dots, $n$,
the tip-to-tip height of the triangle at the origin should be four times
\begin{equation}
\label{stddevw}
\sigma
= \frac{\sqrt{\sum_{k=1}^n (W_{i_k})^2 \tilde{R}_{i_k} (1-\tilde{R}_{i_k})}}
       {\sum_{k=1}^n W_{i_k}},
\end{equation}
where $\tilde{R}_{i_k}$ is defined in~(\ref{averagew});
that is, we replace~(\ref{stddev}) with~(\ref{stddevw}).
When $R_i$ can take on more than two possible values,
the figures instead use the empirical variance
of the members from the full population for each bin from $B_{k-1}$ to $B_k$,
that is, we replace~(\ref{empiricalstddev}) and~(\ref{stddevw}) with
\begin{equation}
\label{empiricalstddevw}
\sigma = \frac{\sqrt{\sum_{k=1}^n (W_{i_k})^2 V_{i_k}}}{\sum_{k=1}^n W_{i_k}},
\end{equation}
where the empirical variance is
\begin{equation}
\label{varestw}
V_{i_k} = \frac{\sum_{i : B_{k-1} < S_i \le B_k} W_i (R_i - \tilde{R}_{i_k})^2}
               {\sum_{i : B_{k-1} < S_i \le B_k} W_i}
\end{equation}
for $k = 1$,~$2$, \dots, $n$,
with $\tilde{R}_{i_k}$ defined in~(\ref{averagew}).
The numerators of~(\ref{stddevw}) and~(\ref{empiricalstddevw})
include the square $(W_{i_k})^2$, unlike the other formulae.

The scalar summary statistics due to Kuiper and to Kolmogorov and Smirnov
of course use the same formulae (\ref{Kolmogorov-Smirnov}) and~(\ref{Kuiper})
as in the unweighted (or uniformly weighted) case,
except for replacing the definition of~$F_k$ from~(\ref{empirical})
with the definition from~(\ref{empiricalw})
and the definition of~$\tilde{F}_k$ from~(\ref{full})
with the definition from~(\ref{fullw}).

\begin{remark}
\label{weightedremark}
We can adapt to the case of weighted sampling
the classical methods discussed in the introduction.
As in the introduction, we choose some partitions of the real line
into $\ell$ disjoint intervals with endpoints $B_1$, $B_2$, \dots, $B_{\ell-1}$
and another (possibly the same) $\ell$ disjoint intervals
with endpoints $\tilde{B}_1$, $\tilde{B}_2$, \dots, $\tilde{B}_{\ell-1}$
such that $B_1 < B_2 < \dots < B_{\ell-1}$
and $\tilde{B}_1 < \tilde{B}_2 < \dots < \tilde{B}_{\ell-1}$,
and then replace~(\ref{subY}) with the weighted averages for the subpopulation
\begin{equation}
\label{subYw}
Y_k = \frac{\sum_{j : B_{k-1} < S_{i_j} \le B_k} W_{i_j} R_{i_j}}
           {\sum_{j : B_{k-1} < S_{i_j} \le B_k} W_{i_j}}
\end{equation}
and replace~(\ref{fullY}) with the weighted averages for the full population
\begin{equation}
\label{fullYw}
\tilde{Y}_k = \frac{\sum_{i : \tilde{B}_{k-1} < S_i \le \tilde{B}_k} W_i R_i}
                   {\sum_{i : \tilde{B}_{k-1} < S_i \le \tilde{B}_k} W_i}
\end{equation}
for $k = 1$, $2$, \dots, $\ell$,
under the convention that $B_0 = \tilde{B}_0 = -\infty$
and $B_{\ell} = \tilde{B}_{\ell} = \infty$.
We also replace~(\ref{subX}) with the weighted averages of the scores
in the bins for the subpopulation
\begin{equation}
\label{subXw}
X_k = \frac{\sum_{j : B_{k-1} < S_{i_j} \le B_k} W_{i_j} S_{i_j}}
           {\sum_{j : B_{k-1} < S_{i_j} \le B_k} W_{i_j}}
\end{equation}
and replace~(\ref{fullX}) with the weighted averages for the full population
\begin{equation}
\label{fullXw}
\tilde{X}_k = \frac{\sum_{i : \tilde{B}_{k-1} < S_i \le \tilde{B}_k} W_i S_i}
                   {\sum_{i : \tilde{B}_{k-1} < S_i \le \tilde{B}_k} W_i}
\end{equation}
for $k = 1$, $2$, \dots, $\ell$,
under the same convention that $B_0 = \tilde{B}_0 = -\infty$
and $B_{\ell} = \tilde{B}_{\ell} = \infty$.
The reliability diagram for assessing the deviation of the subpopulation
from the full population is then the scatterplot of the pairs
$(X_1, Y_1)$, $(X_2, Y_2)$, \dots, $(X_{\ell}, Y_{\ell})$ in black
and the pairs $(\tilde{X}_1, \tilde{Y}_1)$, $(\tilde{X}_2, \tilde{Y}_2)$,
\dots, $(\tilde{X}_{\ell}, \tilde{Y}_{\ell})$ in gray.
Comparing the black plotted points (possibly connected with black lines)
to the gray plotted points (possibly connected with gray lines)
gives an indication of deviation of the subpopulation from the full population.
Two natural choices of the bins whose endpoints are
$B_1$, $B_2$, \dots, $B_{\ell-1}$ (similar choices pertain to the bins
whose endpoints are $\tilde{B}_1$, $\tilde{B}_2$, \dots, $\tilde{B}_{\ell-1}$)
include \{1\} have $B_1$, $B_2$, \dots, $B_{\ell-1}$ be equispaced, and
\{2\} choose $B_1$, $B_2$, \dots, $B_{\ell-1}$ such that
\begin{equation}
\label{subpopulationnorms}
U_k = \frac{\sqrt{\sum_{j : B_{k-1} < S_{i_j} \le B_k} (W_{i_j})^2}}
           {\sum_{j : B_{k-1} < S_{i_j} \le B_k} W_{i_j}}
\end{equation}
has a similar value for all $k = 1$,~$2$, \dots, $\ell$.
Remark~\ref{equierrs} below details the procedure we followed
in the second case; the plots entitled,
``reliability diagram ($\|W\|_2 / \|W\|_1$ is similar for every bin),''
display this second possible choice of bins.
The plots entitled simply, ``reliability diagram,''
display the first possible choice of bins.
In the special case that the weights are uniform, that is,
$W_1 = W_2 = \dots = W_m$, the second choice of bins results
in every bin containing about the same number of scores,
with $U_1 \approx U_2 \approx \dots \approx U_{\ell} \approx \sqrt{\ell/n}$,
and similarly $\tilde{U}_1 \approx \tilde{U}_2 \approx \dots
\approx \tilde{U}_{\ell} \approx \sqrt{\ell/m}$, where
\begin{equation}
\label{fullnorms}
\tilde{U}_k
= \frac{\sqrt{\sum_{i : \tilde{B}_{k-1} < S_i \le \tilde{B}_k} (W_i)^2}}
       {\sum_{i : \tilde{B}_{k-1} < S_i \le \tilde{B}_k} W_i}
\end{equation}
for $k = 1$,~$2$, \dots, $\ell$.
\end{remark}

\begin{remark}
\label{equierrs}
In the case of weighted sampling, the most useful reliability diagrams
are usually those entitled,
``reliability diagram ($\|W\|_2/\|W\|_1$ is similar for every bin).''
These diagrams construct $\ell$ bins with endpoints
$B_0$,~$B_1$, \dots, $B_{\ell}$ such that
$U_1 \approx U_2 \approx \dots \approx U_{\ell}$,
where $U_k$ is defined in~(\ref{subpopulationnorms}).
These diagrams also construct $\tilde{\ell}$ bins with endpoints
$\tilde{B}_0$,~$\tilde{B}_1$, \dots, $\tilde{B}_{\tilde{\ell}}$ such that
$\tilde{U}_1 \approx \tilde{U}_2 \approx \dots
\approx \tilde{U}_{\tilde{\ell}}$,
where $\tilde{U}_k$ is defined in~(\ref{fullnorms}).
The algorithmic details are as follows:
Given a value $U$ for which hopefully $U_k \approx U$
for all $k = 1$,~$2$, \dots, $\ell$,
we set $B_0 = -\infty$ and, iterating from $k = 1$ to $k = \ell$,
incrementally increase $B_k$ to the least value greater than $B_{k-1}$
such that $U_k \le U$.
If this causes the bin $\ell$ (the bin for the highest scores)
to contain less than half as many subpopulation observations as bin $\ell-1$,
that is, $\#\{j : B_{\ell-1} < S_{i_j} \le B_{\ell}\}
< \#\{j : B_{\ell-2} < S_{i_j} \le B_{\ell-1}\} / 2$,
then we merge bin $\ell$ with bin $\ell-1$.
In the aforementioned algorithm, we computed $U$ via the formula
\begin{equation}
\label{heuristic}
U = \frac{\sqrt{\sum_{j = 1}^{\lfloor n / \bar{\ell} \rfloor} (W_{p_j})^2}}
         {\sum_{j = 1}^{\lfloor n / \bar{\ell} \rfloor} W_{p_j}},
\end{equation}
where $p_1$,~$p_2$, \dots, $p_n$ are a uniformly random permutation
of the integers $i_1$,~$i_2$, \dots, $i_n$,
and $\bar{\ell}$ is the desired number of bins.
Calculating $U$ via the heuristic~(\ref{heuristic}) worked well
for all examples reported below,
producing $U_1 \approx U_2 \approx \dots \approx U_\ell$,
with $\ell$ close to $\bar{\ell}$; the procedure yielding $\tilde{U}_1
\approx \tilde{U}_2 \approx \dots \approx \tilde{U}_{\tilde{\ell}}$ is similar
(in fact, the implementation is identical, viewing the full population
as a subpopulation of itself).
\end{remark}

\section{Results and discussion}
\label{results}

This section illustrates via numerous examples
the previous section's methods,
together with the traditional plots --- so-called ``reliability diagrams'' ---
discussed in the introduction.\footnote{Permissively licensed
open-source software for reproducing all figures
and statistics reported here is available at
\url{https://github.com/facebookresearch/fbcdgraph}}
Subsection~\ref{synthetic} presents several (hopefully insightful)
toy examples.
Subsection~\ref{imagenetex} analyzes a popular, unweighted data set of images,
ImageNet.
Subsection~\ref{census} analyzes a weighted data set,
the most recent (year 2019) American Community Survey
of the United States Census Bureau.
Subsection~\ref{outlook} proposes directions for future developments.

The figures display the classical calibration plots (``reliability diagrams'')
as well as both the plots of cumulative differences and the exact expectations
in the absence of noise from random sampling when the exact expectations
are known (as with synthetically generated data).
The captions of the figures discuss the numerical results depicted.

The title, ``subpopulation deviation is the slope as a function of $k/n$,''
labels a plot of $F_k-\tilde{F}_k$
from~(\ref{empirical}) and~(\ref{full}) as a function of $k/n$.
In each such plot, the upper axis specifies $k/n$,
while the lower axis specifies $S_{i_k}$ for the corresponding value of $k$.
The title, ``subpopulation deviation is the slope as a function of $A_k$,''
labels a plot of $F_k-\tilde{F}_k$
from~(\ref{empiricalw}) and~(\ref{fullw}) versus 
the cumulative weight $A_k$ from~(\ref{abscissae}).
In each such plot, the major ticks on the upper axis specify $k/n$,
while the major ticks on the lower axis specify $S_{i_k}$
for the corresponding value of $k$; the points in the plot
are the ordered pairs $(A_k, F_k-\tilde{F}_k)$ for $k = 1$,~$2$, \dots, $n$,
with $A_k$ being the abscissa and $F_k-\tilde{F}_k$ being the ordinate.

The titles, ``reliability diagram,''
``reliability diagram (equal number of subpopulation scores per bin),''
and ``reliability diagram ($\|W\|_2/\|W\|_1$ is similar for every bin),''
label a plot of the pairs
$(X_1, Y_1)$,~$(X_2, Y_2)$, \dots, $(X_{\ell}, Y_{\ell})$ and
$(\tilde{X}_1, \tilde{Y}_1)$,~$(\tilde{X}_2, \tilde{Y}_2)$, \dots,
$(\tilde{X}_{\ell}, \tilde{Y}_{\ell})$
from~(\ref{subY}), (\ref{fullY}), (\ref{subX}), and~(\ref{fullX})
or from~(\ref{subYw}), (\ref{fullYw}), (\ref{subXw}), and~(\ref{fullXw})
in the case of weighted sampling; the subpopulation's pairs
$(X_1, Y_1)$,~$(X_2, Y_2)$, \dots, $(X_{\ell}, Y_{\ell})$
are in black, while the full population's pairs
$(\tilde{X}_1, \tilde{Y}_1)$,~$(\tilde{X}_2, \tilde{Y}_2)$, \dots,
$(\tilde{X}_{\ell}, \tilde{Y}_{\ell})$ are in gray.

To give a sense of the uncertainties in the classical, binned plots,
we vary the number of bins and observe how the plotted values vary.
Displaying the bin frequencies is another way to indicate uncertainties,
as suggested, for example, by~\cite{murphy-winkler}.
Still other possibilities could use kernel density estimation,
as suggested, for example, by~\cite{brocker} and~\cite{wilks}.
Such uncertainty estimates require selecting widths for the bins
or kernel smoothing; varying the widths (as done in the present paper)
avoids having to make what would otherwise be a rather arbitrary choice.
A thorough survey of the various possibilities is available
in Chapter~8 of~\cite{wilks}.

As noted in the introduction, there are two canonical choices for the bins
in the case of unweighted (or uniformly weighted) sampling:
\{1\} make the average of $S_{i_k}$ (or $S_i$) in each bin
be approximately equidistant from the average of $S_{i_k}$ (or $S_i$)
in each neighboring bin or
\{2\} make the number of $S_{i_k}$ (or $S_i$) in every bin
(except perhaps for the last) be the same.
The figures label the first, more conventional possibility
with the short title, ``reliability diagram,'' and the second possibility
with the longer title,
``reliability diagram (equal number of subpopulation scores per bin).''
As noted in Remark~\ref{weightedremark}, there are two natural choices
for the bins in the case of weighted sampling:
\{1\} make the weighted average of $S_{i_k}$ (or $S_i$) in each bin
be approximately equidistant from the weighted average of $S_{i_k}$ (or $S_i$)
in each neighboring bin or
\{2\} follow Remark~\ref{equierrs} above.
The figures label the first possibility with the short title,
``reliability diagram,'' and the second possibility
with the longer title,
``reliability diagram ($\|W\|_2/\|W\|_1$ is similar for every bin).''

Setting the number of bins together with any of these choices
fully specifies the bins. As discussed earlier, we vary the number of bins,
since there is no perfect setting --- using fewer bins offers higher-confidence
estimates, yet limits the resolution for detecting deviations
and for assessing how the deviations vary as a function of $S_{i_k}$.

\subsection{Synthetic}
\label{synthetic}

\begin{figure}
\begin{centering}

\parbox{\imsize}{\includegraphics[width=\imsize]
                {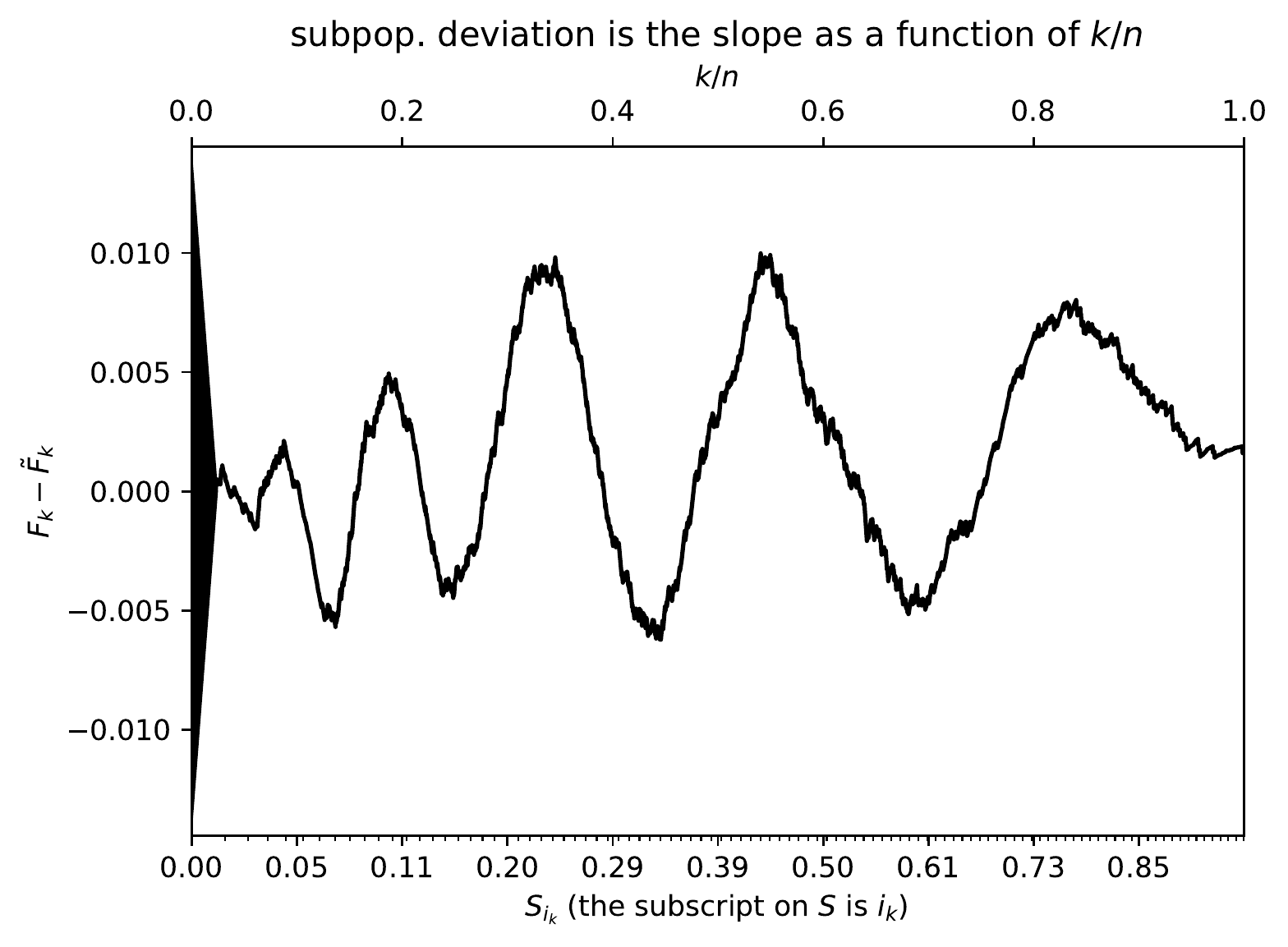}}
\quad\quad
\parbox{\imsize}{\includegraphics[width=\imsize]
                {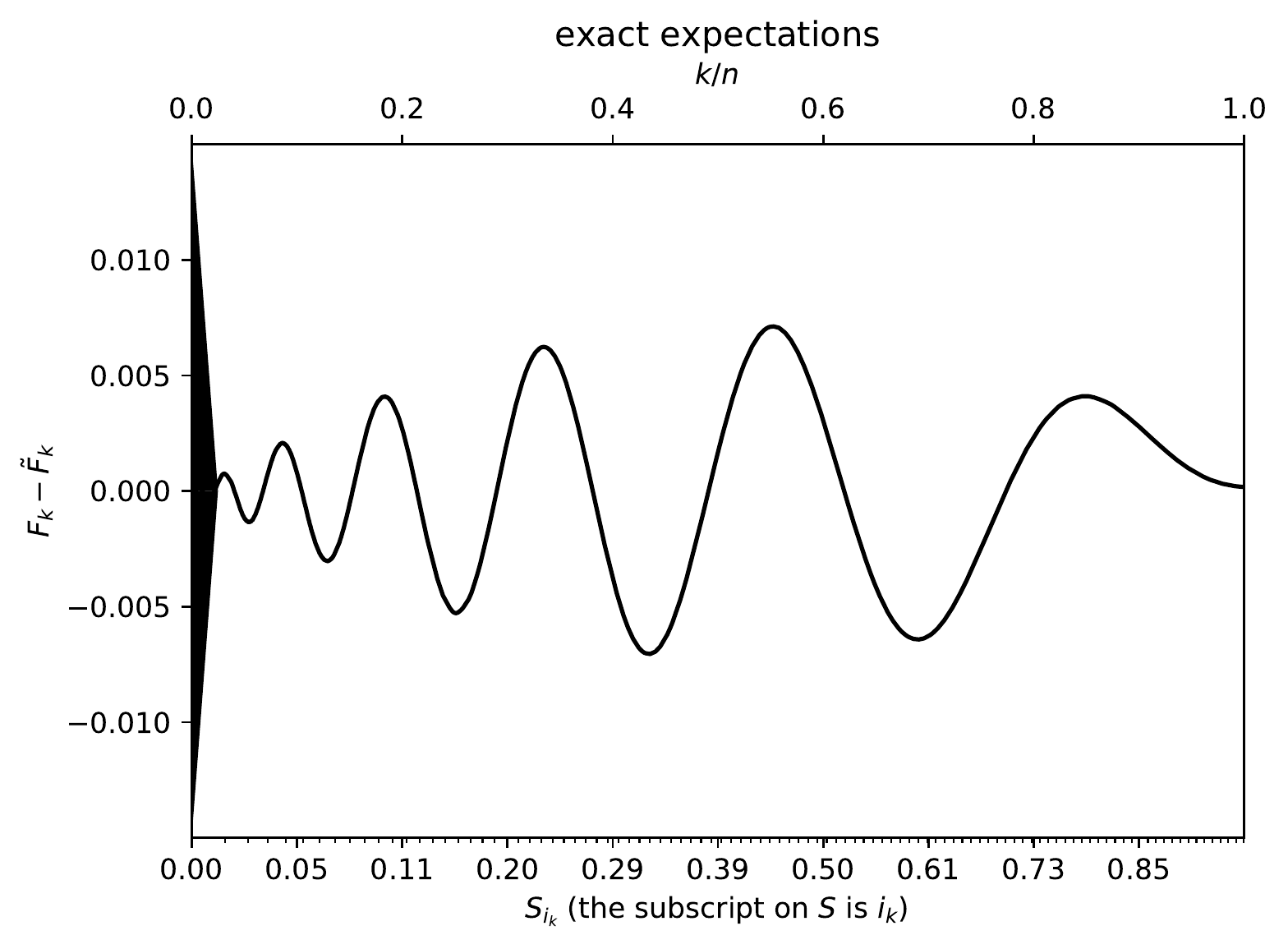}}

\vspace{\vertsep}

\parbox{\imsize}{\includegraphics[width=\imsize]
                {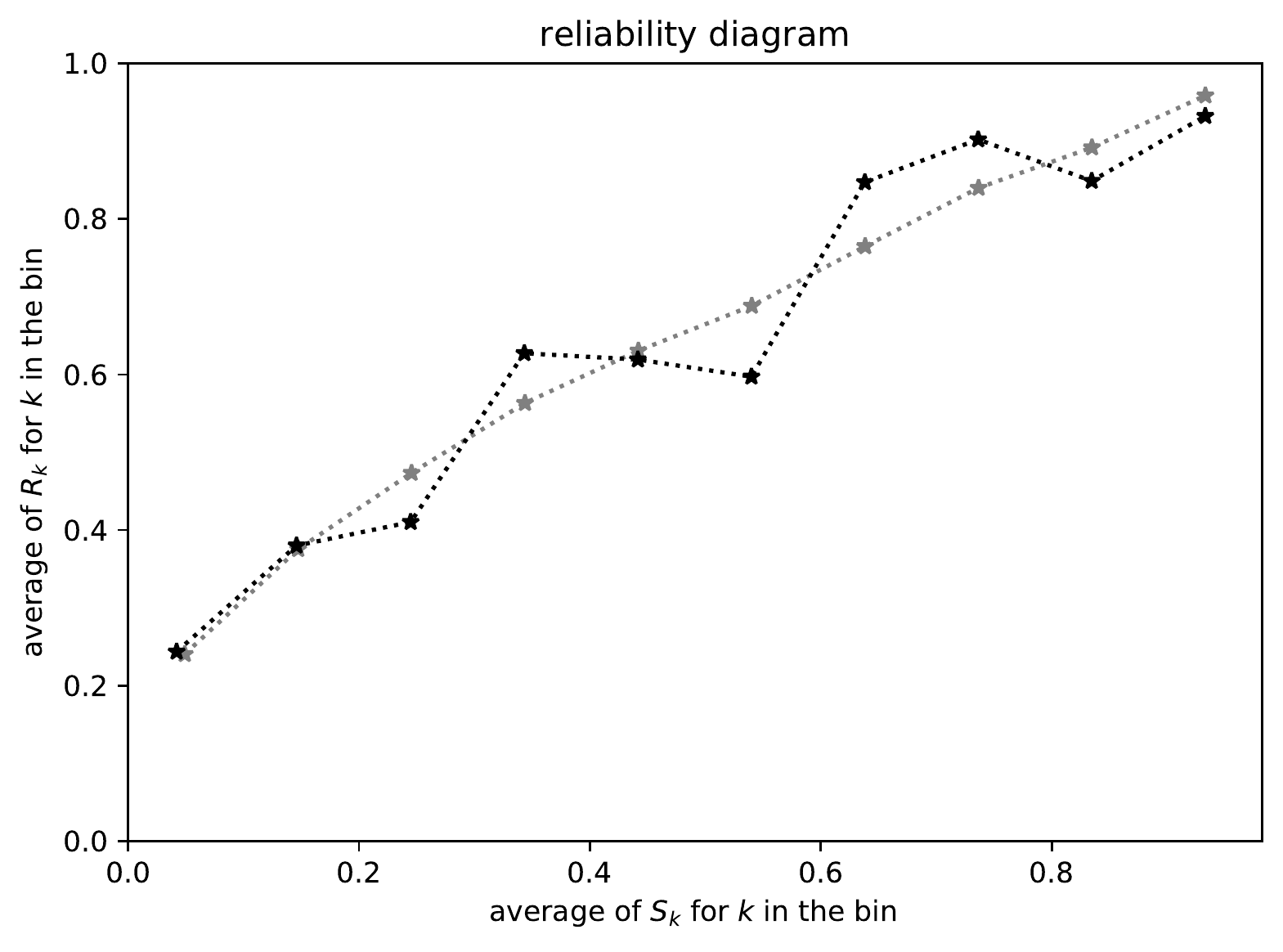}}
\quad\quad
\parbox{\imsize}{\includegraphics[width=\imsize]
                {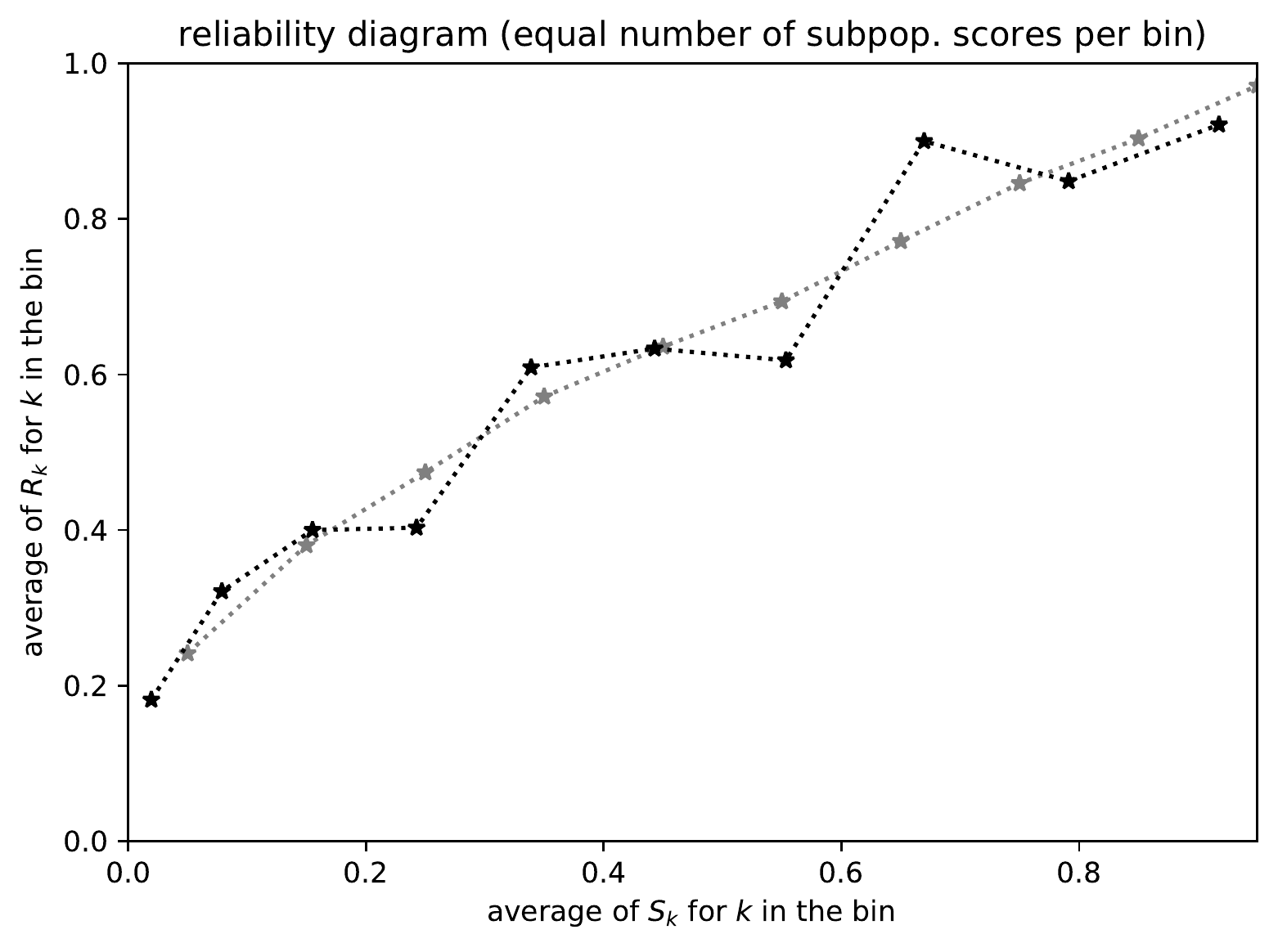}}

\vspace{\vertsep}

\parbox{\imsize}{\includegraphics[width=\imsize]
                {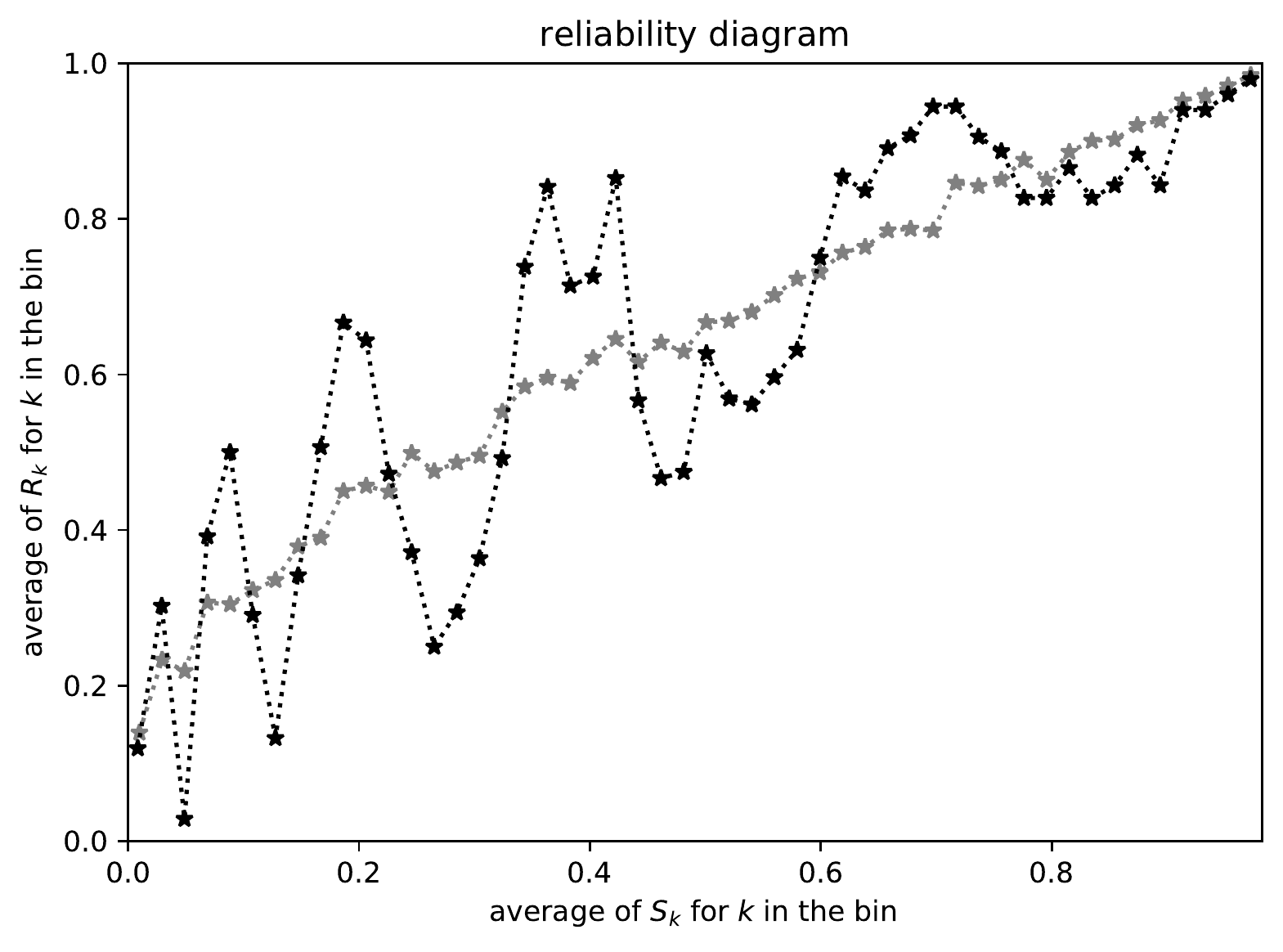}}
\quad\quad
\parbox{\imsize}{\includegraphics[width=\imsize]
                {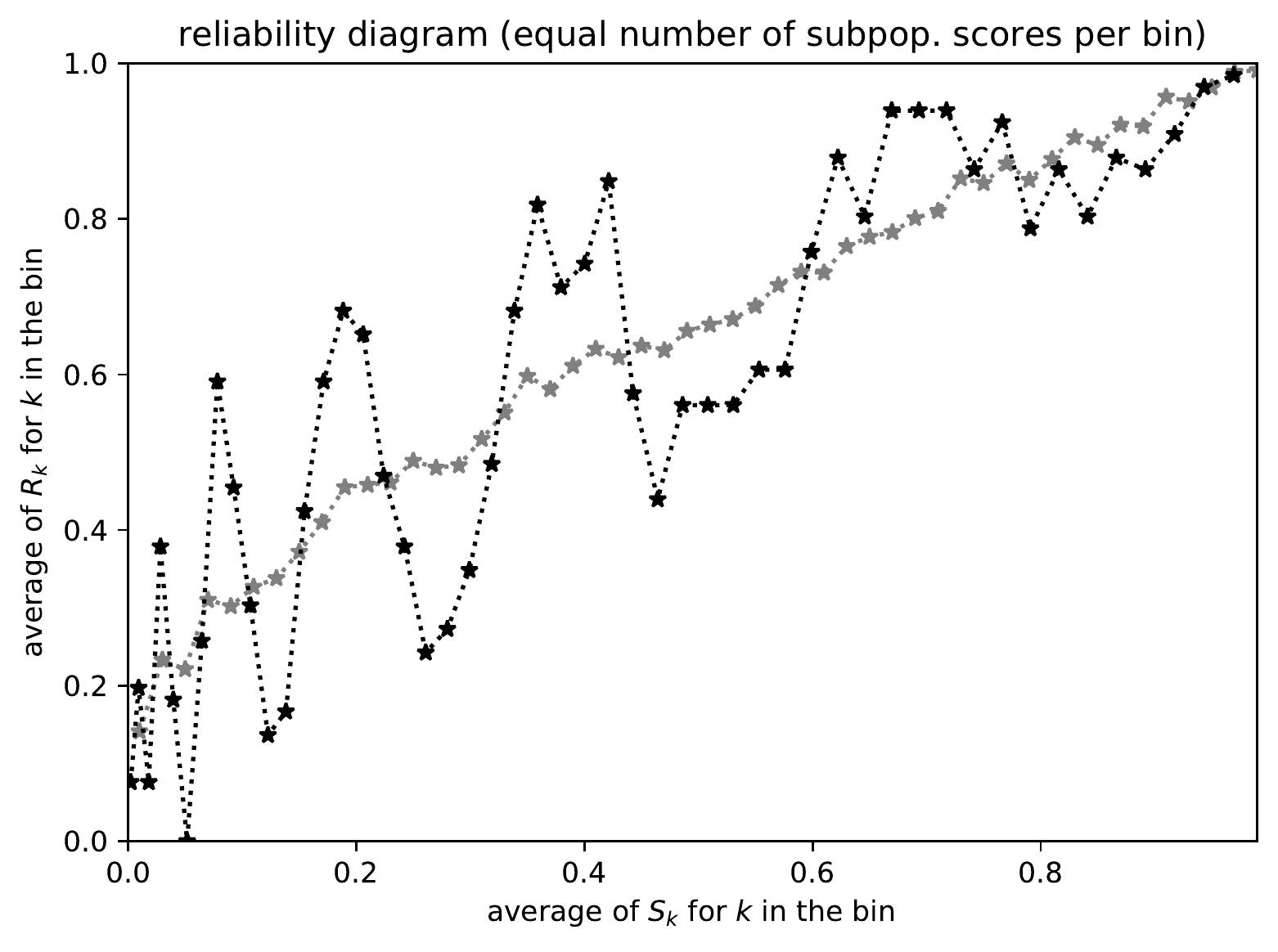}}

\end{centering}
\caption{$n =$ 3,300;
         Kuiper's statistic is $0.01621 / \sigma = 2.243$,
         Kolmogorov's and Smirnov's is $0.009983 / \sigma = 1.381$.
Figure~\ref{3300e} displays the ground-truth reliability diagram.
Distinguishing random fluctuations from real variations
is difficult in the reliability diagrams with 50 bins each.
The reliability diagrams that each have only 10 bins could be misleading,
as the depicted variations in the subpopulation's outcomes are grossly
lower than the actual underlying variations as a function of score.
The plot of cumulative deviation is far from perfect, yet captures
the exact expectations quite well qualitatively
and tolerably well quantitatively.
The scalar summary statistics have trouble
detecting the significant deviation of the subpopulation
from the full population.
This illustrates a blind spot in the Kolmogorov-Smirnov and Kuiper statistics,
namely, they have a hard time detecting oscillatory discrepancies
that average away upon summation. Neither the Kolmogorov-Smirnov metric
nor the Kuiper metric is very sensitive to high-frequency deviations
whose mean is small.
}
\label{3300}
\end{figure}

\begin{figure}
\begin{centering}

\parbox{\imsize}{\includegraphics[width=\imsize]
                {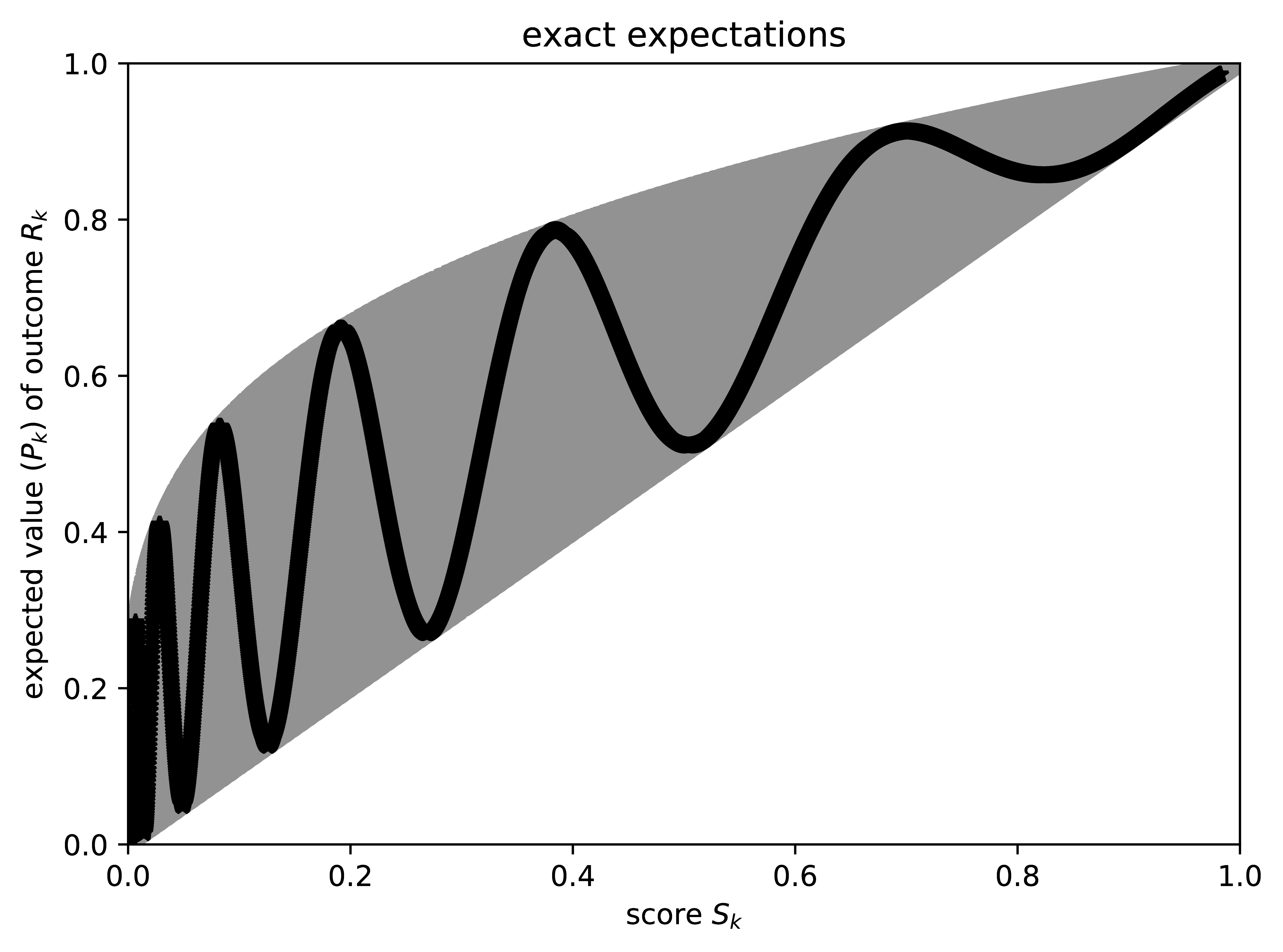}}

\end{centering}
\caption{Ground-truth reliability diagram for Figure~\ref{3300}}
\label{3300e}
\end{figure}

\begin{figure}
\begin{centering}

\parbox{\imsize}{\includegraphics[width=\imsize]
                {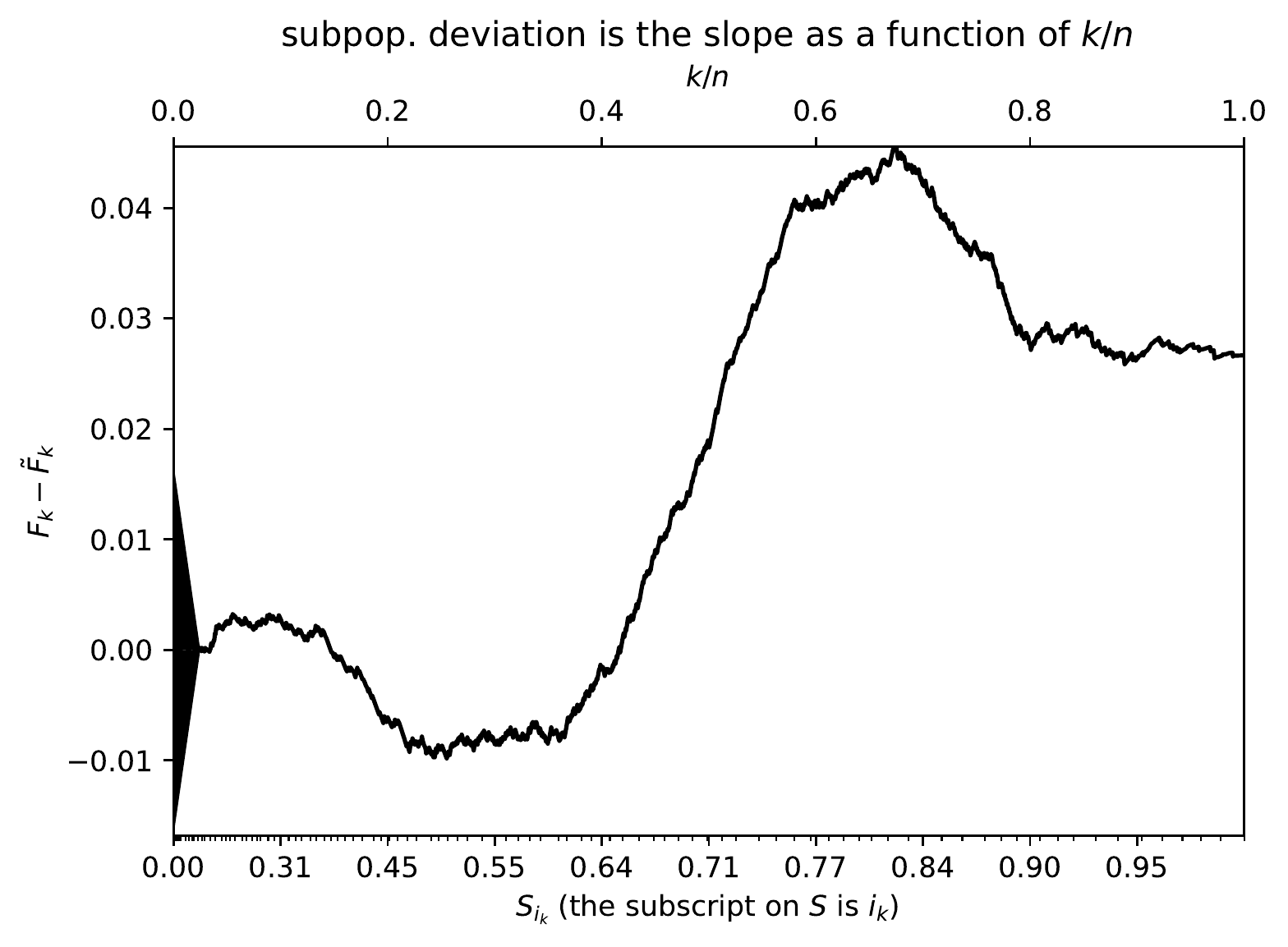}}
\quad\quad
\parbox{\imsize}{\includegraphics[width=\imsize]
                {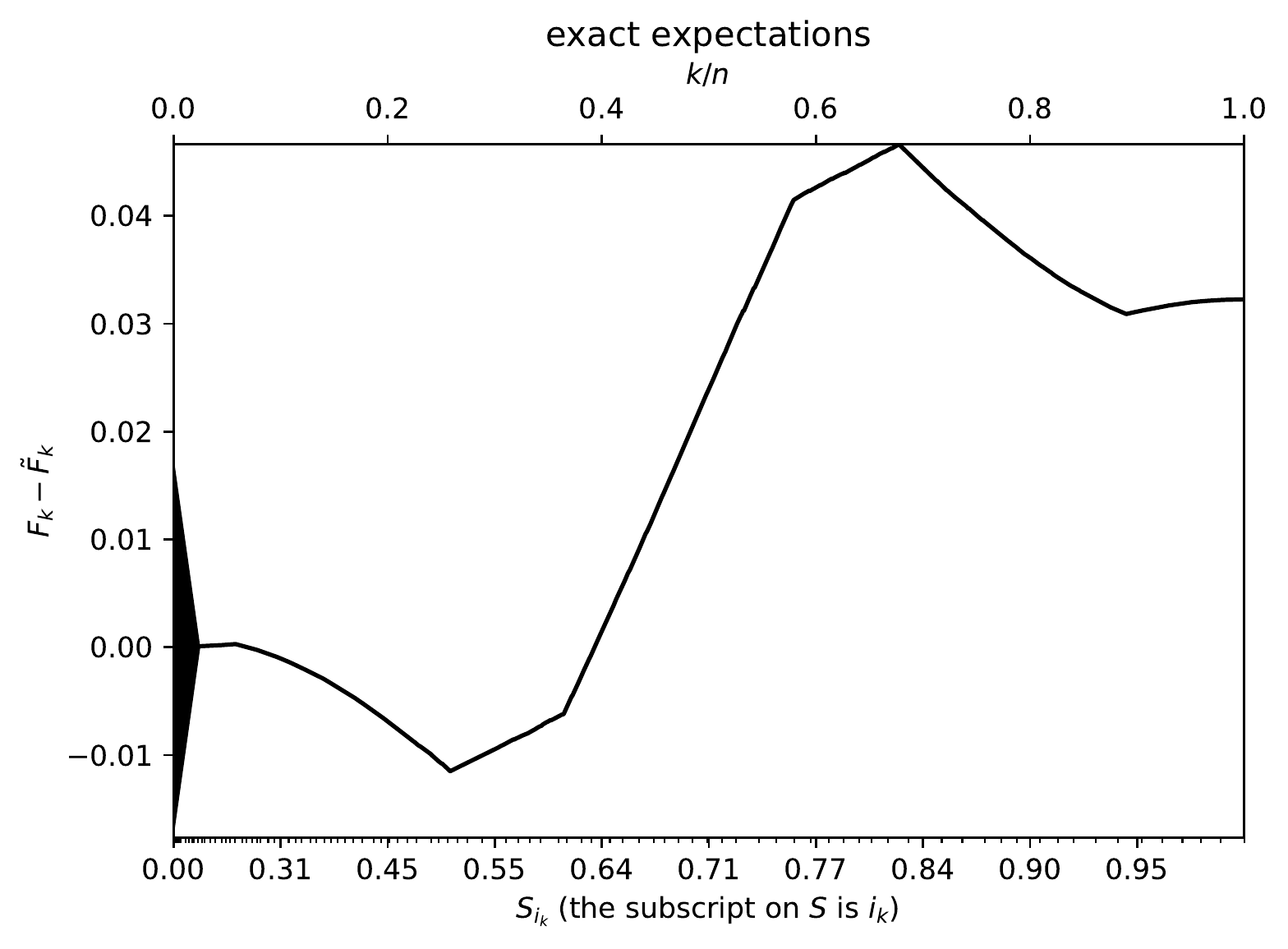}}

\vspace{\vertsep}

\parbox{\imsize}{\includegraphics[width=\imsize]
                {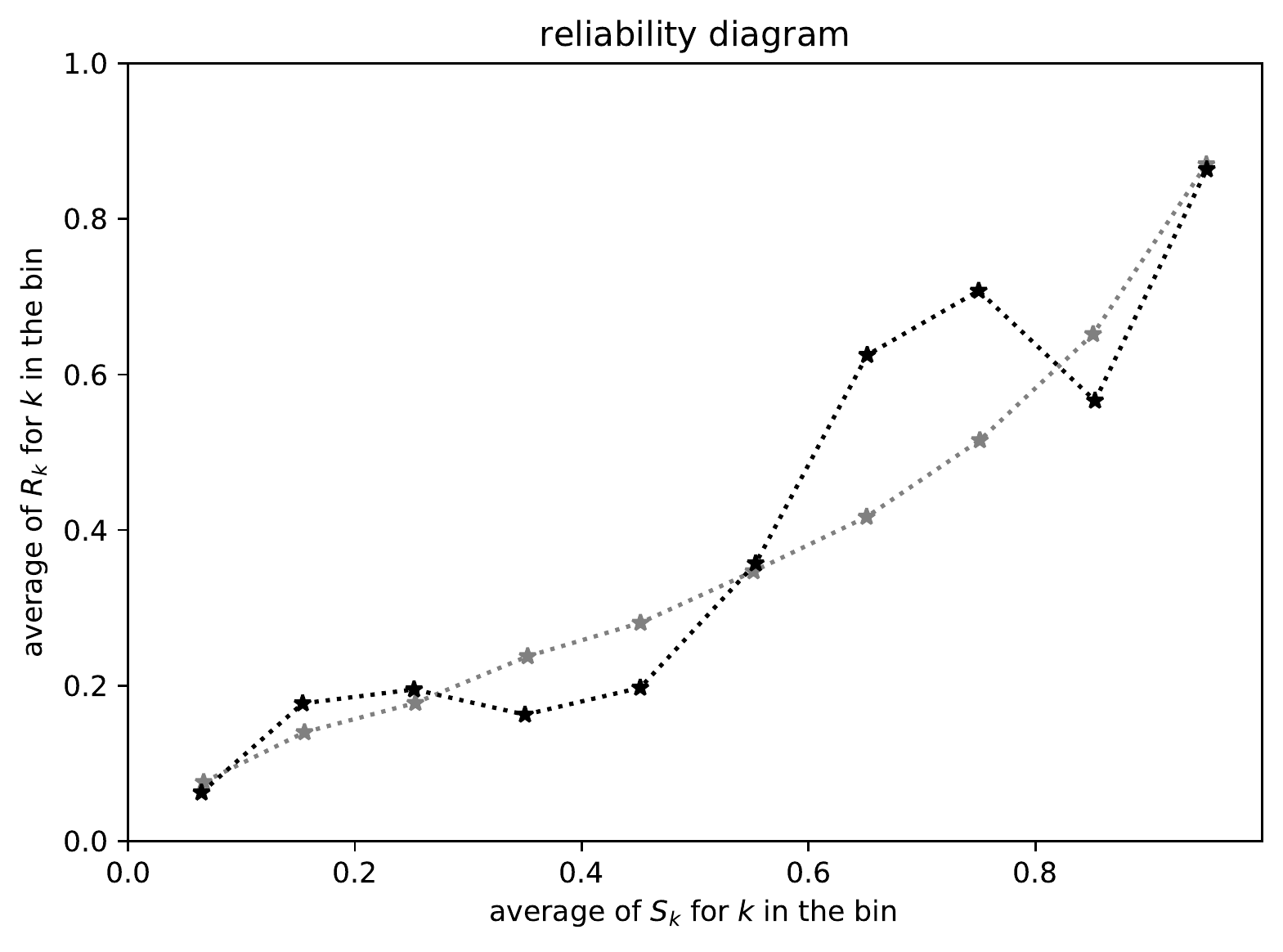}}
\quad\quad
\parbox{\imsize}{\includegraphics[width=\imsize]
                {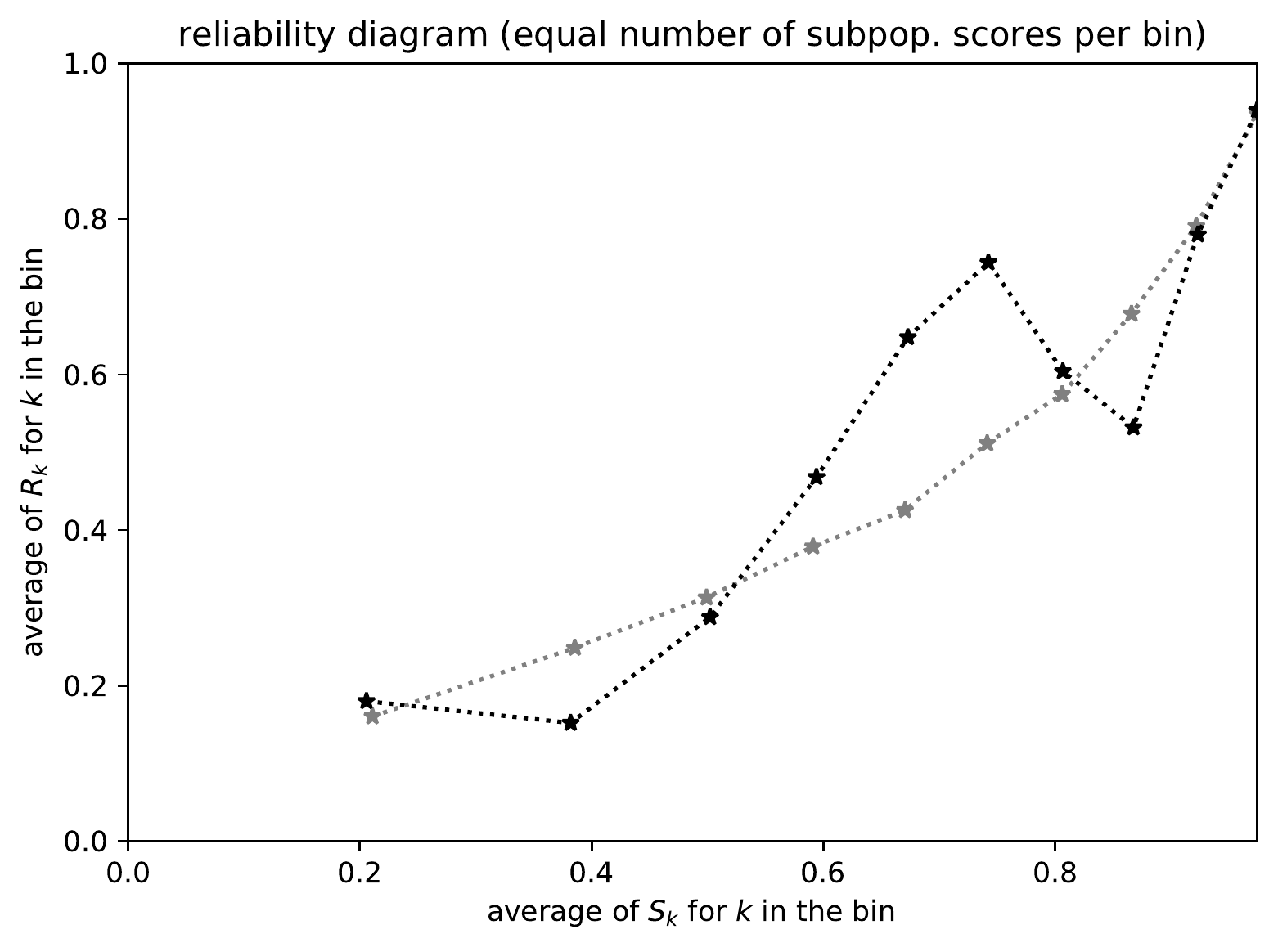}}

\vspace{\vertsep}

\parbox{\imsize}{\includegraphics[width=\imsize]
                {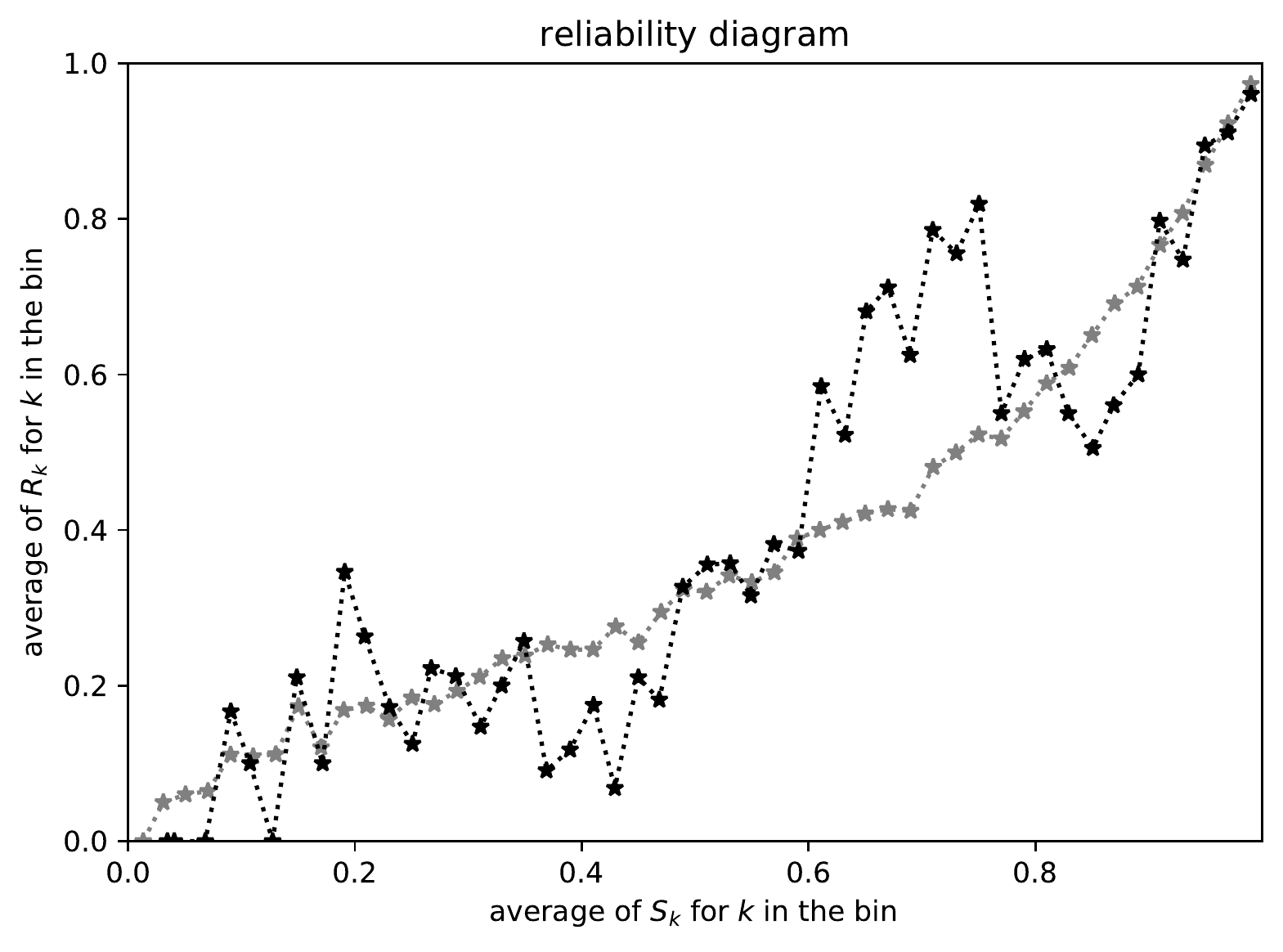}}
\quad\quad
\parbox{\imsize}{\includegraphics[width=\imsize]
                {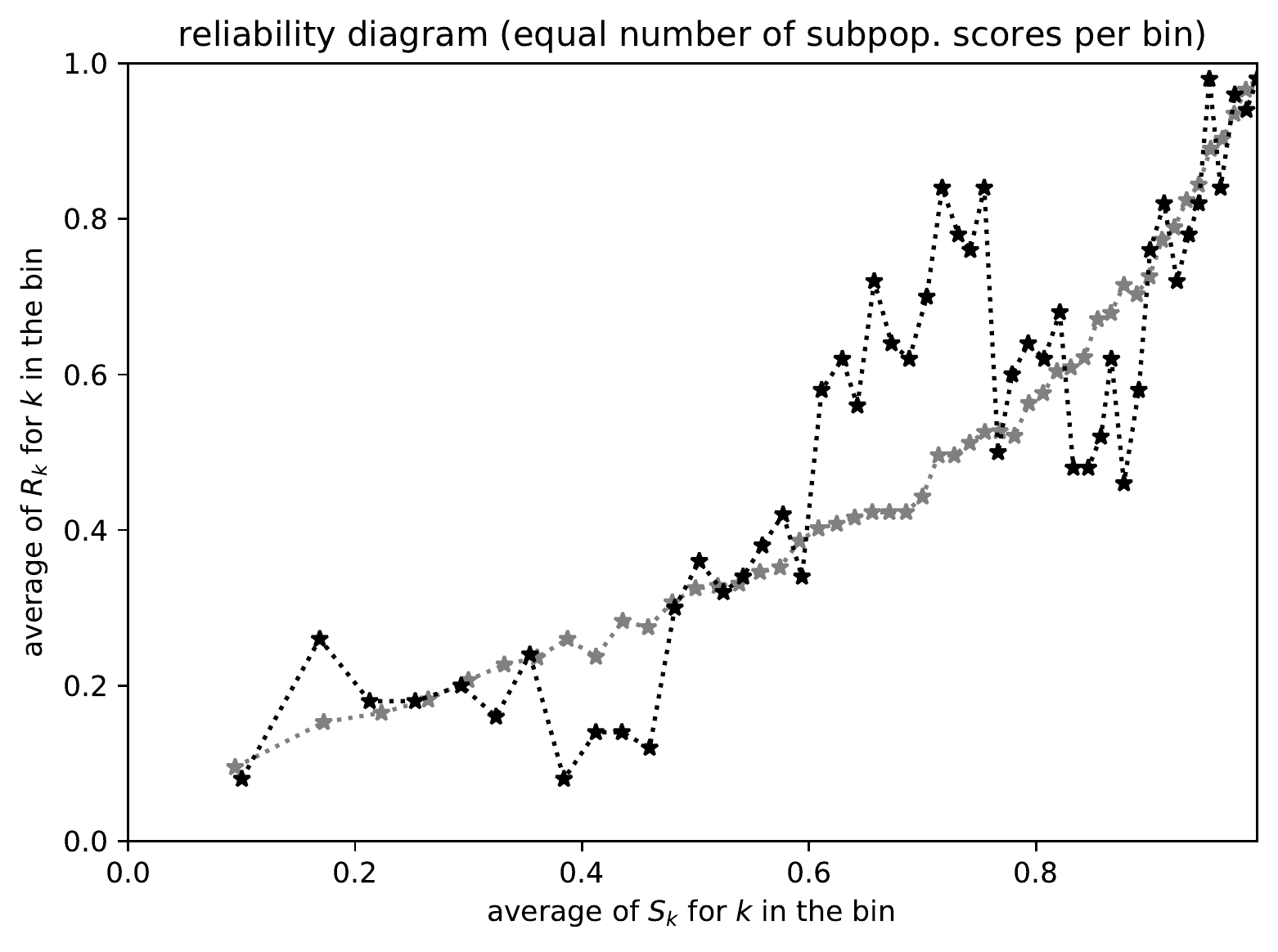}}

\end{centering}
\caption{$n =$ 2,500;
         Kuiper's statistic is $0.05536 / \sigma = 6.586$,
         Kolmogorov's and Smirnov's is $0.04554 / \sigma = 5.418$.
Figure~\ref{2500e} displays the ground-truth reliability diagram.
The observed reliability diagrams fail to depict
the underlying discontinuous jumps in the subpopulation's expected outcomes
as a function of the score. The plot of cumulative deviation succeeds
in resolving some of the corresponding cusps, but does exhibit significant
random fluctuations nearly as high as a quarter of the height of the triangle
at the origin.
The scalar summary statistics successfully
detect the statistically significant deviation
of the subpopulation from the full population.
}
\label{2500}
\end{figure}

\begin{figure}
\begin{centering}

\parbox{\imsize}{\includegraphics[width=\imsize]
                {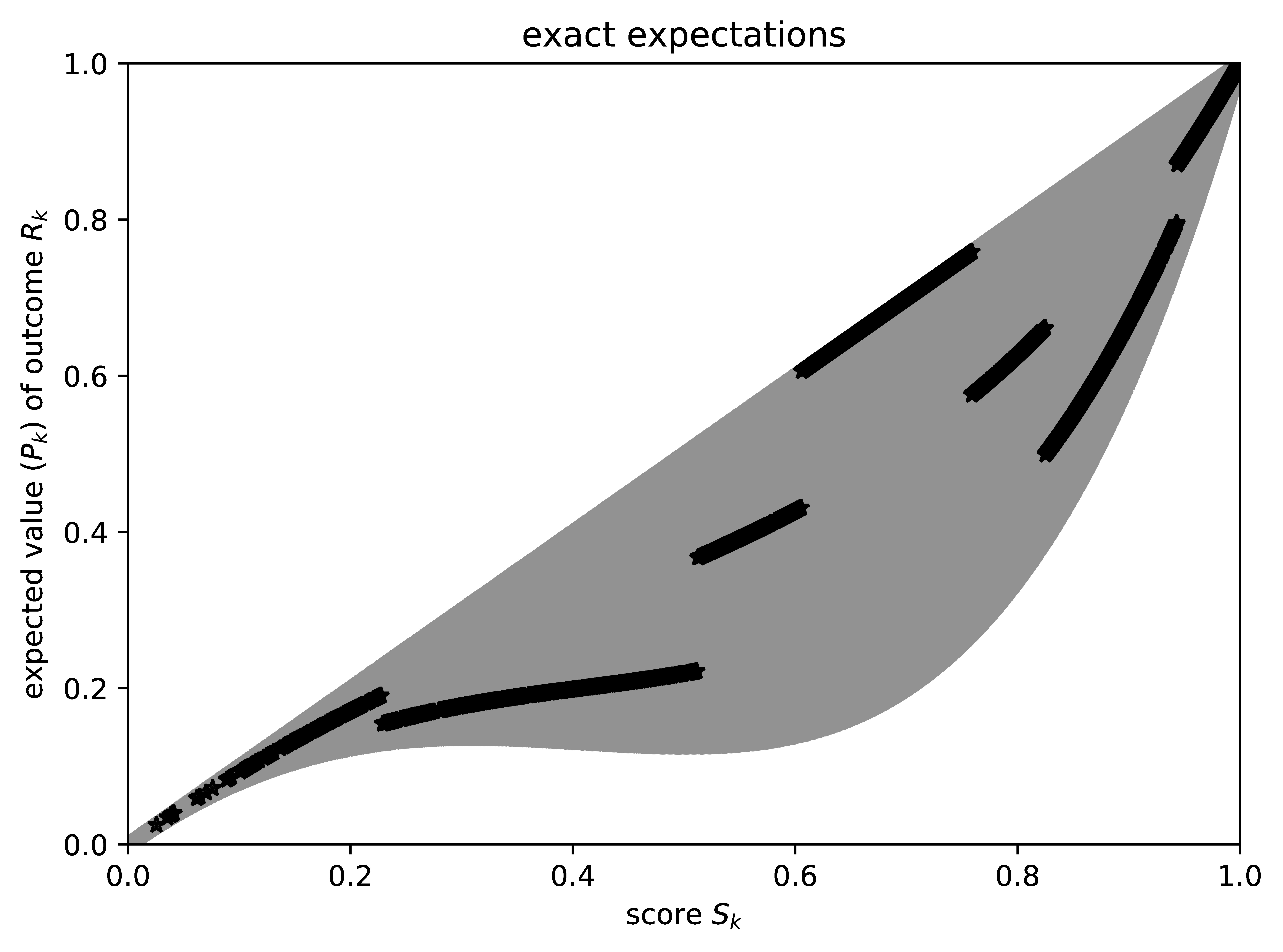}}

\end{centering}
\caption{Ground-truth reliability diagram for Figure~\ref{2500}}
\label{2500e}
\end{figure}

\begin{figure}
\begin{centering}

\parbox{\imsize}{\includegraphics[width=\imsize]
                {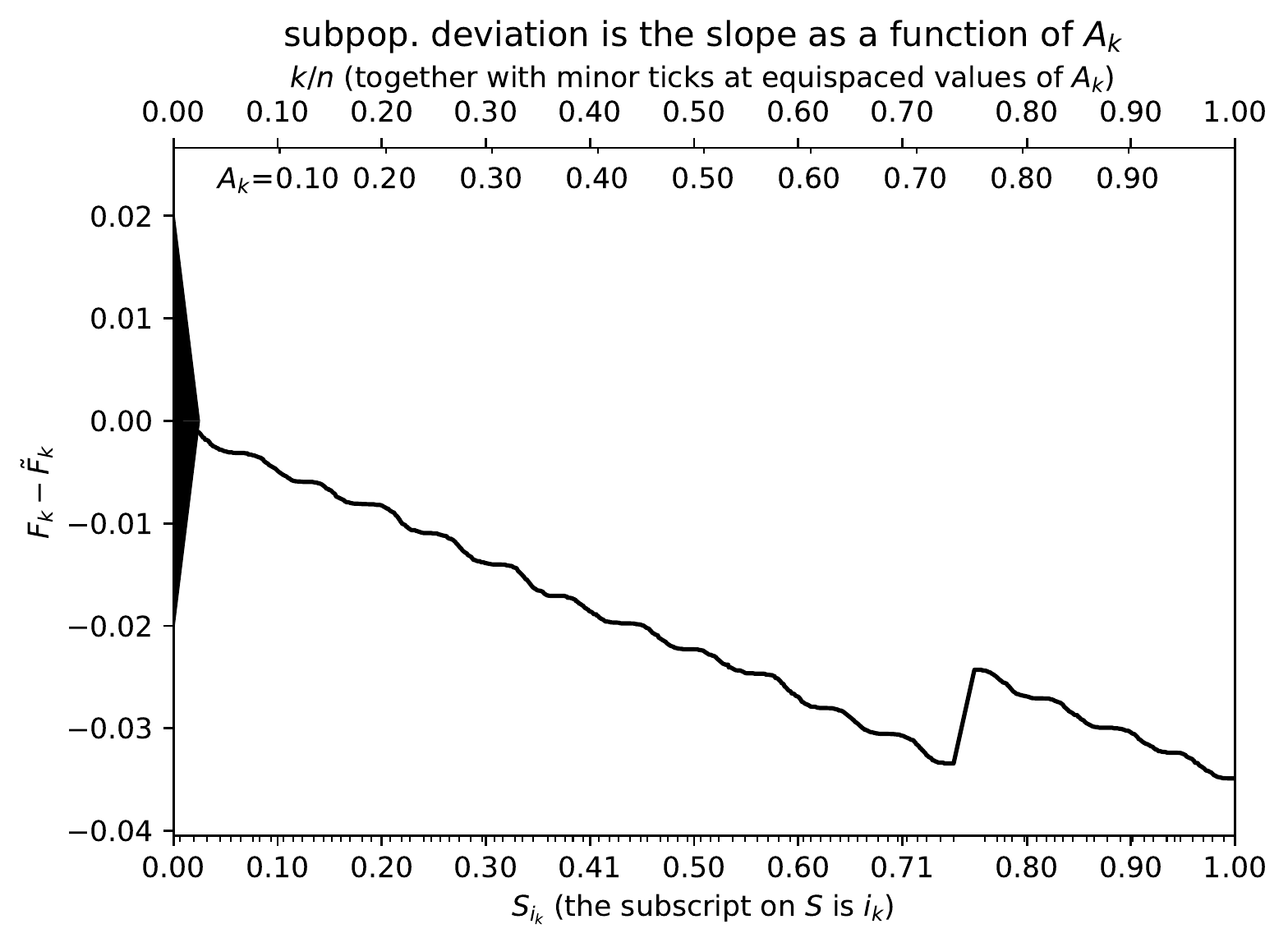}}
\quad\quad
\parbox{\imsize}{\includegraphics[width=\imsize]
                {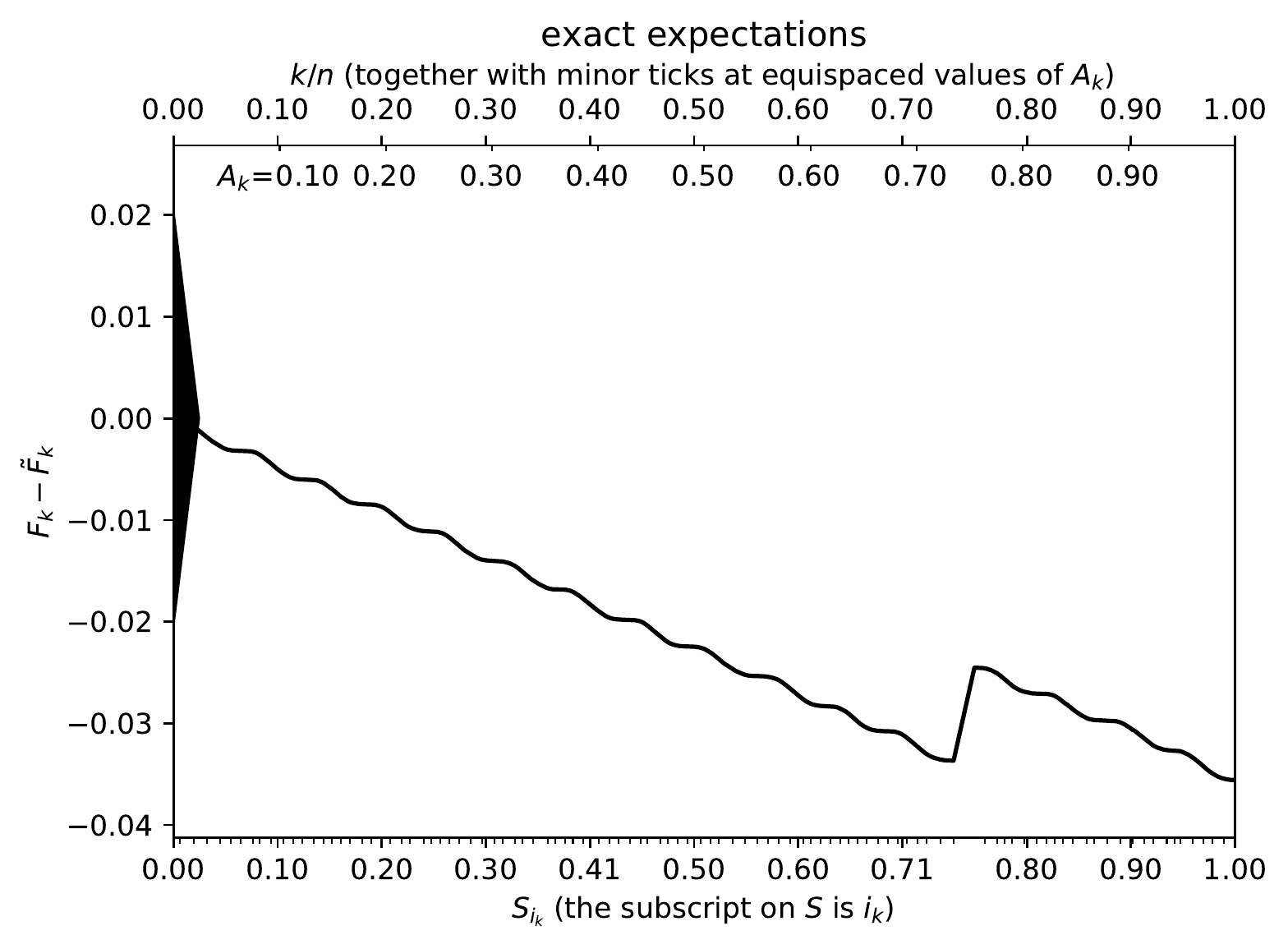}}

\vspace{\vertsep}

\parbox{\imsize}{\includegraphics[width=\imsize]
                {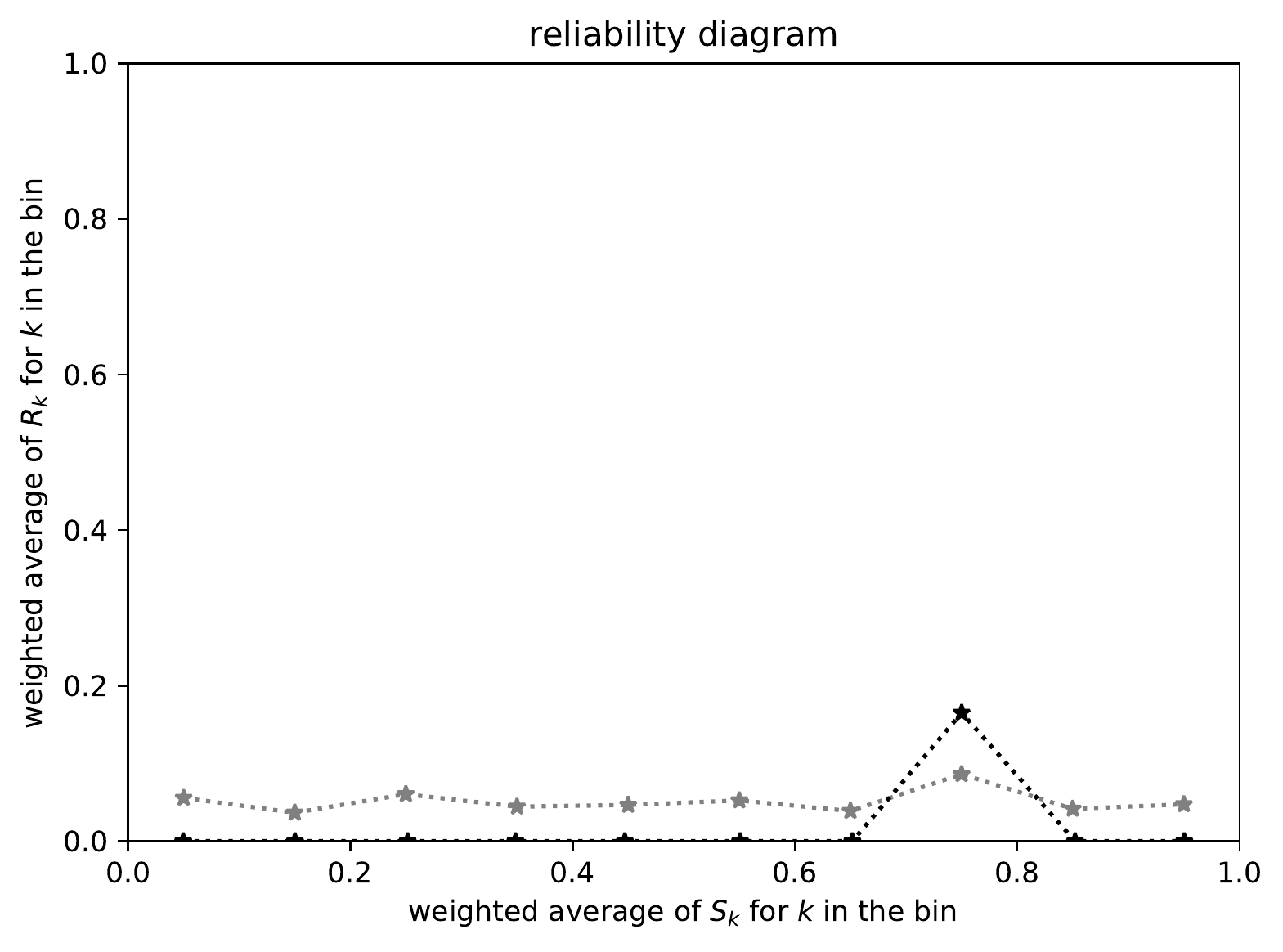}}
\quad\quad
\parbox{\imsize}{\includegraphics[width=\imsize]
                {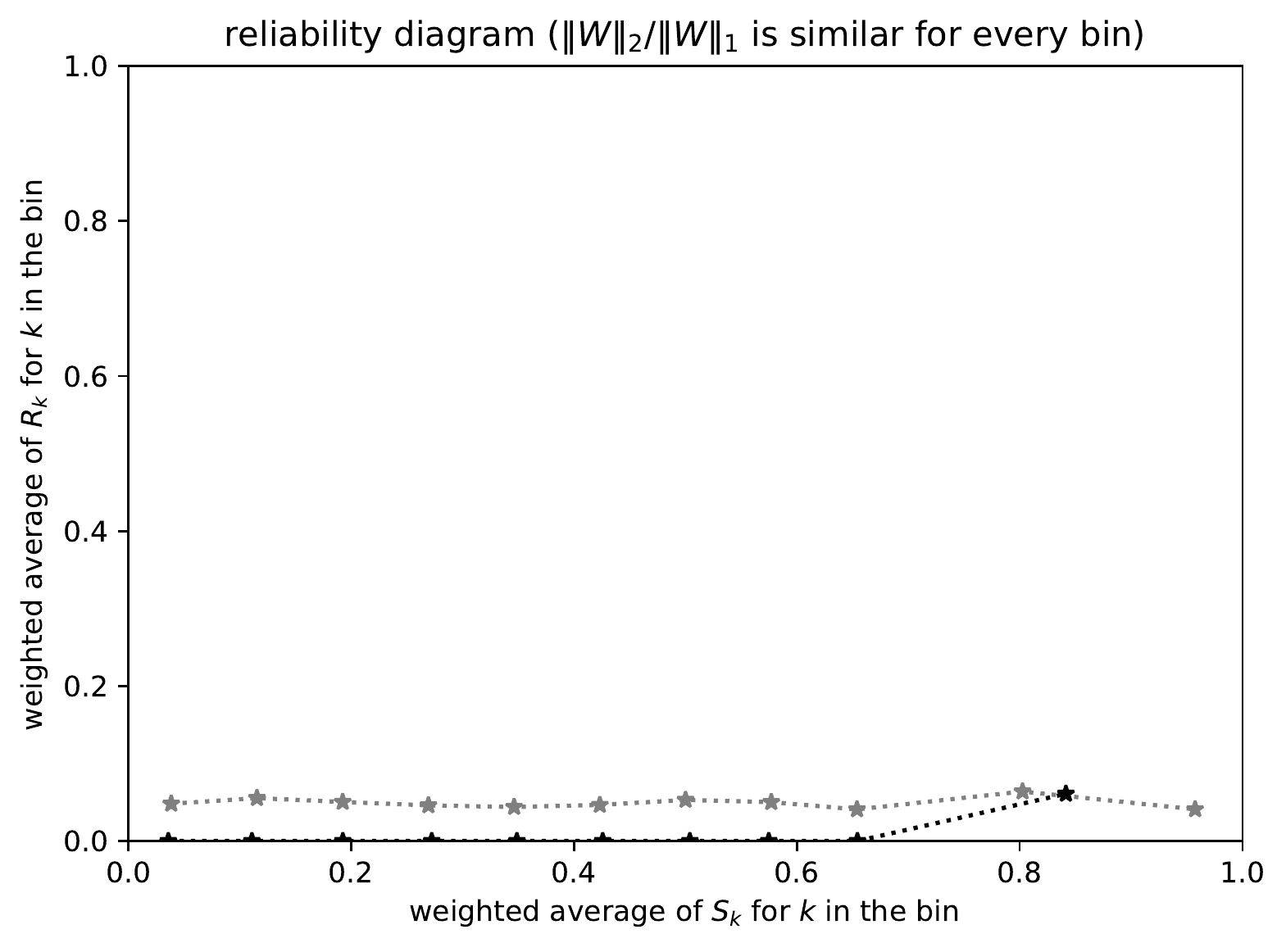}}

\vspace{\vertsep}

\parbox{\imsize}{\includegraphics[width=\imsize]
                {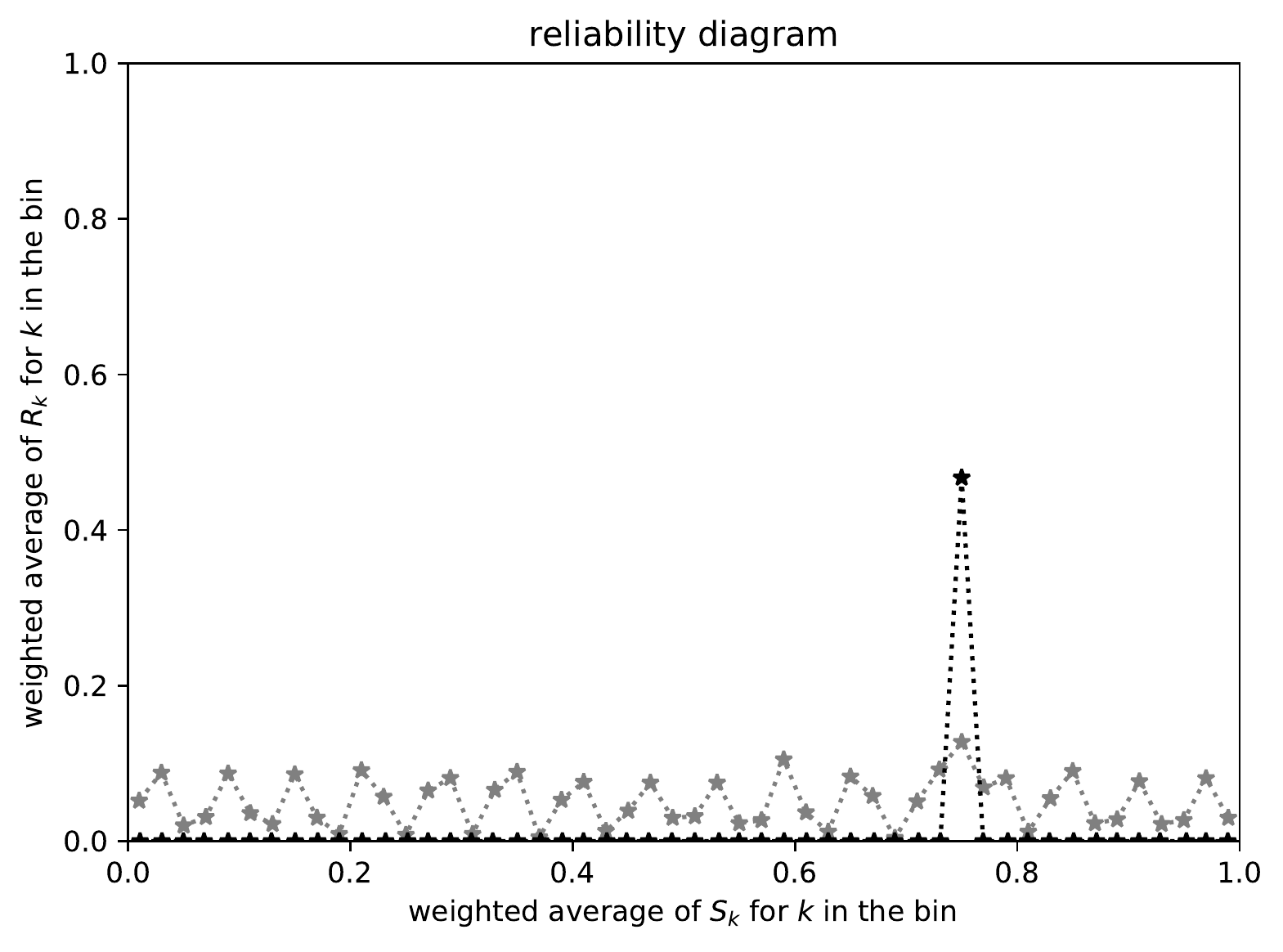}}
\quad\quad
\parbox{\imsize}{\includegraphics[width=\imsize]
                {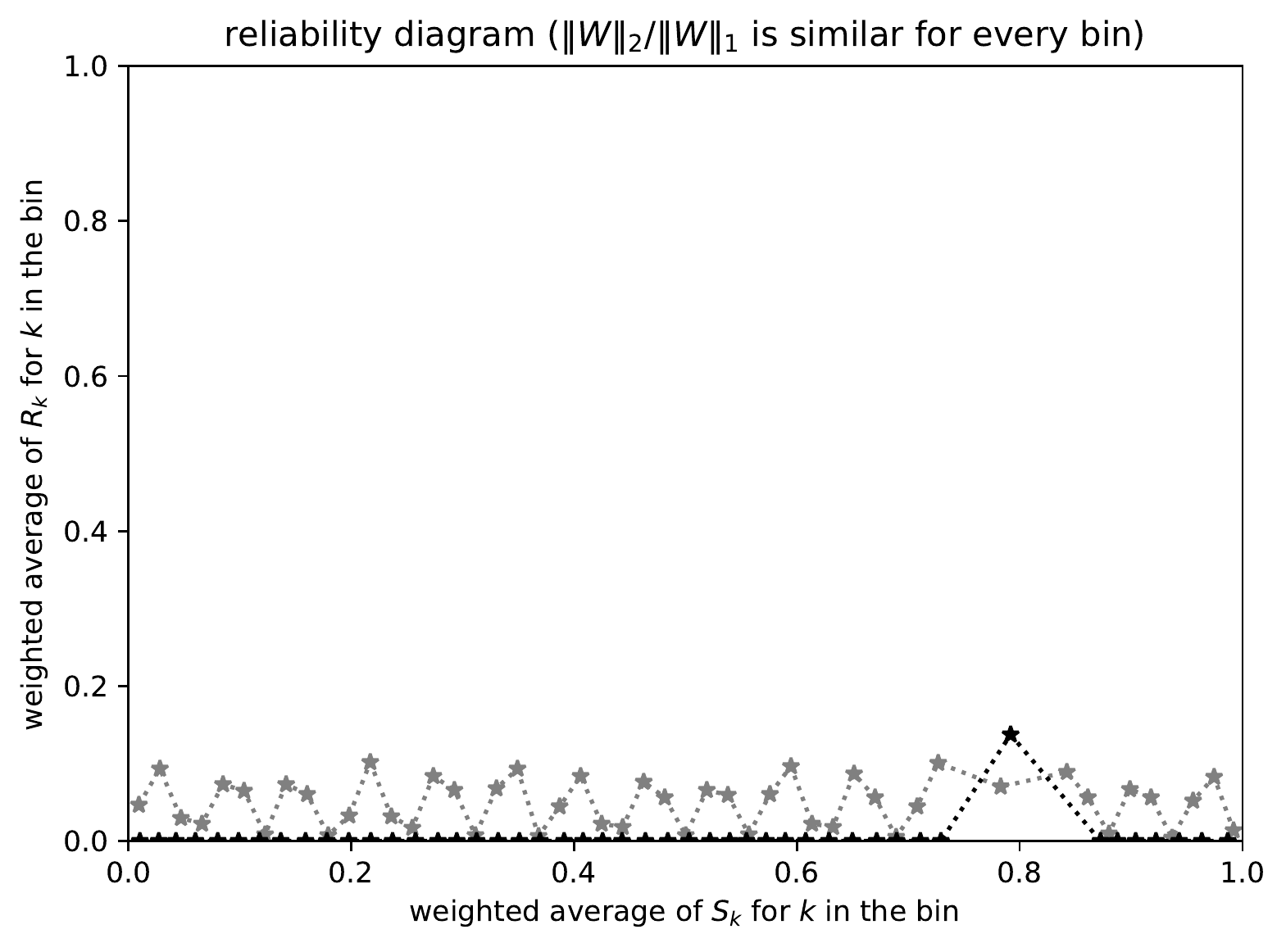}}

\end{centering}
\caption{$n =$ 2,500, with weighted sampling;
         Kuiper's statistic is $0.03490 / \sigma = 3.318$,
         Kolmogorov's and Smirnov's is $0.03490 / \sigma = 3.318$.
Figure~\ref{2500we} displays the ground-truth reliability diagram.
The cumulative plot displays the distinguished observation
from the subpopulation as a straight, steep jump at its score around 0.75;
the constant slope of that steep jump shows that the corresponding
high deviation between the subpopulation and the full population
is due to a single highly weighted observation. This single observation has
no effect on the slopes in the rest of the cumulative plot,
whereas the few highly weighted observations dramatically
(perhaps misleadingly?)\
alter the bins around scores of 0.75 in the observed reliability diagrams.
The scalar summary statistics report
statistically significant deviation of the subpopulation
from the full population, though the steep jump in the cumulative plot
reduces the effectiveness of the scalar statistics.
}
\label{2500w}
\end{figure}

\begin{figure}
\begin{centering}

\parbox{\imsize}{\includegraphics[width=\imsize]
                {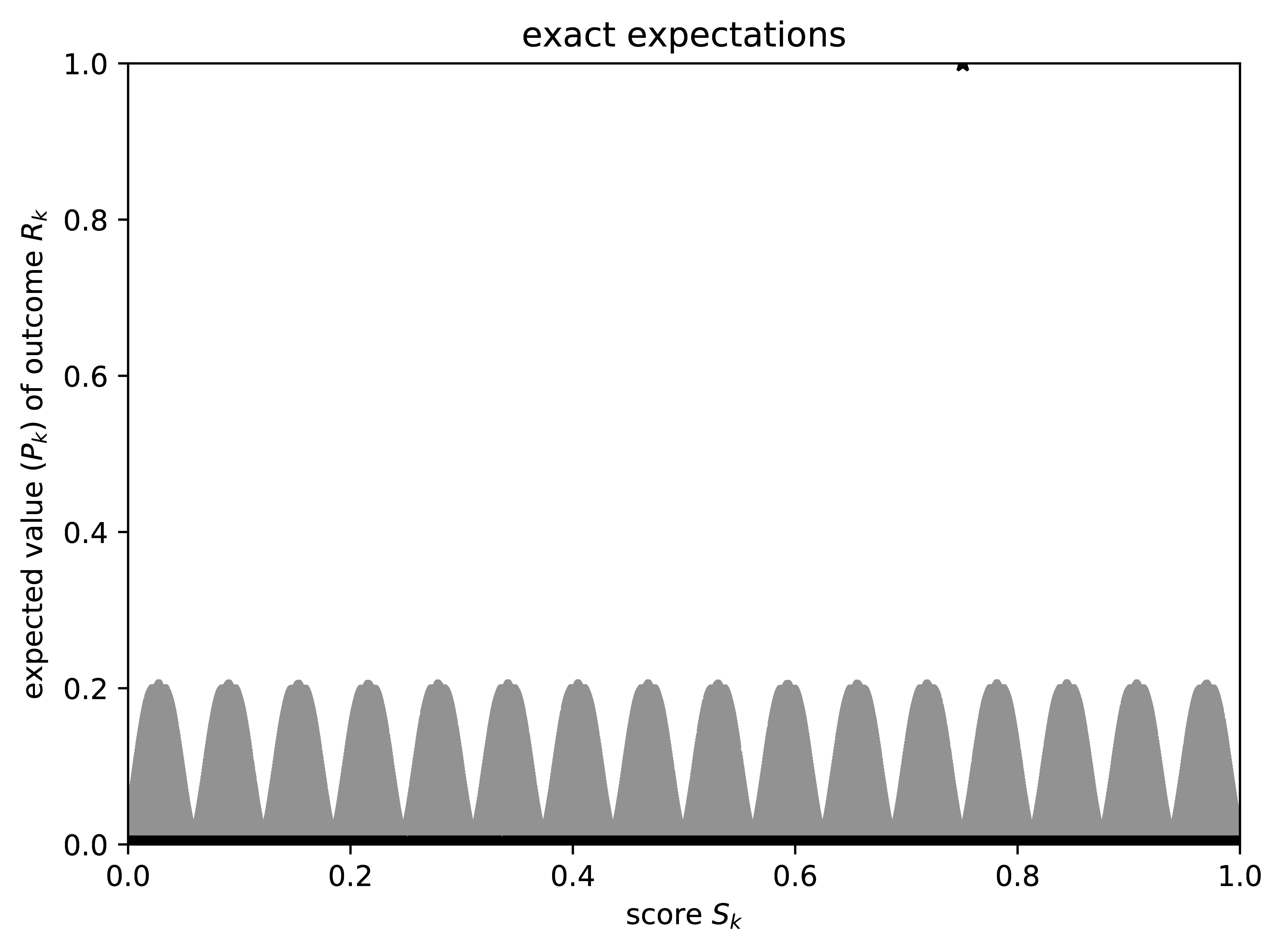}}

\end{centering}
\caption{Ground-truth reliability diagram for Figure~\ref{2500w}}
\label{2500we}
\end{figure}

In this subsection, the examples draw observations at random
from various statistical models so that the underlying ``ground-truth''
is available. To generate the corresponding figures,
Figures~\ref{5000}--\ref{2500we}, we specify values for the scores
$S_1$,~$S_2$, \dots, $S_m$ and for the indices $i_1$,~$i_2$, \dots, $i_n$.
We also select probabilities $P_1$,~$P_2$, \dots, $P_m$
and then independently draw the outcomes $R_1$,~$R_2$, \dots, $R_m$
from the Bernoulli distributions with parameters $P_1$,~$P_2$, \dots, $P_m$,
respectively. The first three examples,
corresponding to Figures~\ref{5000}--\ref{2500e}, use unweighted
(or, equivalently, uniformly weighted) data;
the fourth example, corresponding to Figures~\ref{2500w} and~\ref{2500we},
uses weighted data, in which each pair of score $S_i$ and result $R_i$ comes
with an additional positive scalar weight $W_i$.
Appendix~\ref{caution} illustrates the degenerate case of when deviation
is absent, via further examples.

The top rows of Figures~\ref{5000}, \ref{3300}, and~\ref{2500} plot
$F_k-\tilde{F}_k$ from~(\ref{empirical}) and~(\ref{full}) as a function
of $k/n$, with the rightmost plot displaying its noiseless expected value
rather than using the random observations $R_1$,~$R_2$, \dots, $R_m$.
The top row of Figure~\ref{2500w} plots $F_k-\tilde{F}_k$
from~(\ref{empiricalw}) and~(\ref{fullw}) versus
the cumulative weight $A_k$ from~(\ref{abscissae}),
with the rightmost plot displaying its noiseless expected value
rather than using the random observations $R_1$,~$R_2$, \dots, $R_m$.
Figures~\ref{5000e}, \ref{3300e}, \ref{2500e}, and~\ref{2500we} plot the pairs
$(S_1, P_1)$,~$(S_2, P_2)$, \dots, $(S_m, P_m)$ in gray, and plot the pairs
$(S_{i_1}, P_{i_1})$,~$(S_{i_2}, P_{i_2})$, \dots, $(S_{i_n}, P_{i_n})$
in black, producing ground-truth diagrams that the lowermost two rows of plots
from the associated Figures~\ref{5000}, \ref{3300}, \ref{2500}, and~\ref{2500w}
are trying to estimate using only the observations $R_1$,~$R_2$, \dots, $R_m$,
without access to the underlying probabilities $P_1$,~$P_2$, \dots, $P_m$.

For the examples, we consider full populations which are mixtures
of many subpopulations, including for each example one specific subpopulation
that we analyze for deviations from the average over the full population.
To make the models realistic, we assign expected outcomes to the members
of the full population such that the expected outcomes span a continuous range
which includes the expected outcomes attained by members of the subpopulation
being analyzed for deviations. The reliability diagrams use gray points
and gray connecting-lines to indicate the full population,
and solid black points and black connecting-lines to indicate
the specific subpopulation under consideration.

For the first example, corresponding to Figures~\ref{5000} and~\ref{5000e},
we consider a mixture of subpopulations with a reasonably wide range
of expected outcomes for most ranges of scores, except for a narrow notch
around scores of 0.25 which lacks the significant subpopulation deviation.
We select for the specific subpopulation analyzed for deviations
a subpopulation with among the highest expected outcomes around every score.
The scores are $S_j = ((j - 0.5) / m)^2$ for $j = 1$,~$2$, \dots, $m$,
thus more concentrated near 0 than near 1.
For the indices $i_1$,~$i_2$, \dots, $i_n$ of the subpopulation, we start
with the integer multiples of 20 and then add three levels of refinement
increasingly focused around $m/2$,
where $m/2$ corresponds to the middle of the notch.
The total number of scores considered is $m =$ 50,000, and
(following the refinement) the number of subpopulation indices is $n =$ 5,000.

For the second example, corresponding to Figures~\ref{3300} and~\ref{3300e},
we consider a mixture of subpopulations that includes one whose expected
outcomes oscillate smoothly between the minimum and the maximum of the
expected outcomes of all subpopulations as a function of the score; we analyze
that particular one subpopulation for deviations from the full population.
The scores are $S_j = (j - 0.5) / m$ for $j = 1$,~$2$, \dots, $m$,
hence equispaced between 0 and 1.
The total number of scores considered is $m =$ 50,000.
For the indices $i_1$,~$i_2$, \dots, $i_n$ of the subpopulation,
we raise to the power $4/3$ each positive integer,
then round to the nearest integer,
and finally retain the lowest $n =$ 3,300 unique resulting integers.

For the third example, corresponding to Figures~\ref{2500} and~\ref{2500e},
we consider a mixture of subpopulations similar to that for the second example,
but this time selecting for analysis a subpopulation
whose expected outcomes oscillate in discrete steps between the minimum
and the maximum of the expected outcomes of all subpopulations as a function
of the score.
The scores are $S_j = \sqrt{(j - 0.5) / m)}$ for $j = 1$,~$2$, \dots, $m$,
thus less concentrated near 0 than near 1.
For the indices $i_1$,~$i_2$, \dots, $i_n$ of the subpopulation,
we generate a random permutation of the integers $1$,~$2$, \dots, $m$,
retain the first $n$, and then sort them.
The total number of scores considered is $m =$ 50,000,
and the number of subpopulation indices is $n =$ 2,500.

For the fourth example, corresponding to Figures~\ref{2500w} and~\ref{2500we},
we consider weighted samples (whereas the three previous examples
all used unweighted or uniformly weighted data,
in which all weights would be equal).
We start by constructing the full population as a mixture of subpopulations
whose probabilities of success oscillate between 0 and 0.2,
and with the subpopulation being analyzed selected uniformly at random
such that the probability of success for each observation is 0.
The scores are $S_j = (j - 0.5) / m$ for $j = 1$,~$2$, \dots, $m$,
hence equispaced between 0 and 1.
The total number of scores considered is $m =$ 50,000,
and the number of subpopulation indices is $n =$ 2,500.
All observations just mentioned we weight equally with weight 1,
and then introduce three outliers near the score 0.75:
one being an observation from the subpopulation,
altering its probability of success to 1 and its weight to $0.02 n$,
and the other two being the two members from the full population
not belonging to the subpopulation whose scores are adjacent to the score
for the observation from the subpopulation, setting the probabilities
of success for these two members to 0 and 1,
both with the same weight $0.002 m$.

The captions of the figures comment on the numerical results displayed.

\begin{remark}
\label{Simpson}
If (as in Figures~\ref{2500w} and~\ref{2500we}) the weight $W_i$ for some
single observation is extraordinarily disproportionately high, then any bin
containing that observation will be strongly biased toward only that individual
observation, blind to the other observations in the bin. Such a phenomenon
can lead to misleading behavior akin to Simpson's Paradox of~\cite{simpson}
in the canonical binned reliability diagrams; for instance,
the subpopulation may very well
attain results $R_{i_1}$, $R_{i_2}$, \dots, $R_{i_n}$ that are {\it less} than
all others in the full population except for the one disproportionately heavily
weighted one, yet if the results $R_{i_1}$, $R_{i_2}$, \dots, $R_{i_n}$
for the subpopulation are greater than the result $R_i$ corresponding
to the one disproportionately large weight $W_i$, then
any bin in reliability diagrams containing that heavily weighted observation
will show that the subpopulation attains {\it greater} results
than the weighted average in the full population.
In contrast, a single heavily weighted observation affects
only one subpopulation observation in the plots of cumulative differences,
merely introducing a jump in the constant offset
at the score corresponding to the one observation
(and the jump will be disproportionately large only in proportion
to the weight of the corresponding observation from the subpopulation).
Figure~\ref{2500w} illustrates related behavior,
and Figure~\ref{humboldt} below exhibits similar behavior
for the data from the U.S. Census Bureau analyzed there.
Comparing the subpopulation against the full population
via cumulative differences is thus similar to subtracting off baseline rates
for calibration (that is, similar to ``de-trending'').
\end{remark}

\subsection{ImageNet}
\label{imagenetex}

\begin{figure}
\begin{centering}

\parbox{\imsize}{\includegraphics[width=\imsize]
                {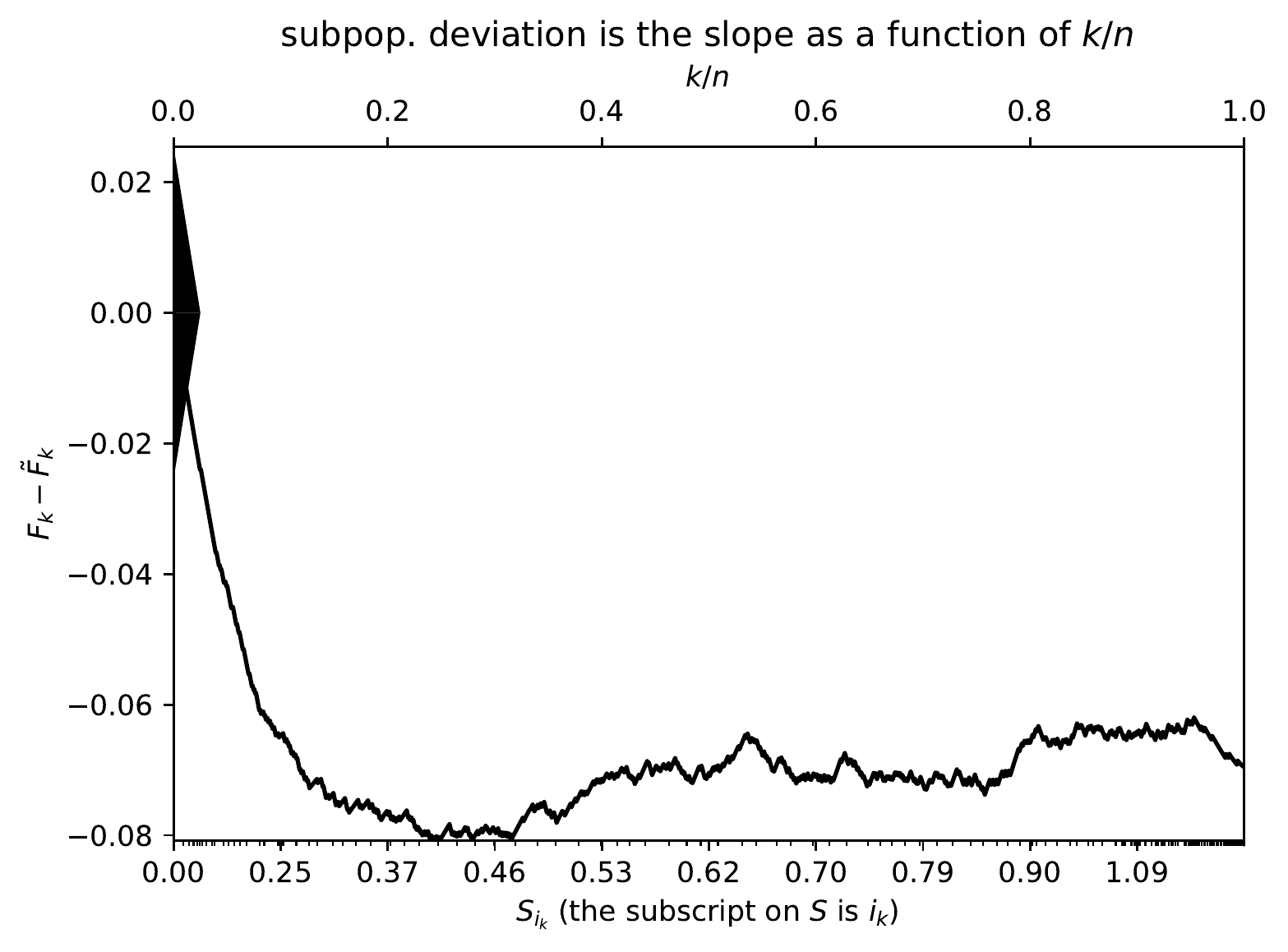}}
\quad\quad
\parbox{\imsize}{\includegraphics[width=\imsize]
                {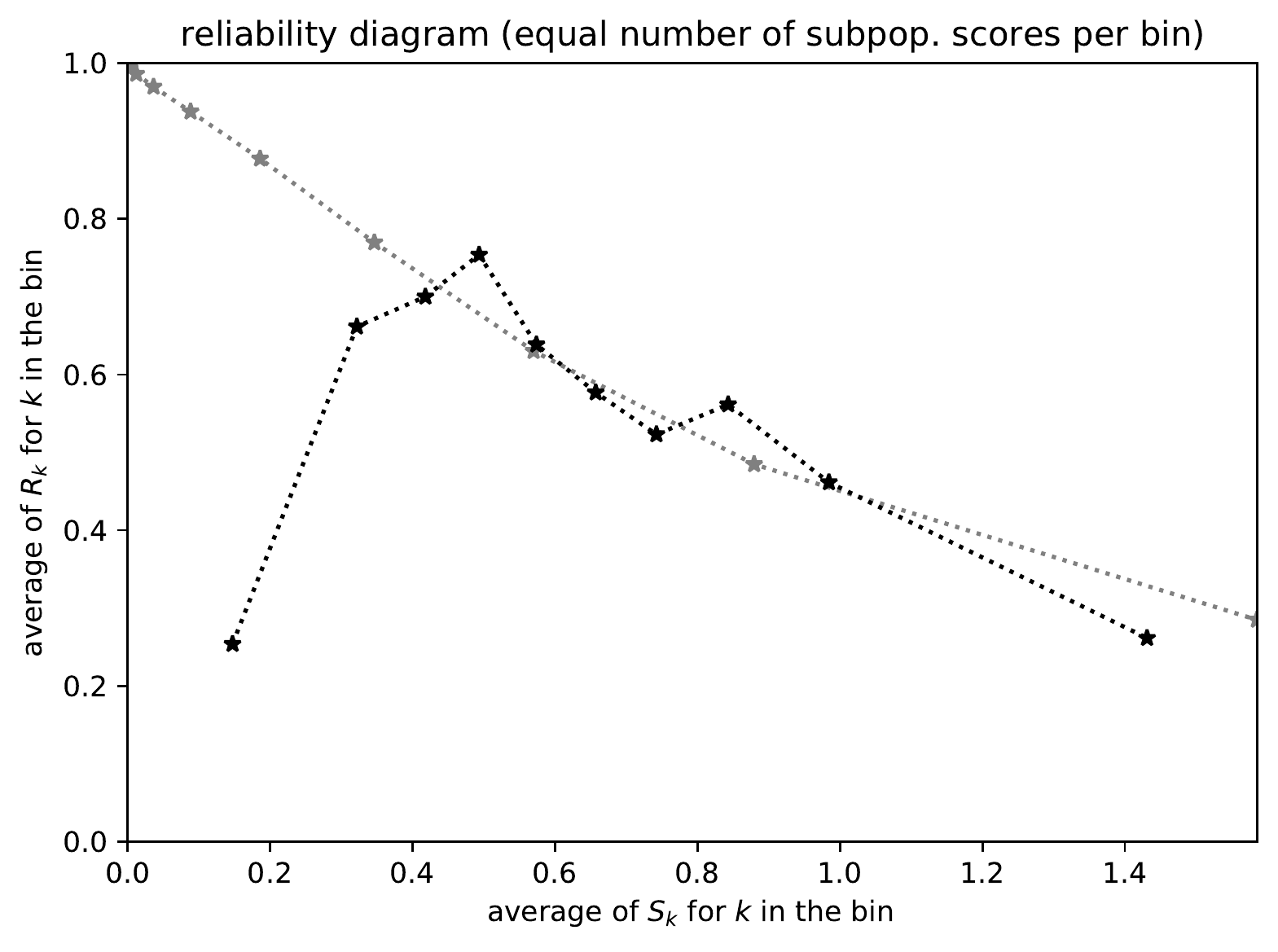}}

\vspace{\vertsep}

\parbox{\imsize}{\includegraphics[width=\imsize]
                {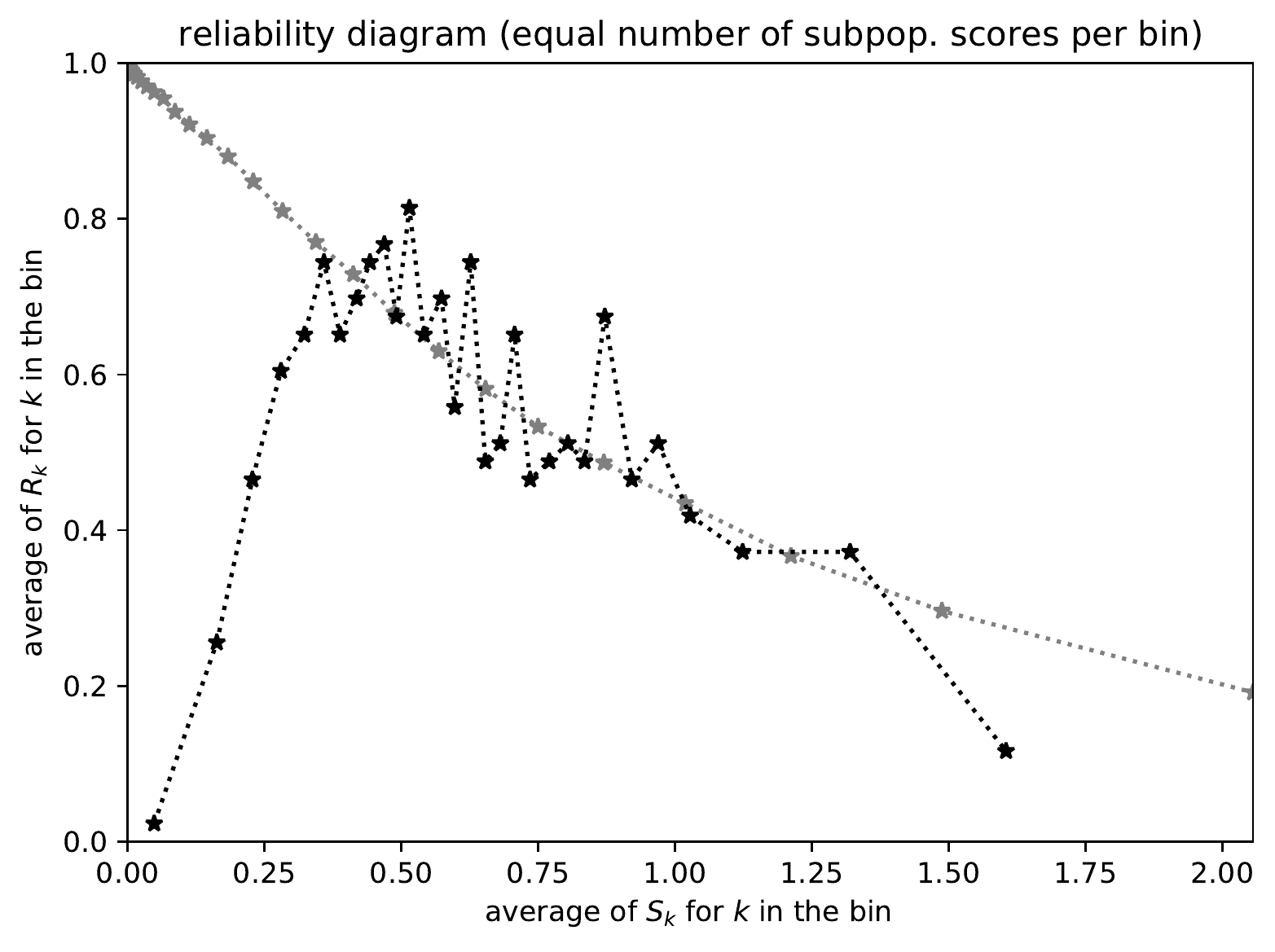}}
\quad\quad
\parbox{\imsize}{\includegraphics[width=\imsize]
                {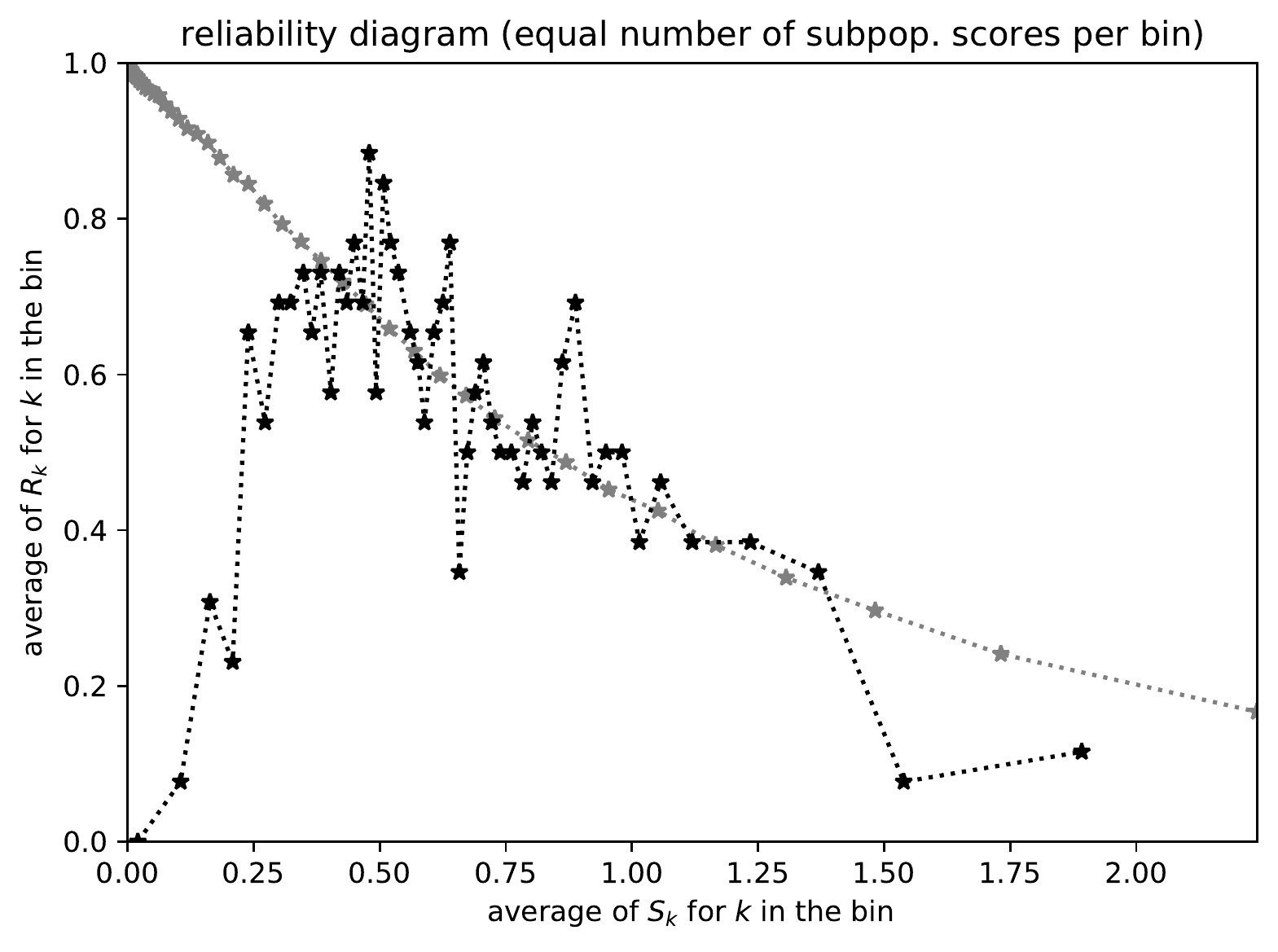}}

\vspace{\vertsep}

\parbox{\imsize}{\includegraphics[width=\imsize]
                {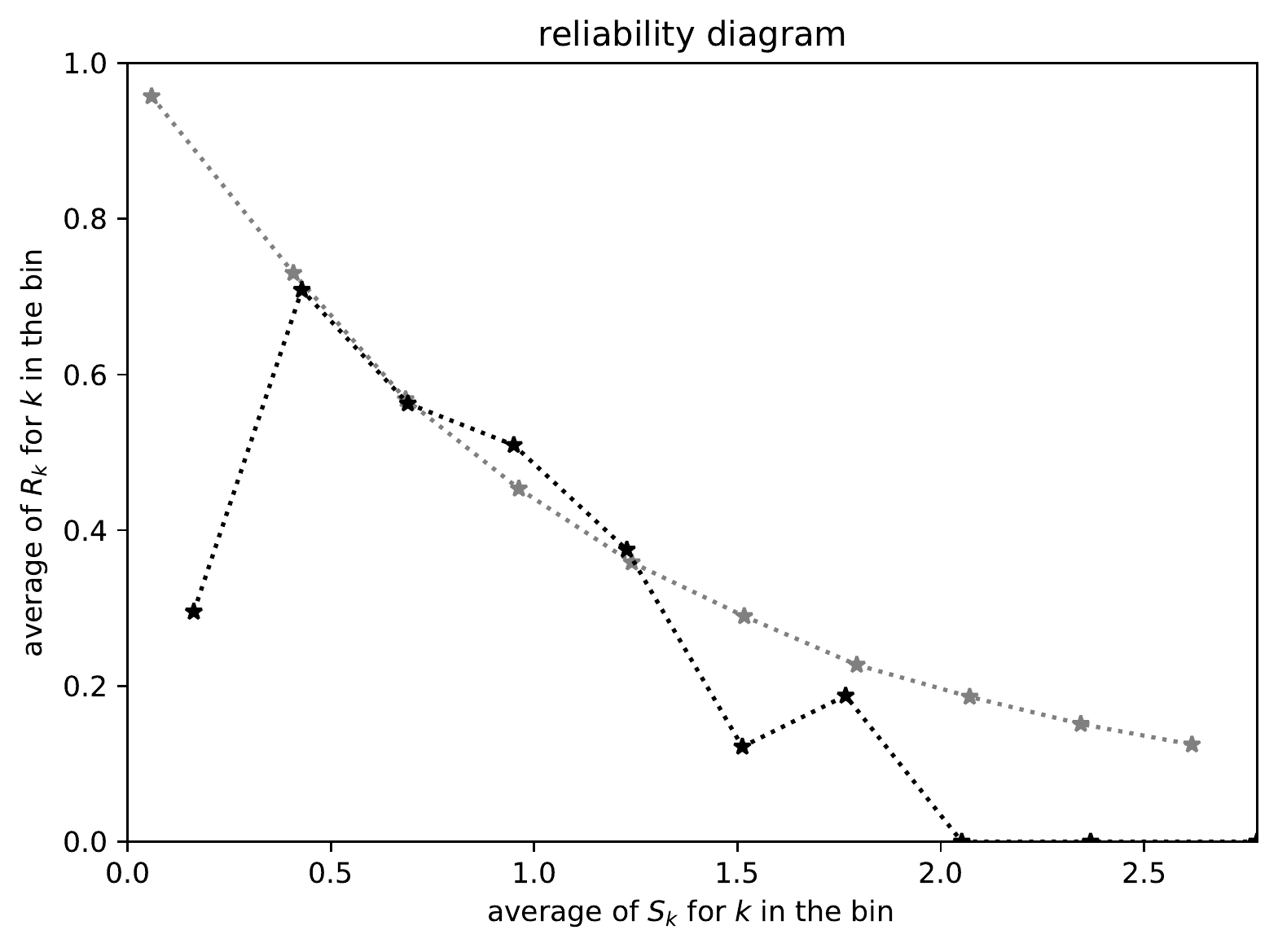}}
\quad\quad
\parbox{\imsize}{\includegraphics[width=\imsize]
                {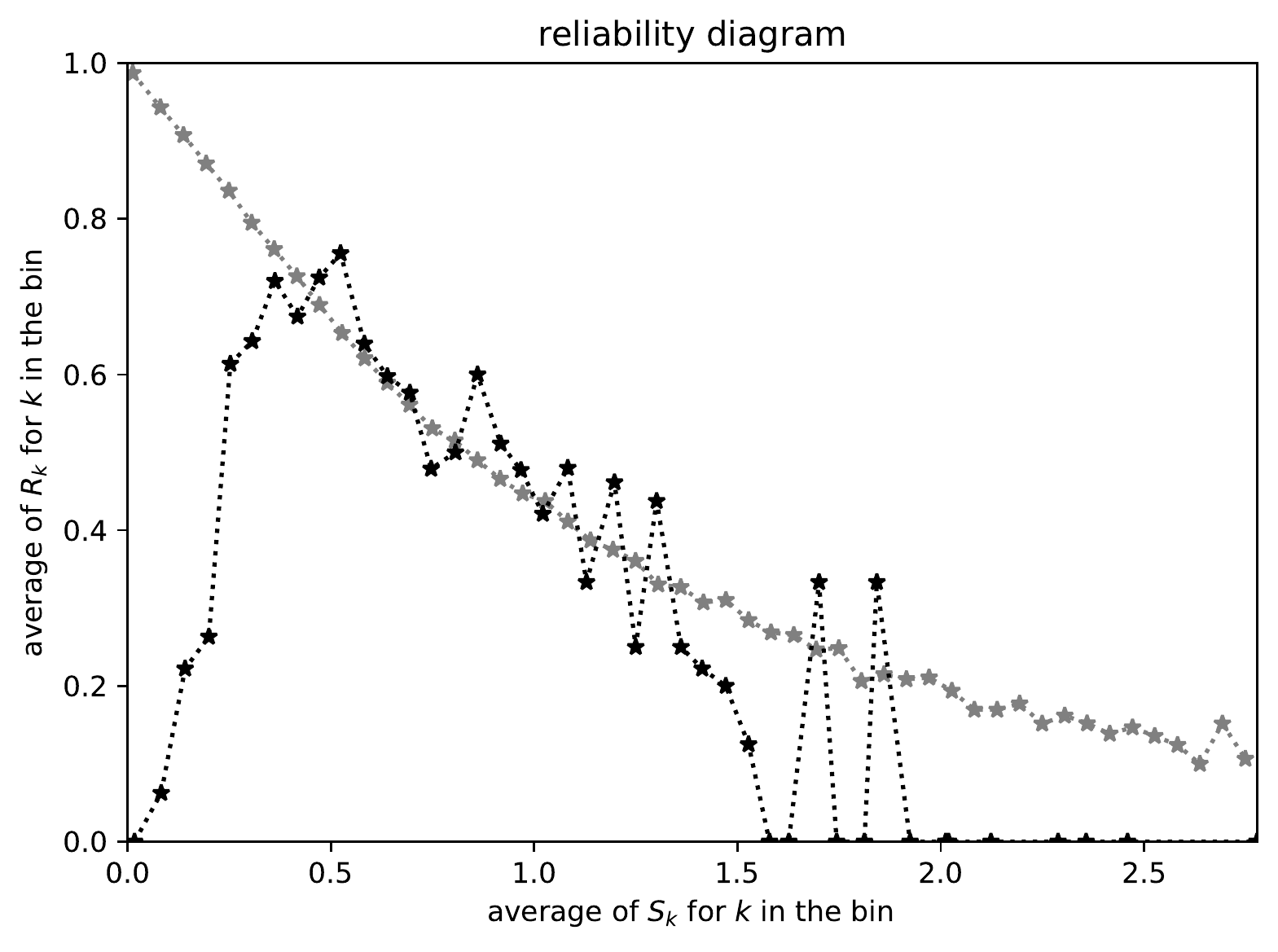}}

\end{centering}
\caption{Eskimo Dog or Husky, with scores ($S_1$,~$S_2$, \dots, $S_m$)
         being the negative log-likelihoods;
         $n =$ 1,300; Kuiper's statistic is $0.08082 / \sigma = 6.363$,
         Kolmogorov's and Smirnov's is $0.08082 / \sigma = 6.363$.
None of the reliability diagrams is able
to smooth away the irrelevant variations while simultaneously capturing
the severe deviation at the lowest negative log-likelihoods.
The scalar summary statistics very successfully
detect the statistically highly significant deviation
of the subpopulation from the full population.
}
\label{eskimo-dog-husky-nll}
\end{figure}

This subsection analyzes the standard training data set ``ImageNet-1000''
of~\cite{imagenet}. Each of the thousand labeled classes in the data set
forms a natural subpopulation to consider; each class considered consists
of $n =$ 1,300 images of a particular noun (such as a
``sidewinder/horned rattlesnake,'' a ``night snake,''
or an ``Eskimo Dog or Husky'').
Some classes in the data set contain fewer than 1,300 images,
so that the total number of members of the data set is $m =$ 1,281,167, but
each subpopulation we analyze below corresponds to a class with 1,300 images.
The particular classes reported below are cherry-picked to illustrate
a variety of problems that can afflict the classical reliability diagrams;
not all classes exhibit such problems, nor are these all the classes
that exhibit problems (many, many others have similar issues).
The images are unweighted (or, equivalently, uniformly weighted),
not requiring the methods of Subsection~\ref{weighted} above.
To generate the corresponding figures,
we calculate the scores $S_1$,~$S_2$, \dots, $S_m$
using the pretrained ResNet18 classifier of~\cite{he-zhang-ren-sun}
from the computer-vision module, ``torchvision,''
in the PyTorch software library of~\cite{pytorch};
the score for an image can be either \{1\} the probability
assigned by the classifier to the class predicted to be most likely
or \{2\} the corresponding negative log-likelihood
(that is, the negative of the natural logarithm of the probability),
with the scores randomly perturbed by about one part in $10^8$ to guarantee
their uniqueness.
For $i = 1$,~$2$, \dots, $m$, the result $R_i$ corresponding to a score $S_i$
is $R_i = 1$ when the class predicted to be most likely is the correct class,
and $R_i = 0$ otherwise.
The figures below omit display of reliability diagrams whose bins are
equispaced with respect to the scores in the typical case
that these reliability diagrams are so noisy as to be useless
(the figures do display the diagrams when they are not too noisy).
Figures~\ref{eskimo-dog-husky-nll}--%
\ref{sidewinder-horned-rattlesnake-Crotalus-cerastes-prob}
present several examples, listing in the captions the associated names
of the classes for the subpopulations.
Figure~\ref{zoom} illustrates the utility of the zooming proposed
in Remark~\ref{zooming} above.
In all figures, the plots of cumulative differences
depict all observed phenomena clearly.

\begin{figure}
\begin{centering}

\parbox{\imsize}{\includegraphics[width=\imsize]
                {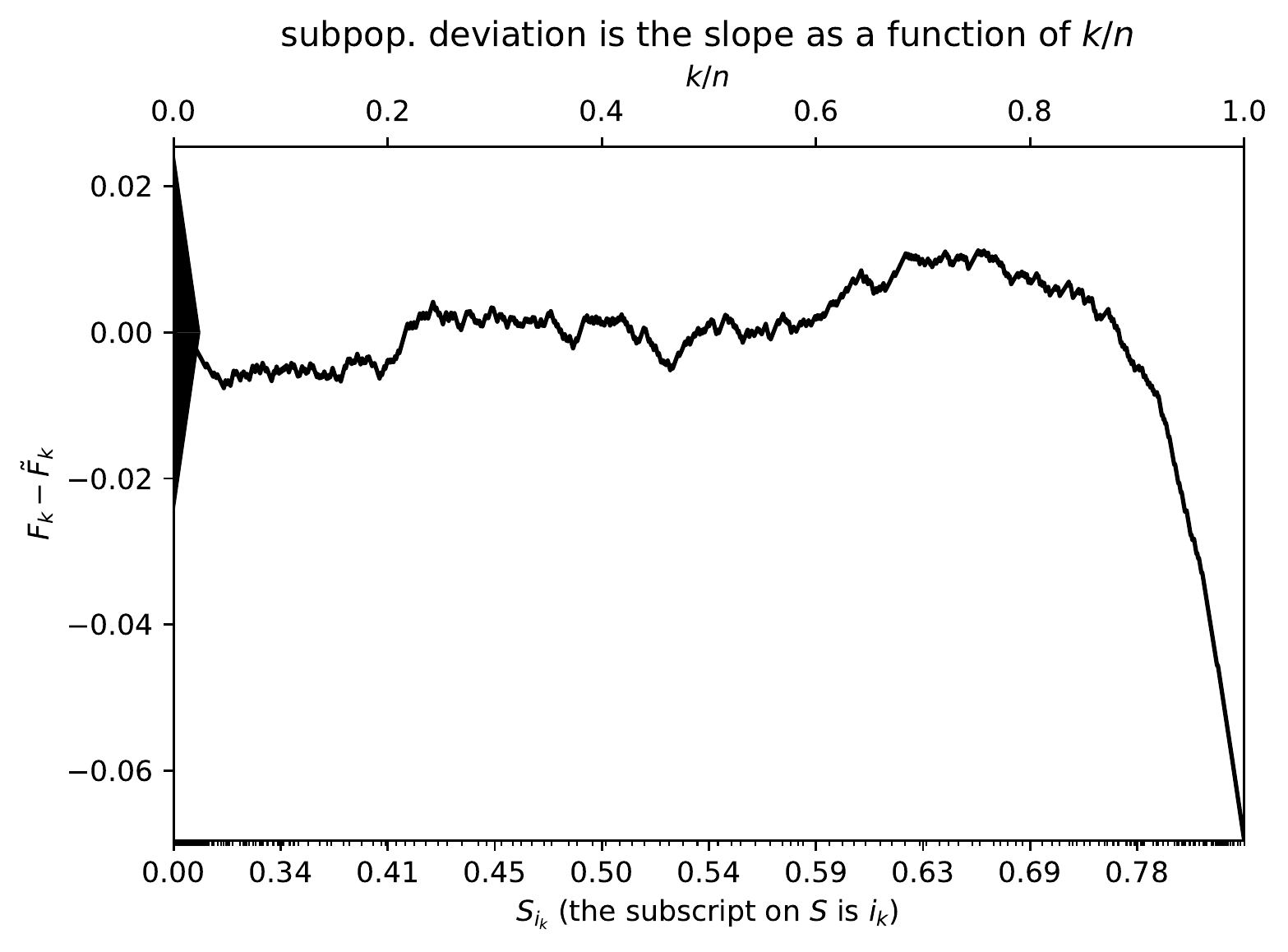}}
\quad\quad
\parbox{\imsize}{\includegraphics[width=\imsize]
                {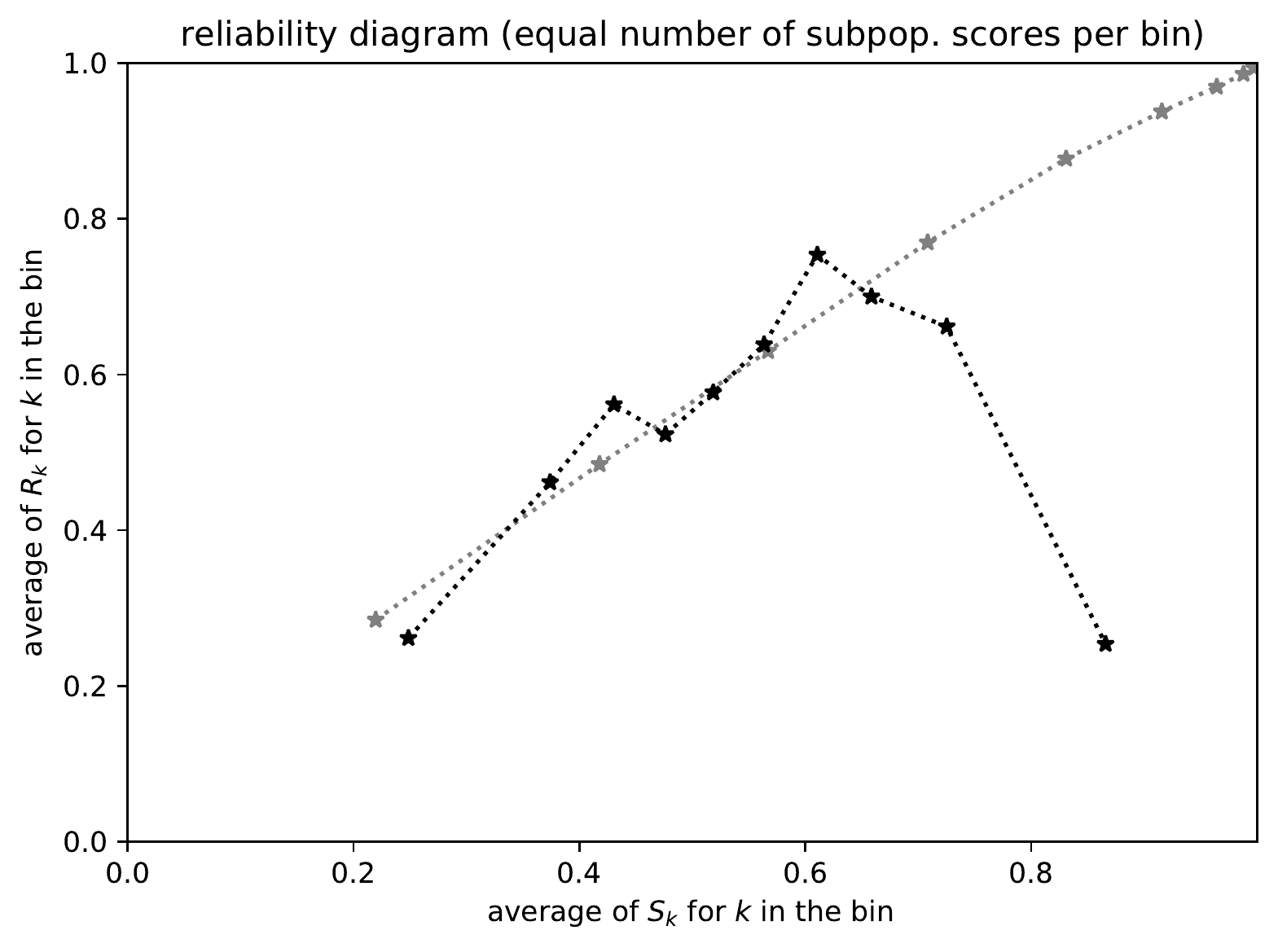}}

\vspace{\vertsep}

\parbox{\imsize}{\includegraphics[width=\imsize]
                {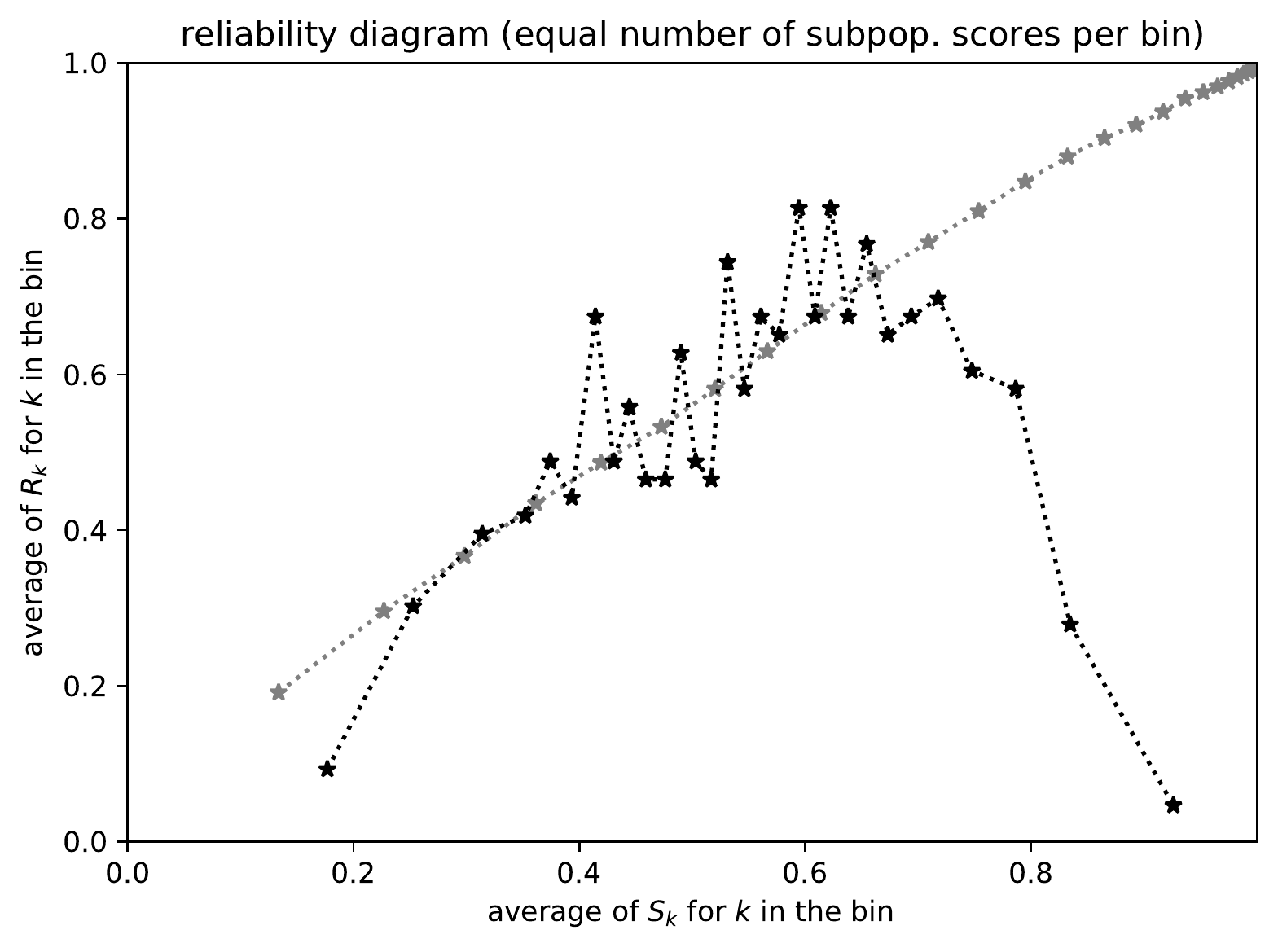}}
\quad\quad
\parbox{\imsize}{\includegraphics[width=\imsize]
                {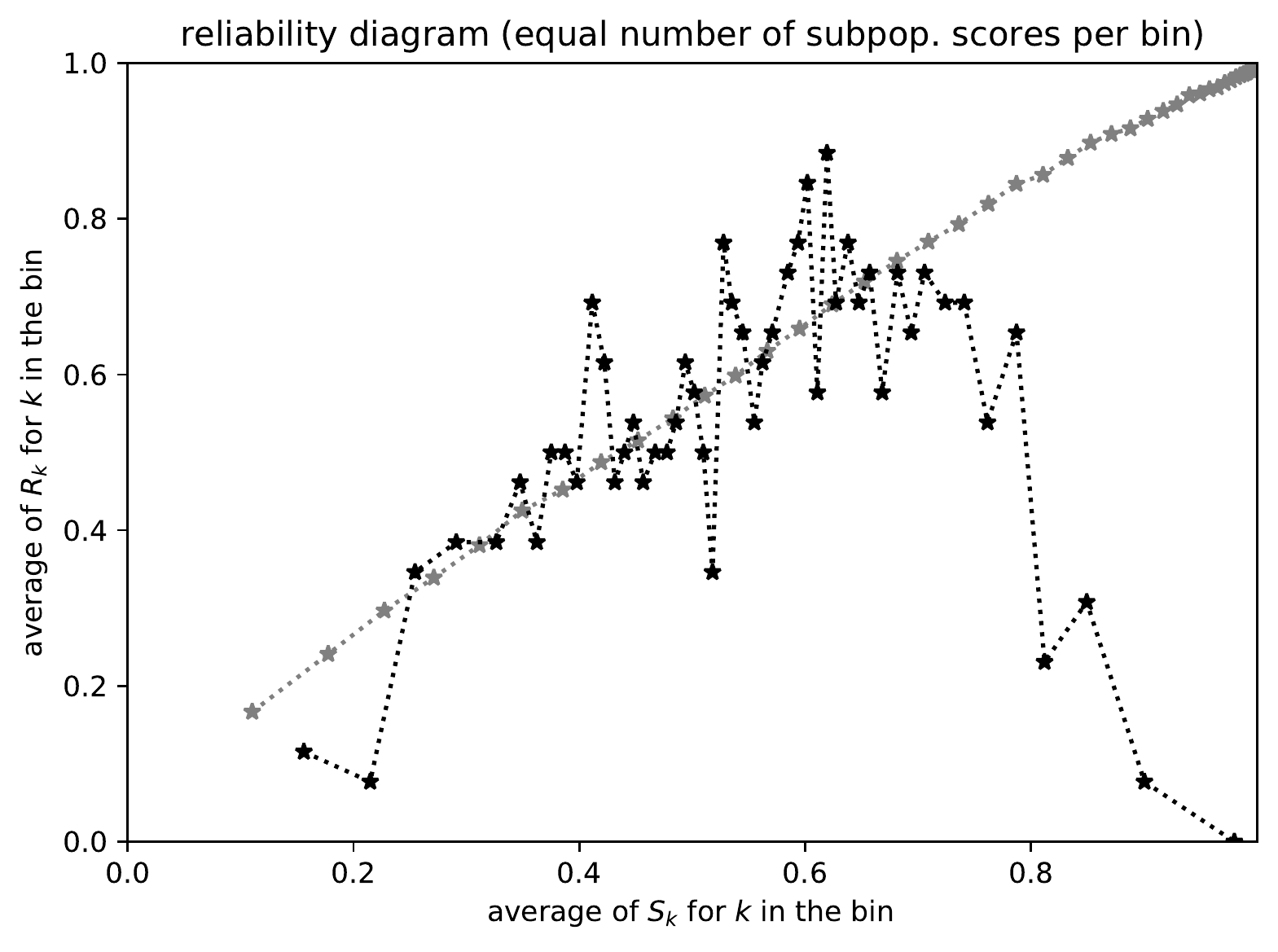}}

\vspace{\vertsep}

\parbox{\imsize}{\includegraphics[width=\imsize]
                {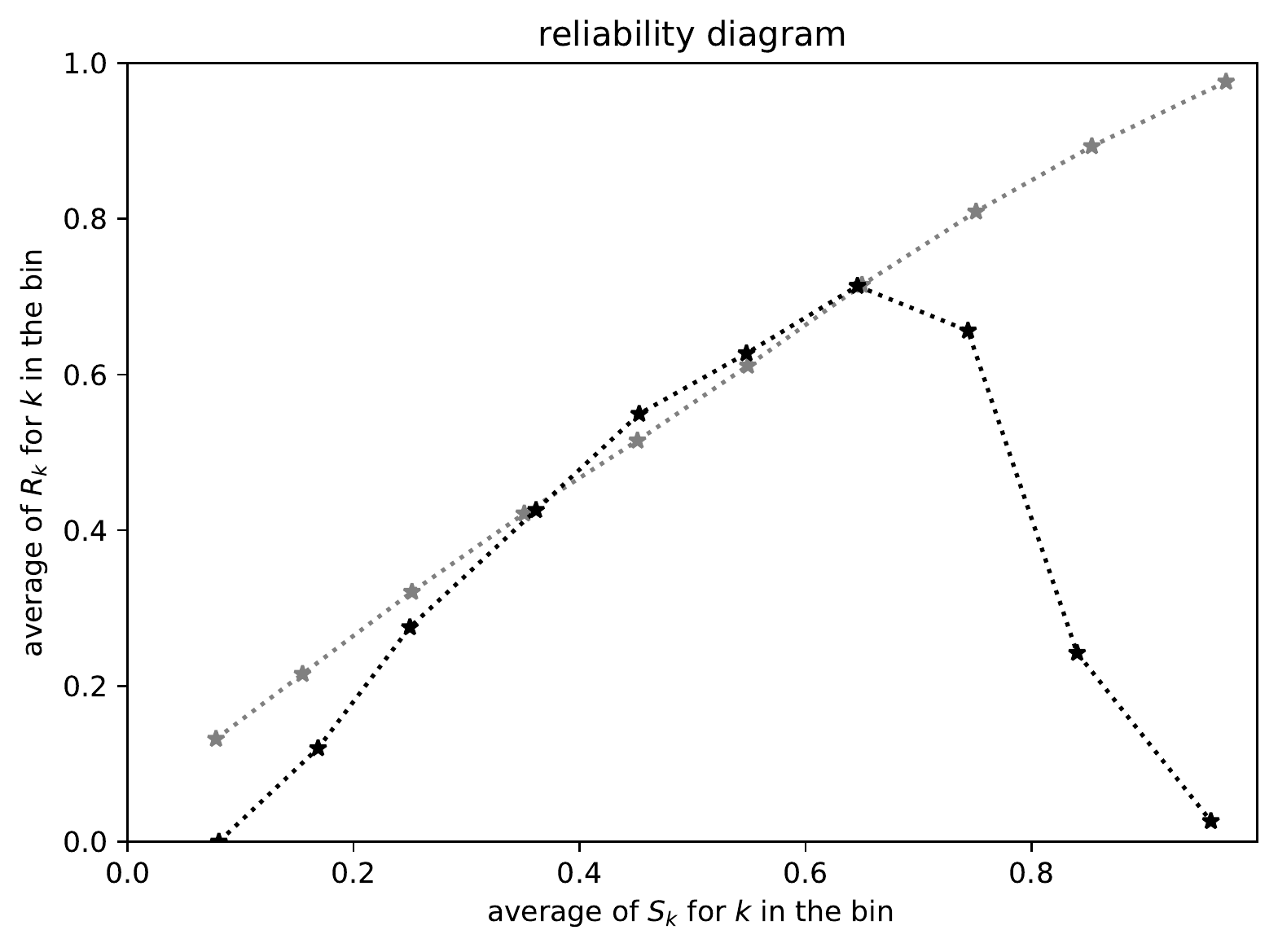}}
\quad\quad
\parbox{\imsize}{\includegraphics[width=\imsize]
                {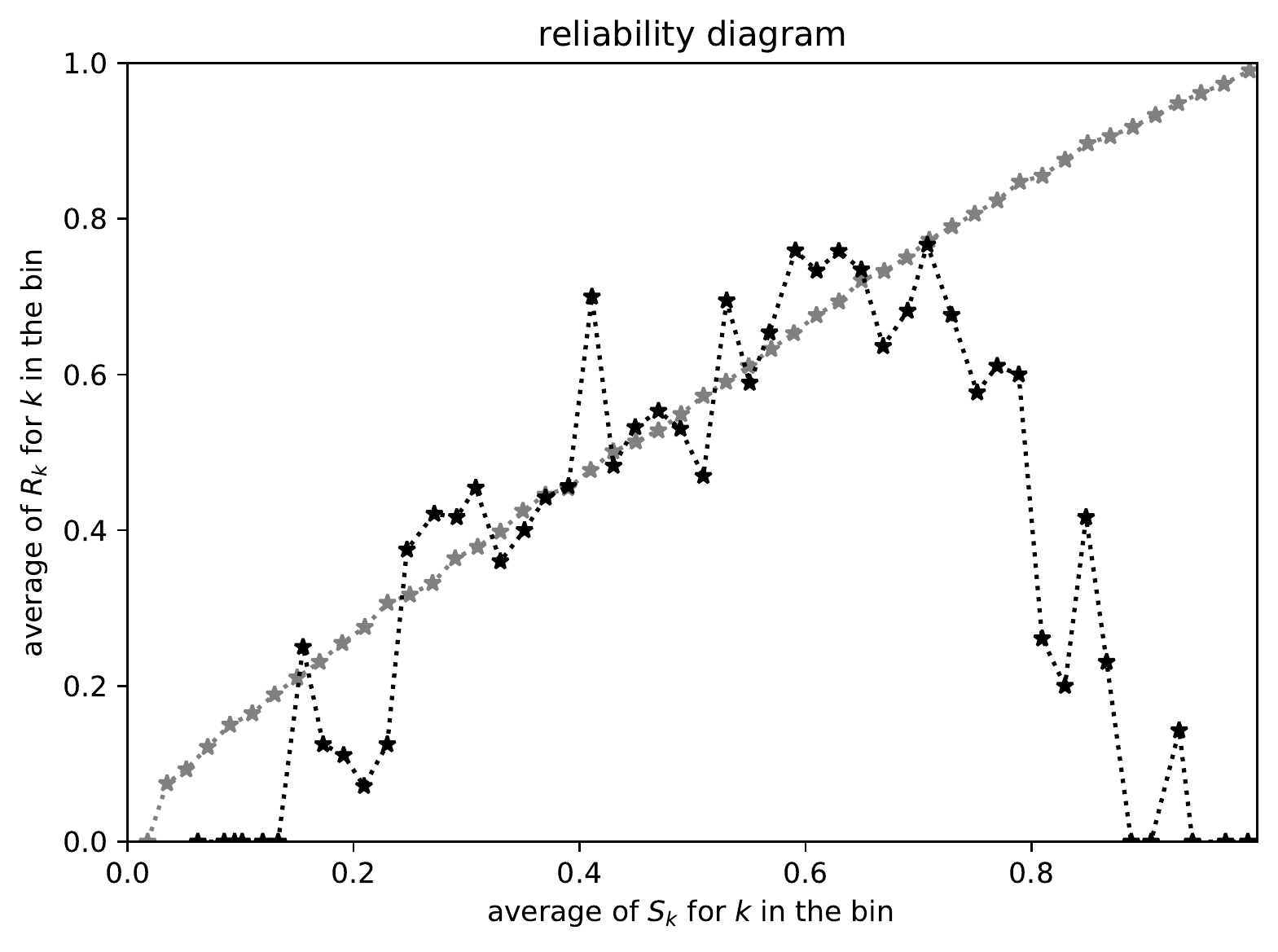}}

\end{centering}
\caption{Eskimo Dog or Husky, with scores being the probabilities;
         $n =$ 1,300; Kuiper's statistic is $0.08082 / \sigma = 6.363$,
         Kolmogorov's and Smirnov's is $0.06959 / \sigma = 5.478$.
As in Figure~\ref{eskimo-dog-husky-nll}, none of the reliability diagrams
with the same number of subpopulation scores per bin is able to smooth away
the irrelevant variations while simultaneously capturing the severe deviation
at the highest probabilities (however, the reliability diagram with 10 bins
that are equispaced in probabilities captures everything nicely).
The scalar summary statistics very successfully
detect the statistically highly significant deviation
of the subpopulation from the full population.
}
\label{eskimo-dog-husky-prob}
\end{figure}

\begin{figure}
\begin{centering}

\parbox{\imsize}{\includegraphics[width=\imsize]
       {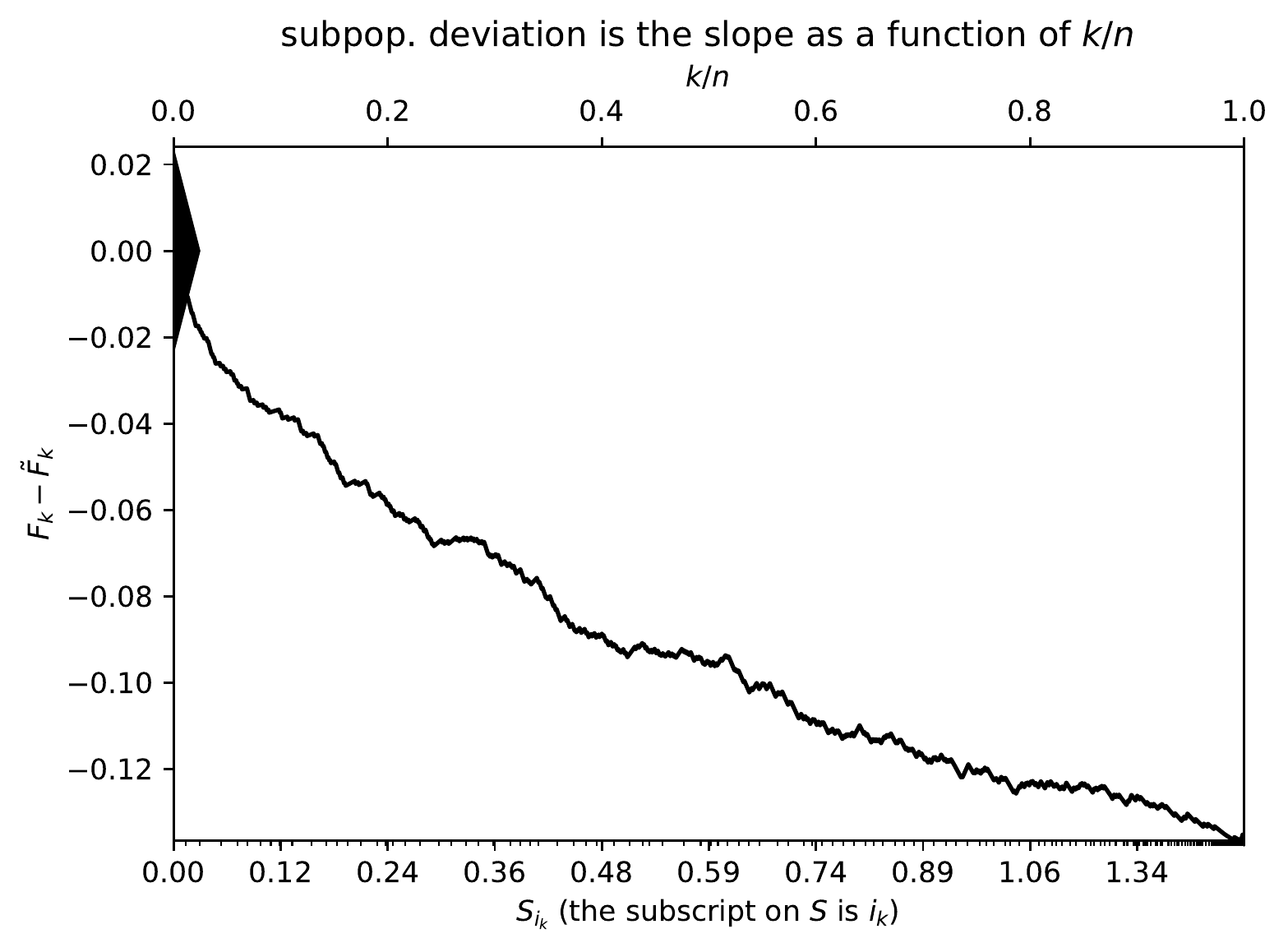}}
\quad\quad
\parbox{\imsize}{\includegraphics[width=\imsize]
       {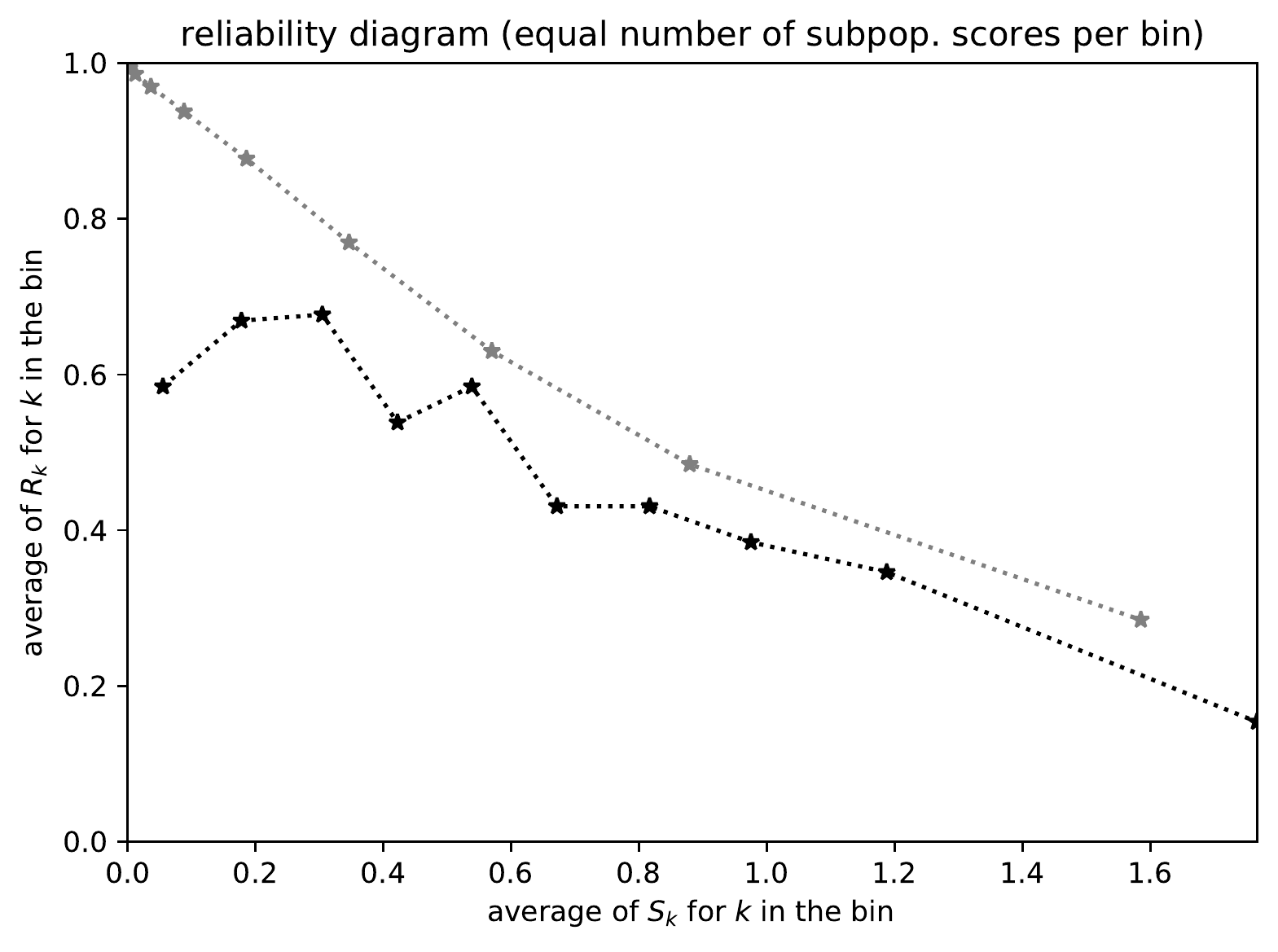}}

\vspace{\vertsep}

\parbox{\imsize}{\includegraphics[width=\imsize]
       {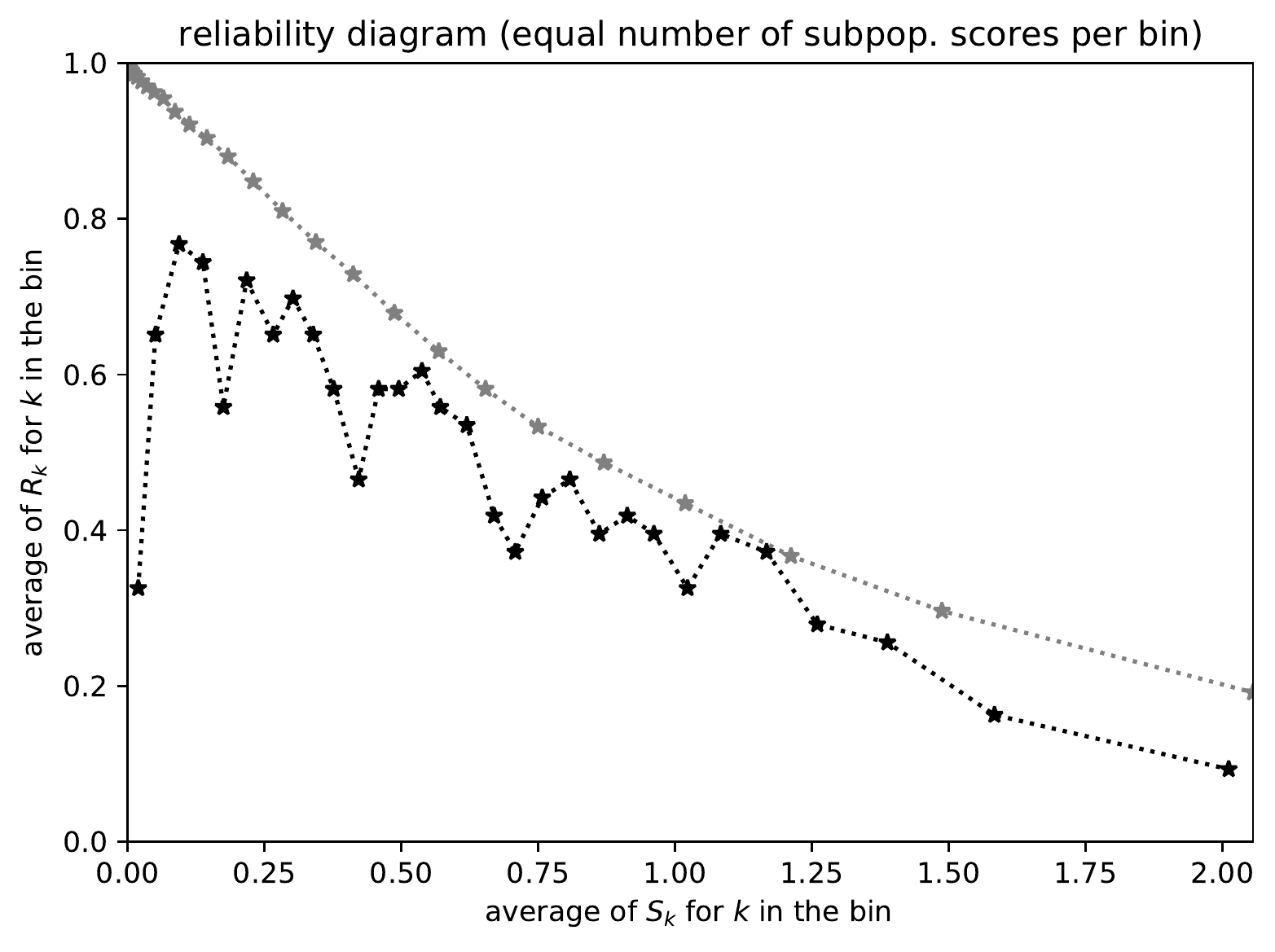}}
\quad\quad
\parbox{\imsize}{\includegraphics[width=\imsize]
       {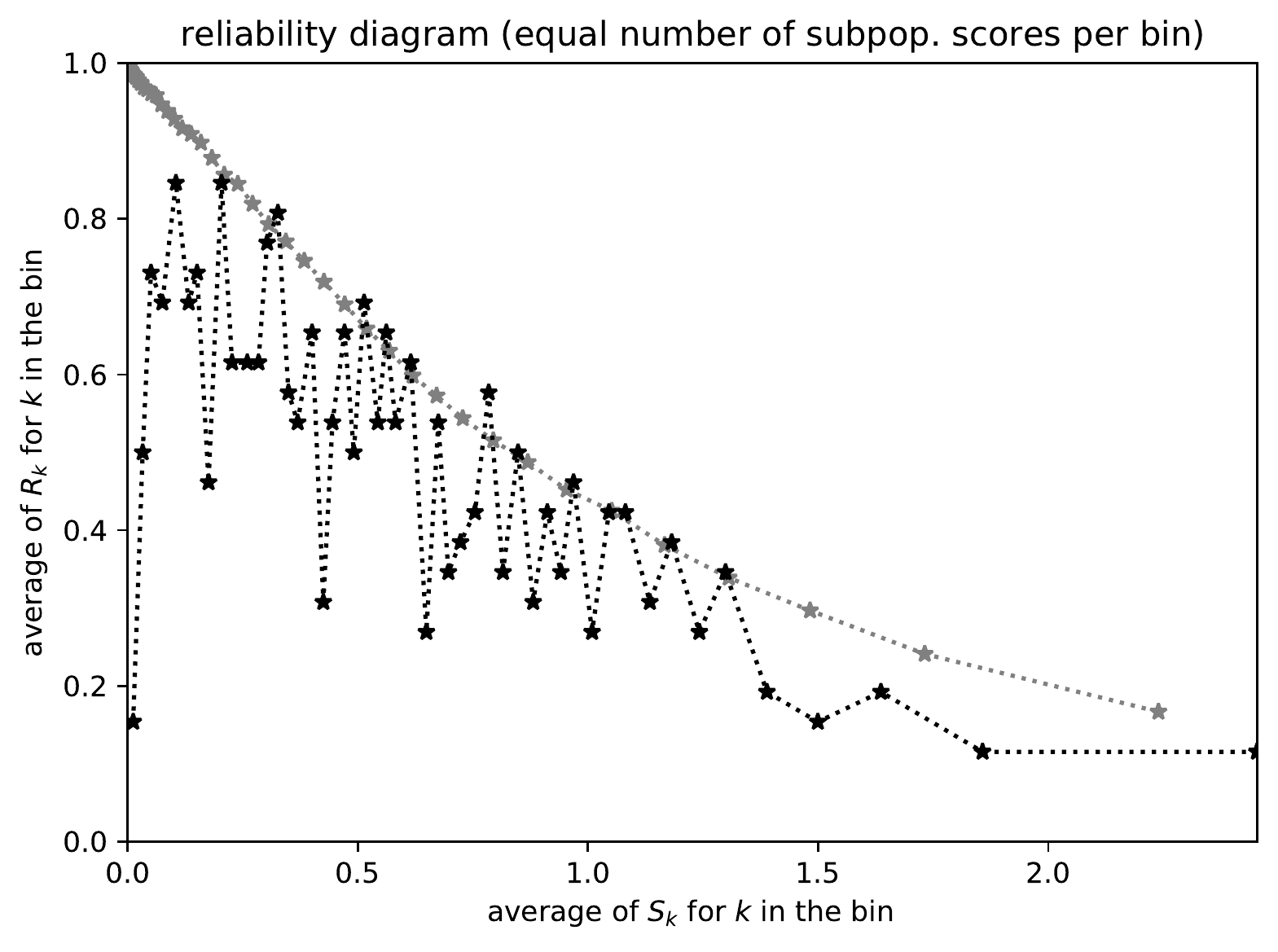}}

\vspace{\vertsep}

\parbox{\imsize}{\includegraphics[width=\imsize]
       {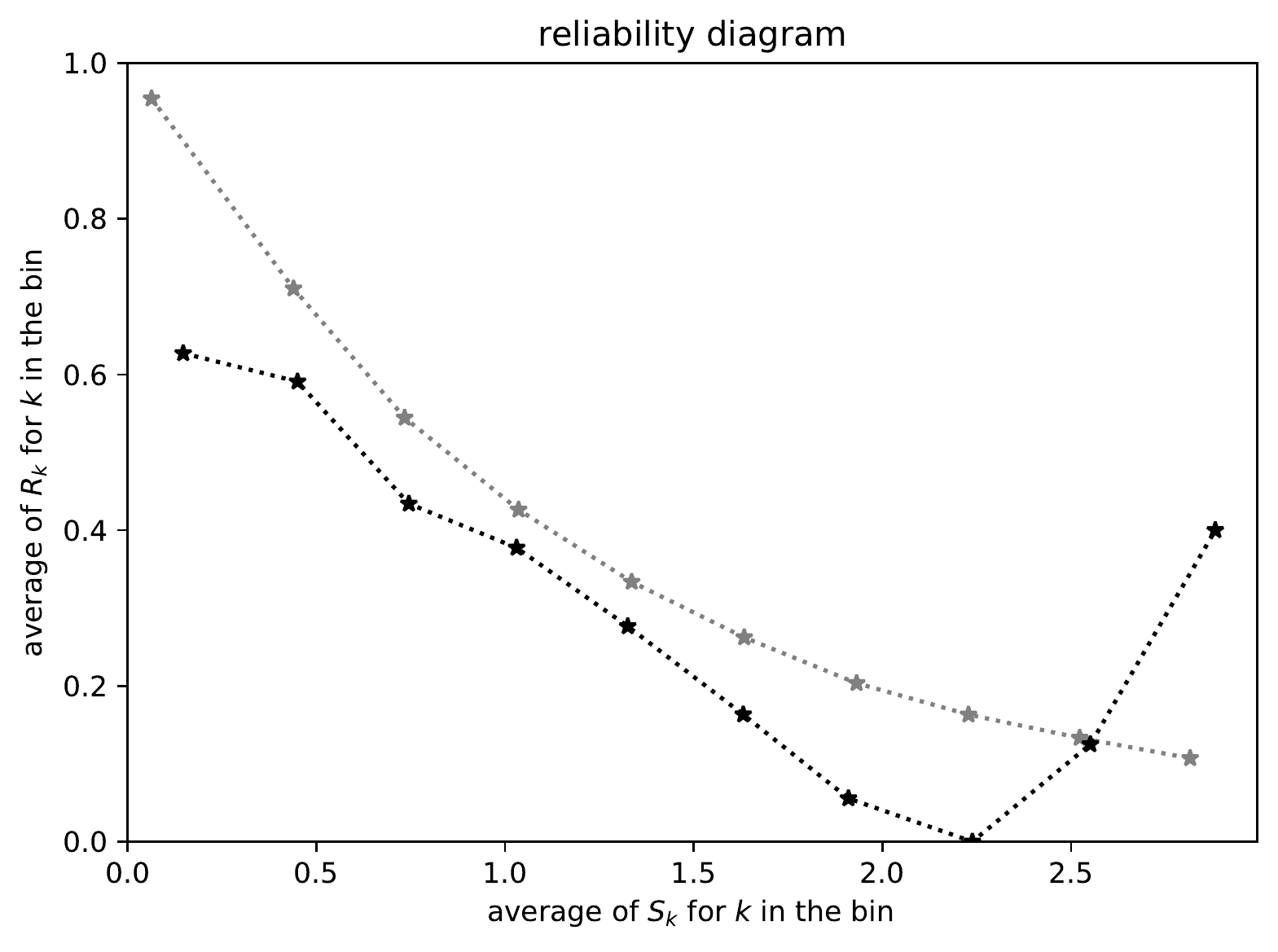}}
\quad\quad
\parbox{\imsize}{\includegraphics[width=\imsize]
       {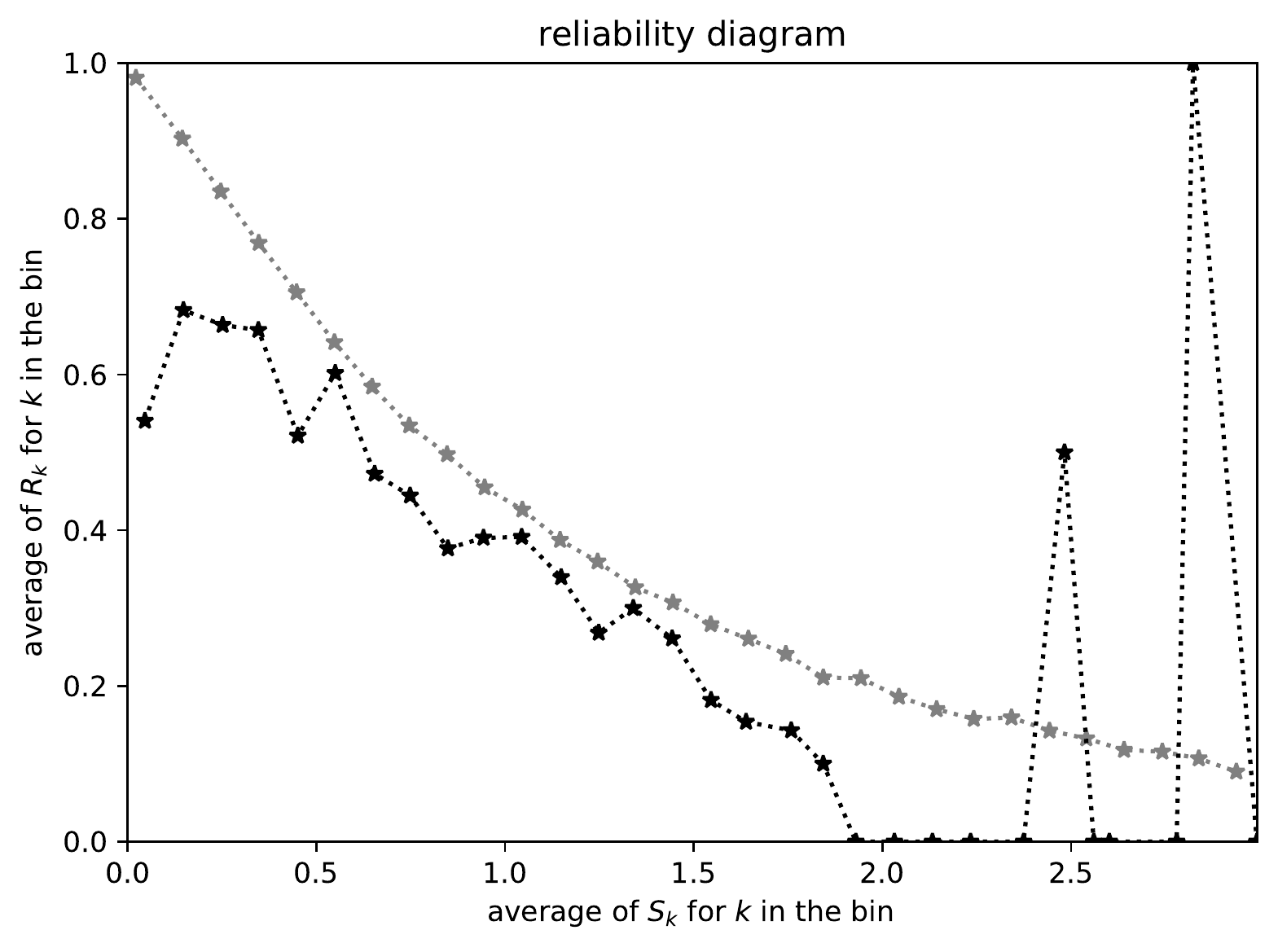}}

\end{centering}
\caption{Night snake (Hypsiglena torquata),
         with scores being the negative log-likelihoods;
         $n =$ 1,300; Kuiper's statistic is $0.1365 / \sigma = 11.35$,
         Kolmogorov's and Smirnov's is $0.1365 / \sigma = 11.35$.
Bins equispaced across either the subpopulation's observations or the scores
(where the scores are negative log-likelihoods in this figure)
cannot resolve the severe deviations at low scores
without being overly noisy elsewhere.
The scalar summary statistics extremely successfully
detect the statistically highly significant deviation
of the subpopulation from the full population.
}
\label{night-snake-Hypsiglena-torquata-nll}
\end{figure}

\begin{figure}
\begin{centering}

\parbox{\imsize}{\includegraphics[width=\imsize]
{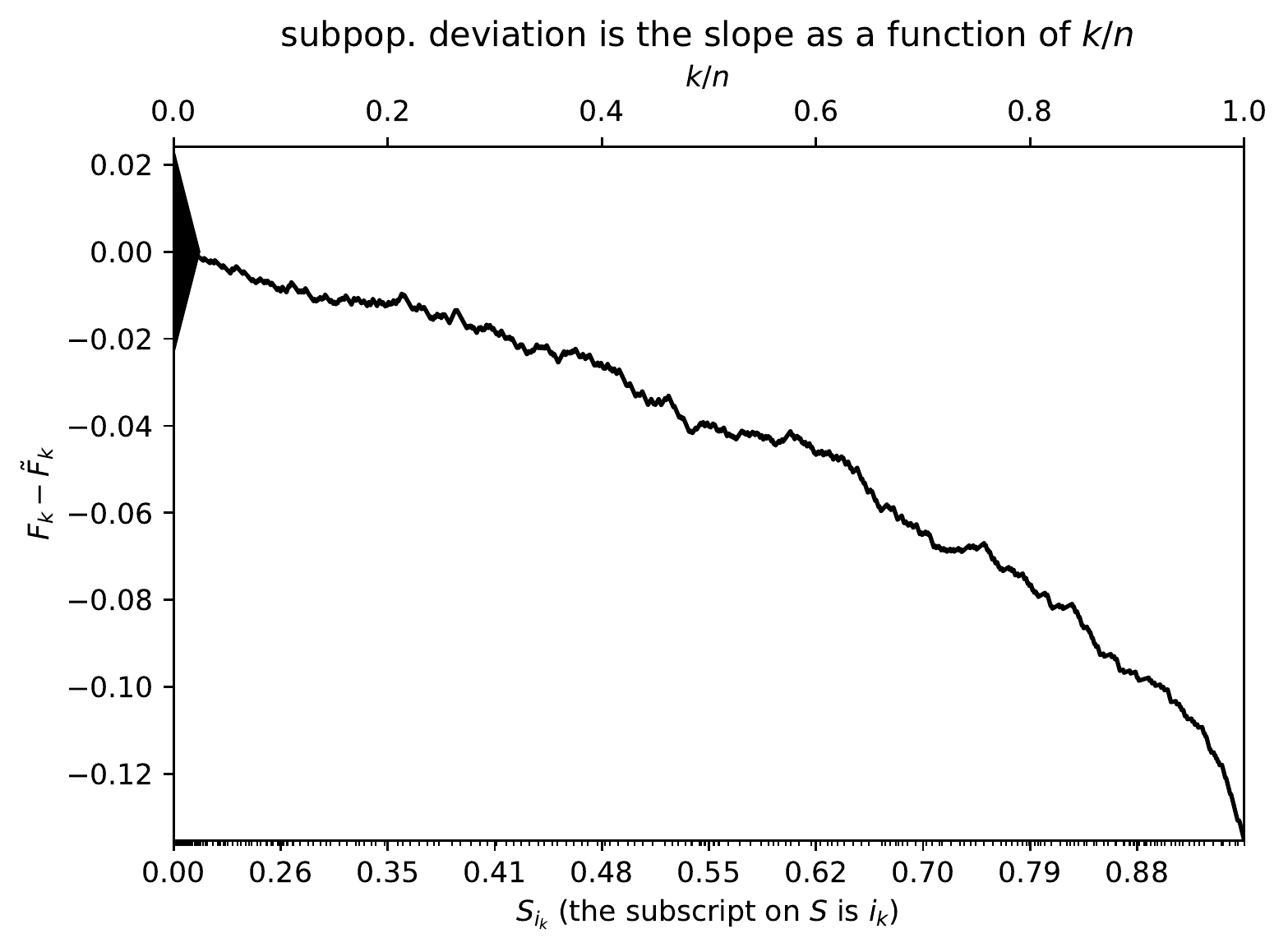}}
\quad\quad
\parbox{\imsize}{\includegraphics[width=\imsize]
{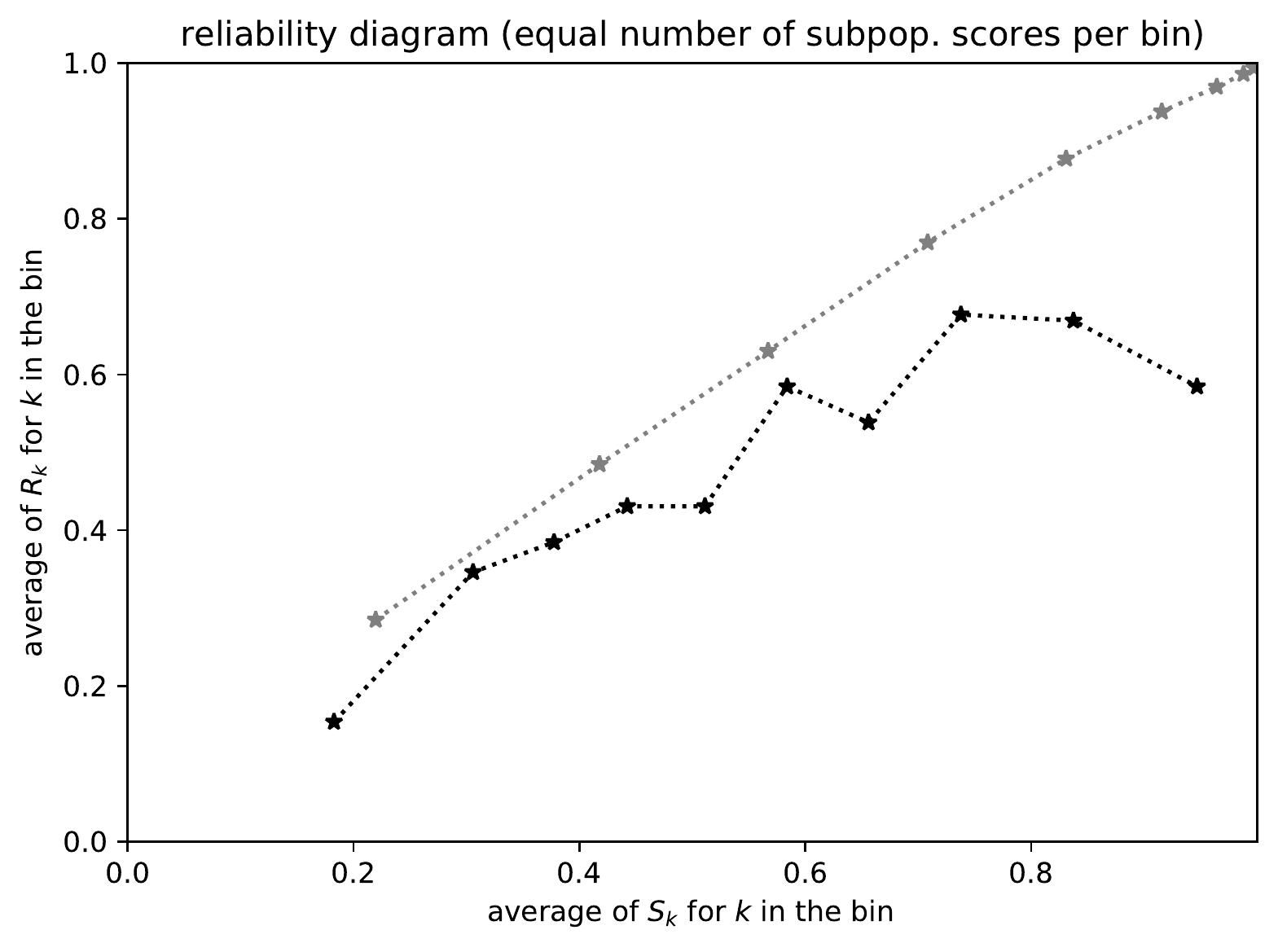}}

\vspace{\vertsep}

\parbox{\imsize}{\includegraphics[width=\imsize]
{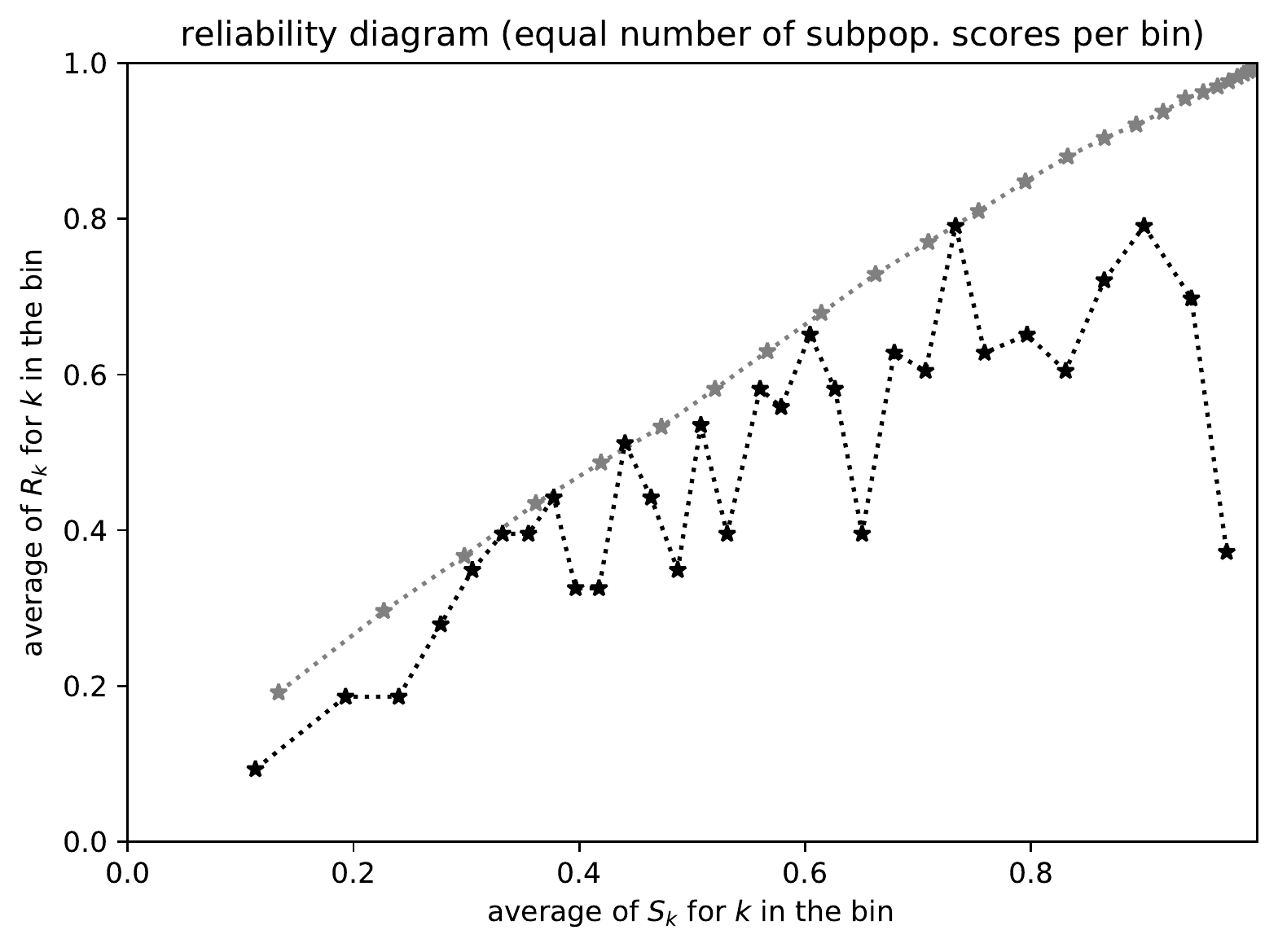}}
\quad\quad
\parbox{\imsize}{\includegraphics[width=\imsize]
{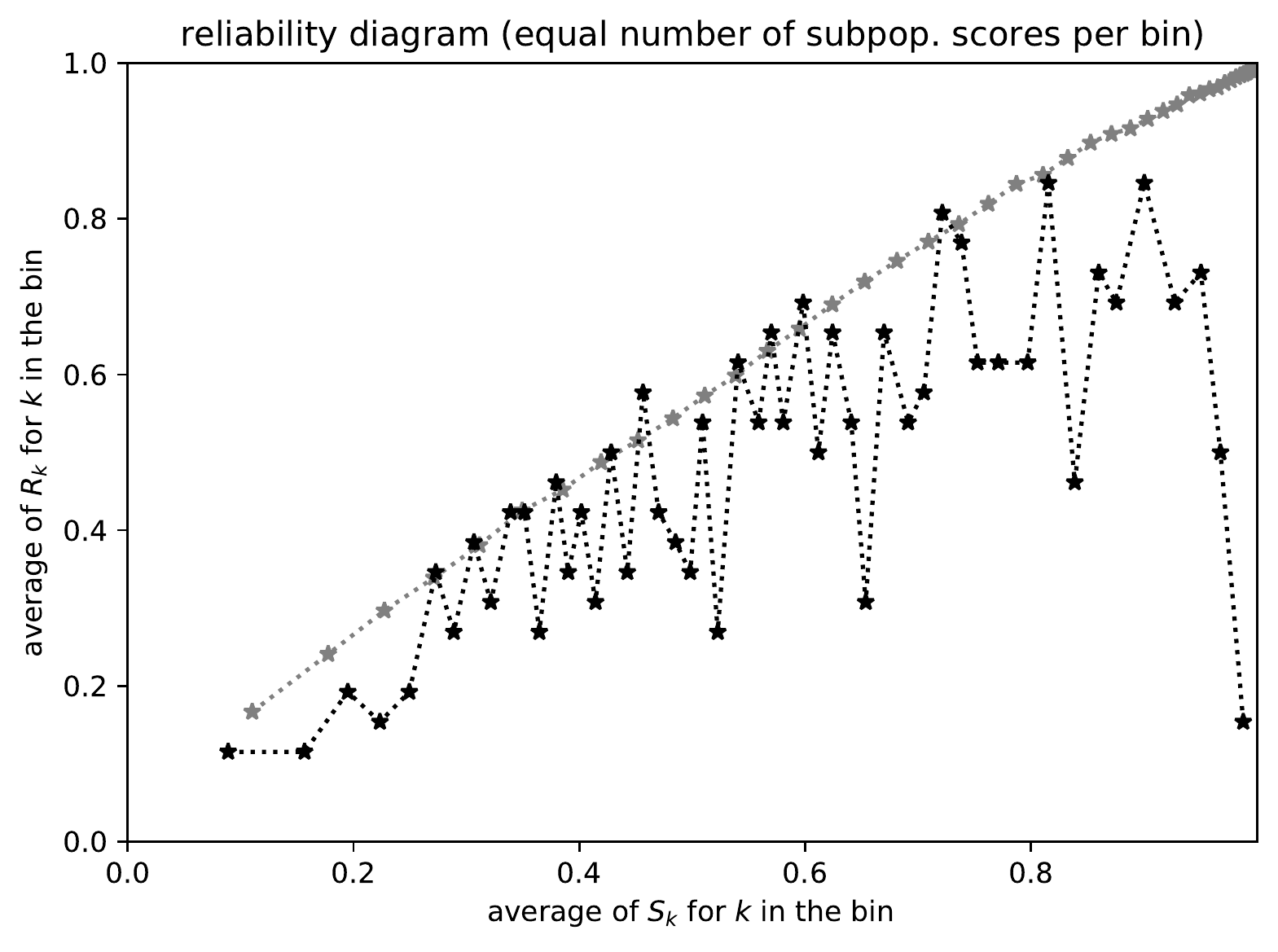}}

\vspace{\vertsep}

\parbox{\imsize}{\includegraphics[width=\imsize]
{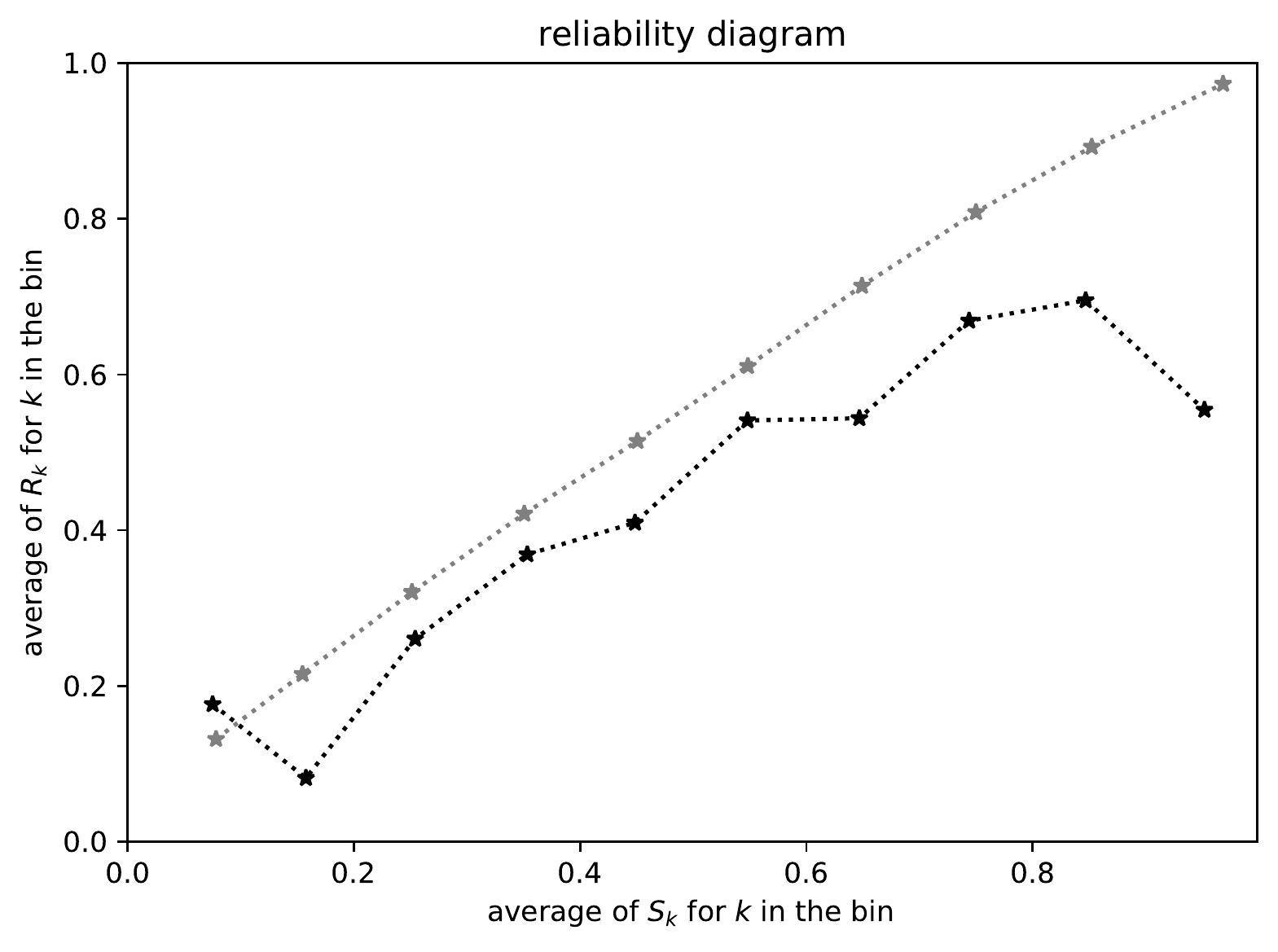}}
\quad\quad
\parbox{\imsize}{\includegraphics[width=\imsize]
{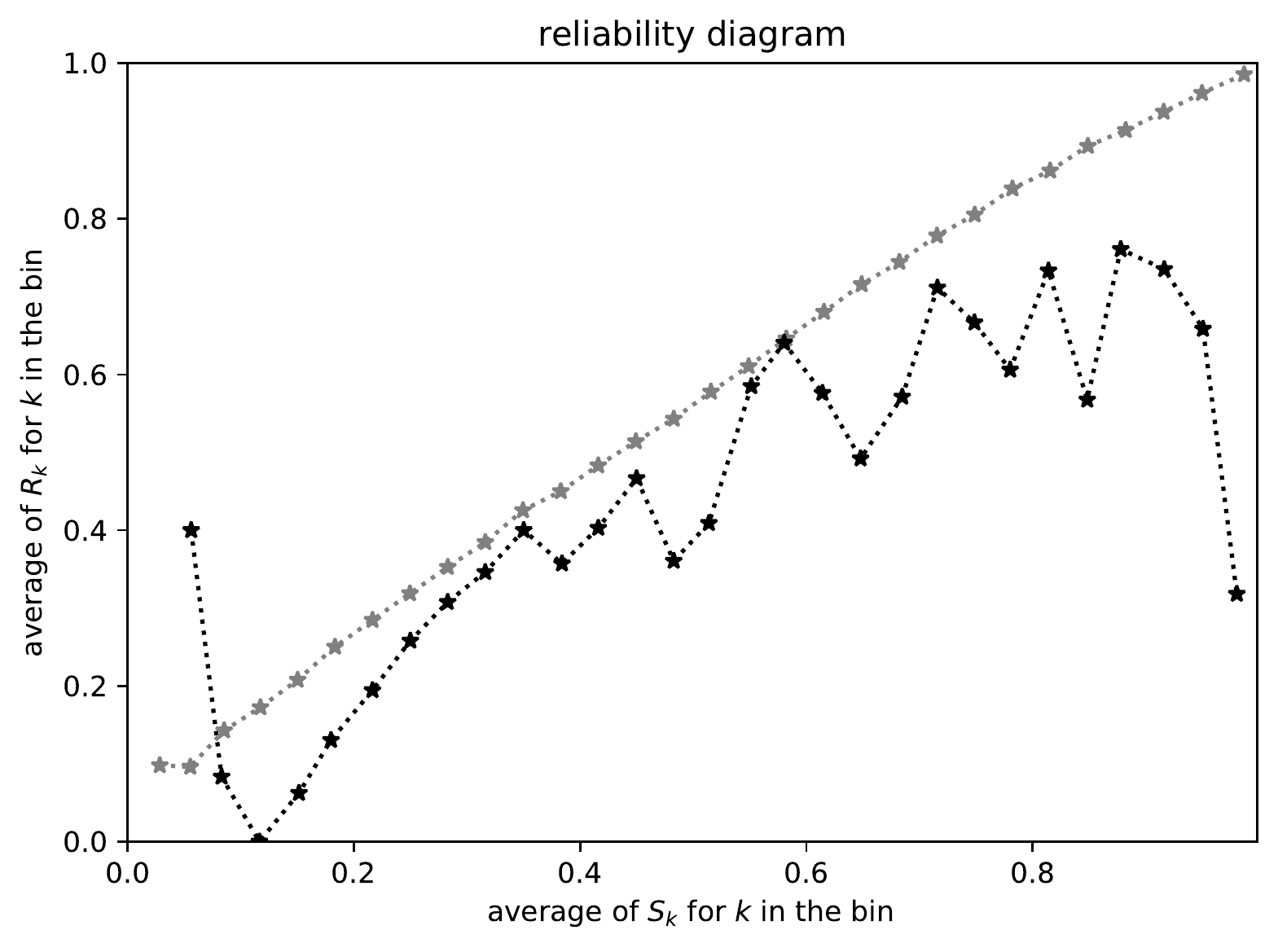}}

\end{centering}
\caption{Night snake (Hypsiglena torquata),
         with scores being the probabilities;
         $n =$ 1,300; Kuiper's statistic is $0.1365 / \sigma = 11.35$,
         Kolmogorov's and Smirnov's is $0.1353 / \sigma = 11.24$.
As in Figure~\ref{night-snake-Hypsiglena-torquata-nll},
uniform resolution in the binned plots
(whether with respect to observations or to scores) is insufficient.
The scalar summary statistics extremely successfully
detect the statistically highly significant deviation
of the subpopulation from the full population.
}
\label{night-snake-Hypsiglena-torquata-prob}
\end{figure}

\begin{figure}
\begin{centering}

\parbox{\imsize}{\includegraphics[width=\imsize]
{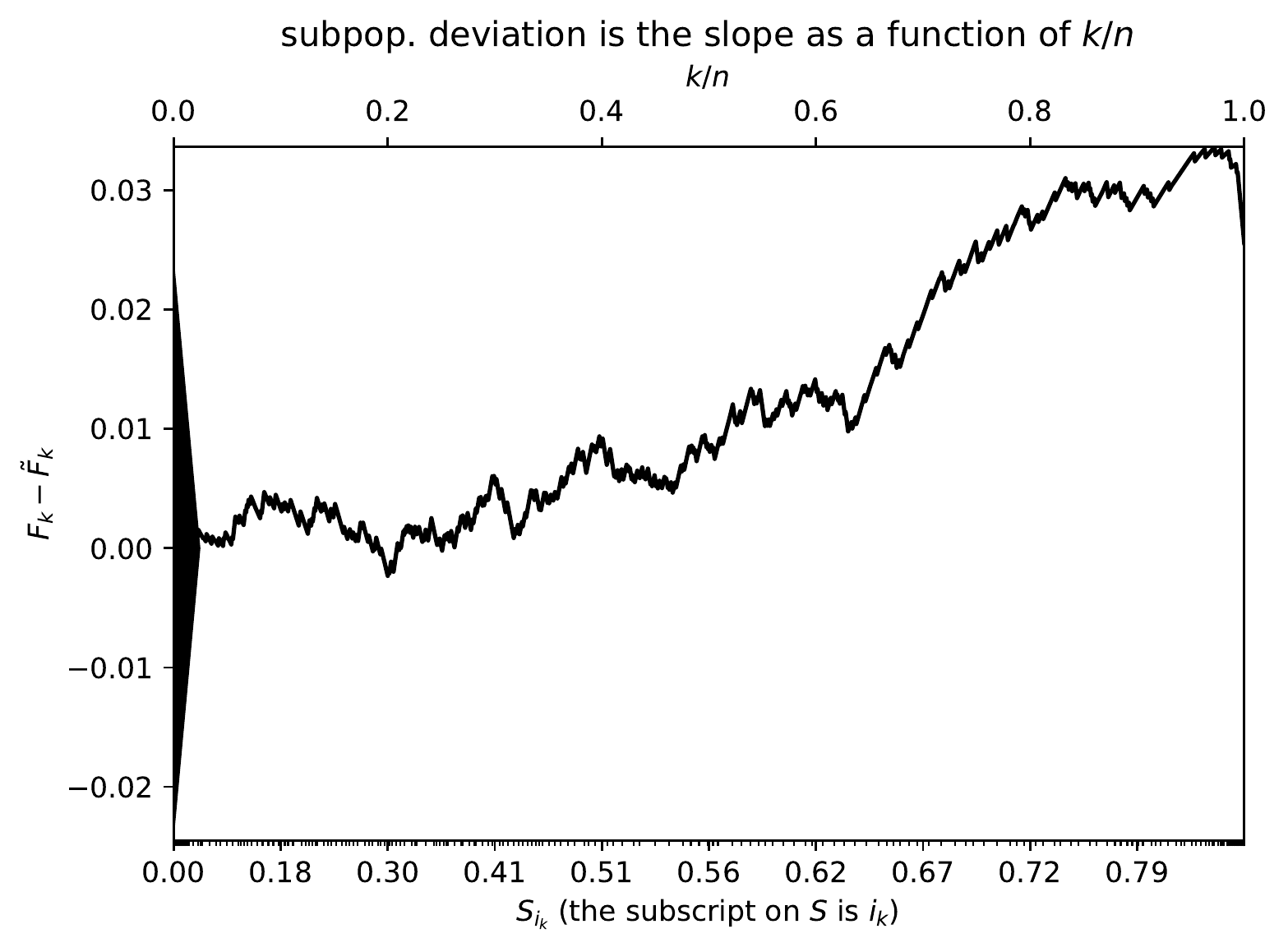}}
\quad\quad
\parbox{\imsize}{\includegraphics[width=\imsize]
{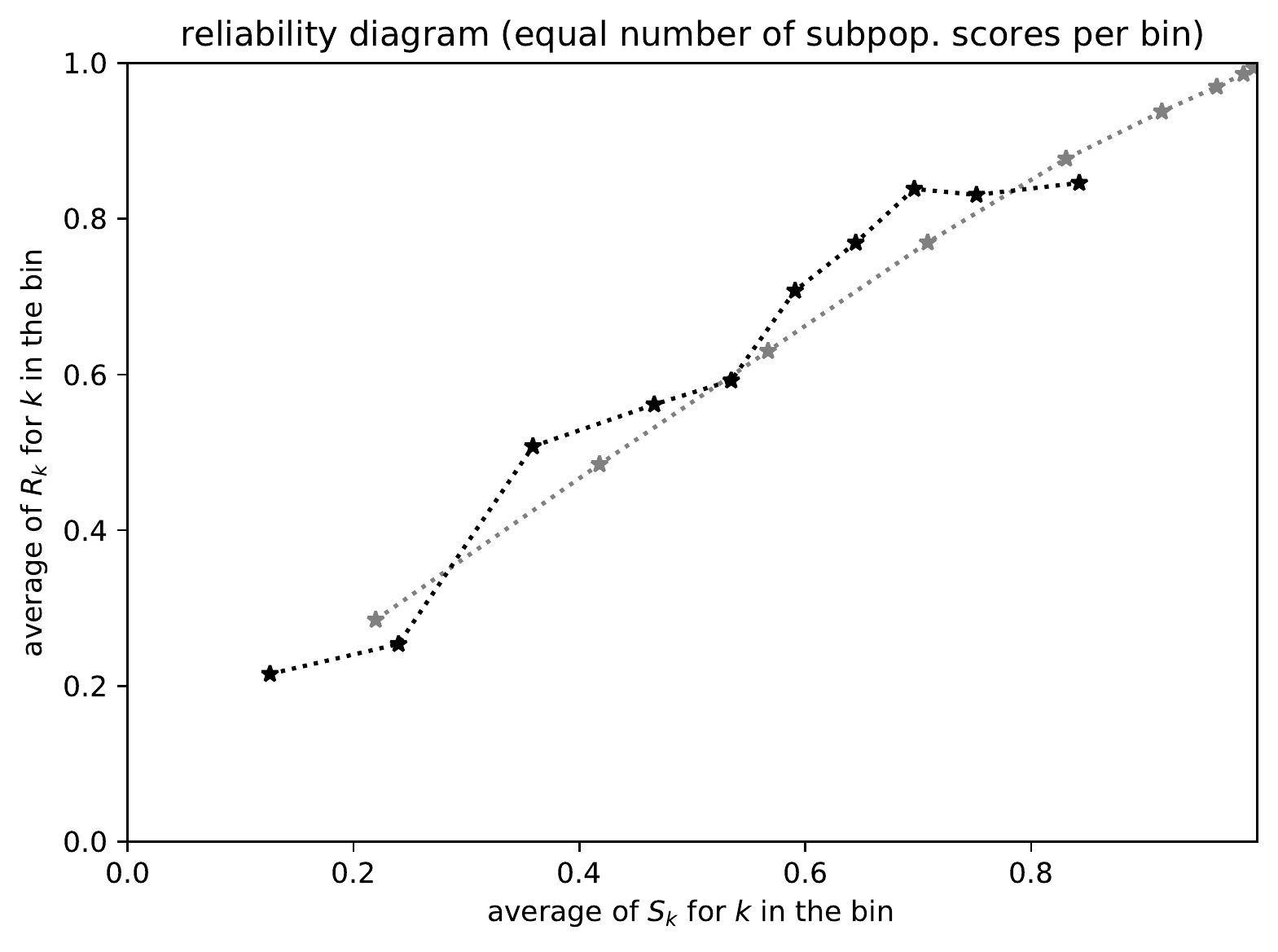}}

\vspace{\vertsep}

\parbox{\imsize}{\includegraphics[width=\imsize]
{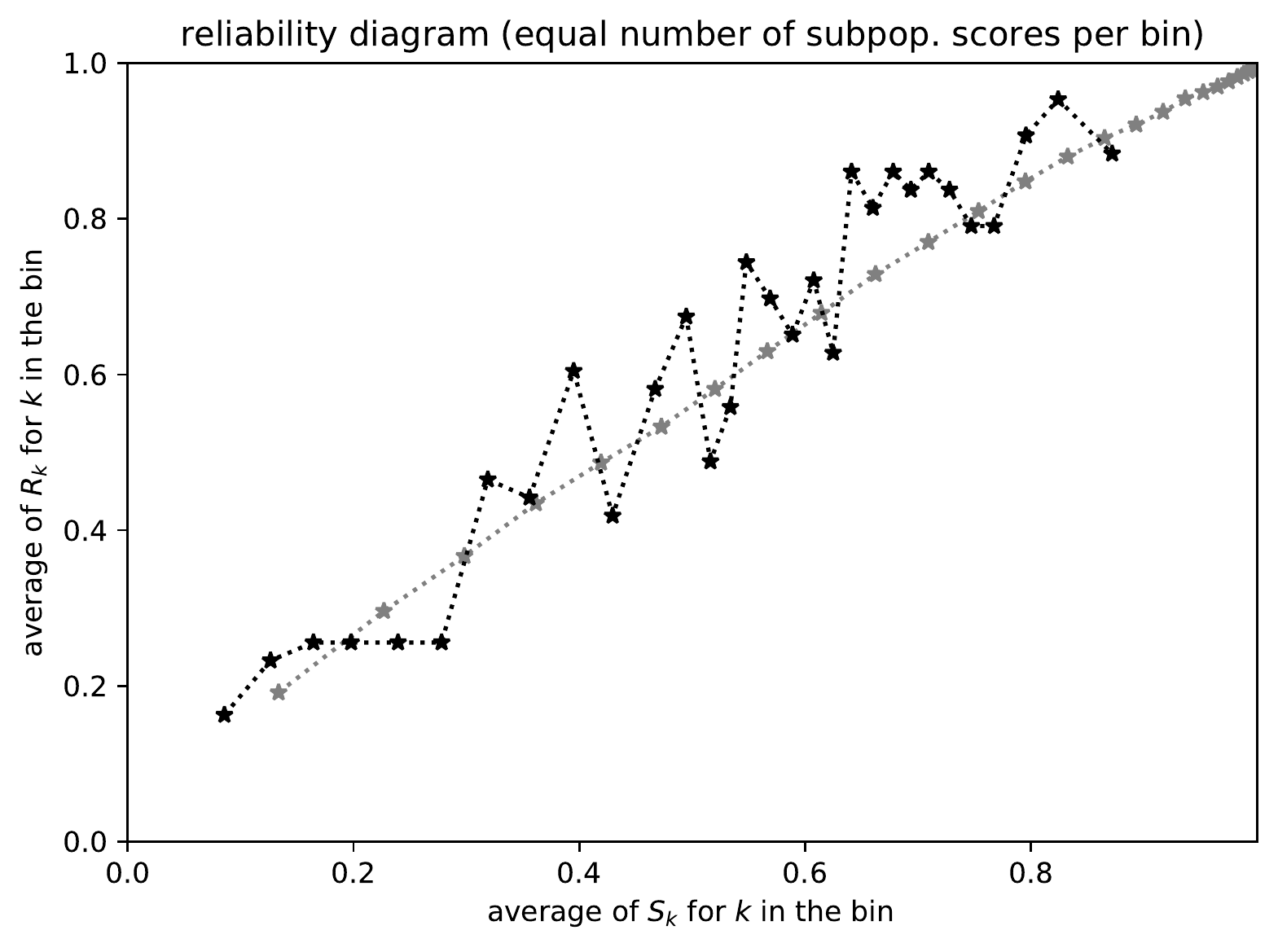}}
\quad\quad
\parbox{\imsize}{\includegraphics[width=\imsize]
{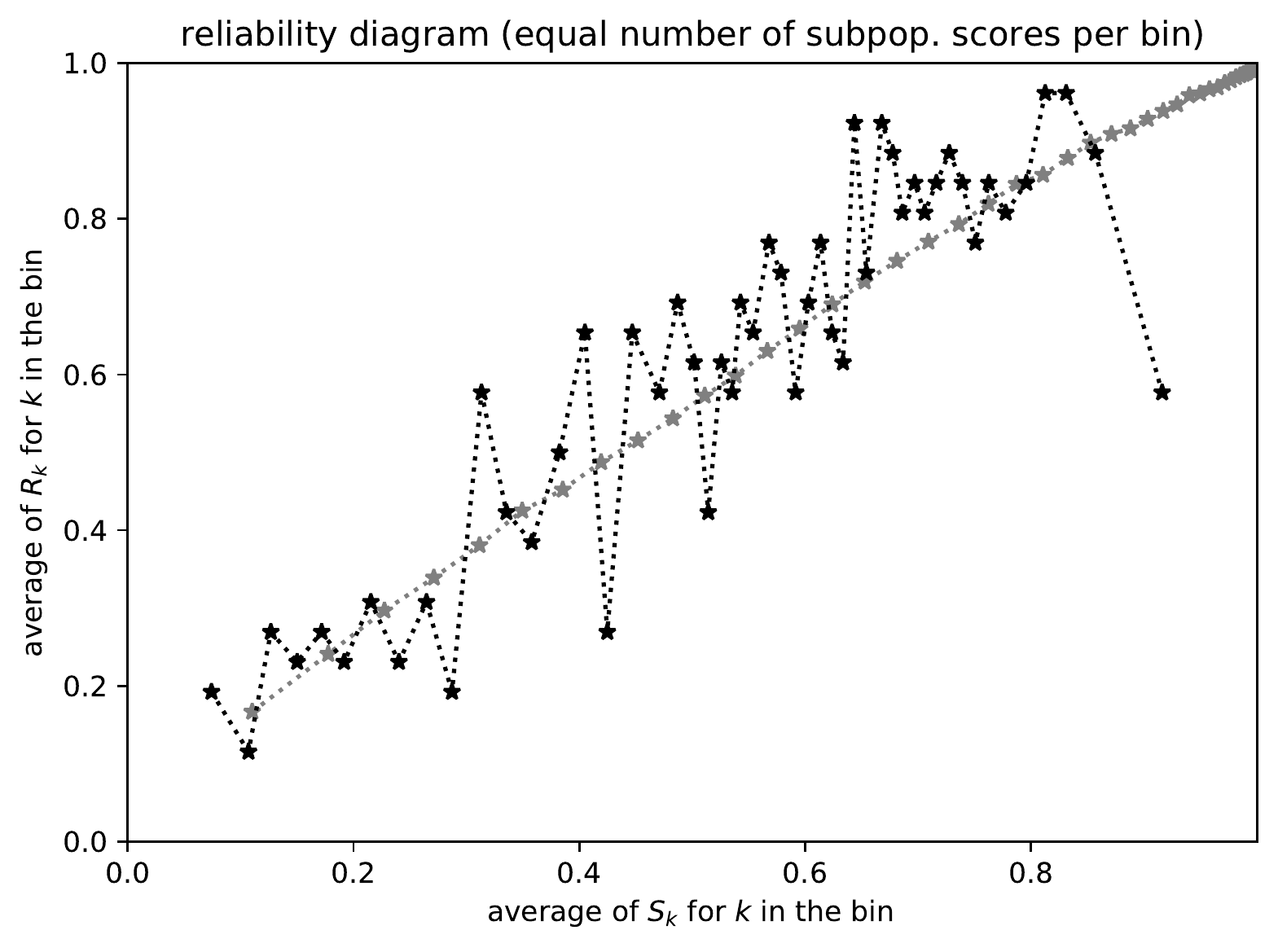}}

\end{centering}
\caption{Sunglasses, with scores being the probabilities;
         $n =$ 1,300; Kuiper's statistic is $0.03597 / \sigma = 2.935$,
         Kolmogorov's and Smirnov's is $0.03365 / \sigma = 2.745$.
The reliability diagrams with only 10 or 30 bins each
fail to resolve the very high deviation
for probabilities greater than 0.9 (unlike the diagram with 50 bins) ---
the smaller numbers of bins average away interesting behavior, without warning.
The scalar summary statistics detect some statistically significant deviation,
yet both are blind to the serious deviation for the highest scores
that the plot of cumulative differences displays prominently;
the steep drop at the highest scores in the cumulative plot
has little to no effect on the Kolmogorov-Smirnov or Kuiper metrics,
unfortunately.
}
\label{sunglasses-dark-glasses-shades-prob}
\end{figure}

\begin{figure}
\begin{centering}

\parbox{\imsize}{\includegraphics[width=\imsize]
       {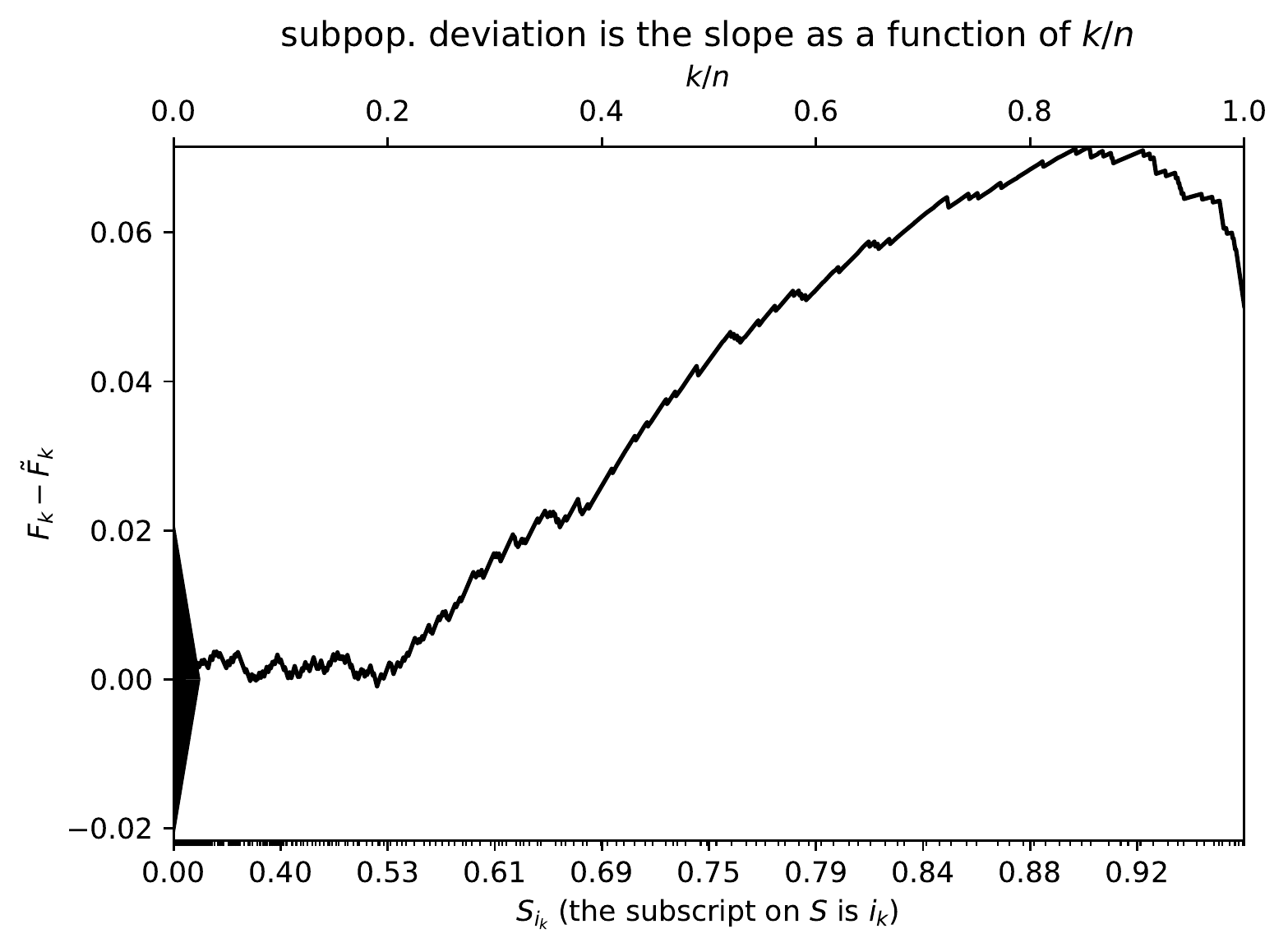}}
\quad\quad
\parbox{\imsize}{\includegraphics[width=\imsize]
       {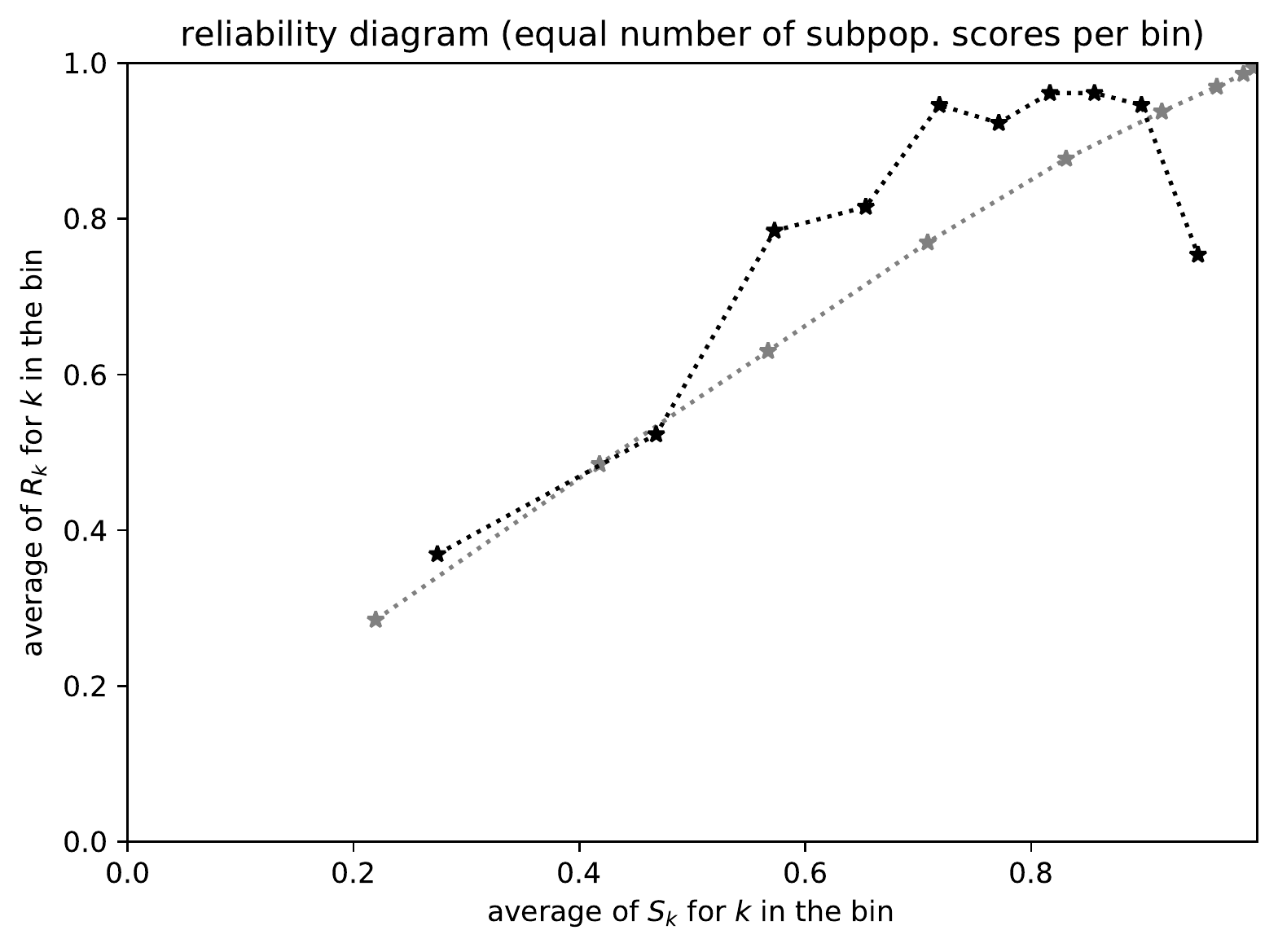}}

\vspace{\vertsep}

\parbox{\imsize}{\includegraphics[width=\imsize]
       {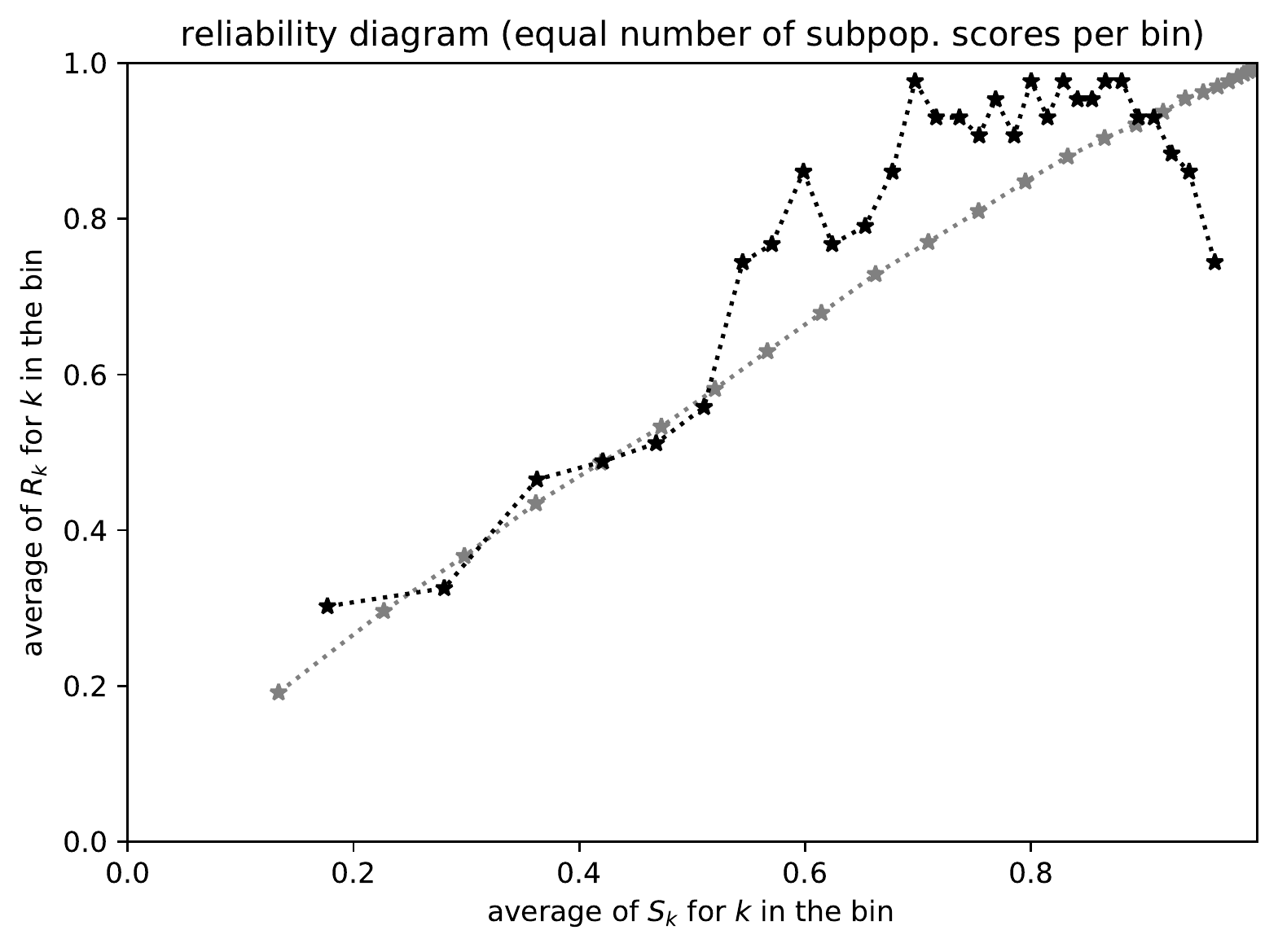}}
\quad\quad
\parbox{\imsize}{\includegraphics[width=\imsize]
       {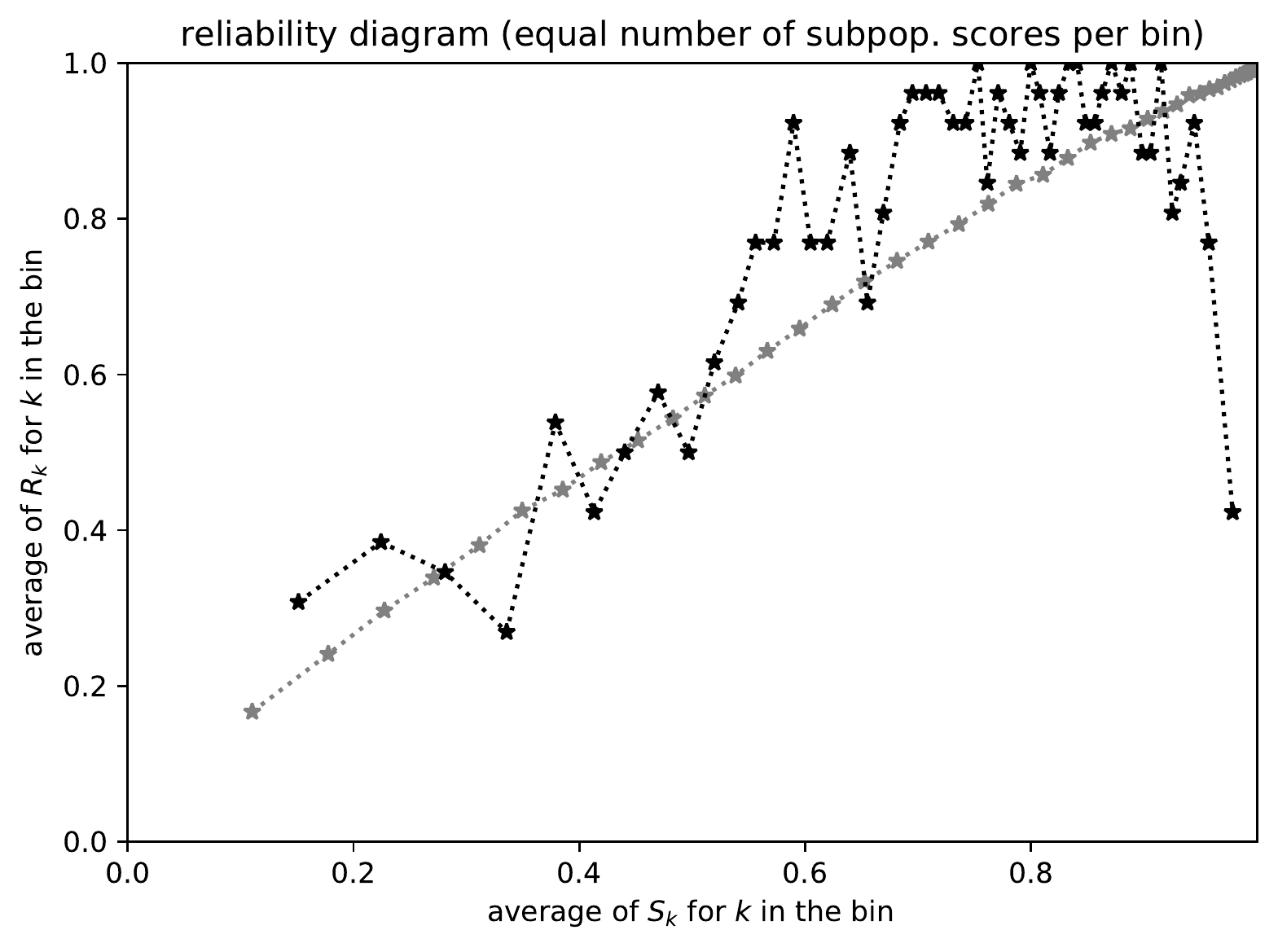}}

\end{centering}
\caption{Wild boar (Sus scrofa), with scores being the probabilities;
         $n =$ 1,300; Kuiper's statistic is $0.07244 / \sigma = 6.688$,
         Kolmogorov's and Smirnov's is $0.07149 / \sigma = 6.600$.
The behavior in the present figure at the highest probabilities
is similar to that in Figure~\ref{sunglasses-dark-glasses-shades-prob}
--- the reliability diagrams with only 10 or 30 bins each
underplay the very high deviation for the highest probabilities,
smoothing away big deviation without warning.
As in Figure~\ref{sunglasses-dark-glasses-shades-prob},
the scalar summary statistics detect statistically significant deviation,
while both are blind to the significant deviation for the highest scores
that the plot of cumulative differences highlights;
the steep drop at the highest scores in the cumulative plot
has essentially no impact on the Kolmogorov-Smirnov or Kuiper metrics,
unfortunately.
}
\label{wild-boar-boar-Sus-scrofa-prob}
\end{figure}

\begin{figure}
\begin{centering}

\parbox{\imsize}{\includegraphics[width=\imsize]
{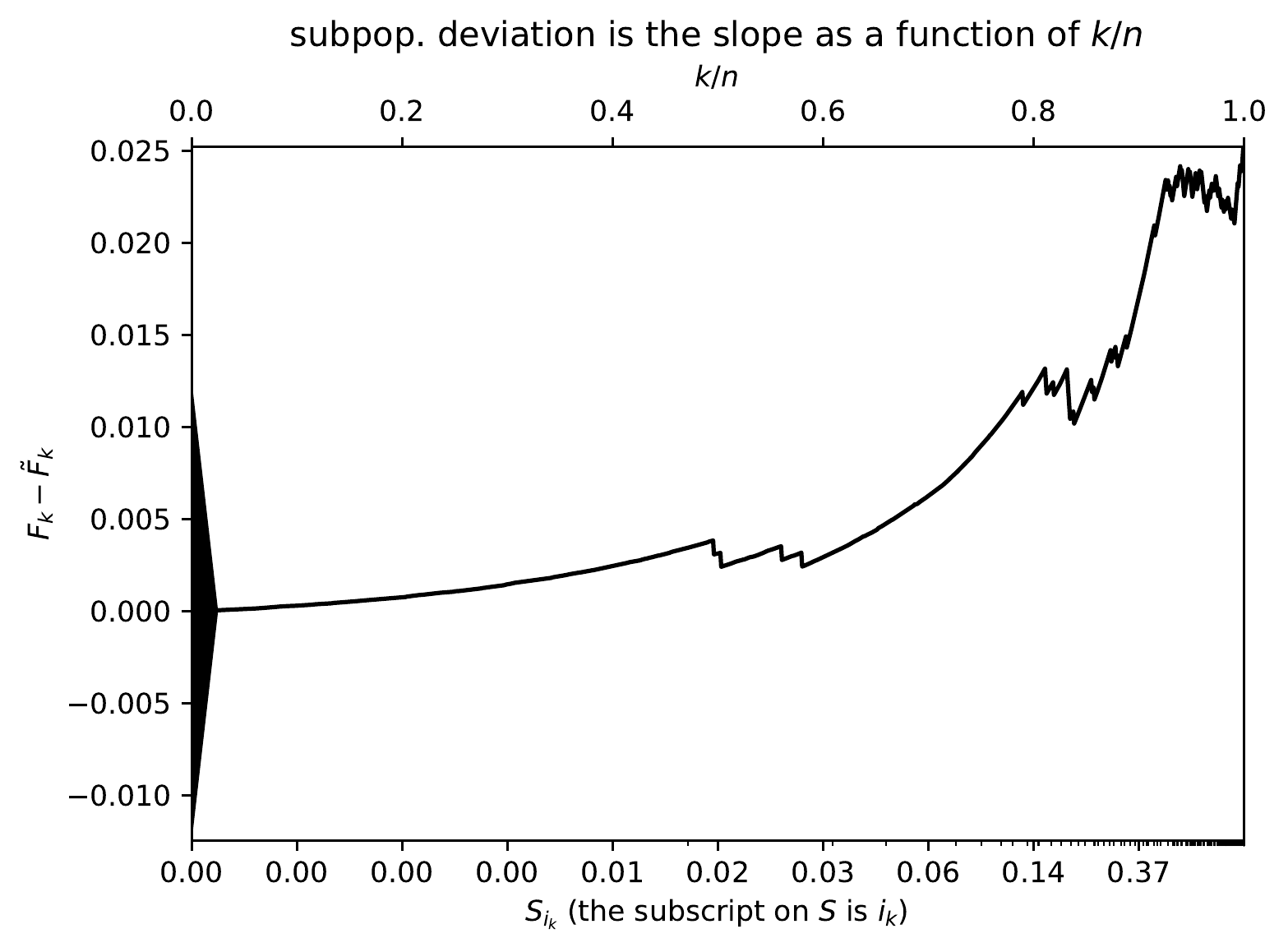}}
\quad\quad
\parbox{\imsize}{\includegraphics[width=\imsize]
{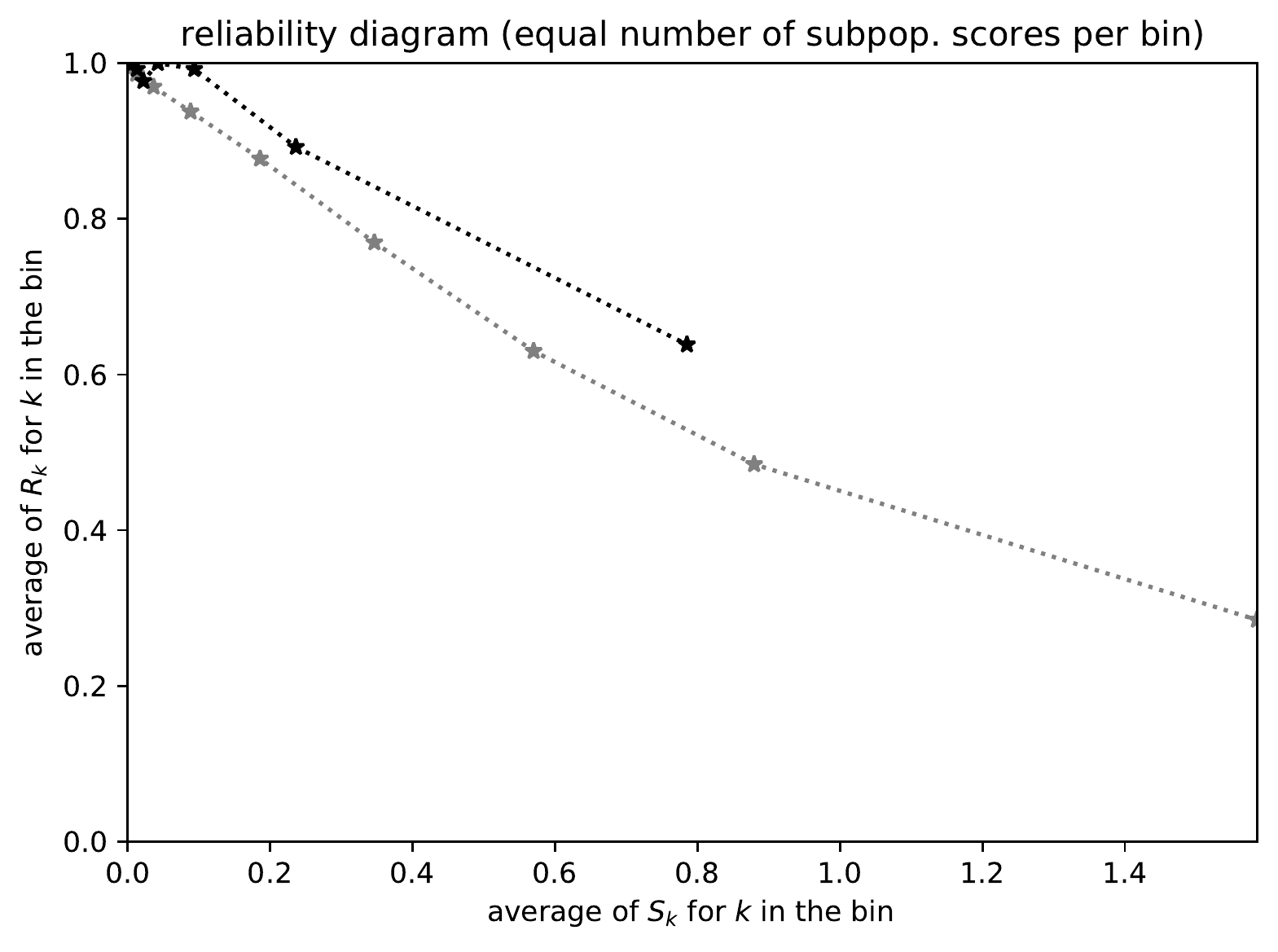}}

\vspace{\vertsep}

\parbox{\imsize}{\includegraphics[width=\imsize]
{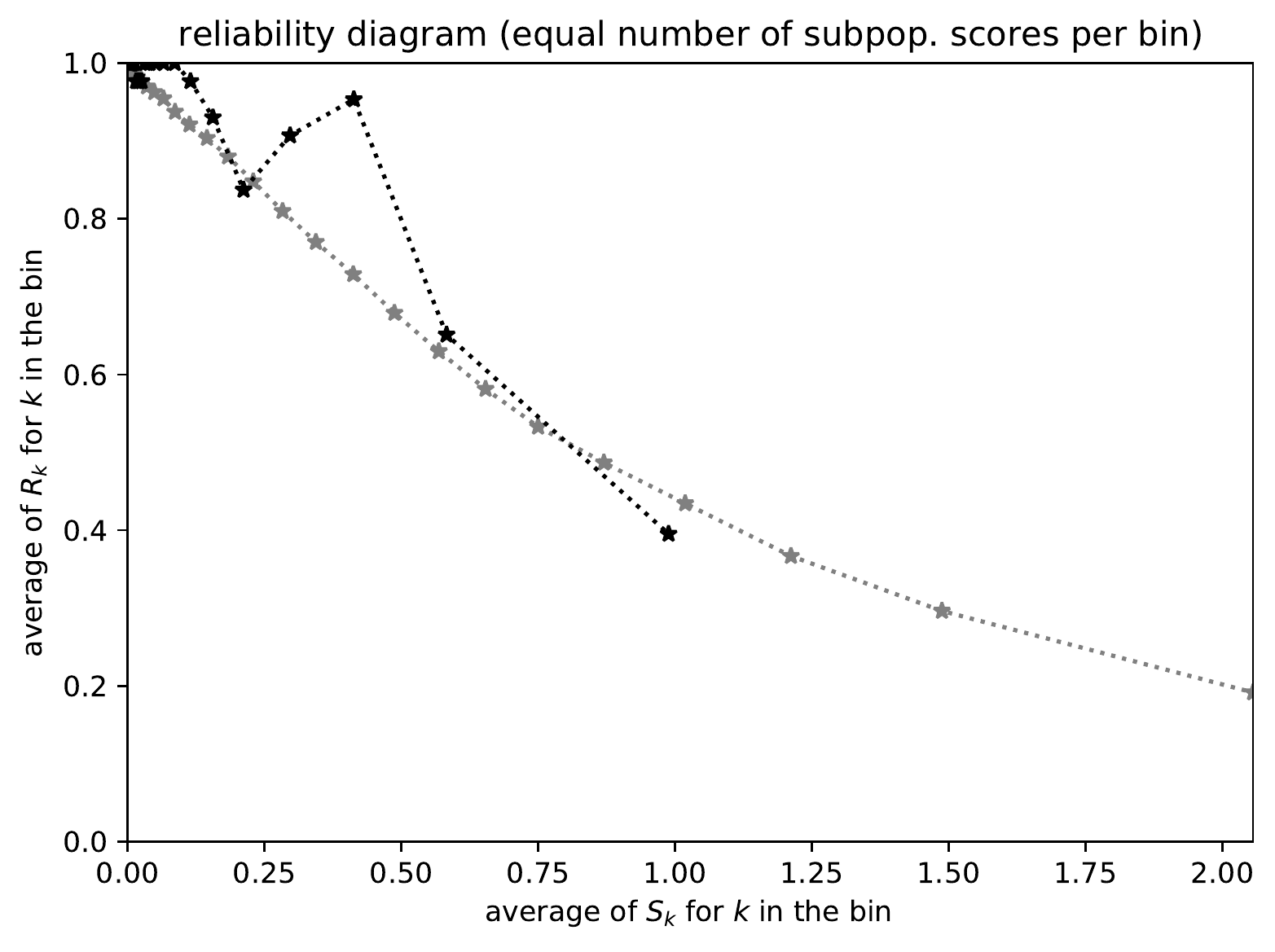}}
\quad\quad
\parbox{\imsize}{\includegraphics[width=\imsize]
{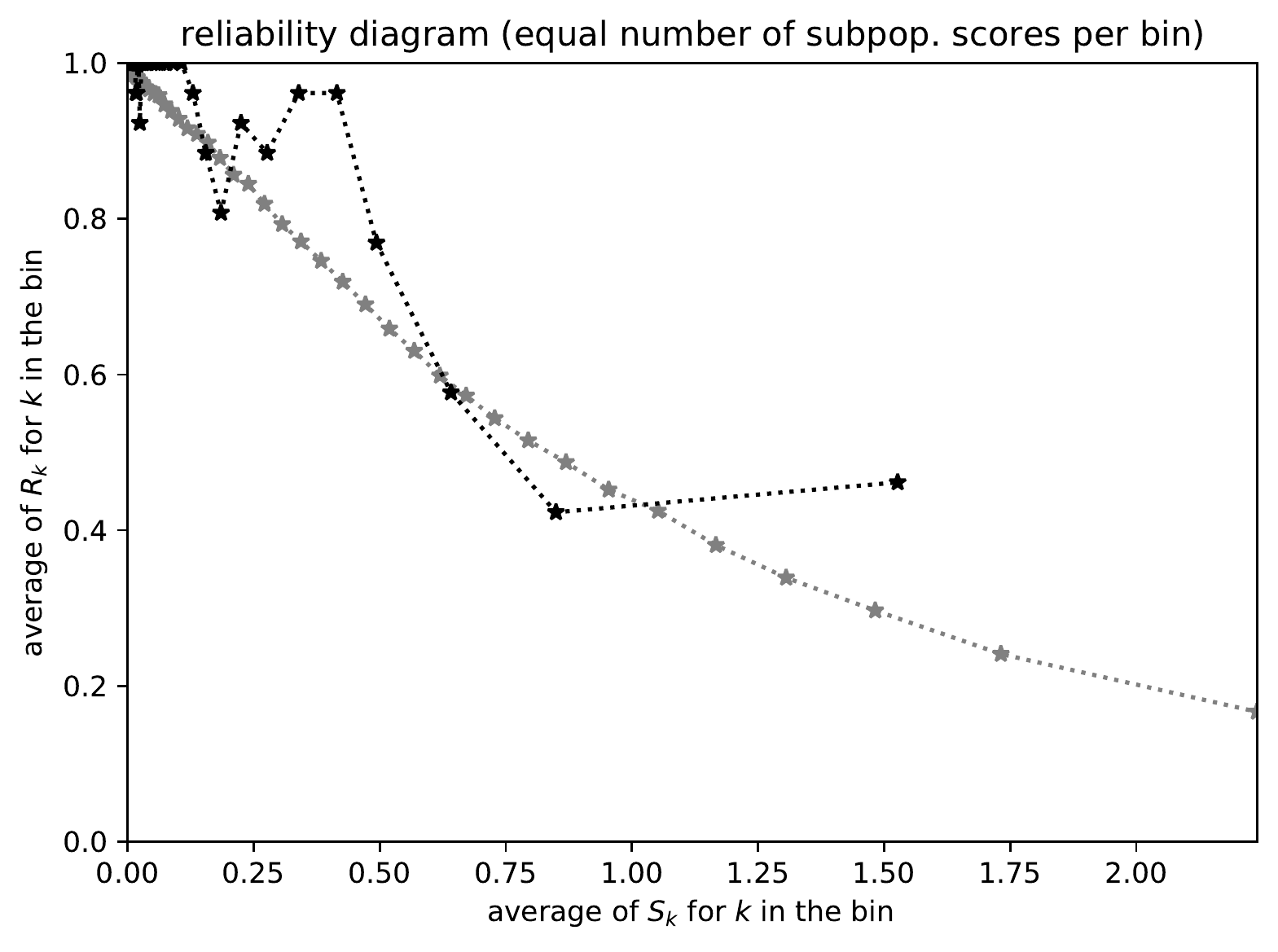}}

\end{centering}
\caption{Cheetah (Acinonyx jubatus),
         with scores being the negative log-likelihoods;
         $n =$ 1,300; Kuiper's statistic is $0.02520 / \sigma = 4.043$,
         Kolmogorov's and Smirnov's is $0.02520 / \sigma = 4.043$.
Only the reliability diagram with 50 bins
detects the high deviation for negative log-likelihoods
greater than 1.25; the diagrams with only 10 or 30 bins give no indication
of the substantial deviation. 30 bins or less are clearly insufficient
to resolve the significant deviation, but that is clear only upon comparison
with the cumulative plot or with diagrams having many bins.
The scalar summary statistics indicate statistically significant deviation
of the subpopulation from the full population, in accord with the plot
of cumulative differences.
}
\label{cheetah-chetah-Acinonyx-jubatus-nll}
\end{figure}

\begin{figure}
\begin{centering}

\parbox{\imsize}{\includegraphics[width=\imsize]
{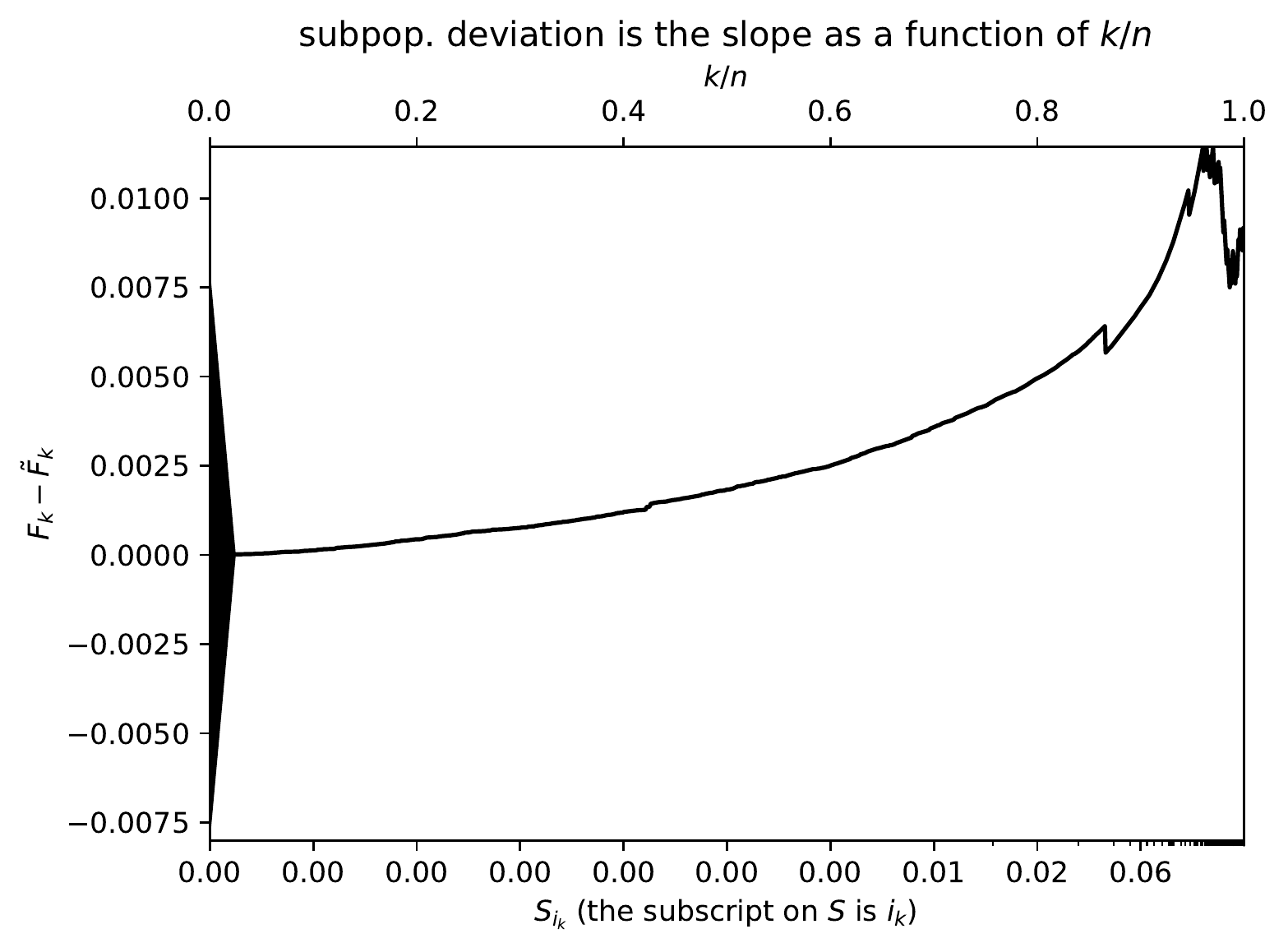}}
\quad\quad
\parbox{\imsize}{\includegraphics[width=\imsize]
{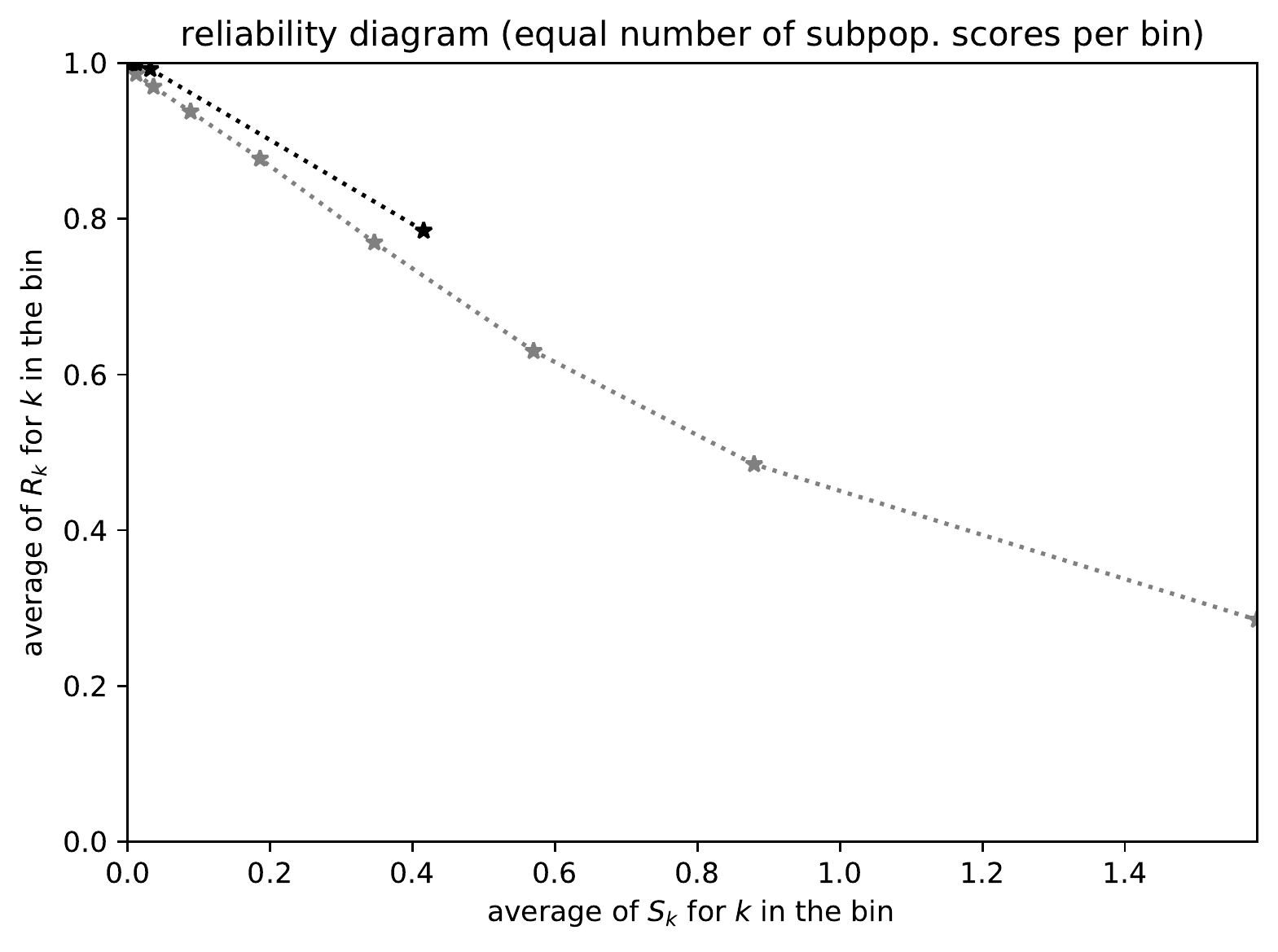}}

\vspace{\vertsep}

\parbox{\imsize}{\includegraphics[width=\imsize]
{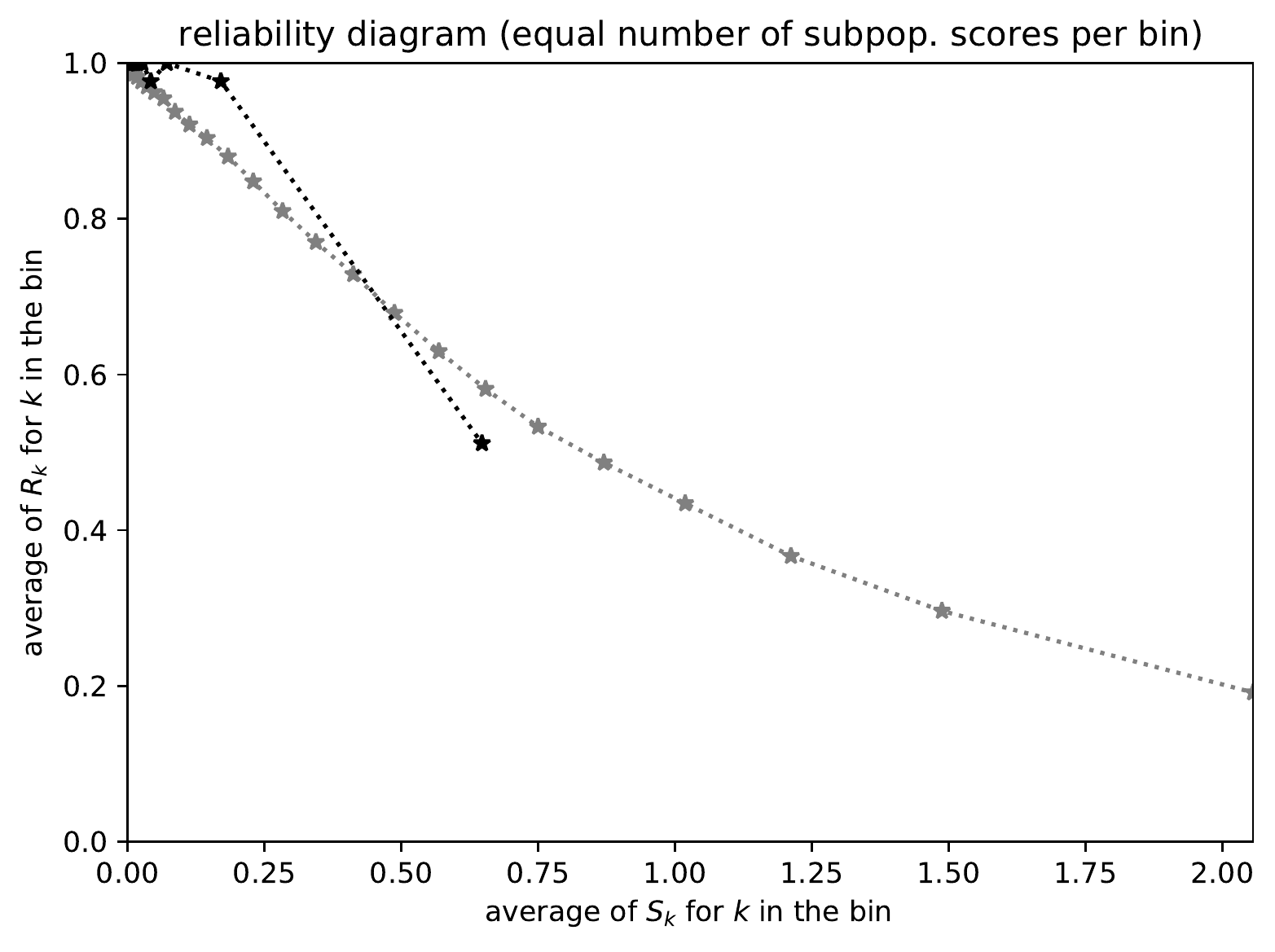}}
\quad\quad
\parbox{\imsize}{\includegraphics[width=\imsize]
{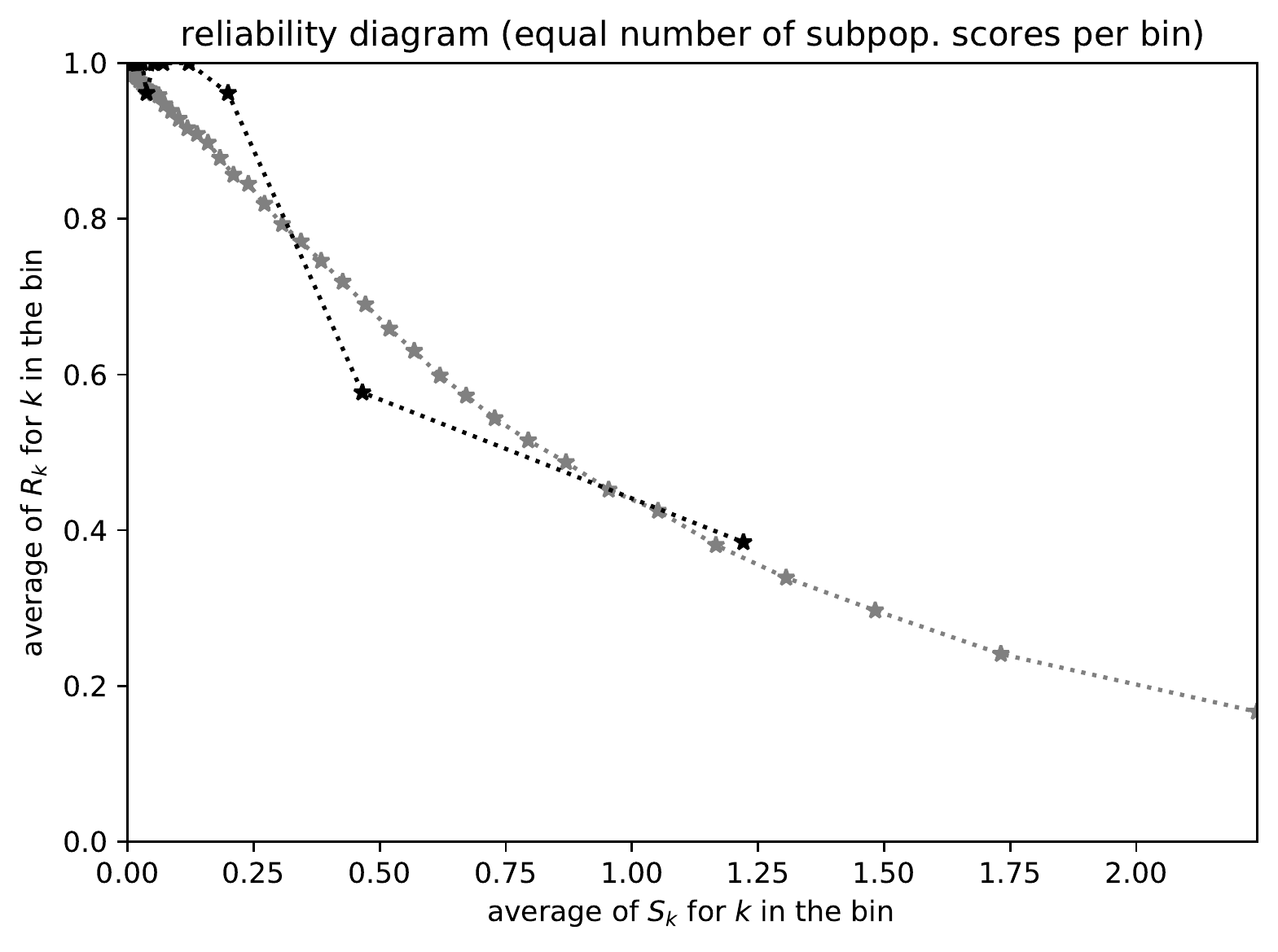}}

\vspace{\vertsep}

\parbox{\imsize}{\includegraphics[width=\imsize]
{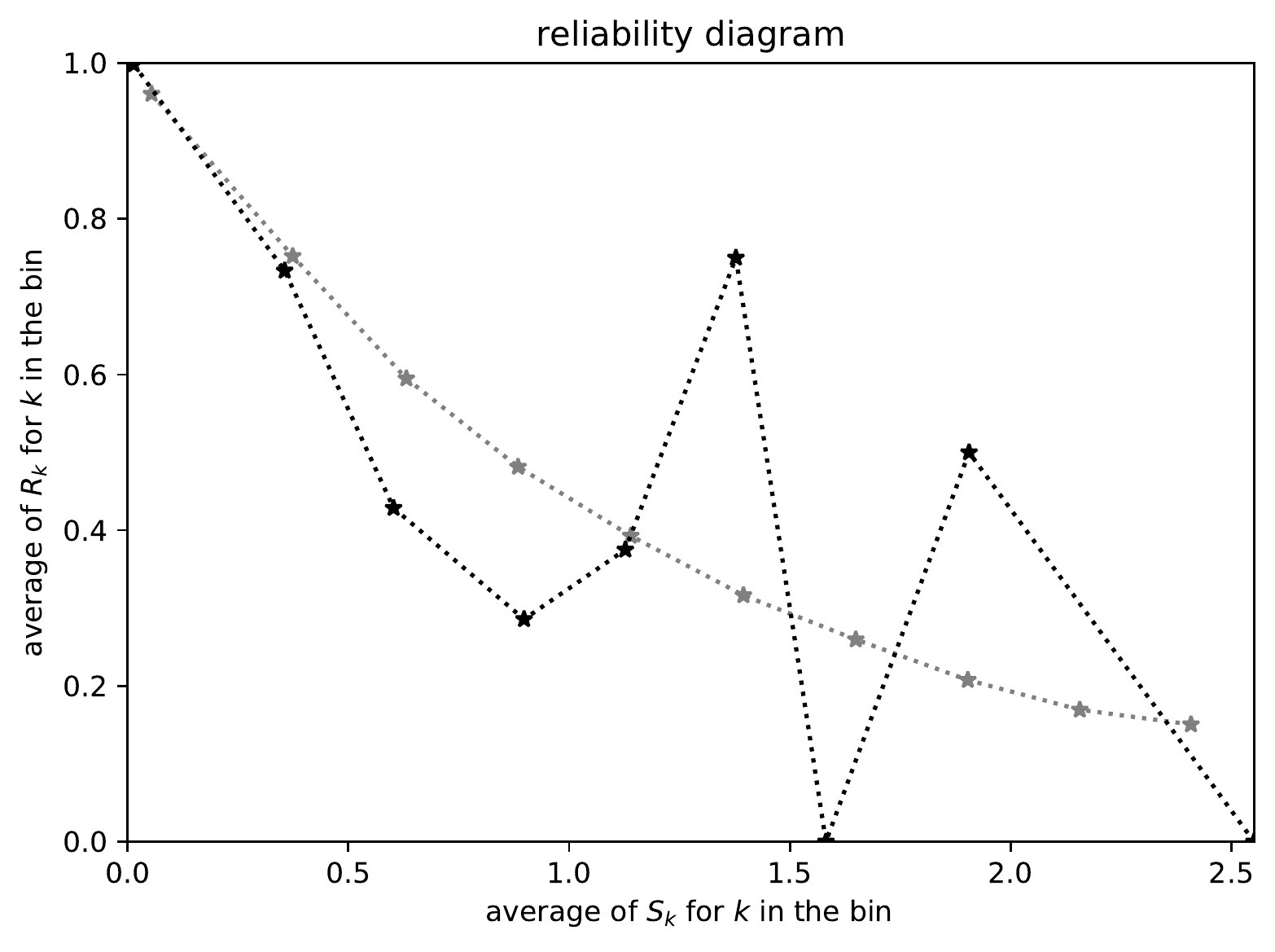}}
\quad\quad
\parbox{\imsize}{\includegraphics[width=\imsize]
{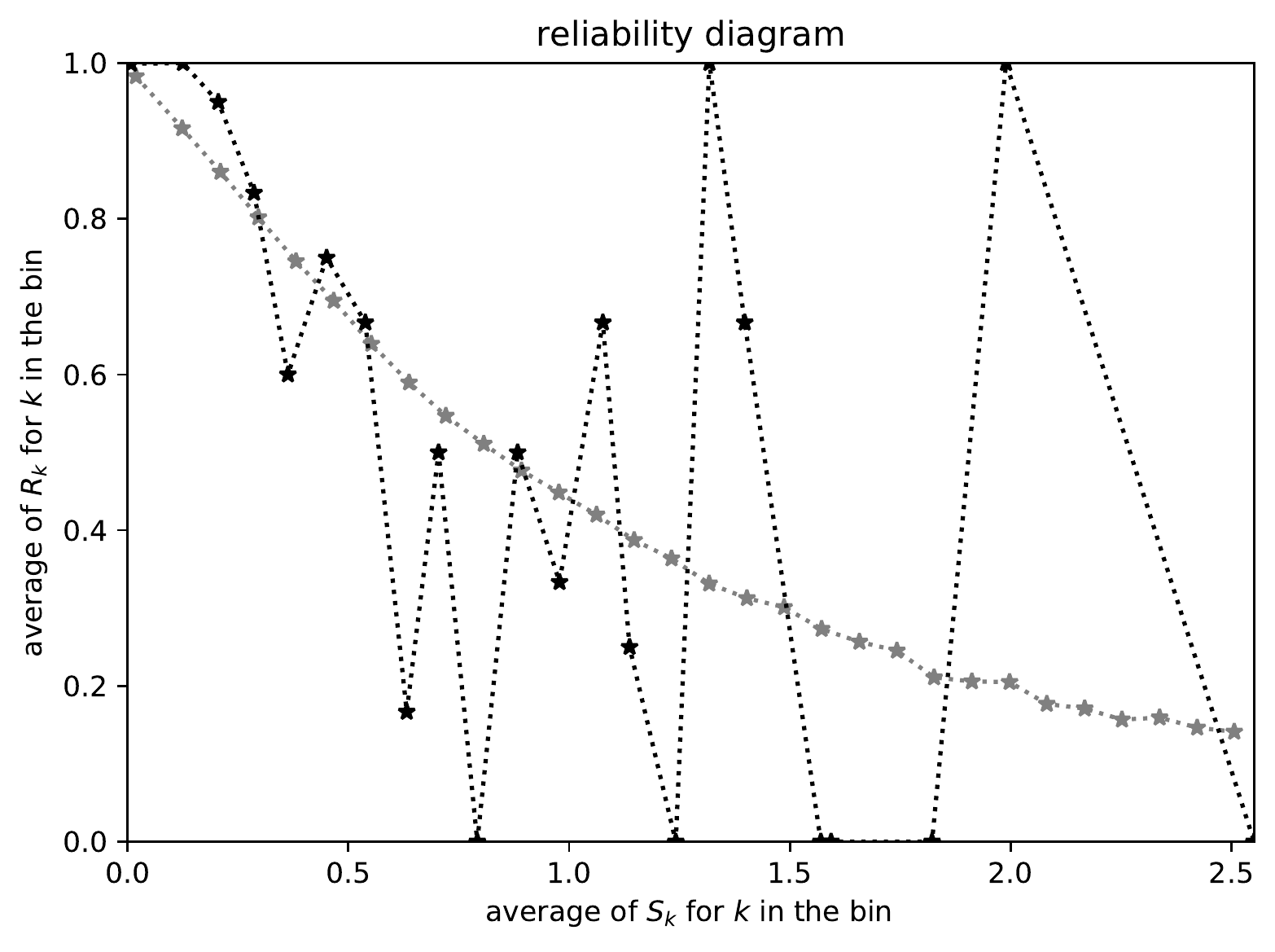}}

\end{centering}
\caption{Monarch (or milkweed) butterfly (Danaus plexippus),
         with scores being the negative log-likelihoods;
         $n =$ 1,300; Kuiper's statistic is $0.01144 / \sigma = 2.856$,
         Kolmogorov's and Smirnov's is $0.01144 / \sigma = 2.856$.
The deviation of the subpopulation below the full population that occurs
for scores greater than 1.5 is apparent only in the cumulative plot
or in the (very noisy) reliability diagrams whose bins
are roughly equispaced along the scores.
The scalar summary statistics detect somewhat significant deviation,
but not as much as would be possible if the Kolmogorov-Smirnov
and Kuiper metrics were to take fully into account the steep drop that occurs
in the plot of cumulative differences.
}
\label{monarch-monarch-butterfly-milkweed-butterfly-Danaus-plexippus-nll}
\end{figure}

\begin{figure}
\begin{centering}

\parbox{\imsize}{\includegraphics[width=\imsize]
{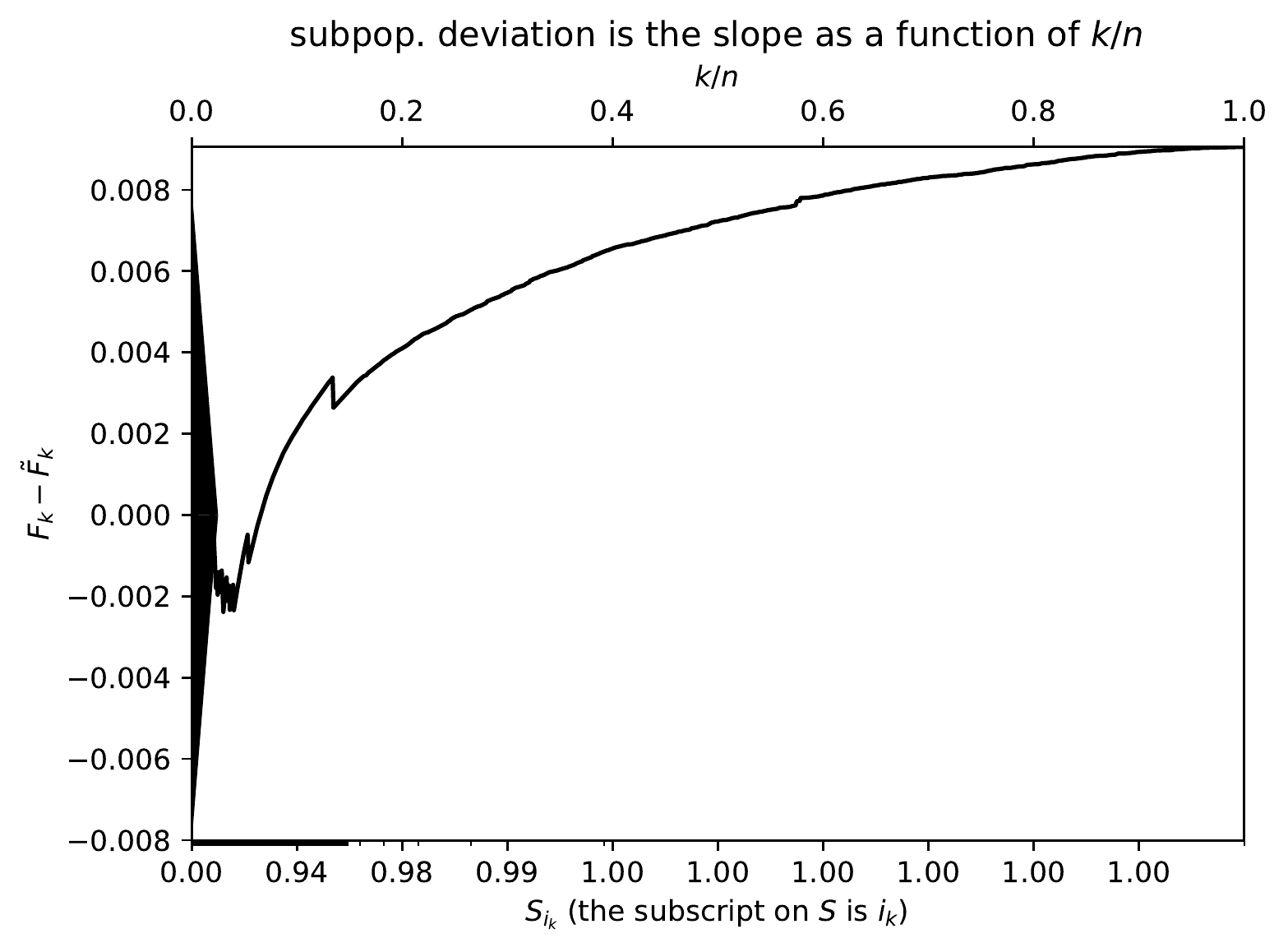}}
\quad\quad
\parbox{\imsize}{\includegraphics[width=\imsize]
{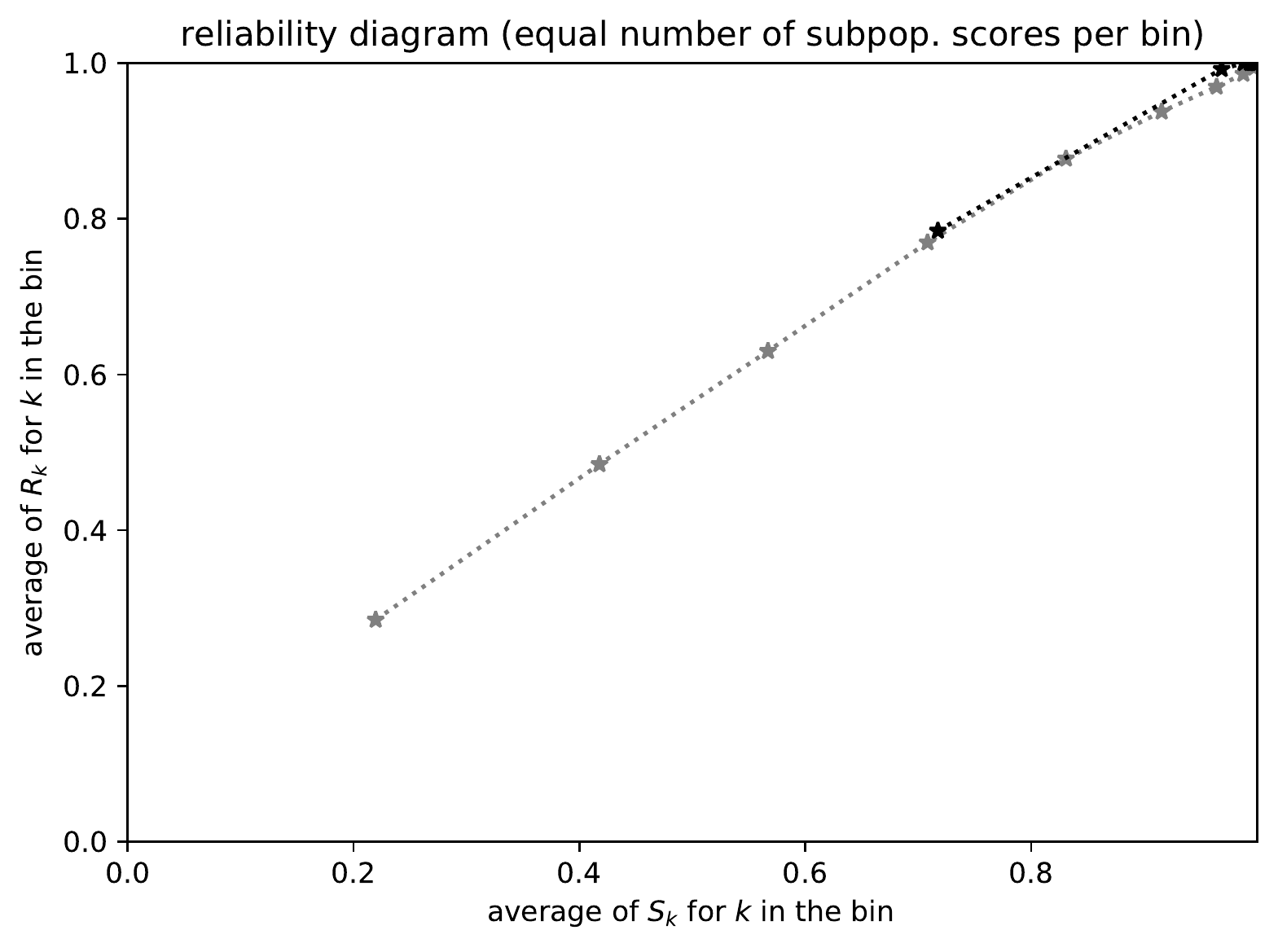}}

\vspace{\vertsep}

\parbox{\imsize}{\includegraphics[width=\imsize]
{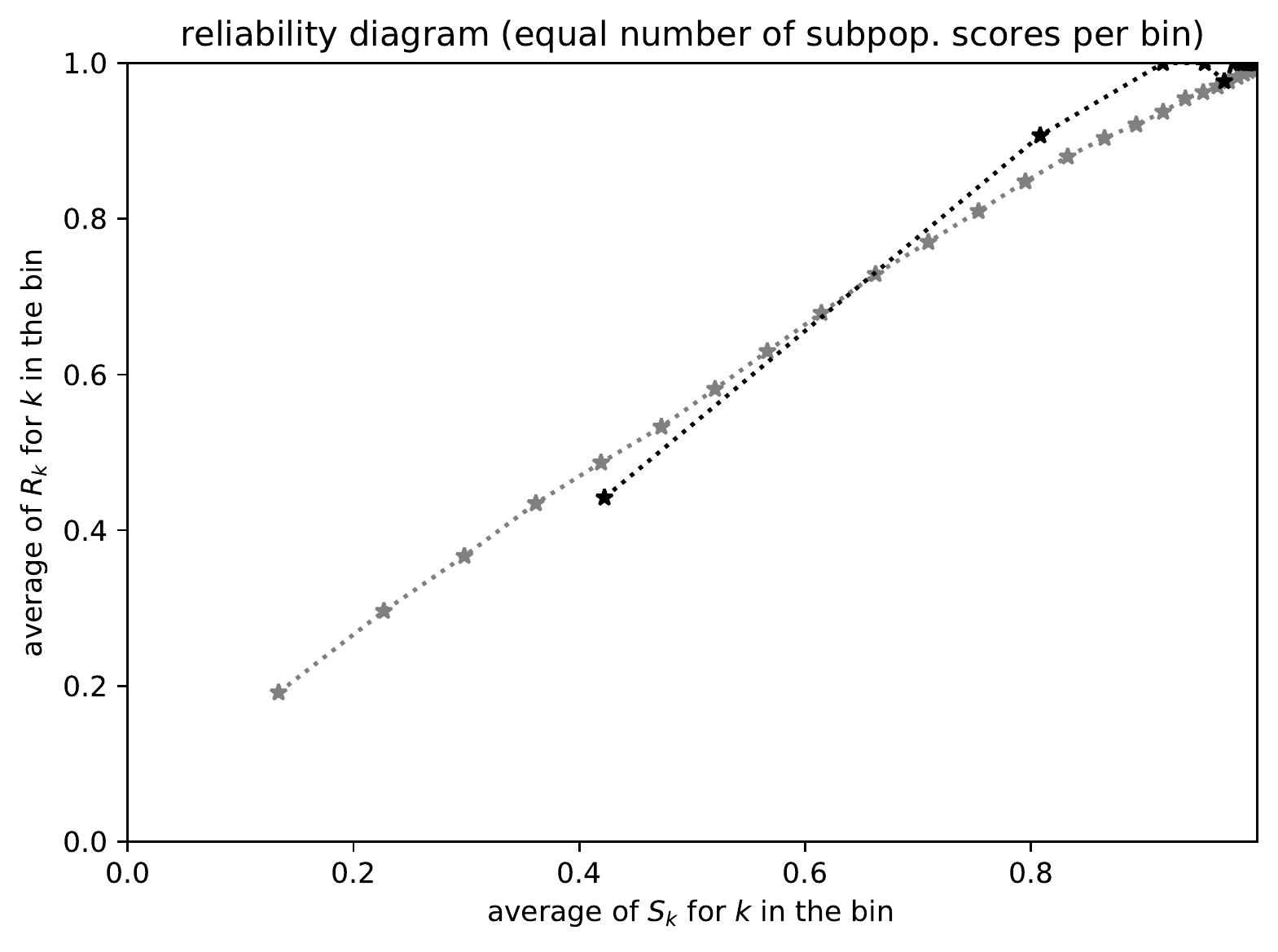}}
\quad\quad
\parbox{\imsize}{\includegraphics[width=\imsize]
{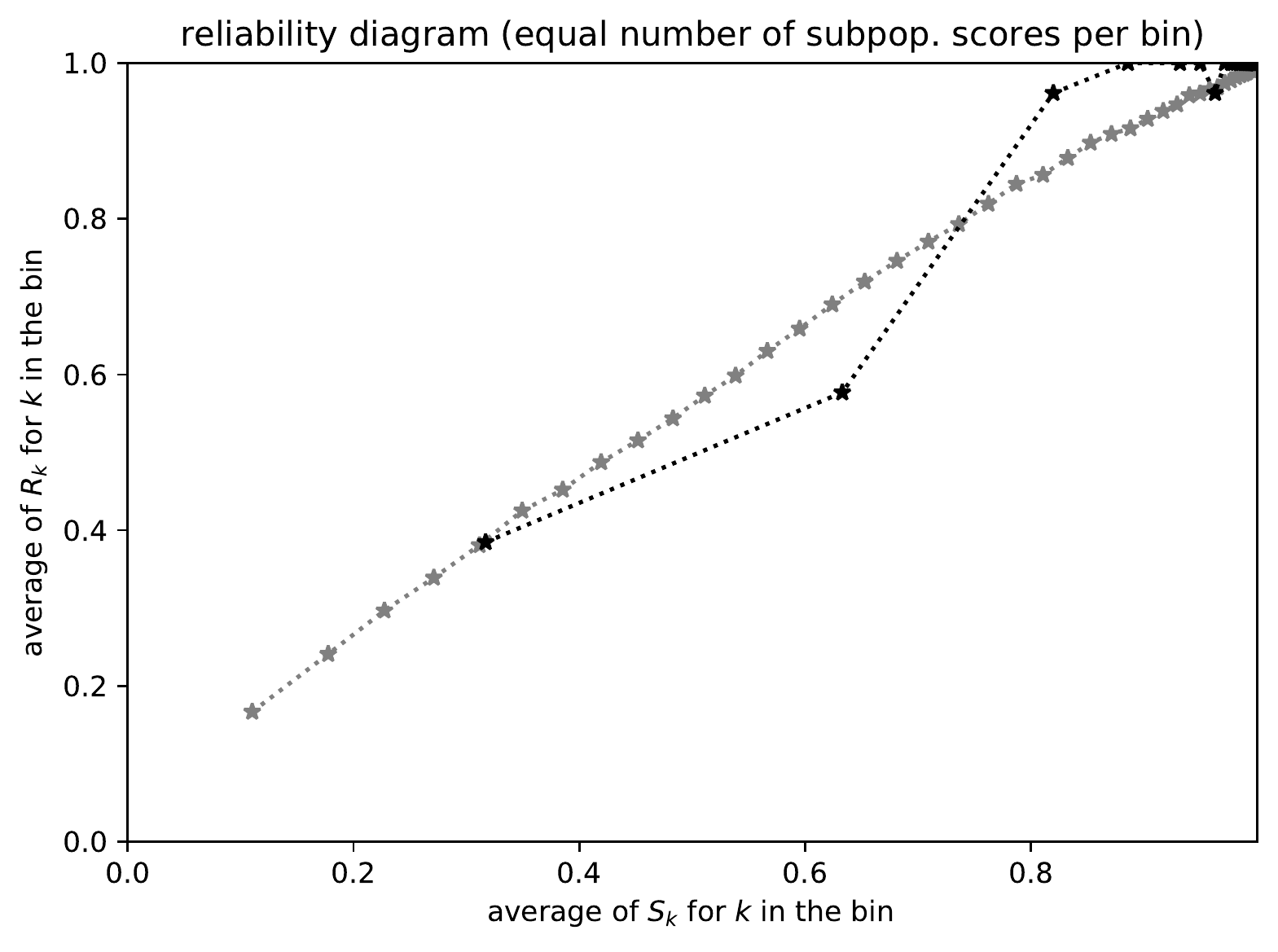}}

\vspace{\vertsep}

\parbox{\imsize}{\includegraphics[width=\imsize]
{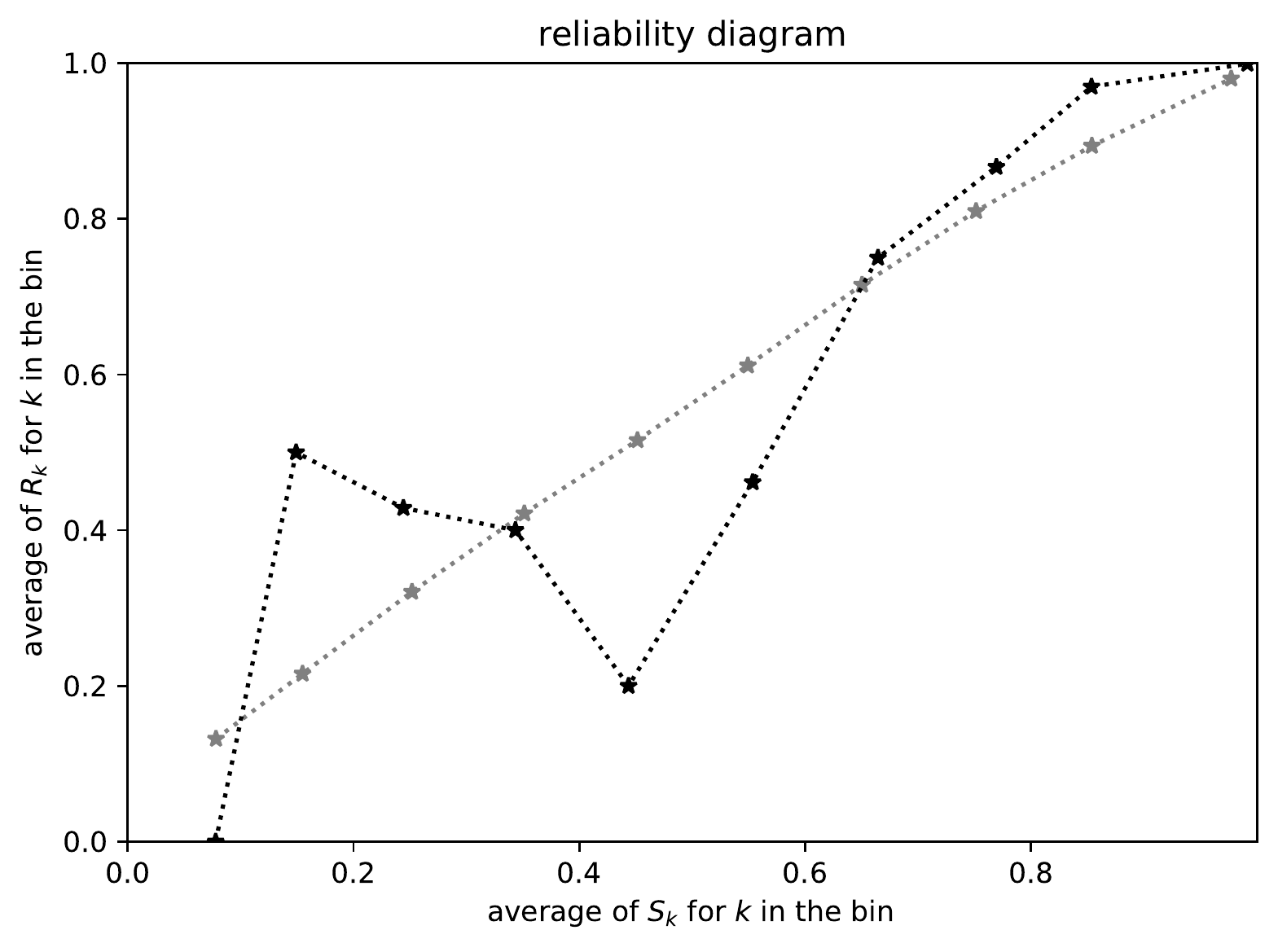}}
\quad\quad
\parbox{\imsize}{\includegraphics[width=\imsize]
{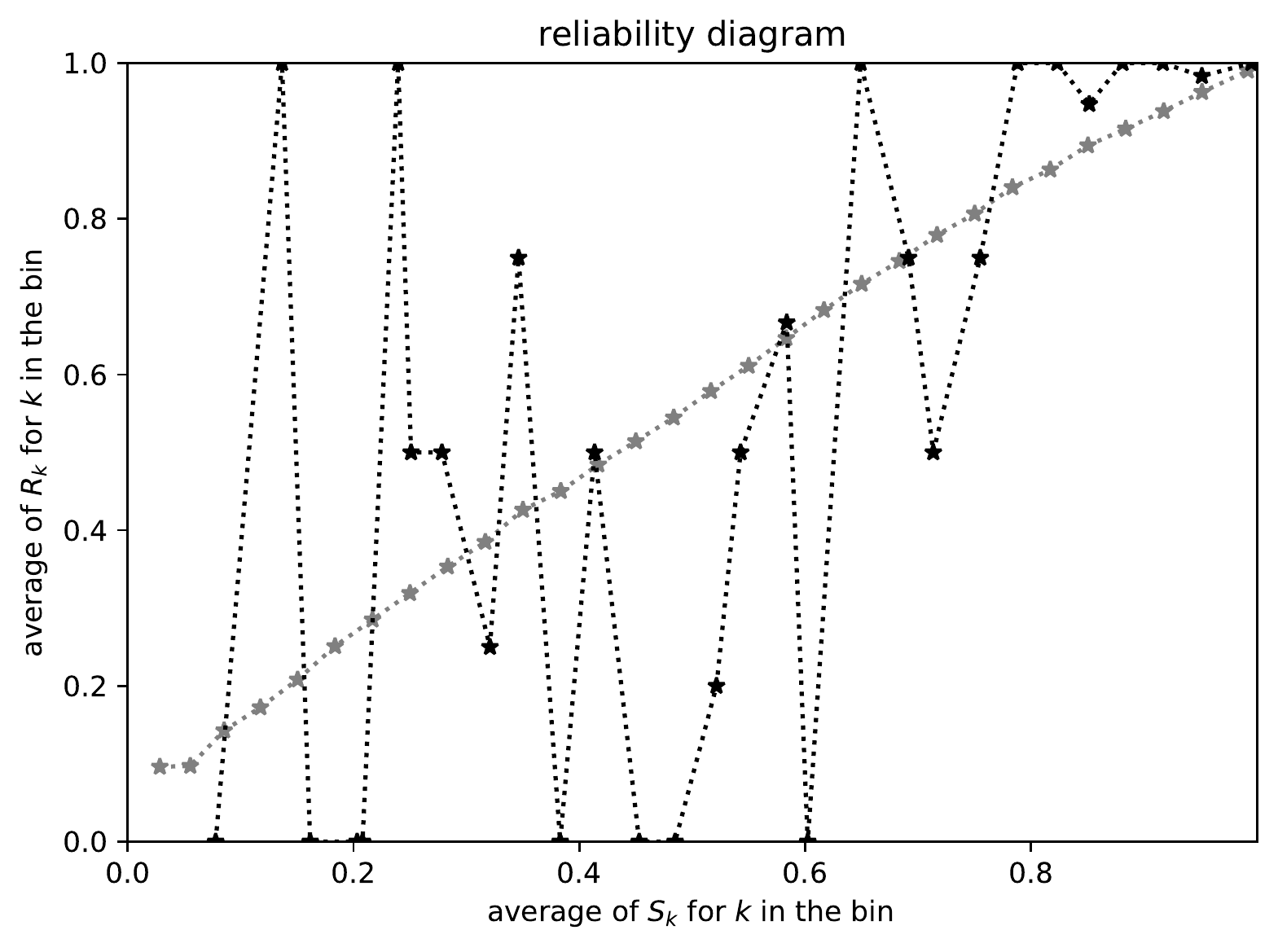}}

\end{centering}
\caption{Monarch (or milkweed) butterfly (Danaus plexippus),
         with scores being the probabilities;
         $n =$ 1,300; Kuiper's statistic is $0.01144 / \sigma = 2.857$,
         Kolmogorov's and Smirnov's is $0.009057 / \sigma = 2.261$.
Much like in
Figure~\ref{monarch-monarch-butterfly-milkweed-butterfly-Danaus-plexippus-nll},
the deviation of the subpopulation below the full population that occurs
for probabilities less than 0.1 is apparent only in the cumulative plot
or in the very noisy reliability diagrams whose bins
are roughly equispaced along the probabilities.
The scalar summary statistics again detect somewhat significant deviation,
though not as much as if the Kolmogorov-Smirnov and Kuiper metrics
could better account for the steep drop that happens
in the plot of cumulative differences.
}
\label{monarch-monarch-butterfly-milkweed-butterfly-Danaus-plexippus-prob}
\end{figure}

\begin{figure}
\begin{centering}

\parbox{\imsize}{\includegraphics[width=\imsize]
{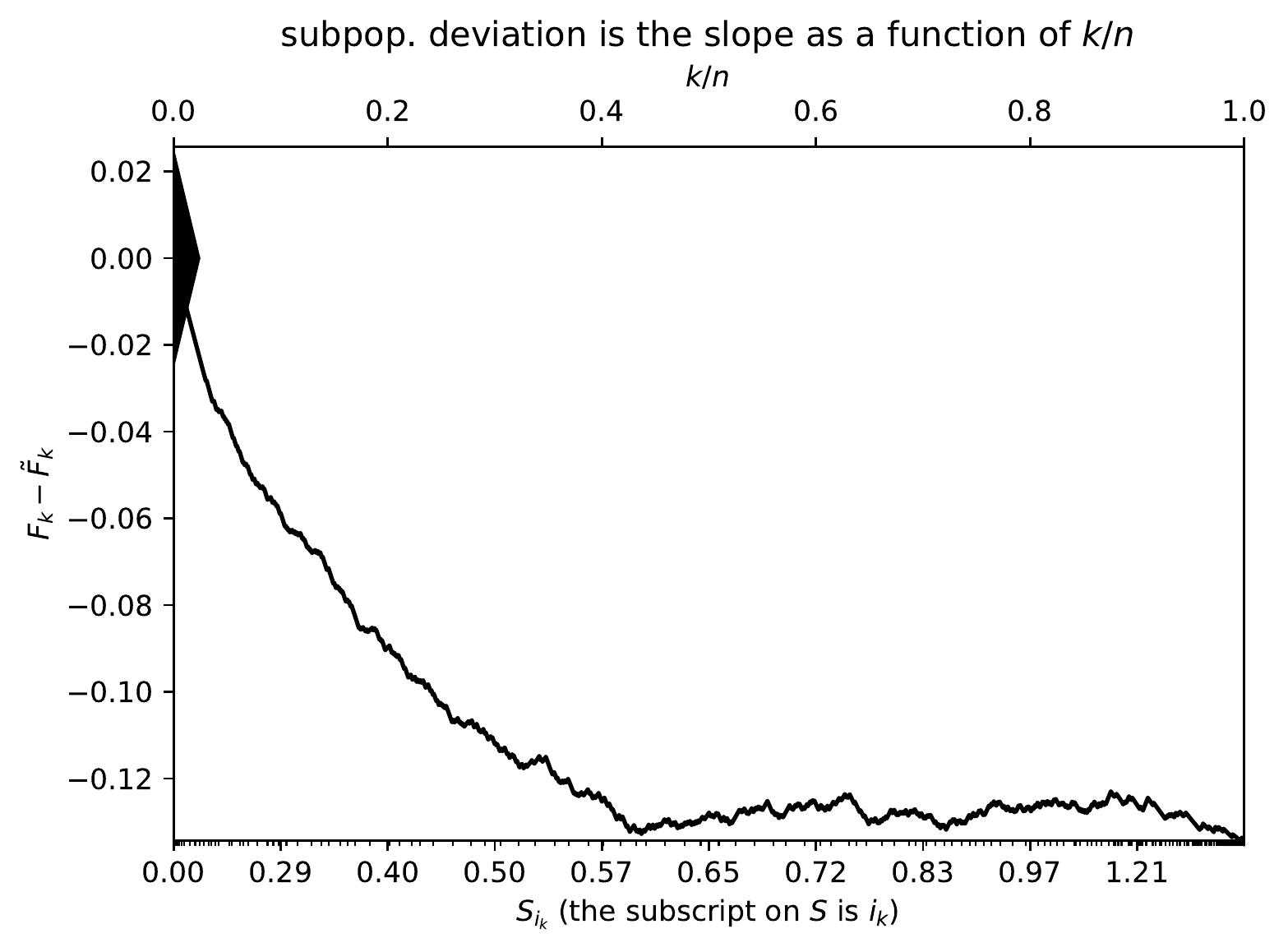}}
\quad\quad
\parbox{\imsize}{\includegraphics[width=\imsize]
{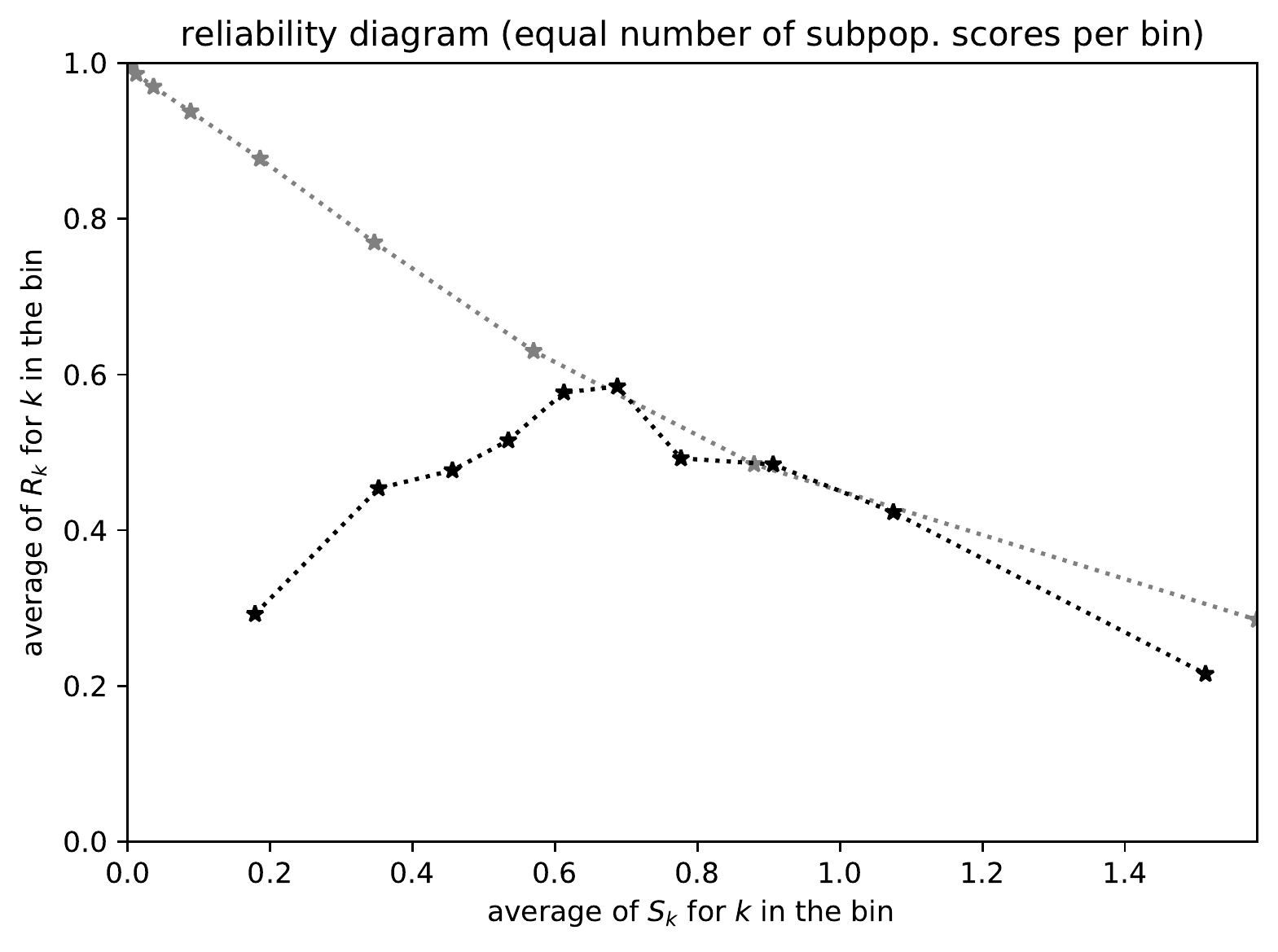}}

\vspace{\vertsep}

\parbox{\imsize}{\includegraphics[width=\imsize]
{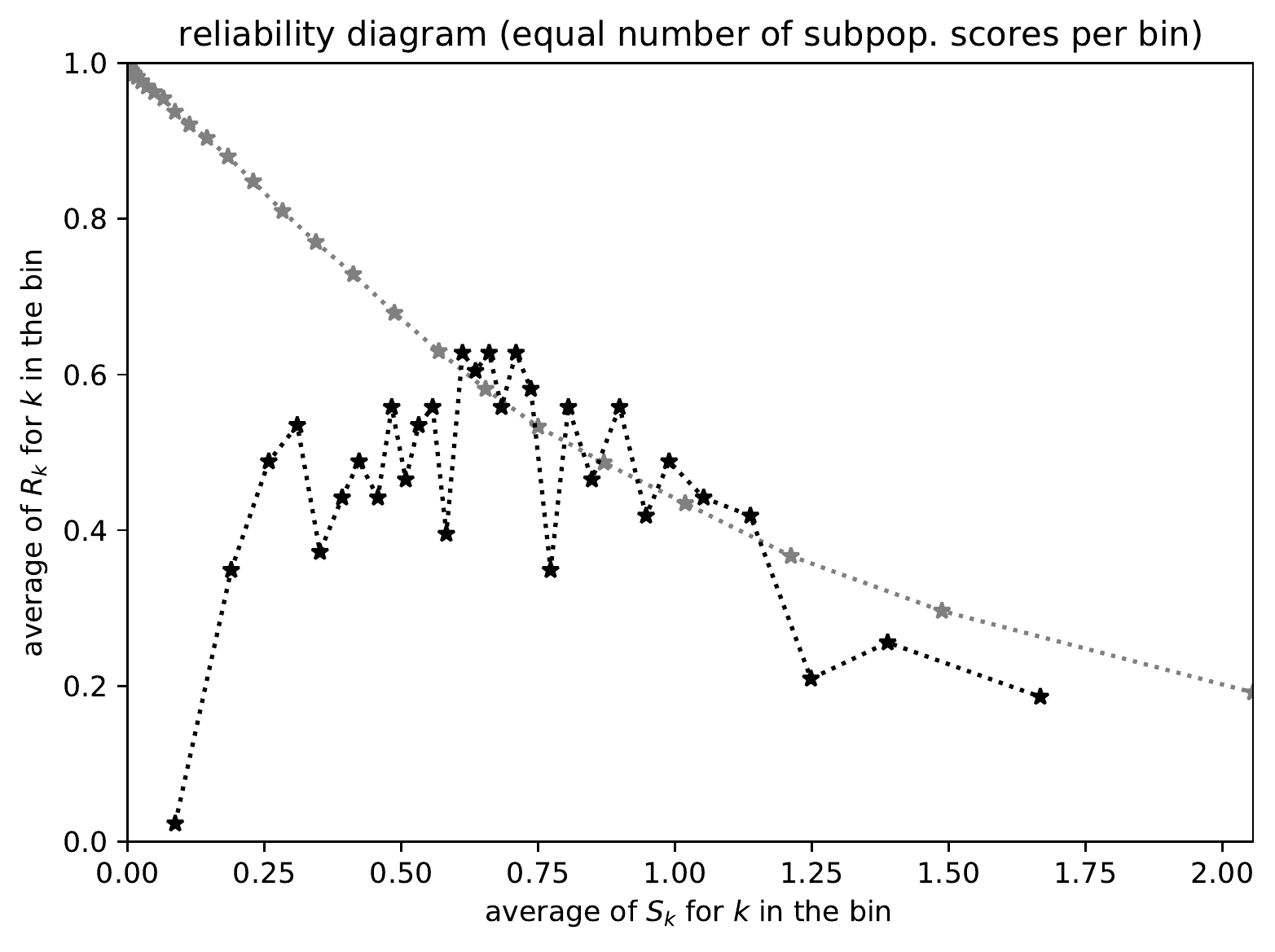}}
\quad\quad
\parbox{\imsize}{\includegraphics[width=\imsize]
{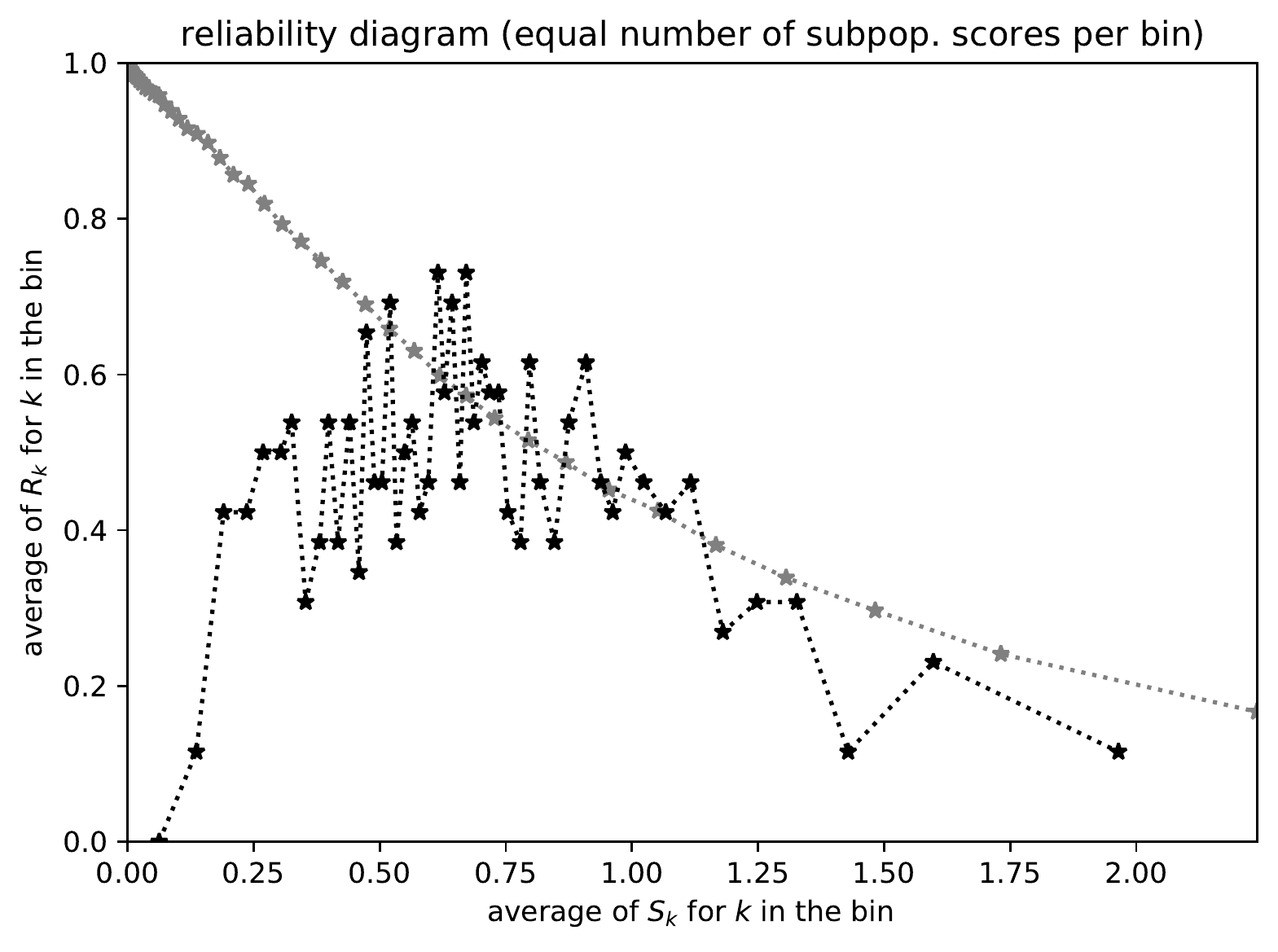}}

\vspace{\vertsep}

\parbox{\imsize}{\includegraphics[width=\imsize]
{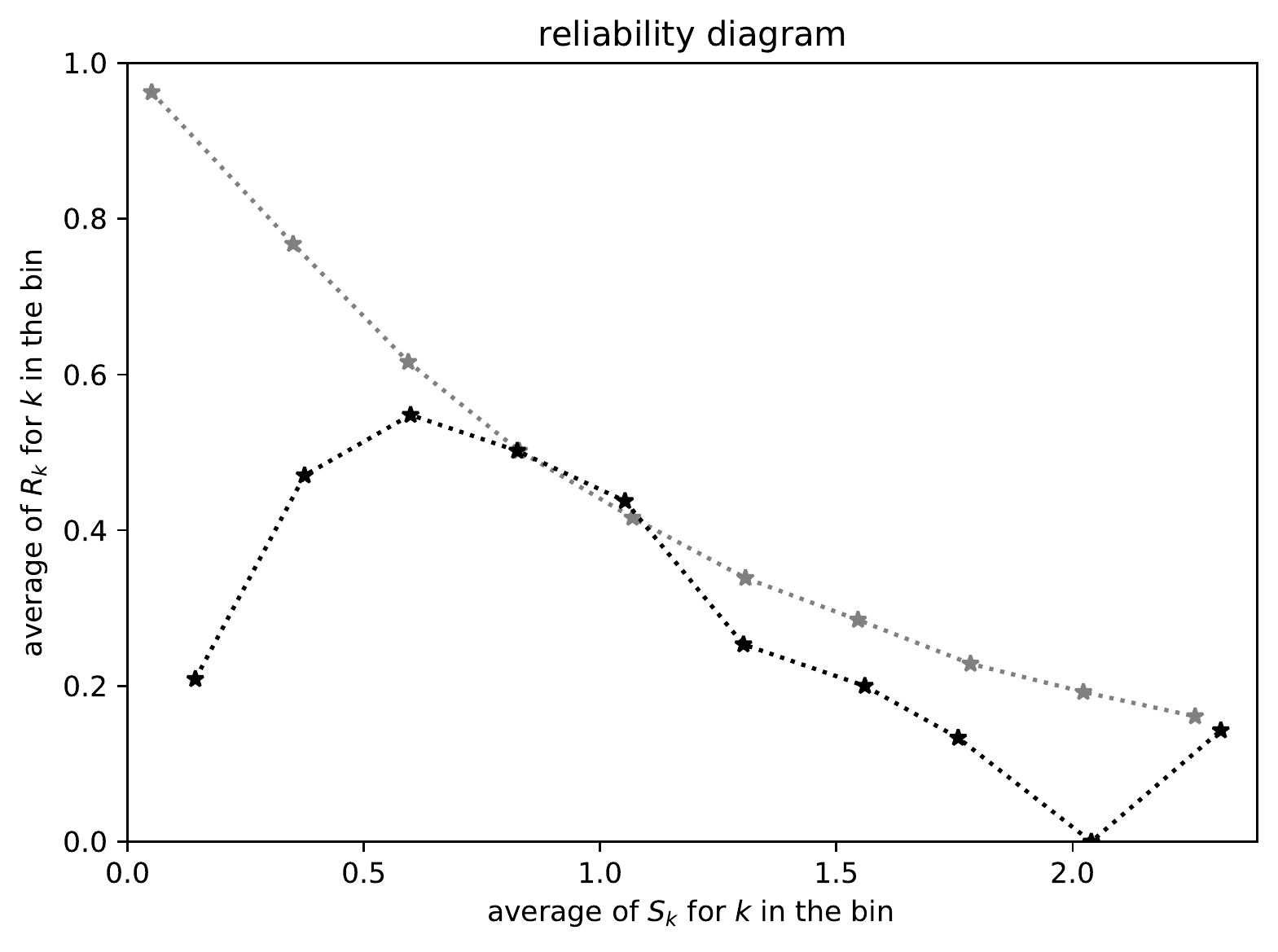}}
\quad\quad
\parbox{\imsize}{\includegraphics[width=\imsize]
{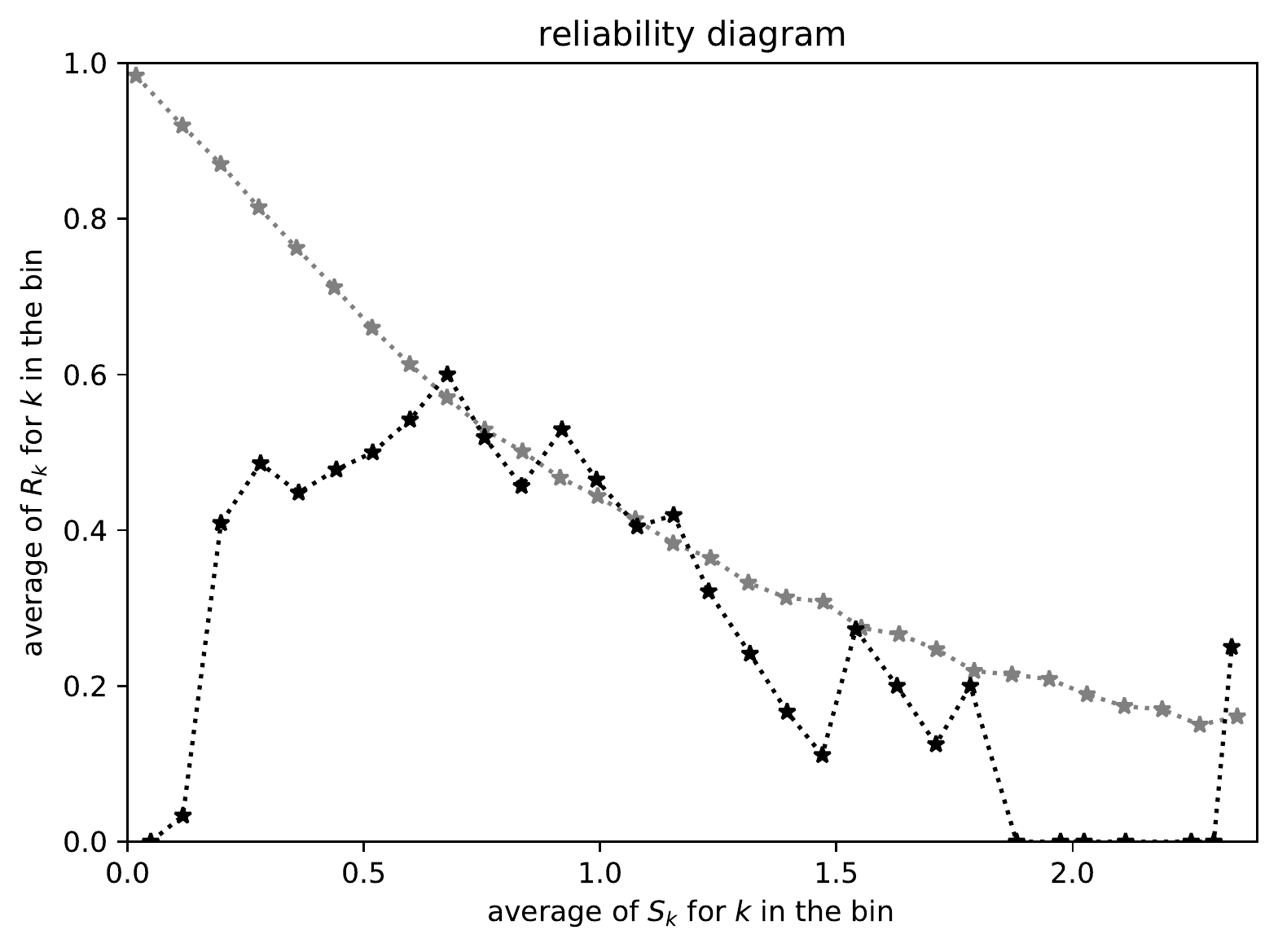}}

\end{centering}
\caption{Sidewinder/horned rattlesnake (Crotalus cerastes),
         with scores being the negative log-likelihoods;
         $n =$ 1,300; Kuiper's statistic is $0.1343 / \sigma = 10.47$,
         Kolmogorov's and Smirnov's is $0.1343 / \sigma = 10.47$.
The reliability diagrams whose bins each contain roughly the same number
of subpopulation scores are either noisier or missing the severest deviations
in comparison with the other reliability diagrams;
so there do exist cases in which choosing bins that are nearly equispaced
along the scores works better than choosing bins that each contain roughly
the same number of subpopulation scores.
The scalar summary statistics extremely successfully
detect the statistically highly significant deviation
of the subpopulation from the full population.
}
\label{sidewinder-horned-rattlesnake-Crotalus-cerastes-nll}
\end{figure}

\begin{figure}
\begin{centering}

\parbox{\imsize}{\includegraphics[width=\imsize]
{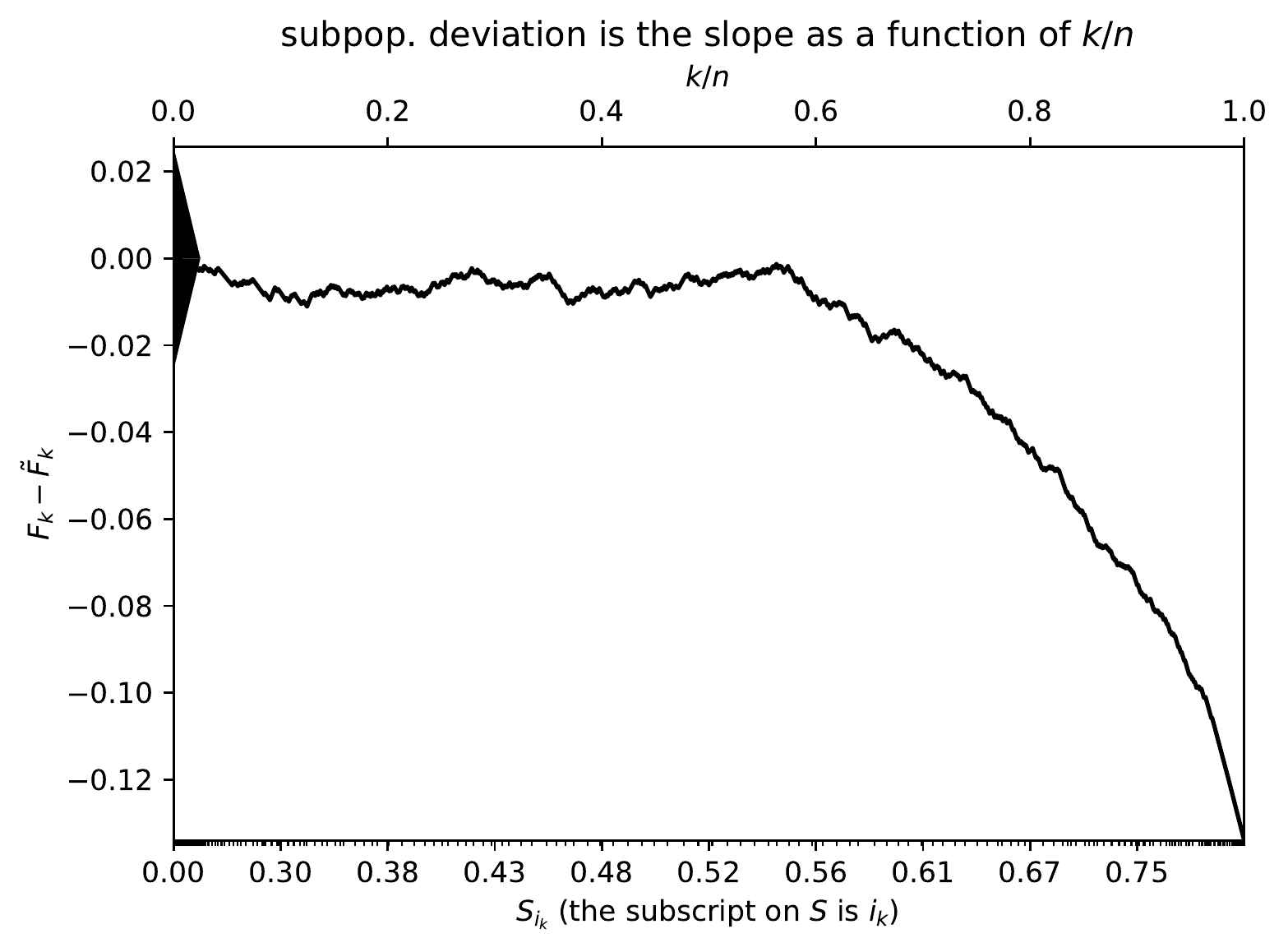}}
\quad\quad
\parbox{\imsize}{\includegraphics[width=\imsize]
{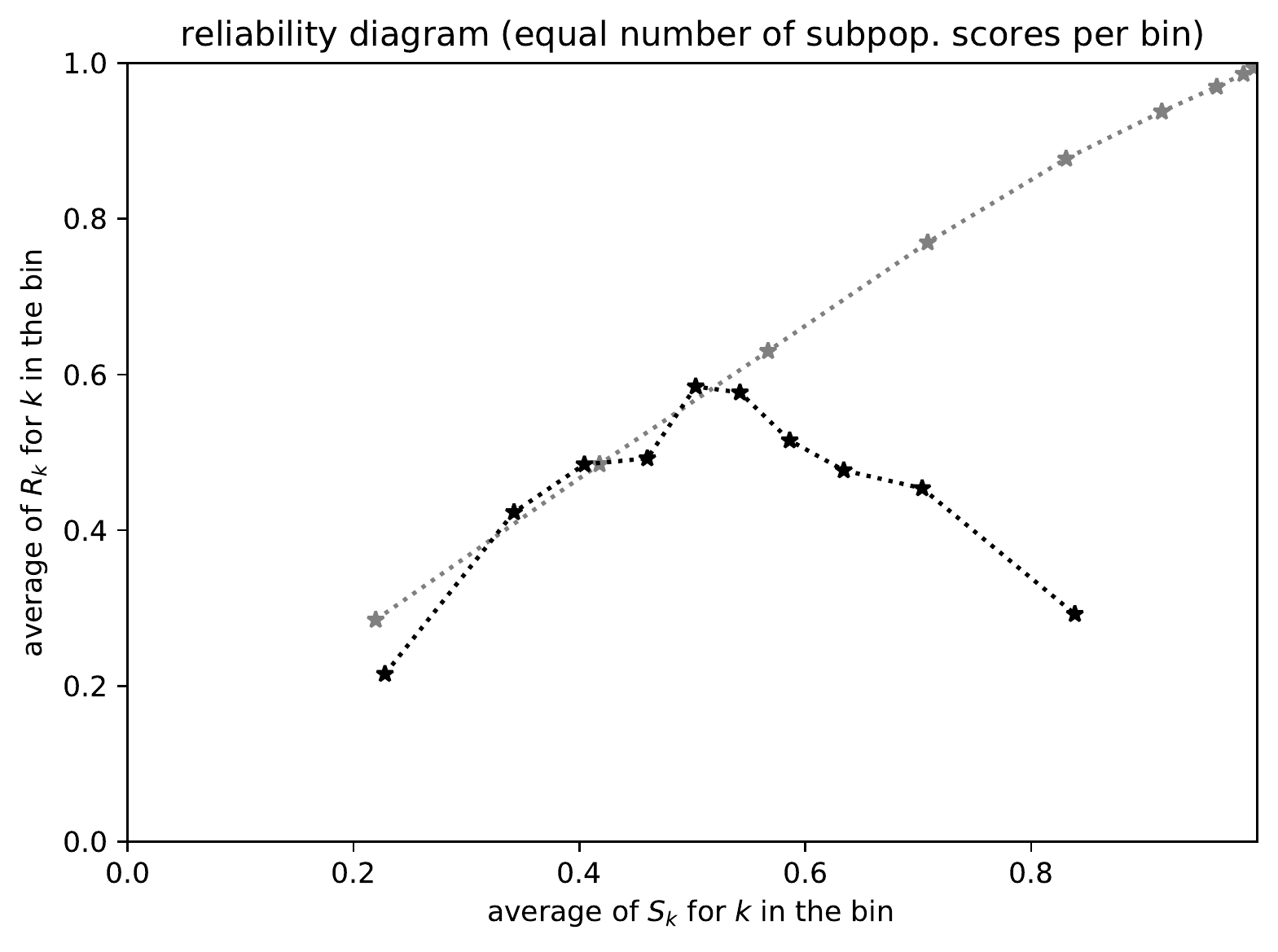}}

\vspace{\vertsep}

\parbox{\imsize}{\includegraphics[width=\imsize]
{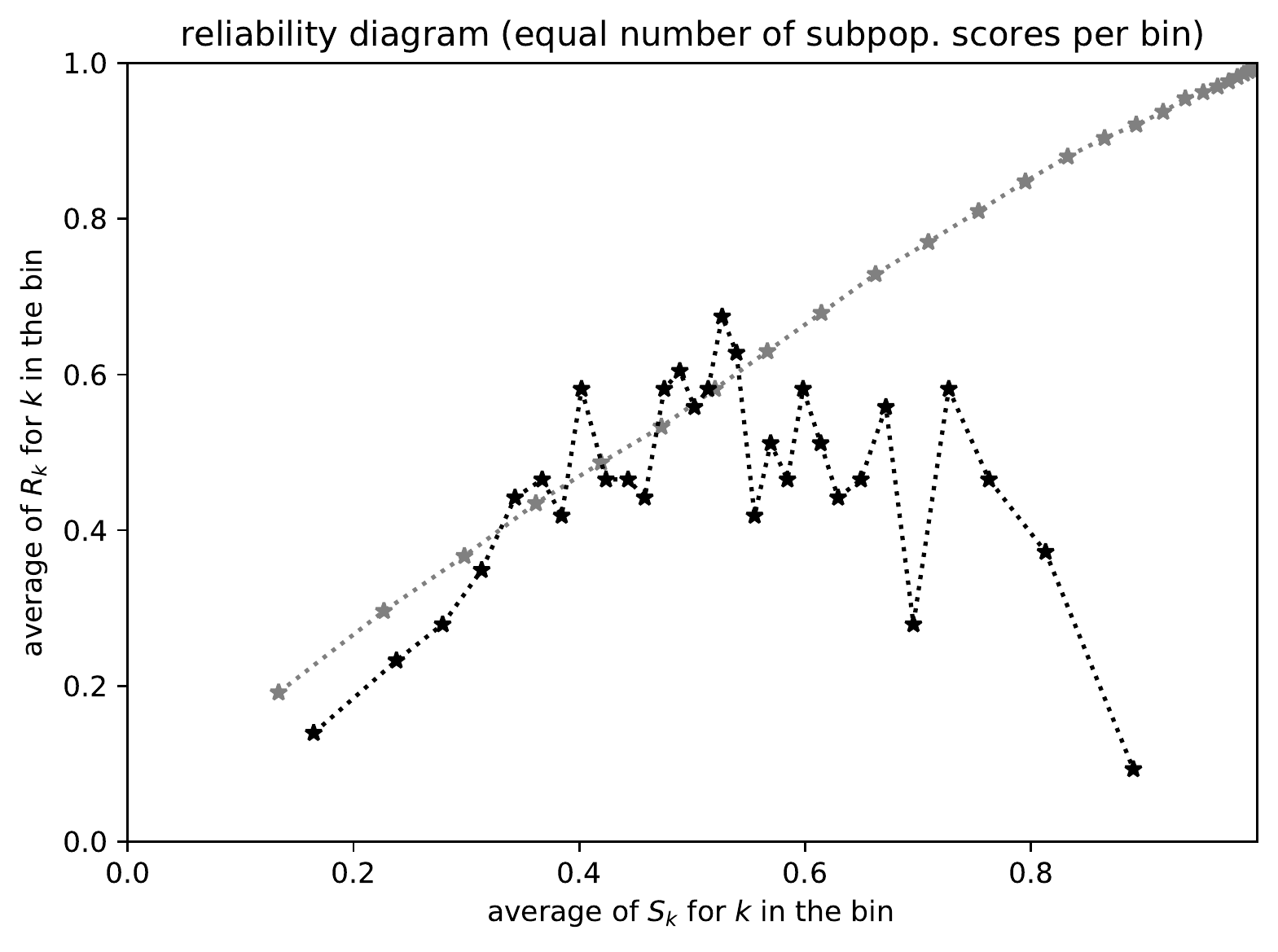}}
\quad\quad
\parbox{\imsize}{\includegraphics[width=\imsize]
{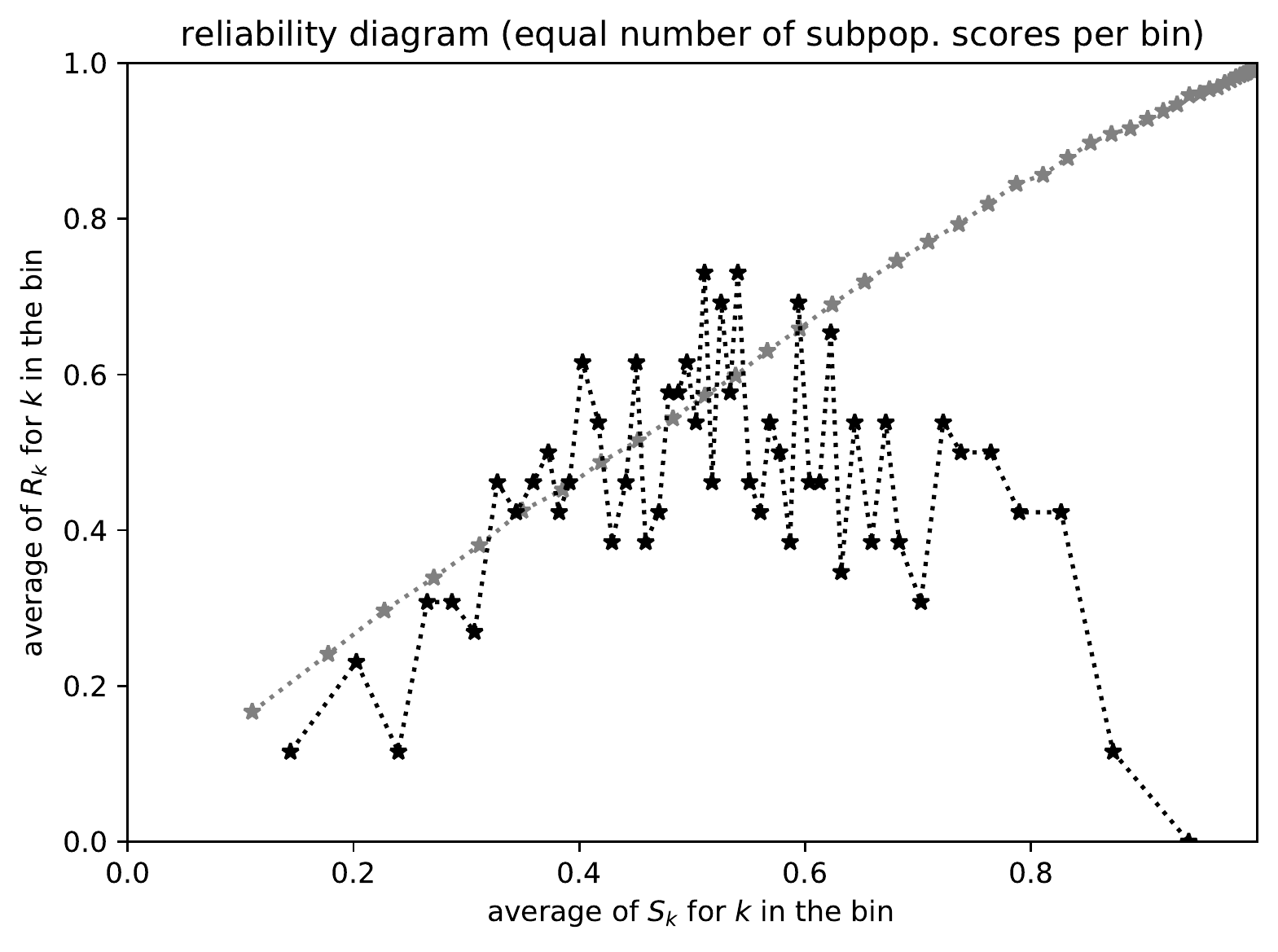}}

\vspace{\vertsep}

\parbox{\imsize}{\includegraphics[width=\imsize]
{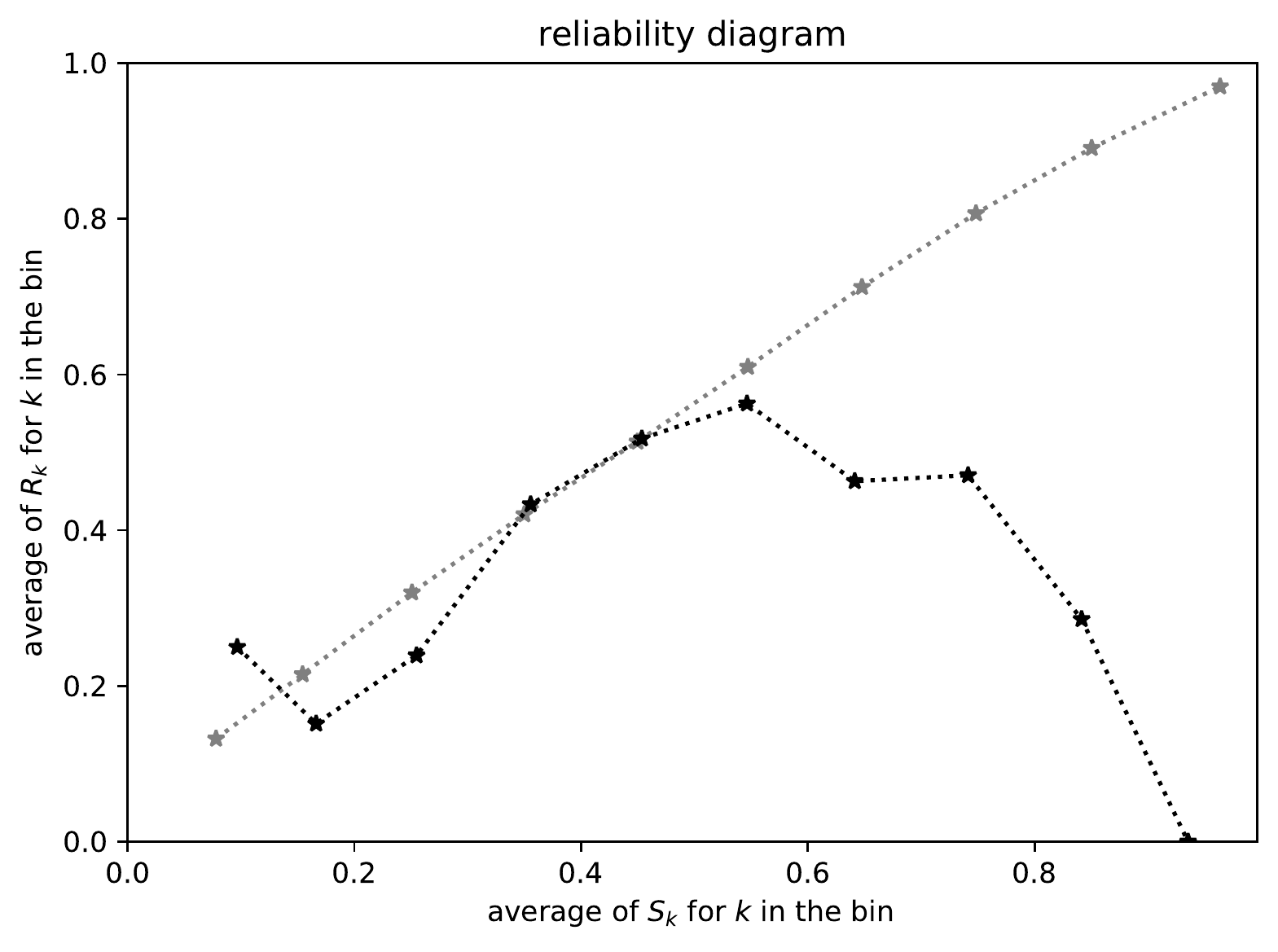}}
\quad\quad
\parbox{\imsize}{\includegraphics[width=\imsize]
{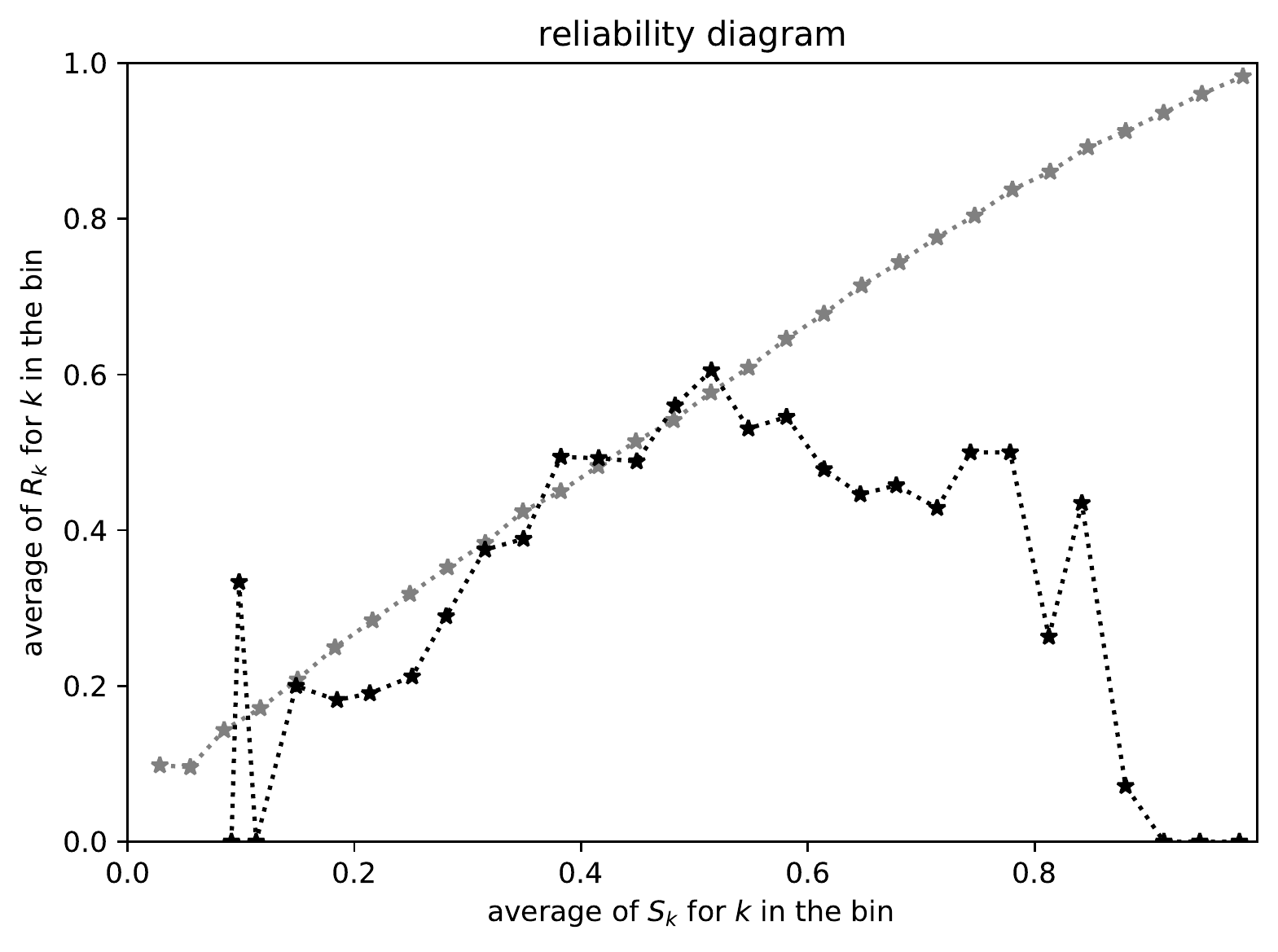}}

\end{centering}
\caption{Sidewinder/horned rattlesnake (Crotalus cerastes),
         with scores being the probabilities;
         $n =$ 1,300; Kuiper's statistic is $0.1343 / \sigma = 10.47$,
         Kolmogorov's and Smirnov's is $0.1340 / \sigma = 10.45$.
As in Figure~\ref{sidewinder-horned-rattlesnake-Crotalus-cerastes-nll},
the reliability diagrams whose bins each contain roughly the same number
of subpopulation scores either miss the severest deviations or are noisier
compared to the other reliability diagrams for this example;
so sometimes choosing bins that are approximately equispaced
along the scores works better than choosing bins that each contain
a similar number of subpopulation scores.
The scalar summary statistics extremely successfully
detect the statistically highly significant deviation
of the subpopulation from the full population.
}
\label{sidewinder-horned-rattlesnake-Crotalus-cerastes-prob}
\end{figure}

\begin{figure}
\begin{centering}

\parbox{\imsize}{\includegraphics[width=\imsize]
       {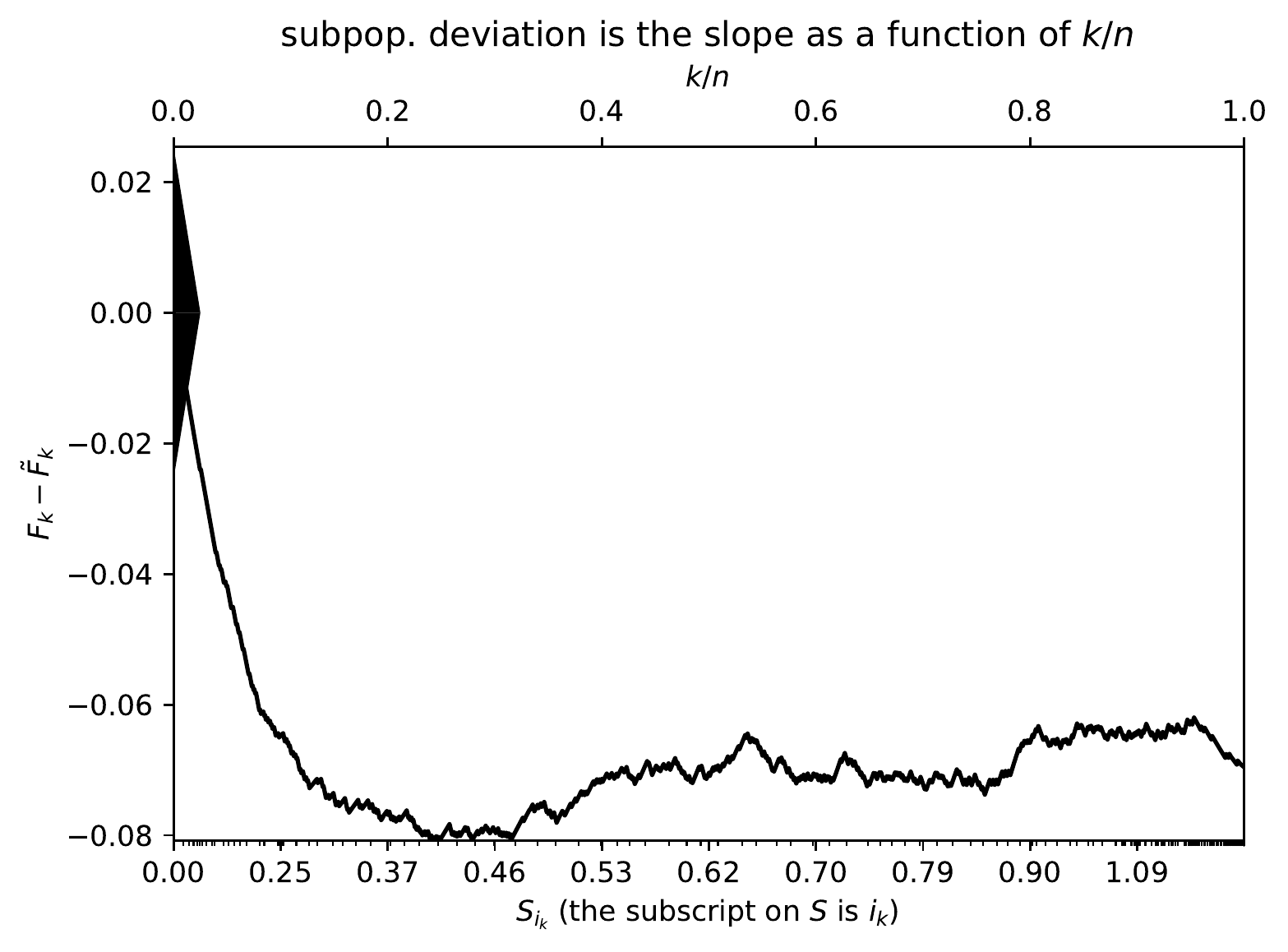}}
\quad\quad
\parbox{\imsize}{\includegraphics[width=\imsize]
       {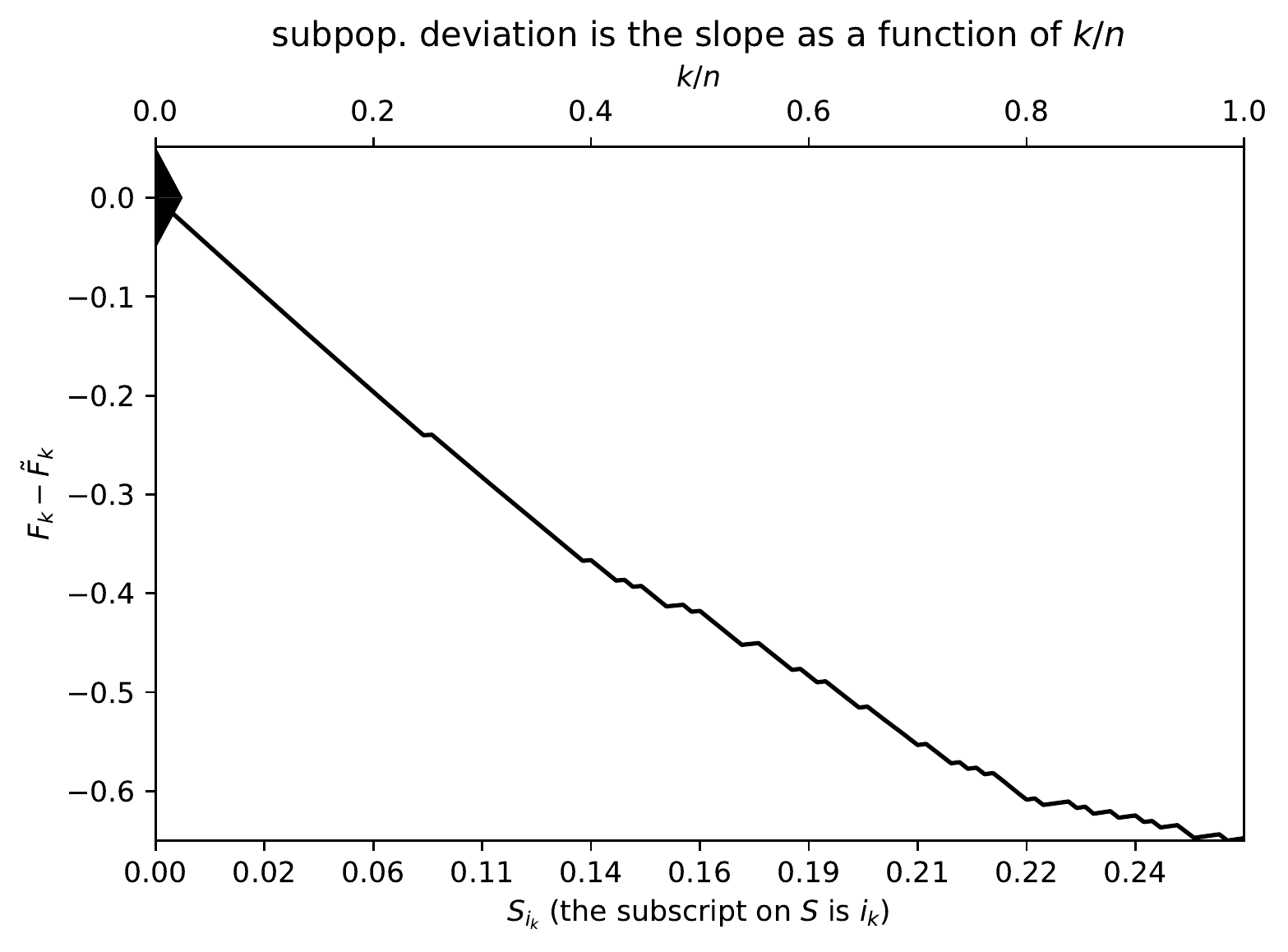}}

\parbox{\imsize}{\centering $n =$ 1,300; Kolmogorov-Smirnov and Kuiper
both $= 0.08082 / \sigma = 6.363$}
\quad\quad
\parbox{\imsize}{\centering $n = 130$;\ \ Kolmogorov-Smirnov and Kuiper
both $= 0.6502 / \sigma = 25.26$}

\vspace{\vertsep}

\parbox{\imsize}{\includegraphics[width=\imsize]
       {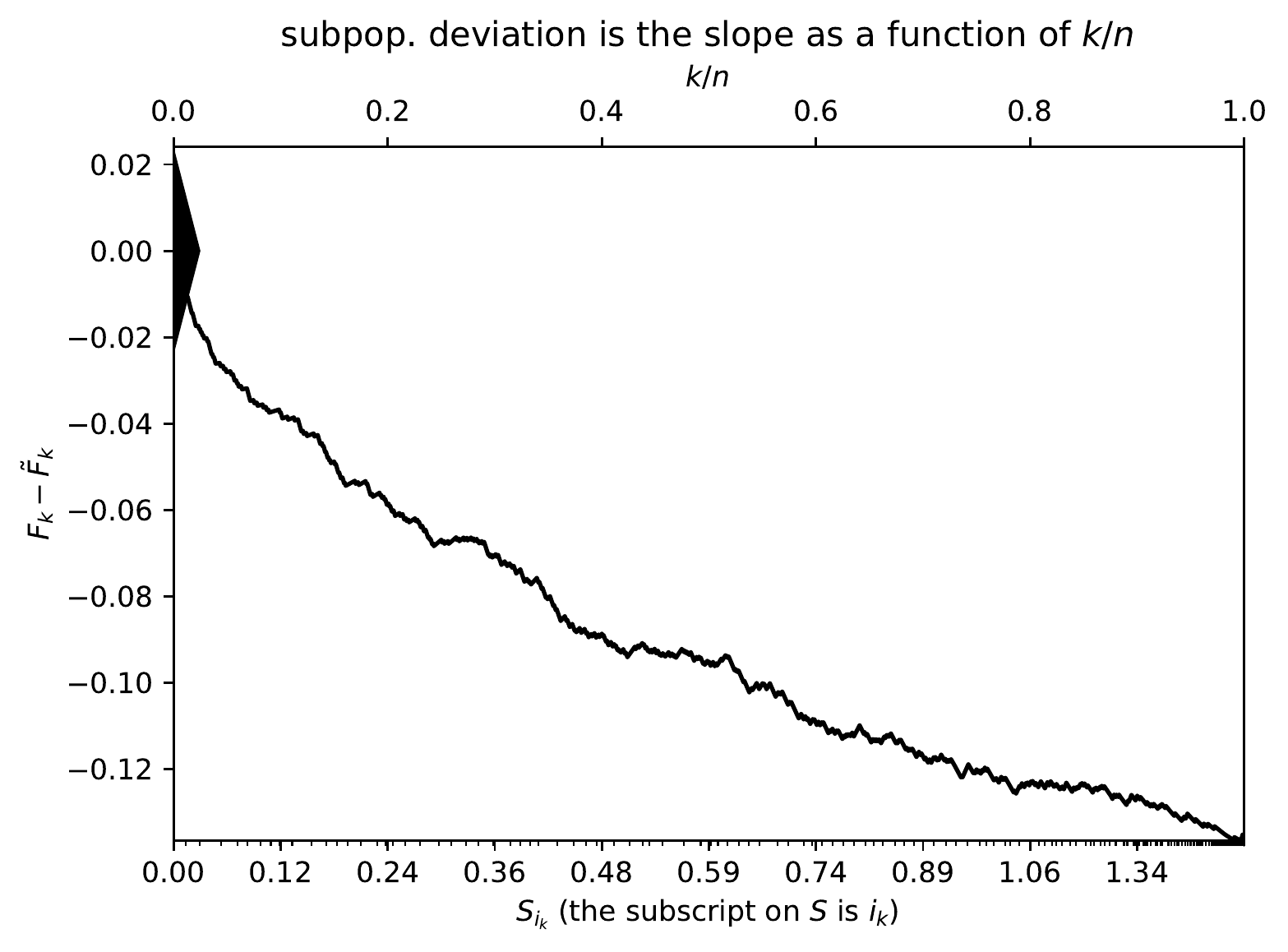}}
\quad\quad
\parbox{\imsize}{\includegraphics[width=\imsize]
       {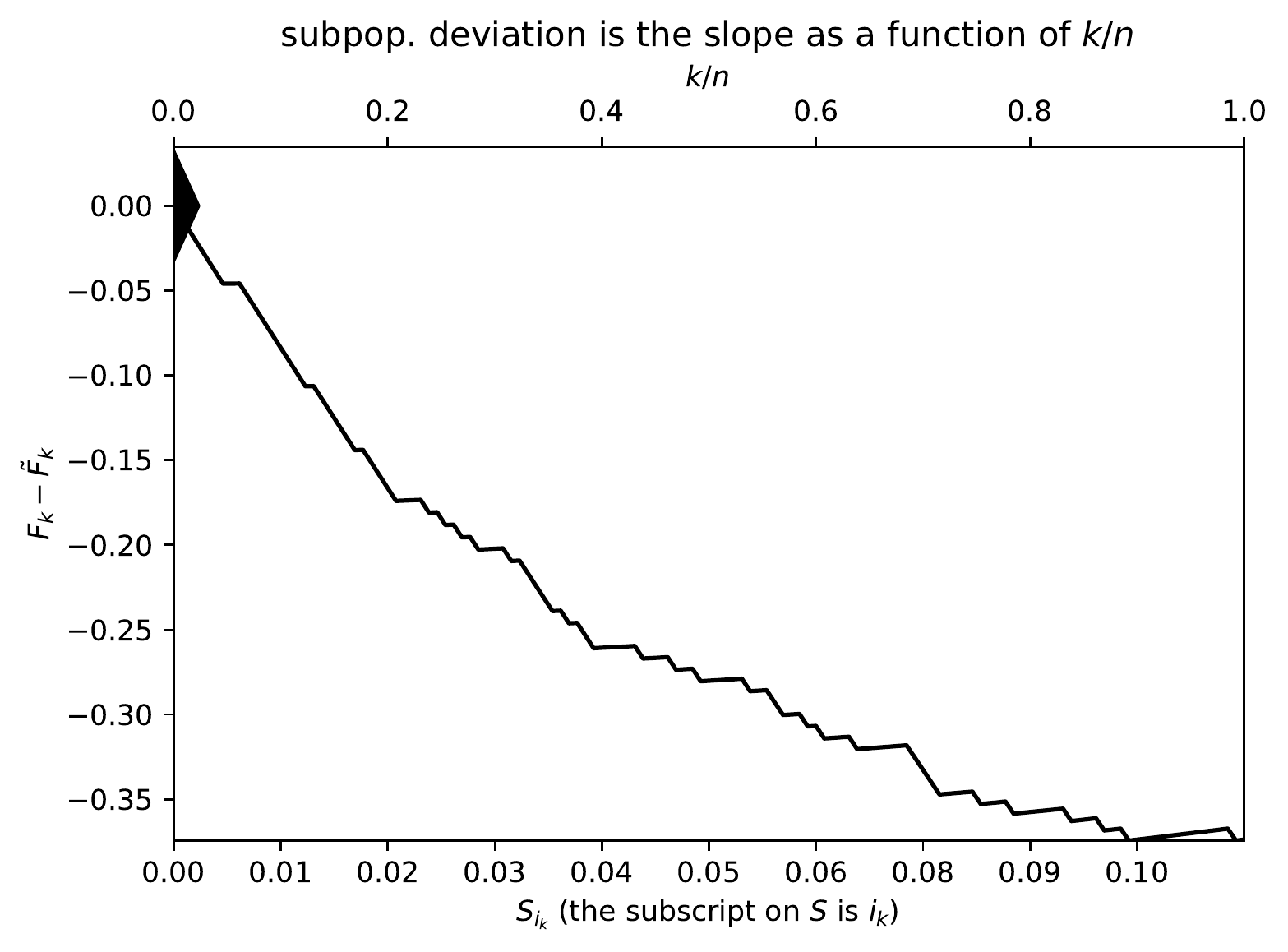}}

\parbox{\imsize}{\centering $n =$ 1,300; Kolmogorov-Smirnov and Kuiper
both $= 0.1365 / \sigma = 11.35$}
\quad\quad
\parbox{\imsize}{\centering $n = 130$;\ \ Kolmogorov-Smirnov and Kuiper
both $= 0.3745 / \sigma = 21.52$}

\vspace{\vertsep}

\parbox{\imsize}{\includegraphics[width=\imsize]
{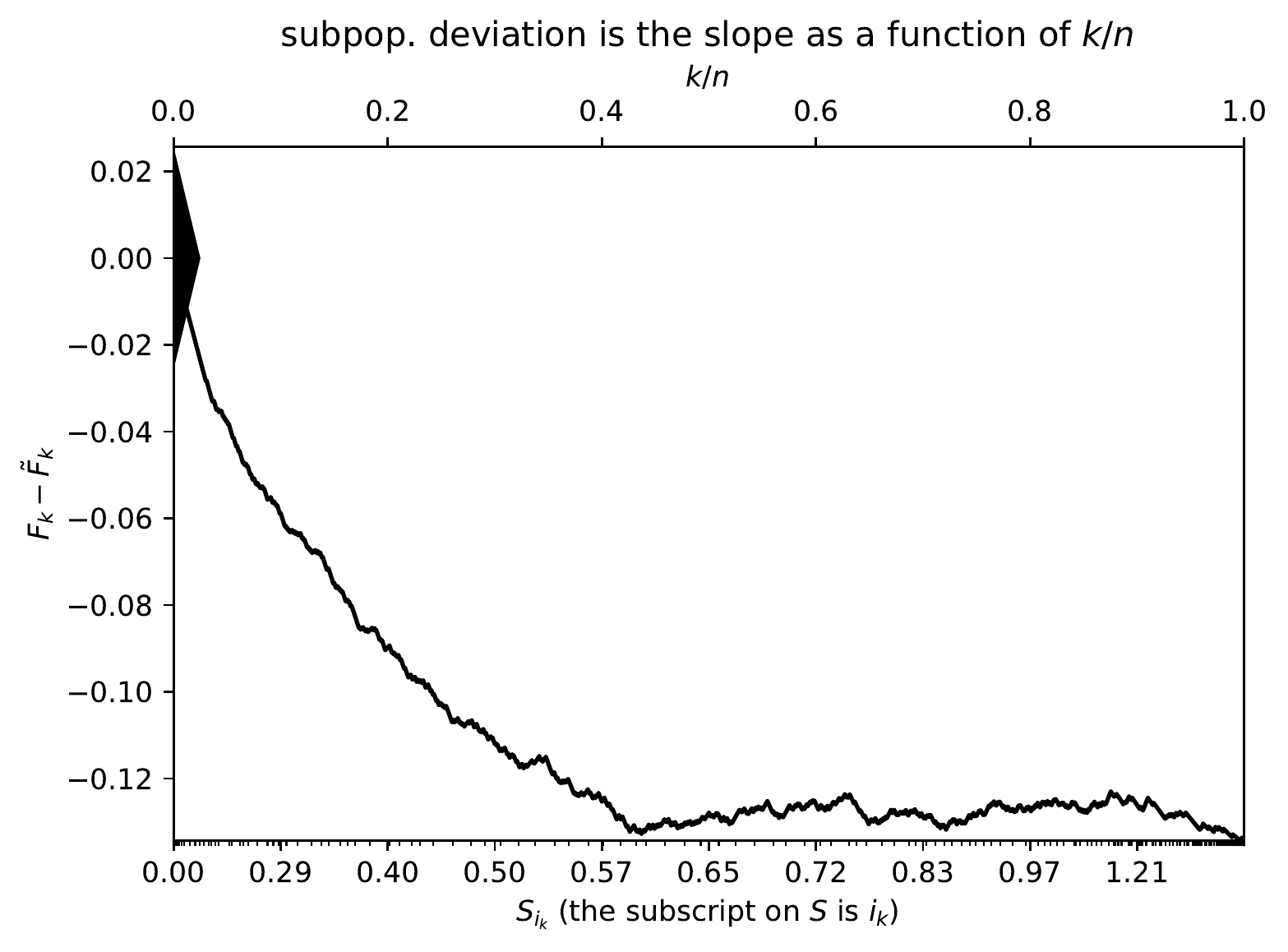}}
\quad\quad
\parbox{\imsize}{\includegraphics[width=\imsize]
{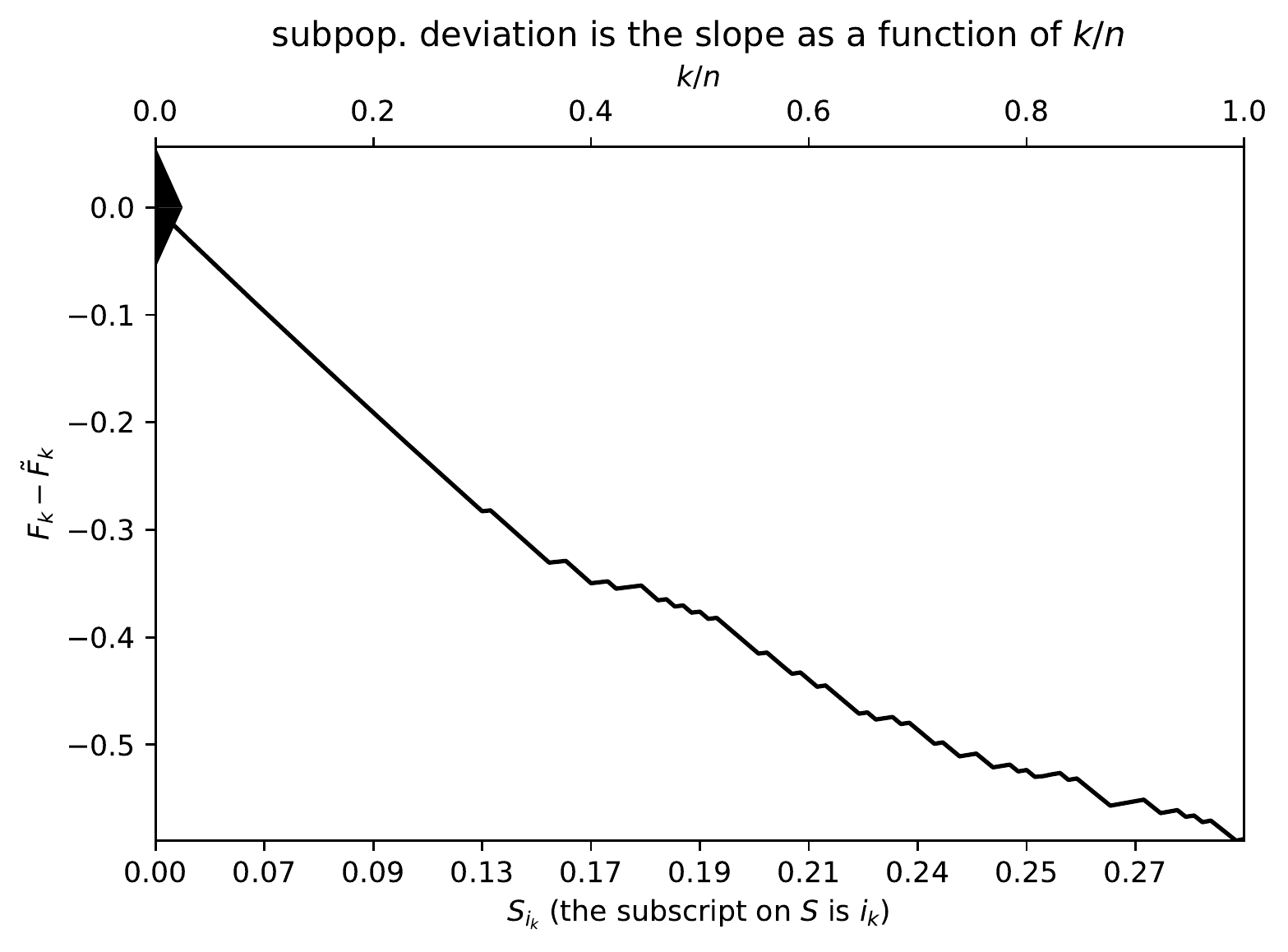}}

\parbox{\imsize}{\centering $n =$ 1,300; Kolmogorov-Smirnov and Kuiper
both $= 0.1343 / \sigma = 10.47$}
\quad\quad
\parbox{\imsize}{\centering $n = 130$;\ \ Kolmogorov-Smirnov and Kuiper
both $= 0.5893 / \sigma = 20.94$}

\end{centering}
\caption{Zooming in on the origin of some examples in which the scores
are the negative log-likelihoods;
the uppermost plots are for the Eskimo Dog or Husky
of Figure~\ref{eskimo-dog-husky-nll},
the middlemost plots are for the night snake (Hypsiglena torquata)
of Figure~\ref{night-snake-Hypsiglena-torquata-nll},
and the lowermost plots are for the sidewinder/horned rattlesnake
(Crotalus cerastes)
of Figure~\ref{sidewinder-horned-rattlesnake-Crotalus-cerastes-nll};
each plot on the right zooms in on the origin of the plot to its left,
adjusting the height of the triangle at the origin,
the number $n$ of observations,
and the Kuiper and Kolmogorov-Smirnov statistics to reflect the smaller range
of scores on the horizontal axis.}
\label{zoom}
\end{figure}

\subsection{American Community Survey of the U.S. Census Bureau}
\label{census}

\begin{figure}
\begin{centering}

\parbox{\imsize}{\includegraphics[width=\imsize]
{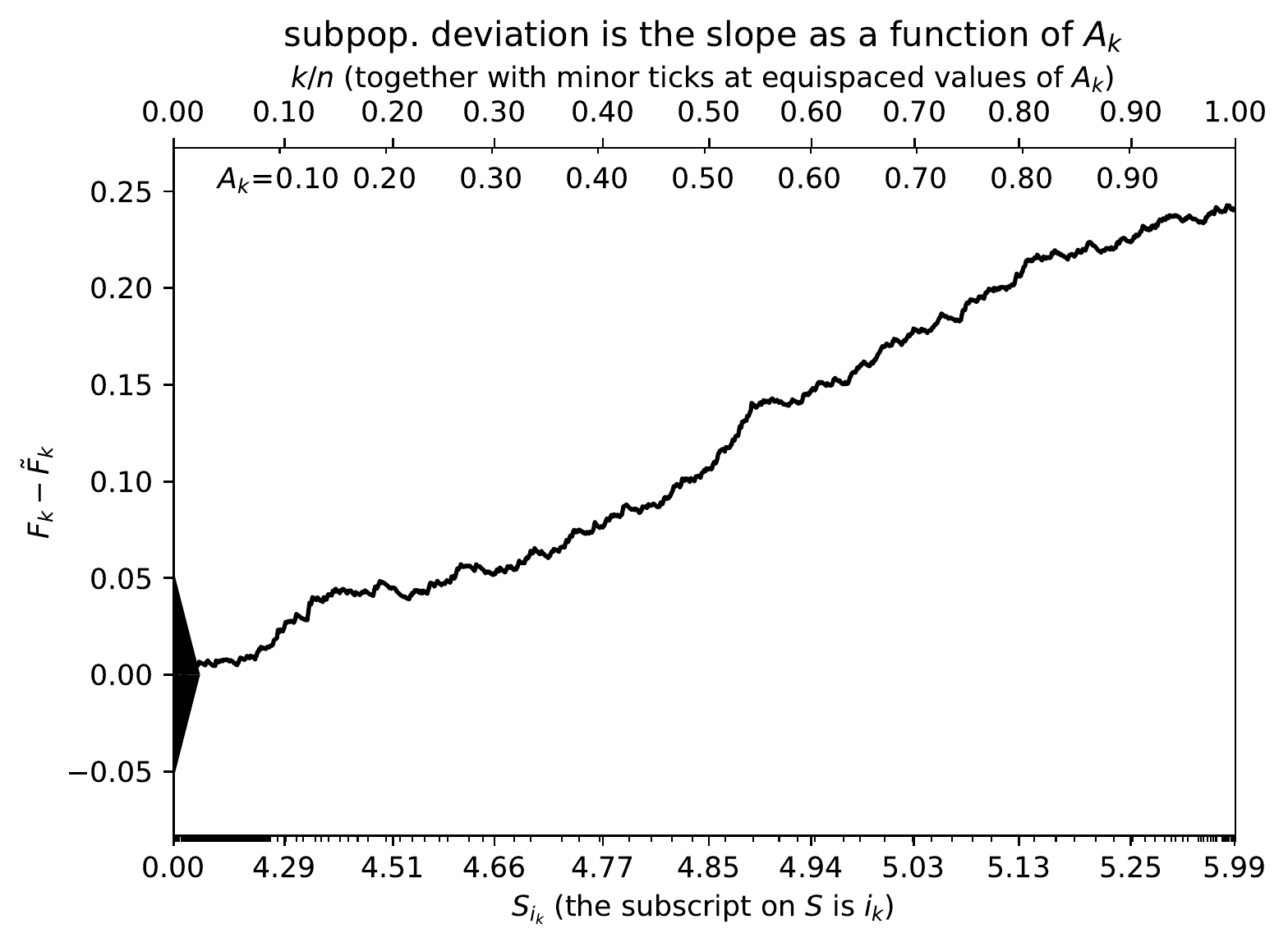}}
\quad\quad
\parbox{\imsize}{\includegraphics[width=\imsize]
{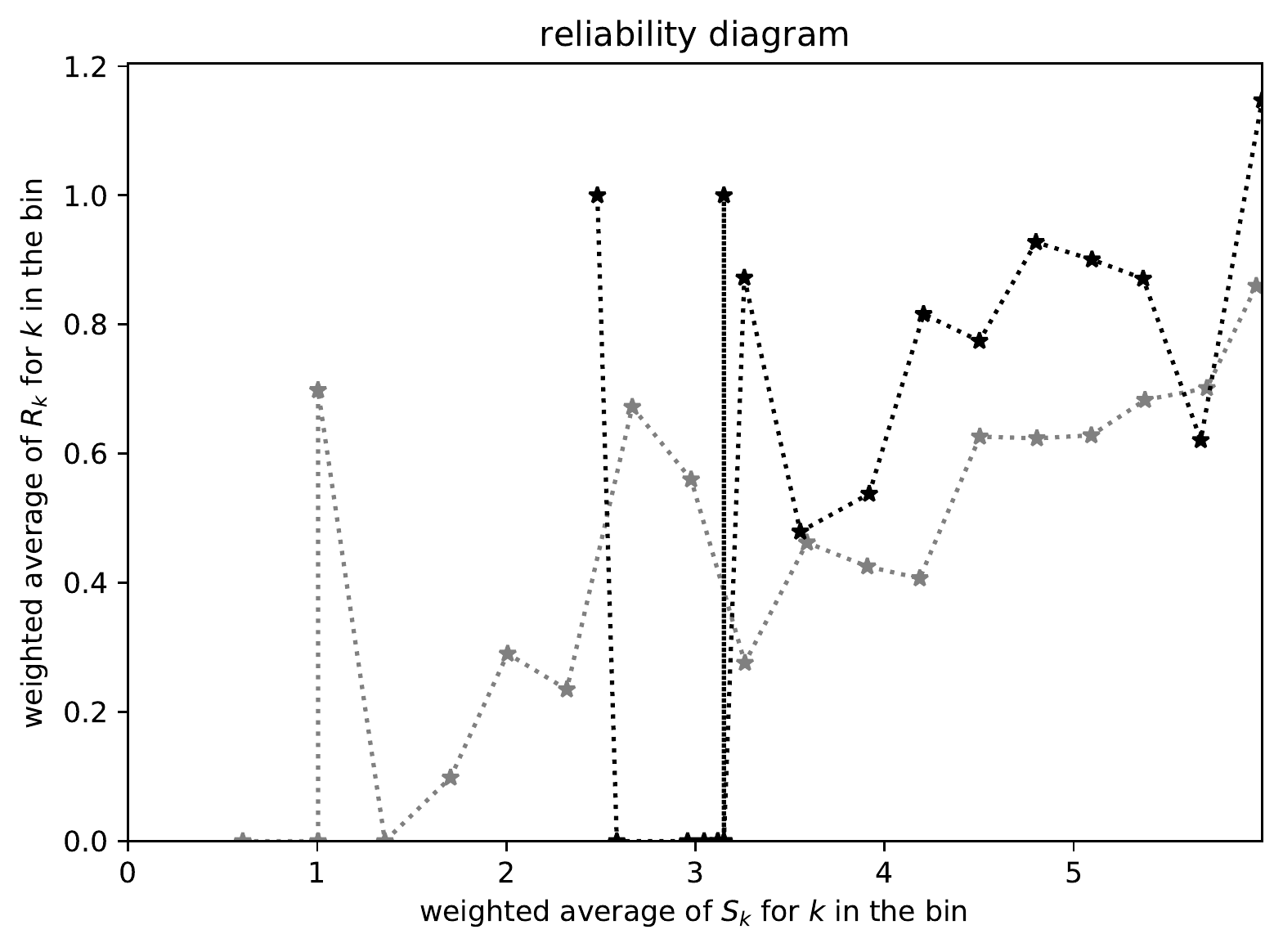}}

\vspace{\vertsep}

\parbox{\imsize}{\includegraphics[width=\imsize]
{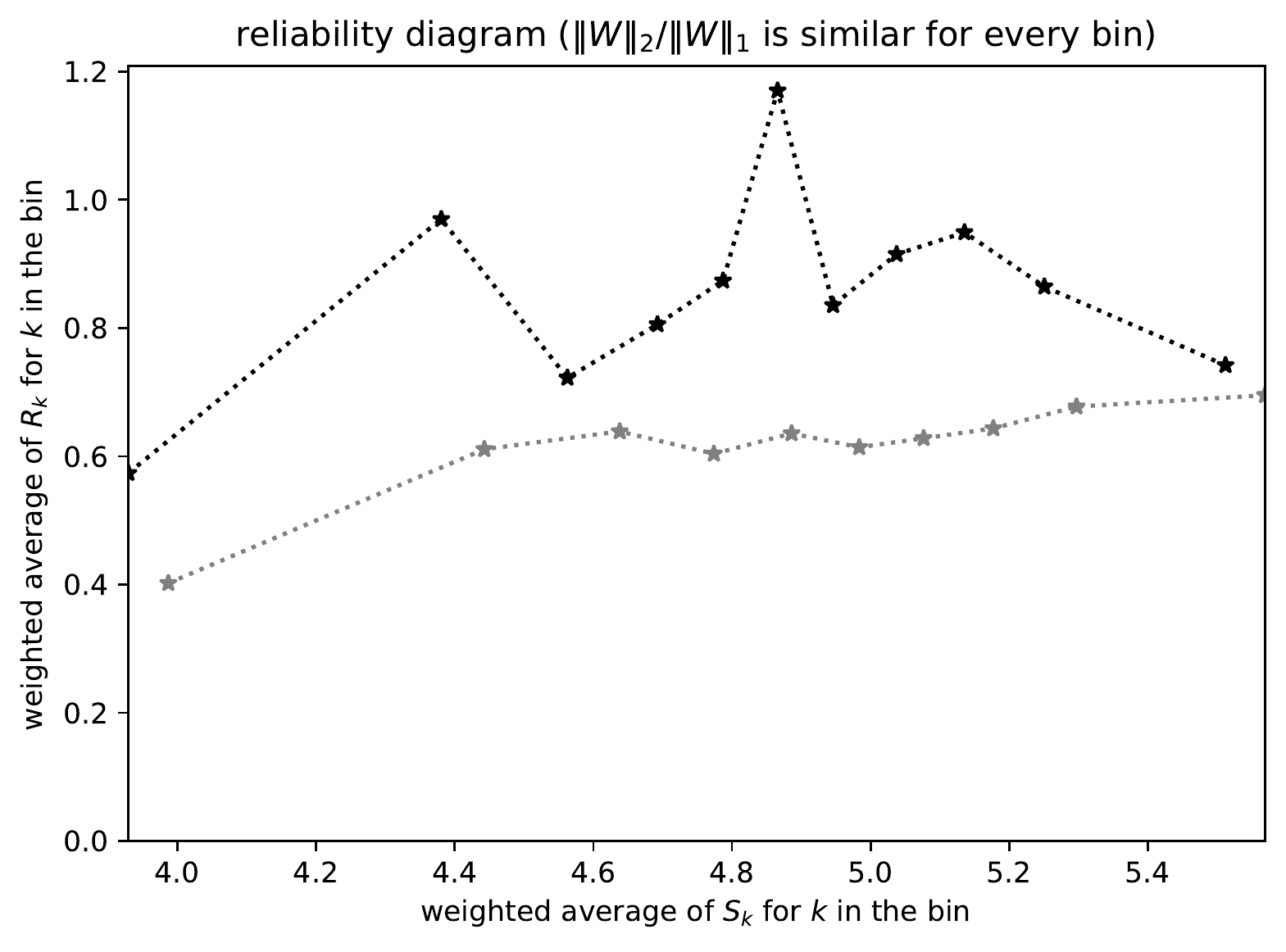}}
\quad\quad
\parbox{\imsize}{\includegraphics[width=\imsize]
{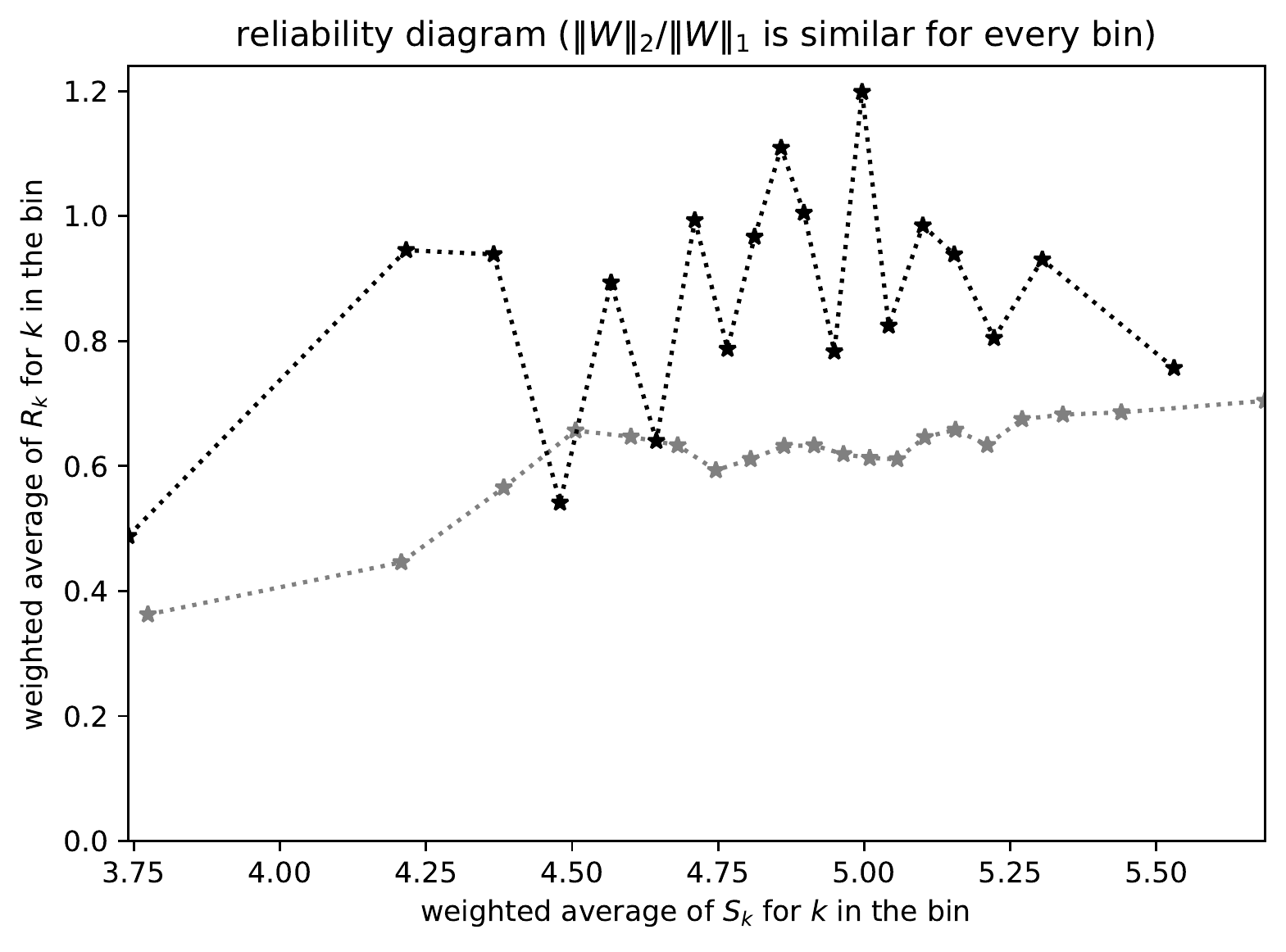}}

\vspace{\vertsep}

\parbox{\imsize}{\includegraphics[width=\imsize]
{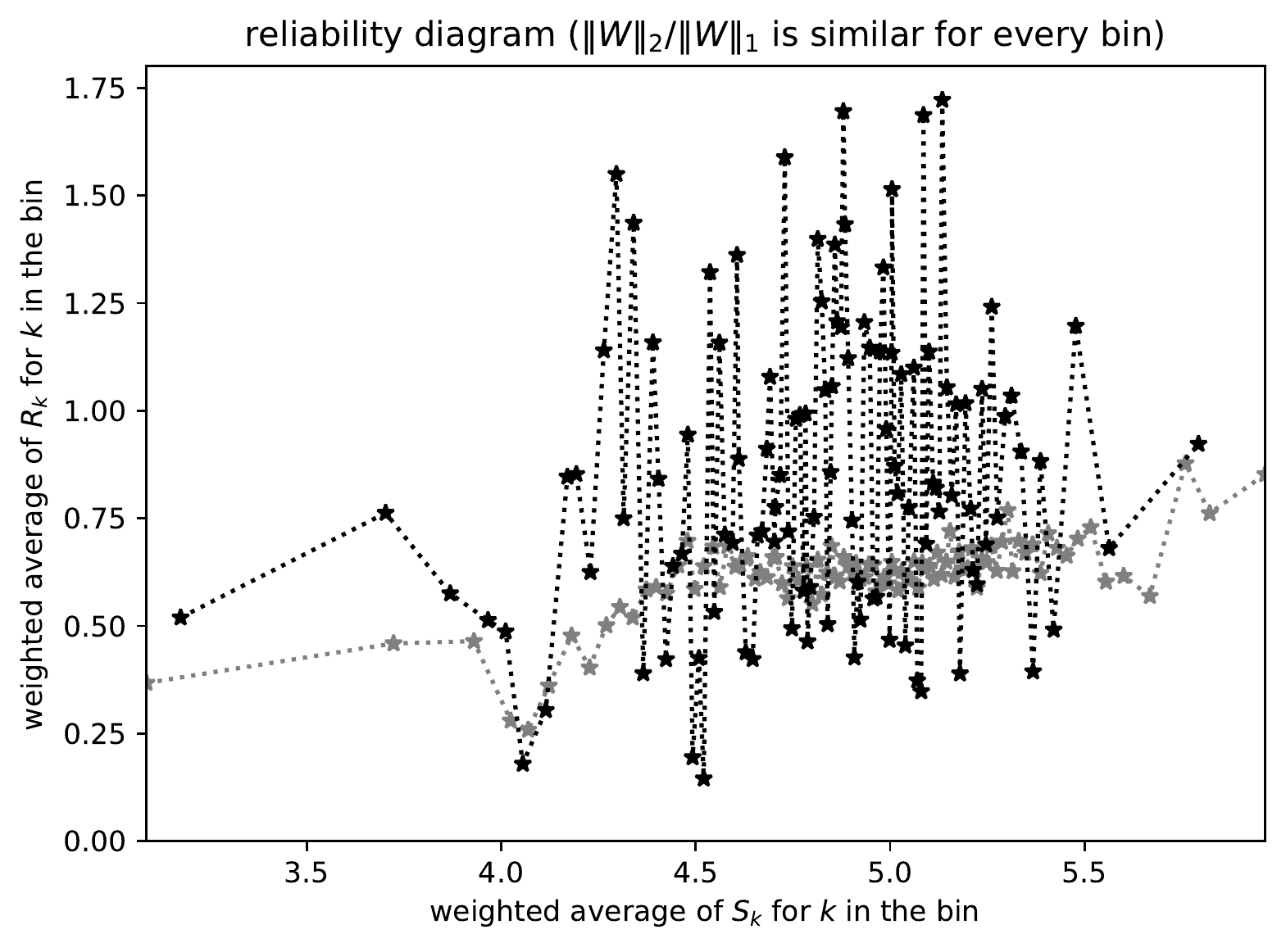}}
\quad\quad
\parbox{\imsize}{\includegraphics[width=\imsize]
{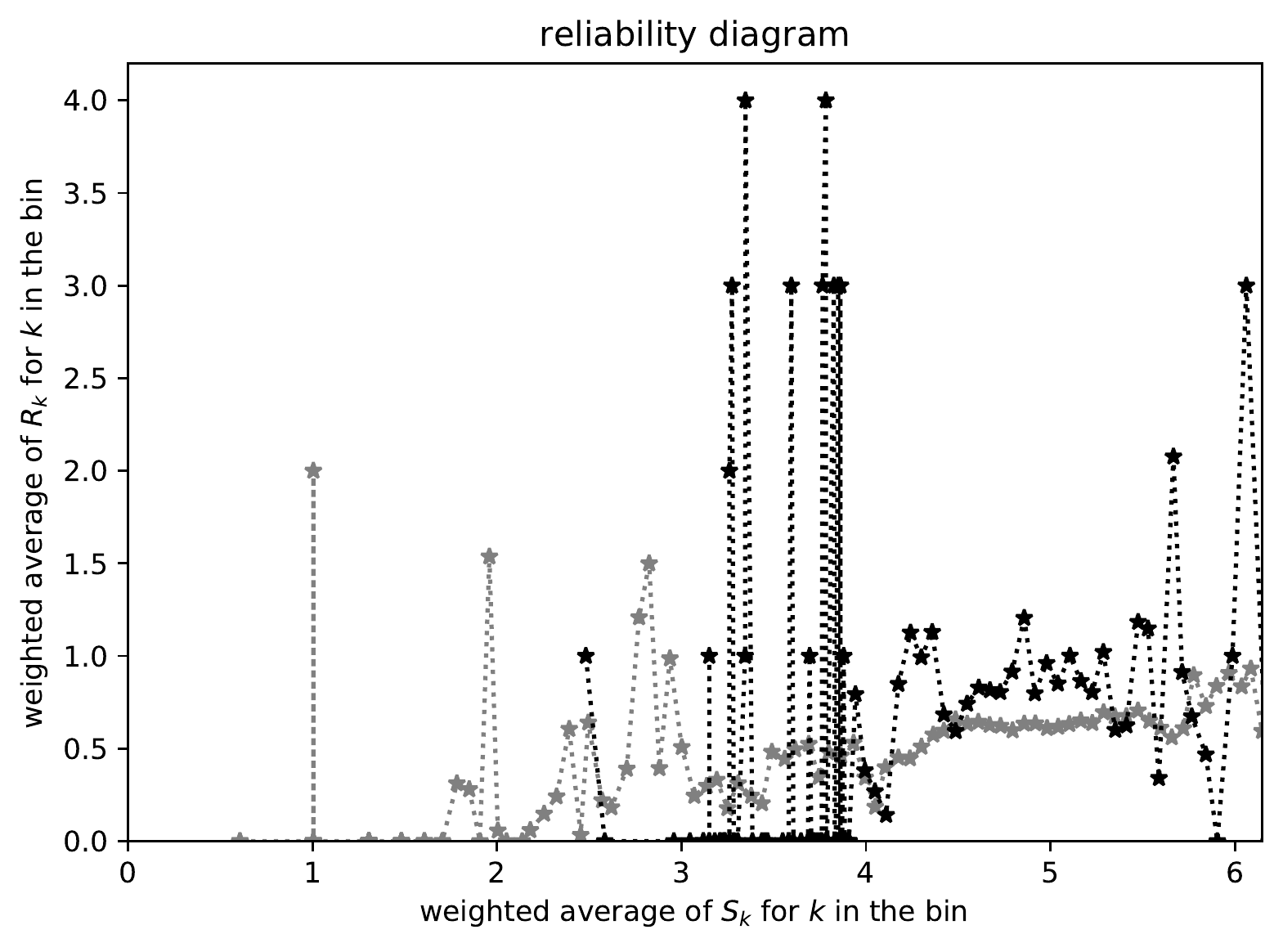}}

\end{centering}
\caption{San Joaquin County, reporting the number of related children
         in the household, with scores being $\log_{10}$
         of the adjusted household income;
         $n =$ 2,282; Kuiper's statistic is $0.2449 / \sigma = 9.120$,
         Kolmogorov's and Smirnov's is $0.2429 / \sigma = 9.045$.
The lack of deviation between the subpopulation
and the full population right near scores of 4.0 is difficult to discern
in the reliability diagrams with 10 or 20 bins each. The reliability diagrams
with around 100 bins each do display the lack of deviation near scores of 4.0,
but the rest of these diagrams is really noisy.
The cumulative plot nicely captures the lack of deviation near scores of 4.0.
Overall, the scalar summary statistics detect
highly statistically significant deviation.
}
\label{san_joaquin}
\end{figure}

\begin{figure}
\begin{centering}

\parbox{\imsize}{\includegraphics[width=\imsize]
{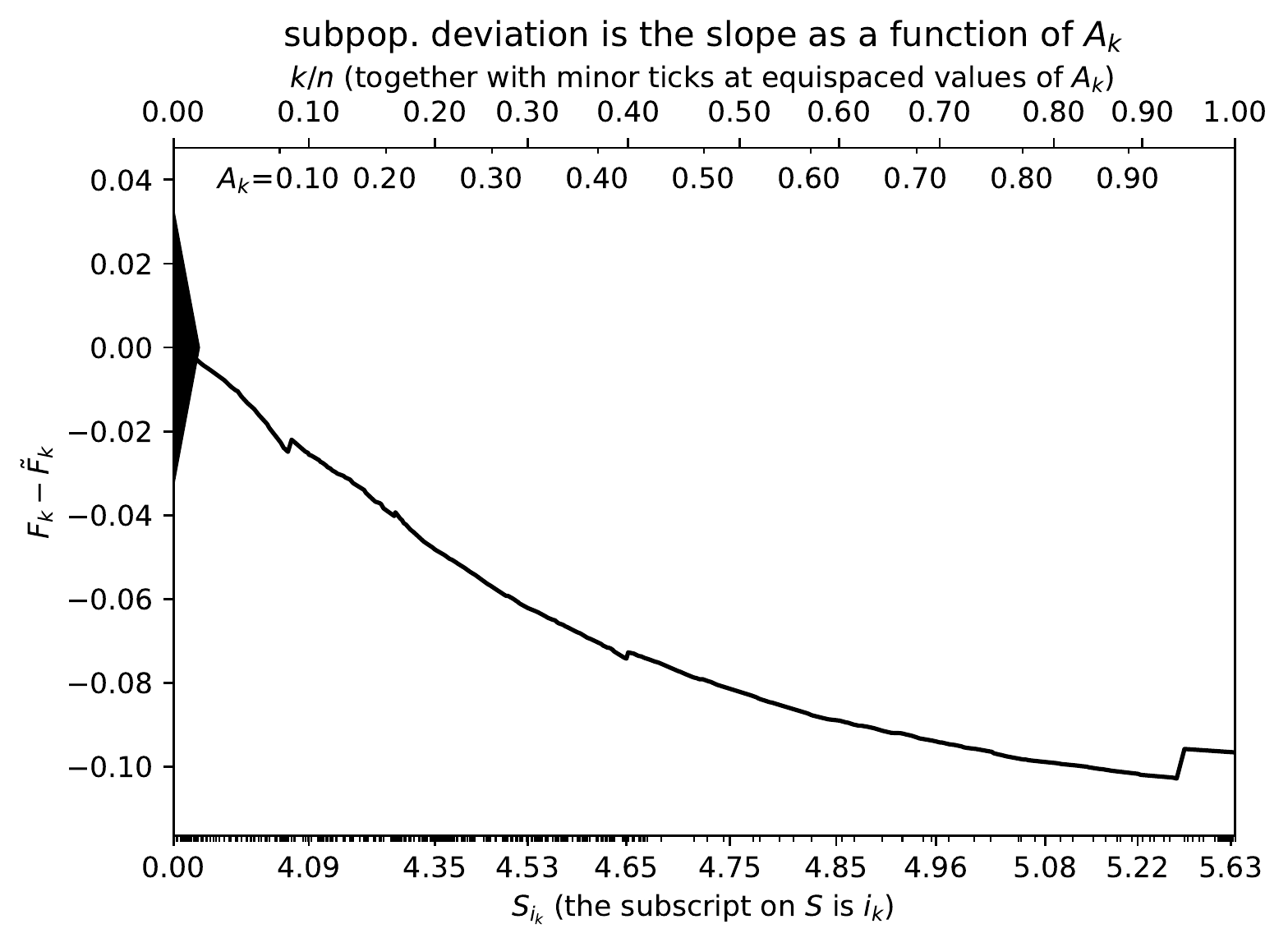}}
\quad\quad
\parbox{\imsize}{\includegraphics[width=\imsize]
{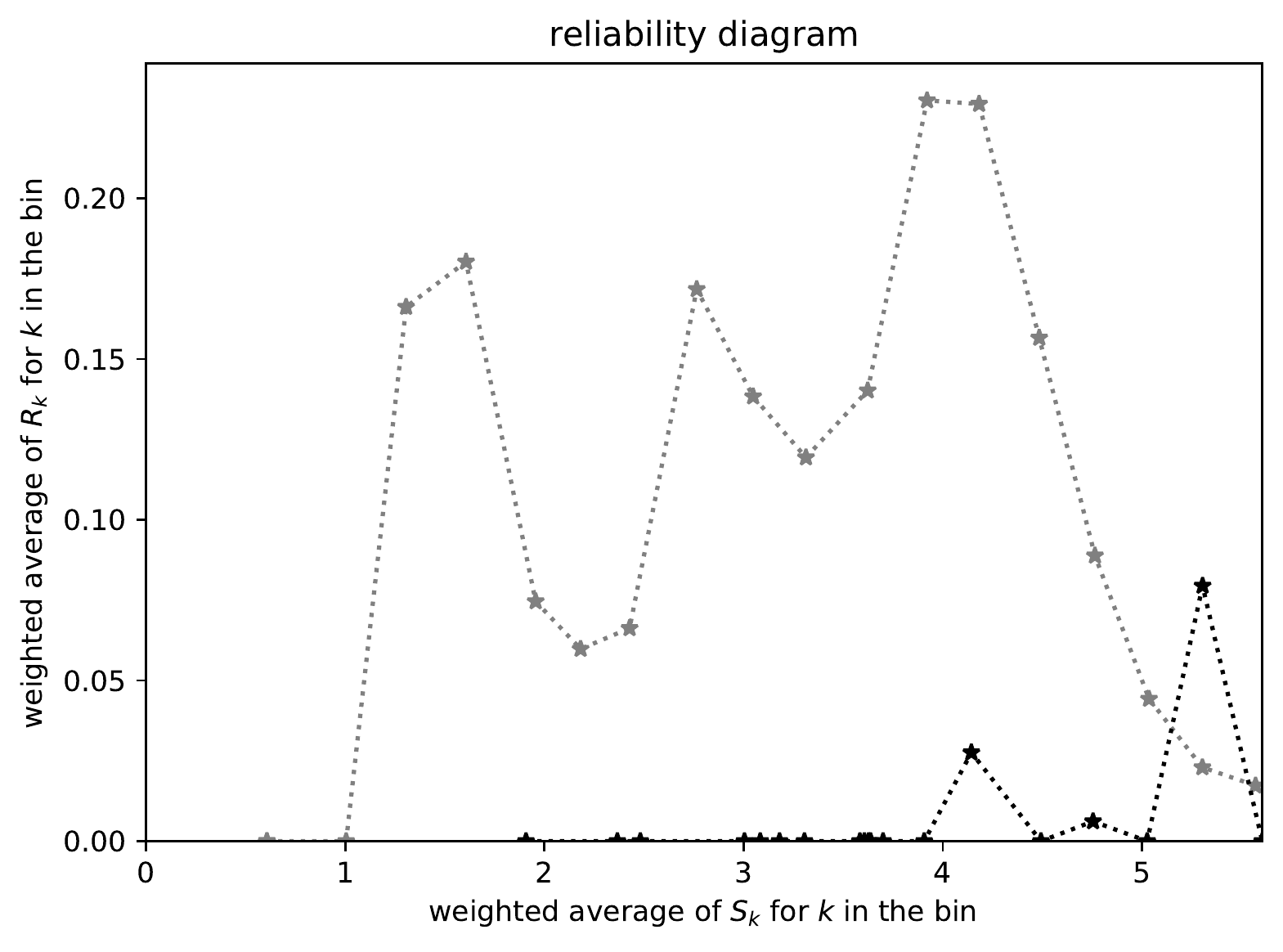}}

\vspace{\vertsep}

\parbox{\imsize}{\includegraphics[width=\imsize]
{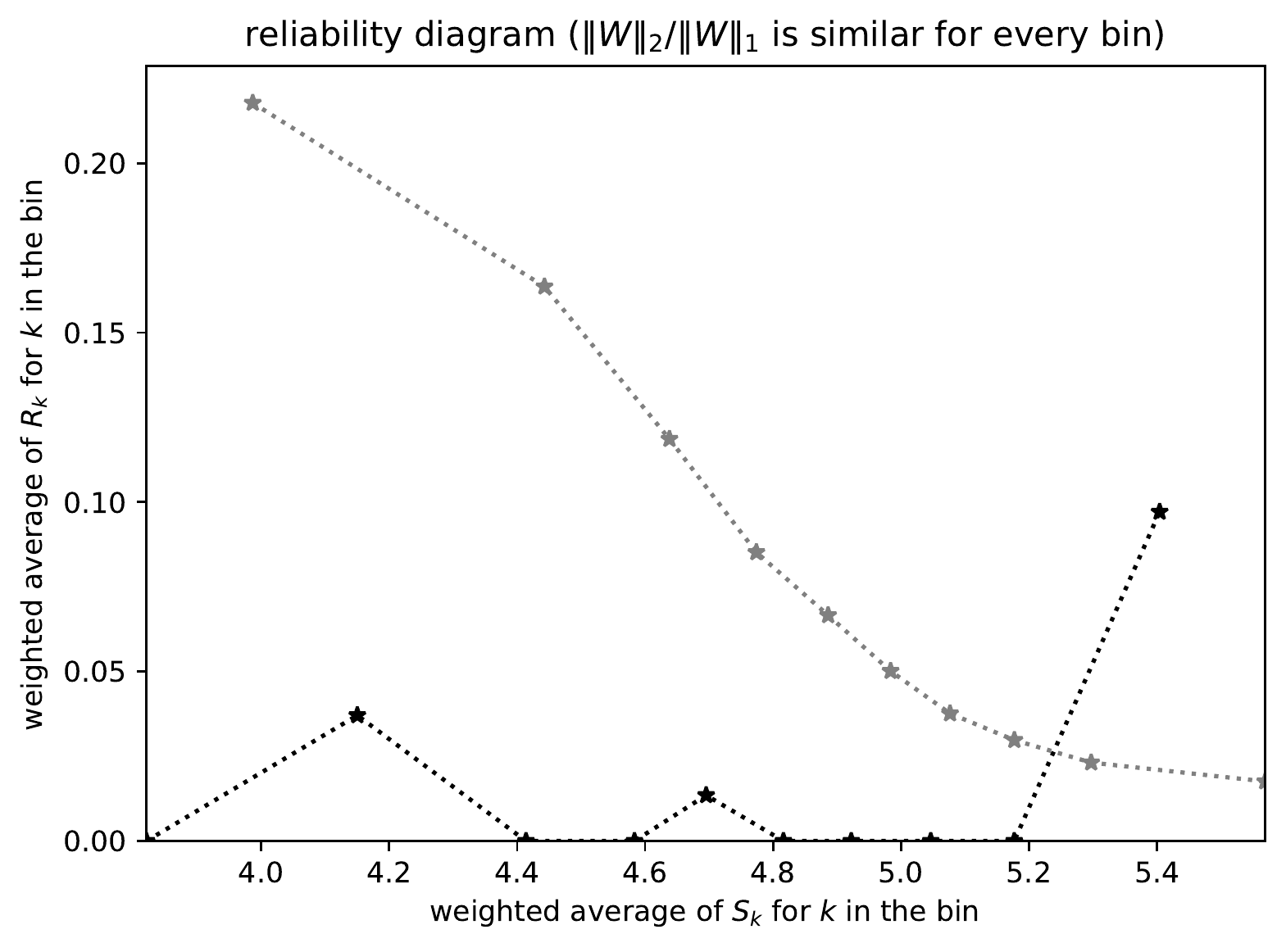}}
\quad\quad
\parbox{\imsize}{\includegraphics[width=\imsize]
{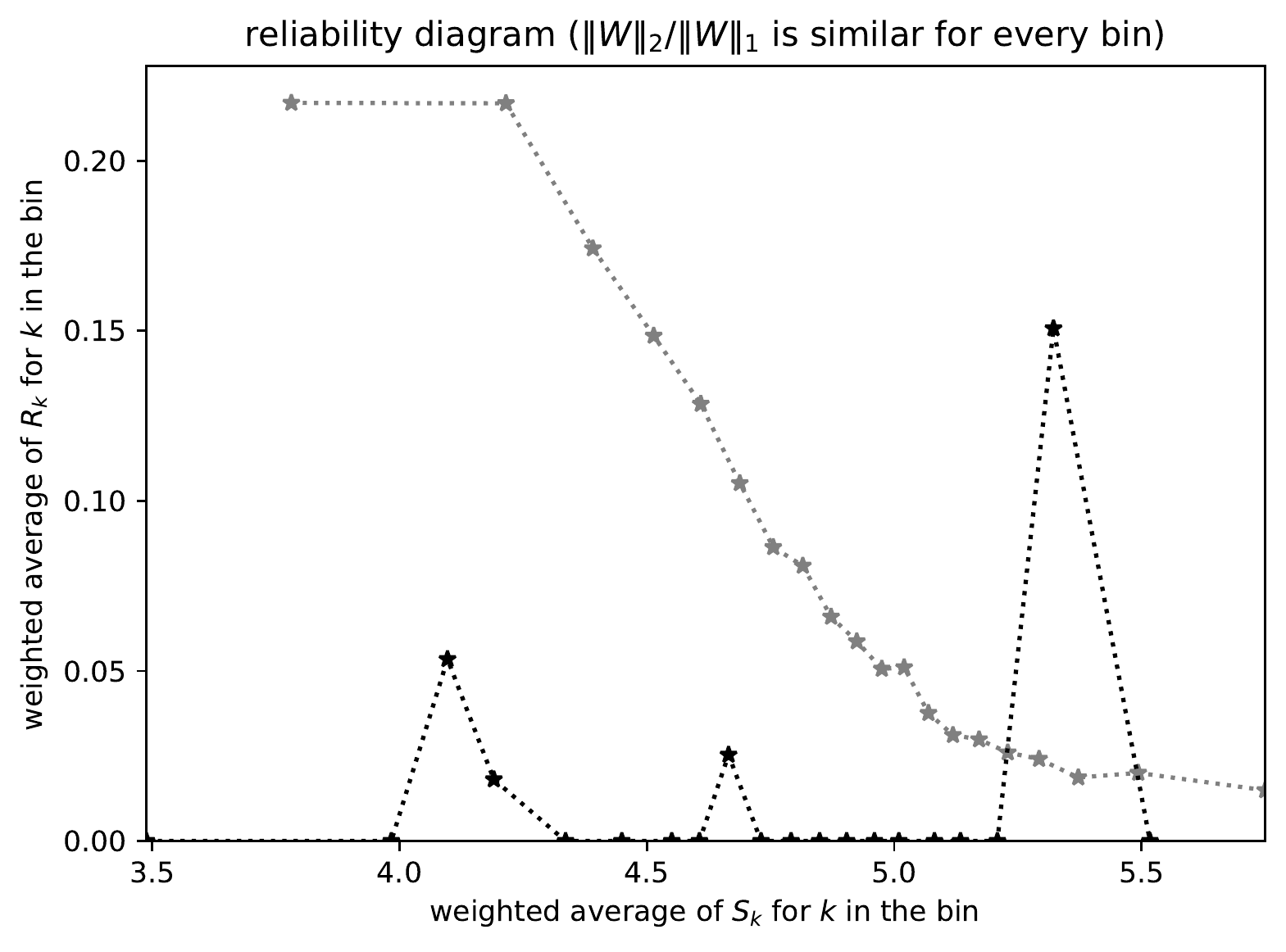}}

\end{centering}
\caption{Humboldt County, reporting whether no one in the household 14 or over
         speaks English very well, with scores being $\log_{10}$
         of the adjusted household income;
         $n =$ 583; Kuiper's statistic is $0.1028 / \sigma = 6.062$,
         Kolmogorov's and Smirnov's is $0.1028 / \sigma = 6.062$.
A single, highly weighted outlying observation
from the subpopulation corrupts a bin at the highest scores
in each reliability diagram. The cumulative plot
includes this outlying observation, too, but displays the observation as
an unmistakable steep jump in the plotted curve; the constant slope
of that steep jump shows that the corresponding high deviation
between the subpopulation and the full population is due
to a single highly weighted observation. This single observation has no effect
on the slopes in the rest of the cumulative plot,
so this problematic observation corrupts only the reliability diagrams
and not the plot of cumulative differences
--- an analogue of Simpson's Paradox of~\cite{simpson}
afflicts the reliability diagrams but not the cumulative plot.
The scalar summary statistics report highly statistically significant deviation
commensurate with all the plots, albeit slightly less due to the steep jump
in the cumulative plot. These behaviors are analogous to those displayed
in Figure~\ref{2500w} above (see also Remark~\ref{Simpson} above).
}
\label{humboldt}
\end{figure}

\begin{figure}
\begin{centering}

\parbox{\imsize}{\includegraphics[width=\imsize]
{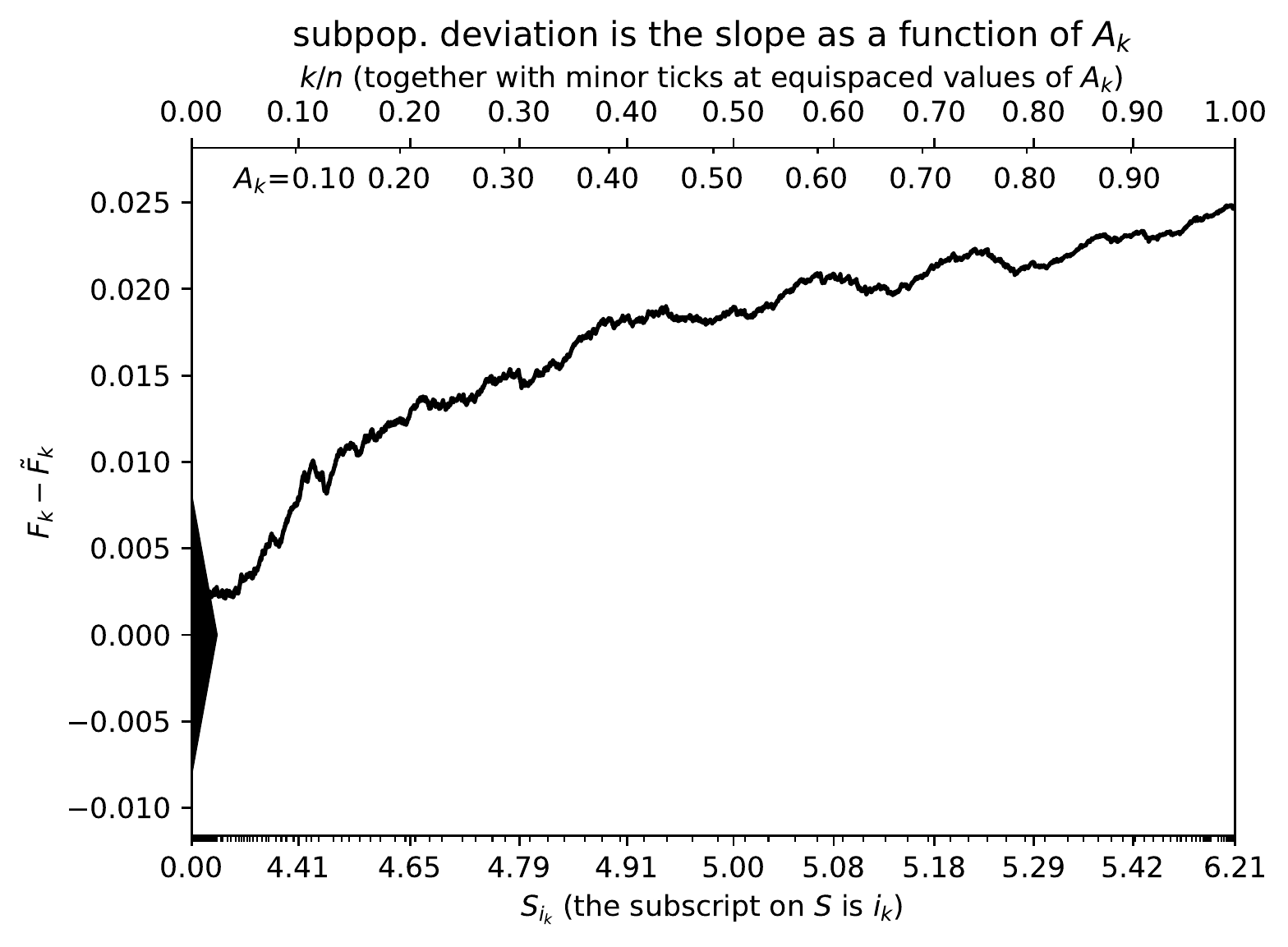}}
\quad\quad
\parbox{\imsize}{\includegraphics[width=\imsize]
{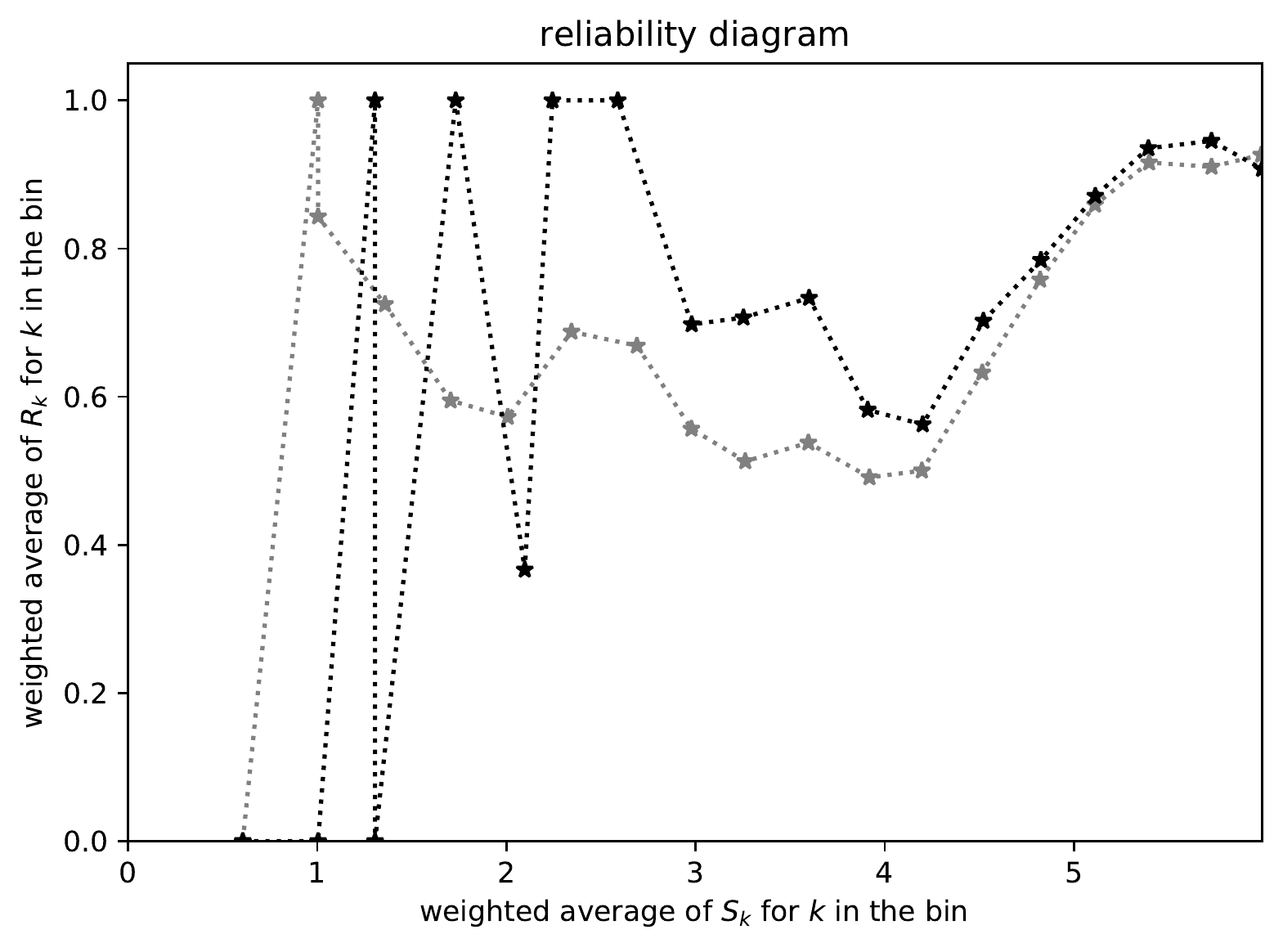}}

\vspace{\vertsep}

\parbox{\imsize}{\includegraphics[width=\imsize]
{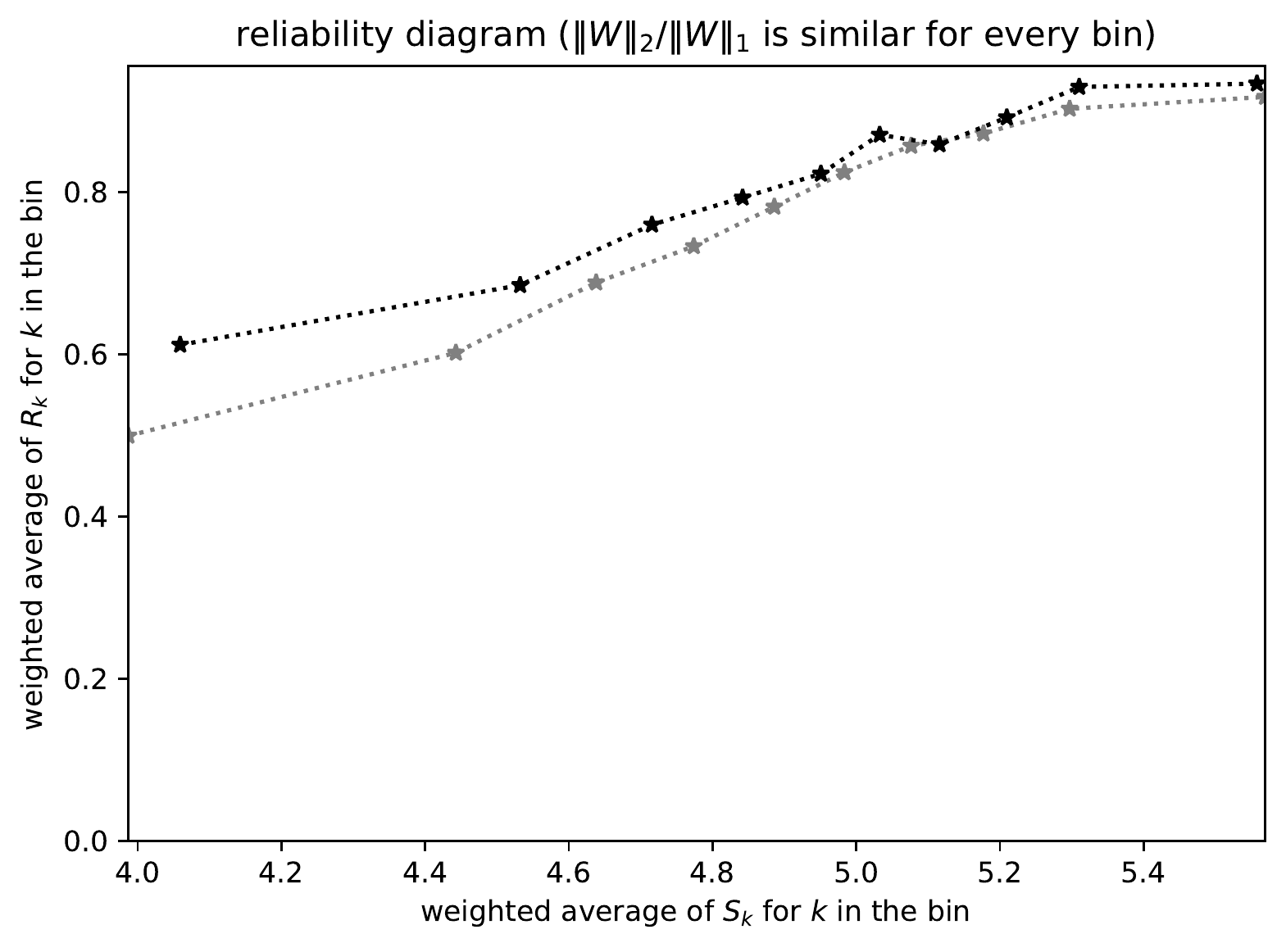}}
\quad\quad
\parbox{\imsize}{\includegraphics[width=\imsize]
{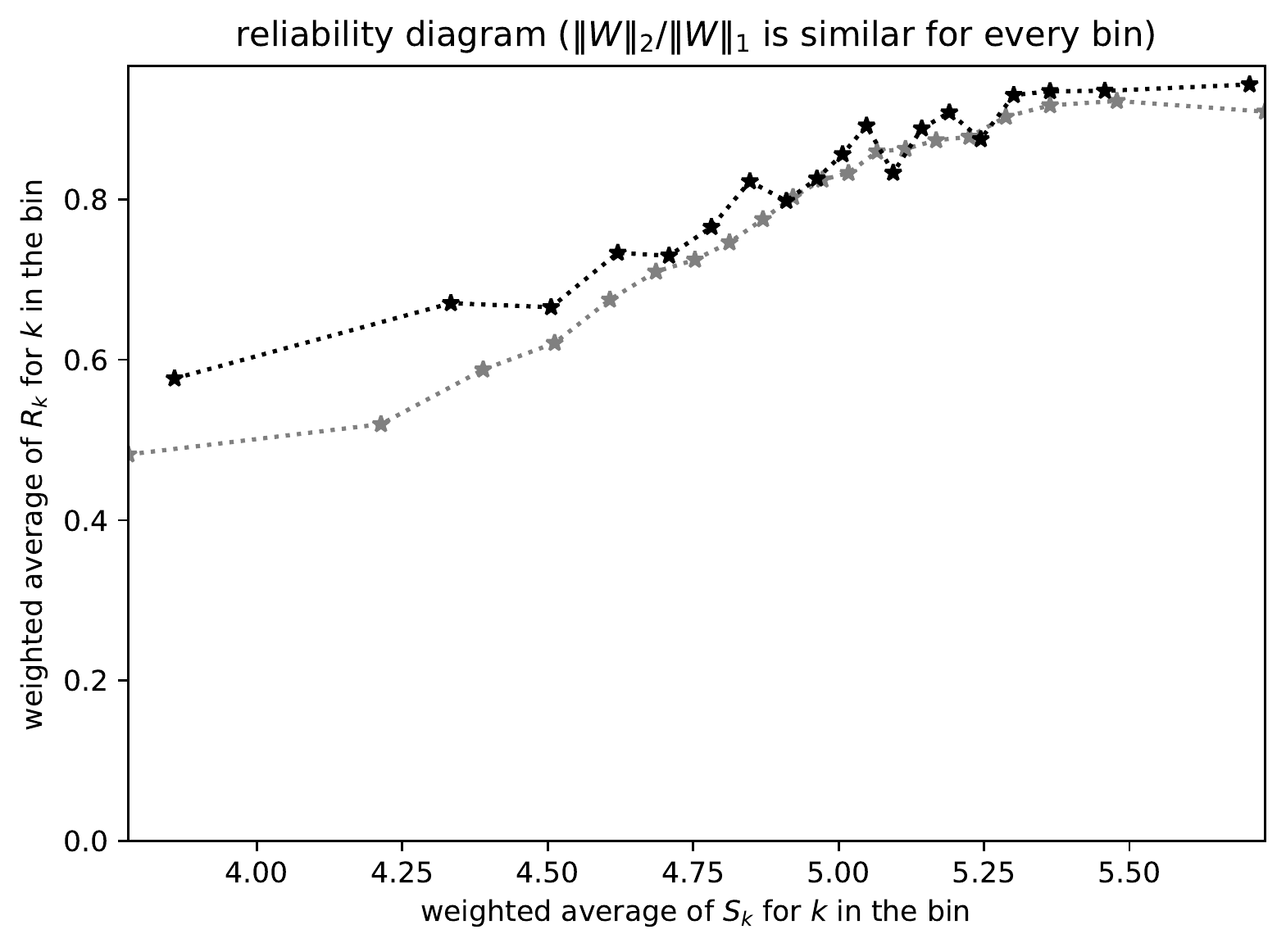}}

\vspace{\vertsep}

\parbox{\imsize}{\includegraphics[width=\imsize]
{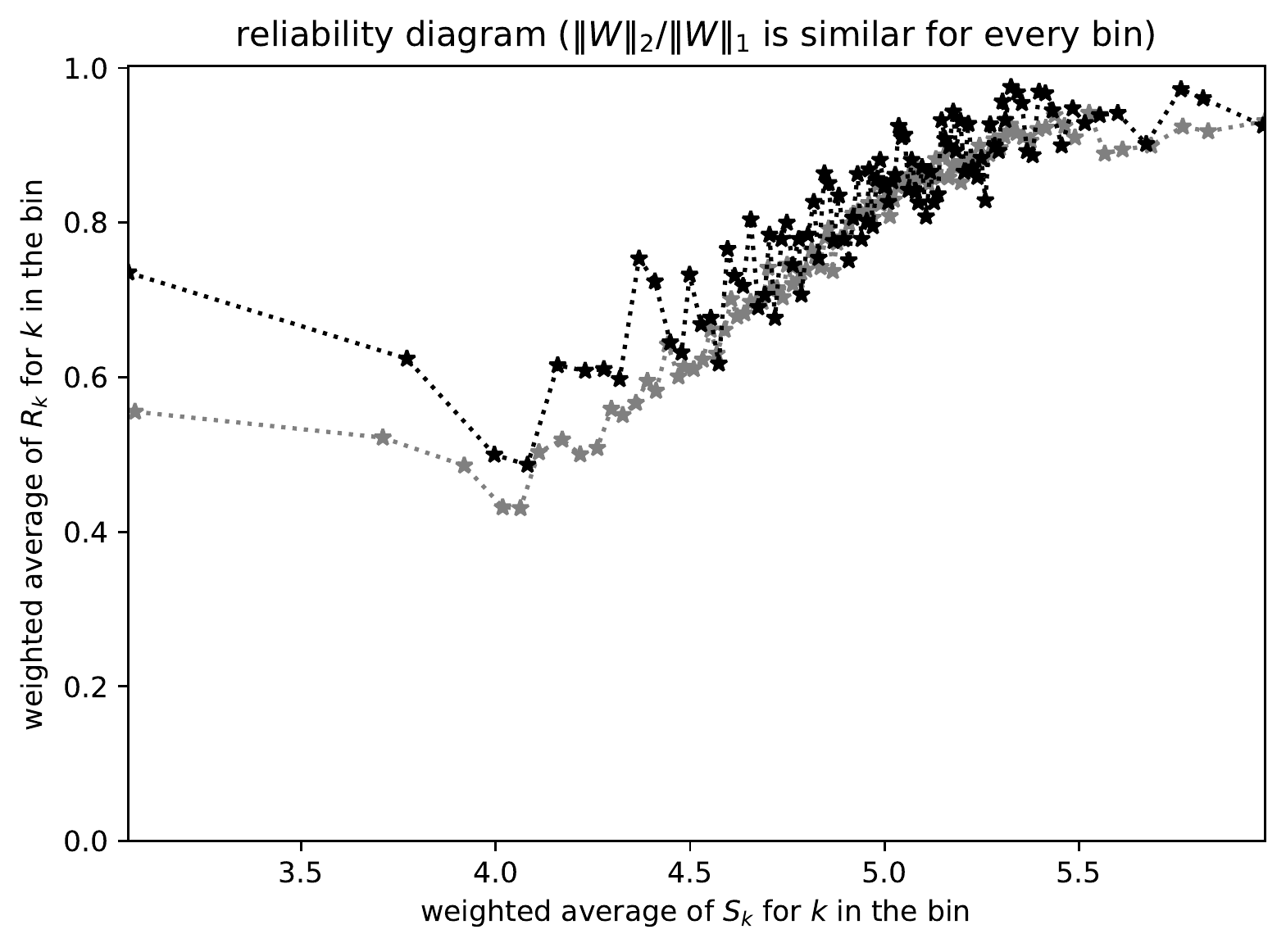}}
\quad\quad
\parbox{\imsize}{\includegraphics[width=\imsize]
{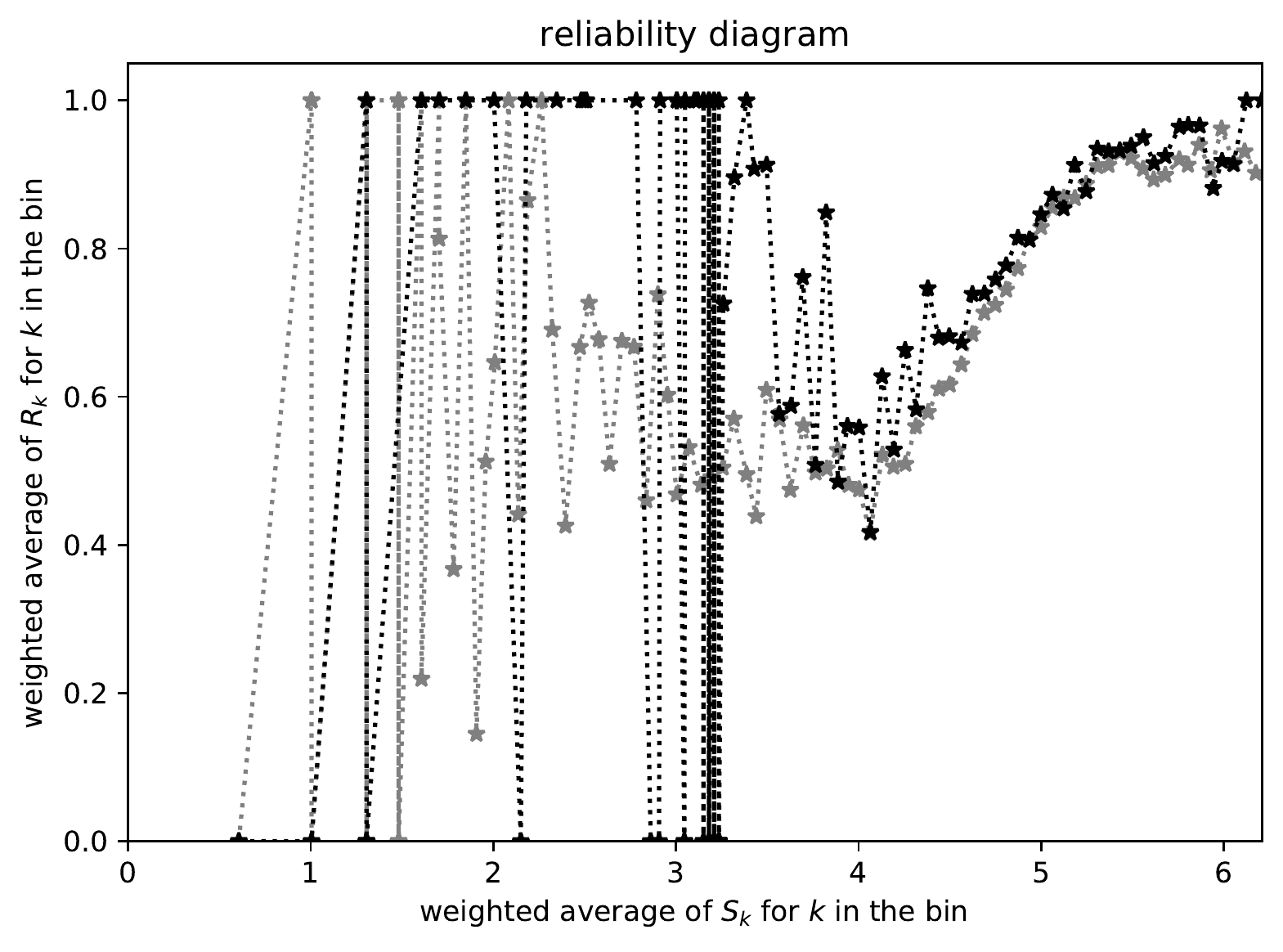}}

\end{centering}
\caption{Orange County, reporting whether the household has high-speed access
         to the internet, with scores being $\log_{10}$
         of the adjusted household income; $n =$ 10,680;
         Kuiper's statistic is $0.02499 / \sigma = 6.013$,
         Kolmogorov's and Smirnov's is $0.02484 / \sigma = 5.979$.
The severe deviation at the very lowest scores
is misleadingly underestimated in the reliability diagrams
with 10 or 20 bins each, even as compared to the diagrams
with around 100 bins each. However, the diagrams with around 100 bins each
are far too noisy for other scores. Only the plot of cumulative differences
resolves the severe deviation at the lowest scores while being informative
at the other scores, too. The scalar summary statistics indicate
that the overall deviation is highly statistically significant.
}
\label{orange}
\end{figure}

\begin{figure}
\begin{centering}

\parbox{\imsize}{\includegraphics[width=\imsize]
{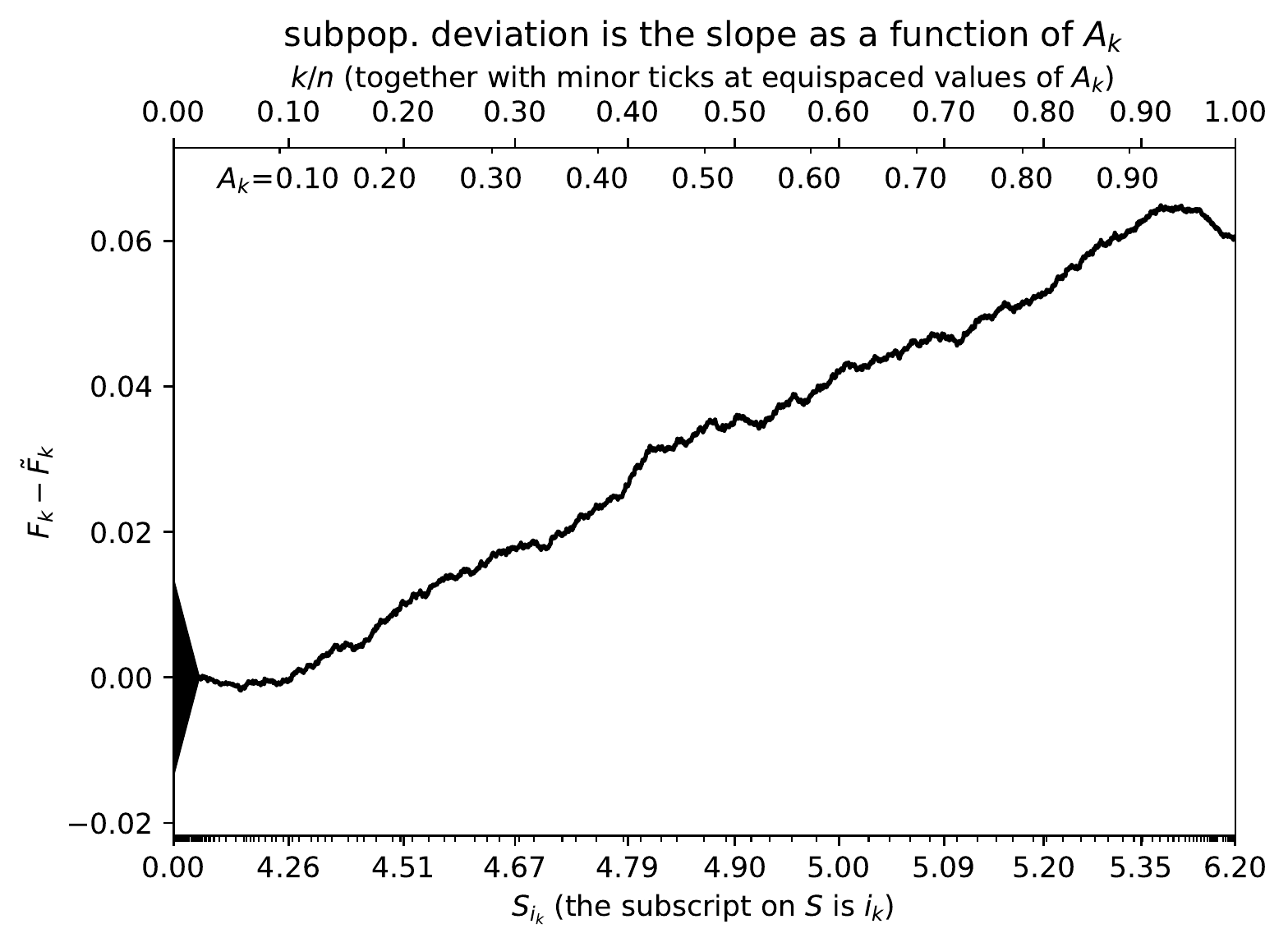}}
\quad\quad
\parbox{\imsize}{\includegraphics[width=\imsize]
{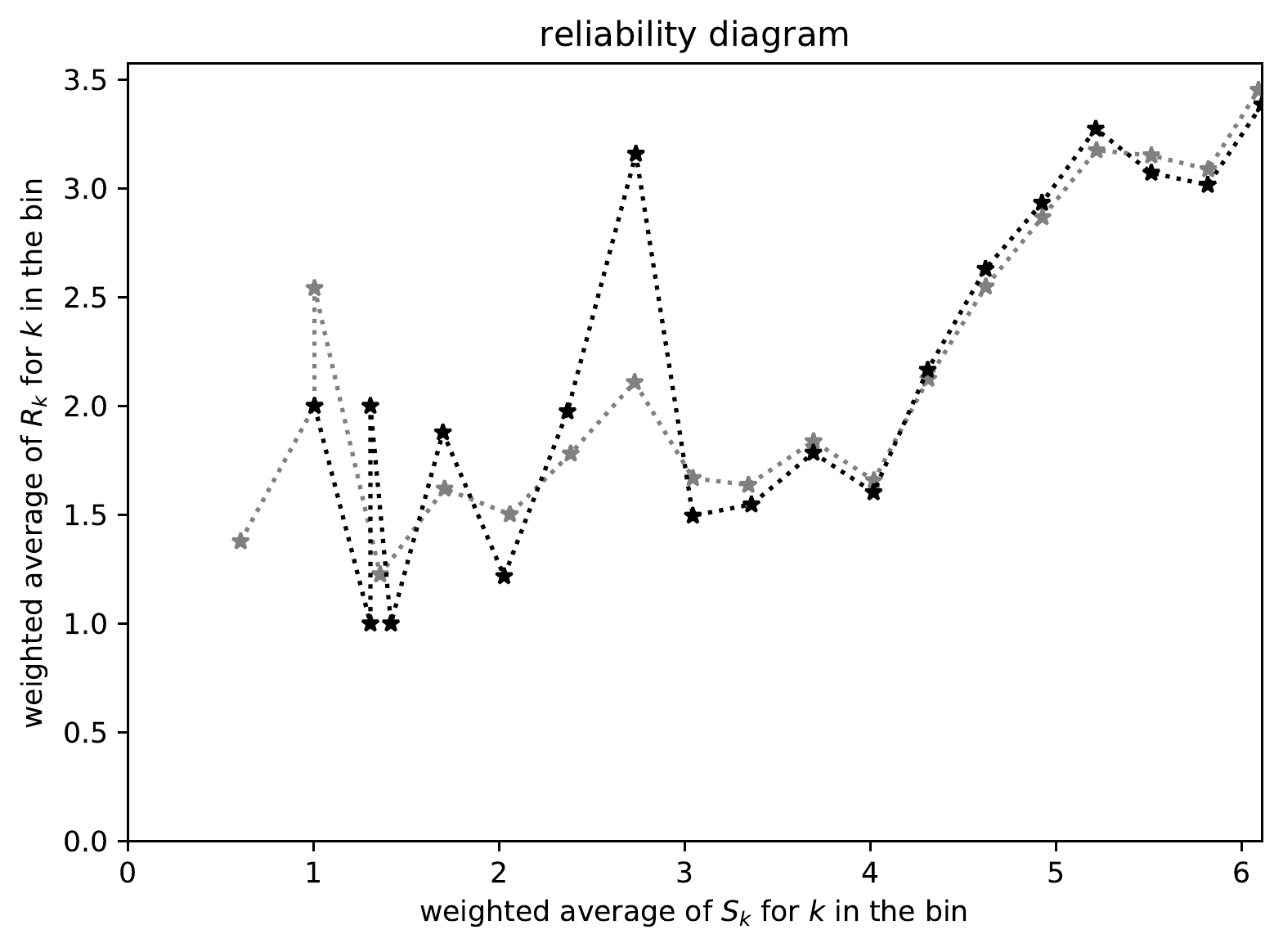}}

\vspace{\vertsep}

\parbox{\imsize}{\includegraphics[width=\imsize]
{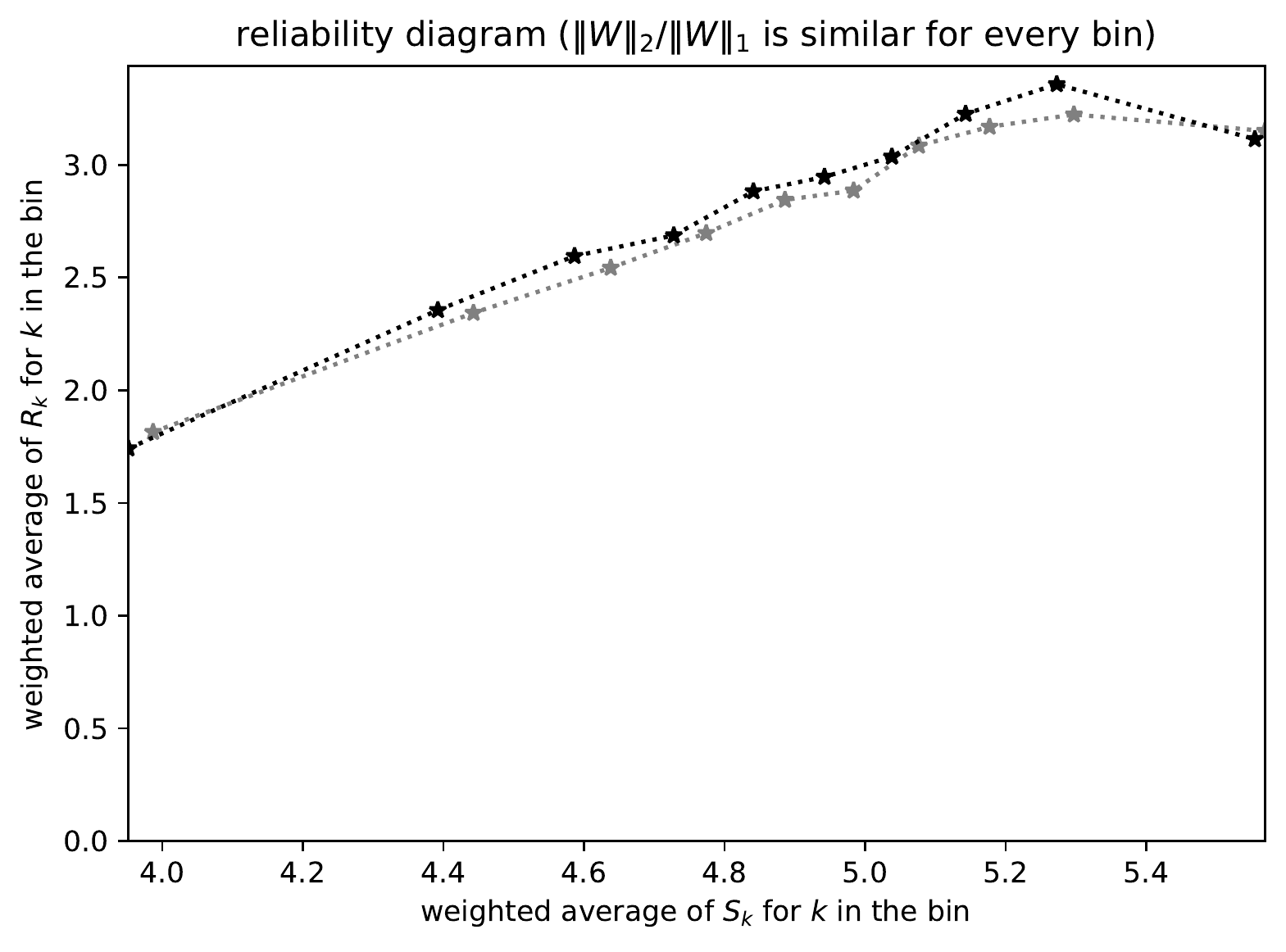}}
\quad\quad
\parbox{\imsize}{\includegraphics[width=\imsize]
{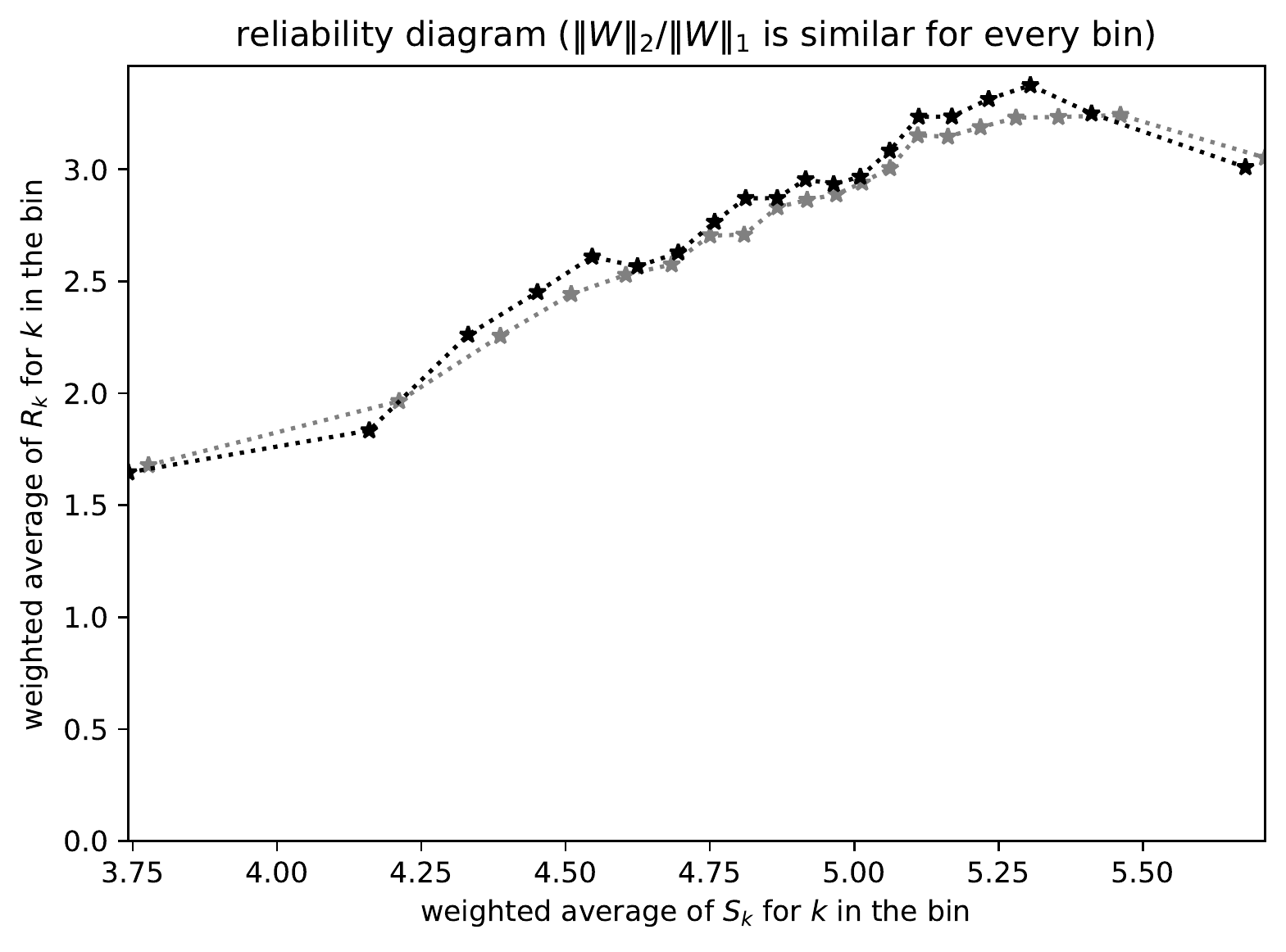}}

\vspace{\vertsep}

\parbox{\imsize}{\includegraphics[width=\imsize]
{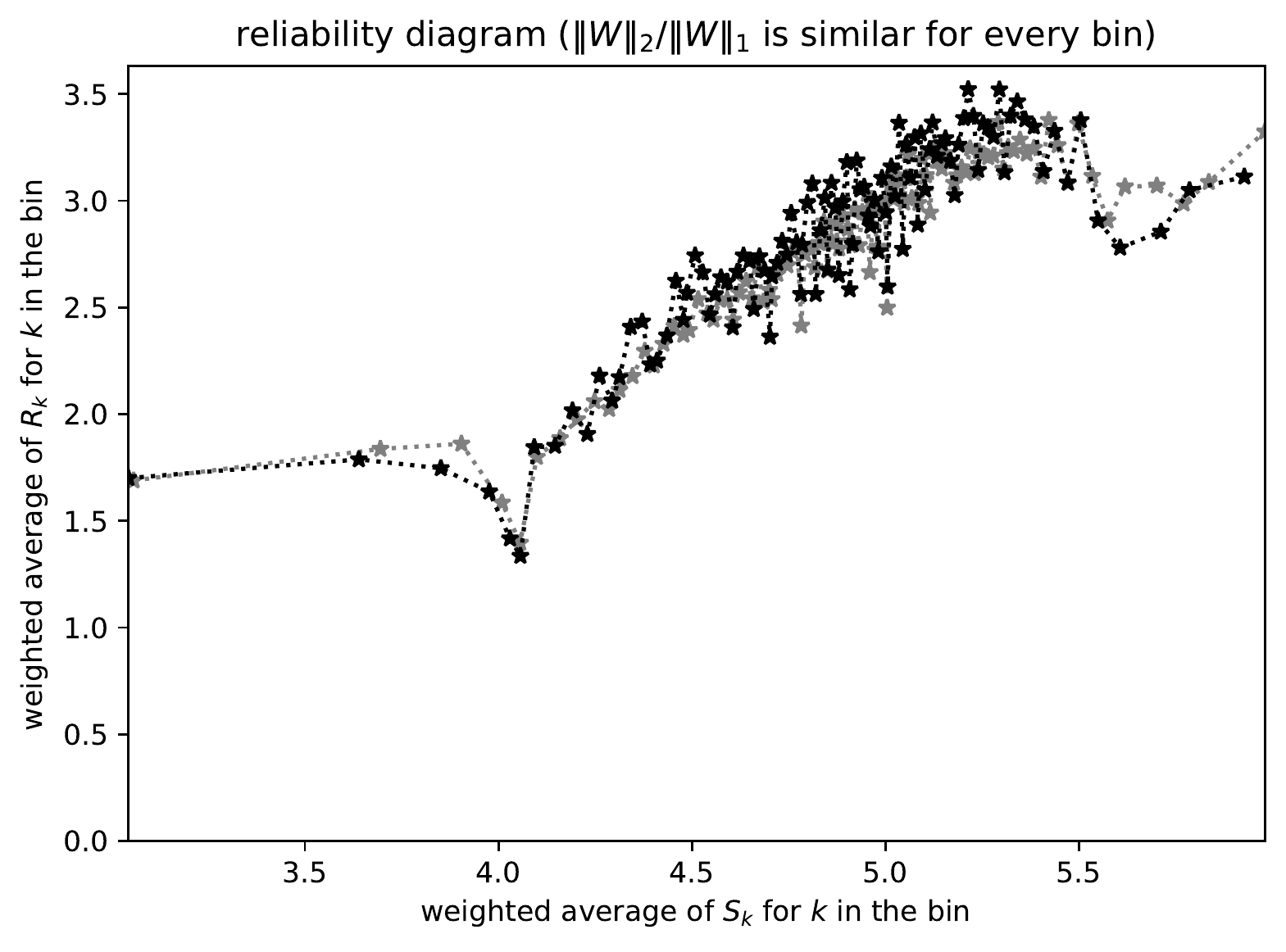}}
\quad\quad
\parbox{\imsize}{\includegraphics[width=\imsize]
{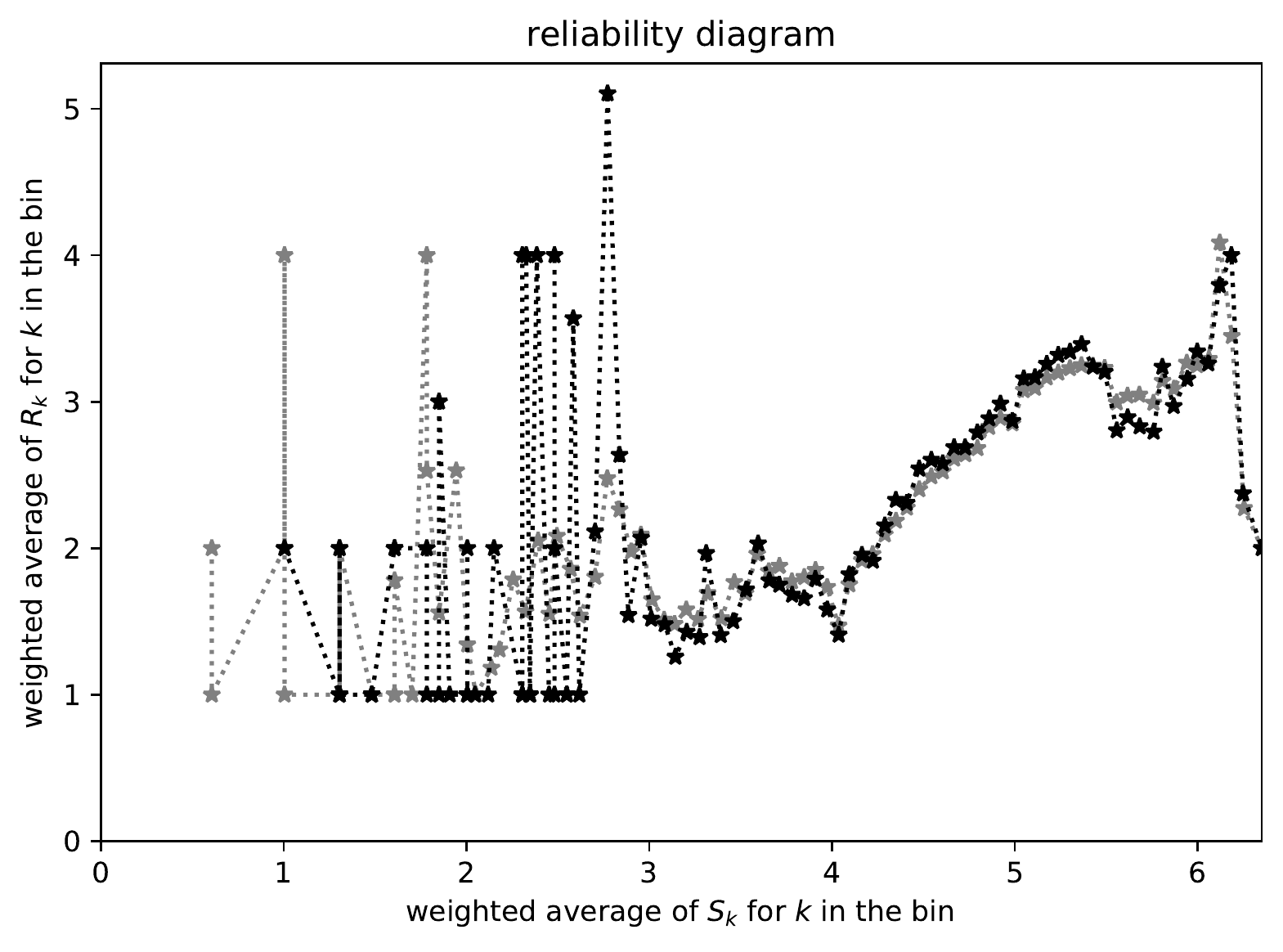}}

\end{centering}
\caption{Los Angeles County, reporting the number of people in the household,
         with scores being $\log_{10}$ of the adjusted household income;
         $n =$ 35,364; Kuiper's statistic is $0.06674 / \sigma = 9.605$,
         Kolmogorov's and Smirnov's is $0.06495 / \sigma = 9.347$.
Discerning the analogue of the dip in the plot
of cumulative differences at the highest scores is possible yet difficult
in the reliability diagrams, while being unmistakable in the cumulative plot;
the reliability diagrams with enough bins do reflect
the corresponding deviation at the highest scores, but are hard to interpret
without the accompanying cumulative plot. The scalar summary statistics
report very highly statistically significant deviation,
largely since the number of observations from this largest county in California
is so large.
}
\label{los_angeles}
\end{figure}

\begin{figure}
\begin{centering}

\parbox{\imsize}{\includegraphics[width=\imsize]
{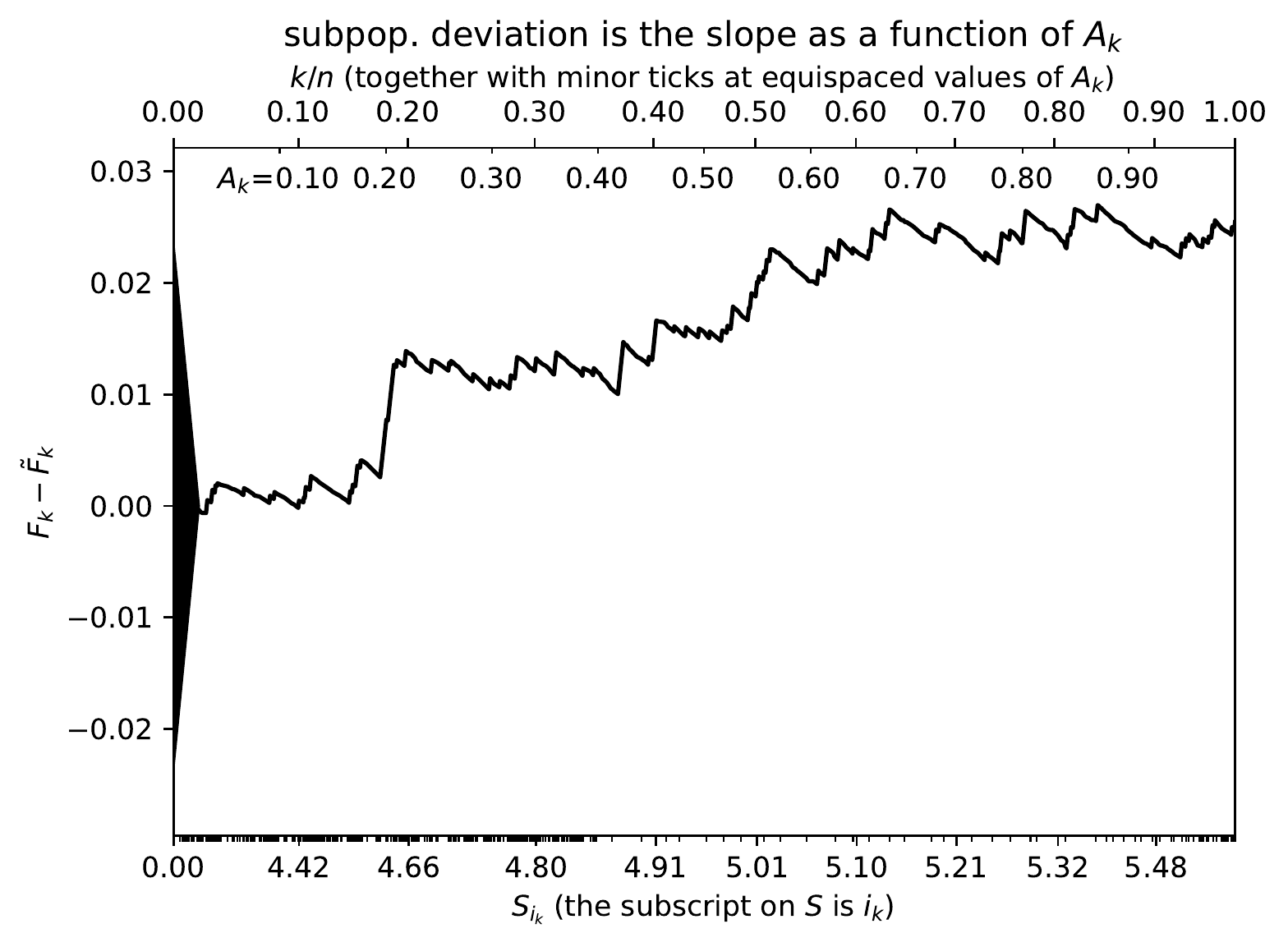}}
\quad\quad
\parbox{\imsize}{\includegraphics[width=\imsize]
{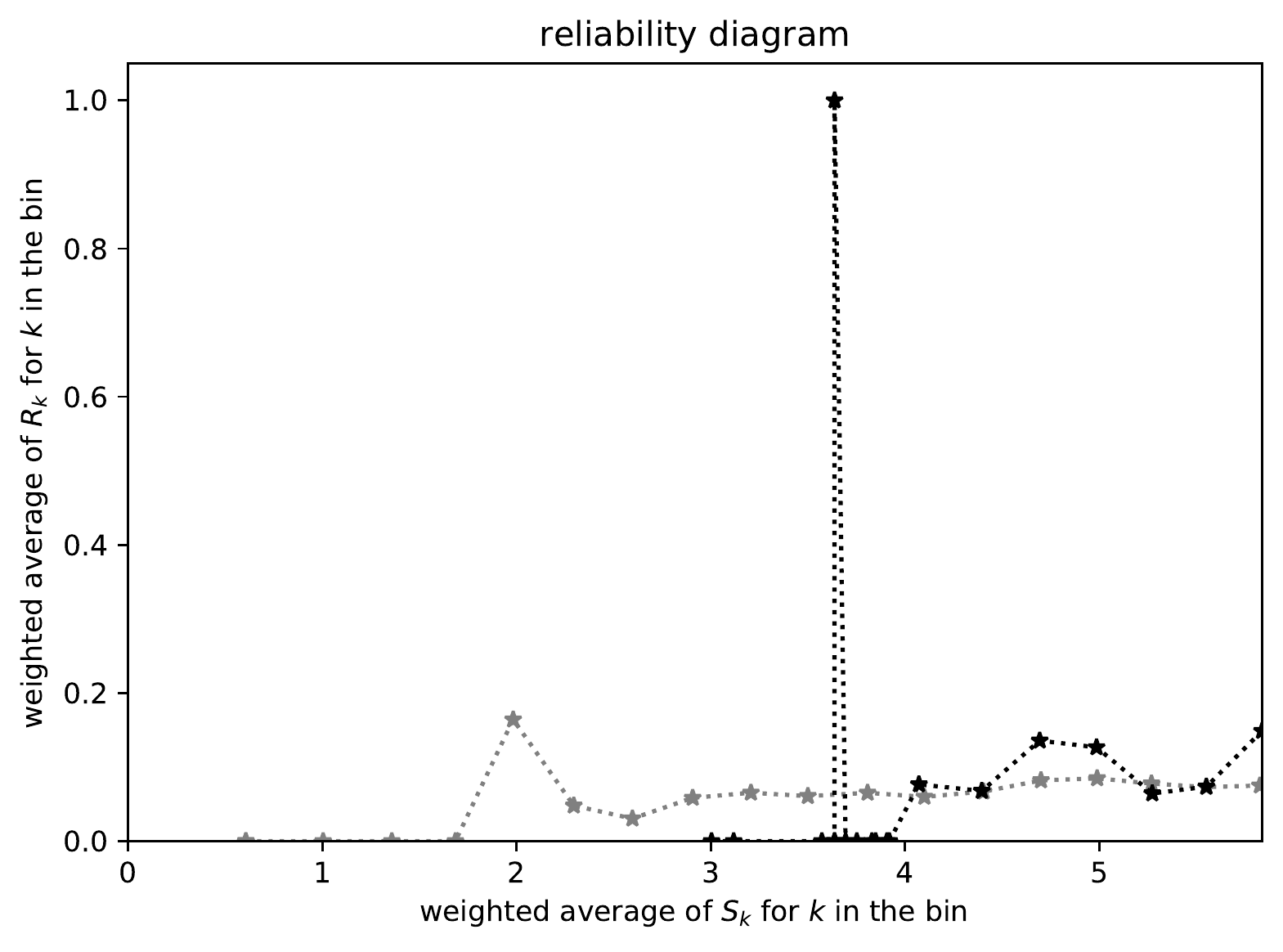}}

\vspace{\vertsep}

\parbox{\imsize}{\includegraphics[width=\imsize]
{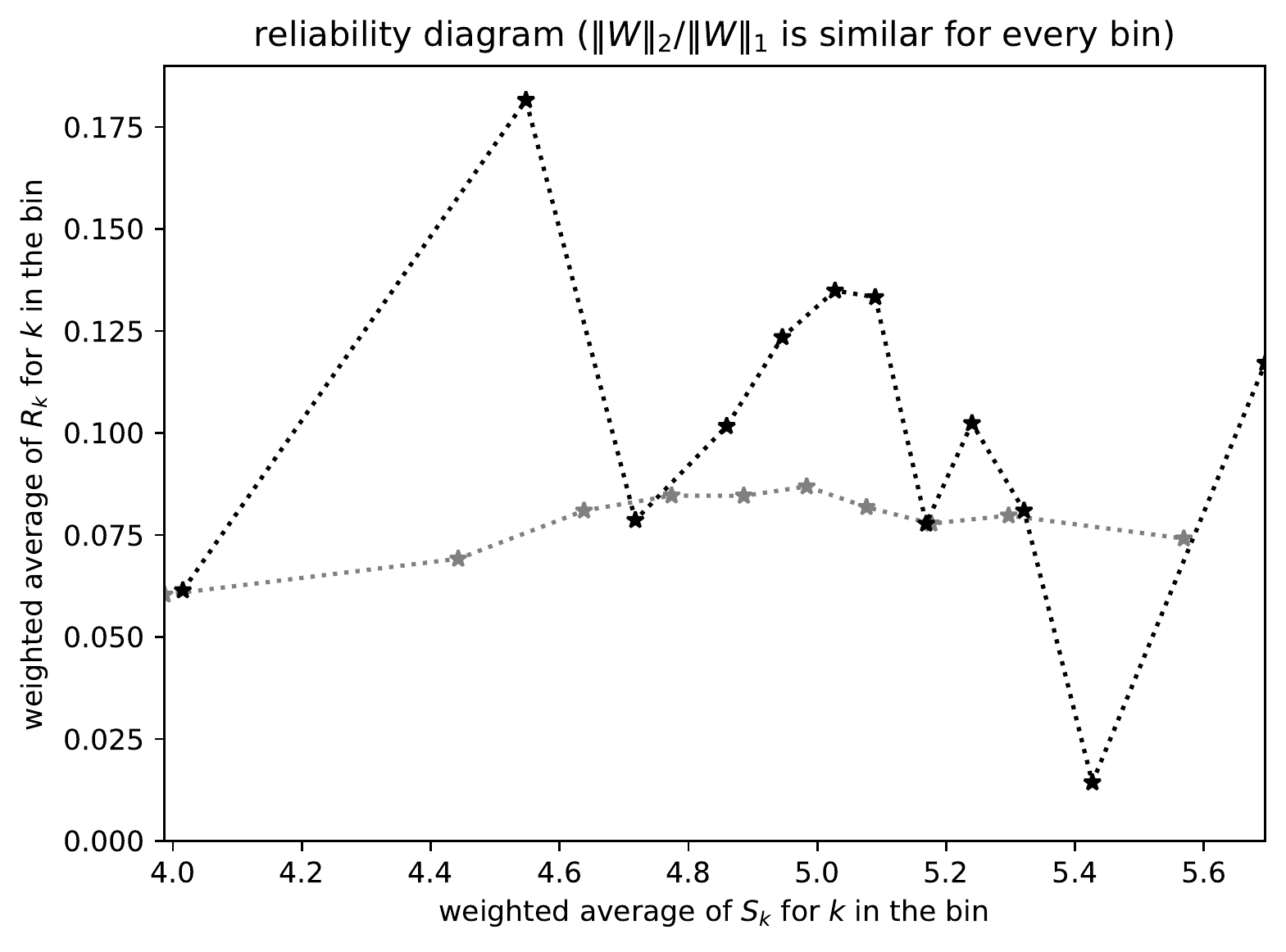}}
\quad\quad
\parbox{\imsize}{\includegraphics[width=\imsize]
{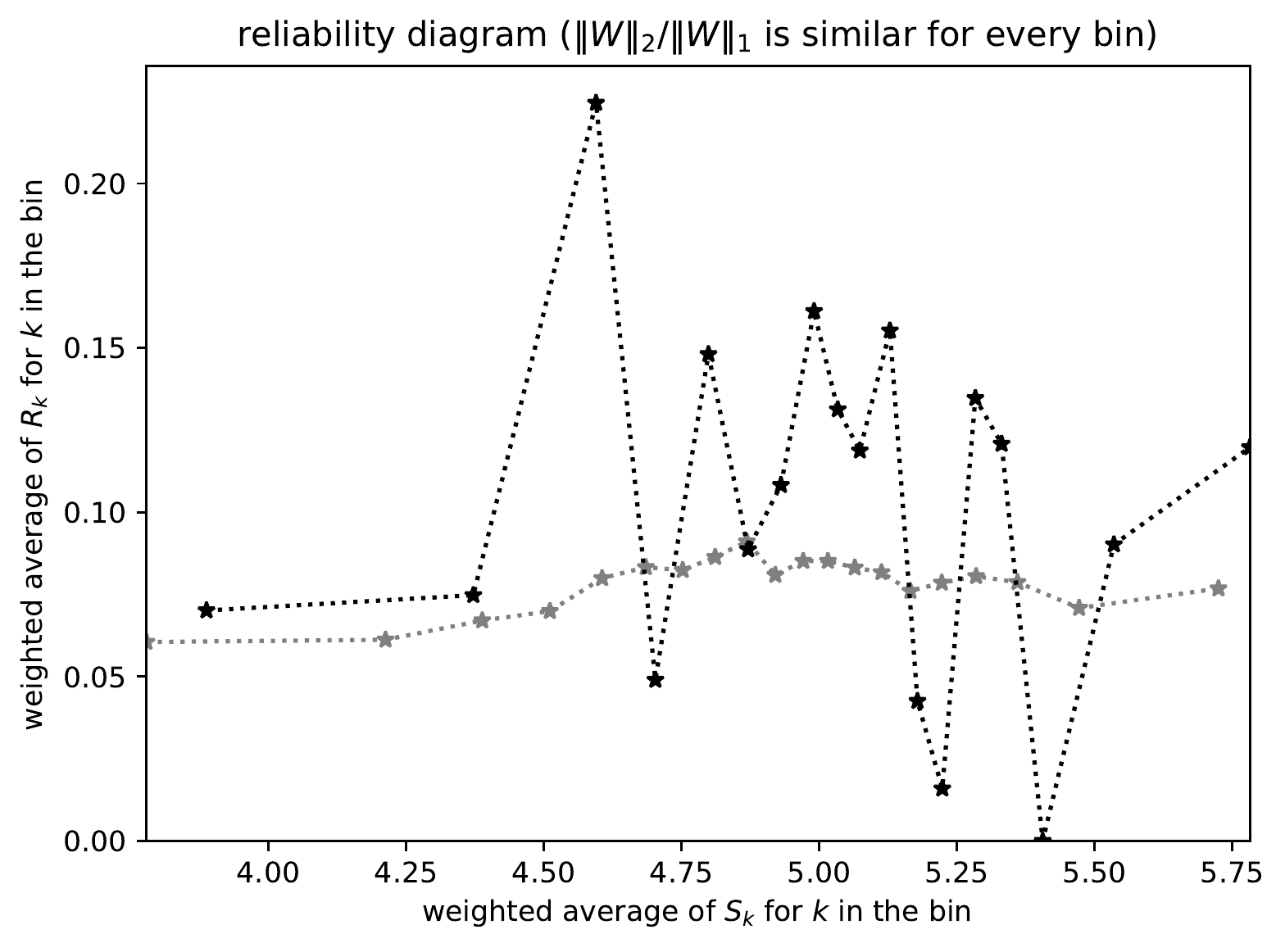}}

\end{centering}
\caption{Napa County, reporting whether the household has access
         via a satellite dish to the internet, with scores being $\log_{10}$
         of the adjusted household income;
         $n =$ 679; Kuiper's statistic is $0.02761 / \sigma = 2.259$,
         Kolmogorov's and Smirnov's is $0.02695 / \sigma = 2.205$.
The intense deviation around scores of 4.6 is apparent
in the reliability diagrams with 10 and 20 bins each,
but the latter resolves the spike much better while being unfortunately
too noisy for many other scores. The plot of cumulative differences
resolves the sharp jump around scores of 4.6 without detracting
from the display at other scores. The scalar summary statistics
report only very mildly statistically significant deviation,
unable to fully account for the high deviation around scores of 4.6.
}
\label{napa}
\end{figure}

\begin{figure}
\begin{centering}

\parbox{\imsize}{\includegraphics[width=\imsize]
{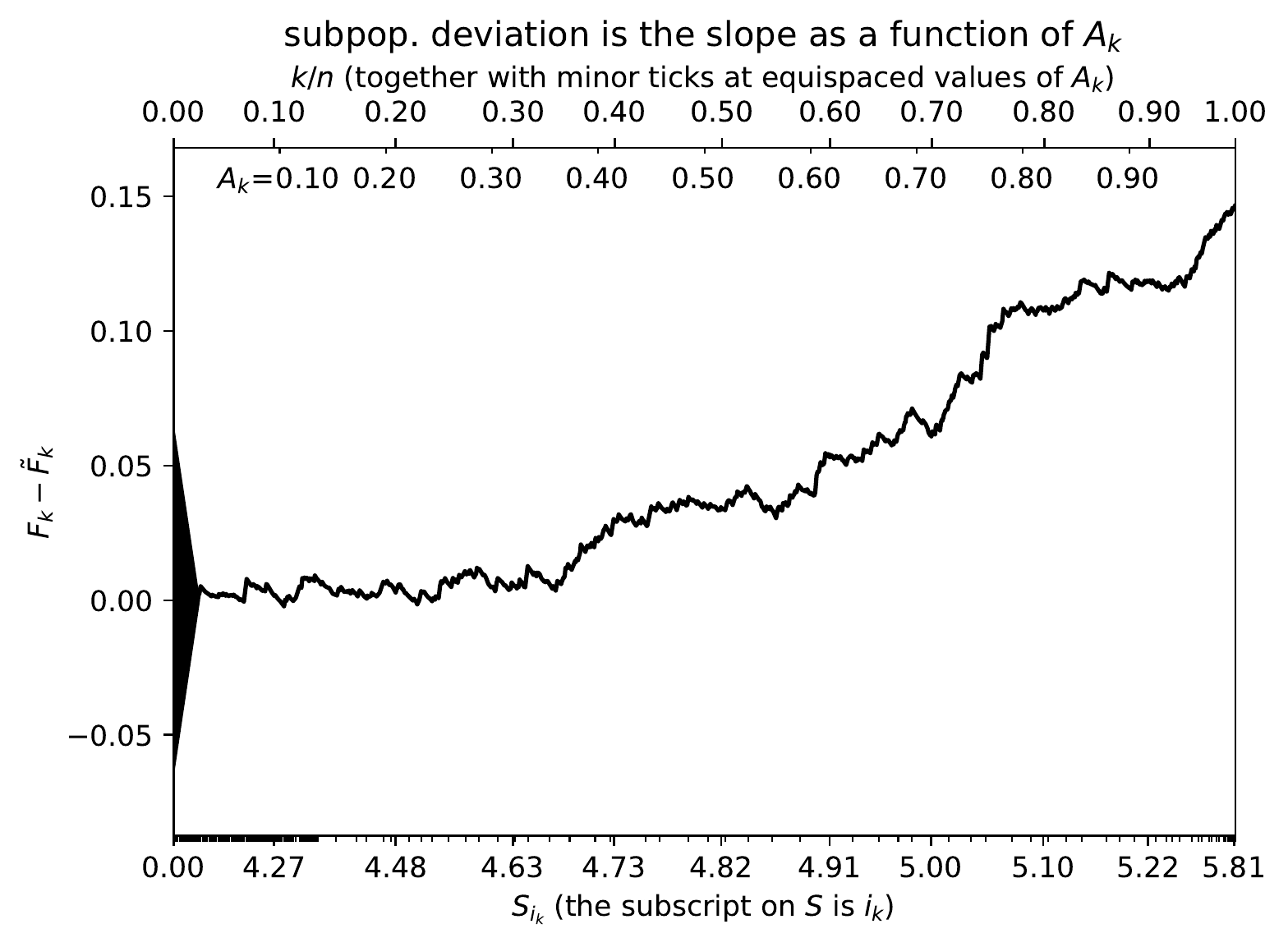}}
\quad\quad
\parbox{\imsize}{\includegraphics[width=\imsize]
{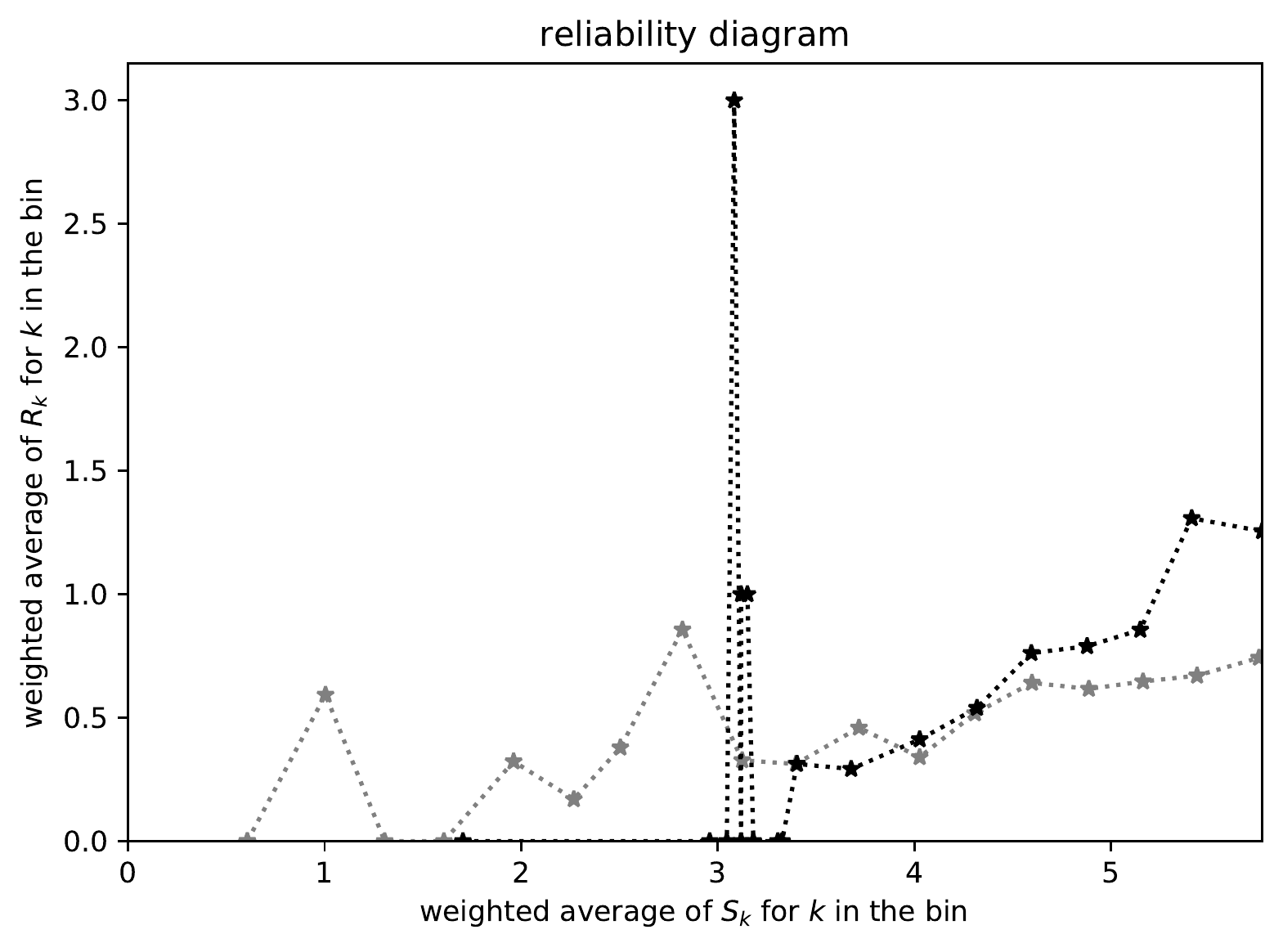}}

\vspace{\vertsep}

\parbox{\imsize}{\includegraphics[width=\imsize]
{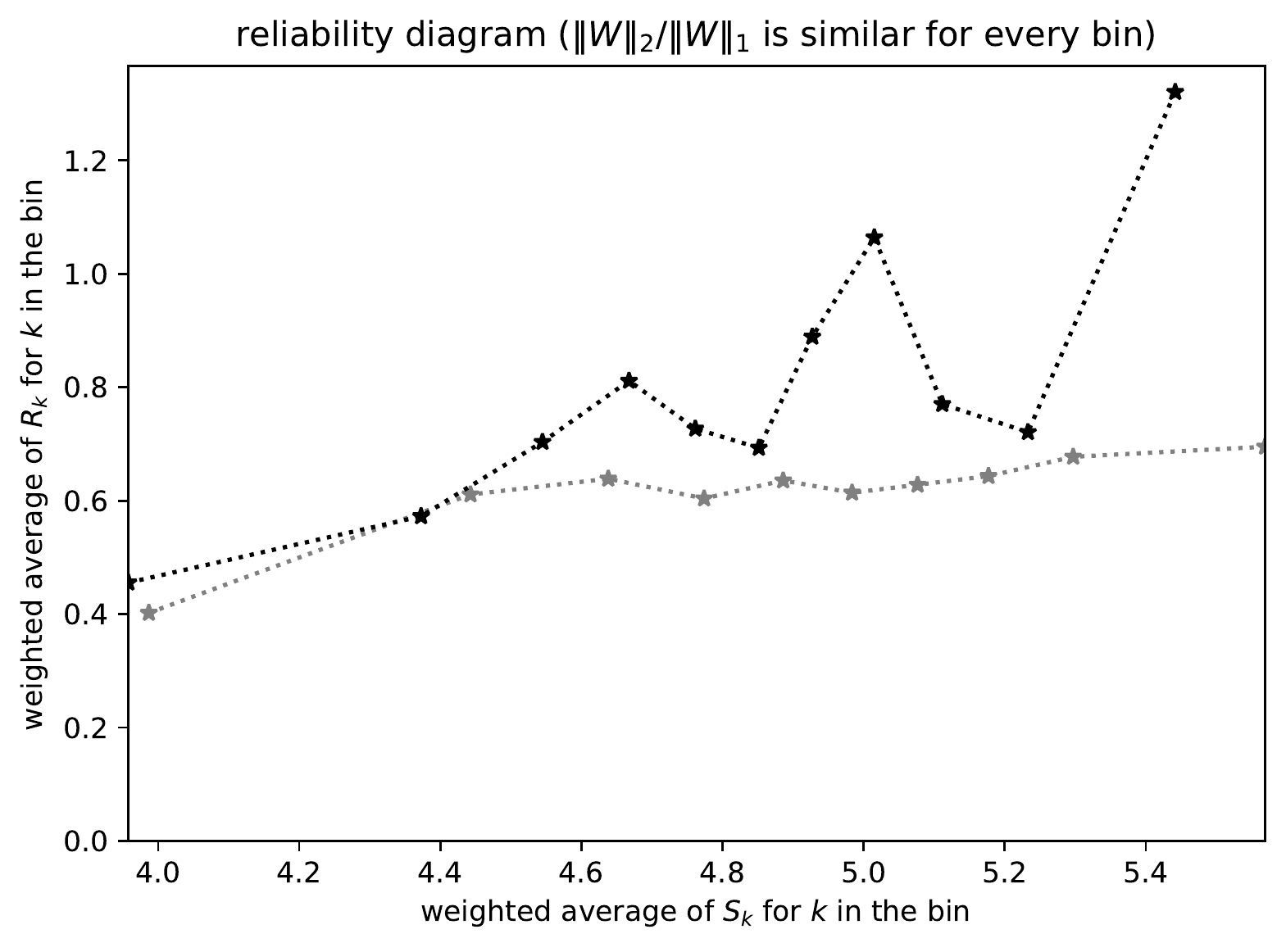}}
\quad\quad
\parbox{\imsize}{\includegraphics[width=\imsize]
{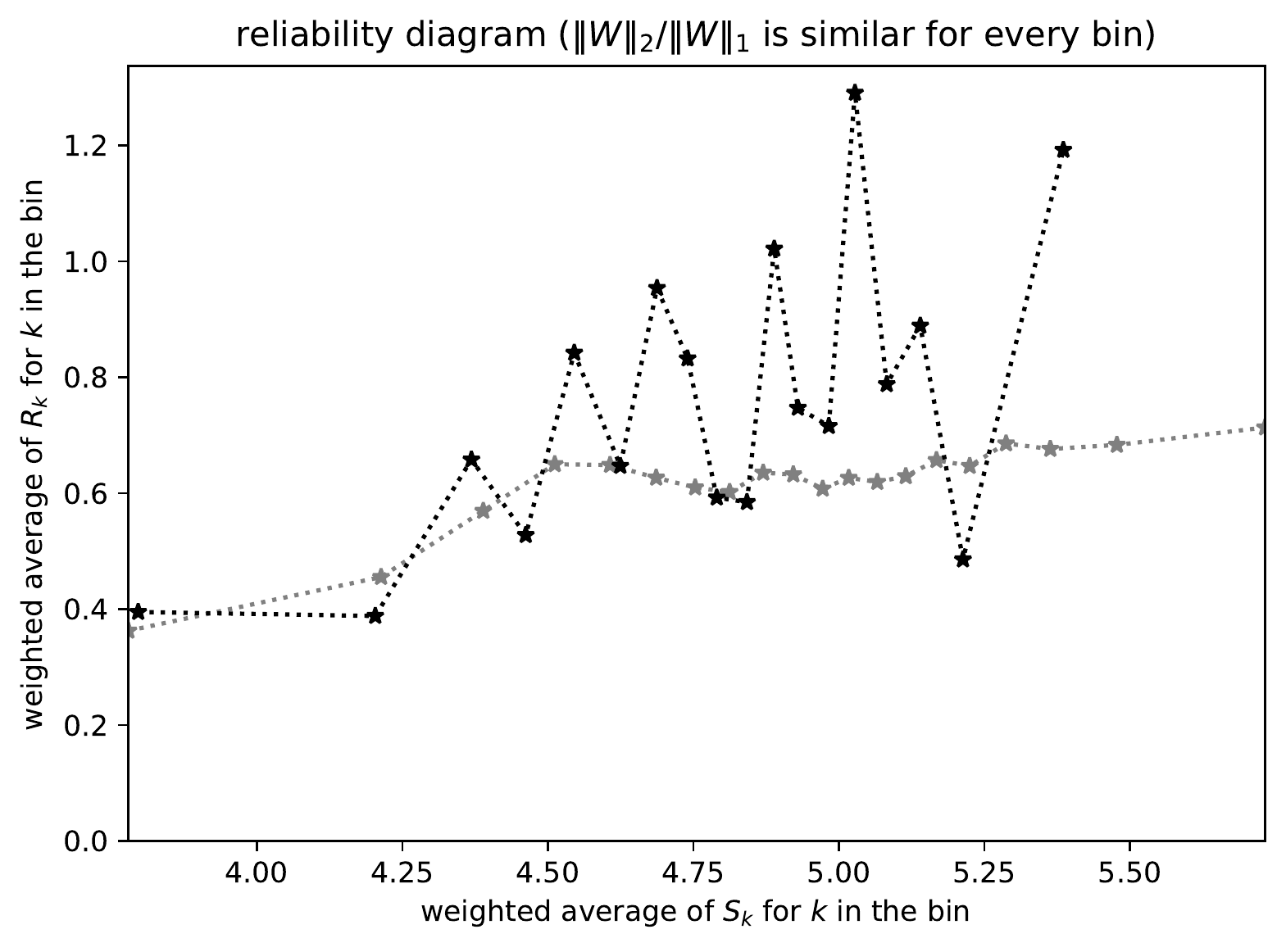}}

\end{centering}
\caption{Stanislaus County, reporting the number of related children
         in the household, with scores being $\log_{10}$
         of the adjusted household income;
         $n =$ 1,624; Kuiper's statistic is $0.1489 / \sigma = 4.500$,
         Kolmogorov's and Smirnov's is $0.1467 / \sigma = 4.433$.
The reliability diagrams are all commensurate with each other,
but that is clear only after careful comparison of them all.
Making sense of the reliability diagrams in this case would seem to require
generating multiple plots, with several different numbers of bins.
The scalar summary statistics have little trouble detecting
statistically significant deviation here.
}
\label{stanislaus}
\end{figure}

This subsection analyzes the latest (year 2019) microdata
from the American Community Survey of the U.S. Census Bureau;\footnote{The data
from the American Community Survey is available
at \url{https://www.census.gov/programs-surveys/acs/microdata.html}}
specifically, we consider the full population to be the members
from all counties in California put together, and consider the subpopulation
to be the observations from an individual county.
The sampling in this survey is weighted;
we retain only those members whose weights (``WGTP'' in the microdata)
are nonzero, and discard any member whose household personal income
(``HINCP'') is zero or for which the adjustment factor to income (``ADJINC'')
is reported as missing.
For the scores, we use the logarithm to base 10
of the adjusted household personal income
(the adjusted income is ``HINCP'' times ``ADJINC,'' divided by one million
when ``ADJINC'' omits its decimal point in the integer-valued microdata),
randomly perturbing the scores by about one part in $10^8$ to guarantee
their uniqueness.
The captions on the figures specify which variables from the data set
we use for the results $R_1$,~$R_2$, \dots, $R_m$
(different figures consider different variables), with $m =$ 134,094.
Figures~\ref{san_joaquin}--\ref{stanislaus} present the examples.
In all figures, the cumulative plots display all significant features clearly.

\subsection{Future outlook}
\label{outlook}

This subsection suggests generalizations suitable for future research.

The present paper plots the cumulative differences
between the results of a subpopulation and the results of the full population,
accumulated as a function of a scalar real-valued score.
If instead the scores to be accumulated are from a Euclidean or Hilbert space
of more than one dimension, a natural generalization is to impose
a total order on the scores via a space-filling curve
such as the Peano or Hilbert curves, effectively replacing
the space of more than one dimension with a one-dimensional ordering.
Space-filling curves constructed via quad-trees in two-dimensional space
or oct-trees in three-dimensional space can be especially efficient.
Google's S2 Geometry Library, which drives Google Maps, Foursquare, MongoDB,
et al., employs a Hilbert curve.\footnote{Google's S2 Geometry Library
is described at \url{https://s2geometry.io/devguide/s2cell_hierarchy.html}}

The present paper focuses on deviations of a subpopulation
from the full population; comparing the subpopulation to the full population
is always legitimate, as binning or interpolating from the scores
of the full population to those of the subpopulation happens at a scale
finer than the subpopulation's sampling, and there is always at least one score
in the full population corresponding to each score in the subpopulation
(namely, the same score). In contrast, assessing deviations
between two different subpopulations that do not have an explicit pairing
for each score can be problematic. Indeed, the ranges of scores
for the subpopulations may not even overlap ---
the scores for one subpopulation can all be significantly
less than all the scores for another subpopulation, for instance.
Binning or interpolating from one subpopulation to another can be ill-posed.
And even when binning or interpolating from one subpopulation
to a second subpopulation is well-posed, binning or interpolating
from that second subpopulation to the first subpopulation can be ill-posed
--- the comparison may not be reflexively well-posed. In general,
a partial ordering governs the comparisons --- we can always analyze deviations
of any subpopulation from any larger population containing the subpopulation,
but cannot always reliably analyze deviations between two different
subpopulations of the same larger population.
Applications demanding direct comparison between different subpopulations
are currently the subject of intensive investigation,
far beyond the scope of the present paper
except in the special case that the different subpopulations are paired
for each score. One possibility for the general case is to bin
both subpopulations being compared to the same finest binning such that
both subpopulations have at least one score in each bin;
however, interpreting significance in this general case can be tricky.
Forthcoming work will treat the special case for which no score
associated with any observation from either subpopulation being compared
is exactly equal to the score for any other observation
from the two subpopulations.

\section{Conclusion}
\label{conclusion}

Plotting the cumulative differences between the outcomes for the subpopulation
and those for the full population binned to the subpopulation's scores
avoids the arbitrary selection of widths for bins or smoothing kernels
that the conventional reliability diagrams and calibration plots require.
The plot of cumulative differences displays deviation directly as the slope
of secant lines for the graph; such slope is easy to perceive independent
of any irrelevant constant offset of a secant line.
The graph of cumulative differences very directly enables detection
and quantification of deviation between the subpopulation and full population,
along with identification of the corresponding ranges of scores.
The cumulative differences estimate the distribution of deviation
fully nonparametrically, letting the data observations speak for themselves
(or nearly for themselves --- the triangle at the origin helps convey the scale
of a driftless random walk's expected random fluctuations).
As seen in the figures, the graph of cumulative differences automatically
adapts its resolving power to the distributions of deviations and scores
for the subpopulation, not imposing any artificial grid of bins
or set-width smoothing kernel, unlike the traditional reliability diagrams
and calibration plots.
The scalar metrics of Kuiper and of Kolmogorov and Smirnov
conveniently summarize the overall statistical significance of deviations
displayed in the graph.

\section*{Declarations}

\subsection*{Ethics approval and consent to participate}

Not applicable.

\subsection*{Consent for publication}

Not applicable.

\subsection*{Availability of data and materials}

The data sets generated during and/or analyzed during the current study
are available in the following repositories:
\begin{itemize}
\item \url{https://github.com/facebookresearch/fbcdgraph}
(for all synthetic data sets)
\item \url{http://image-net.org/download-images}
(for ImageNet)
\item \url{https://www2.census.gov/programs-surveys/acs/data/pums/2019/1-Year}
(for California households file {\tt csv\_hca.zip}
--- which includes the file {\tt psam\_h06.csv} that our software processes ---
from the 2019 American Community Survey of the U.S. Census Bureau)
\end{itemize}

\noindent MIT-licensed open-source codes in Python 3 and shell scripts
that automatically reproduce all figures and statistics of the present paper
are publicly available in the repository {\tt fbcdgraph} at
\url{https://github.com/facebookresearch/fbcdgraph}

\subsection*{Competing interests}

M.T. is employed by and holds stock in Facebook.

\subsection*{Funding}

M.T. is employed by Facebook and conducted this research
as part of his job responsibilities.

\subsection*{Author's contributions}

M.T. is the sole author.

\subsection*{Acknowledgements}

We would like to thank Tiffany Cai, Joaquin Qui\~nonero Candela,
Sam Corbett-Davies, Kenneth Hung, Imanol Arrieta Ibarra, Mike Rabbat,
Jonathan Tannen, Edmund Tong, and the anonymous reviewers and editor.

\subsection*{Author's information}

\noindent Mark Tygert, 786 Coleman Ave., Apt. L, Menlo Park, CA 94025-2440

\noindent {\tt mark@tygert.com}

\noindent \url{http://tygert.com}

\clearpage

\appendix
\section{Calibration}
\label{calibration}

\subsection{Introduction to assessing calibration}
\label{aintro}

This appendix treats calibration, specifically,
we consider $n$ observations $R_1$,~$R_2$, \dots, $R_n$ of the outcomes
of independent Bernoulli trials, and would like to test the hypothesis
that $R_j$ is drawn from a Bernoulli distribution whose probability of success
is $S_j$, that is,
\begin{equation}
\label{null}
R_j \sim \Bernoulli(S_j)
\end{equation}
for all $j = 1$, $2$, \dots, $n$, where the corresponding probabilities
of success are $S_1$, $S_2$, \dots, $S_n$;
following the usual conventions, $R_j = 1$ when the outcome is a success
and $R_j = 0$ when the outcome is a failure.
We view $R_1$,~$R_2$, \dots, $R_n$ (but not $S_1$,~$S_2$, \dots, $S_n$)
as random. Without loss of generality, we order the success probabilities
(preserving the pairing of $R_j$ with $S_j$ for every $j$)
such that $S_1 \le S_2 \le \dots \le S_n$, ordering any ties at random,
perturbed so that $S_1 < S_2 < \dots < S_n$.

As in Section~\ref{intro}, the classical methods 
require partitioning the unit interval $[0, 1]$ into $\ell$ disjoint intervals
with endpoints $B_1$, $B_2$, \dots, $B_{\ell}$ such that
$0 < B_1 < B_2 < \dots < B_{\ell-1} < B_{\ell} = 1$.
We can then form the averages
\begin{equation}
\label{avgY}
Y_k = \frac{\sum_{j : B_{k-1} < S_j \le B_k} R_j}
           {\#\{j : B_{k-1} < S_j \le B_k\}}
    = \frac{\#\{j : B_{k-1} < S_j \le B_k \hbox{ and } R_j = 1\}}
           {\#\{j : B_{k-1} < S_j \le B_k\}}
\end{equation}
for $k = 1$, $2$, \dots, $\ell$, under the convention that $B_0 < 0$.
We also calculate the average success probabilities in the bins
\begin{equation}
\label{avgX}
X_k = \frac{\sum_{j : B_{k-1} < S_j \le B_k} S_j}
           {\#\{j : B_{k-1} < S_j \le B_k\}}
\end{equation}
for $k = 1$, $2$, \dots, $\ell$, under the same convention that $B_0 < 0$.
A graphical method for assessing calibration is then to scatterplot the pairs
$(X_1, Y_1)$, $(X_2, Y_2)$, \dots, $(X_{\ell}, Y_{\ell})$,
along with the line connecting the origin $(0, 0)$ to the point $(1, 1)$;
a pair $(X_k, Y_k)$ lies on that line when its calibration is ideal,
as then $Y_k = X_k$.
This graphical method for assessing the calibration or reliability
of probabilistic predictions is known as a ``reliability diagram''
or ``calibration plot,'' as reviewed, for example, by~\cite{tygert}.
Copious examples of such reliability diagrams are available
in the figures below, as well as in the works
of~\cite{corbett-davies-pierson-feller-goel-huq},
\cite{crowson-atkinson-therneau}, \cite{murphy-winkler}, \cite{brocker},
\cite{wilks}, \cite{brocker-smith}, \cite{gneiting-balabdaoui-raftery},
\cite{guo-pleiss-sun-weinberger},
\cite{vaicenavicius-widmann-andersson-lindsten-roll-schoen},
and many others; those works consider applications ranging
from weather forecasting to medical prognosis to fairness in criminal justice
to quantifying the uncertainty in predictions of artificial neural networks.
An approach closely related to reliability diagrams is to smooth over
the binning using kernel density estimation, as discussed by~\cite{brocker},
\cite{wilks}, and others.

As in Section~\ref{intro}, two common choices of the bins
whose endpoints are $B_1$, $B_2$, \dots, $B_{\ell}$ are
\{1\} to make $B_1$, $B_2$, \dots, $B_{\ell}$ be equispaced, and
\{2\} to make the number of success probabilities that fall in the $k$th bin,
that is, $\#\{j : B_{k-1} < S_j \le B_k\}$, be the same for all $k$
(aside from the rightmost bin, that for $k = \ell$,
if $n$ is not perfectly divisible by $\ell$).
In contrast to the classical approach,
the methods of the following subsections avoid binning.
Recently, \cite{gupta-rahimi-ajanthan-mensink-sminchisescu-hartley}
and~\cite{roelofs-cain-shlens-mozer} have forcefully reiterated
serious problems with existing binned methodologies.

There have been a number of proposals to measure calibration
via Kolmogorov-Smirnov statistics, without binning, including
by~\cite{gupta-rahimi-ajanthan-mensink-sminchisescu-hartley} and~\cite{tygert}.
Section~3.2 of~\cite{gneiting-balabdaoui-raftery},
Chapter 8 of~\cite{wilks}, and Figure~1
of~\cite{gupta-rahimi-ajanthan-mensink-sminchisescu-hartley}
also point to the utility of cumulative reliability diagrams and plots
somewhat similar to those in the present paper.
The particular plots proposed below focus on calibration specifically,
encoding miscalibration directly as the slopes of secant lines for the graphs.
Such plots lucidly depict miscalibration
with significant quantitative precision.
Popular graphical methods for assessing calibration appear not to leverage
the key to the approach advocated below, namely that slope is easy to assess
visually even when the constant offset of the graph (or portion of the graph
under consideration) is arbitrary and meaningless.

\begin{figure}
\begin{centering}

\parbox{\imsize}{\includegraphics[width=\imsize]
                {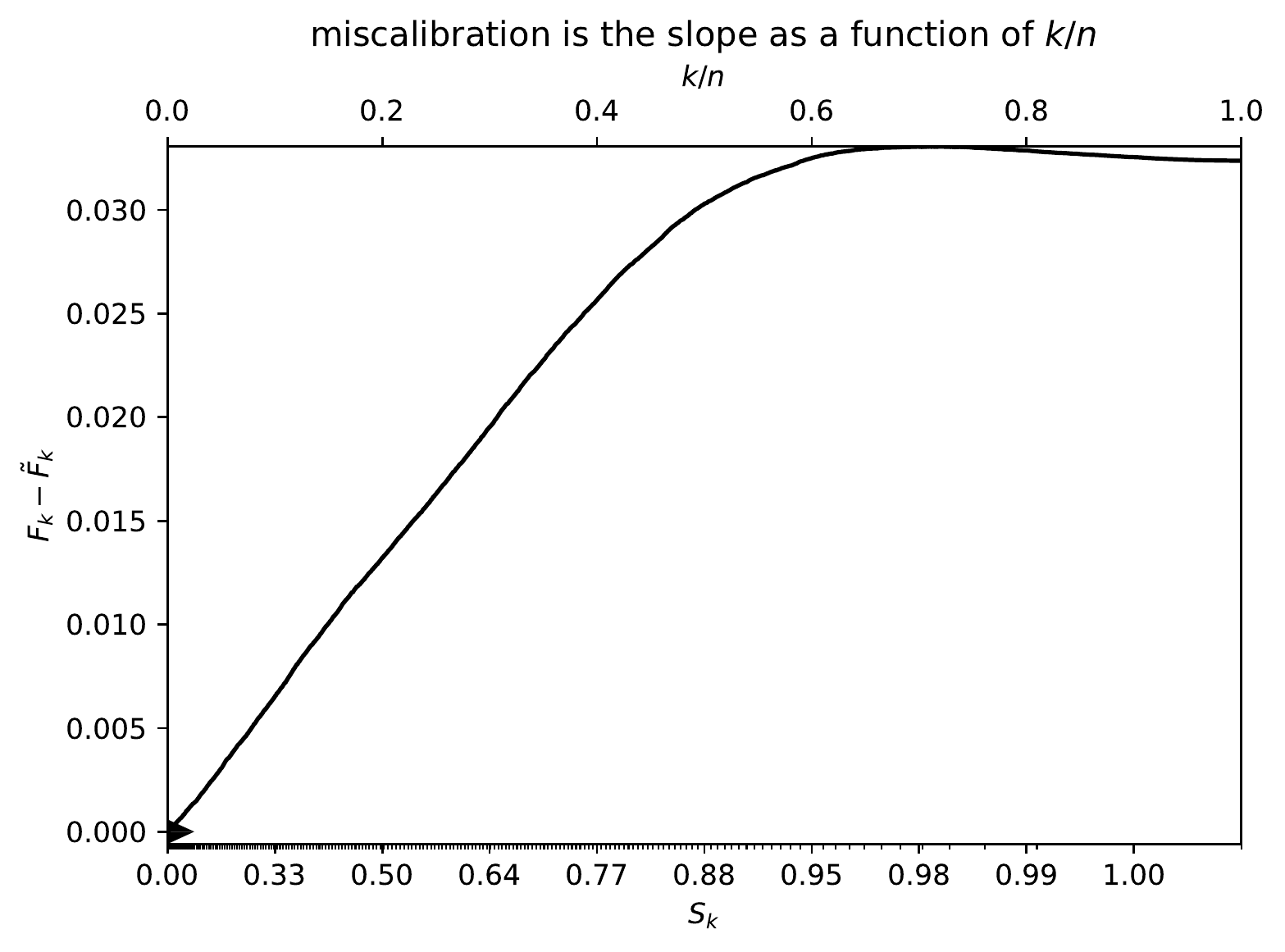}}

\vspace{\vertsep}

\parbox{\imsize}{\includegraphics[width=\imsize]
                {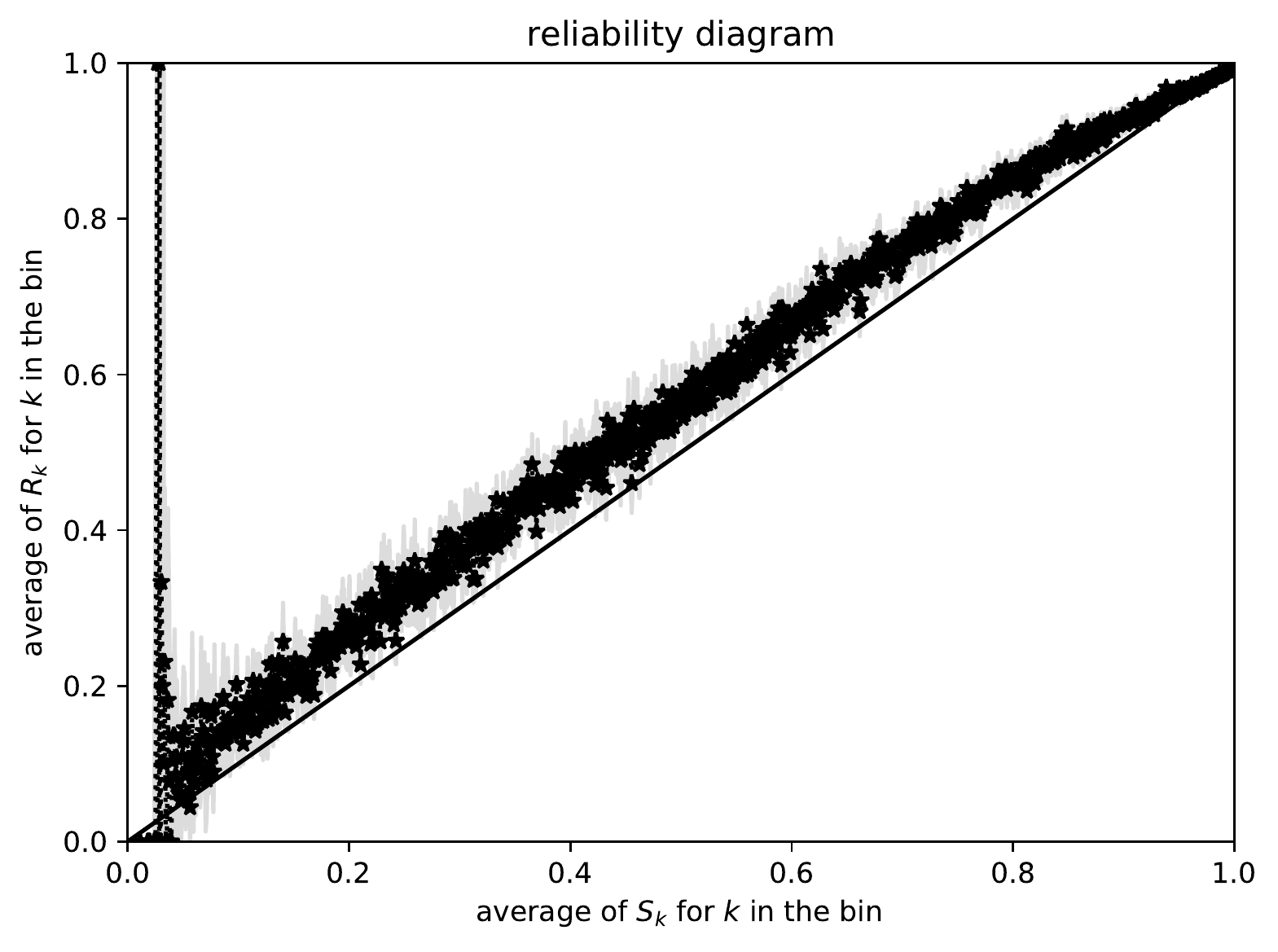}}
\quad\quad
\parbox{\imsize}{\includegraphics[width=\imsize]
                {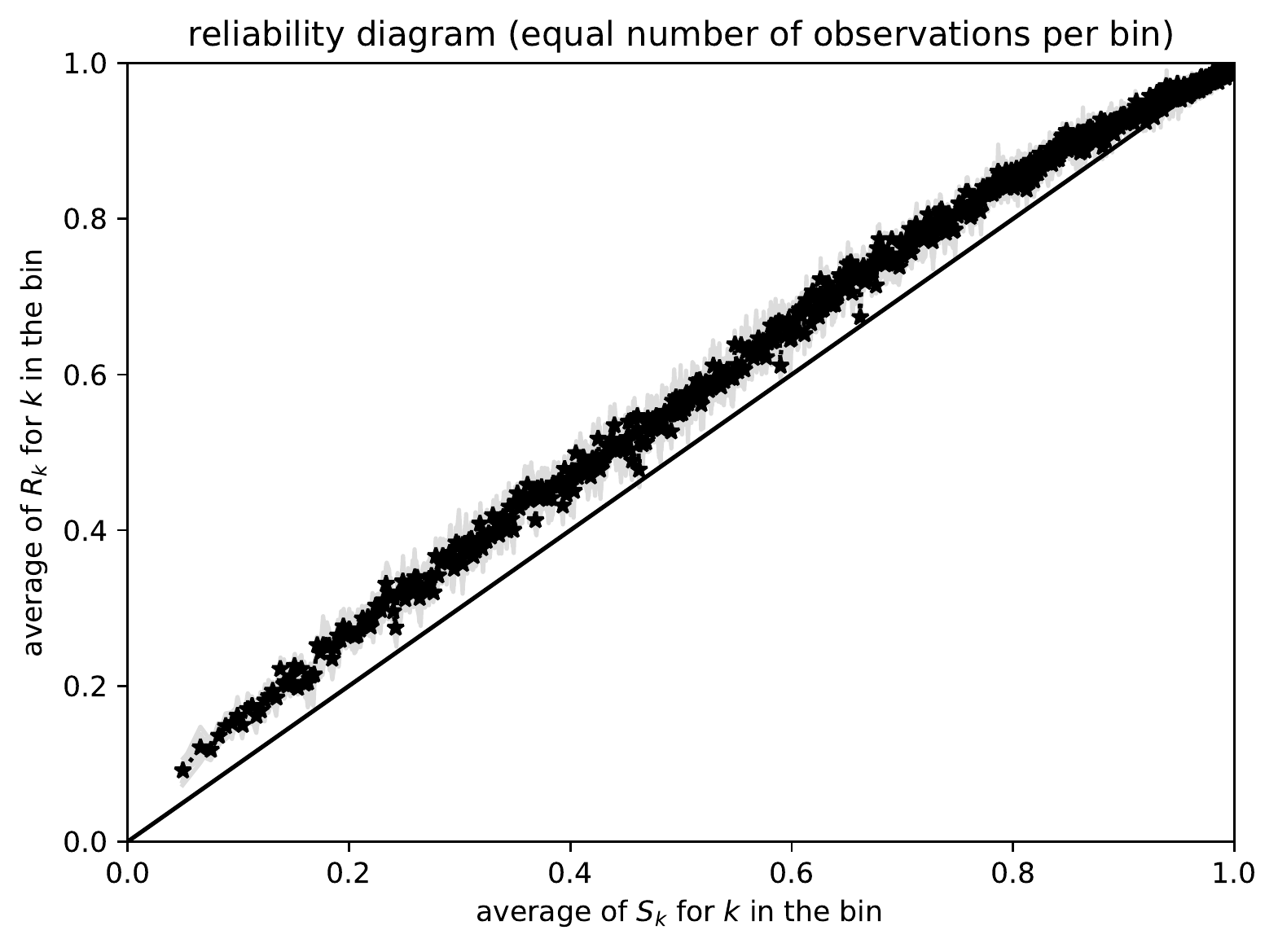}}

\vspace{\vertsep}

\parbox{\imsize}{\includegraphics[width=\imsize]
                {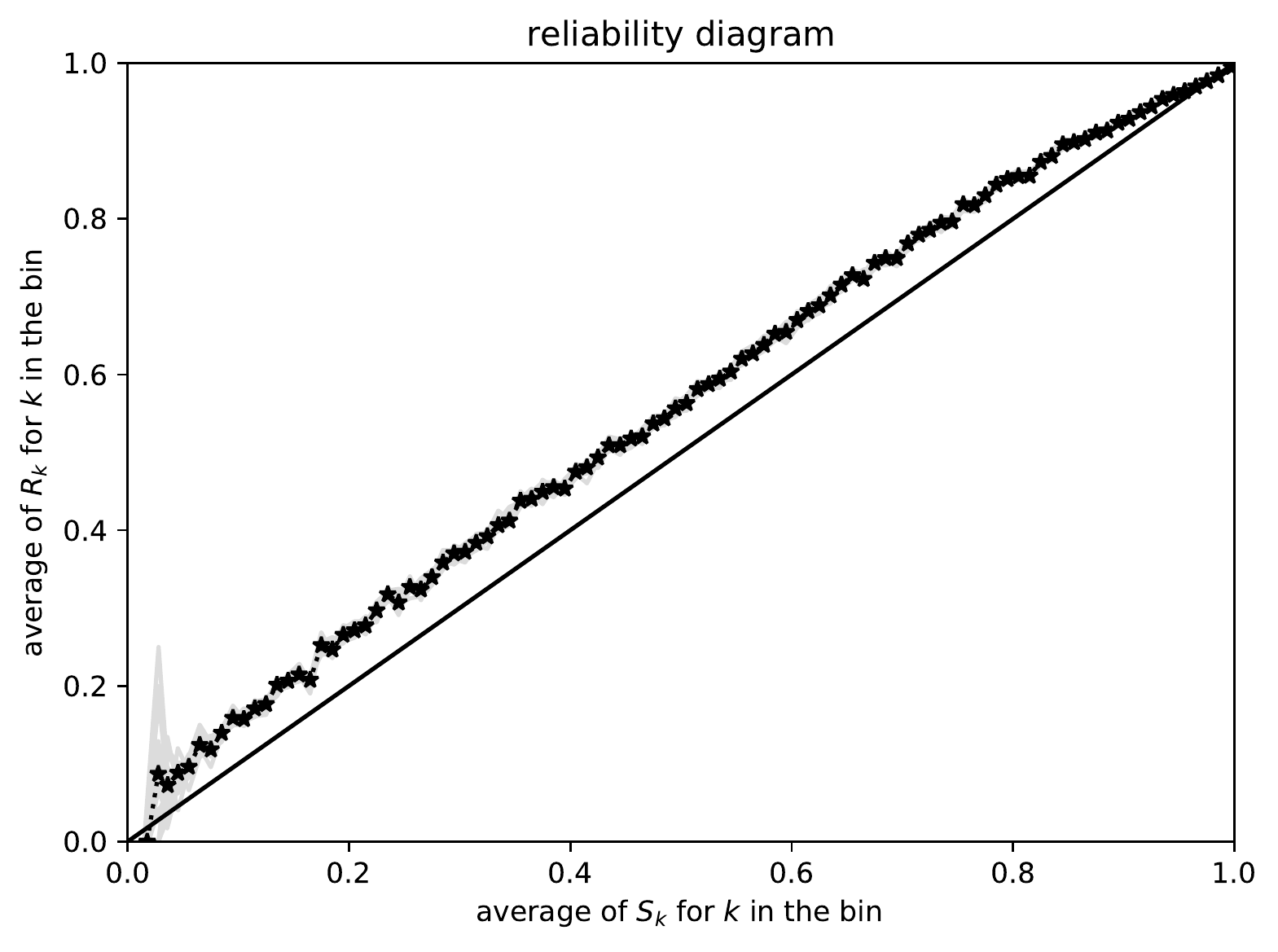}}
\quad\quad
\parbox{\imsize}{\includegraphics[width=\imsize]
                {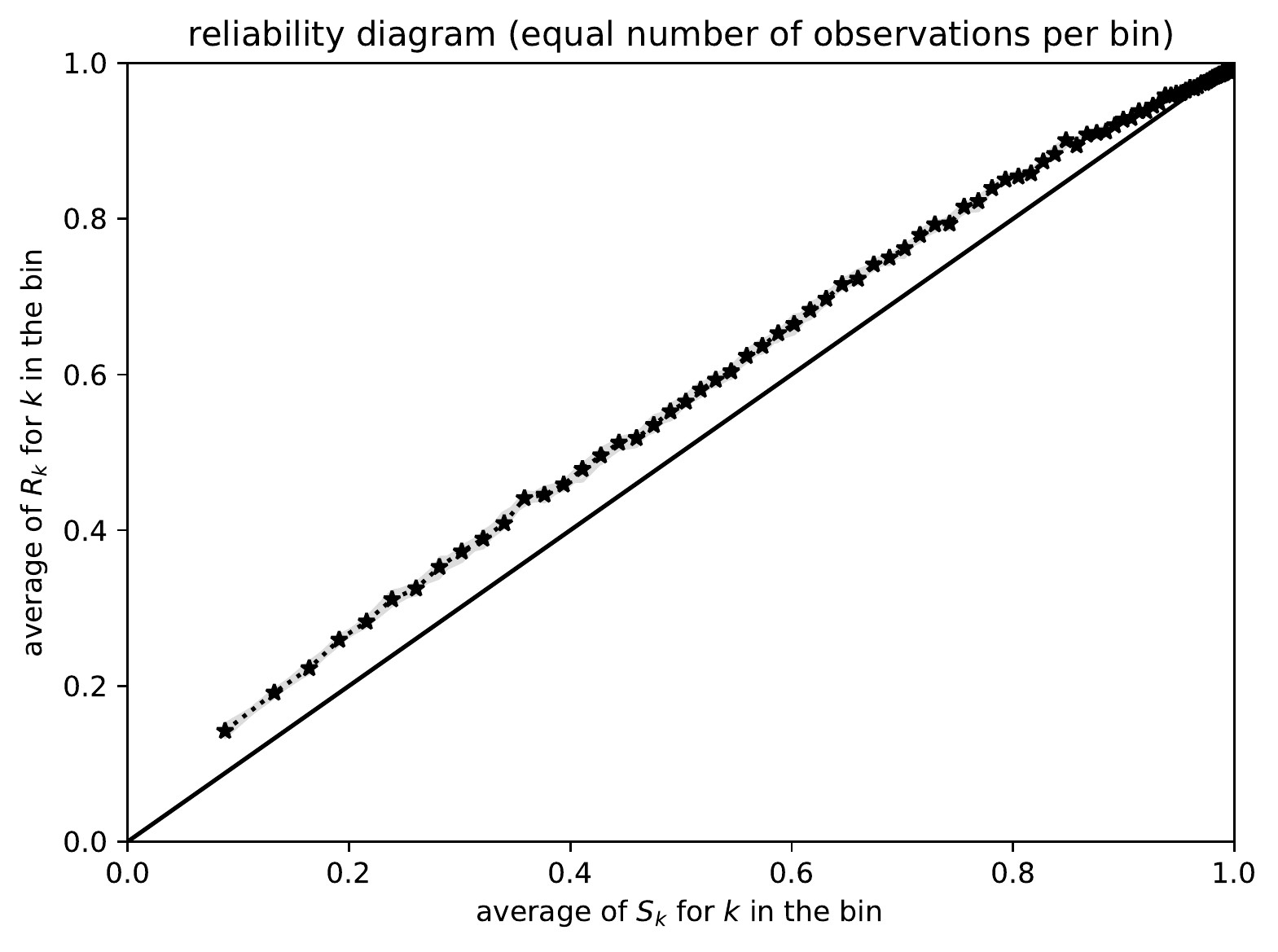}}

\end{centering}
\caption{Calibration of the full ImageNet-1000 training data set
         (the scores are the probabilities); $n =$ 1,281,167;
         Kuiper's statistic is $0.03306 / \sigma = 111.7$,
         Kolmogorov's and Smirnov's is $0.03306 / \sigma = 111.7$.
The reliability diagram with 100 bins,
each with roughly the same number of observations, looks best
among the reliability diagrams; the reliability diagrams with 1,000 bins each
display unreal stochastic variations.
Only the cumulative plot conveniently reveals that a third of all observations
(specifically, those with probabilities of at least 0.97) are well-calibrated.
The scalar summary statistics report
profoundly statistically significant miscalibration,
courtesy of the large number of observations
(the actual effect size is more modest, as seen by the values
without dividing by $\sigma$).
}
\label{imagenetcal}
\end{figure}

To illustrate with an example,
Figure~\ref{imagenetcal} displays both the classical reliability diagrams
as well as the plot of cumulative differences proposed below.
Subsubsection~\ref{aimagenetex} below describes the figure in detail.
The lowermost two rows of Figure~\ref{imagenetcal} are the classical diagrams,
with $n =$ 1,281,167;
there are $\ell =$ 1,000 bins for each diagram in the middlemost row
and $\ell =$ 100 for each diagram in the lowermost row.
The bins are equispaced along the probabilities
in the leftmost reliability diagrams,
whereas each bin contains the same number of probabilities
from $S_1$,~$S_2$, \dots, $S_n$ in the rightmost reliability diagrams.
The light gray lines indicate ``error bars'' constructed via bootstrapping,
as elaborated in Subsection~\ref{aresults} below.

The topmost row of Figure~\ref{imagenetcal} displays the cumulative plot.
Leaving elucidation of the cumulative plots and their construction
to Subsection~\ref{amethods} below,
we just point out here that miscalibration across an interval
is equal to the slope of the secant line for the graph over that interval,
aside from expected stochastic fluctuations
detailed in Subsubsection~\ref{asignificance} below.
Steep slopes correspond to substantial miscalibration across the ranges
of probabilities where the slopes are steep, with the miscalibration
over an interval exactly equal to the expected value of the slope
of the secant line for the graph over that interval.
In Figure~\ref{imagenetcal},
the cumulative plot reveals a curious change-point in calibration
that is difficult to notice in the conventional reliability diagrams,
and the Kolmogorov-Smirnov and Kuiper metrics conveniently summarize
the statistical significance of the overall deviation across all scores,
in accordance
with Subsubsections~\ref{ascalarstats} and~\ref{asignificance} below.

The structure of the remainder of this appendix is as follows:
Subsection~\ref{amethods} details the cumulative methods,
and Subsection~\ref{aresults} then illustrates them via several examples.
Table~\ref{anotation} gives a glossary of the notation used
in the appendices, while Table~\ref{notation} gives a glossary
of the notation used prior to the appendices.

\subsection{Methods for assessing calibration}
\label{amethods}

This subsection provides a detailed mathematical formulation
of the cumulative methods,
with Subsubsection~\ref{ahigh-level} outlining an approach
to large-scale analysis.
Subsubsection~\ref{agraphical} details the graphical methods.
Subsubsection~\ref{ascalarstats} details the scalar summary metrics.
Subsubsection~\ref{asignificance} treats statistical significance
for both the graphical methods and the summary statistics.

\begin{table}
\caption{Notational conventions for the appendices
(Table~\ref{notation} summarizes the notation used before the appendices.
The symbols in the tables are in alphabetical order.)}
\label{anotation}
\begin{center}
\begin{tabular}{lll}
\hline
symbol & meaning & equation \\\hline
$D$ & Kuiper statistic & (\ref{aKuiper}) \\
$\Delta_k$ & expected slope of $F_j - \tilde{F}_j$ from $j = k-1$ to $j = k$ &
(\ref{adelta}) \\
$F_k$ & cumulative response & (\ref{cumresponse}) \\
$\tilde{F}_k$ &
cumulative perfectly calibrated response (cumulative success probability) &
(\ref{expresponse}) \\
$G$ & Kolmogorov-Smirnov statistic & (\ref{aKolmogorov-Smirnov}) \\
$P_k$ & actual probability of success for a Bernoulli trial in synthetic data
& (Subsection~\ref{aresults}) \\
$R_k$ & response --- (random) dependent variable, outcome, or result &
(\ref{null}) \\
$S_k$ & success probability score --- (non-random) independent variable &
(\ref{null}) \\
$\sigma$ &
scale of random fluctuations over the full range of success probabilities &
(\ref{astddev}) \\
$X_k$ & abscissa of the observations for reliability diagrams & (\ref{avgX}) \\
$Y_k$ & ordinate of the observations for reliability diagrams & (\ref{avgY}) \\
\hline
\end{tabular}
\end{center}
\end{table}

\subsubsection{High-level strategy for assessing calibration}
\label{ahigh-level}

This subsubsection proposes a process for large-scale data analysis.

As in Subsection~\ref{high-level}, we can take a two-step approach
when there are many data sets to analyze:
\begin{enumerate}
\item A screening stage assigning a single scalar summary statistic
to each data set (where the size of the statistic measures miscalibration).
\item A detailed drill-down for each data set
whose scalar summary statistic is large, drilling down into
the miscalibration's variation as a function of the probability of success.
\end{enumerate}

The drill-down relies on graphical display of miscalibration;
the scalar statistic for the first stage simply summarizes
the overall miscalibration across all success probabilities,
as either the maximum absolute deviation of the graph from the ideal
or the size of the range of deviations.
Thus, for each data set, both stages are based on a graph;
the first stage collapses the graph into a single scalar summary statistic.
The following subsubsection constructs the graph.

\subsubsection{Graphical method for assessing calibration}
\label{agraphical}

This subsubsection details the construction of cumulative plots
for calibration.

The cumulative response is
\begin{equation}
\label{cumresponse}
F_k = \frac{1}{n} \sum_{j=1}^k R_j
\end{equation}
for $k = 1$,~$2$, \dots, $n$.

Under the null hypothesis~(\ref{null}), the expected cumulative response
(or, just as well, the cumulative expected response) is
\begin{equation}
\label{expresponse}
\tilde{F}_k = \frac{1}{n} \sum_{j=1}^k S_j
\end{equation}
for $k = 1$,~$2$, \dots, $n$.

A plot of $F_k-\tilde{F}_k$ as a function of $k$ displays miscalibration
directly as slopes that deviate significantly from 0;
indeed, the increment in the expected difference $F_j-\tilde{F}_j$
from $j = k-1$ to $j = k$ is
\begin{equation}
\E[ (F_k-\tilde{F}_k) - (F_{k-1}-\tilde{F}_{k-1}) ]
= \frac{\E[ R_k ] - S_k}{n};
\end{equation}
thus, on a plot with the values for $k$ spaced $1/n$ apart,
the slope from $j = k-1$ to $j = k$ is
\begin{equation}
\label{adelta}
\Delta_k = \E[ R_k ] - S_k
\end{equation}
for $k = 1$,~$2$, \dots, $n$.
The miscalibration for success probabilities near $S_k$ is substantial
when $\Delta_k$ is significantly nonzero, that is, when the slope
of the plot of $F_k-\tilde{F}_k$ deviates significantly from horizontal
over a significantly long range.

To emphasize: {\it miscalibration over a contiguous range of $S_k$
is the slope of the secant line for the plot of $F_k-\tilde{F}_k$
as a function of $\frac{k}{n}$ over that range,
aside from expected stochastic fluctuations}.
The following subsubsection reviews two metrics that summarize
how much the graph deviates from 0 (needless to say,
if the slopes of the secant lines are all nearly 0,
then the whole graph cannot deviate much from 0).
Considering these metrics, Subsubsection~\ref{asignificance} then discusses
the expected random fluctuations.
Many examples of plots are available in Subsection~\ref{aresults}.

\subsubsection{Scalar summary statistics for assessing calibration}
\label{ascalarstats}

This subsubsection details the construction of summary statistics
which collapse the plots introduced in the previous subsubsection
into scalars. The captions of the figures report the values of the statistics
for the corresponding examples.

Two standard metrics for miscalibration over the full range
of success probabilities that account for expected random fluctuations
are that due to Kolmogorov and Smirnov, the maximum absolute deviation
\begin{equation}
\label{aKolmogorov-Smirnov}
G = \max_{1 \le k \le n} |F_k-\tilde{F}_k|,
\end{equation}
and that due to Kuiper, the size of the range of the deviations
\begin{equation}
\label{aKuiper}
D = \max_{0 \le k \le n} (F_k-\tilde{F}_k)
  - \min_{0 \le k \le n} (F_k-\tilde{F}_k),
\end{equation}
where $F_0 = 0 = \tilde{F}_0$; Remark~\ref{zero}
of Subsection~\ref{scalarstats}
explains the reason for including $F_0$ and $\tilde{F}_0$.
Under the null hypothesis~(\ref{null}),
the distributions of $G$ and $D$ are known,
and can form the basis for tests of statistical significance,
as described, for example,
by Section~14.3.4 of~\cite{press-teukolsky-vetterling-flannery}.
The distributions are easy to calculate under the null hypothesis~(\ref{null}),
as then the sequence $F_1$, $F_2$, \dots, $F_n$ defined in~(\ref{cumresponse})
is a random walk with fully specified transition probabilities
$S_1$, $S_2$, \dots, $S_n$.
For assessing statistical significance (rather than overall effect size),
$G$ and $D$ should be divided by $\sigma$,
where $\sigma$ is $1/n$ times the standard deviation
of the sum of independent Bernoulli variates whose probabilities of success
are $S_1$, $S_2$, \dots, $S_n$, that is,
\begin{equation}
\label{astddev}
\sigma = \frac{1}{n} \sqrt{\sum_{j=1}^n S_j (1-S_j)};
\end{equation}
the following remark explains why.

\begin{remark}
\label{proofs}
Under the null hypothesis~(\ref{null}) that assumes that the outcomes
$R_1$, $R_2$, \dots, $R_n$ are drawn independently from Bernoulli distributions
whose probabilities of success are $S_1$, $S_2$, \dots, $S_n$,
the expected value of $G/\sigma$ is less than or equal to
the expected value of the maximum (over a subset of the unit interval $[0, 1]$)
of the absolute value of the standard Brownian motion
over the unit interval $[0, 1]$, in the limit $n \to \infty$.
As discussed by~\cite{masoliver} (with $x = 0$ and $D = 1$ in Formula~46
from the associated arXiv publication\footnote{A freely available preprint
of~\cite{masoliver} is available at \url{https://arxiv.org/pdf/1401.4939.pdf}}
\dots\ or see immediately before Remark~I
on the relevant StackExchange thread\footnote{The associated discussion
on StackExchange is available at
{\expandafter\def\expandafter\UrlBreaks\expandafter{\UrlBreaks\do\/\do-}
\url{https://math.stackexchange.com/questions/3251957/calculating-the-expecation-of-the-supremum-of-absolute-value-of-a-brownian-motion}}}),
the expected value of the maximum of the absolute value
of the standard Brownian motion over $[0, 1]$ is $\sqrt{\pi/2} \approx 1.25$.
The discussion by~\cite{masoliver} immediately following Formula~44
from the associated arXiv publication\footnote{A freely available preprint
of~\cite{masoliver} is available at \url{https://arxiv.org/pdf/1401.4939.pdf}}
shows that the probability distribution of the maximum of the absolute value
of the standard Brownian motion over $[0, 1]$ is sub-Gaussian,
decaying past its mean $\sqrt{\pi/2} \approx 1.25$.
Values of $G/\sigma$ much greater than 1.25 imply serious miscalibration,
while values of $G/\sigma$ close to 0 imply that $G$ did not detect
any statistically significant miscalibration.
Needless to say, similar remarks pertain to $D$.
\end{remark}

\subsubsection{Significance of stochastic fluctuations
for assessing calibration}
\label{asignificance}

This subsubsection discusses statistical significance
both for the graphical methods of Subsubsection~\ref{agraphical}
and for the summary statistics of Subsubsection~\ref{ascalarstats}.

The plot of $F_k-\tilde{F}_k$ as a function of $k/n$ automatically
includes some ``error bars'' courtesy of the discrepancy
$F_k-\tilde{F}_k$ fluctuating randomly as the index $k$ increments.
Of course, the standard deviation of a Bernoulli variate
whose expected value is $S_j$ is $\sqrt{S_j (1-S_j)}$ ---
smaller both for $S_j$ near 0 and for $S_j$ near 1.
To indicate the size of the fluctuations, the plots should include
a triangle centered at the origin whose height above the origin is $2/n$
times the standard deviation of the sum of independent Bernoulli variates
with success probabilities $S_1$, $S_2$, \dots, $S_n$;
thus, the height of the triangle above the origin 
(where the triangle itself is centered at the origin) is
$2 \sqrt{\sum_{j=1}^n S_j (1-S_j)} / n$.
The expected deviation from 0 of $|F_k-\tilde{F}_k|$
(at any specified value for $k$)
is no greater than this height, under the assumption that the responses
$R_1$,~$R_2$, \dots, $R_n$ are draws
from independent Bernoulli distributions with the correct success probabilities
$S_1$, $S_2$, \dots, $S_n$, that is, under the null hypothesis~(\ref{null}).
The triangle is similar to the classic confidence bands
around an empirical cumulative distribution function
given by Kolmogorov and Smirnov, as reviewed by~\cite{doksum}.

\subsection{Results and discussion for assessing calibration}
\label{aresults}

This subsection illustrates via several examples
the previous subsection's methods,
together with the traditional plots --- so-called ``reliability diagrams'' ---
discussed in Subsection~\ref{aintro}.\footnote{Permissively licensed
open-source software for reproducing all figures and statistics displayed here
is available at \url{https://github.com/facebookresearch/fbcdgraph}}
Subsubsection~\ref{asynthetic} presents toy examples.
Subsubsection~\ref{aimagenetex} analyzes a popular data set of images,
ImageNet.

Figures~\ref{imagenetcal}--\ref{100_1e} display the classical calibration plots
as well as both the plots of cumulative differences and the exact expectations
in the absence of noise from random sampling when the exact expectations
are available (as in Subsubsection~\ref{asynthetic}).
The captions of the figures discuss the numerical results depicted.

To generate the figures in Subsubsection~\ref{asynthetic},
we specify values for $S_1$,~$S_2$, \dots, $S_n$
and for $P_1$,~$P_2$, \dots, $P_n$ differing from $S_1$,~$S_2$, \dots, $S_n$,
then independently draw $R_1$,~$R_2$, \dots, $R_n$
from the Bernoulli distributions with parameters $P_1$,~$P_2$, \dots, $P_n$,
respectively. Ideally the plots would show how and where 
$P_1$,~$P_2$, \dots, $P_n$ differs from $S_1$,~$S_2$, \dots, $S_n$.
Appendix~\ref{caution} considers the case in which $P_k = S_k$
for all $k = 1$,~$2$, \dots, $n$.

The top rows of the figures with three rows plot $F_k-\tilde{F}_k$
from~(\ref{cumresponse}) and~(\ref{expresponse}) as a function of $k/n$,
with the rightmost plot displaying its noiseless expected value
rather than using the observations $R_1$,~$R_2$, \dots, $R_n$
(Figure~\ref{imagenetcal} omits the rightmost plot since the expected values
are unknown in that case).
In each of these plots,
the upper axis specifies $k/n$, while the lower axis specifies $S_k$
for the corresponding value of $k$.
The lowermost two rows of the figures with three rows plot the pairs
$(X_1, Y_1)$,~$(X_2, Y_2)$, \dots, $(X_{\ell}, Y_{\ell})$
from~(\ref{avgY}) and~(\ref{avgX}),
with the rightmost plots using an equal number of observations per bin.
The dark black lines and points of the left and right plots in the lowermost
two rows of Figures~\ref{10000}, \ref{1000}, and~\ref{100} are in fact
identical, since $S_1$,~$S_2$, \dots, $S_n$ are equispaced
for those examples (so equally wide bins contain equal numbers
of observations). The figures with only a single diagram
plot pairs $(X_1, Y_1)$,~$(X_2, Y_2)$, \dots, $(X_n, Y_n)$
from~(\ref{avgY}) and~(\ref{avgX}),
but this time using their noiseless expected values
$(S_1, P_1)$,~$(S_2, P_2)$, \dots, $(S_n, P_n)$
instead of using the random observations $R_1$,~$R_2$, \dots, $R_n$.

Perhaps the simplest, most straightforward method to gauge uncertainty
in the binned plots is to vary the number of bins
and observe how the plotted values vary.
All figures displayed employ this method, with the number of bins increased in
the second rows of plots beyond the number of bins in the third rows of plots.
The figures also include the ``error bars'' resulting
from one of the bootstrap resampling schemes proposed by~\cite{brocker-smith},
obtained by drawing $n$ observations independently and uniformly at random
with replacement from $(S_1, R_1)$, $(S_2, R_2)$, \dots, $(S_n, R_n)$
and then plotting (in light gray) the corresponding reliability diagram,
and repeating for a total of 20 times (thus displaying 20 gray lines per plot).
The chance that all 20 lines are unrepresentative
of the expected statistical variations would be roughly $1/20 = 5$\%,
so plotting these 20 lines corresponds to approximately 95\% confidence.
An alternative is to display the bin frequencies as suggested,
for example, by~\cite{murphy-winkler}.
Other possibilities often involve kernel density estimation,
as suggested, for example, by~\cite{brocker} and~\cite{wilks}.
All such methods require selecting widths for the bins or kernel smoothing;
avoiding having to make what is a necessarily somewhat arbitrary choice
is possible by varying the widths, as done in the plots of the present paper.
Chapter~8 of~\cite{wilks} comprehensively reviews the extant literature.

We may set the widths of the bins such that either
\{1\} the average of $S_k$ for $k$ in each bin is approximately equidistant
from the average of $S_k$ for $k$ in each neighboring bin or
\{2\} the range of $k$ for every bin has the same width.
Both options are natural; the first is the canonical choice,
whereas the second ensures that error bars would be similarly sized
for every bin. The figures display both possibilities,
with the first on the left and the second on the right.
Setting the number of bins together with either of these choices
fully specifies the bins. As discussed earlier, we vary the number of bins
since there is no perfect setting --- using fewer bins offers estimates
with higher confidence yet limits the resolution for detecting miscalibration
and for assessing the dependence of calibration as a function of $S_k$.

\subsubsection{Synthetic examples for assessing calibration}
\label{asynthetic}

In this subsubsection, the examples are sampled randomly
from various statistical models so that the underlying ``ground-truth''
is known explicitly.

\begin{figure}
\begin{centering}

\parbox{\imsize}{\includegraphics[width=\imsize]
                {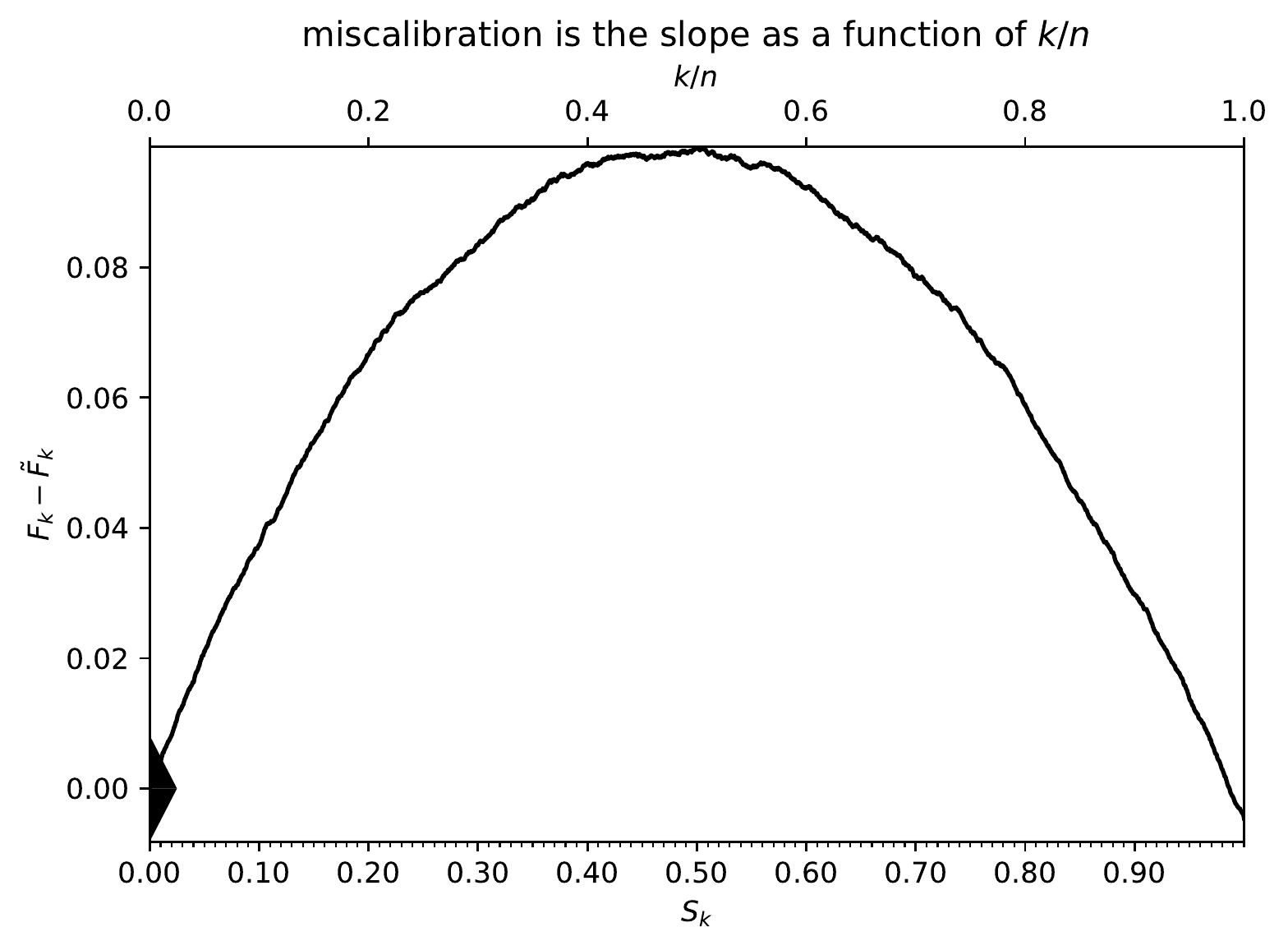}}
\quad\quad
\parbox{\imsize}{\includegraphics[width=\imsize]
                {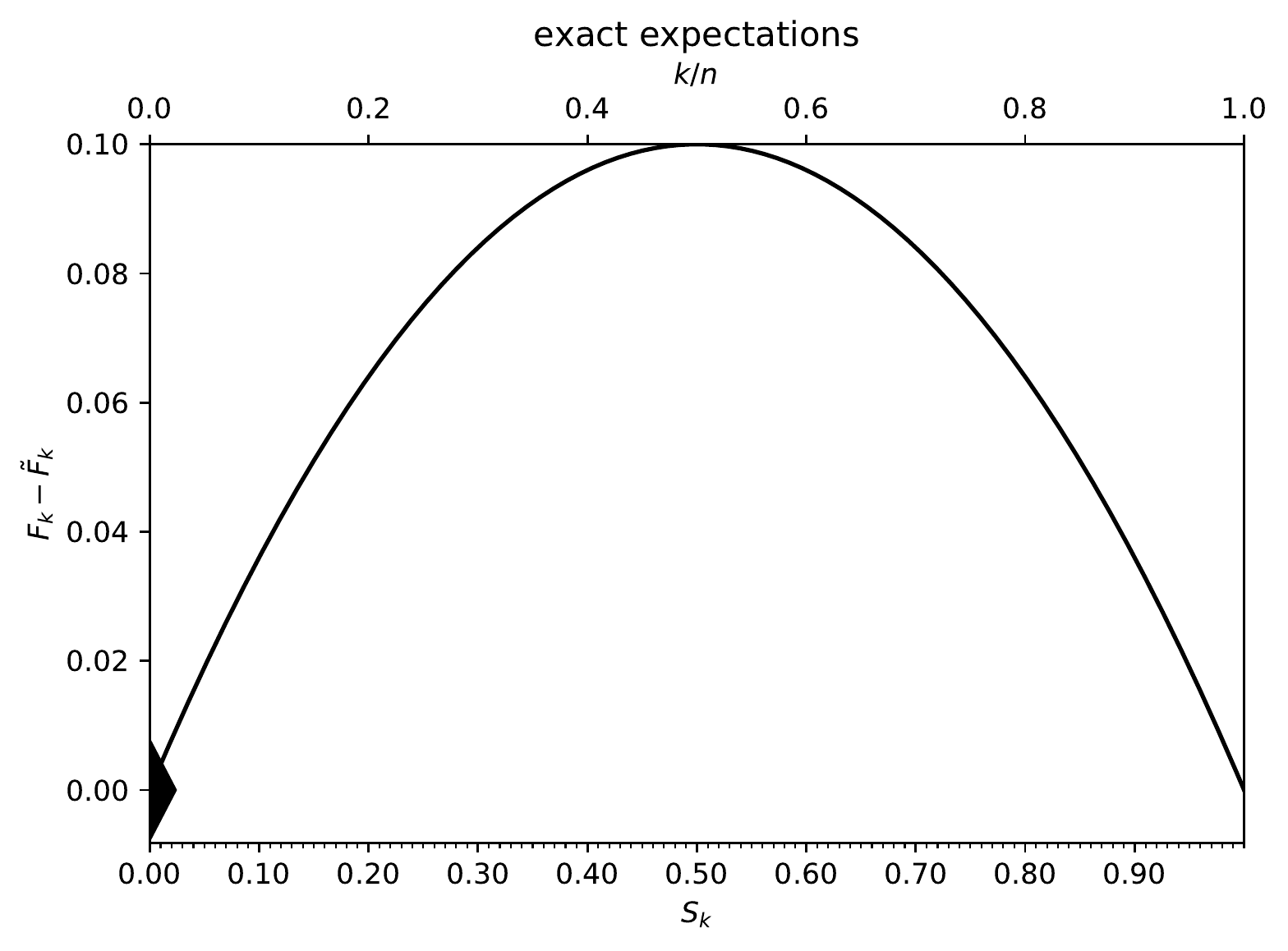}}

\vspace{\vertsep}

\parbox{\imsize}{\includegraphics[width=\imsize]
                {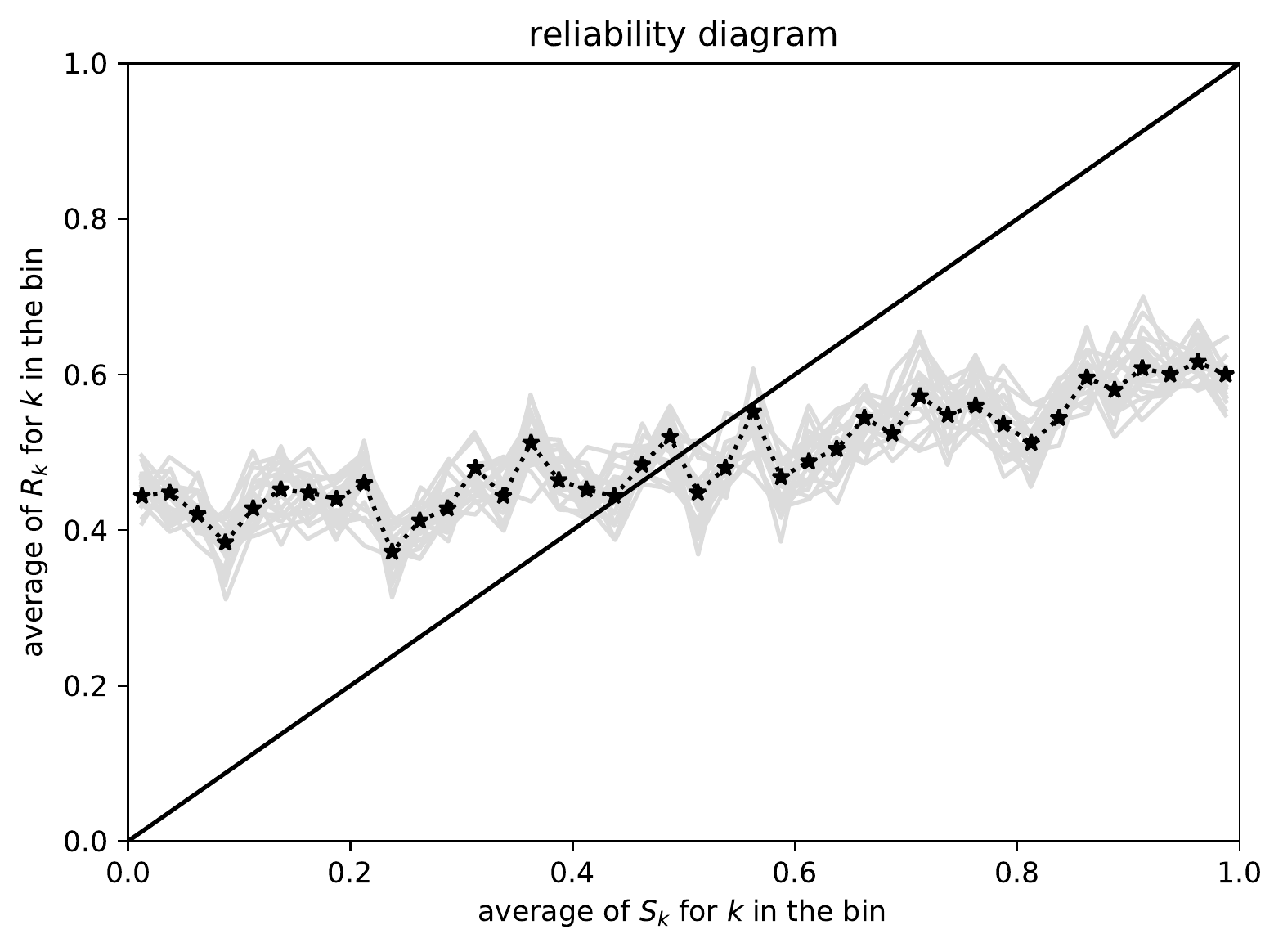}}
\quad\quad
\parbox{\imsize}{\includegraphics[width=\imsize]
                {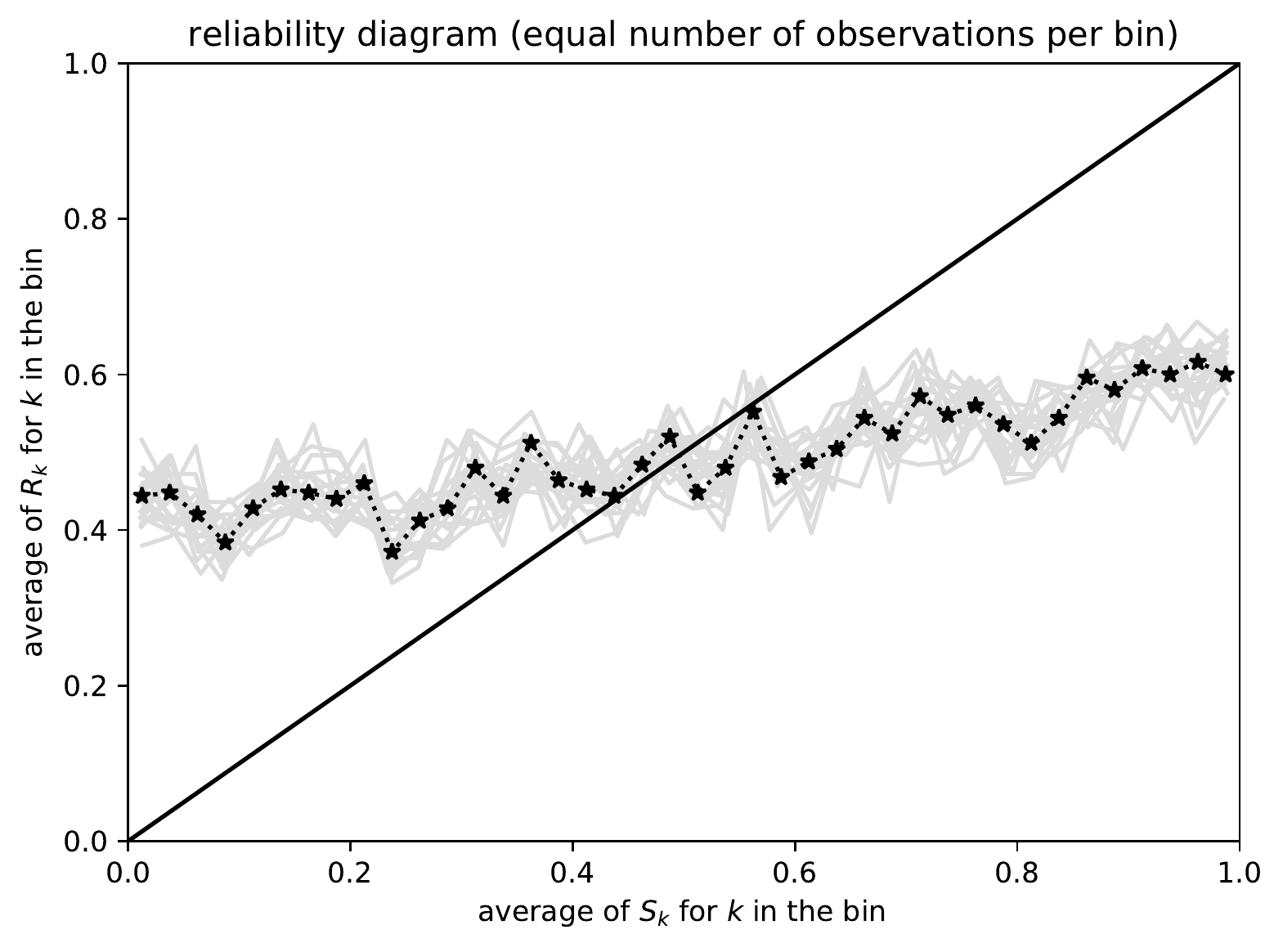}}

\vspace{\vertsep}

\parbox{\imsize}{\includegraphics[width=\imsize]
                {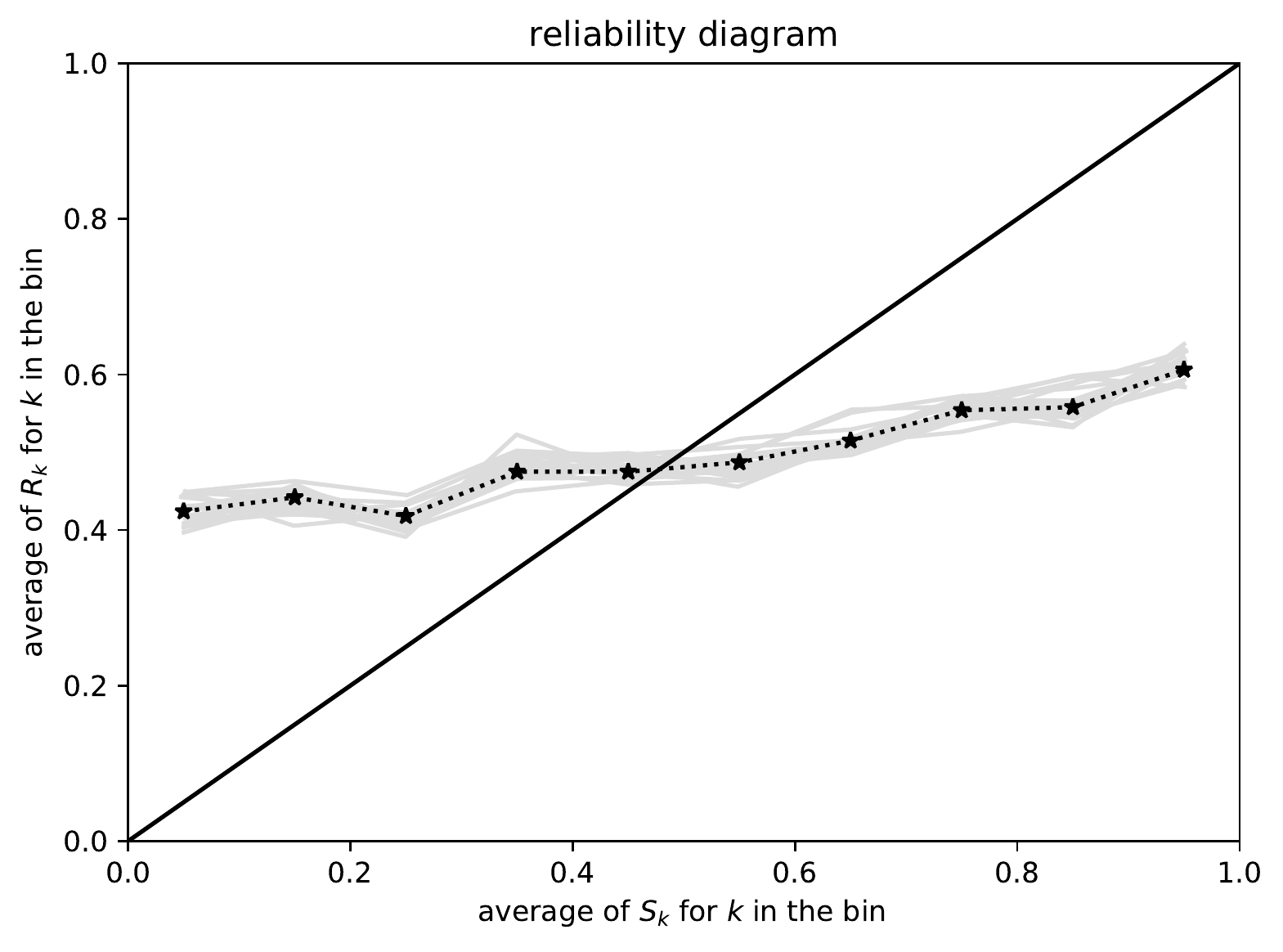}}
\quad\quad
\parbox{\imsize}{\includegraphics[width=\imsize]
                {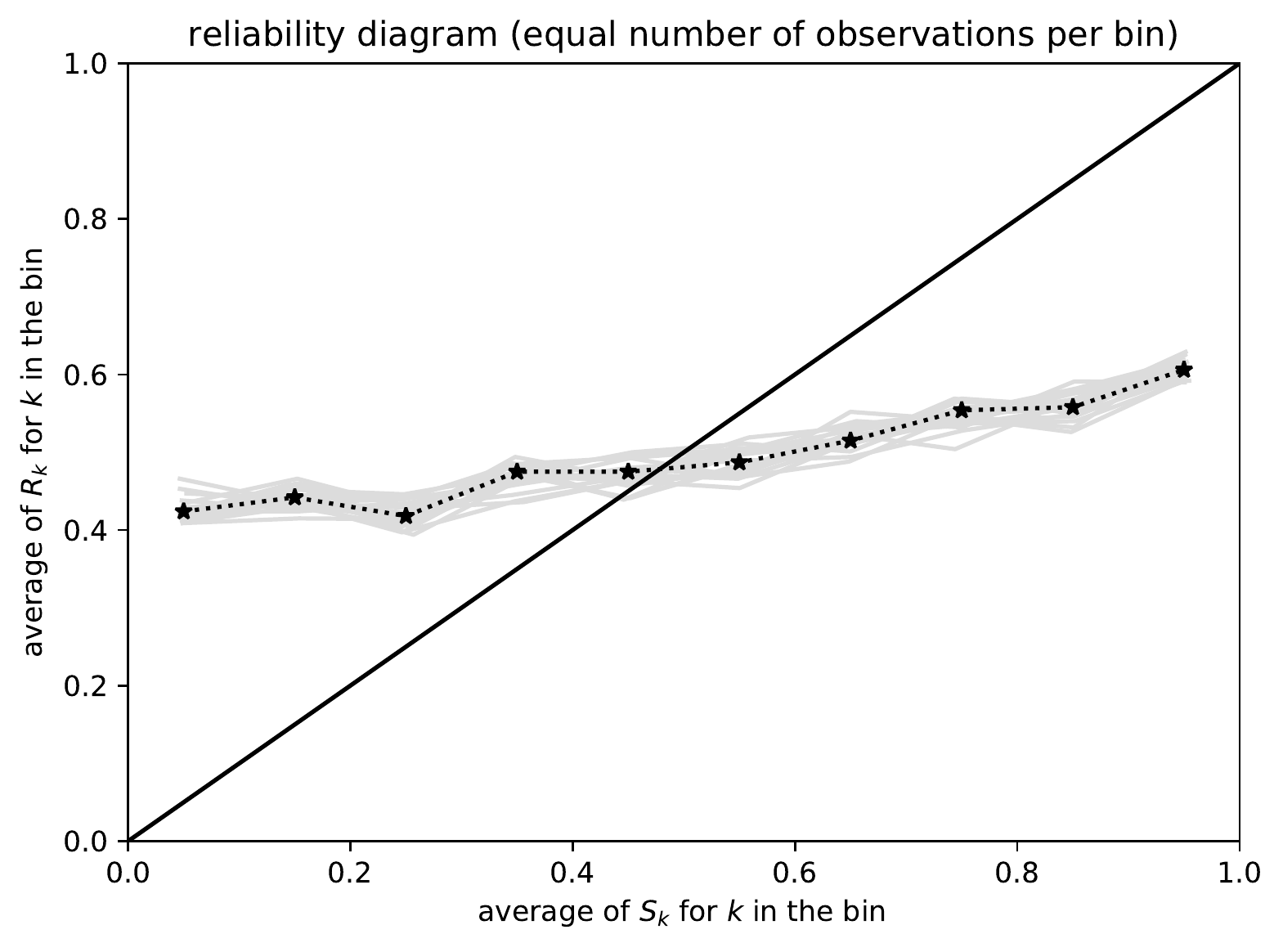}}

\end{centering}
\caption{$n =$ 10,000; $S_1$, $S_2$, \dots, $S_n$ are equispaced;
         Kuiper's statistic is $0.1031 / \sigma = 25.25$,
         Kolmogorov's and Smirnov's is $0.09849 / \sigma = 24.12$.
Figure~\ref{10000e} displays the ground-truth reliability diagram.
All plots, whether cumulative or conventional, appear to work well.
}
\label{10000}
\end{figure}

\begin{figure}
\begin{centering}

\parbox{\imsize}{\includegraphics[width=\imsize]
                {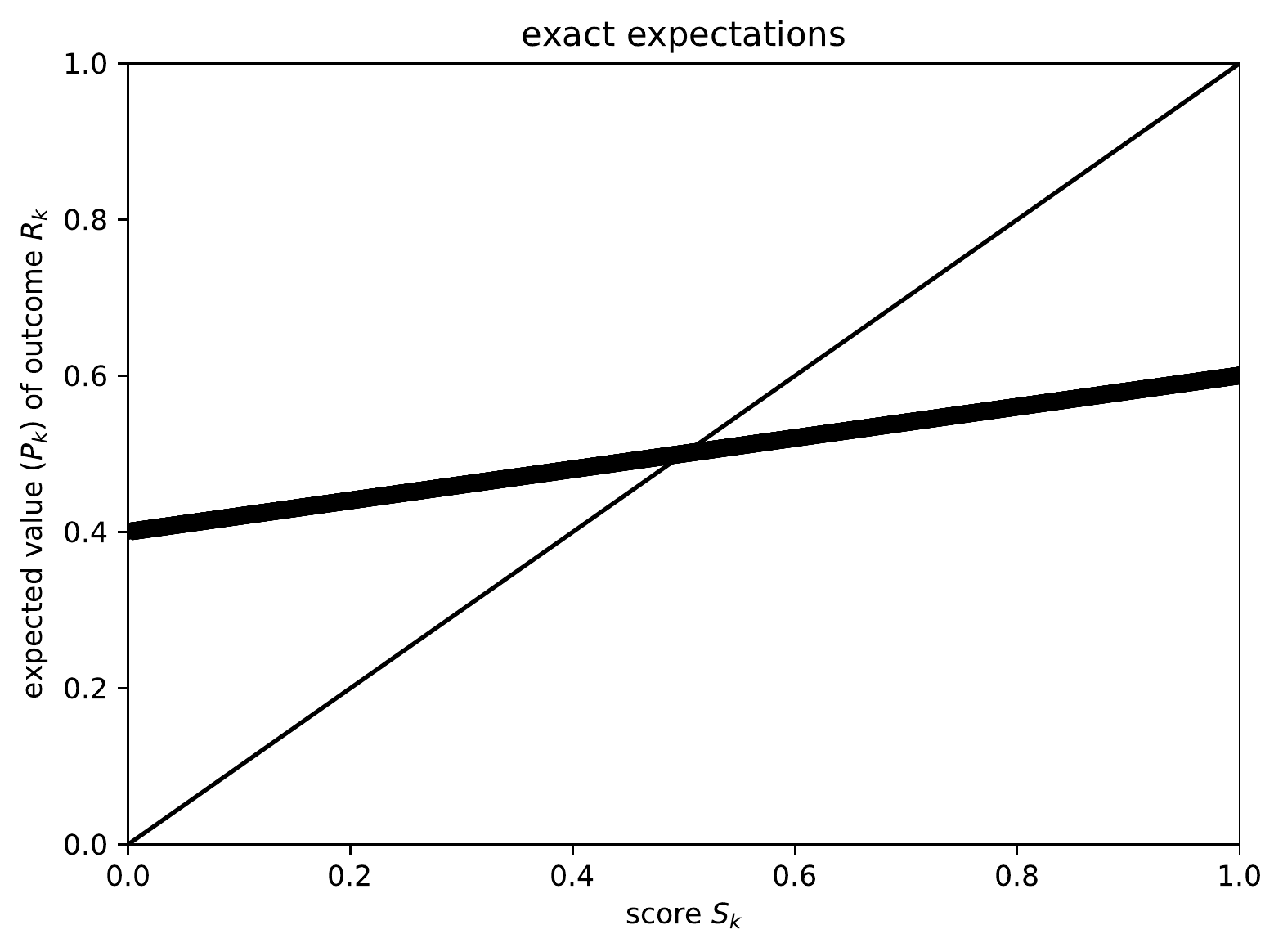}}

\end{centering}
\caption{Ground-truth reliability diagram for Figure~\ref{10000}}
\label{10000e}
\end{figure}

\begin{figure}
\begin{centering}

\parbox{\imsize}{\includegraphics[width=\imsize]
                {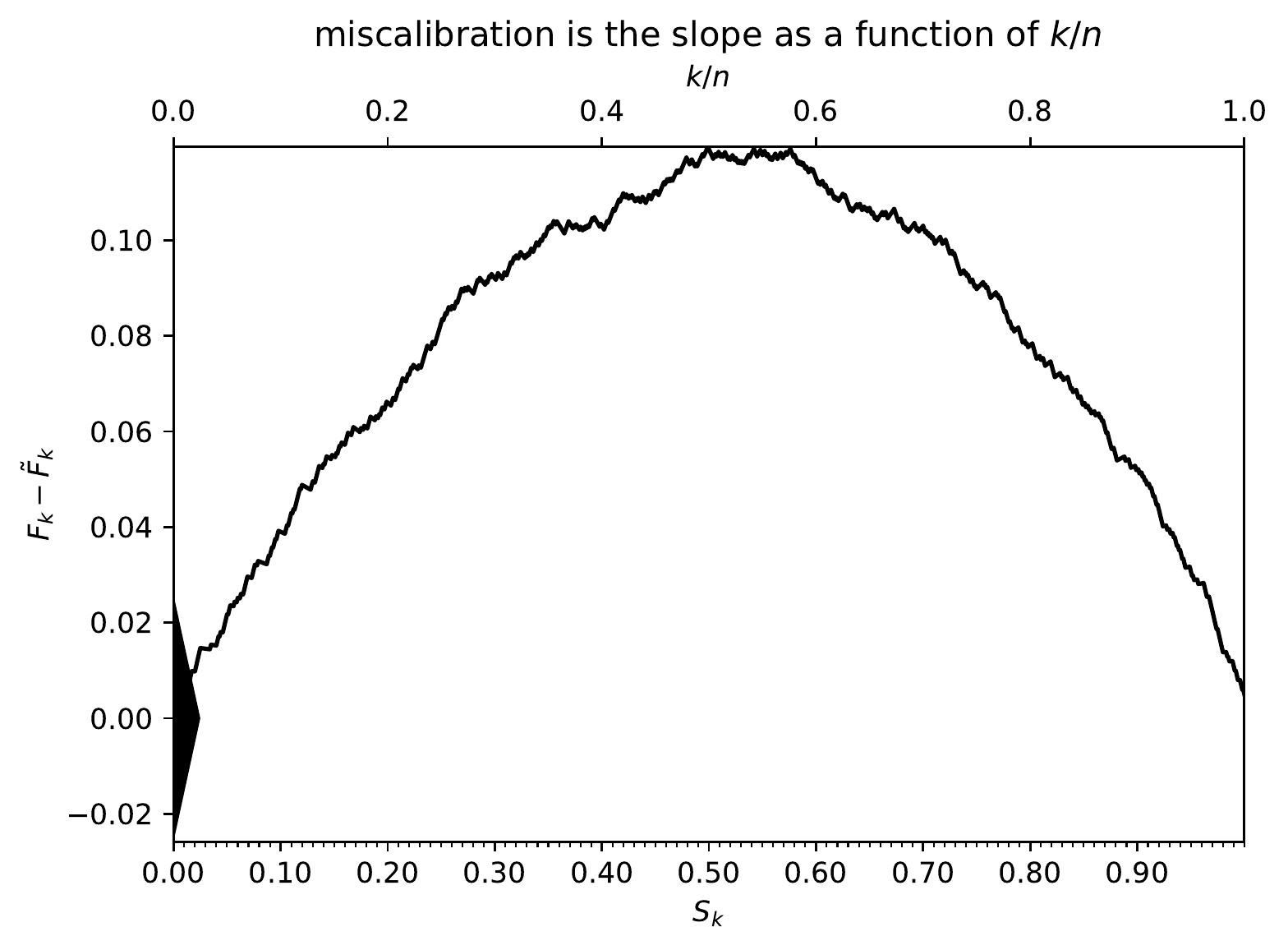}}
\quad\quad
\parbox{\imsize}{\includegraphics[width=\imsize]
                {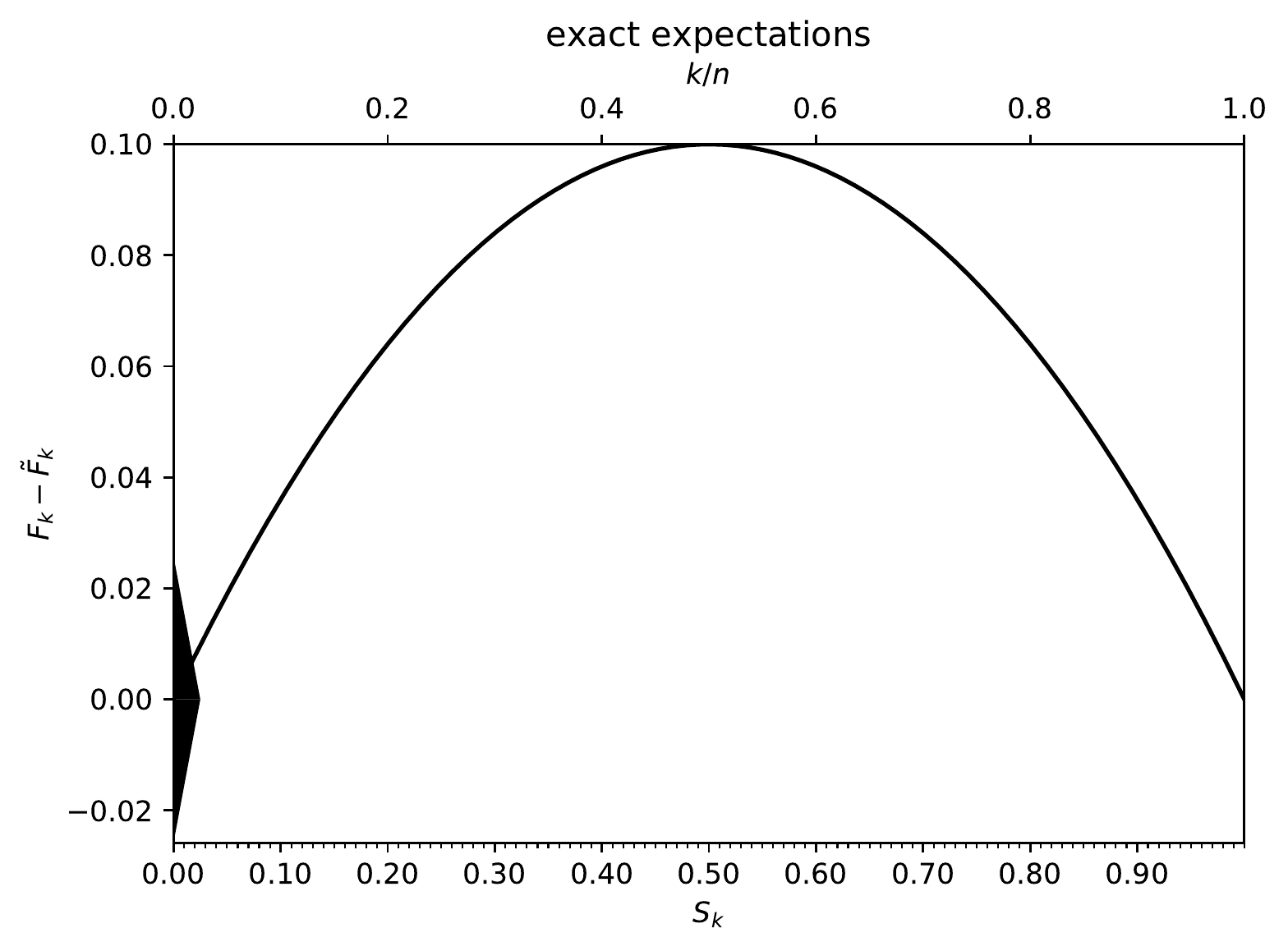}}

\vspace{\vertsep}

\parbox{\imsize}{\includegraphics[width=\imsize]
                {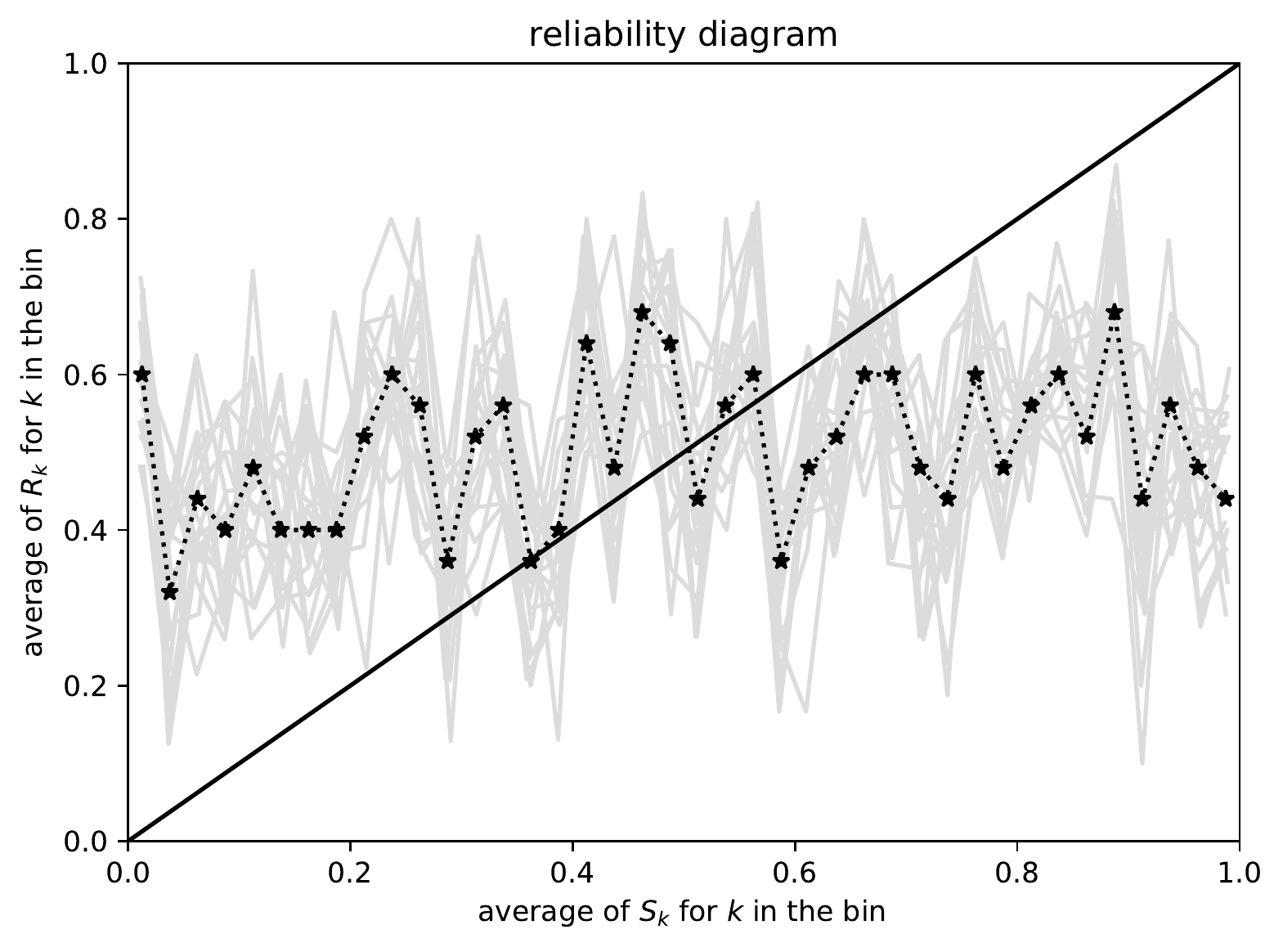}}
\quad\quad
\parbox{\imsize}{\includegraphics[width=\imsize]
                {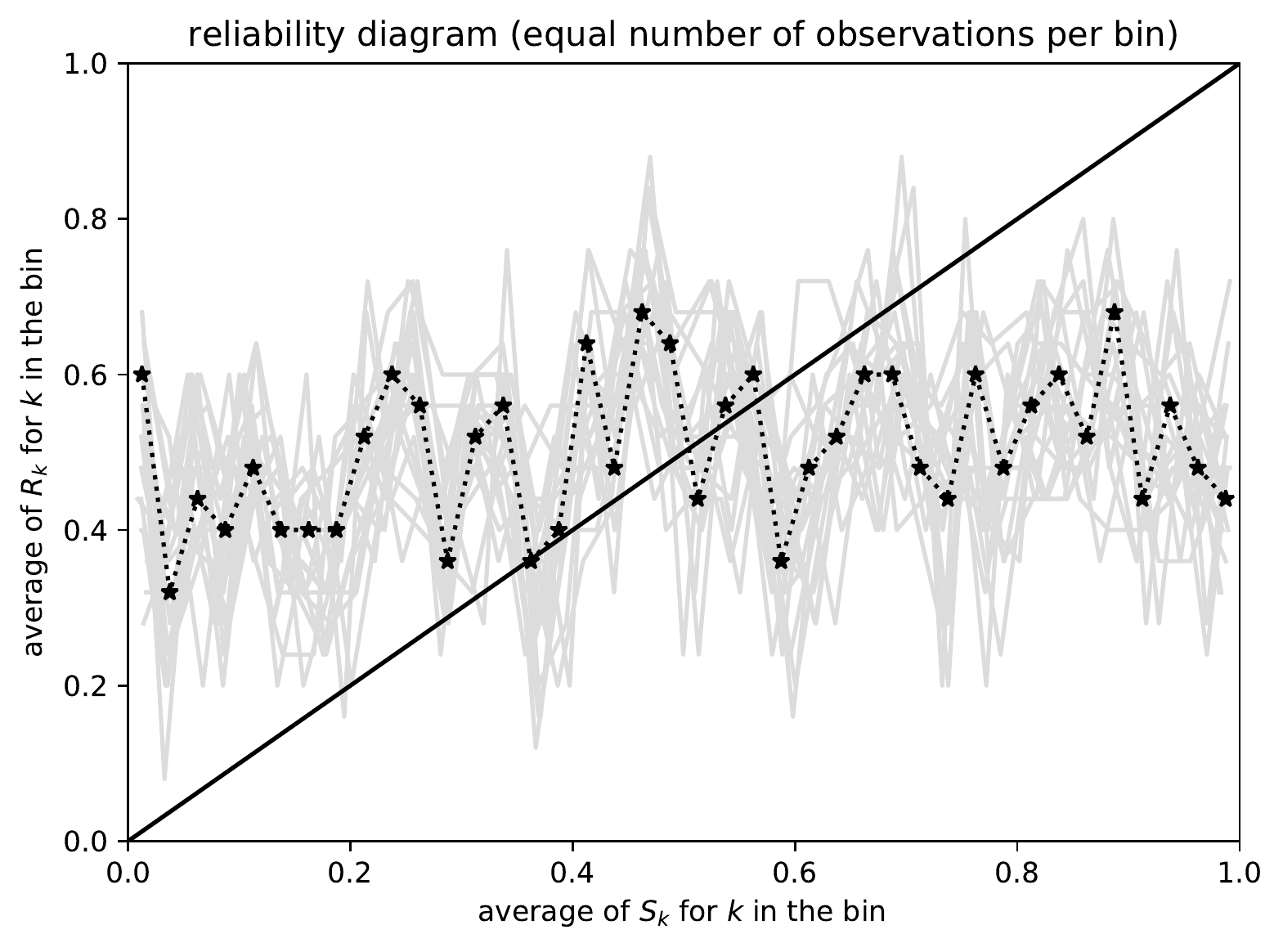}}

\vspace{\vertsep}

\parbox{\imsize}{\includegraphics[width=\imsize]
                {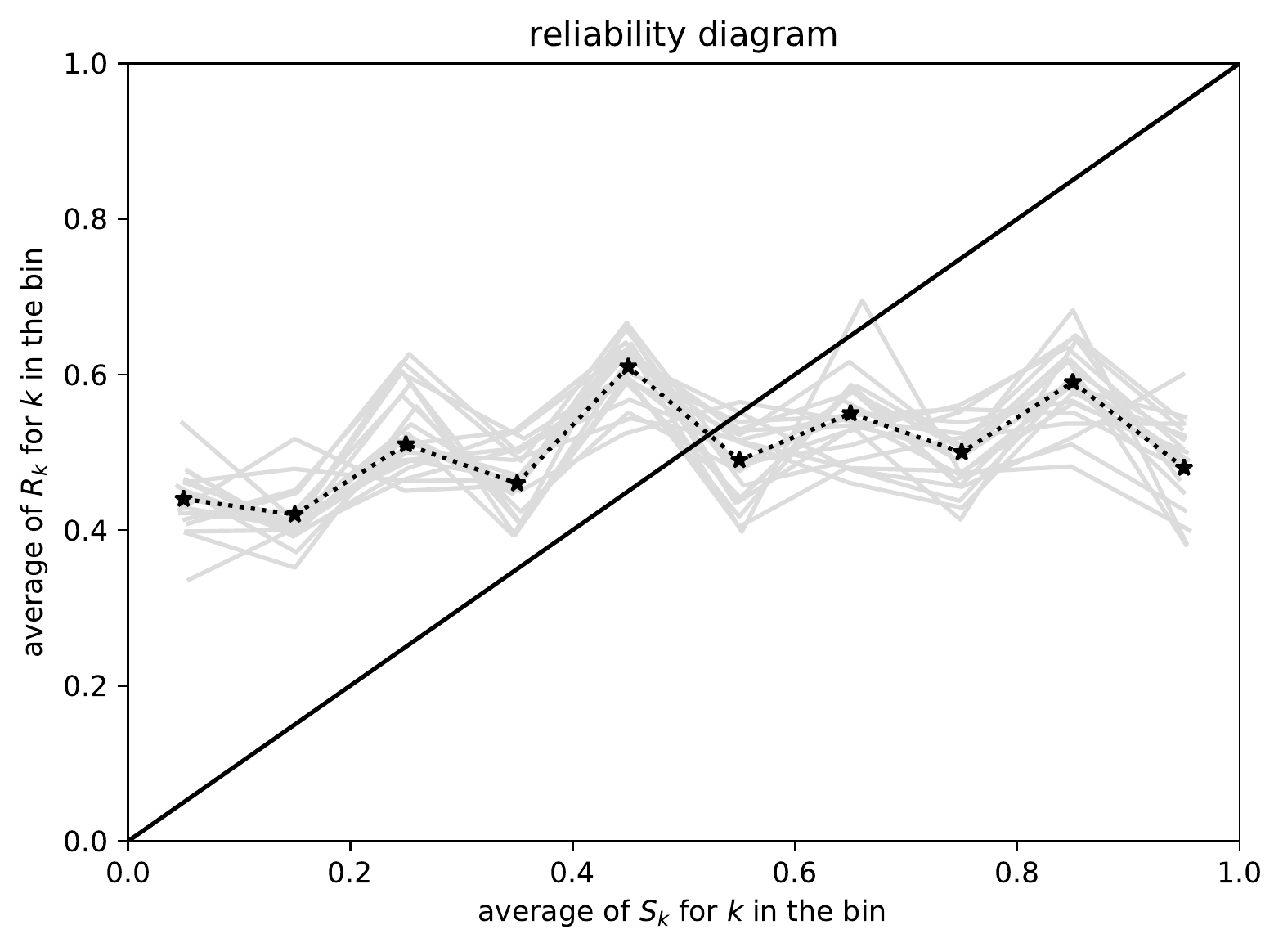}}
\quad\quad
\parbox{\imsize}{\includegraphics[width=\imsize]
                {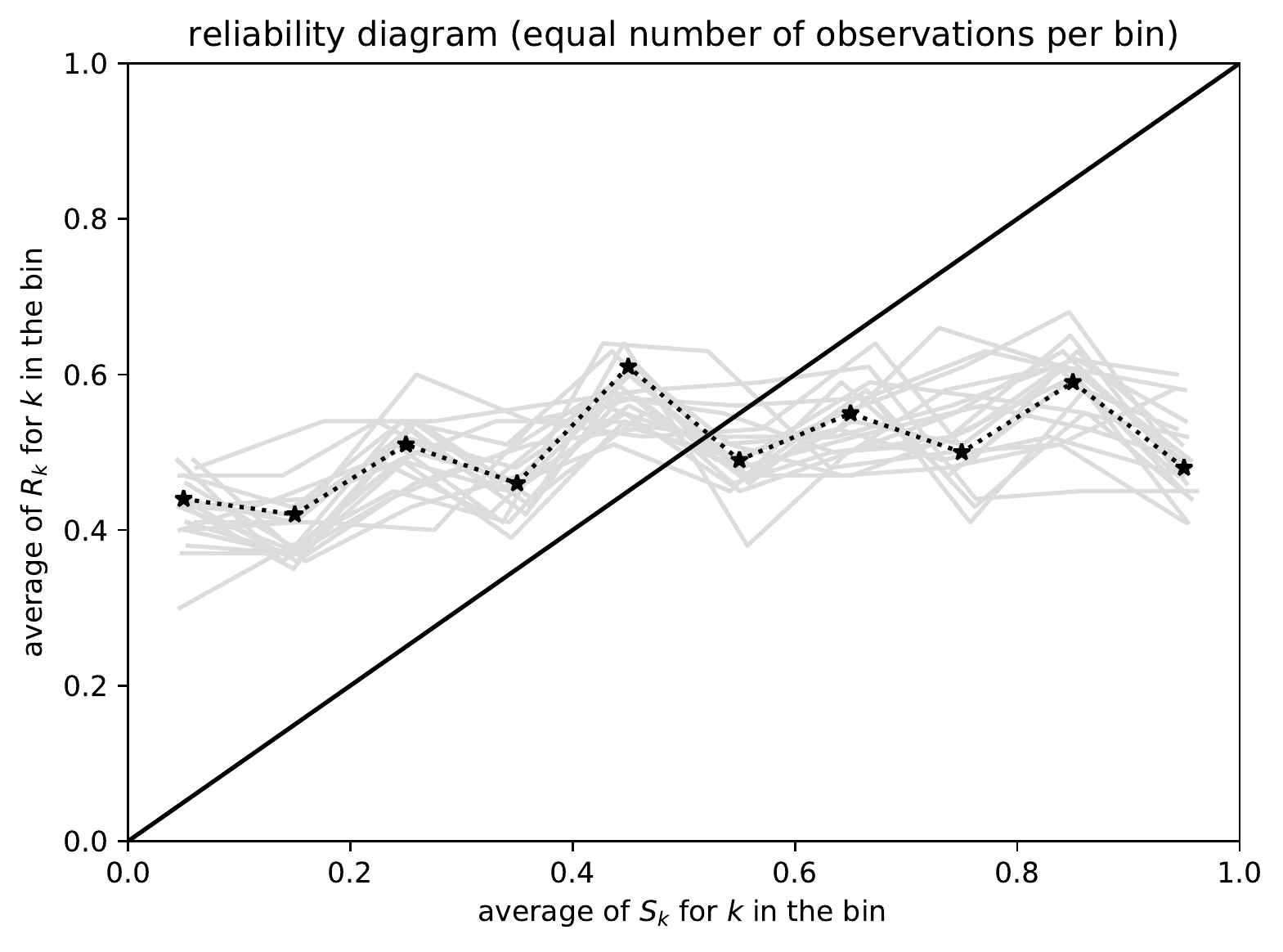}}

\end{centering}
\caption{$n =$ 1,000; $S_1$, $S_2$, \dots, $S_n$ are equispaced;
         Kuiper's statistic is $0.1195 / \sigma = 9.256$,
         Kolmogorov's and Smirnov's is $0.1195 / \sigma = 9.256$.
Figure~\ref{1000e} displays the ground-truth reliability diagram.
All plots, whether cumulative or conventional, appear to work well.
}
\label{1000}
\end{figure}

\begin{figure}
\begin{centering}

\parbox{\imsize}{\includegraphics[width=\imsize]
                {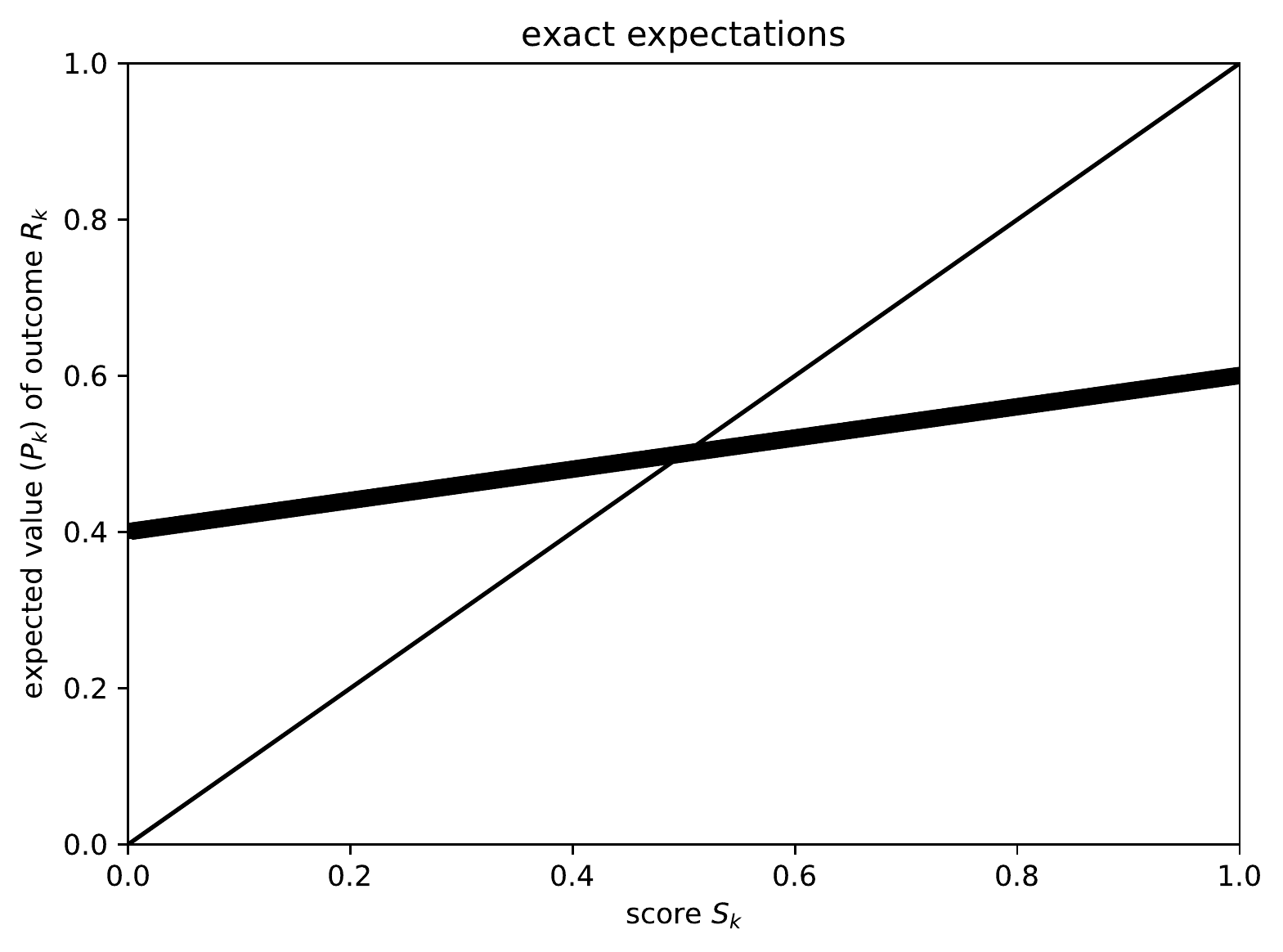}}

\end{centering}
\caption{Ground-truth reliability diagram for Figure~\ref{1000}}
\label{1000e}
\end{figure}

\begin{figure}
\begin{centering}

\parbox{\imsize}{\includegraphics[width=\imsize]
                {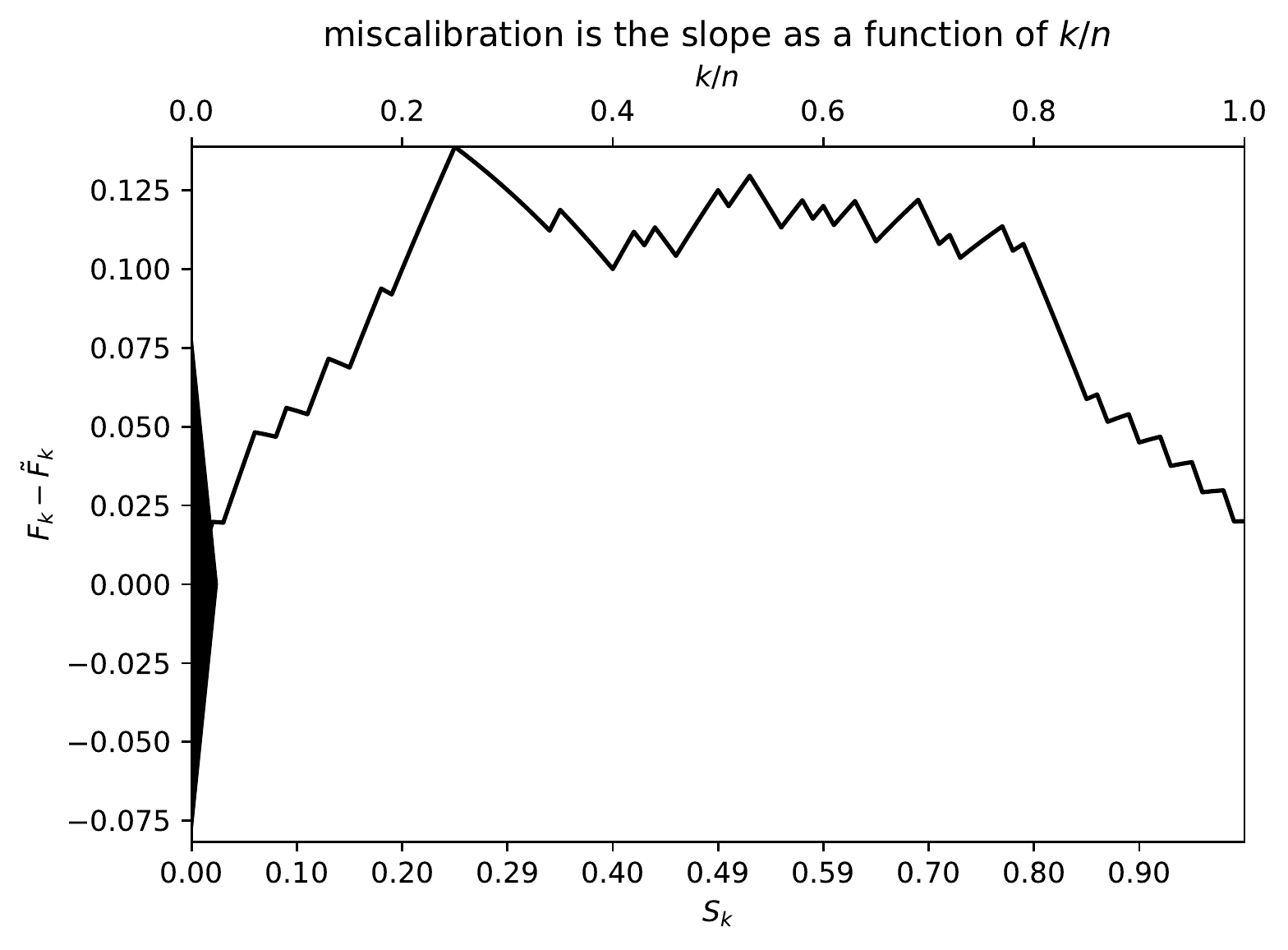}}
\quad\quad
\parbox{\imsize}{\includegraphics[width=\imsize]
                {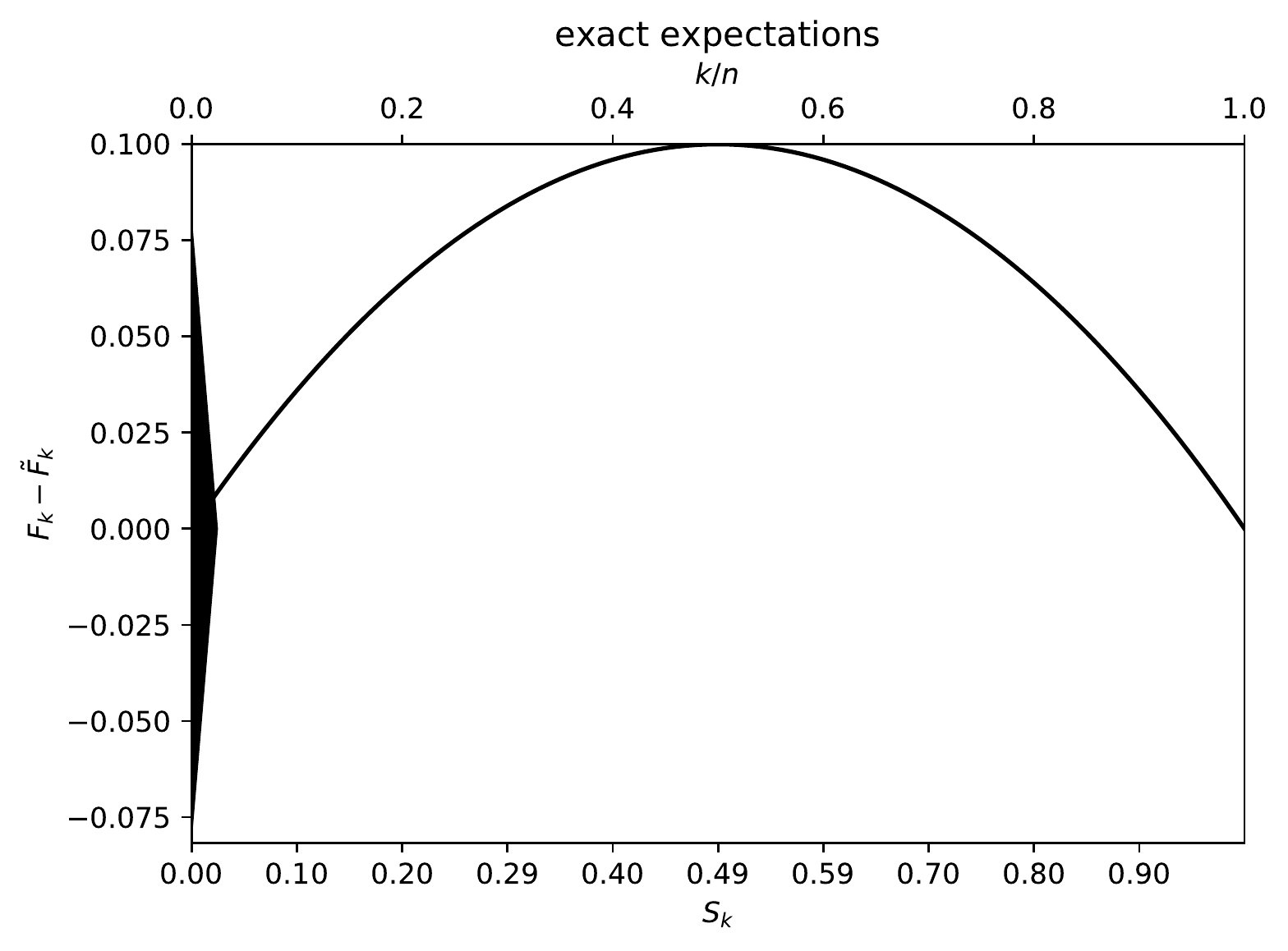}}

\vspace{\vertsep}

\parbox{\imsize}{\includegraphics[width=\imsize]
                {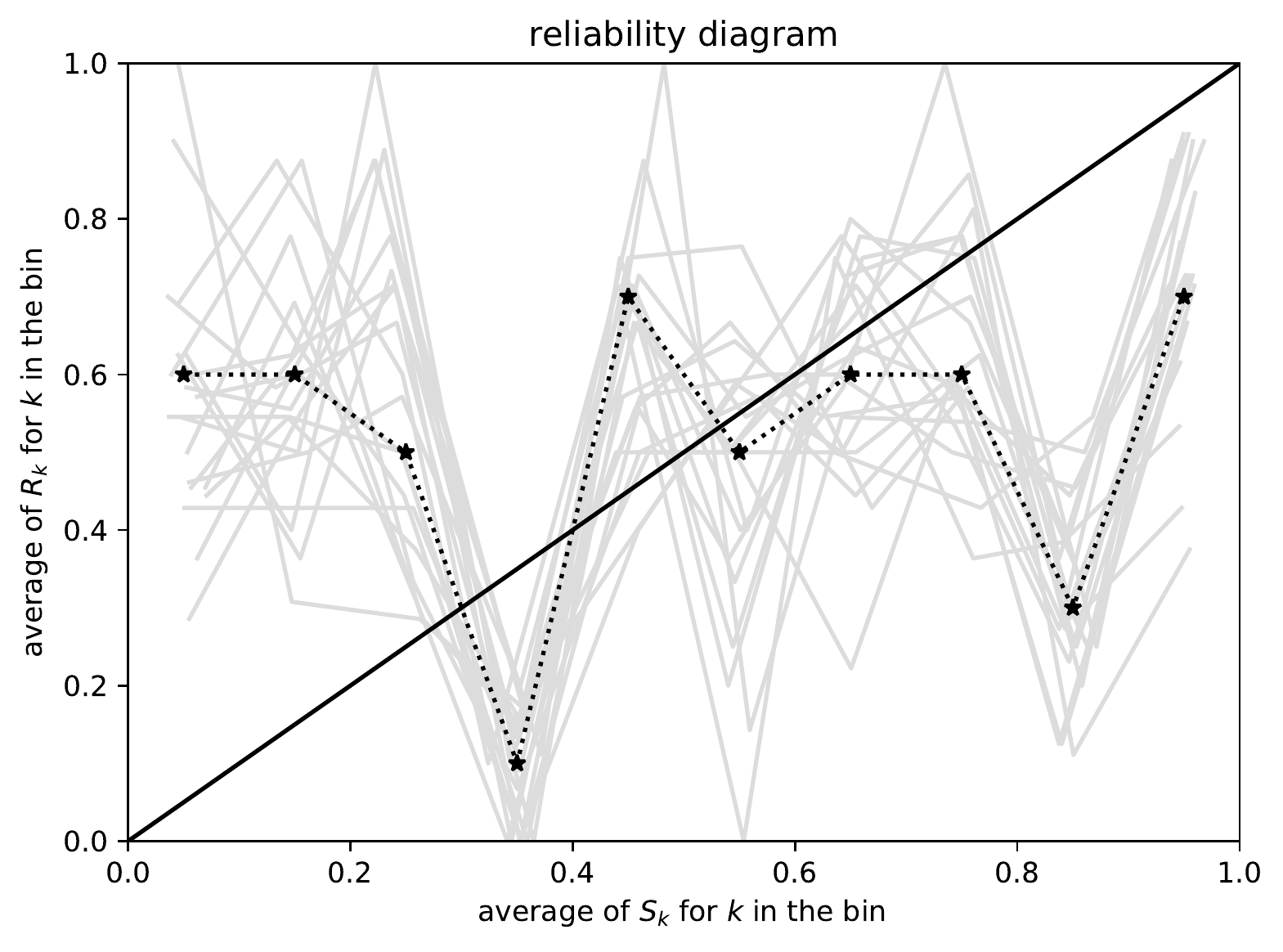}}
\quad\quad
\parbox{\imsize}{\includegraphics[width=\imsize]
                {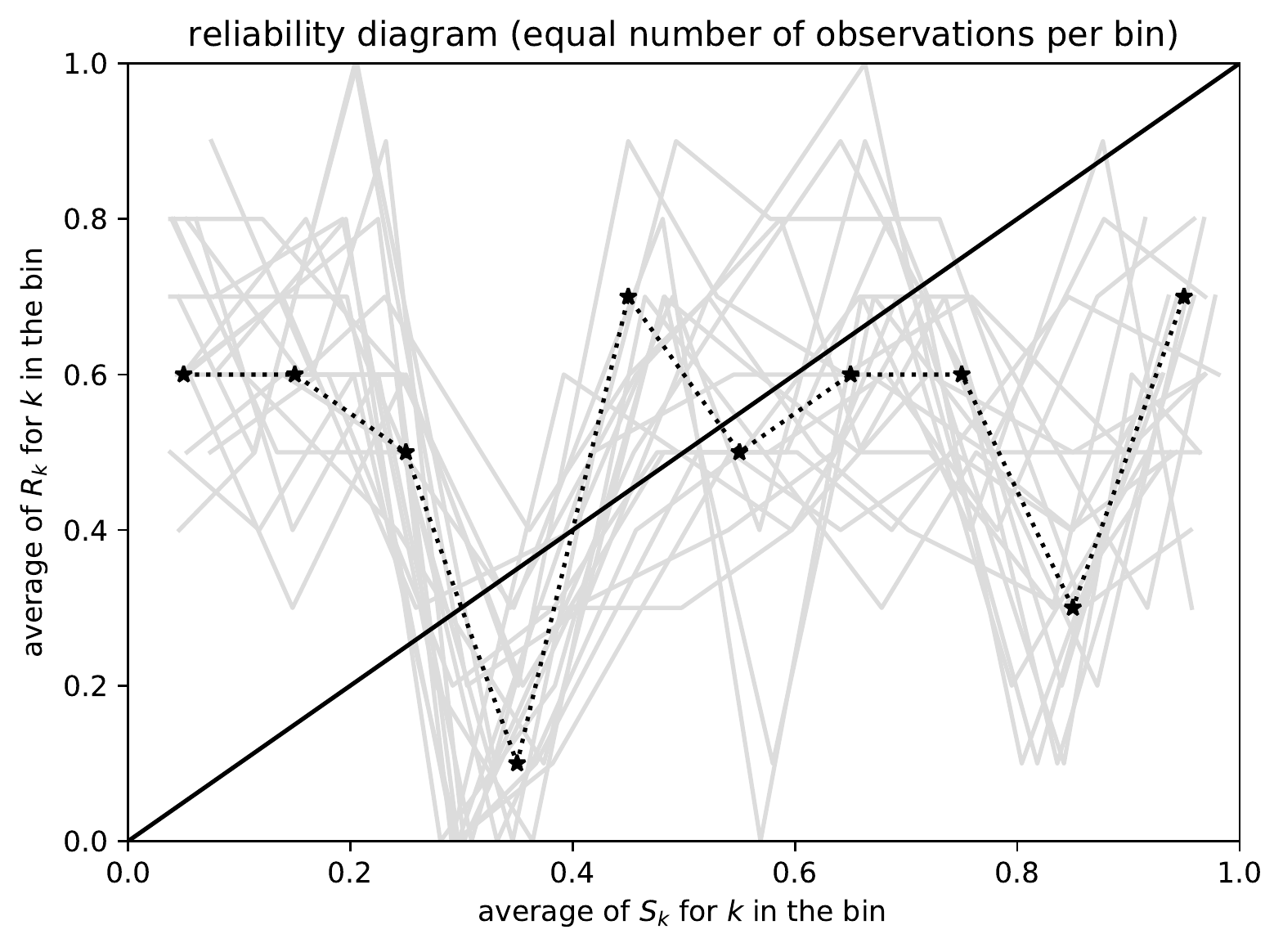}}

\vspace{\vertsep}

\parbox{\imsize}{\includegraphics[width=\imsize]
                {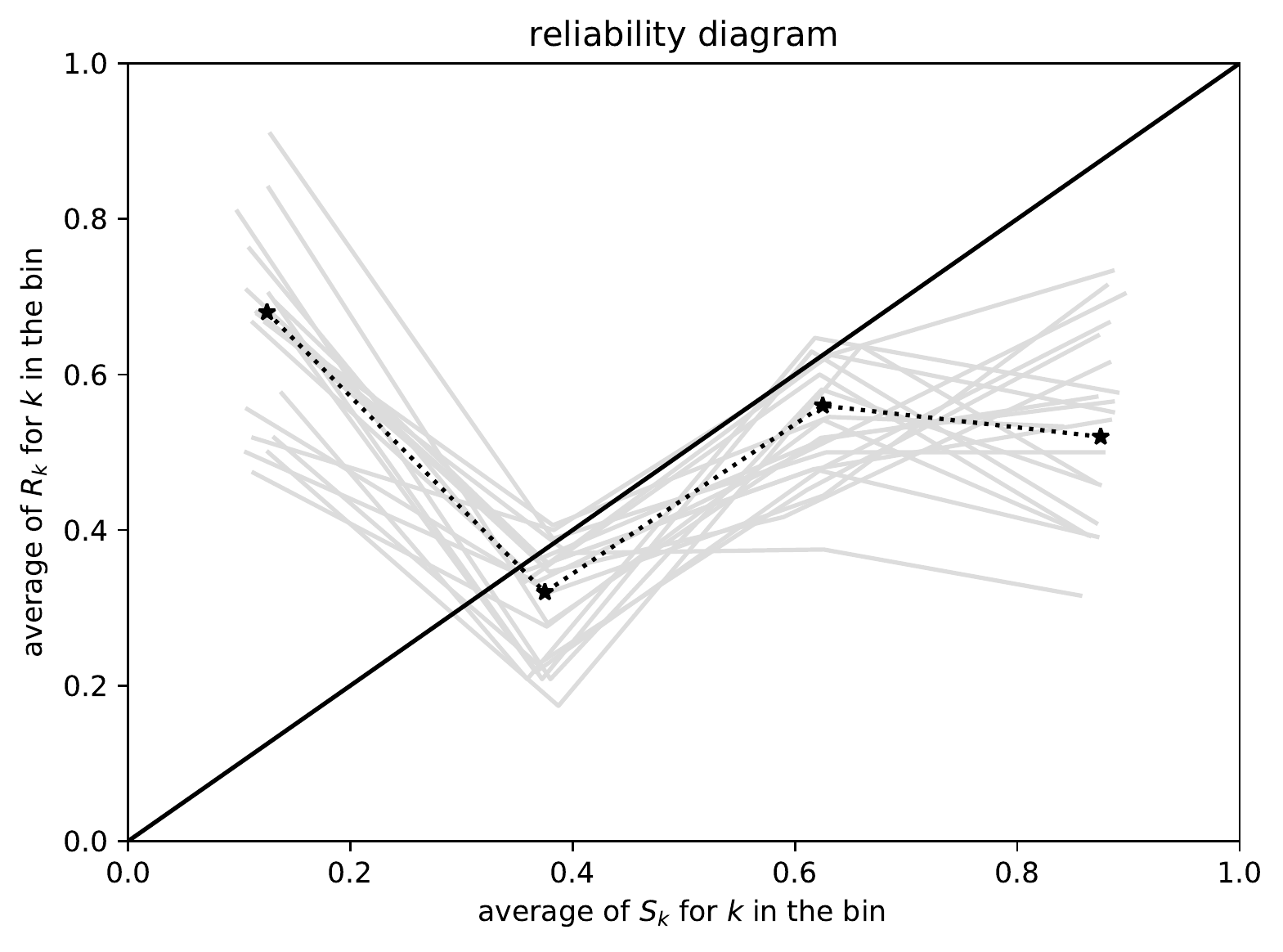}}
\quad\quad
\parbox{\imsize}{\includegraphics[width=\imsize]
                {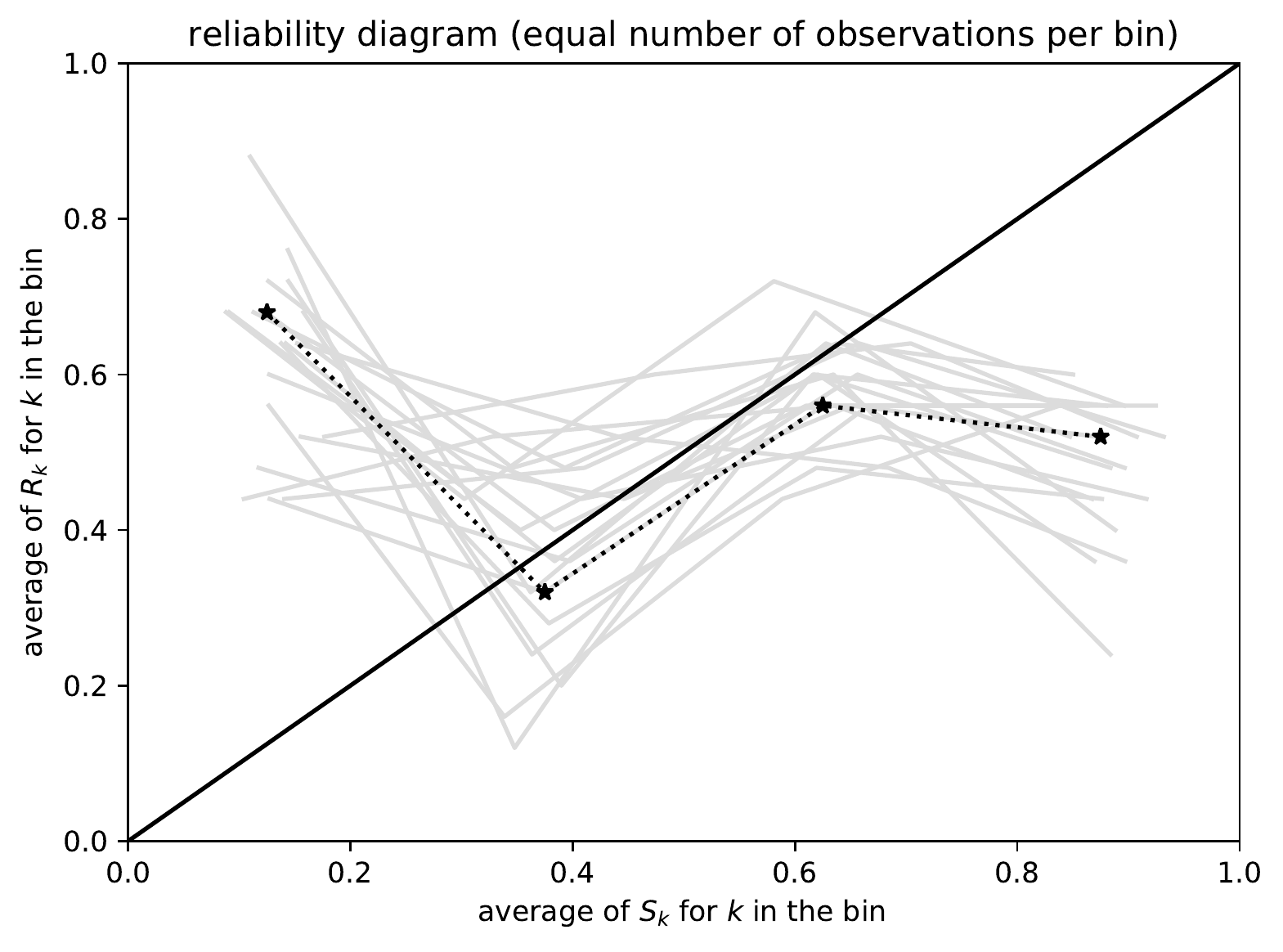}}

\end{centering}
\caption{$n =$ 100; $S_1$, $S_2$, \dots, $S_n$ are equispaced;
         Kuiper's statistic is $0.1388 / \sigma = 3.399$,
         Kolmogorov's and Smirnov's is $0.1388 / \sigma = 3.399$.
Figure~\ref{100e} displays the ground-truth reliability diagram.
The conventional plots become increasingly problematic as $n$ reduces to 100
from 10,000 and 1,000 in Figures~\ref{10000} and~\ref{1000},
whereas the cumulative plot still detects roughly the right level
of miscalibration for $0 \lesssim S_k \lesssim 0.2$
and $0.8 \lesssim S_k \lesssim 1$; the cumulative plot
indicates that too little data is available for $0.2 \lesssim S_k \lesssim 0.8$
to detect any statistically significant miscalibration in that range of $S_k$
(note the size of the triangle centered at the origin of the plot).
}
\label{100}
\end{figure}

\begin{figure}
\begin{centering}

\parbox{\imsize}{\includegraphics[width=\imsize]
                {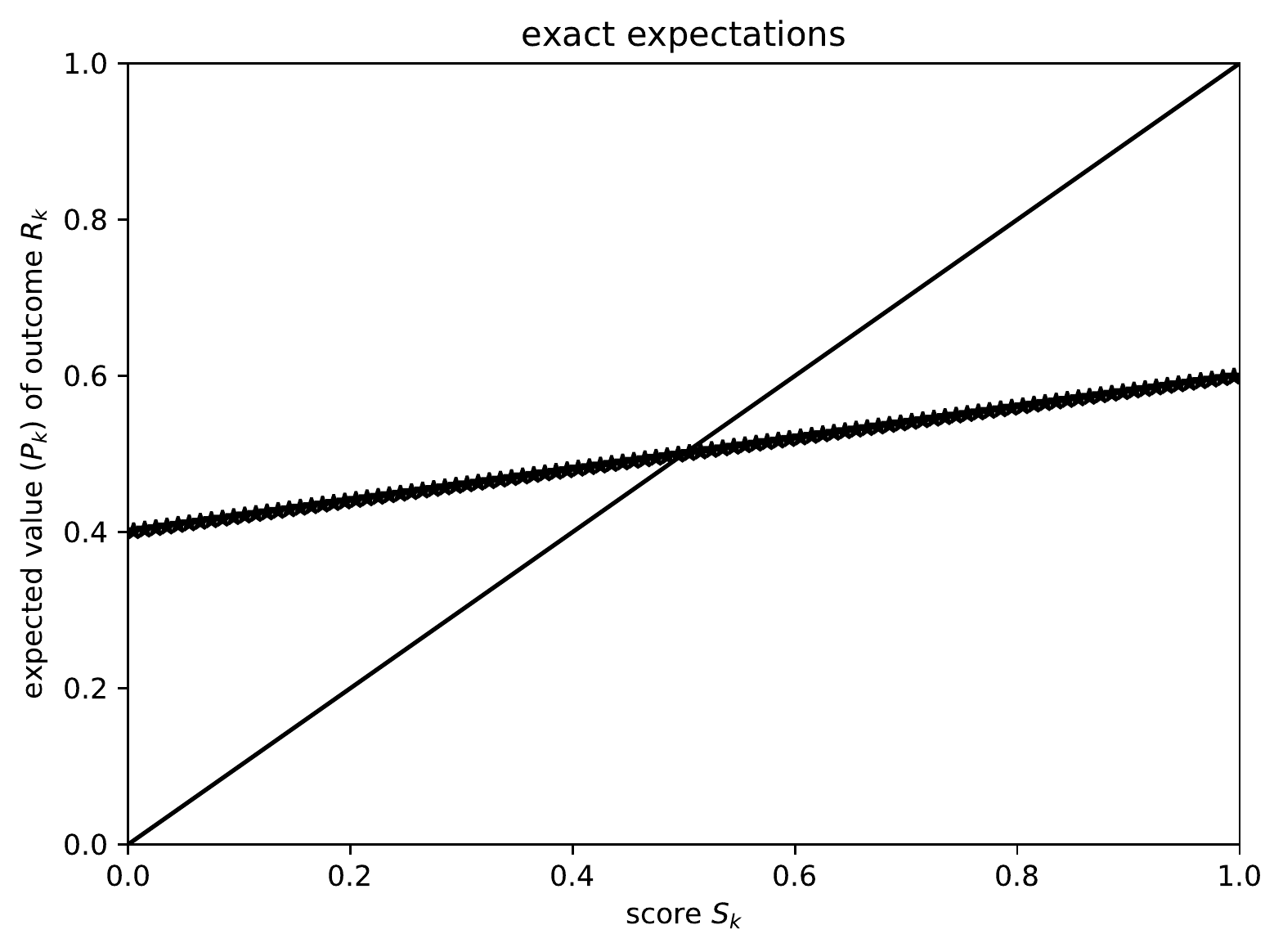}}

\end{centering}
\caption{Ground-truth reliability diagram for Figure~\ref{100}}
\label{100e}
\end{figure}

Figures~\ref{10000}--\ref{100e} all draw from the same underlying distribution
that deviates linearly as a function of $k$ from the distribution of $S_k$,
and $S_1$, $S_2$, \dots, $S_n$ are equispaced;
Figures~\ref{10000} and~\ref{10000e} set $n =$ 10,000,
Figures~\ref{1000} and~\ref{1000e} set $n =$ 1,000,
and Figures~\ref{100} and~\ref{100e} set $n =$ 100.
Overall, the cumulative plots seem more informative
(or at least easier to interpret) in Figures~\ref{10000}--\ref{100e},
but only mildly.

\begin{figure}
\begin{centering}

\parbox{\imsize}{\includegraphics[width=\imsize]
                {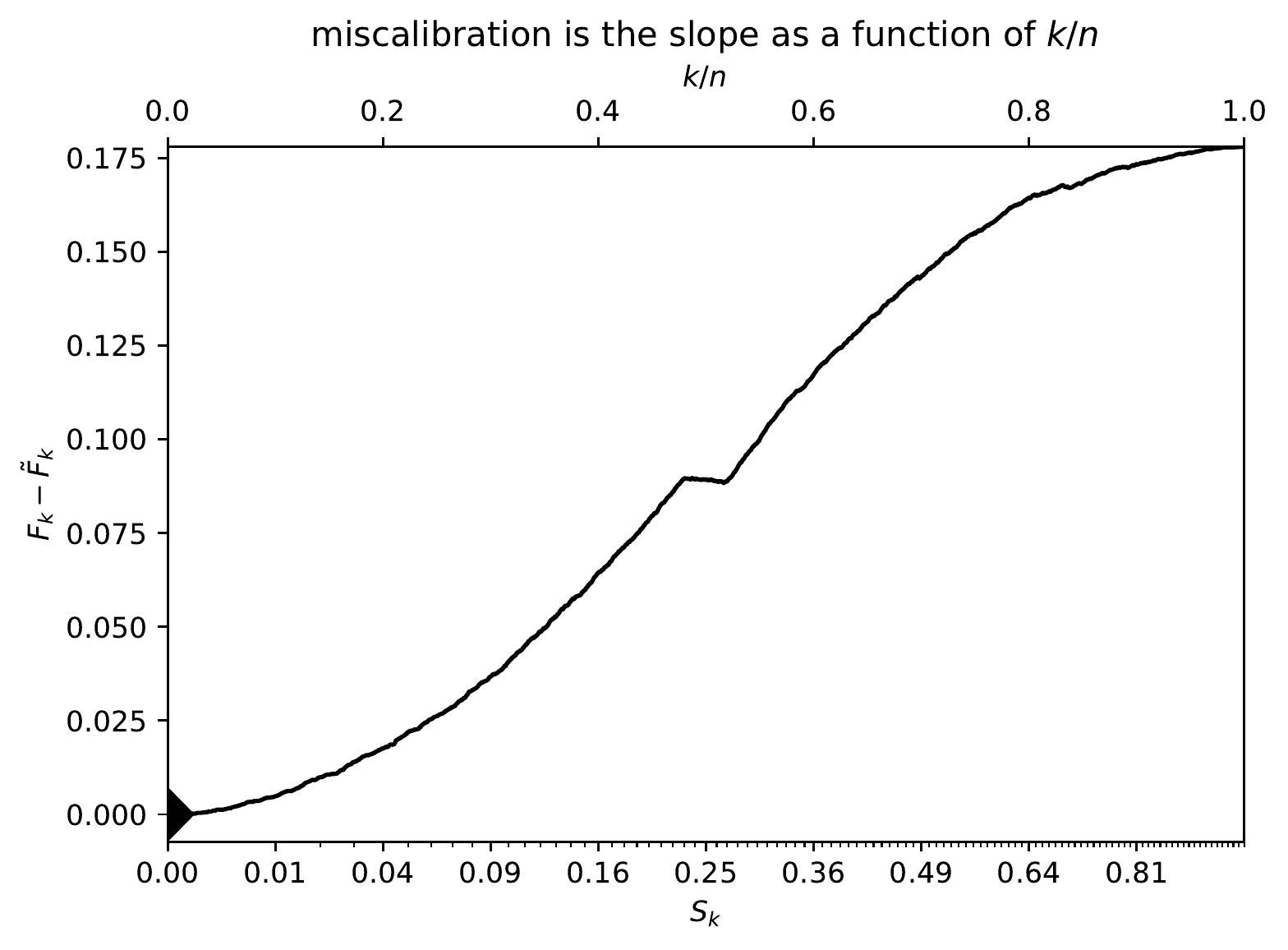}}
\quad\quad
\parbox{\imsize}{\includegraphics[width=\imsize]
                {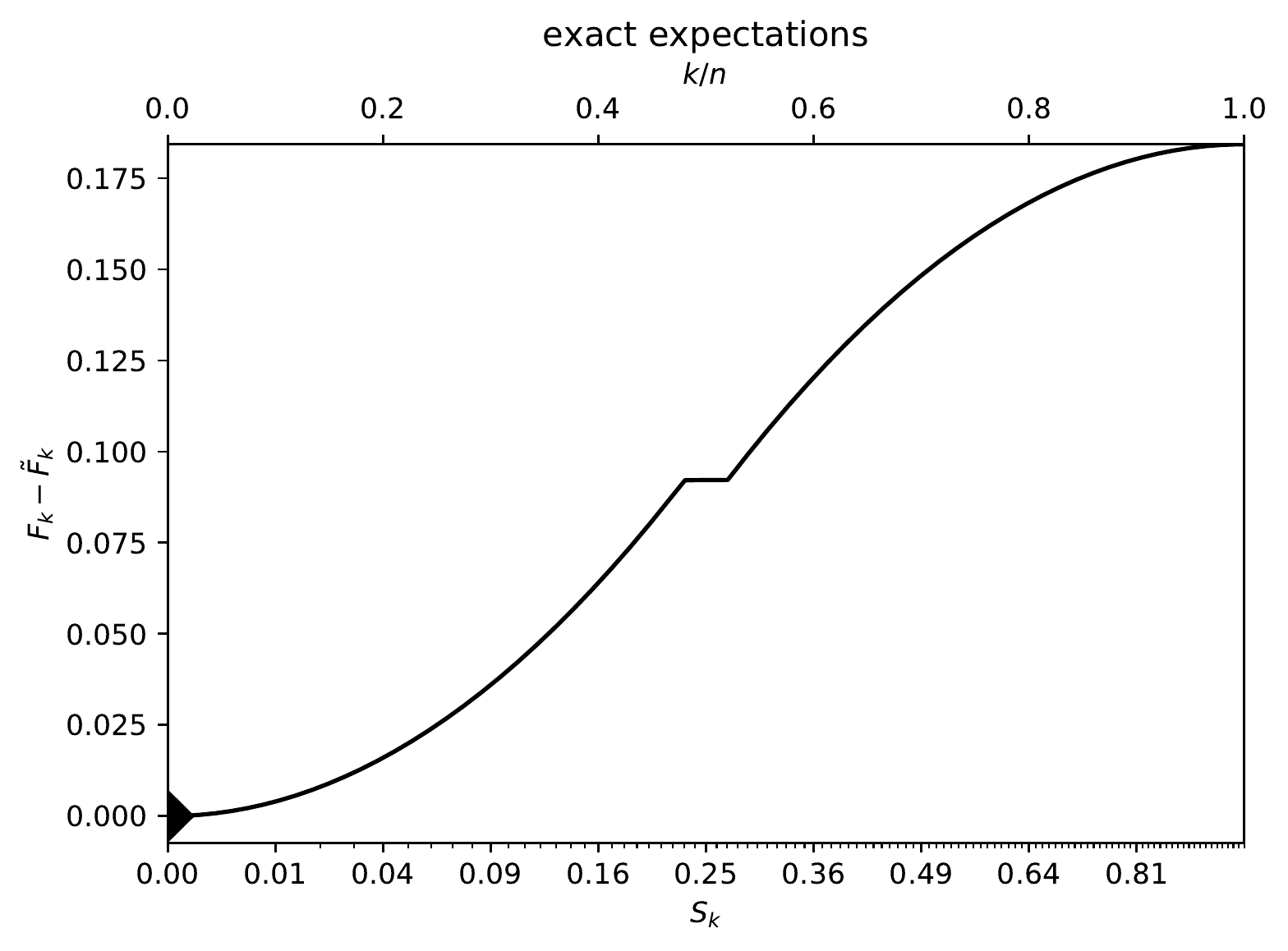}}

\vspace{\vertsep}

\parbox{\imsize}{\includegraphics[width=\imsize]
                {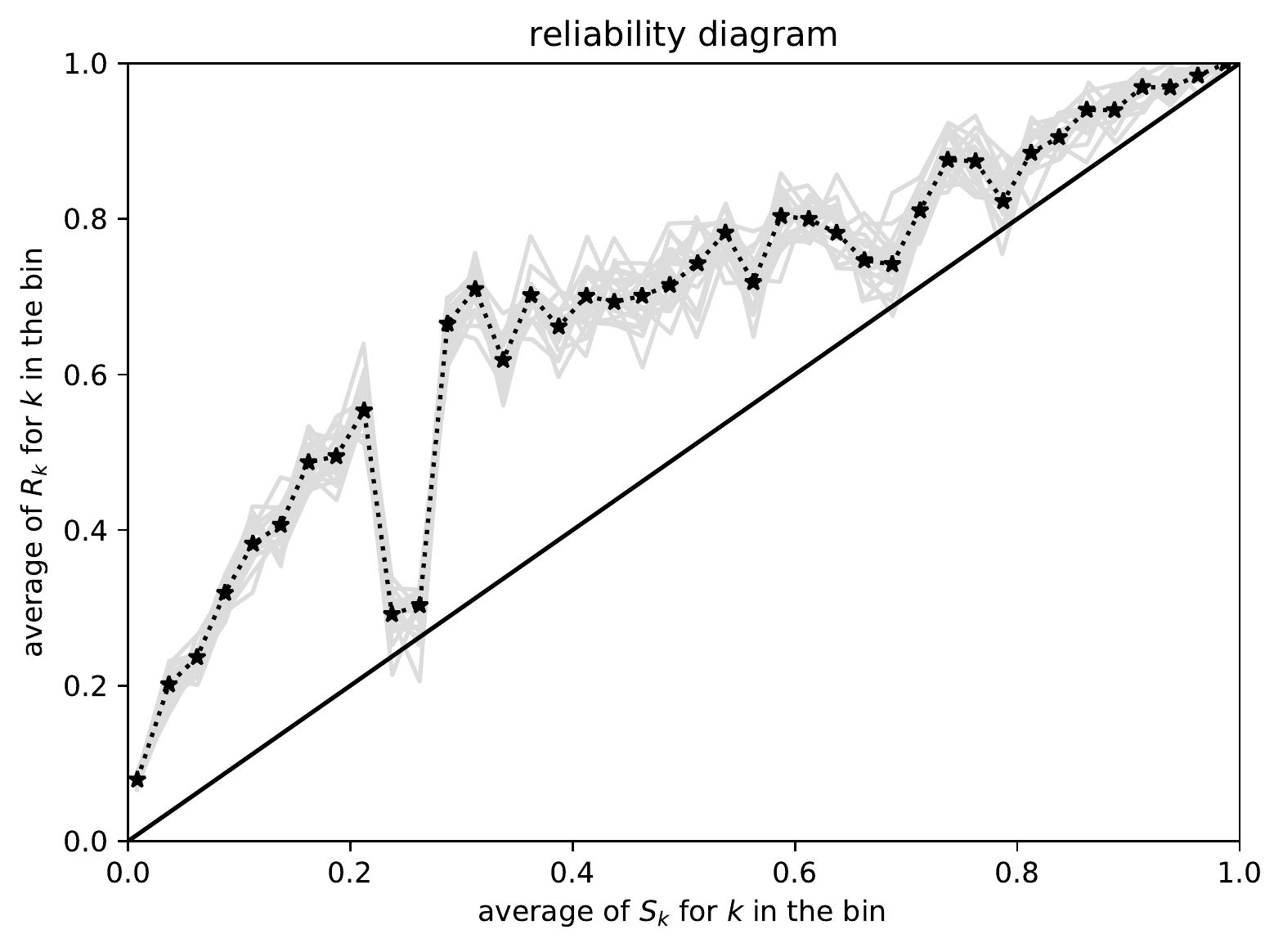}}
\quad\quad
\parbox{\imsize}{\includegraphics[width=\imsize]
                {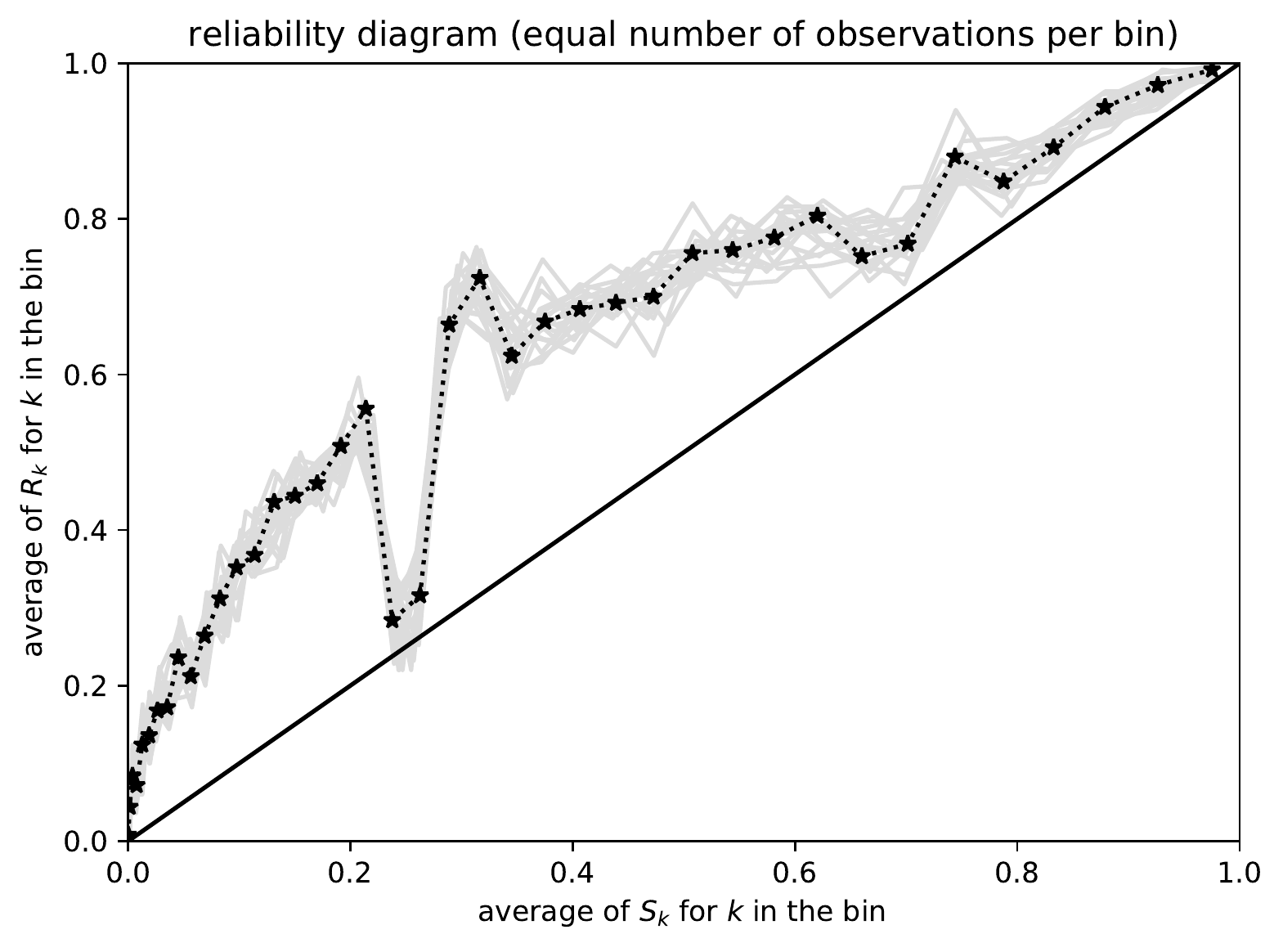}}

\vspace{\vertsep}

\parbox{\imsize}{\includegraphics[width=\imsize]
                {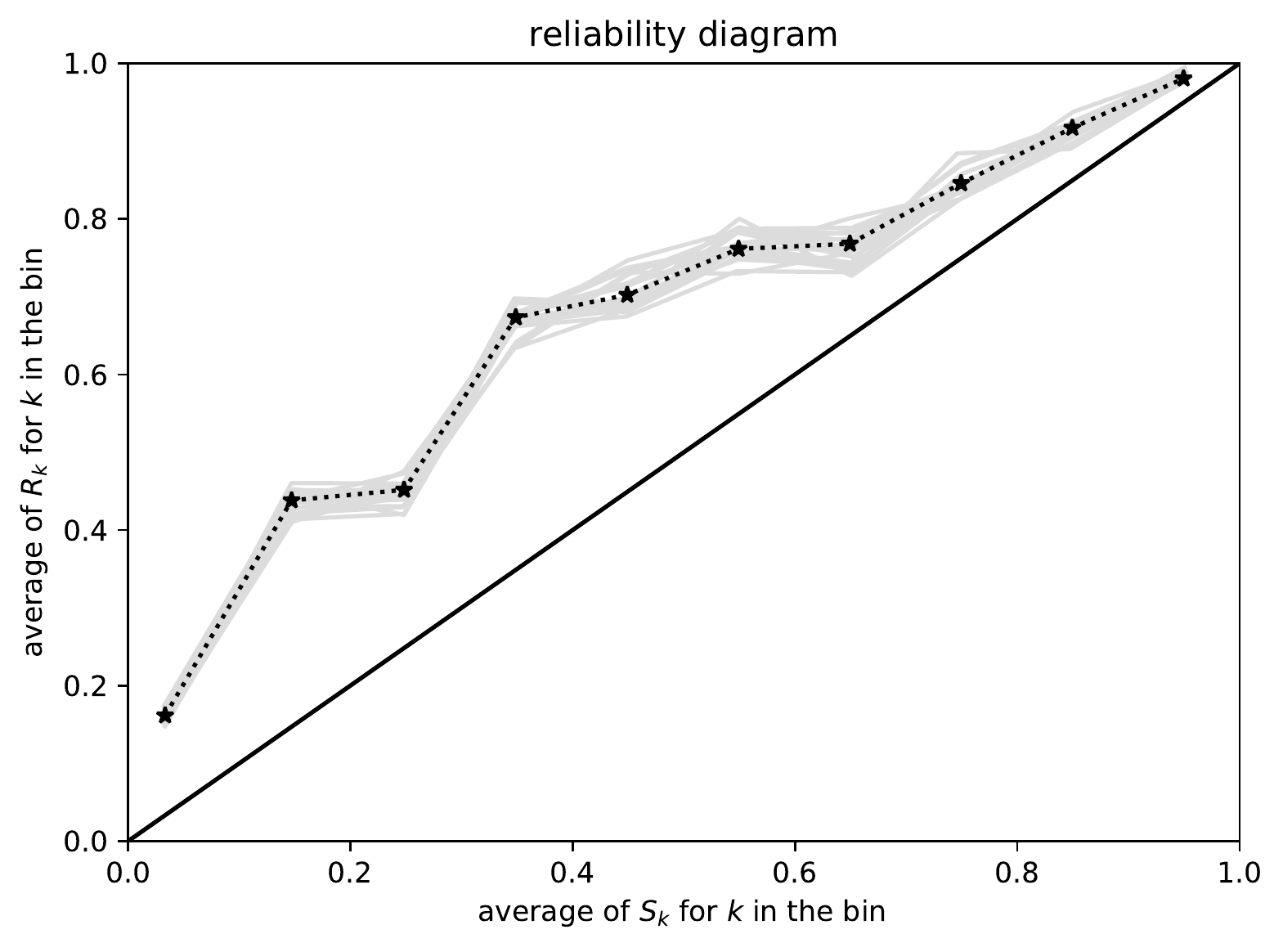}}
\quad\quad
\parbox{\imsize}{\includegraphics[width=\imsize]
                {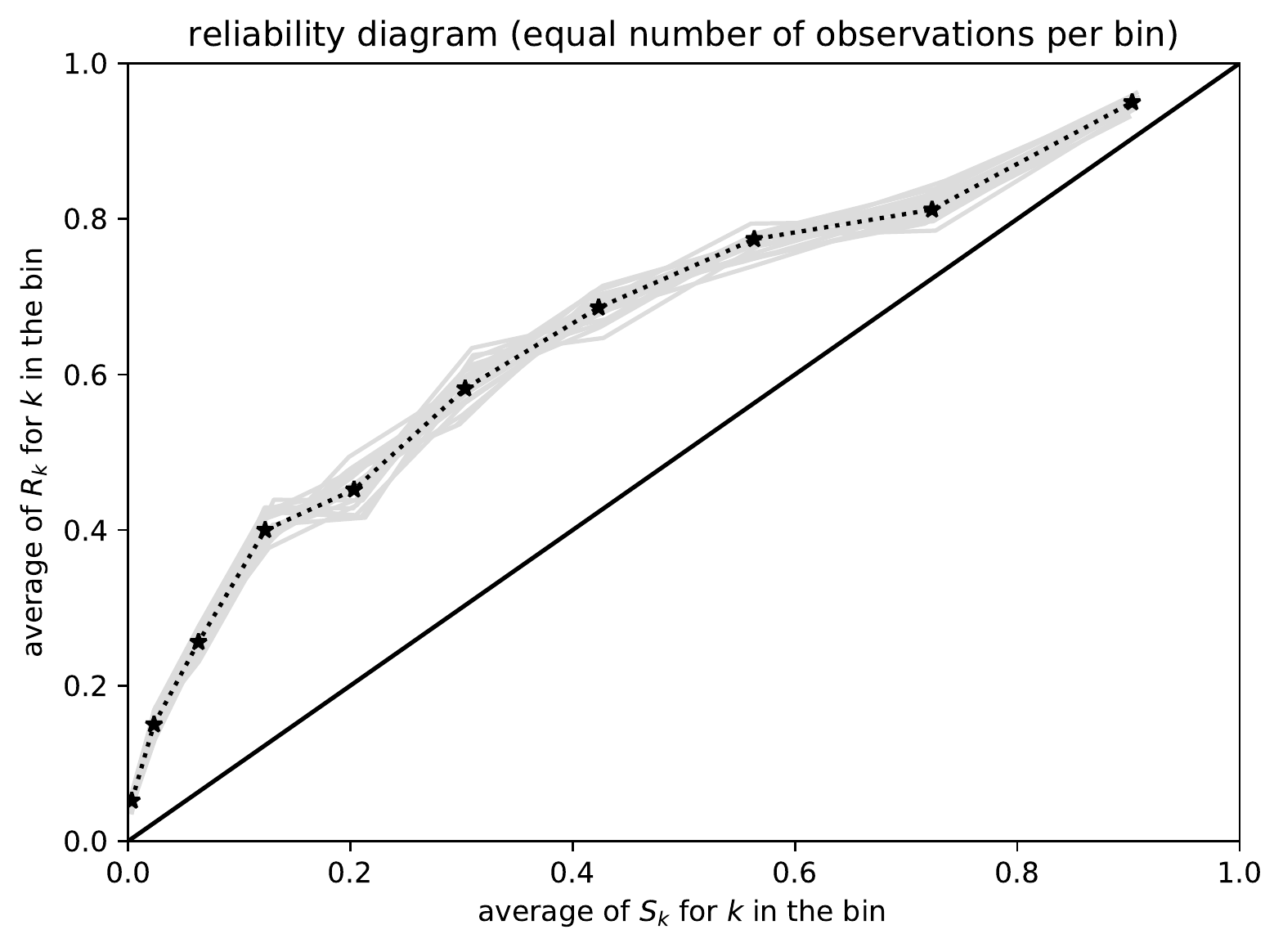}}

\end{centering}
\caption{$n =$ 10,000; $S_1$, $S_2$, \dots, $S_n$ are denser near 0;
         Kuiper's statistic is $0.1781 / \sigma = 48.77$,
         Kolmogorov's and Smirnov's is $0.1781 / \sigma = 48.77$.
Figure~\ref{10000_0e} displays the ground-truth reliability diagram.
All plots, whether cumulative or conventional, work well enough,
though the reliability diagrams
might be mistakenly misleading relative to the exact expectations,
at least without diligent attention to the significant variation
with the number of bins.
}
\label{10000_0}
\end{figure}

\begin{figure}
\begin{centering}

\parbox{\imsize}{\includegraphics[width=\imsize]
                {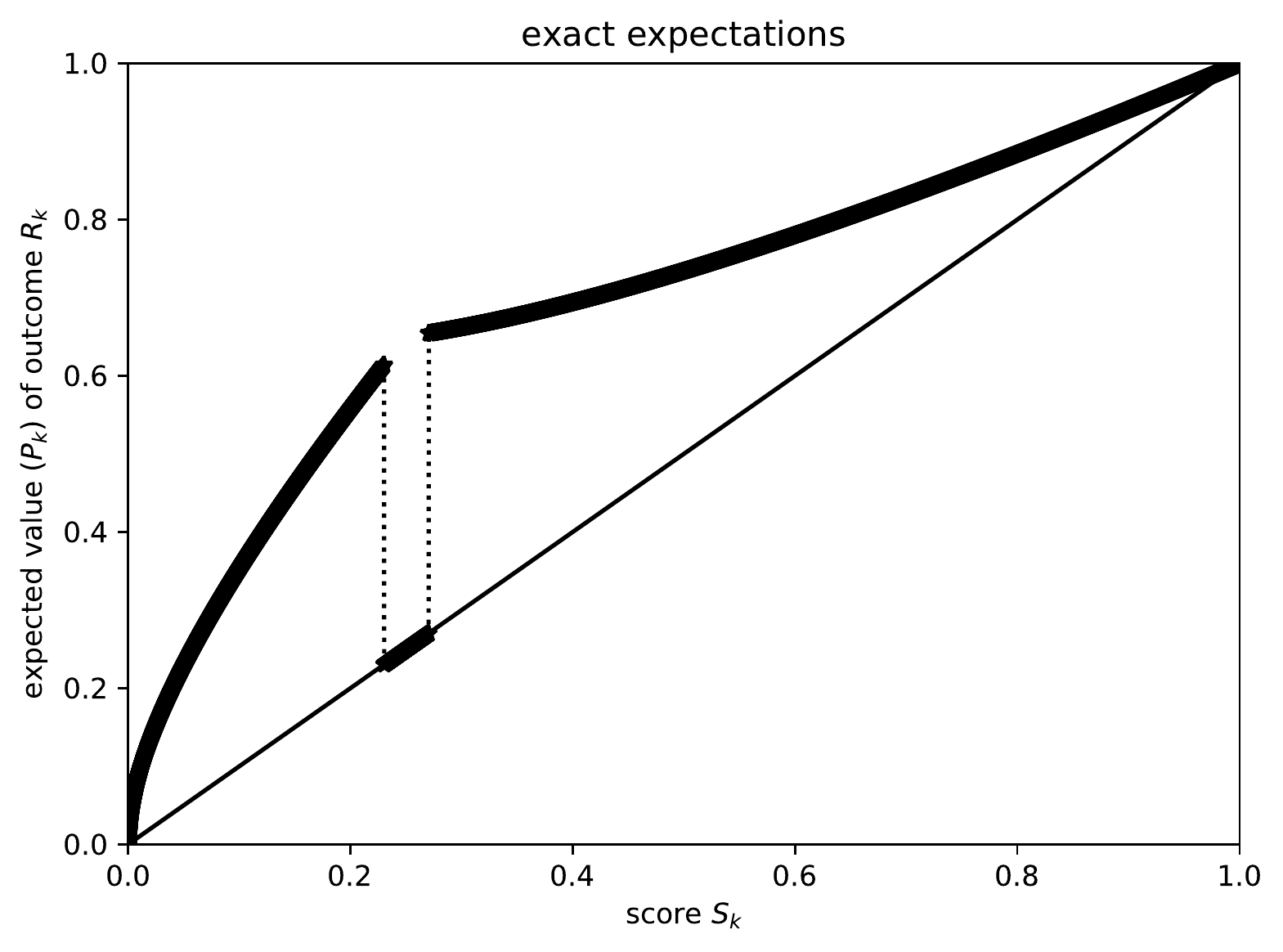}}

\end{centering}
\caption{Ground-truth reliability diagram for Figure~\ref{10000_0}}
\label{10000_0e}
\end{figure}

\begin{figure}
\begin{centering}

\parbox{\imsize}{\includegraphics[width=\imsize]
                {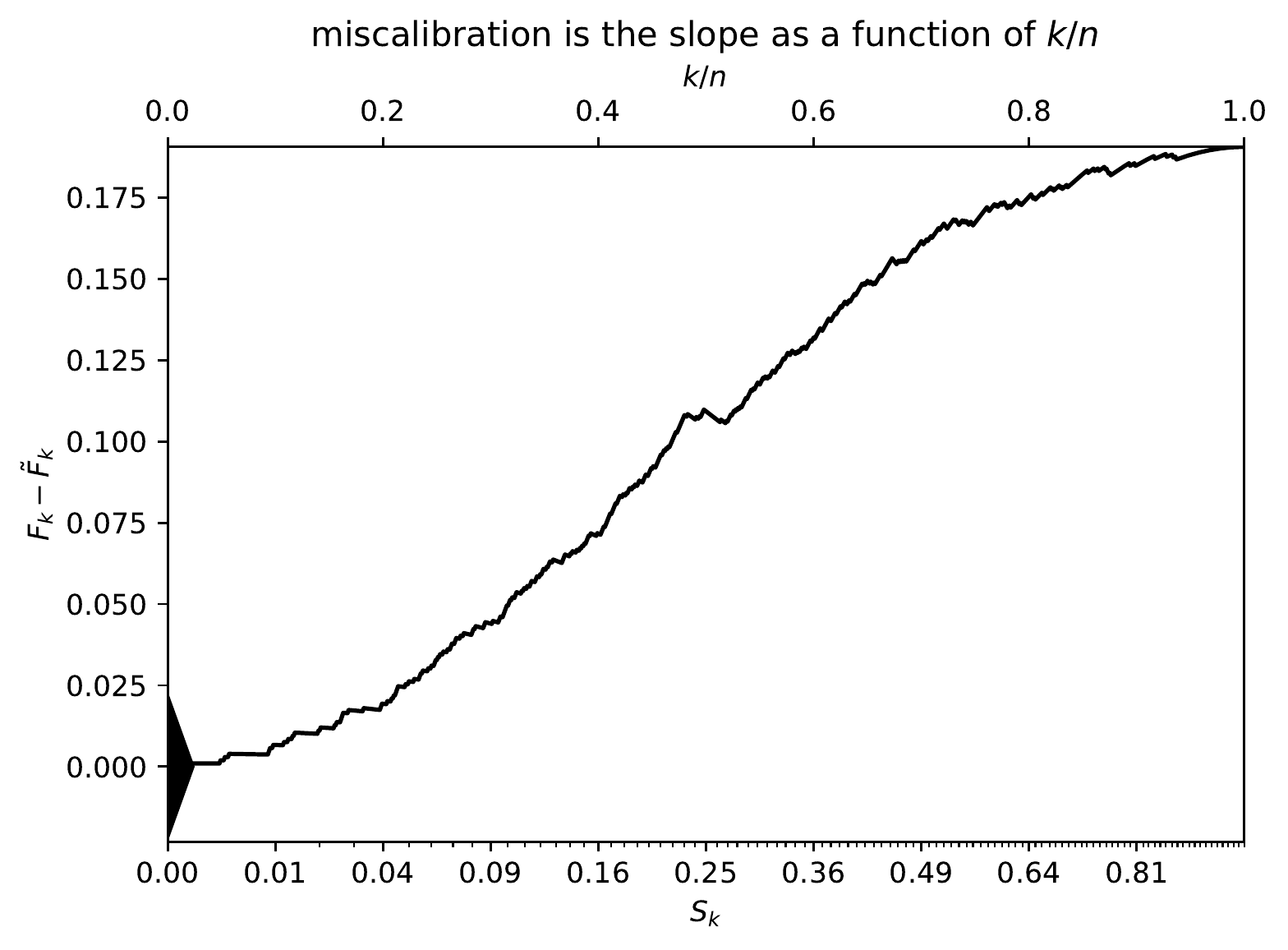}}
\quad\quad
\parbox{\imsize}{\includegraphics[width=\imsize]
                {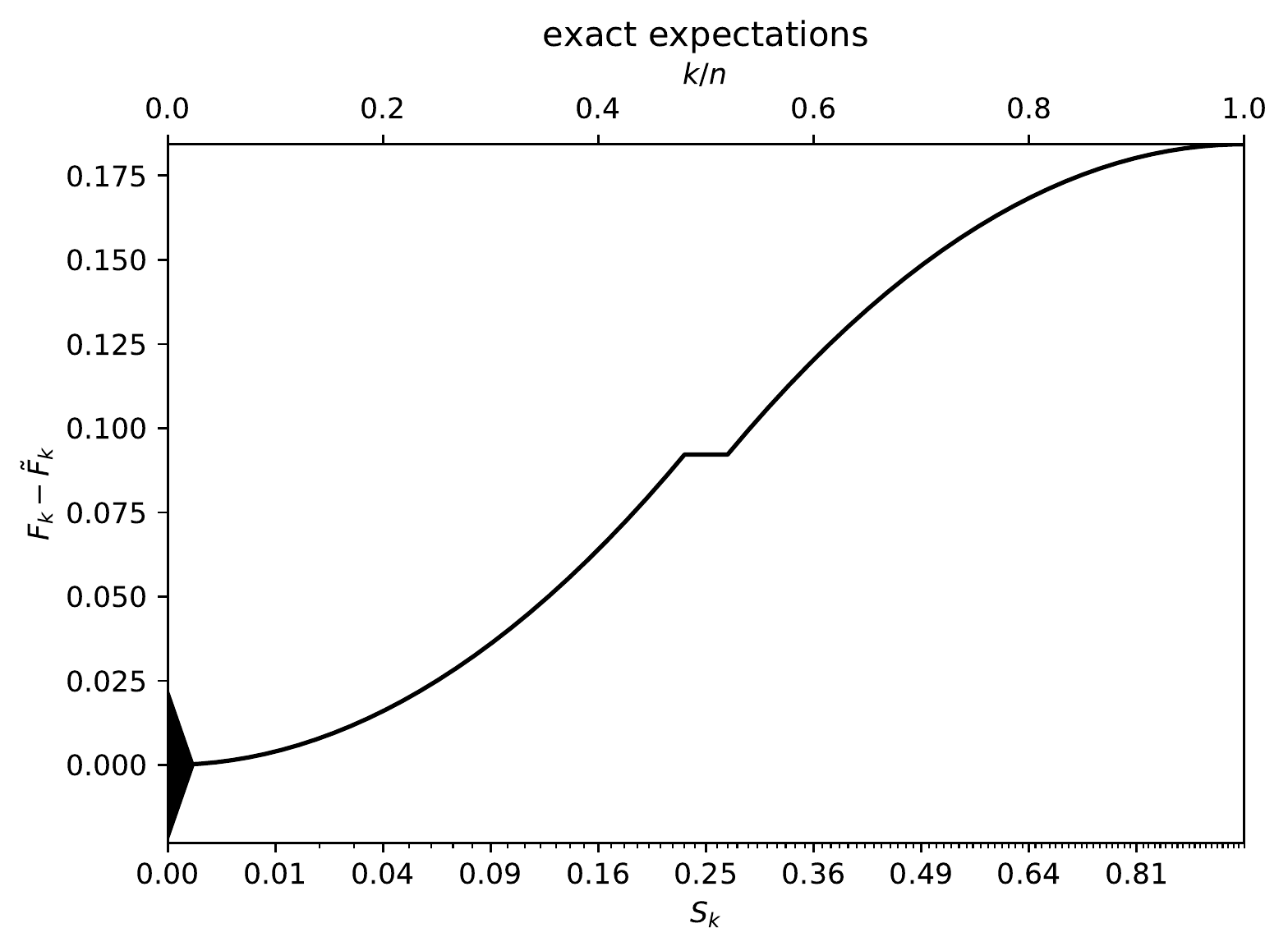}}

\vspace{\vertsep}

\parbox{\imsize}{\includegraphics[width=\imsize]
                {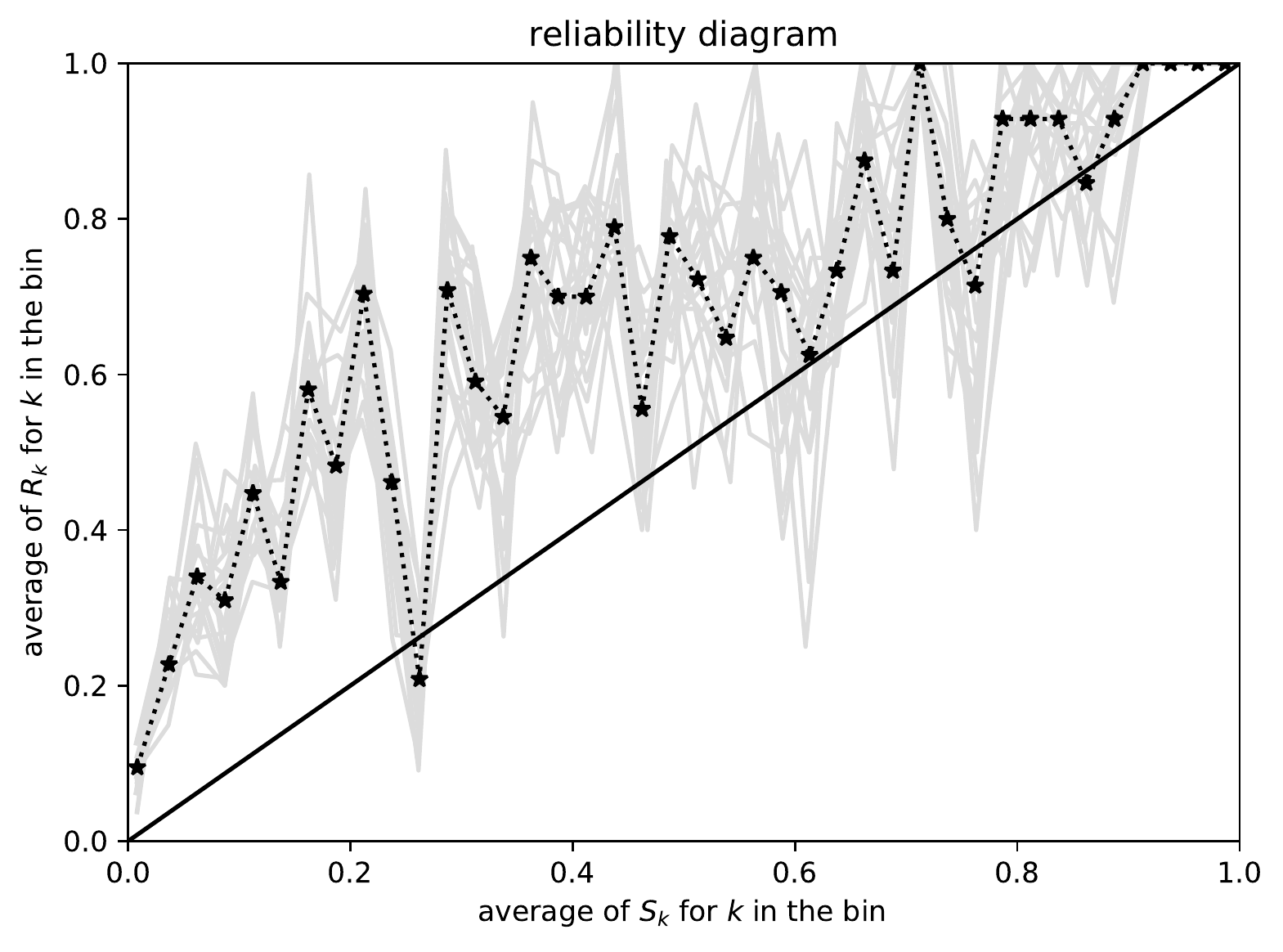}}
\quad\quad
\parbox{\imsize}{\includegraphics[width=\imsize]
                {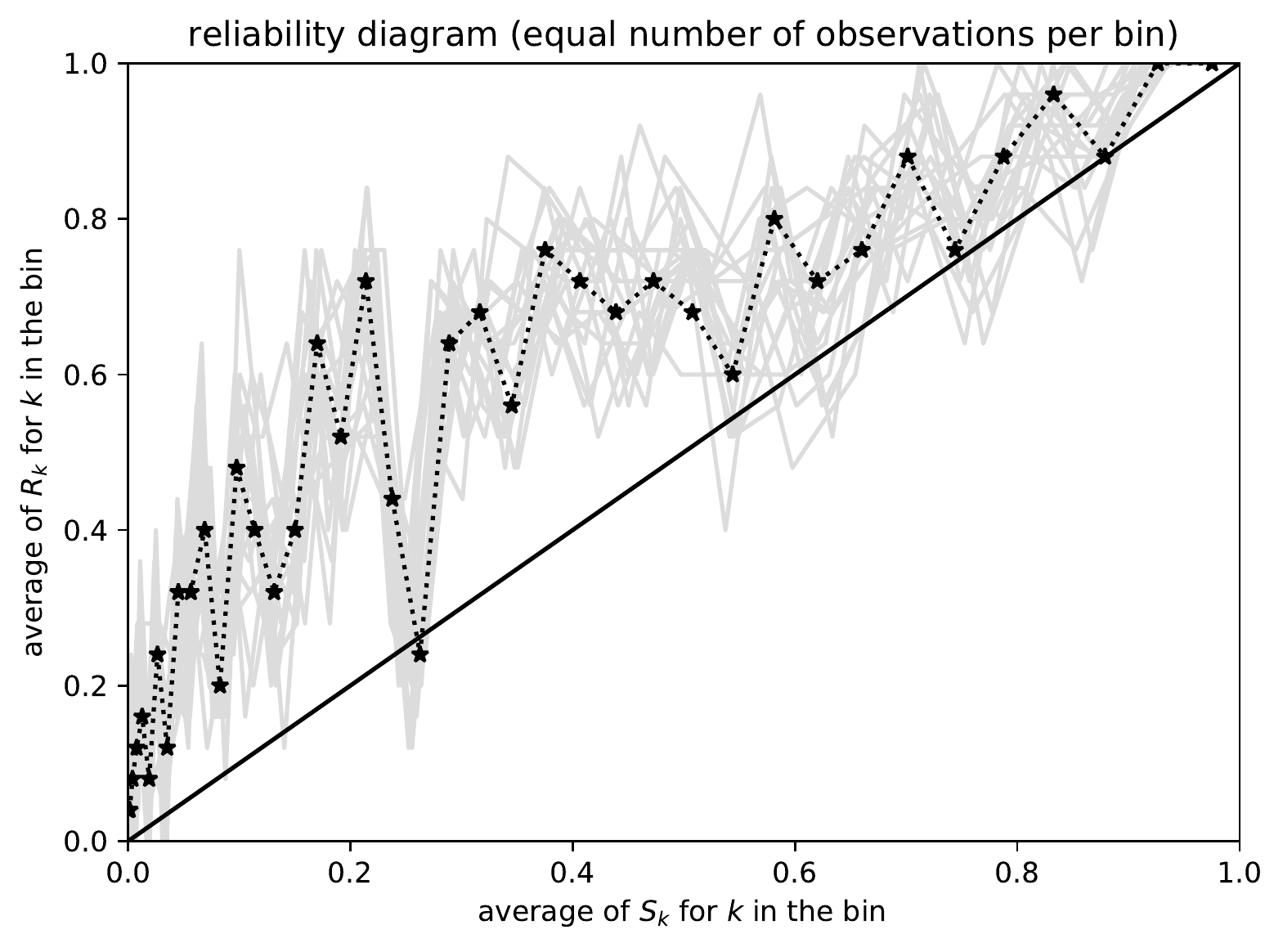}}

\vspace{\vertsep}

\parbox{\imsize}{\includegraphics[width=\imsize]
                {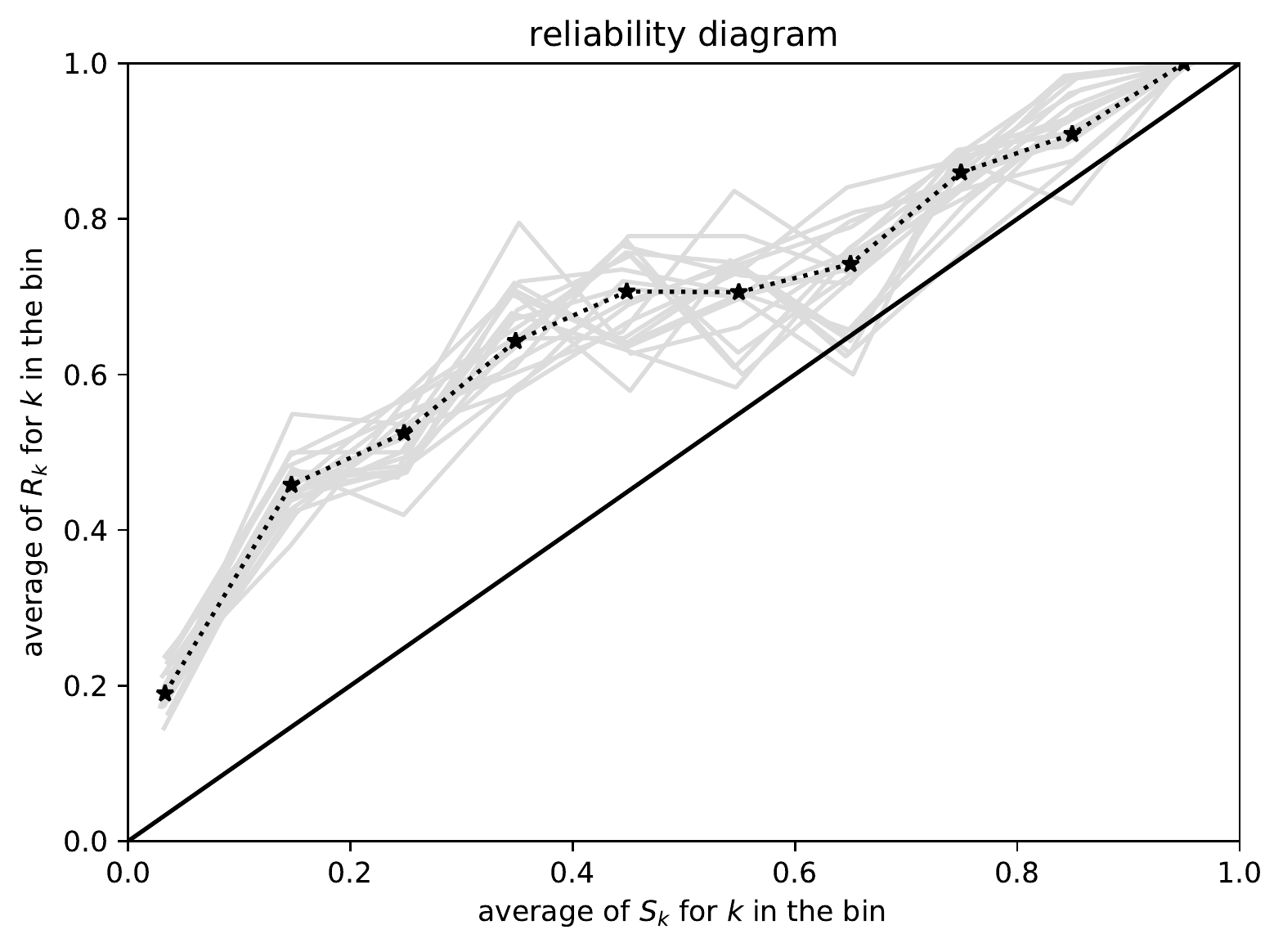}}
\quad\quad
\parbox{\imsize}{\includegraphics[width=\imsize]
                {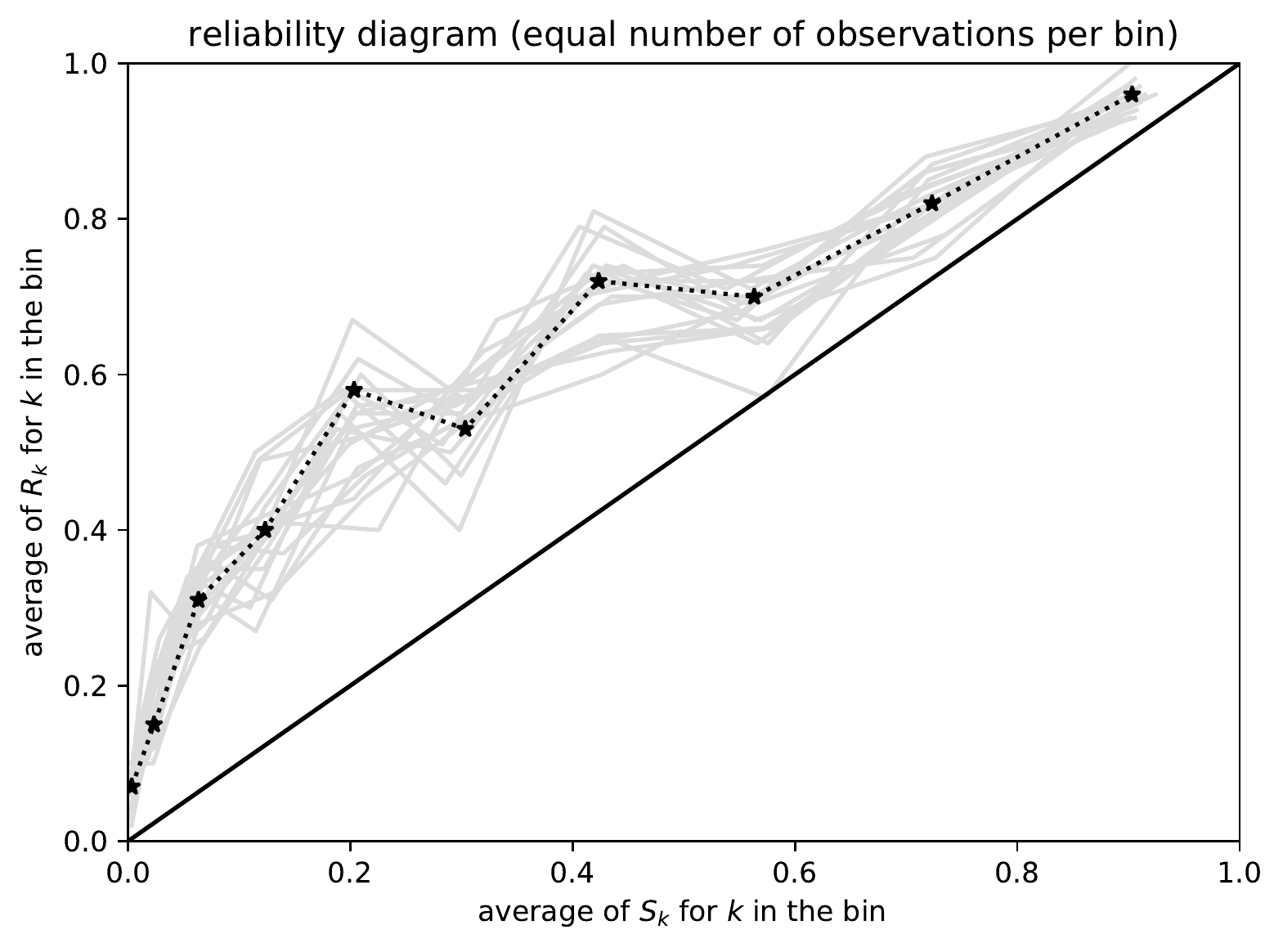}}

\end{centering}
\caption{$n =$ 1,000; $S_1$, $S_2$, \dots, $S_n$ are denser near 0;
         Kuiper's statistic is $0.1907 / \sigma = 16.51$,
         Kolmogorov's and Smirnov's is $0.1907 / \sigma = 16.51$.
Figure~\ref{1000_0e} displays the ground-truth reliability diagram.
As in Figure~\ref{10000_0}, all plots work sufficiently well,
though the observed reliability diagrams may be misleading
relative to the exact expectations,
at least without careful attention to the variation with the number of bins.
}
\label{1000_0}
\end{figure}

\begin{figure}
\begin{centering}

\parbox{\imsize}{\includegraphics[width=\imsize]
                {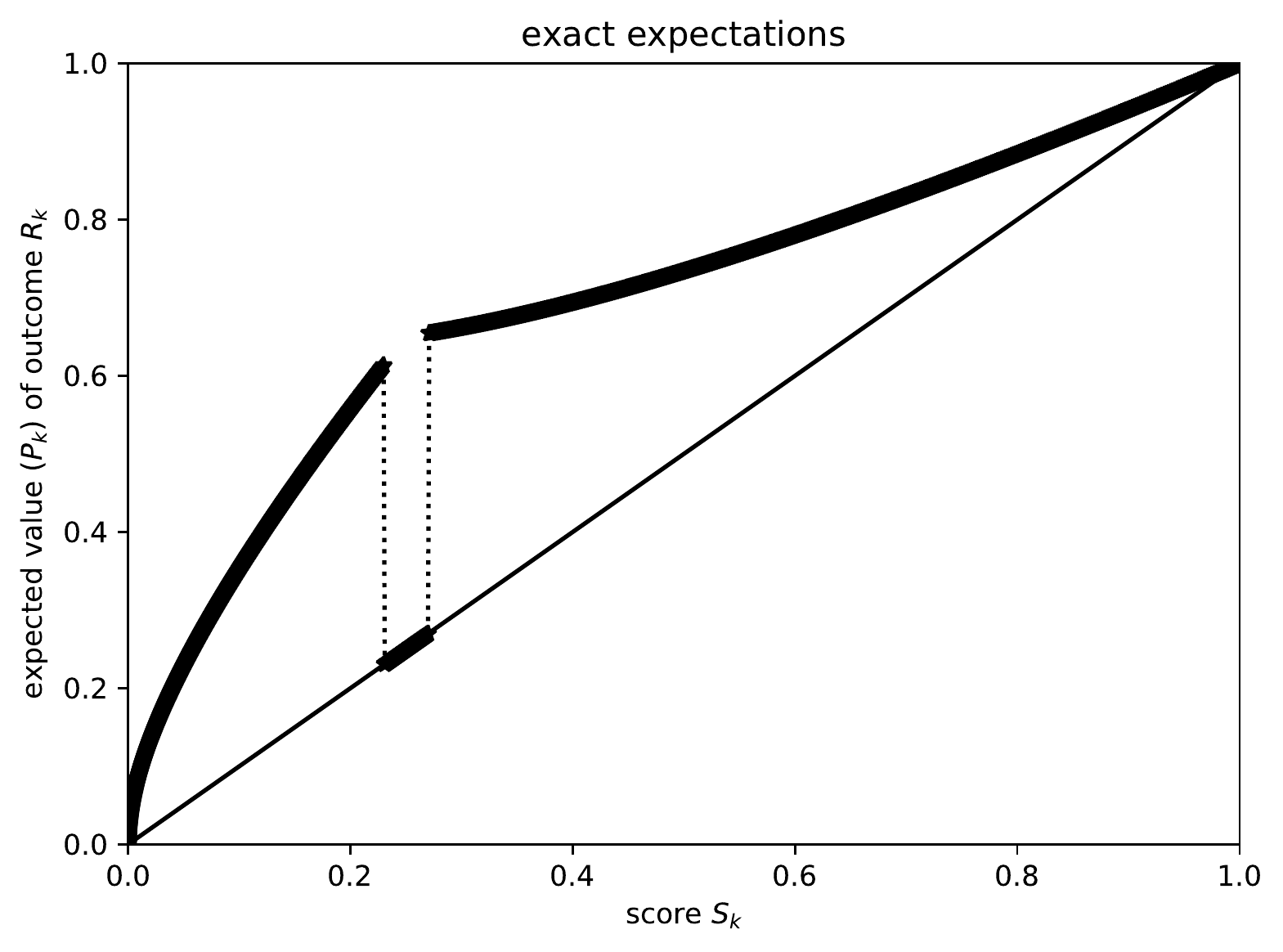}}

\end{centering}
\caption{Ground-truth reliability diagram for Figure~\ref{1000_0}}
\label{1000_0e}
\end{figure}

\begin{figure}
\begin{centering}

\parbox{\imsize}{\includegraphics[width=\imsize]
                {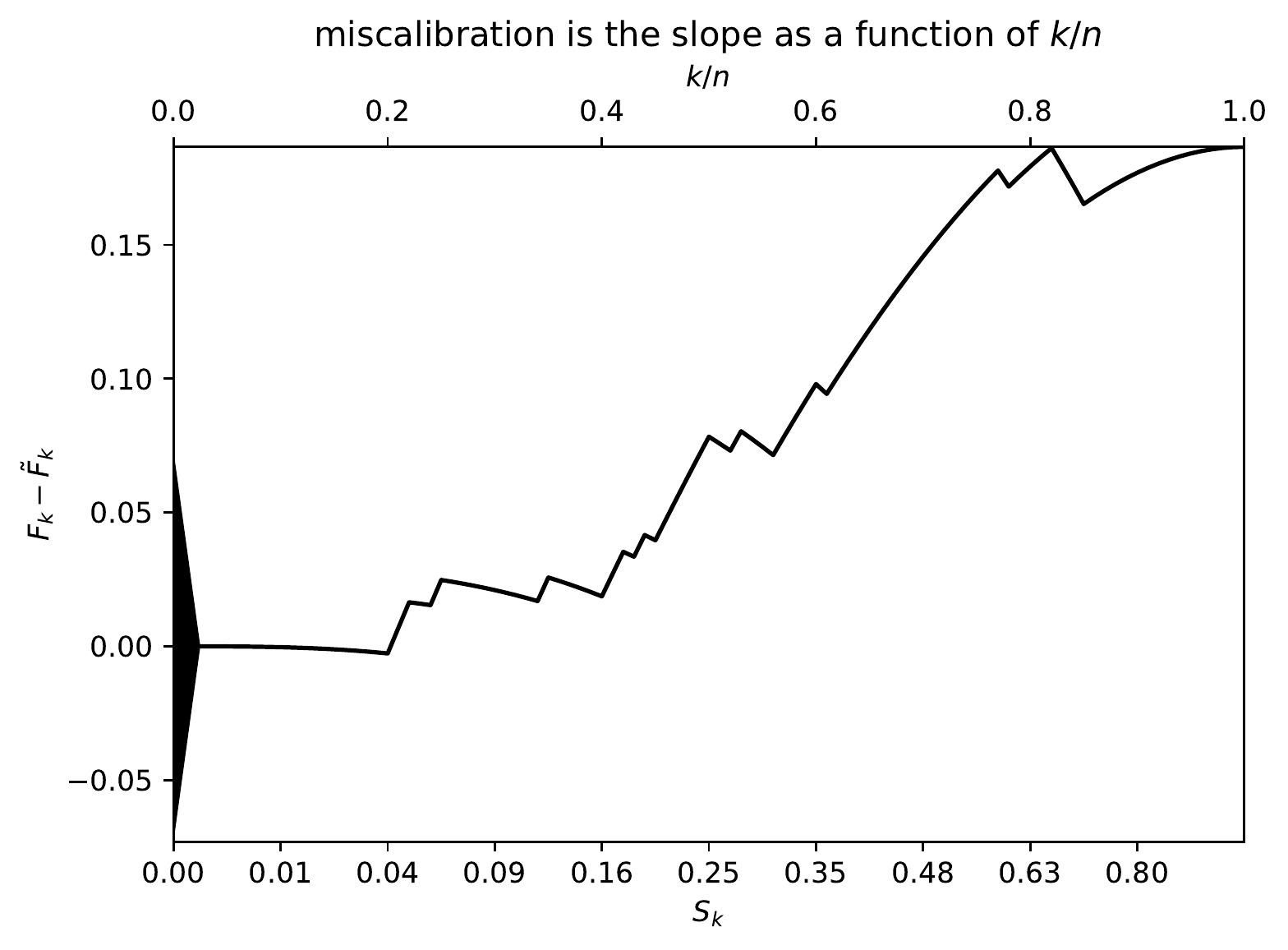}}
\quad\quad
\parbox{\imsize}{\includegraphics[width=\imsize]
                {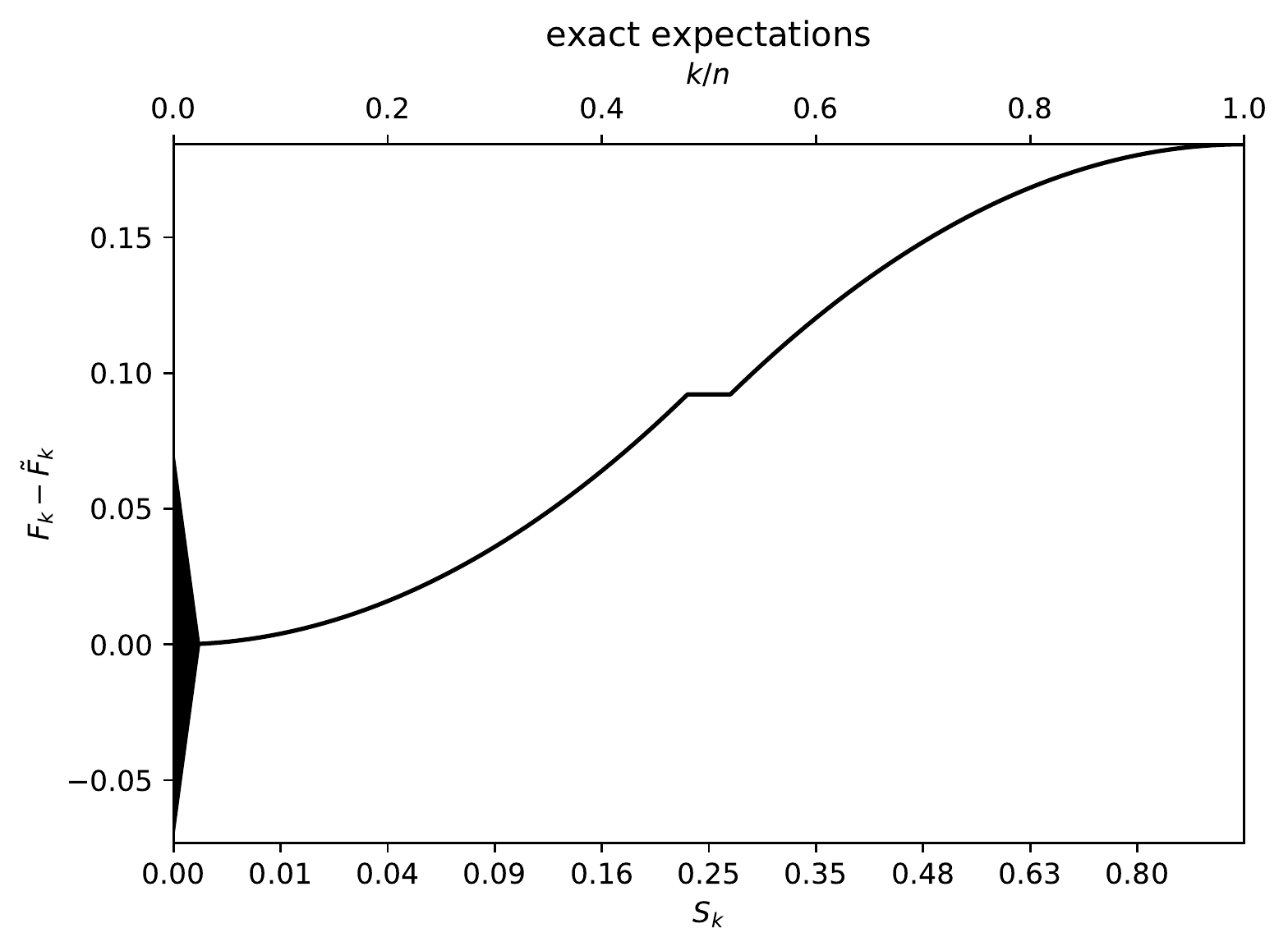}}

\vspace{\vertsep}

\parbox{\imsize}{\includegraphics[width=\imsize]
                {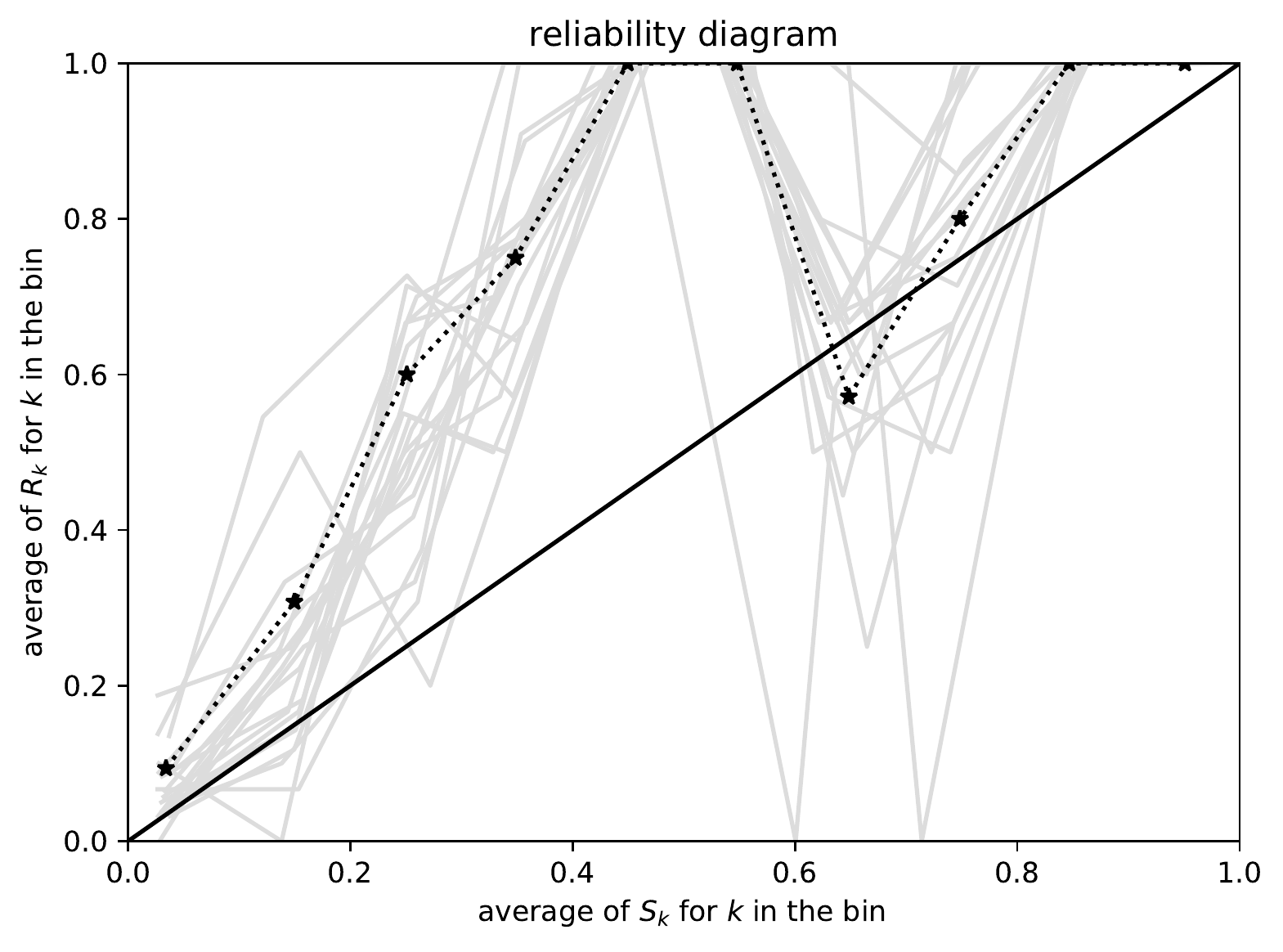}}
\quad\quad
\parbox{\imsize}{\includegraphics[width=\imsize]
                {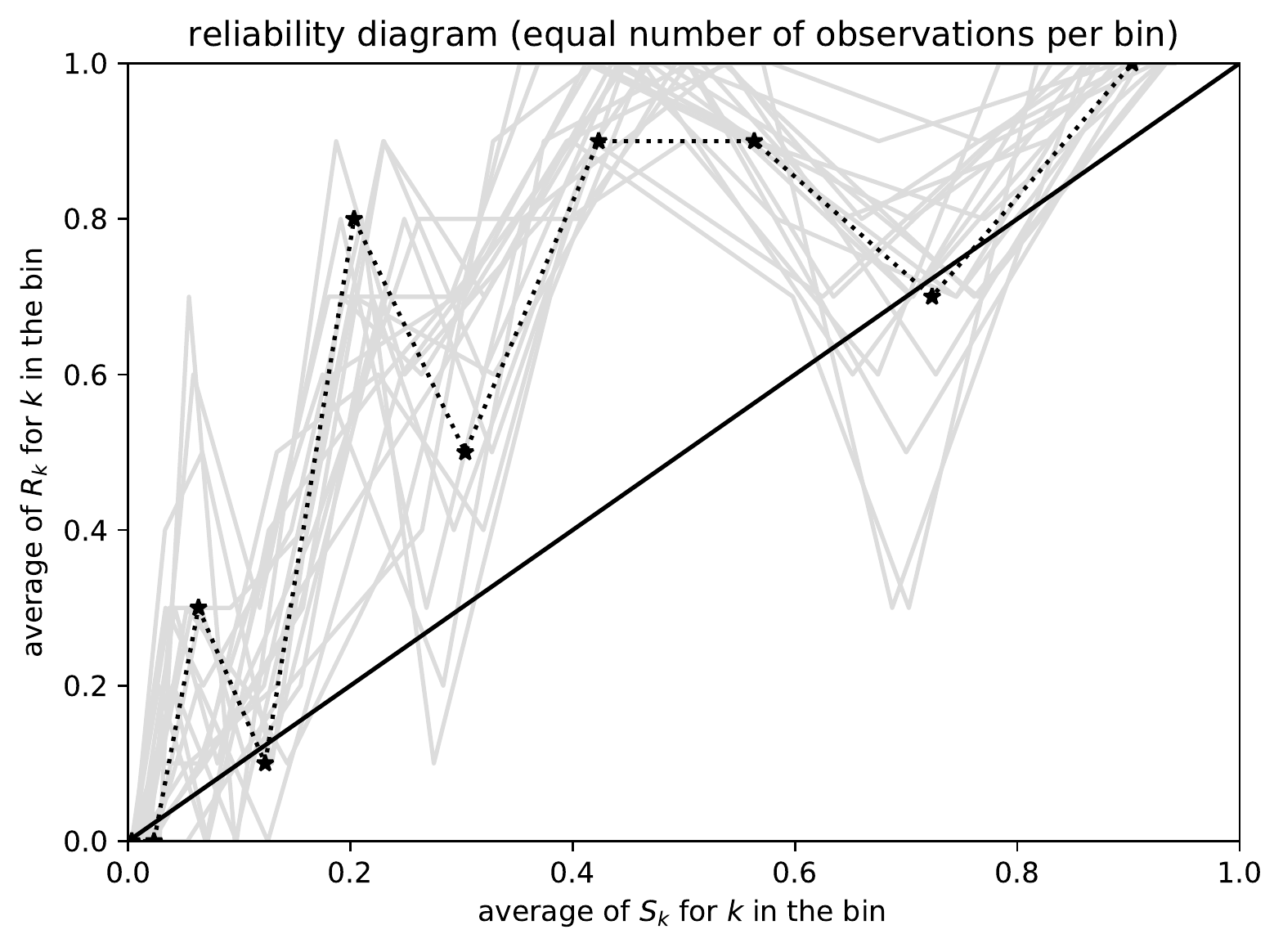}}

\vspace{\vertsep}

\parbox{\imsize}{\includegraphics[width=\imsize]
                {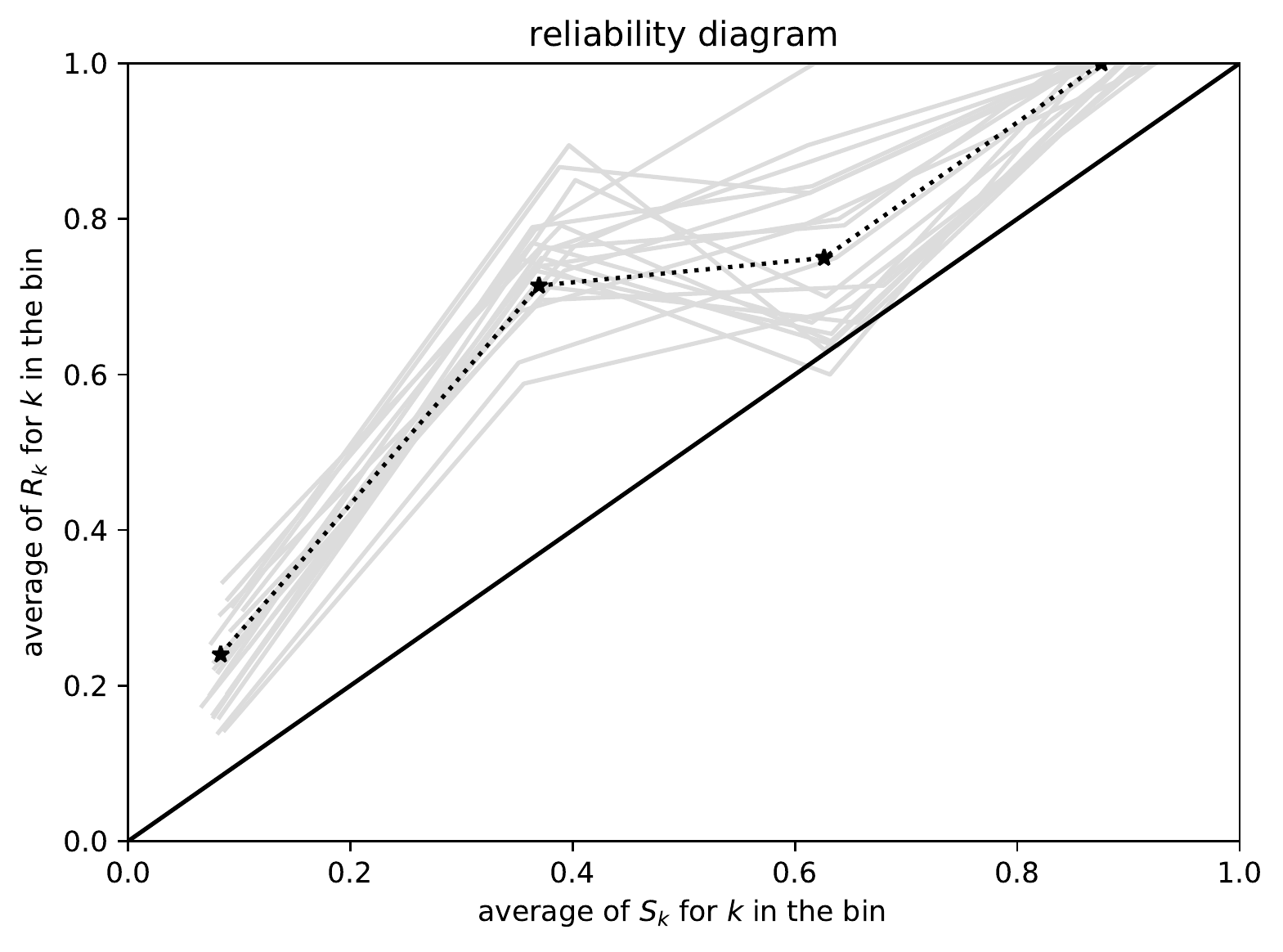}}
\quad\quad
\parbox{\imsize}{\includegraphics[width=\imsize]
                {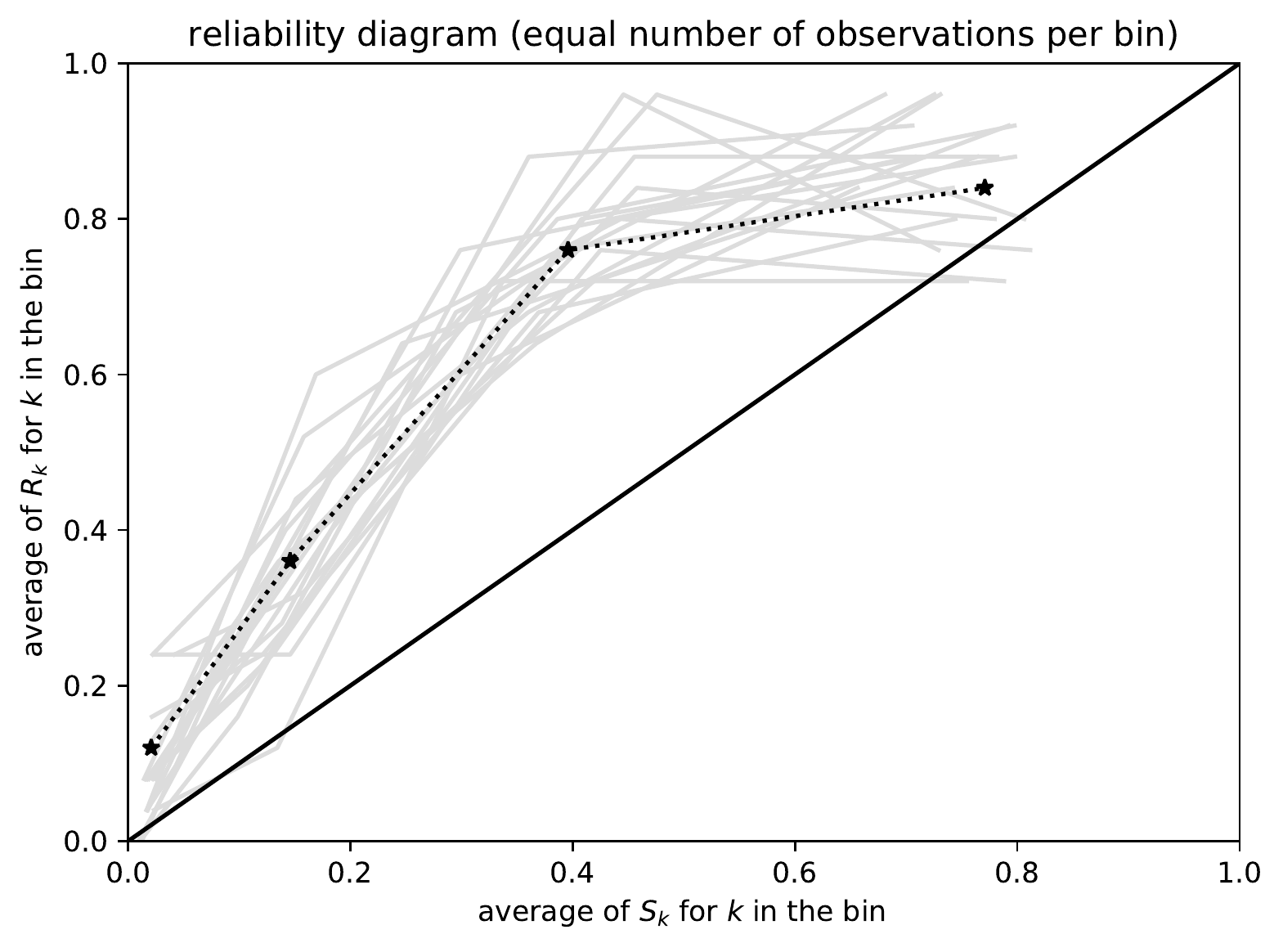}}

\end{centering}
\caption{$n =$ 100; $S_1$, $S_2$, \dots, $S_n$ are denser near 0;
         Kuiper's statistic is $0.1893 / \sigma = 5.185$,
         Kolmogorov's and Smirnov's is $0.1867 / \sigma = 5.112$.
Figure~\ref{100_0e} displays the ground-truth reliability diagram.
The plots, whether cumulative or conventional, reveal similar information here,
though the reliability diagrams with an equal number of observations per bin
provide more reliable estimates than the other reliability diagrams.
The cumulative plot is perhaps the easiest to interpret:
the miscalibration is significant for $0.1 \lesssim S_k \lesssim 0.23$
and $0.27 \lesssim S_k \lesssim 0.6$,
with about the correct amount of miscalibration
(the amount is correct since the secant lines have the expected slopes).
}
\label{100_0}
\end{figure}

\begin{figure}
\begin{centering}

\parbox{\imsize}{\includegraphics[width=\imsize]
                {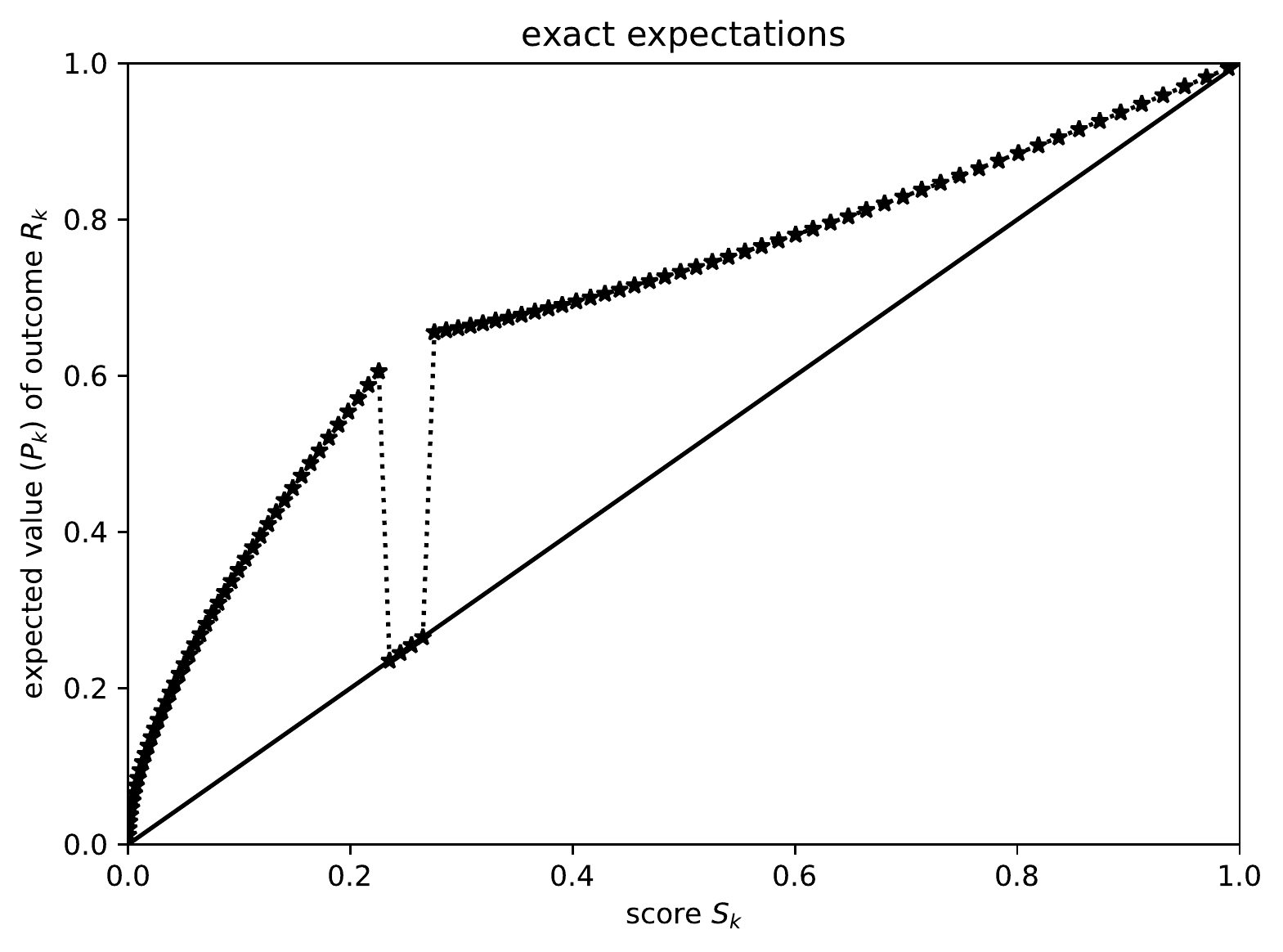}}

\end{centering}
\caption{Ground-truth reliability diagram for Figure~\ref{100_0}}
\label{100_0e}
\end{figure}

Figures~\ref{10000_0}--\ref{100_0e} all draw
from the same underlying distribution that is overconfident
(lying above the perfectly calibrated ideal),
with the overconfidence peaking for $S_k$ around $0.25$
(aside from a perfectly calibrated notch right around $0.25$),
where $S_k$ is proportional to $(k-0.5)^2$;
Figures~\ref{10000_0} and~\ref{10000_0e} set $n =$ 10,000,
Figures~\ref{1000_0} and~\ref{1000_0e} set $n =$ 1,000,
and Figures~\ref{100_0} and~\ref{100_0e} set $n =$ 100.
The cumulative plots look to work better.

\begin{figure}
\begin{centering}

\parbox{\imsize}{\includegraphics[width=\imsize]
                {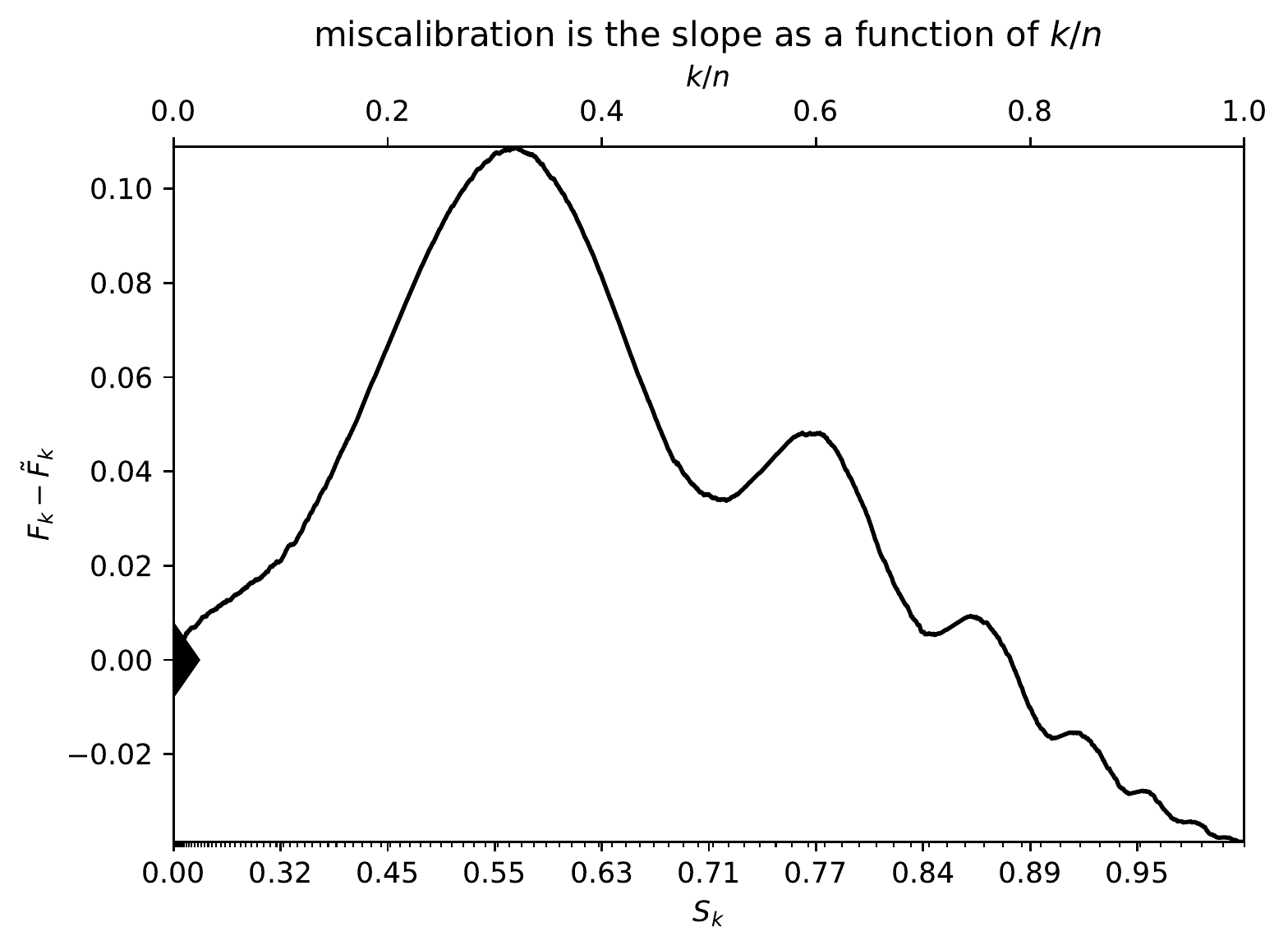}}
\quad\quad
\parbox{\imsize}{\includegraphics[width=\imsize]
                {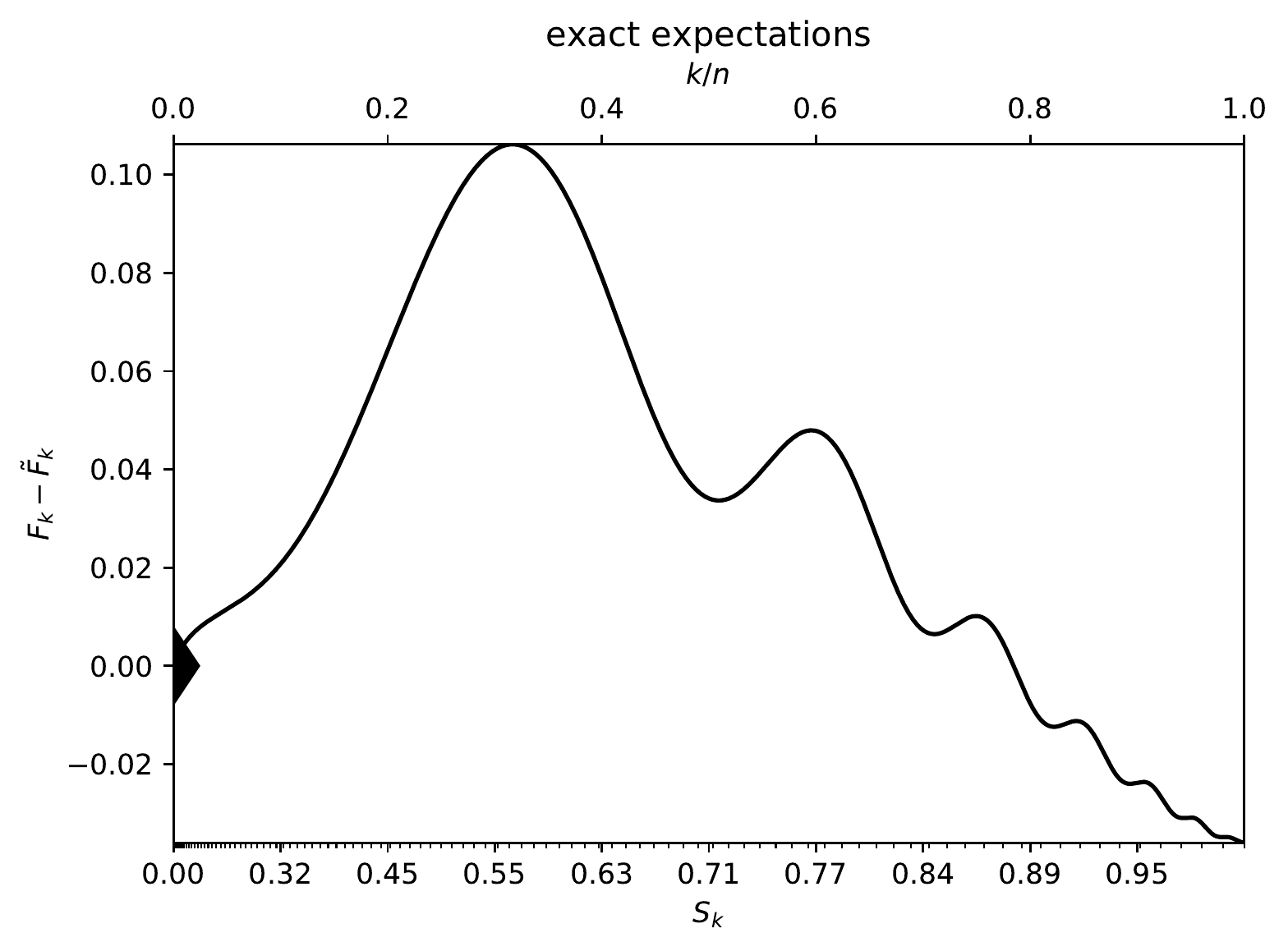}}

\vspace{\vertsep}

\parbox{\imsize}{\includegraphics[width=\imsize]
                {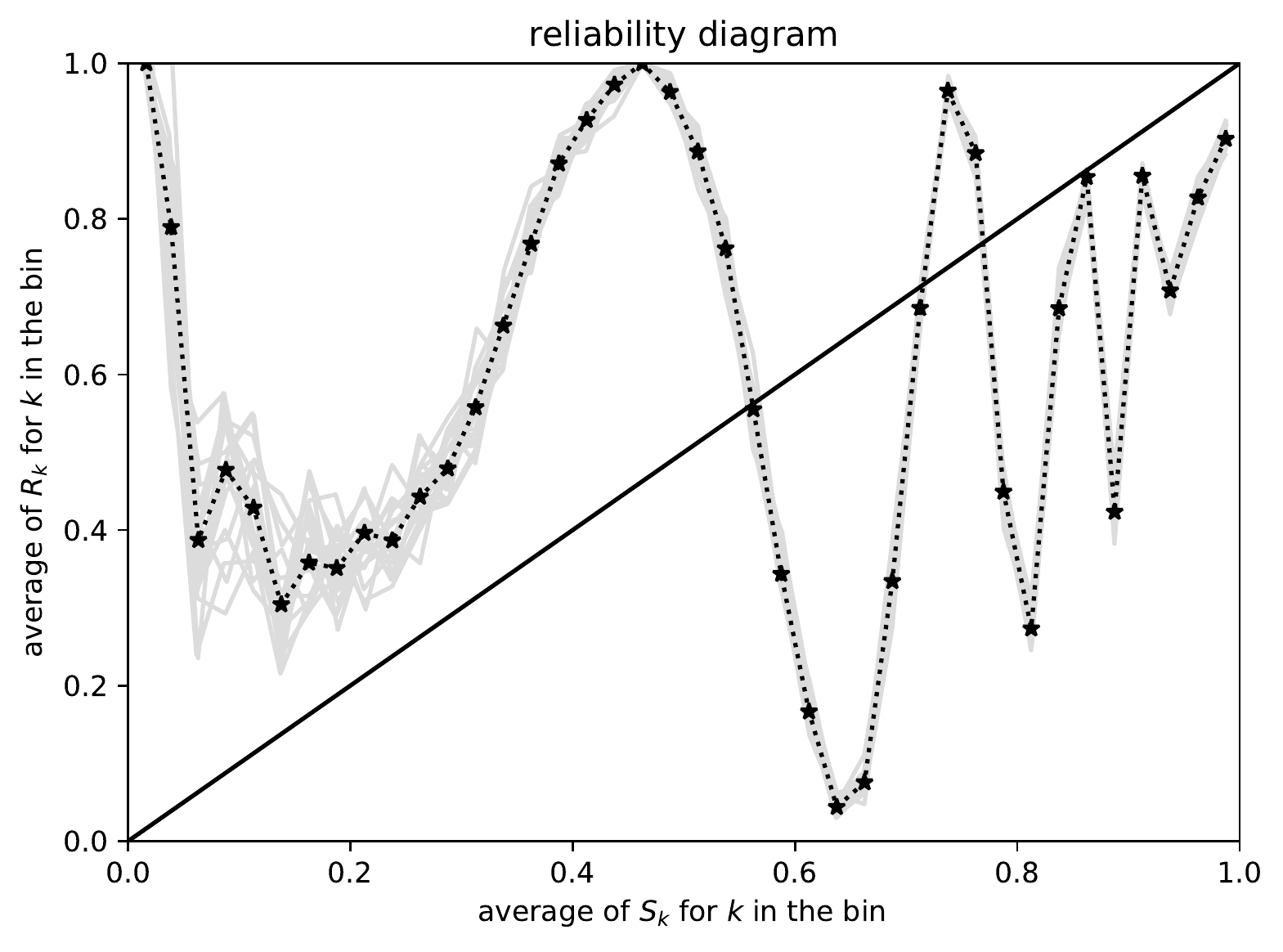}}
\quad\quad
\parbox{\imsize}{\includegraphics[width=\imsize]
                {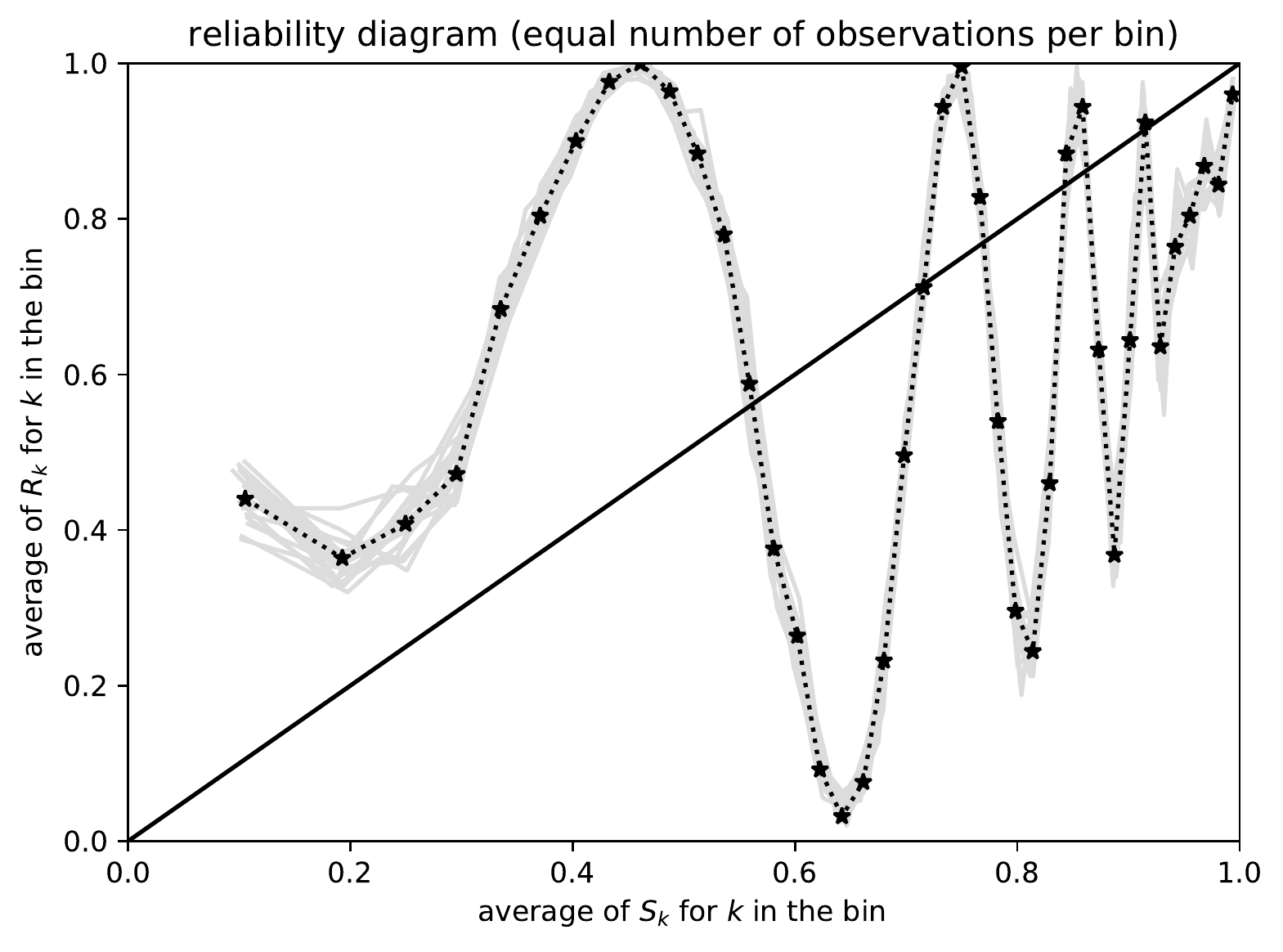}}

\vspace{\vertsep}

\parbox{\imsize}{\includegraphics[width=\imsize]
                {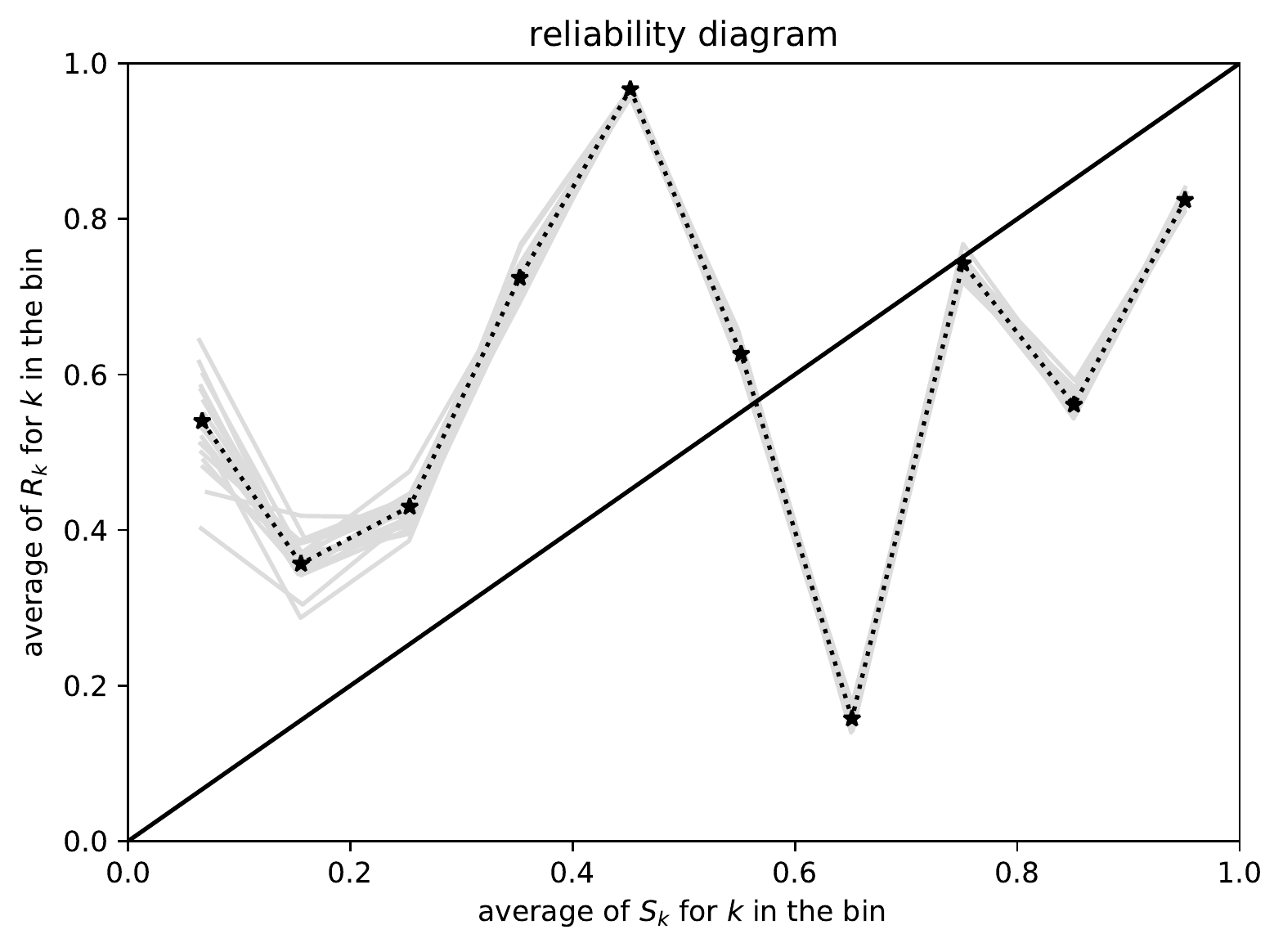}}
\quad\quad
\parbox{\imsize}{\includegraphics[width=\imsize]
                {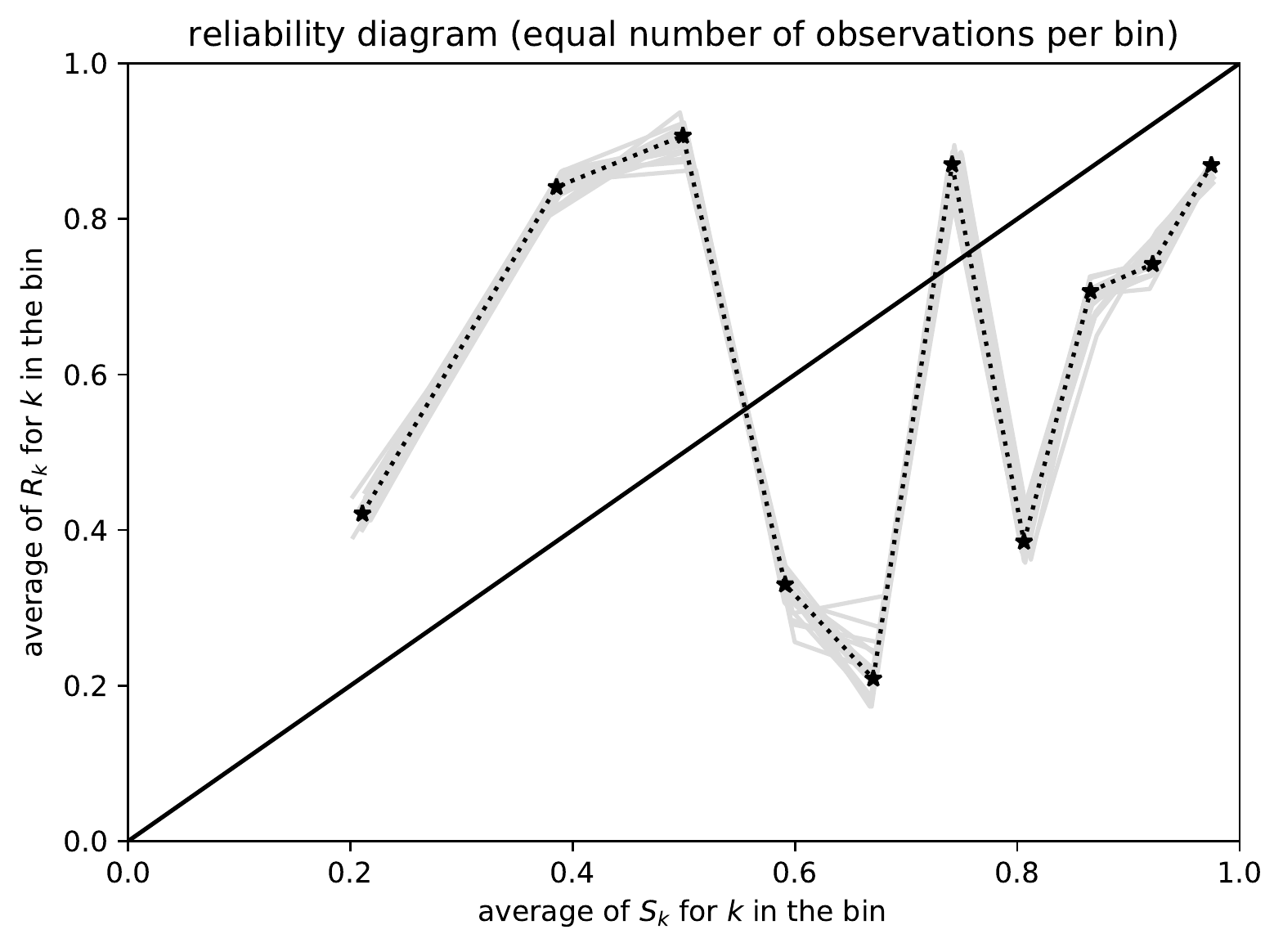}}

\end{centering}
\caption{$n =$ 10,000; $S_1$, $S_2$, \dots, $S_n$ are denser near 1;
         Kuiper's statistic is $0.1475 / \sigma = 36.12$,
         Kolmogorov's and Smirnov's is $0.1089 / \sigma = 26.67$.
Figure~\ref{10000_1e} displays the ground-truth reliability diagram.
The cumulative plot captures more of the oscillations in the miscalibration,
as do to some extent the reliability diagrams with an equal number
of observations per bin; however, the variations in the reliability diagrams
could be difficult to interpret without access
to the ground-truth exact expectations.
}
\label{10000_1}
\end{figure}

\begin{figure}
\begin{centering}

\parbox{\imsize}{\includegraphics[width=\imsize]
                {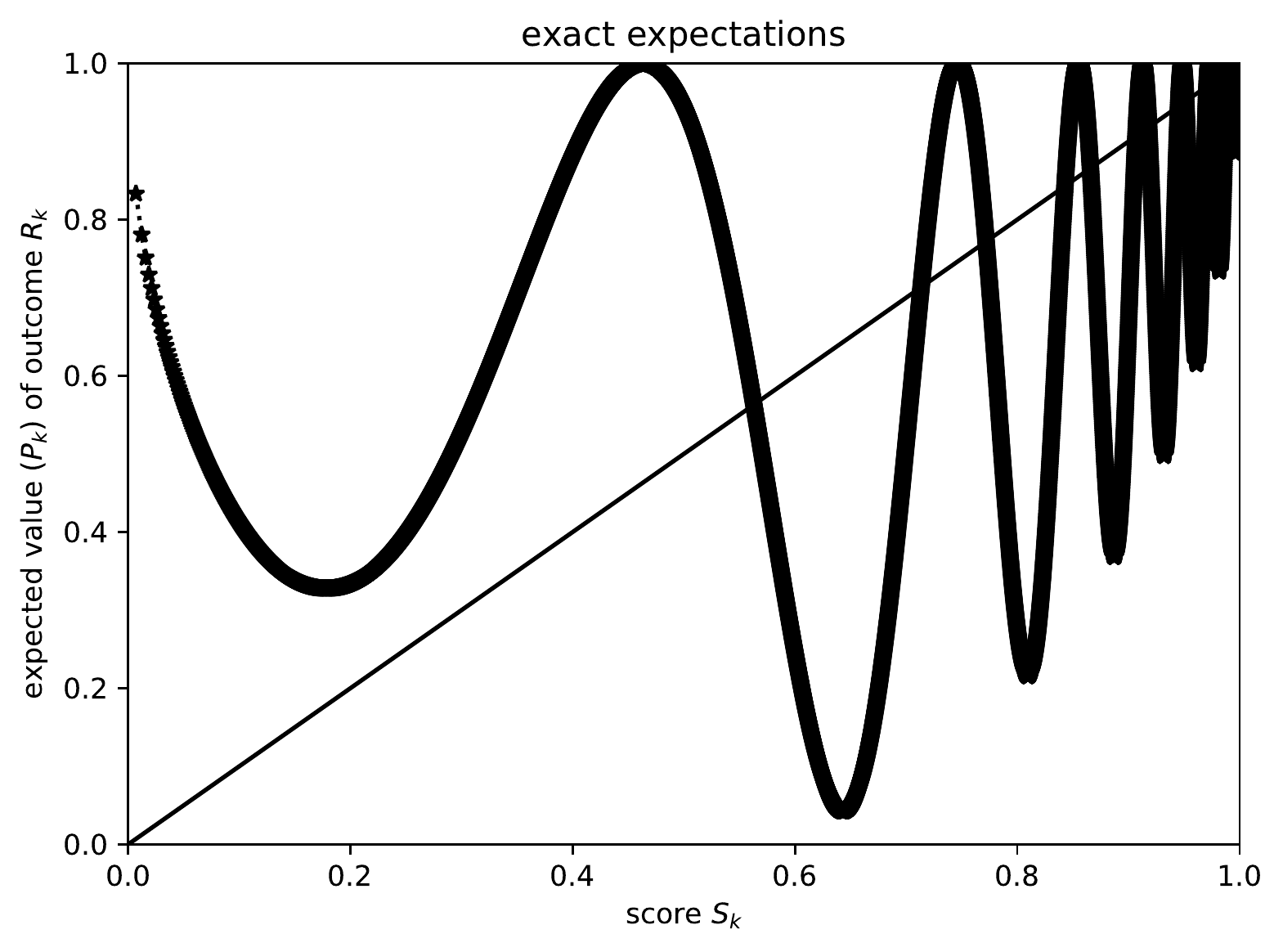}}

\end{centering}
\caption{Ground-truth reliability diagram for Figure~\ref{10000_1}}
\label{10000_1e}
\end{figure}

\begin{figure}
\begin{centering}

\parbox{\imsize}{\includegraphics[width=\imsize]
                {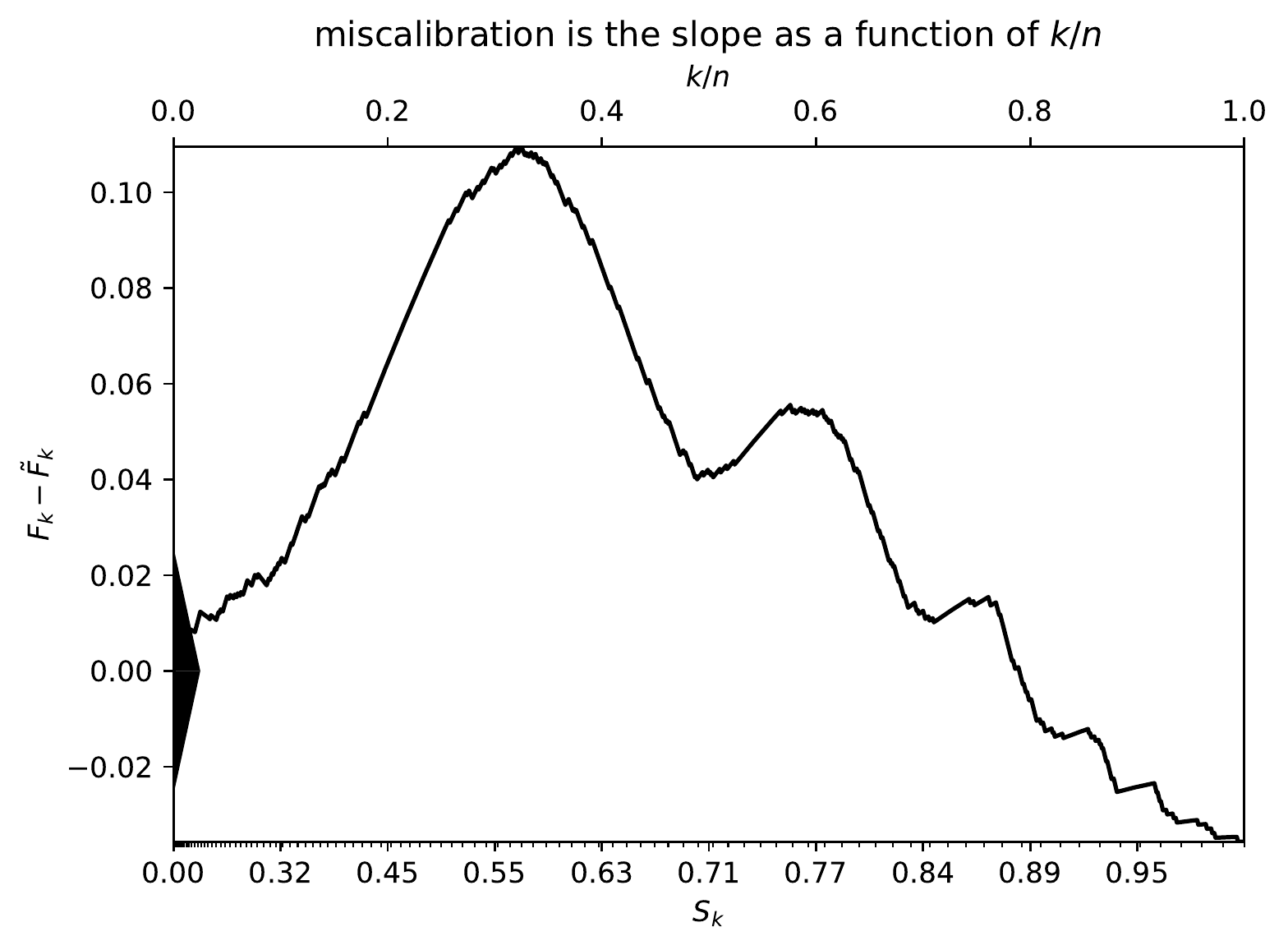}}
\quad\quad
\parbox{\imsize}{\includegraphics[width=\imsize]
                {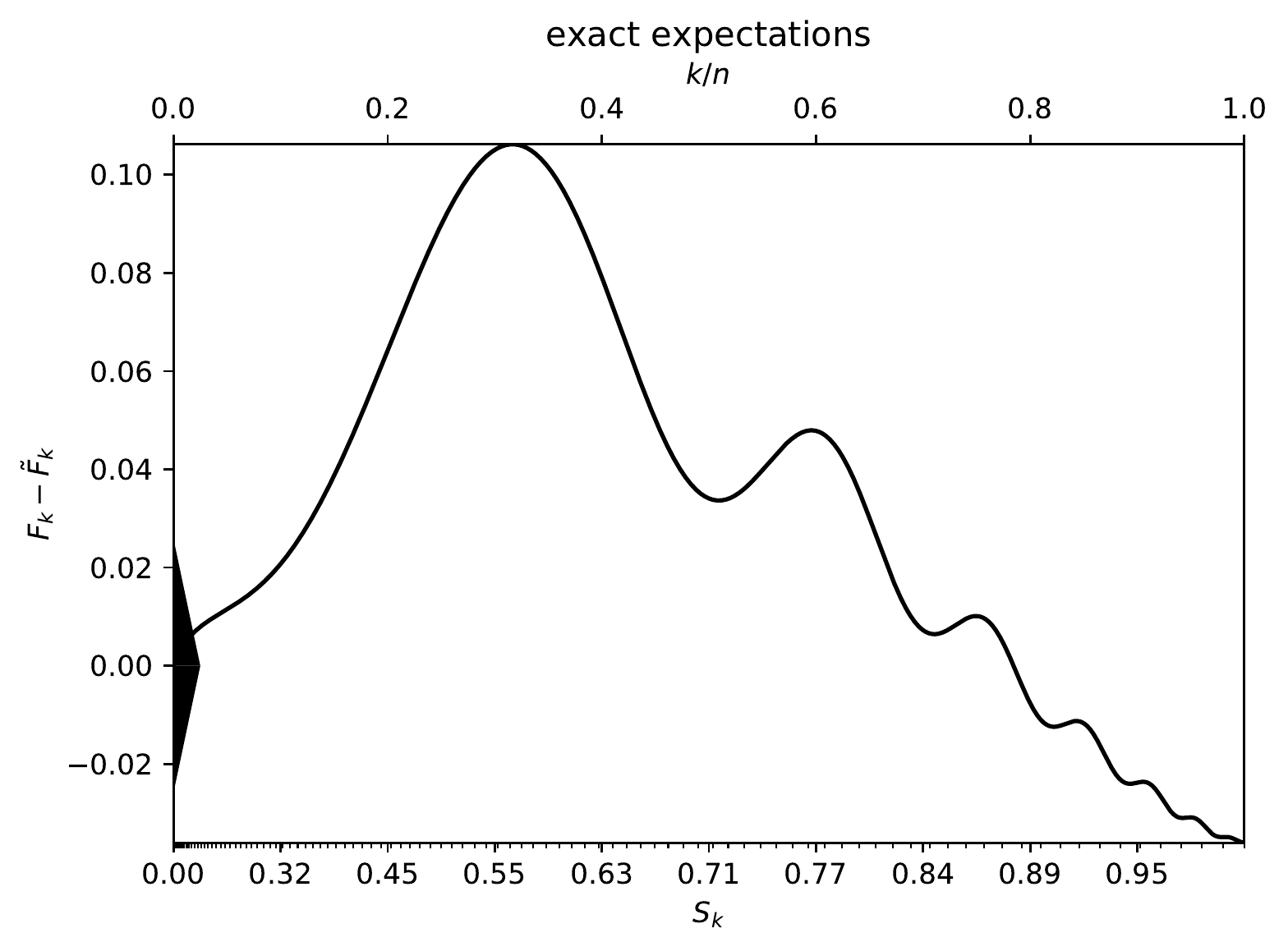}}

\vspace{\vertsep}

\parbox{\imsize}{\includegraphics[width=\imsize]
                {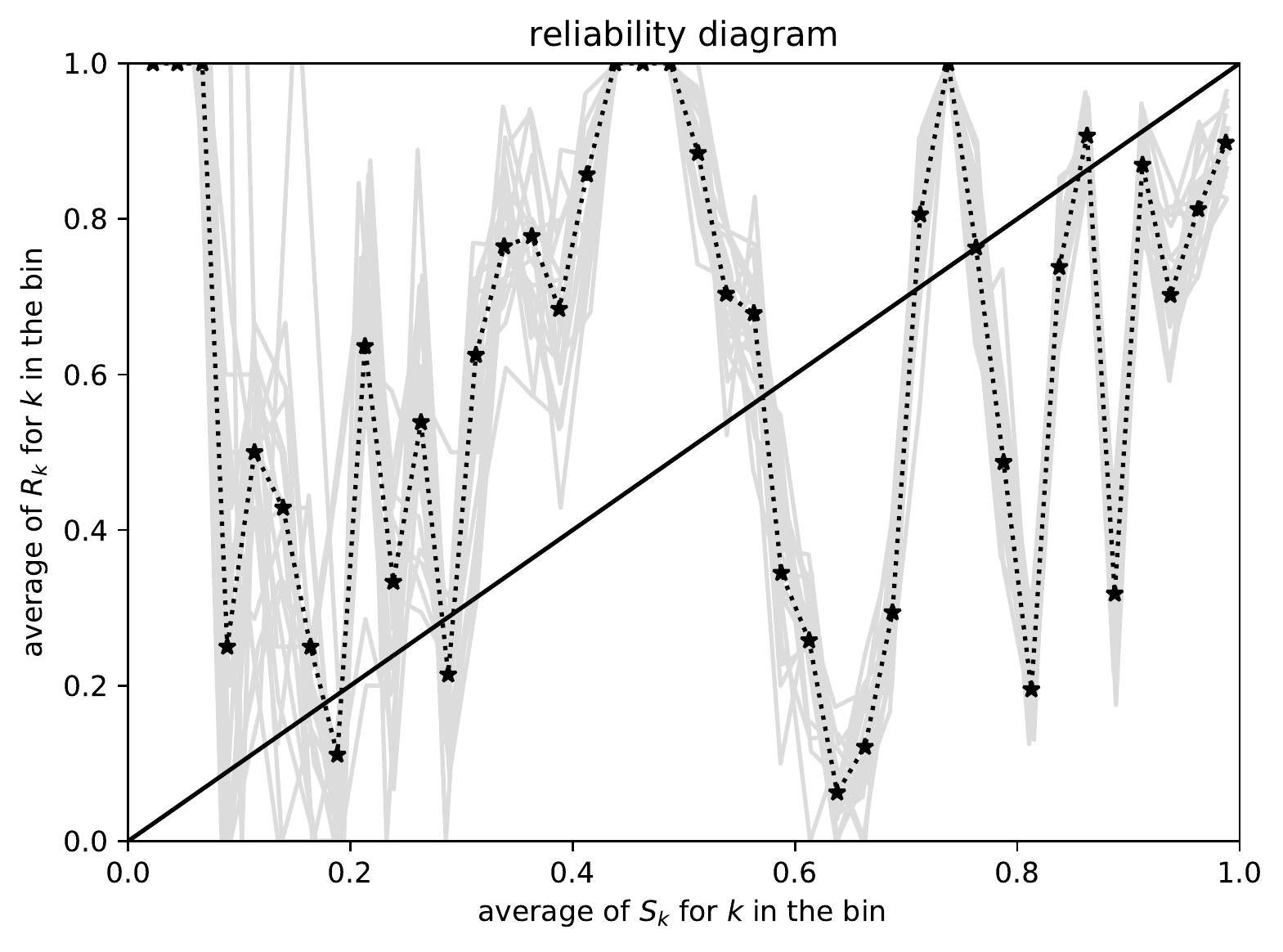}}
\quad\quad
\parbox{\imsize}{\includegraphics[width=\imsize]
                {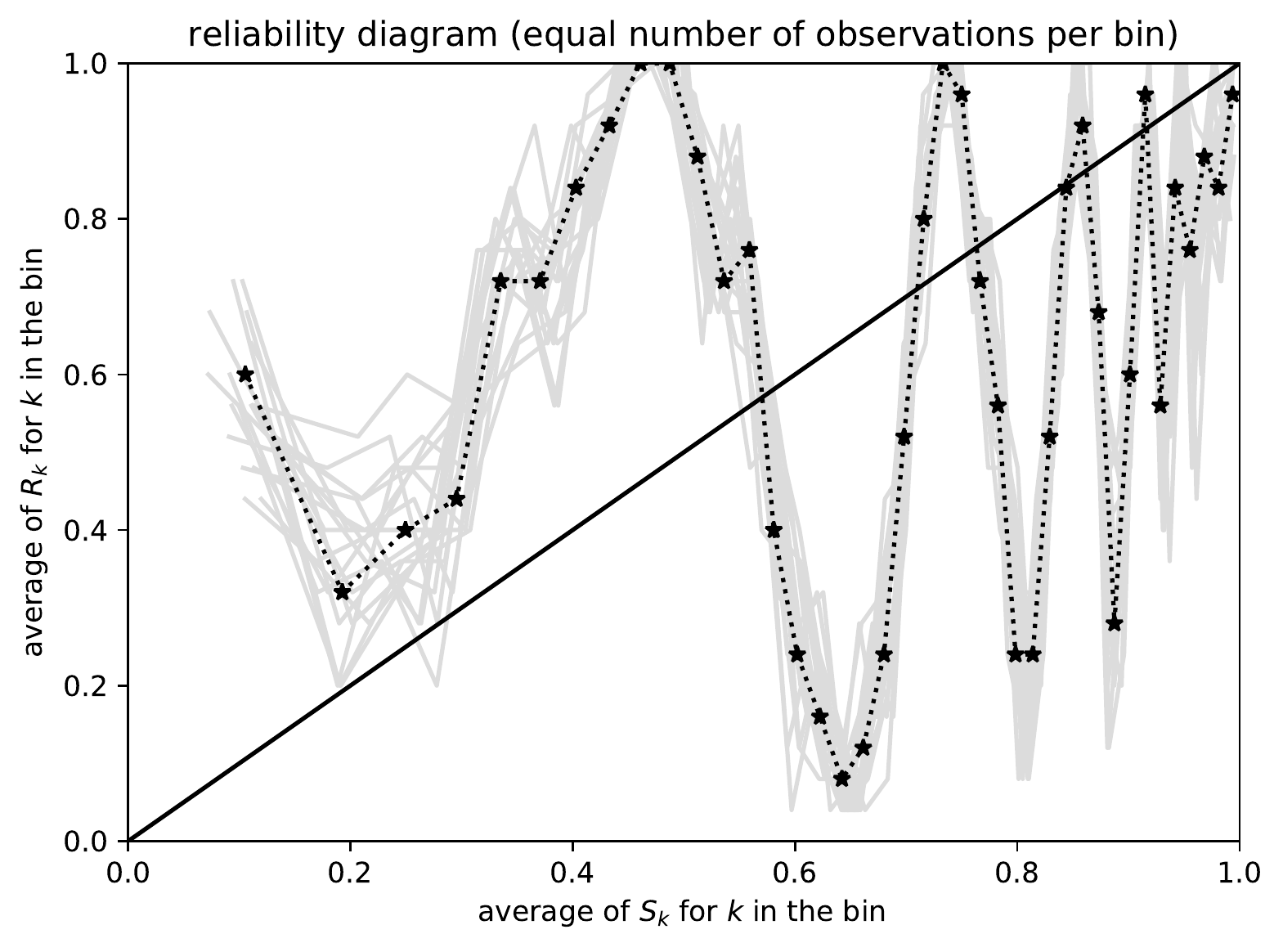}}

\vspace{\vertsep}

\parbox{\imsize}{\includegraphics[width=\imsize]
                {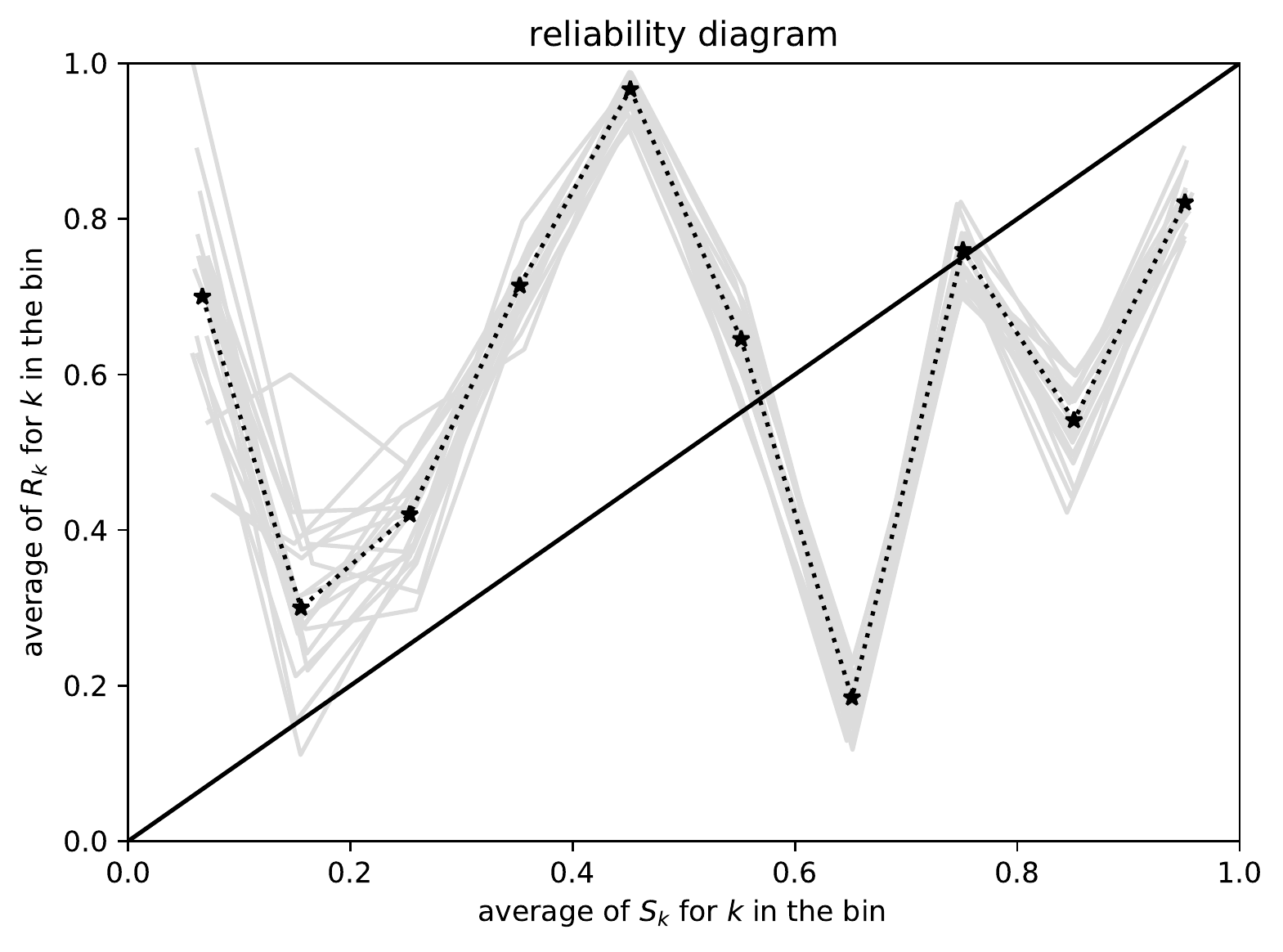}}
\quad\quad
\parbox{\imsize}{\includegraphics[width=\imsize]
                {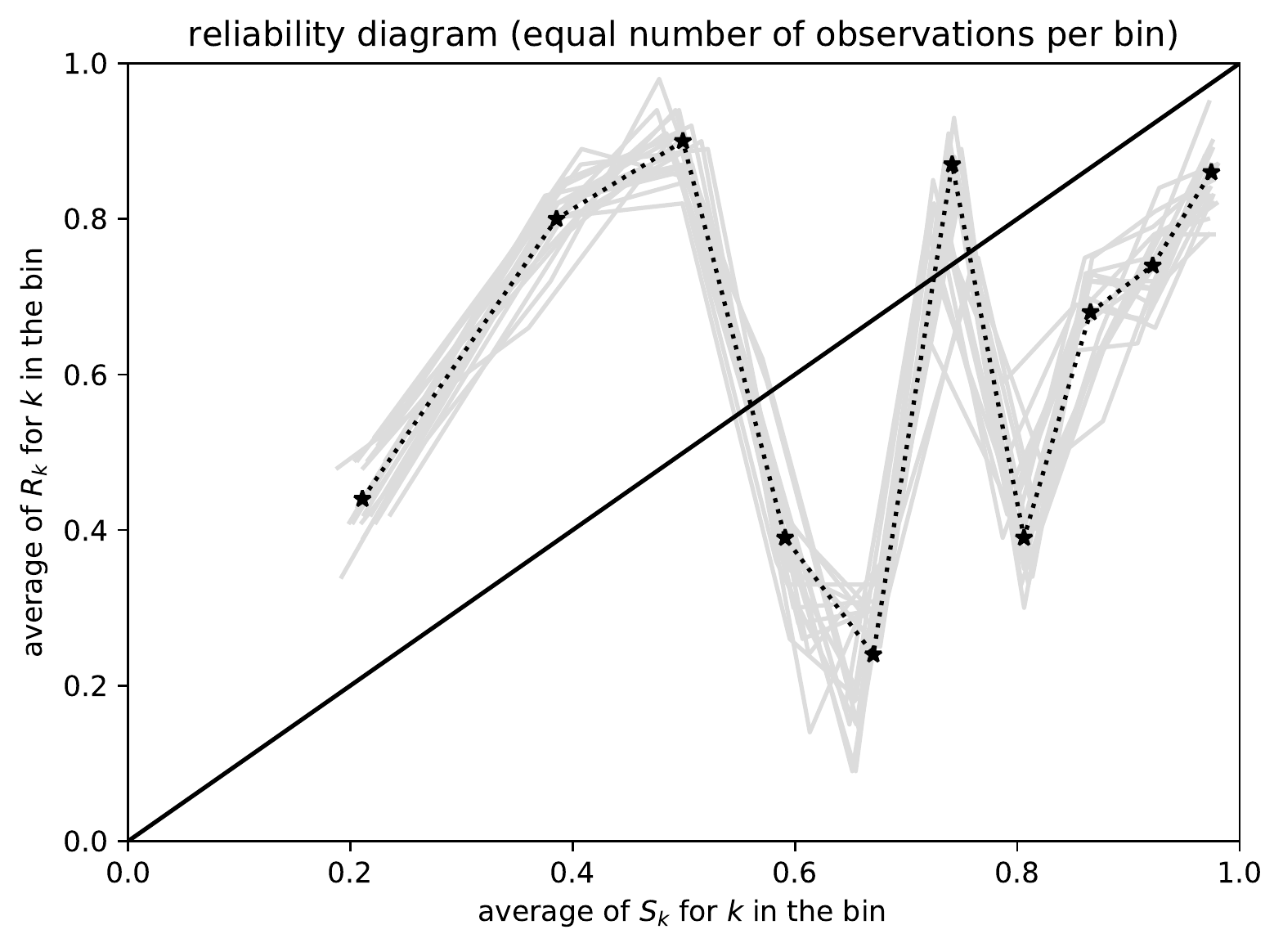}}

\end{centering}
\caption{$n =$ 1,000; $S_1$, $S_2$, \dots, $S_n$ are denser near 1;
         Kuiper's statistic is $0.1452 / \sigma = 11.24$,
         Kolmogorov's and Smirnov's is $0.1095 / \sigma = 8.480$.
Figure~\ref{1000_1e} displays the ground-truth reliability diagram.
As in Figure~\ref{10000_1},
the cumulative plot resolves more of the oscillations in the miscalibration,
as do somewhat the reliability diagrams with an equal number
of observations per bin; that said, the variations in the reliability diagrams
could be hard to interpret without access to the exact expectations.
}
\label{1000_1}
\end{figure}

\begin{figure}
\begin{centering}

\parbox{\imsize}{\includegraphics[width=\imsize]
                {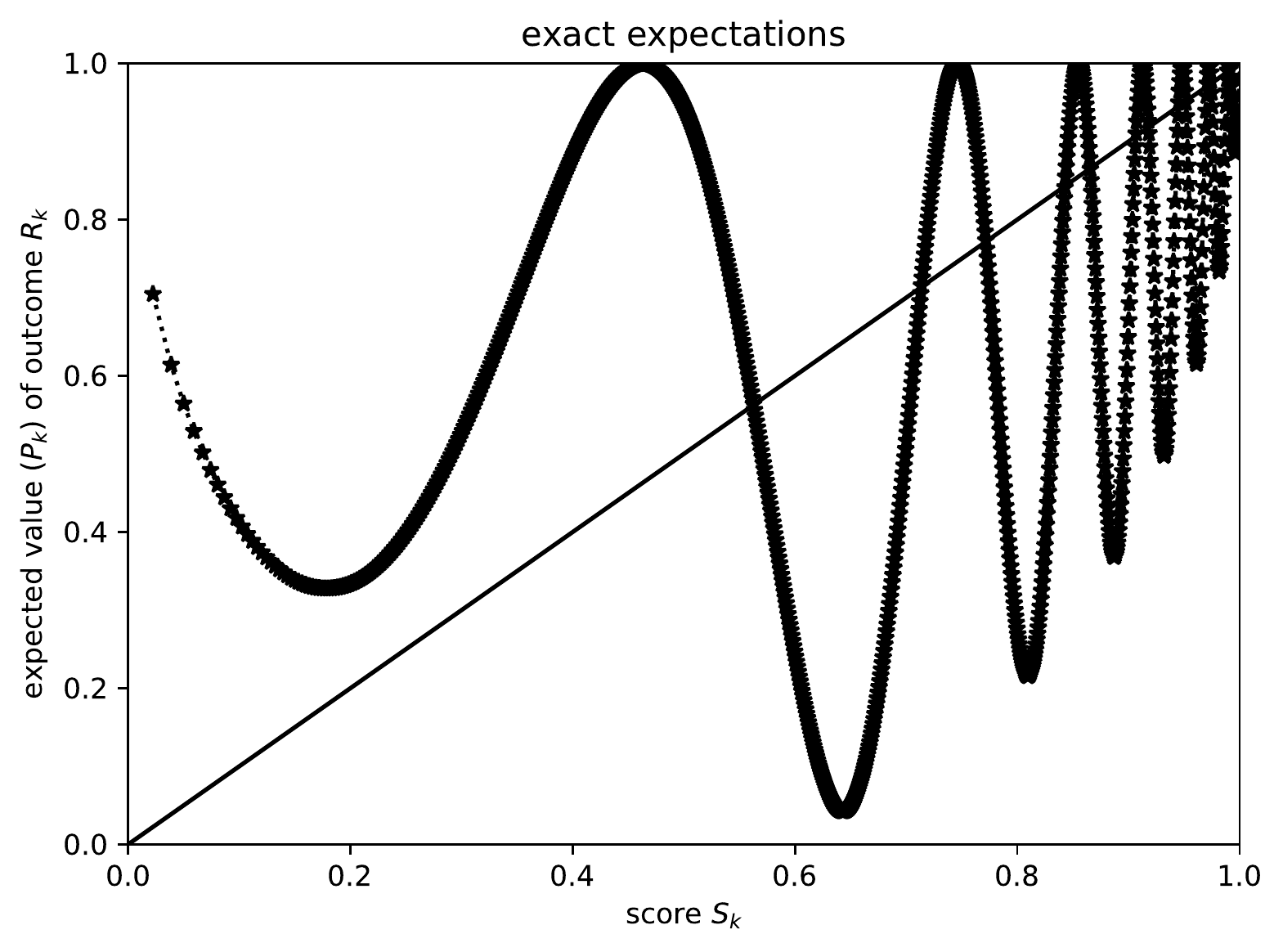}}

\end{centering}
\caption{Ground-truth reliability diagram for Figure~\ref{1000_1}}
\label{1000_1e}
\end{figure}

\begin{figure}
\begin{centering}

\parbox{\imsize}{\includegraphics[width=\imsize]
                {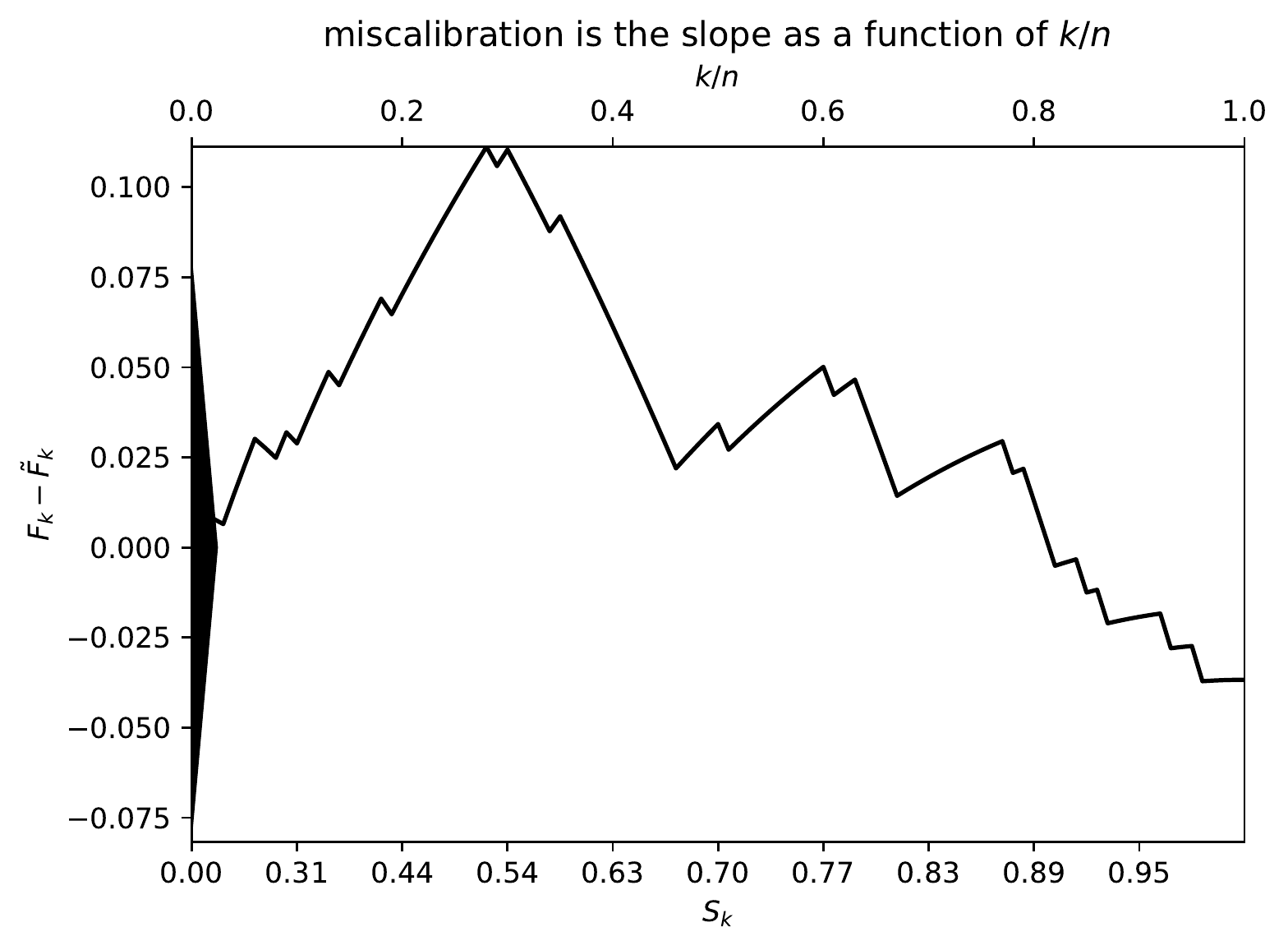}}
\quad\quad
\parbox{\imsize}{\includegraphics[width=\imsize]
                {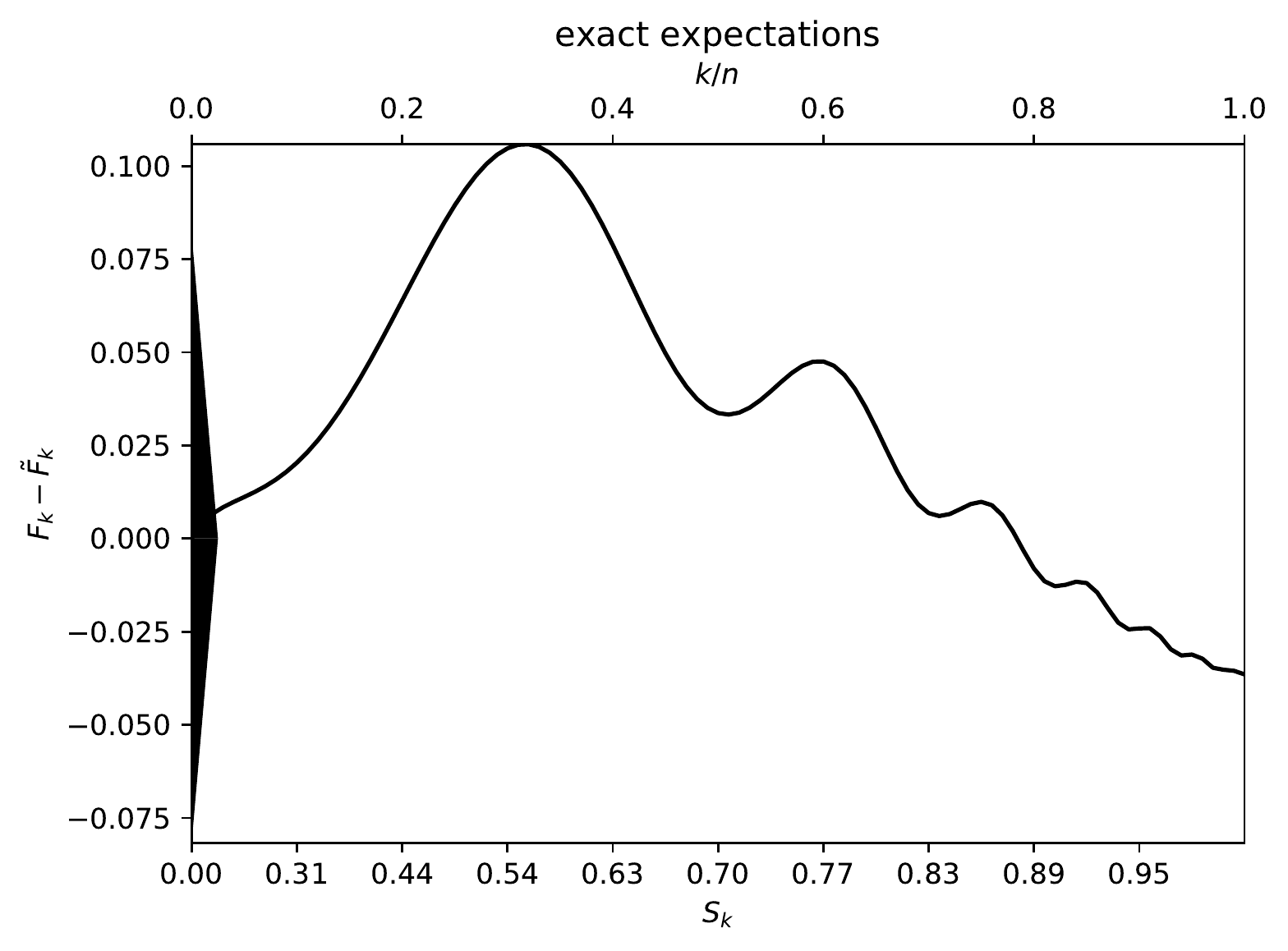}}

\vspace{\vertsep}

\parbox{\imsize}{\includegraphics[width=\imsize]
                {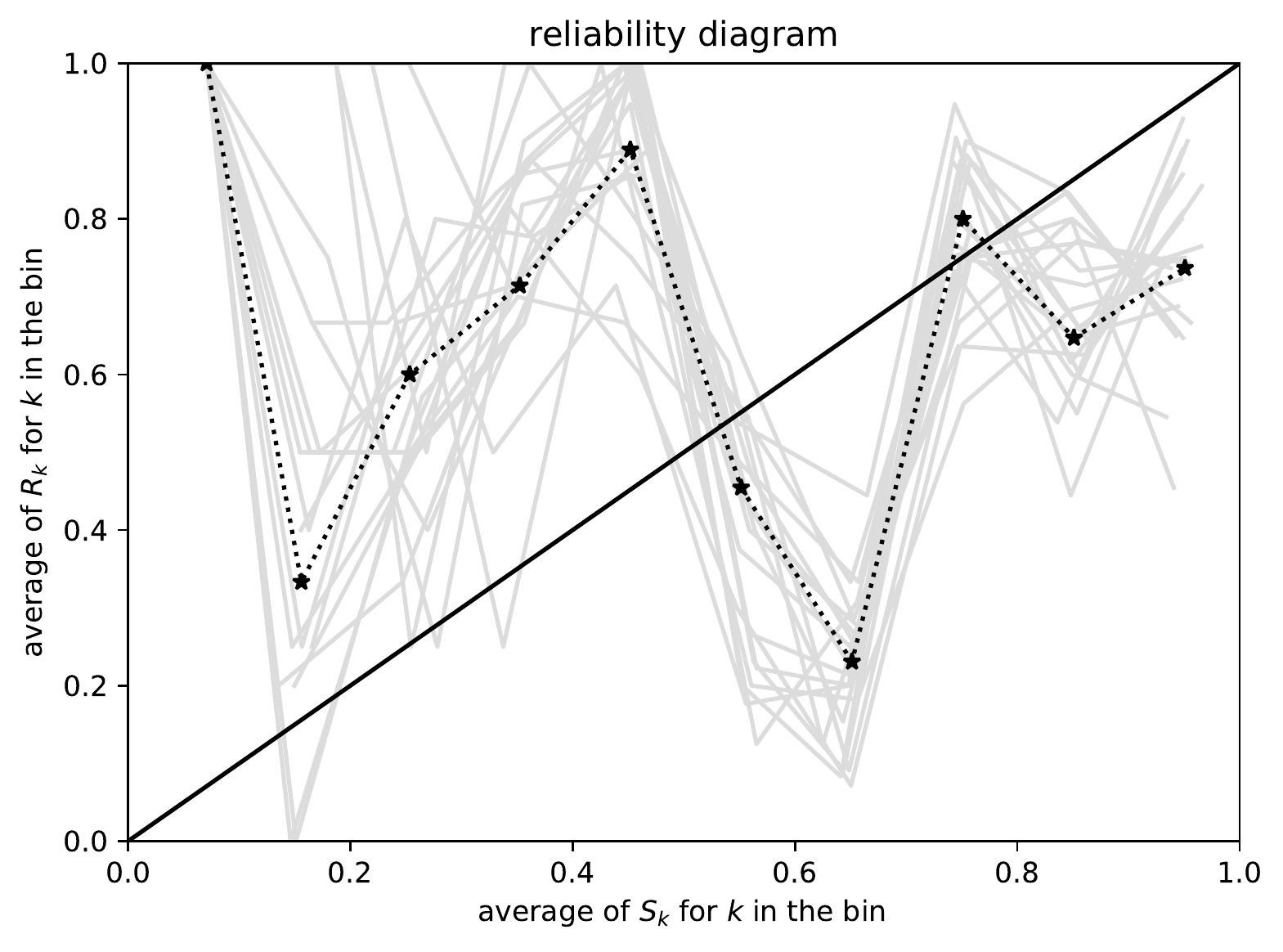}}
\quad\quad
\parbox{\imsize}{\includegraphics[width=\imsize]
                {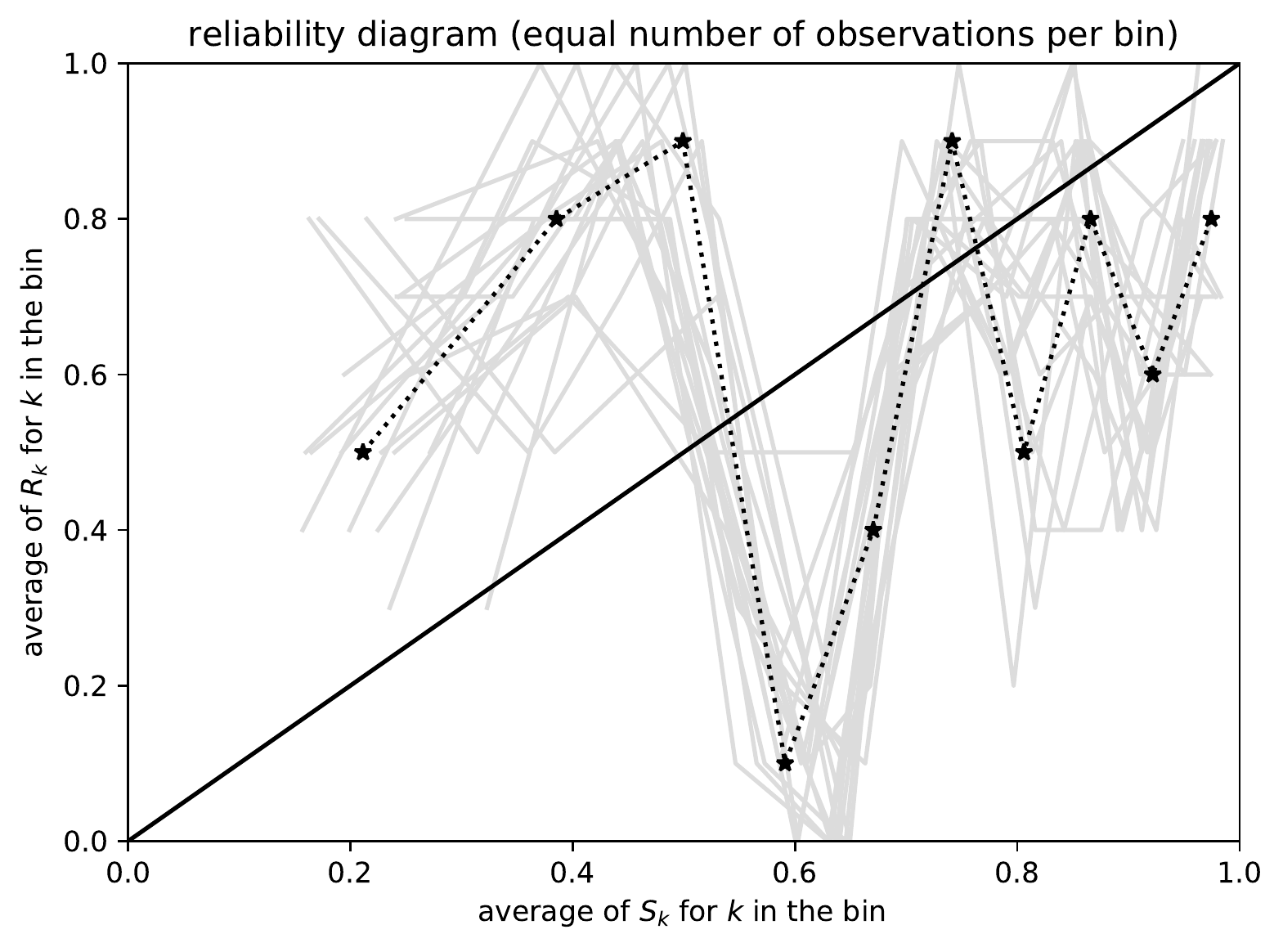}}

\vspace{\vertsep}

\parbox{\imsize}{\includegraphics[width=\imsize]
                {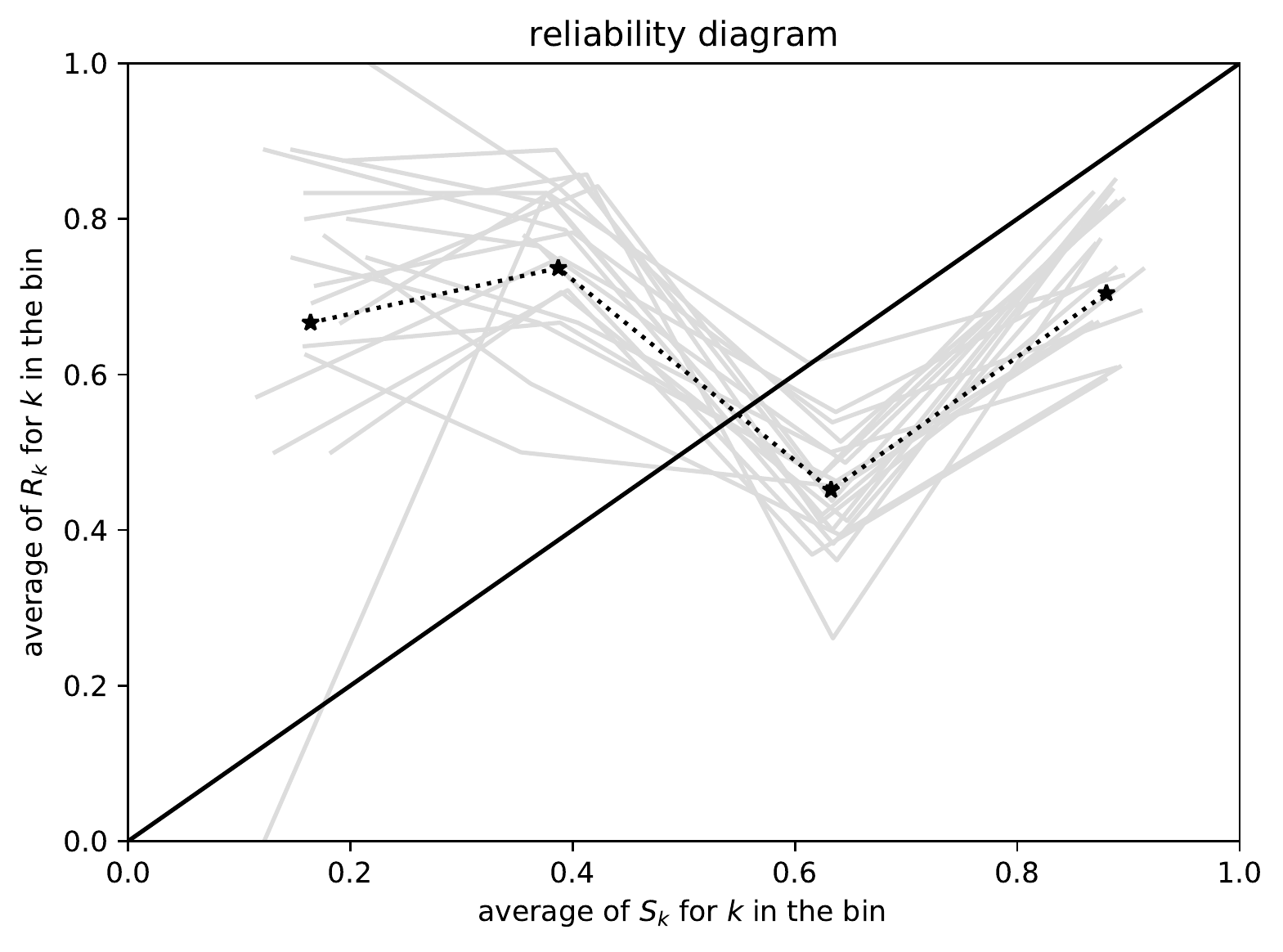}}
\quad\quad
\parbox{\imsize}{\includegraphics[width=\imsize]
                {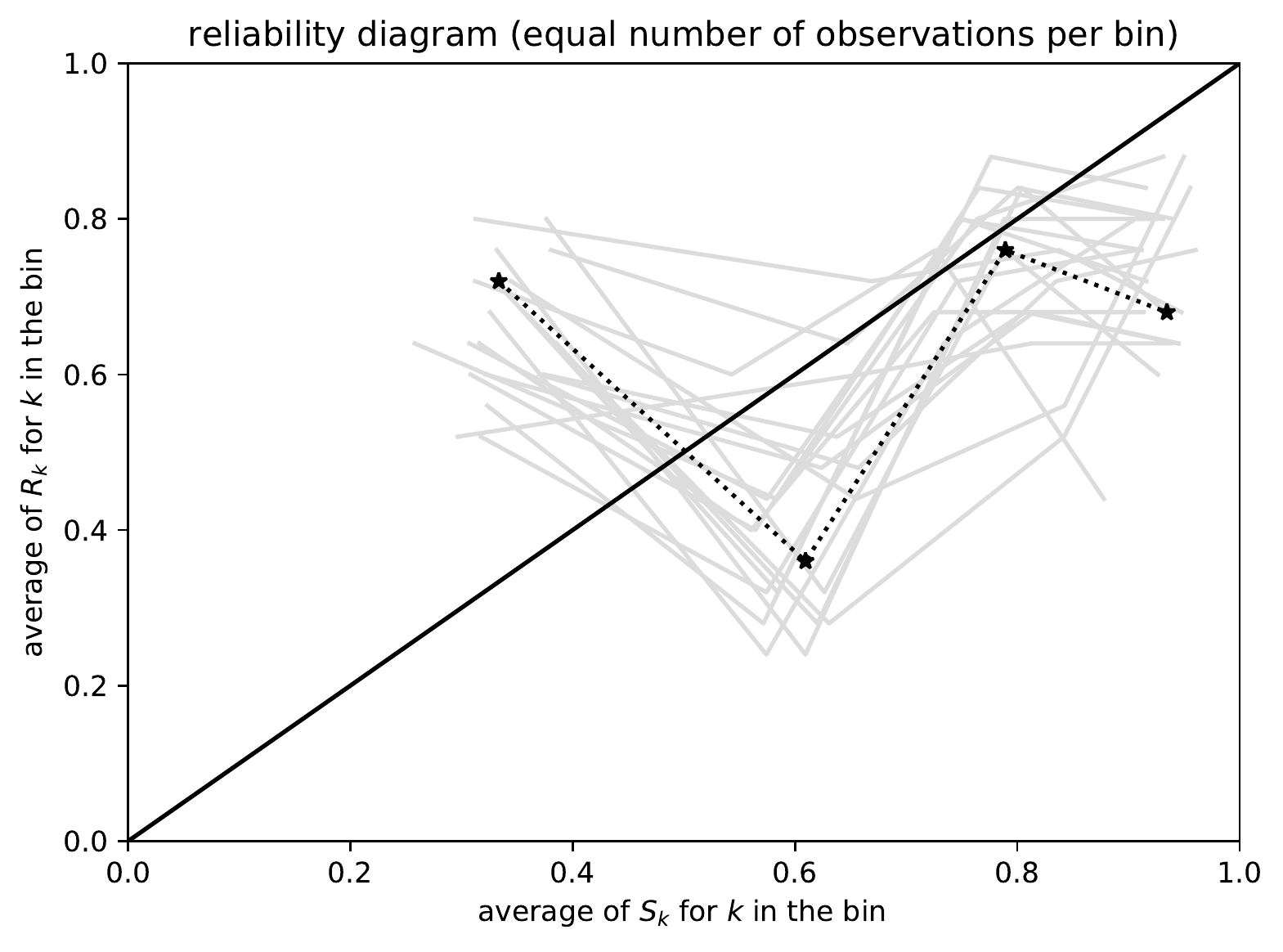}}

\end{centering}
\caption{$n =$ 100; $S_1$, $S_2$, \dots, $S_n$ are denser near 1;
         Kuiper's statistic is $0.1483 / \sigma = 3.632$,
         Kolmogorov's and Smirnov's is $0.1112 / \sigma = 2.723$.
Figure~\ref{100_1e} displays the ground-truth reliability diagram.
Just like in Figures~\ref{10000_1} and~\ref{1000_1},
the cumulative plot captures more of the oscillations in the miscalibration,
as do to some extent the reliability diagrams with an equal number
of observations per bin; even so, the variations in the reliability diagrams
are difficult to interpret without access to the exact expectations.
}
\label{100_1}
\end{figure}

\begin{figure}
\begin{centering}

\parbox{\imsize}{\includegraphics[width=\imsize]
                {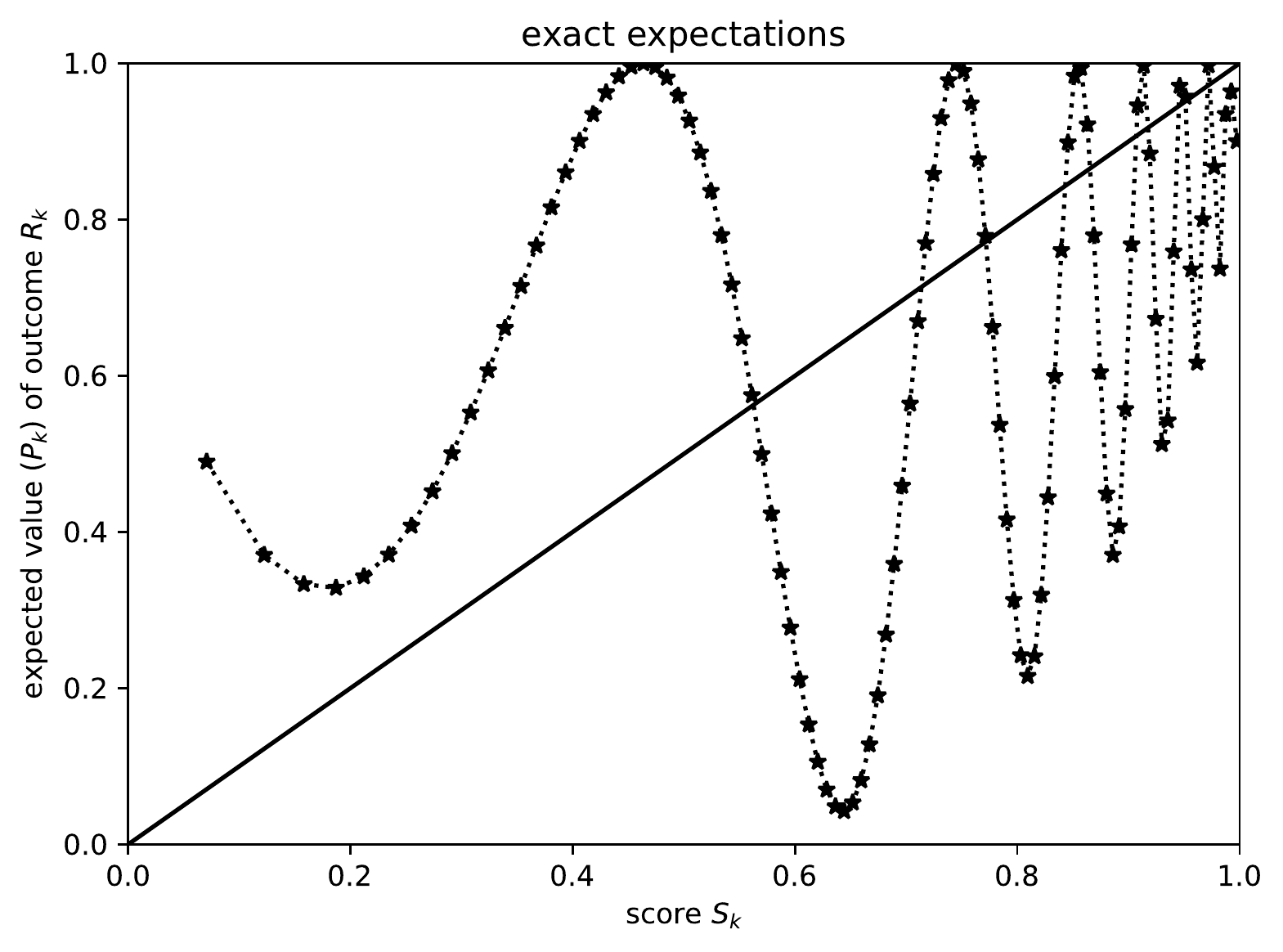}}

\end{centering}
\caption{Ground-truth reliability diagram for Figure~\ref{100_1}}
\label{100_1e}
\end{figure}

Figures~\ref{10000_1}--\ref{100_1e} all draw
from the same, relatively complicated underlying distribution,
with $S_k$ being proportional to $\sqrt{k-0.5}$;
Figures~\ref{10000_1} and~\ref{10000_1e} set $n =$ 10,000,
Figures~\ref{1000_1} and~\ref{1000_1e} set $n =$ 1,000,
and Figures~\ref{100_1} and~\ref{100_1e} set $n =$ 100.
The cumulative plots appear advantageous.

In all cases with $n =$ 10,000, that is, in Figures~\ref{10000}, \ref{10000_0},
and~\ref{10000_1}, the scalar summary statistics detect
extremely statistically significant miscalibration.
Moreover, in all cases with $n =$ 1,000, that is, in Figures~\ref{1000},
\ref{1000_0}, and~\ref{1000_1}, the scalar summary statistics detect
highly statistically significant miscalibration.
In all cases with $n =$ 100, that is, in Figures~\ref{100}, \ref{100_0},
and~\ref{100_1}, the scalar summary statistics detect
some statistically significant miscalibration,
though not with nearly as much confidence as when $n =$ 1,000
or (even more starkly) as when $n =$ 10,000.
Naturally, the Kolmogorov-Smirnov and Kuiper statistics get larger
for miscalibration that is solely overcalibration
(and the same would be true of solely undercalibrated data),
such as that in Figures~\ref{10000_0}--\ref{100_0e}.

\subsubsection{ImageNet examples for assessing calibration}
\label{aimagenetex}

This subsubsection analyzes the standard training data set ``ImageNet-1000''
of~\cite{imagenet}. Each of the thousand labeled classes in the data set
consists of about 1,300 images of a particular noun,
so that the total number of members of the data set is $n =$ 1,281,167.
To generate the corresponding plots, displayed in Figure~\ref{imagenetcal},
we calculate the scores $S_1$,~$S_2$, \dots, $S_n$
using the pretrained ResNet18 classifier of~\cite{he-zhang-ren-sun}
from the computer-vision module, ``torchvision,''
in the PyTorch software library of~\cite{pytorch};
the score for an image is the probability assigned by the classifier
to the class predicted to be most likely,
with the scores randomly perturbed by about one part in $10^8$ to guarantee
their uniqueness.
For $j = 1$,~$2$, \dots, $n$, the result $R_j$ corresponding to a score $S_j$
is $R_j = 1$ when the class predicted to be most likely is the correct class,
and $R_j = 0$ otherwise. The cumulative plot works nicely,
as discussed in the caption of Figure~\ref{imagenetcal}.

\clearpage

\section{Cautions}
\label{caution}

In addition to noting the size of the triangle at the origin,
interpreting plots of cumulative differences
requires careful attention to one caveat:
avoid hallucination of minor deviations!
The sample paths of random walks and Brownian motion can look surprisingly
non-random (drifting?)\ quite often for short stints.
The most trustworthy detections of deviation are long ranges
(as a function of $k/n$ in the unweighted case
or of $A_k$ in the weighted case) of steep slopes for $F_k-\tilde{F}_k$.
The triangles centered at the origins of the plots give a sense
of the length scale for variations that are statistically significant.

For all plots, whether cumulative or classical,
bear in mind that even at 95\% confidence, one in twenty detections
is likely to be false. So, if there are a hundred bins,
each with a 95\% confidence interval, the reality is likely to violate
around 5 of those confidence intervals.
Beware when conducting multiple tests of significance
(or be sure to adjust the confidence level accordingly).

Figures~\ref{10000_00}--\ref{100_00e} illustrate these cautions;
these figures are analogous to those presented in Appendix~\ref{calibration},
but with the observations drawn from the same predicted probabilities
of success used to generate the graphs, so that the discrepancy
from perfect calibration should be statistically insignificant.
More precisely, Figures~\ref{10000_00}--\ref{100_00e}
all set $S_k$ to be proportional to $(k-0.5)^2$
and draw $R_1$,~$R_2$, \dots, $R_n$ from independent Bernoulli distributions
with expected success probabilities $S_1$,~$S_2$, \dots, $S_n$, respectively;
this corresponds to setting $P_k = S_k$ for all $k = 1$,~$2$, \dots, $n$,
in the numerical experiments of Appendix~\ref{calibration}.
Figures~\ref{10000_00}, \ref{1000_00}, and~\ref{100_00}
consider $n =$ 10,000, $n =$ 1,000, and $n =$ 100, respectively.
Please note that the ranges of the vertical axes
for the top rows of plots are tiny.
The leftmost topmost plots in Figures~\ref{10000_00}, \ref{1000_00},
and~\ref{100_00} look like driftless random walks;
in fact, they really are driftless random walks.
The variations of the graphs are comparable to the heights
of the triangles centered at the origins.
Comparing the second rows with the third rows
shows that the deviations from perfect calibration
are consistent with expected random fluctuations.
Indeed, all plots in this appendix depict only small deviations
from perfect calibration, as expected (and as desired).
None of the scalar summary statistics significantly exceeds
its expected value under the null hypothesis of~(\ref{null}).

\begin{figure}
\begin{centering}

\parbox{\imsize}{\includegraphics[width=\imsize]
                {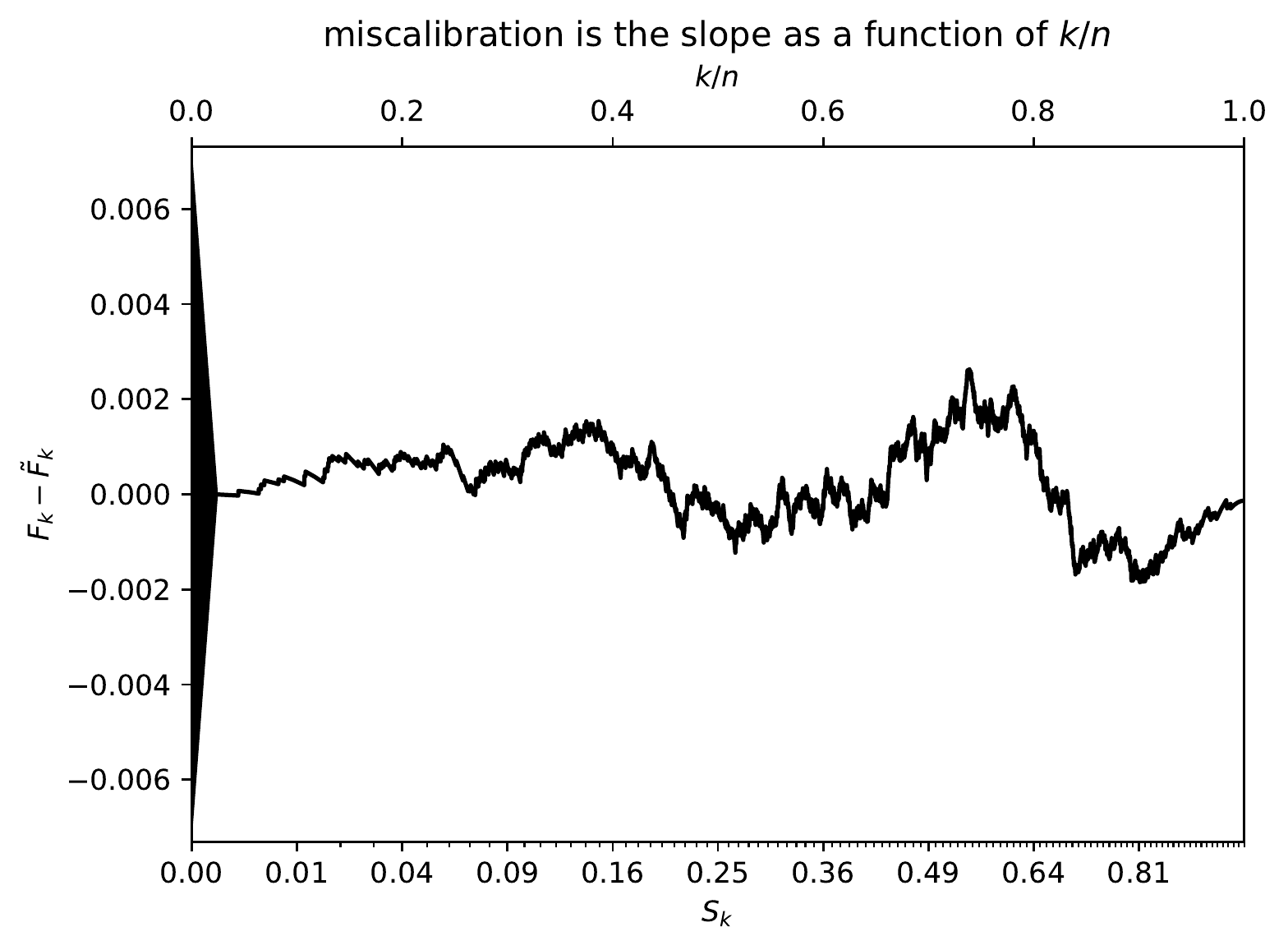}}
\quad\quad
\parbox{\imsize}{\includegraphics[width=\imsize]
                {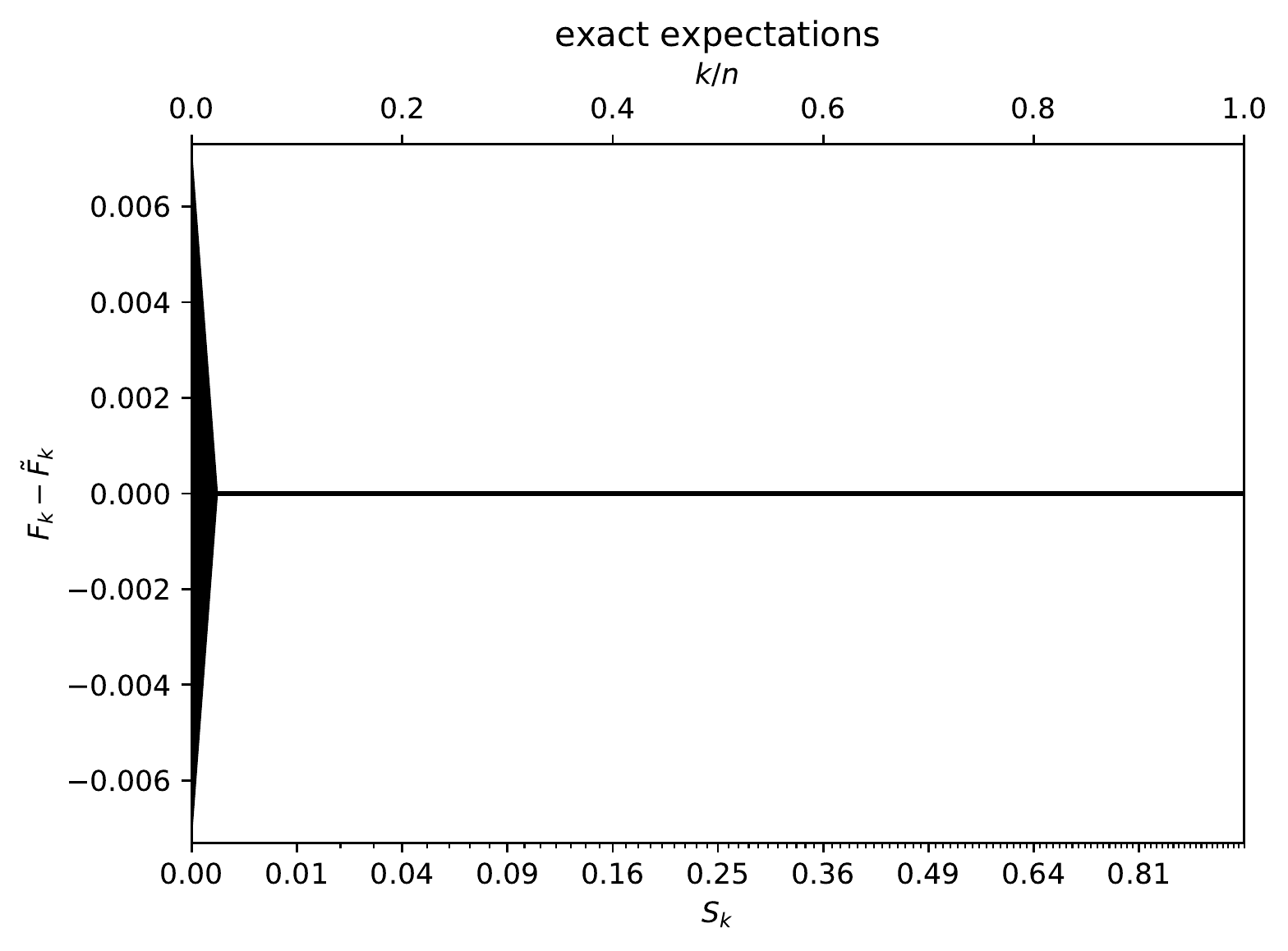}}

\vspace{\vertsep}

\parbox{\imsize}{\includegraphics[width=\imsize]
                {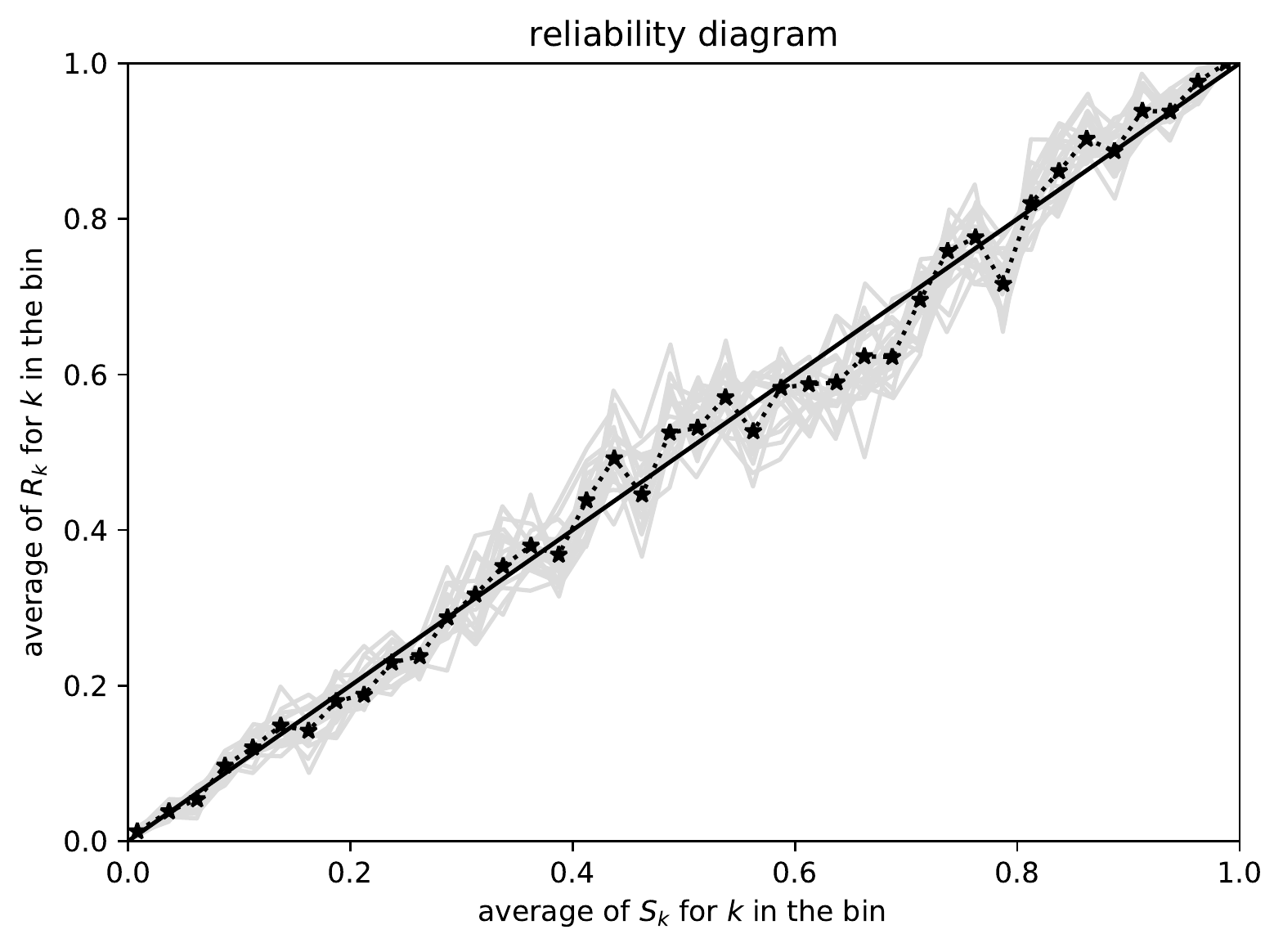}}
\quad\quad
\parbox{\imsize}{\includegraphics[width=\imsize]
                {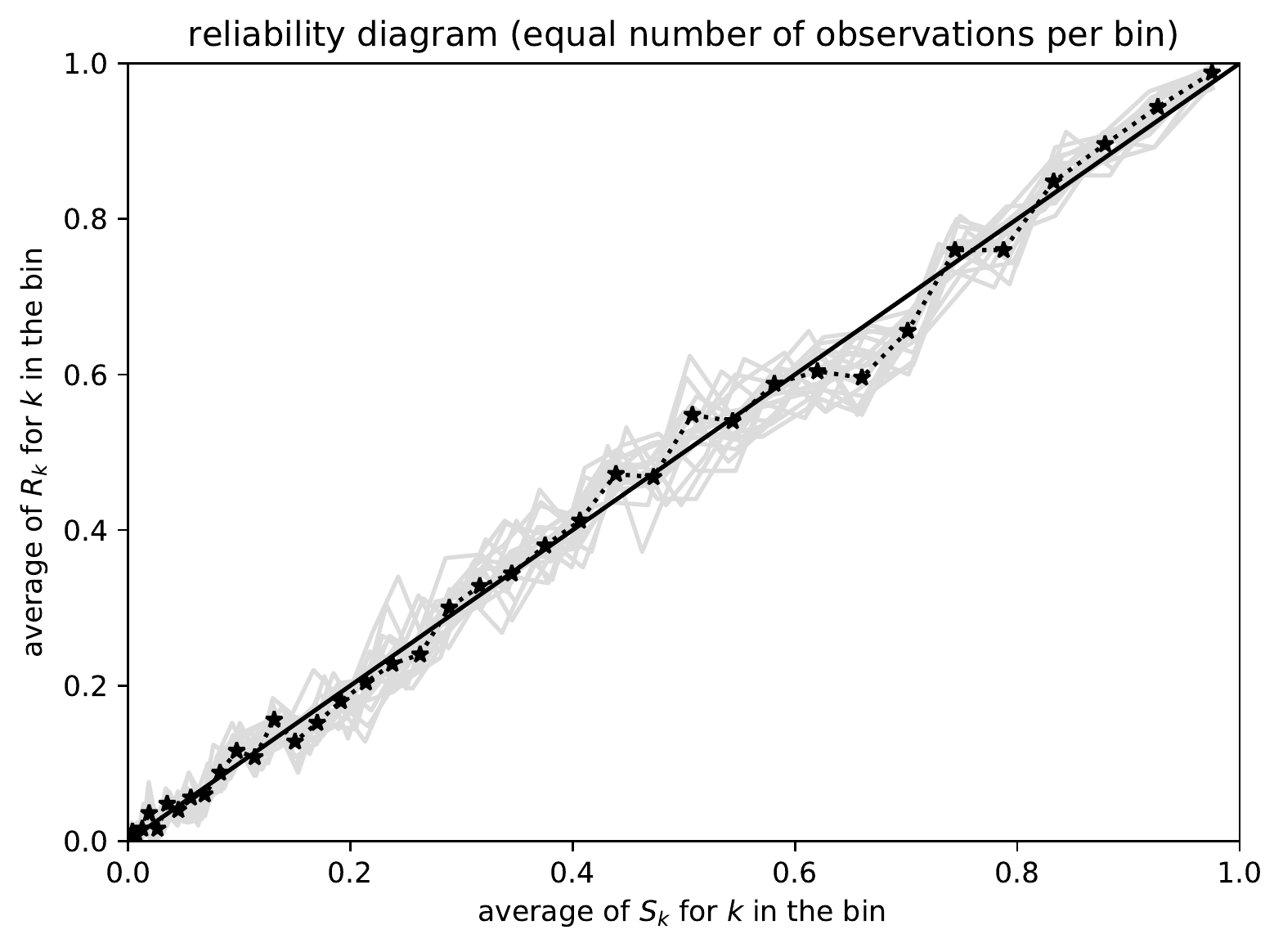}}

\vspace{\vertsep}

\parbox{\imsize}{\includegraphics[width=\imsize]
                {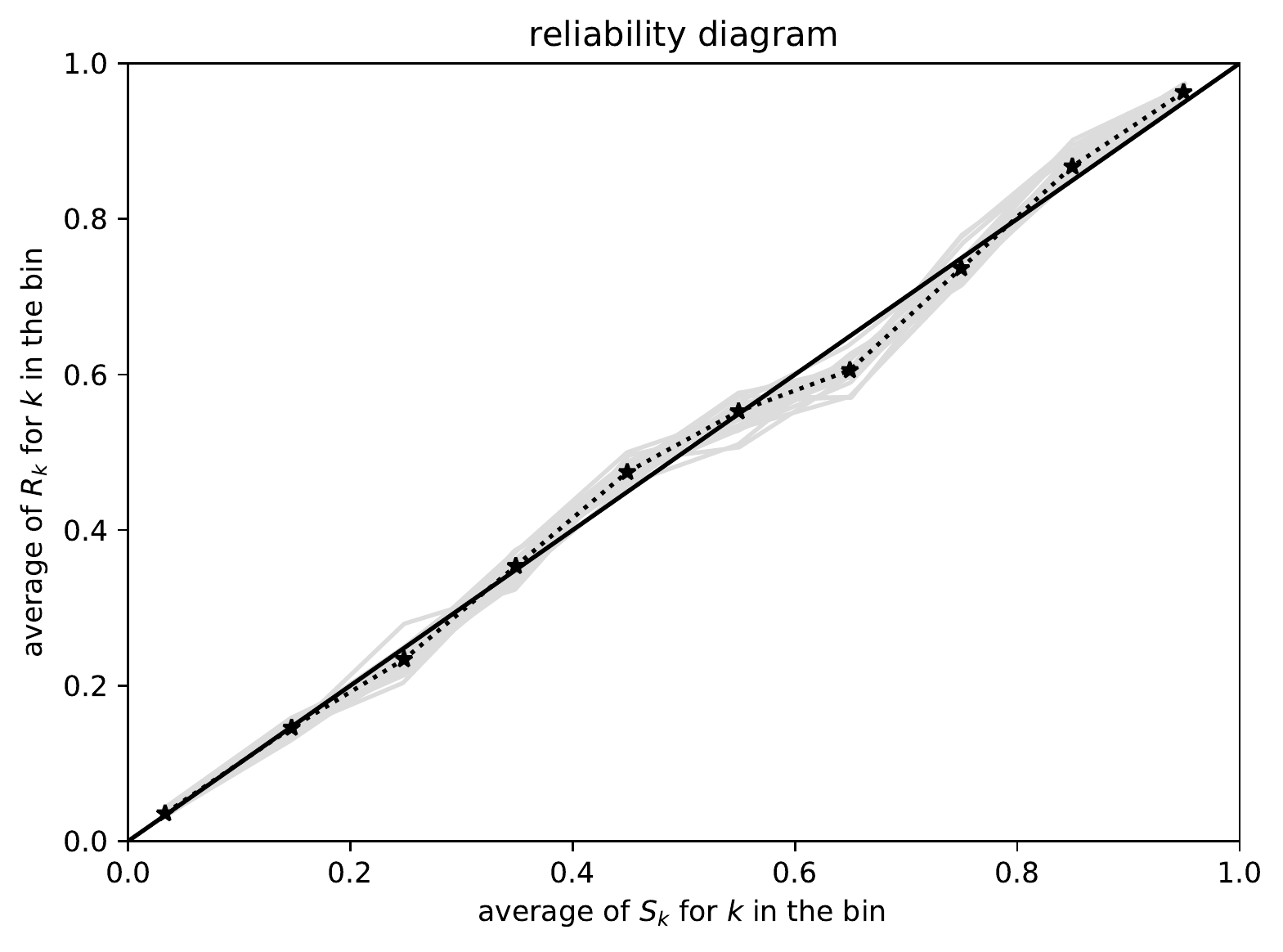}}
\quad\quad
\parbox{\imsize}{\includegraphics[width=\imsize]
                {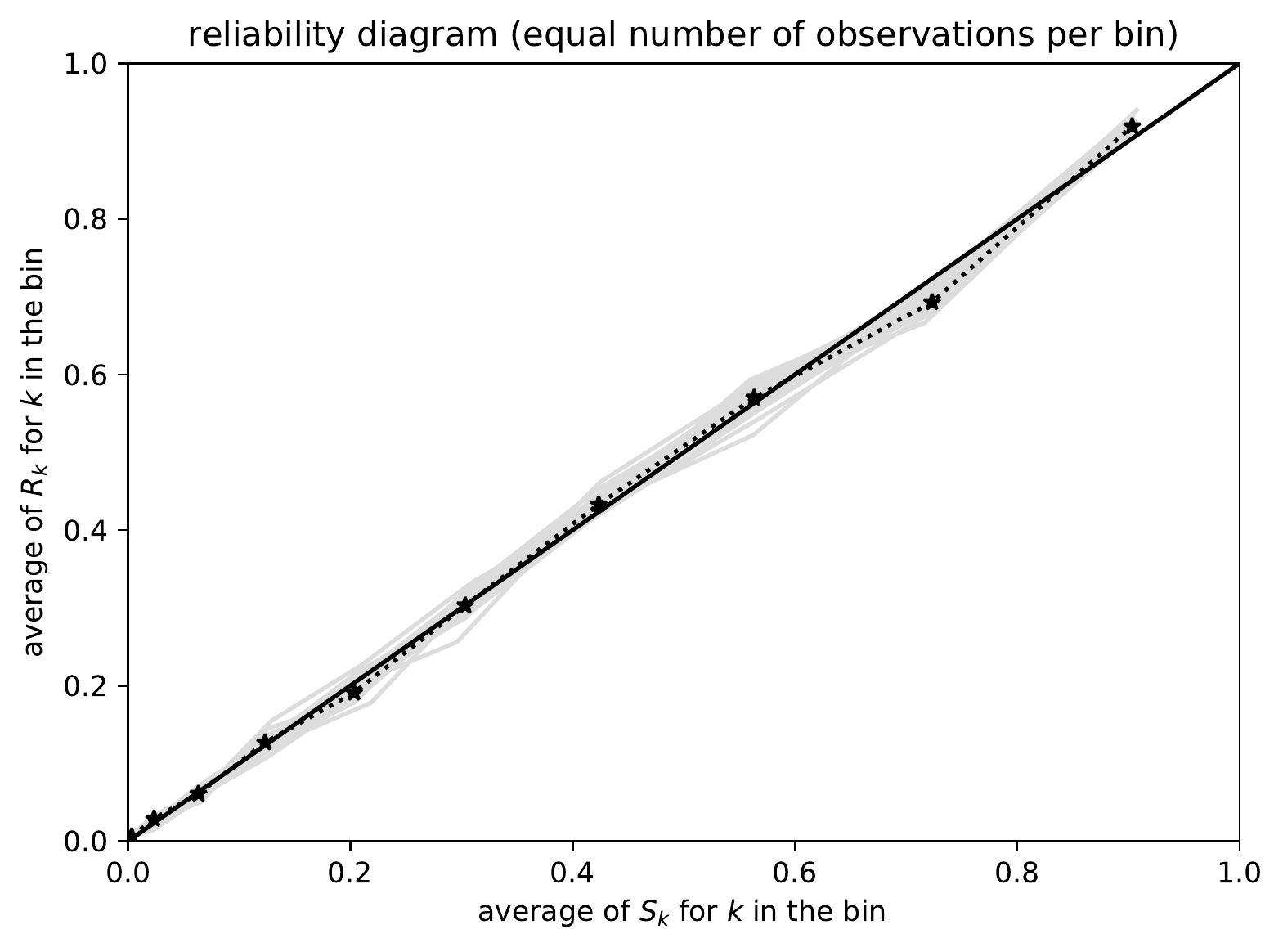}}

\end{centering}
\caption{$n =$ 10,000; $S_1$, $S_2$, \dots, $S_n$ are denser near 0;
         Kuiper's statistic is $0.004475 / \sigma = 1.226$,
         Kolmogorov's and Smirnov's is $0.002627 / \sigma = 0.7194$.
Figure~\ref{10000_00e} displays the ground-truth reliability diagram.
}
\label{10000_00}
\end{figure}

\begin{figure}
\begin{centering}

\parbox{\imsize}{\includegraphics[width=\imsize]
                {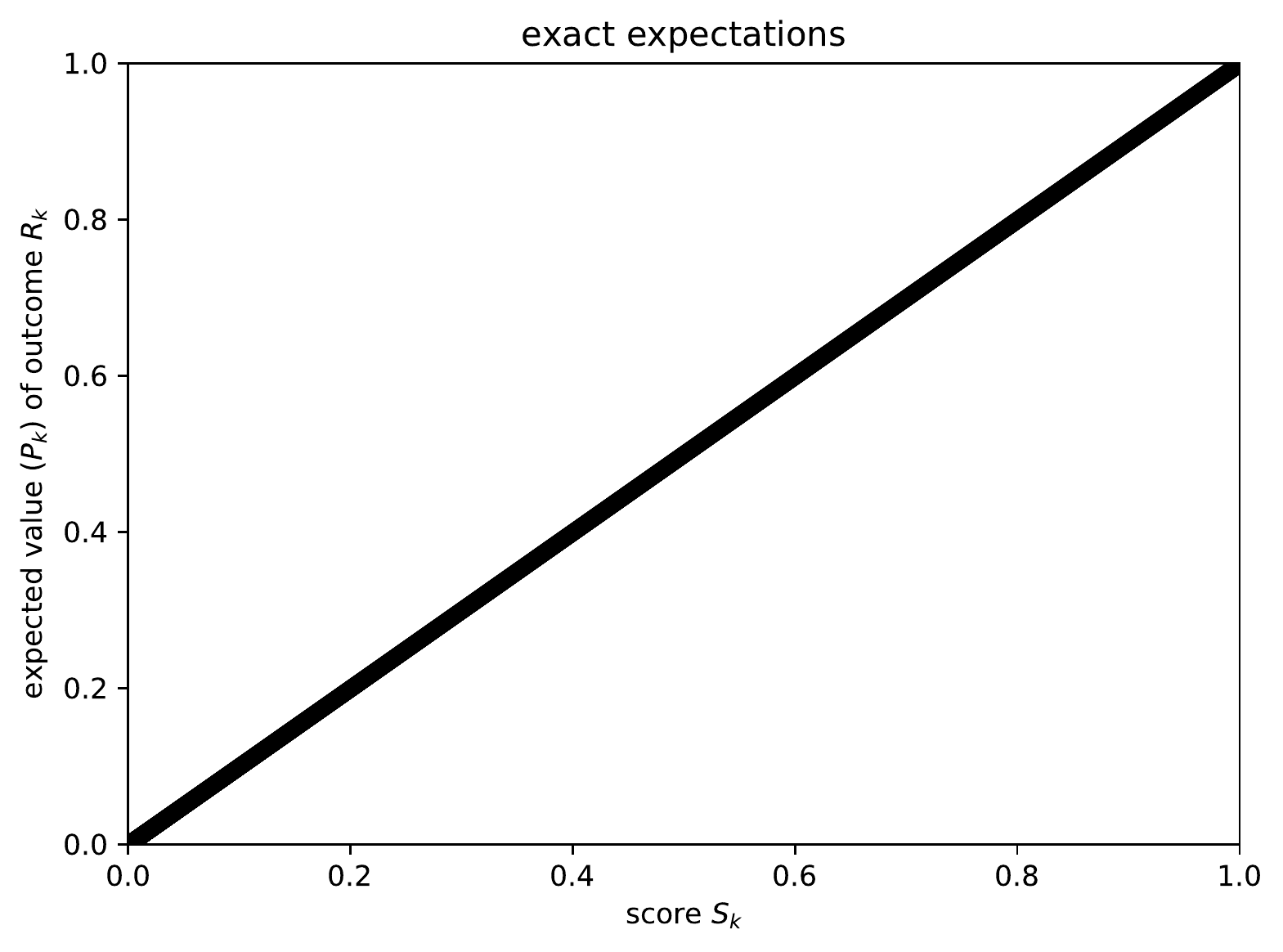}}

\end{centering}
\caption{Ground-truth reliability diagram for Figure~\ref{10000_00}}
\label{10000_00e}
\end{figure}

\begin{figure}
\begin{centering}

\parbox{\imsize}{\includegraphics[width=\imsize]
                {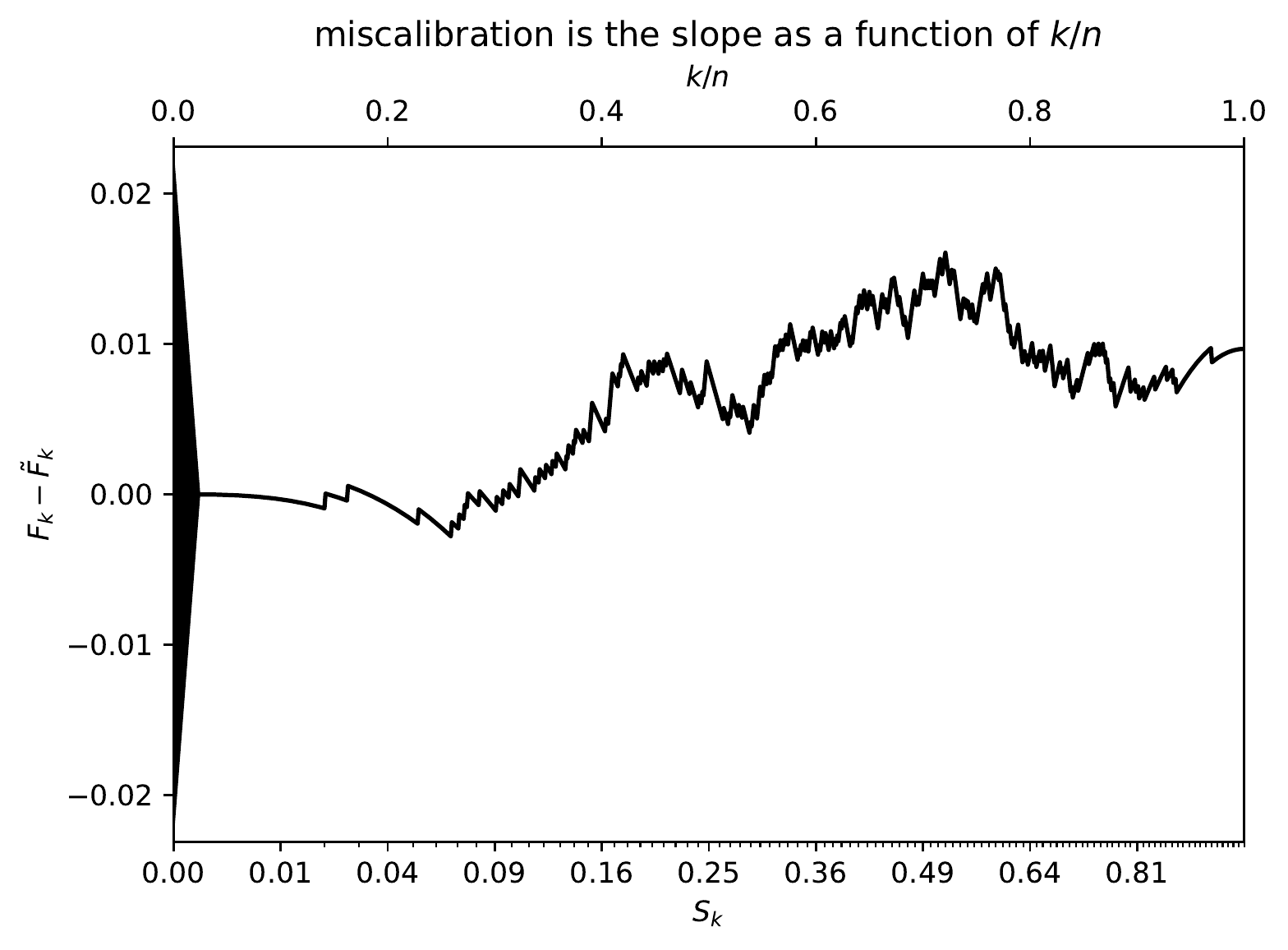}}
\quad\quad
\parbox{\imsize}{\includegraphics[width=\imsize]
                {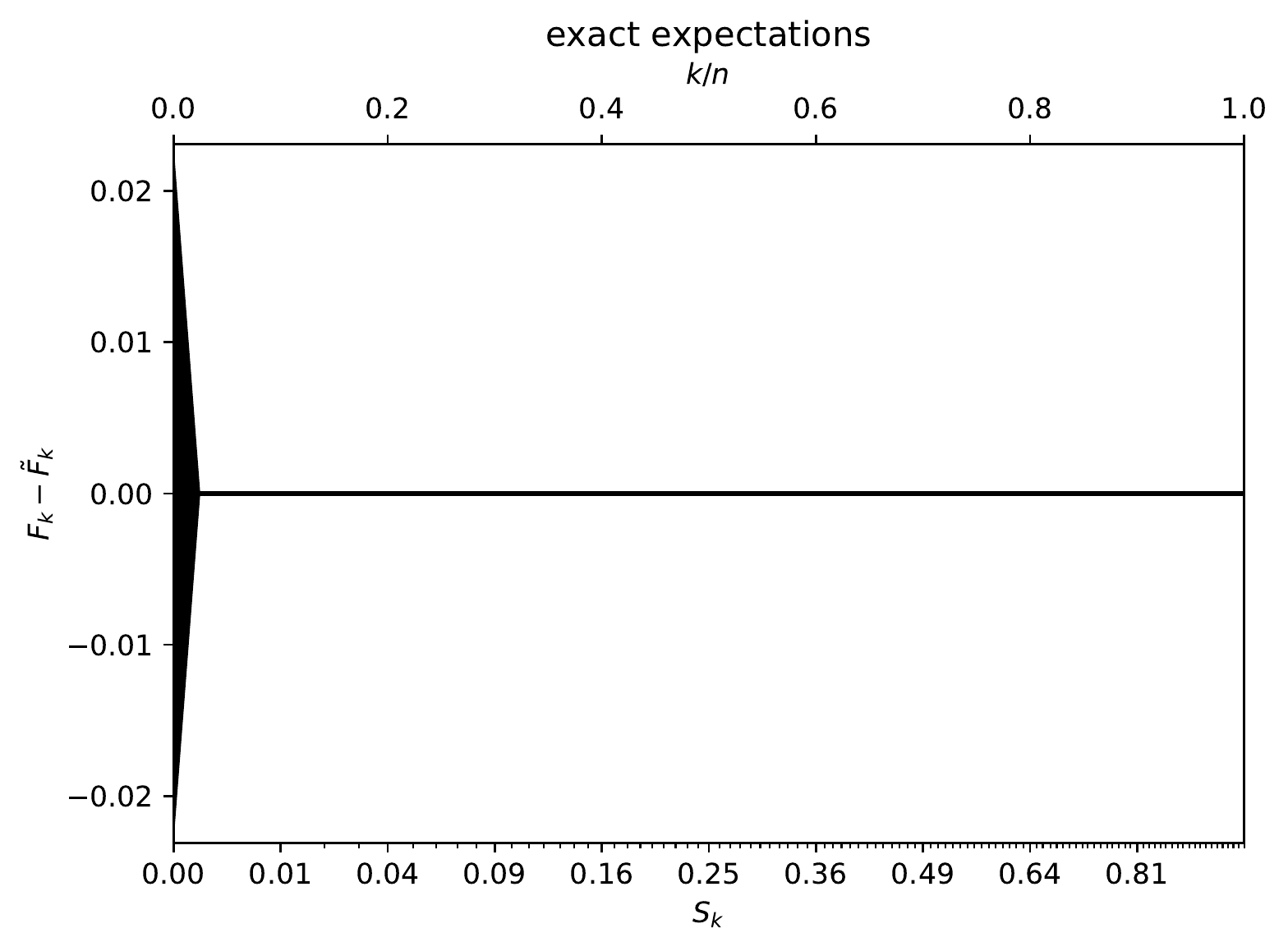}}

\vspace{\vertsep}

\parbox{\imsize}{\includegraphics[width=\imsize]
                {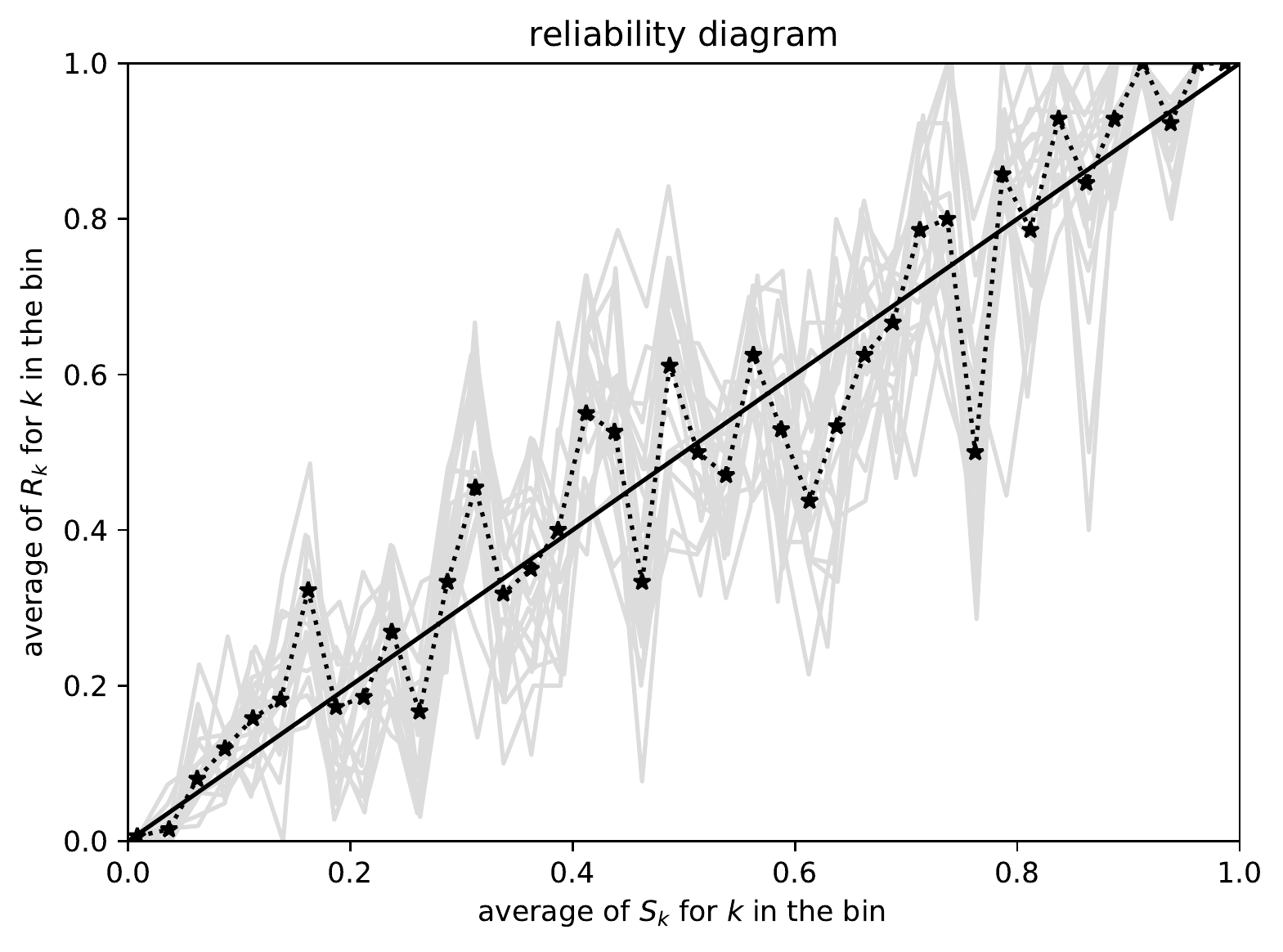}}
\quad\quad
\parbox{\imsize}{\includegraphics[width=\imsize]
                {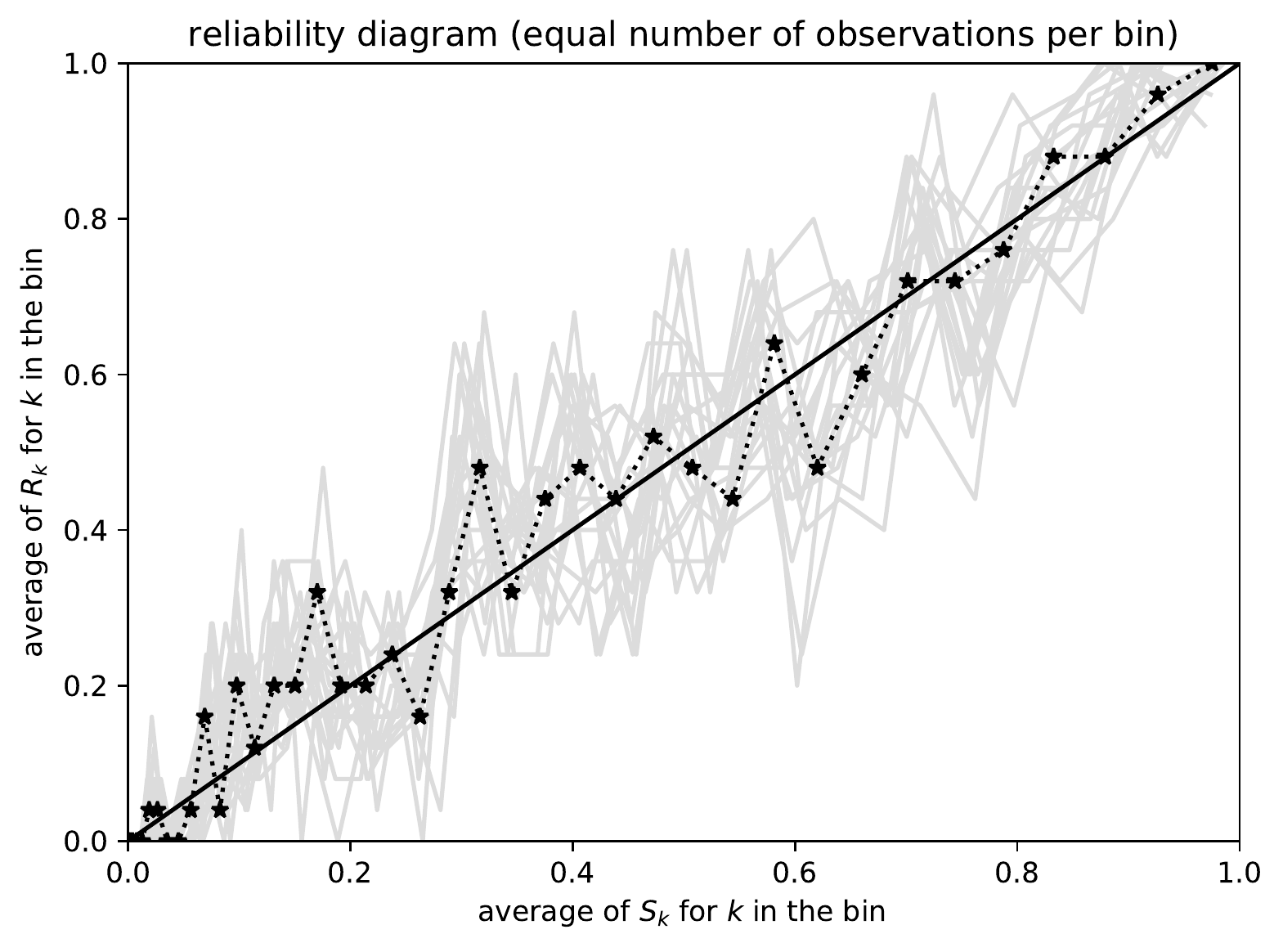}}

\vspace{\vertsep}

\parbox{\imsize}{\includegraphics[width=\imsize]
                {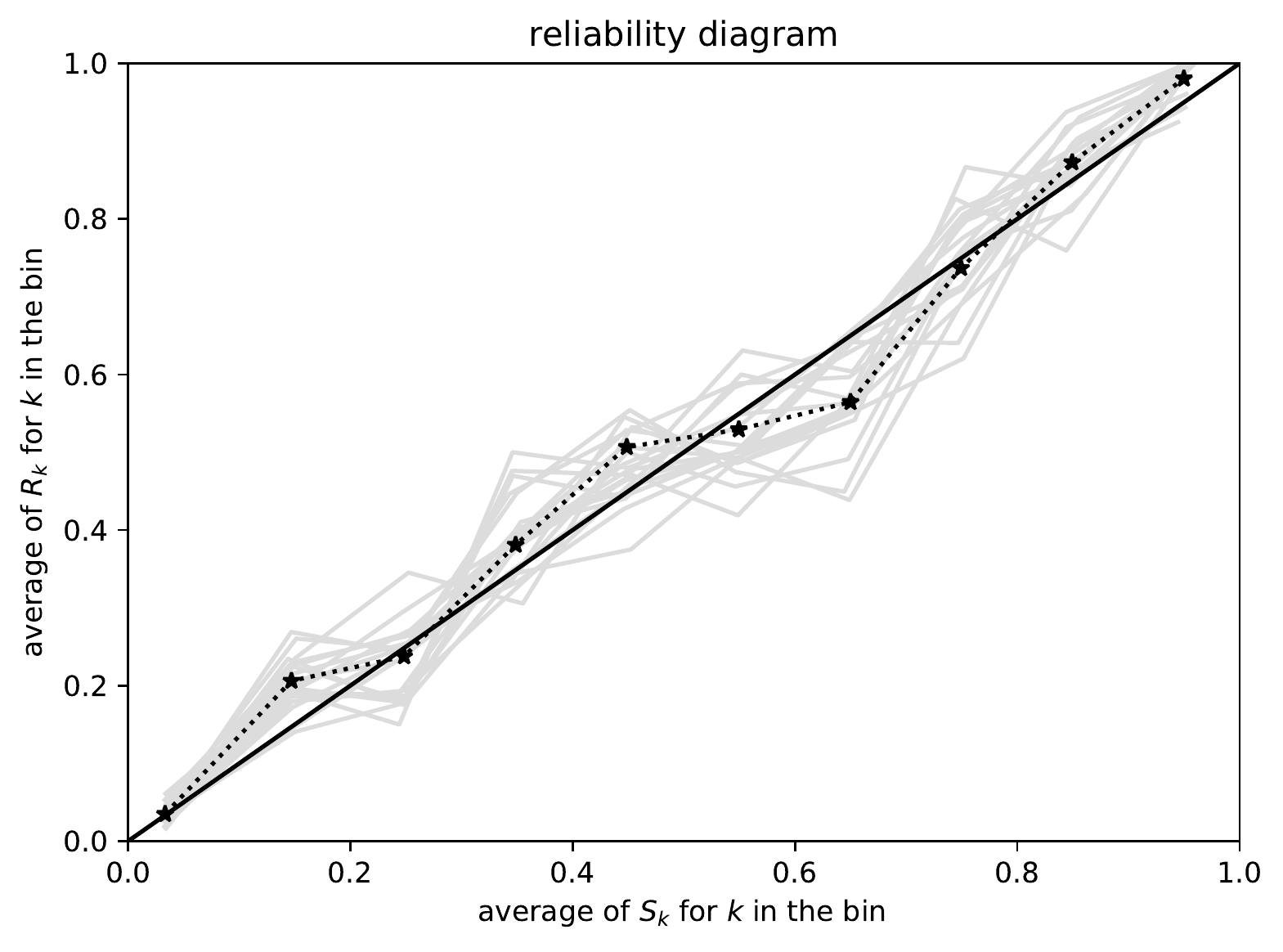}}
\quad\quad
\parbox{\imsize}{\includegraphics[width=\imsize]
                {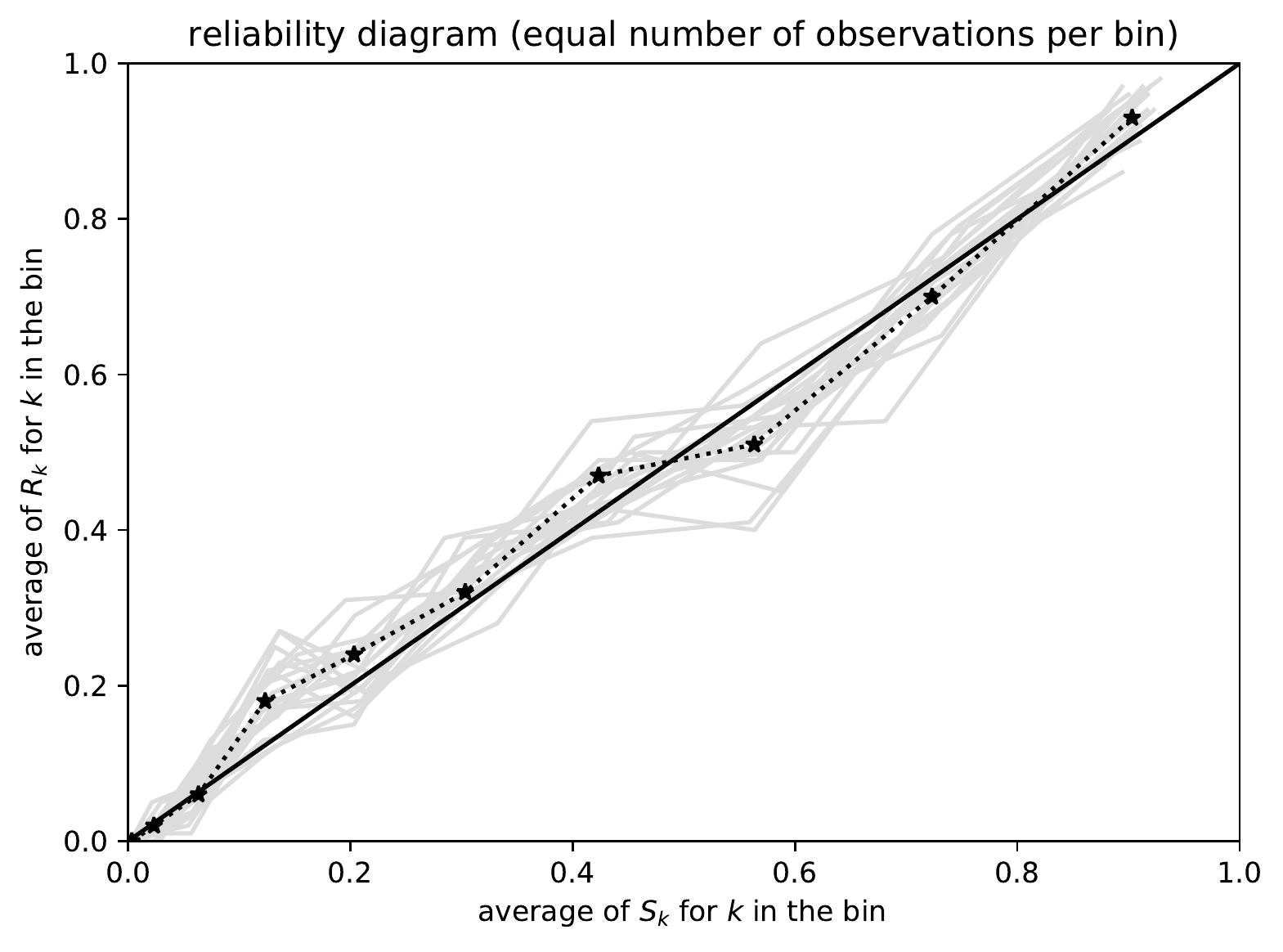}}

\end{centering}
\caption{$n =$ 1,000; $S_1$, $S_2$, \dots, $S_n$ are denser near 0;
         Kuiper's statistic is $0.01886 / \sigma = 1.633$,
         Kolmogorov's and Smirnov's is $0.01606 / \sigma = 1.391$.
Figure~\ref{1000_00e} displays the ground-truth reliability diagram.
}
\label{1000_00}
\end{figure}

\begin{figure}
\begin{centering}

\parbox{\imsize}{\includegraphics[width=\imsize]
                {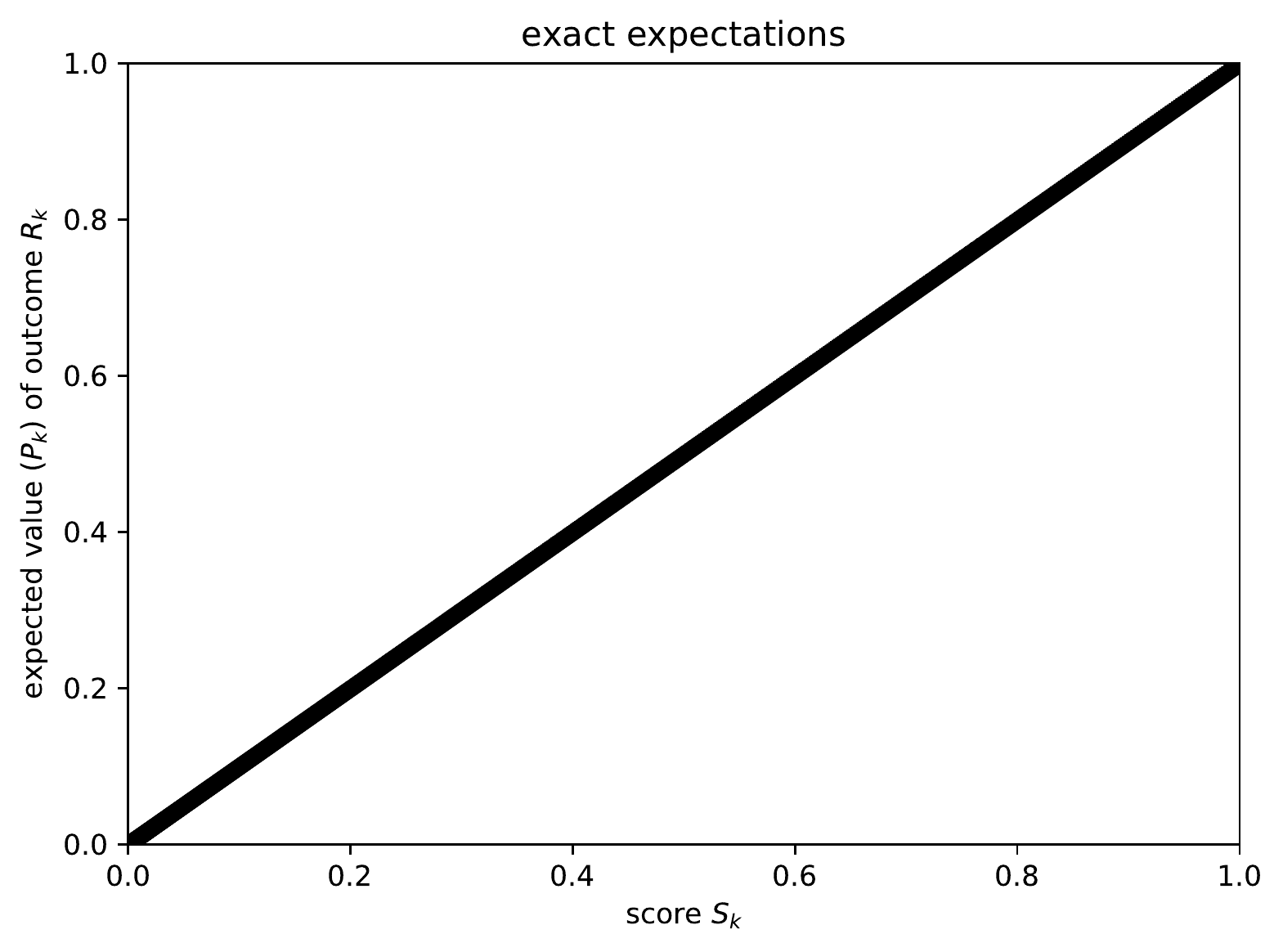}}

\end{centering}
\caption{Ground-truth reliability diagram for Figure~\ref{1000_00}}
\label{1000_00e}
\end{figure}

\begin{figure}
\begin{centering}

\parbox{\imsize}{\includegraphics[width=\imsize]
                {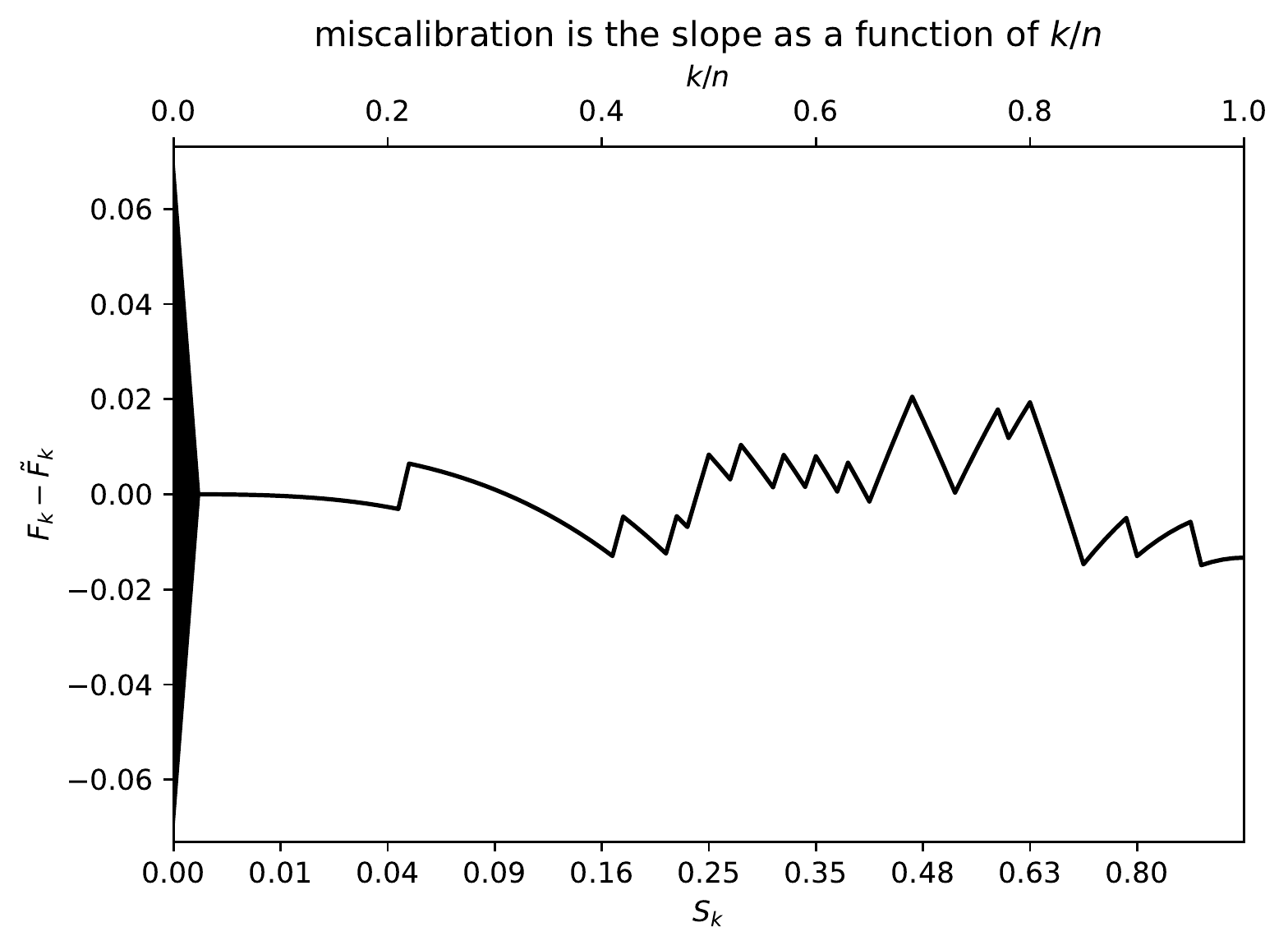}}
\quad\quad
\parbox{\imsize}{\includegraphics[width=\imsize]
                {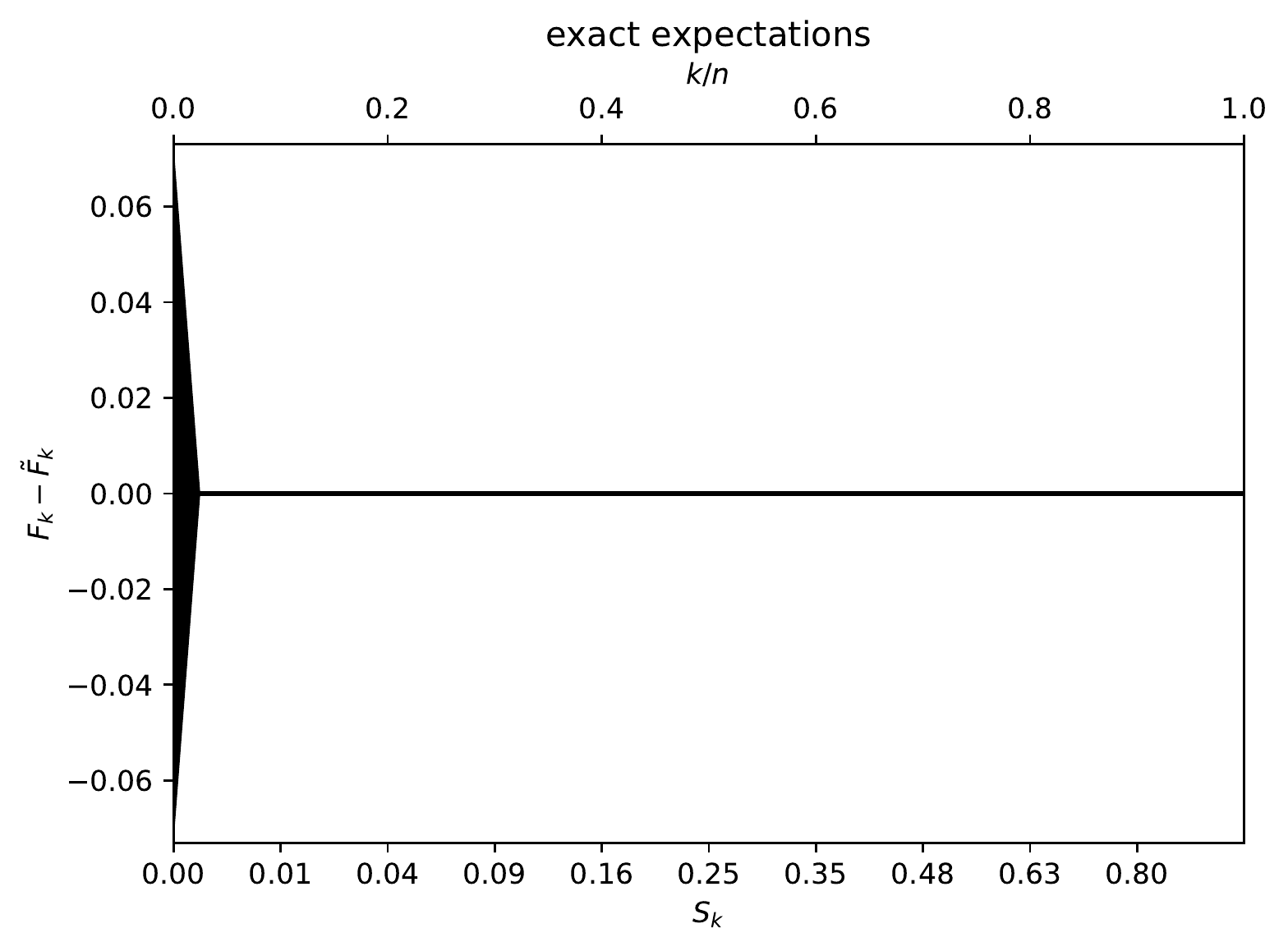}}

\vspace{\vertsep}

\parbox{\imsize}{\includegraphics[width=\imsize]
                {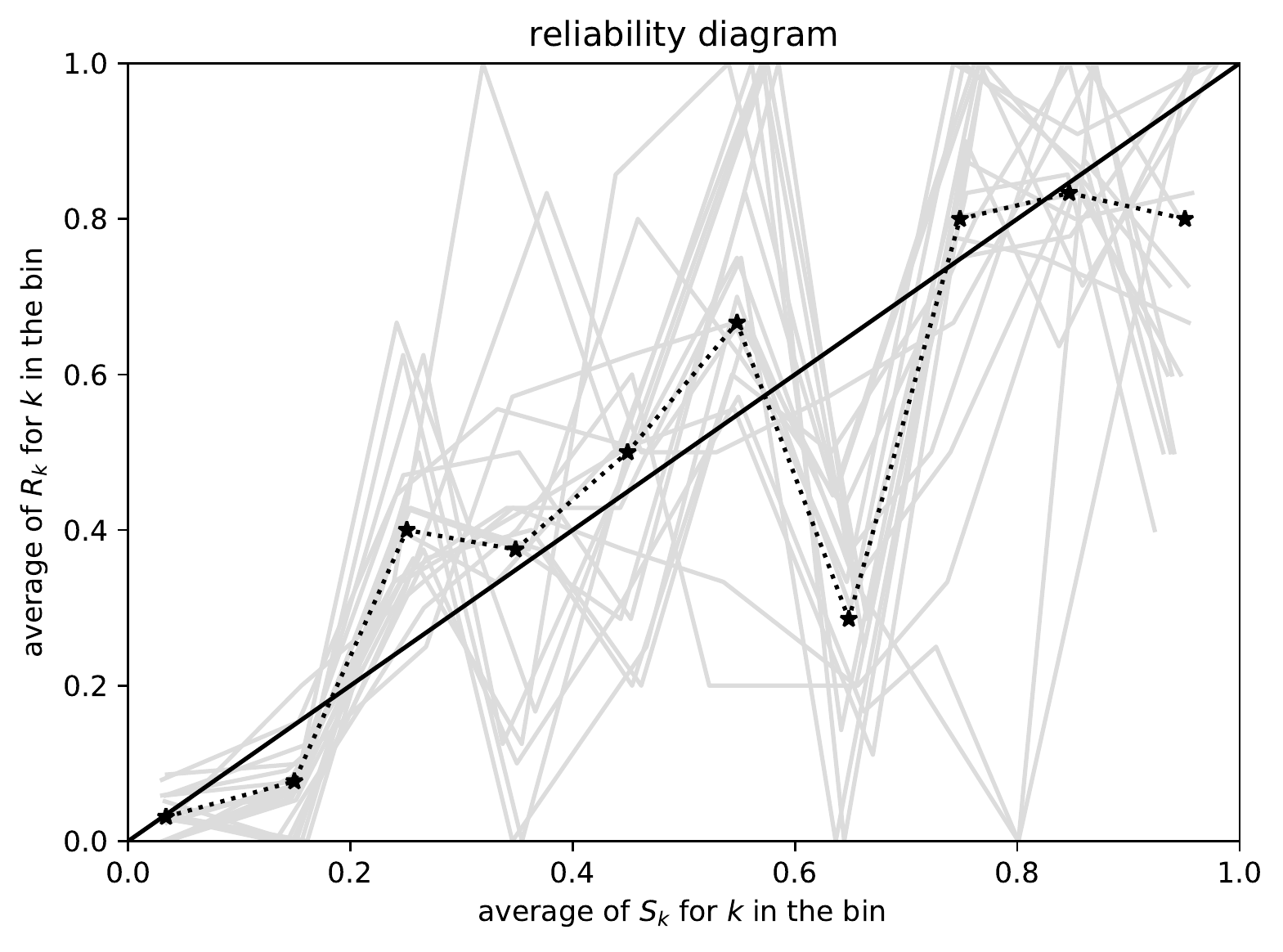}}
\quad\quad
\parbox{\imsize}{\includegraphics[width=\imsize]
                {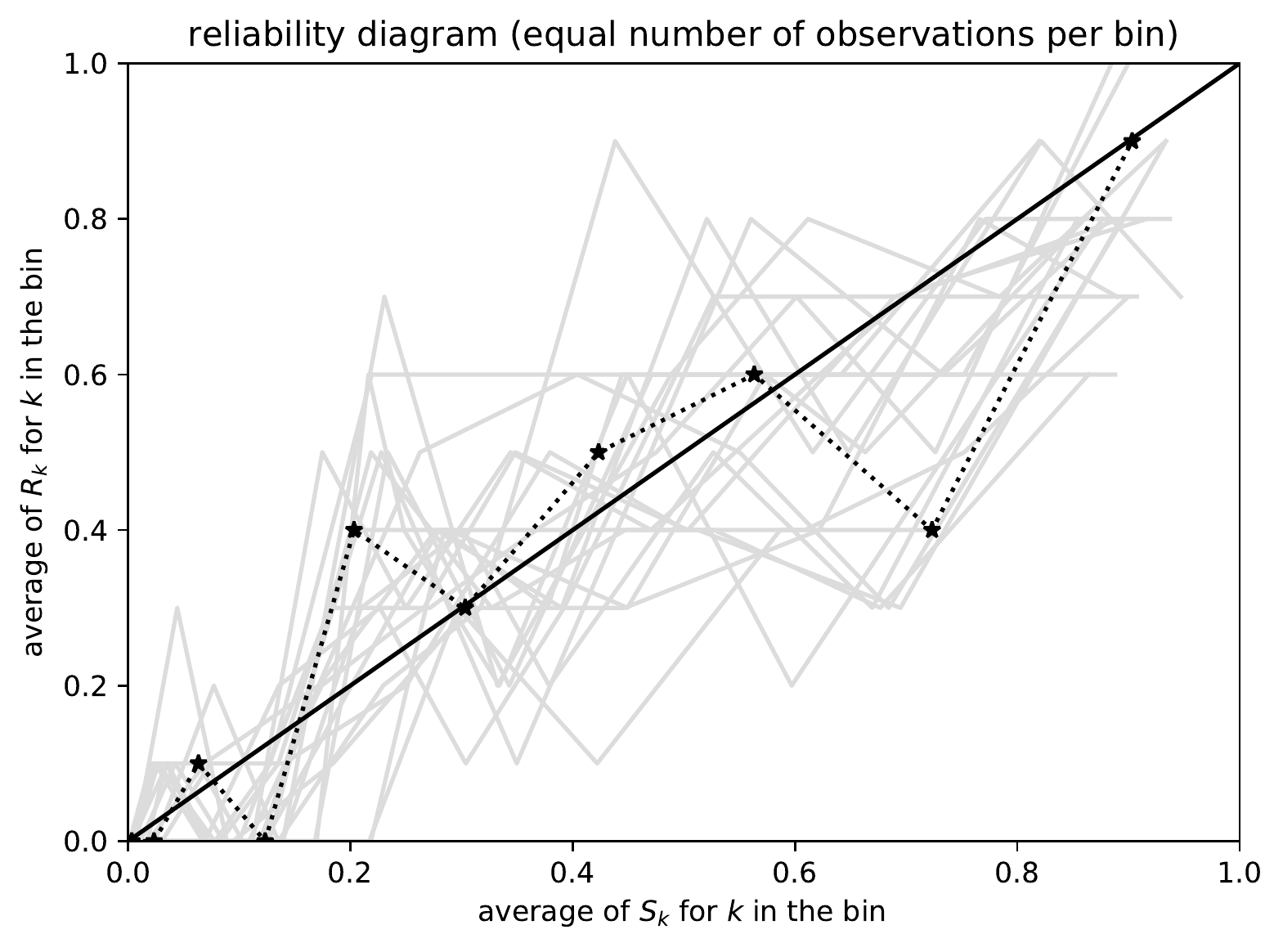}}

\vspace{\vertsep}

\parbox{\imsize}{\includegraphics[width=\imsize]
                {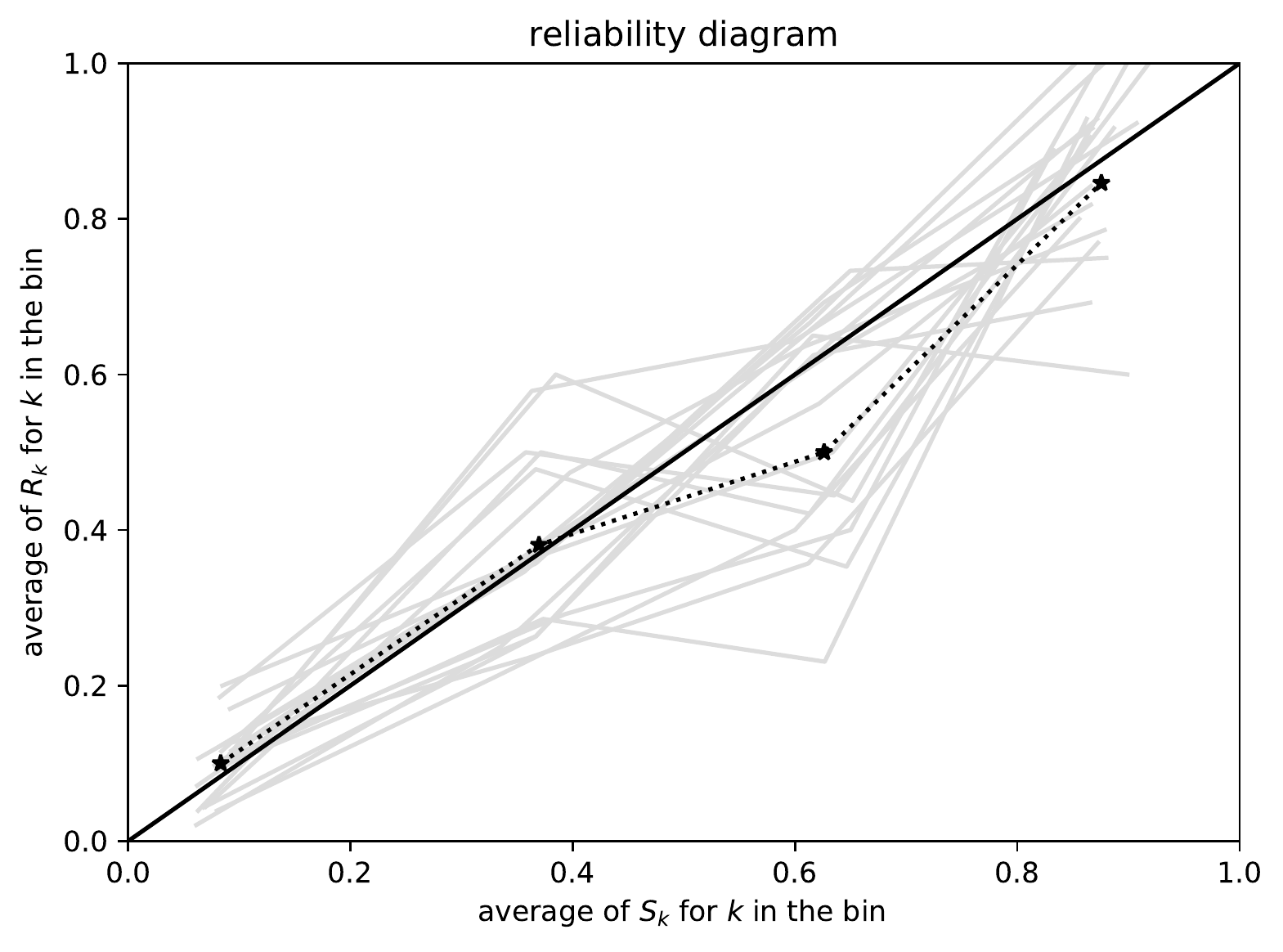}}
\quad\quad
\parbox{\imsize}{\includegraphics[width=\imsize]
                {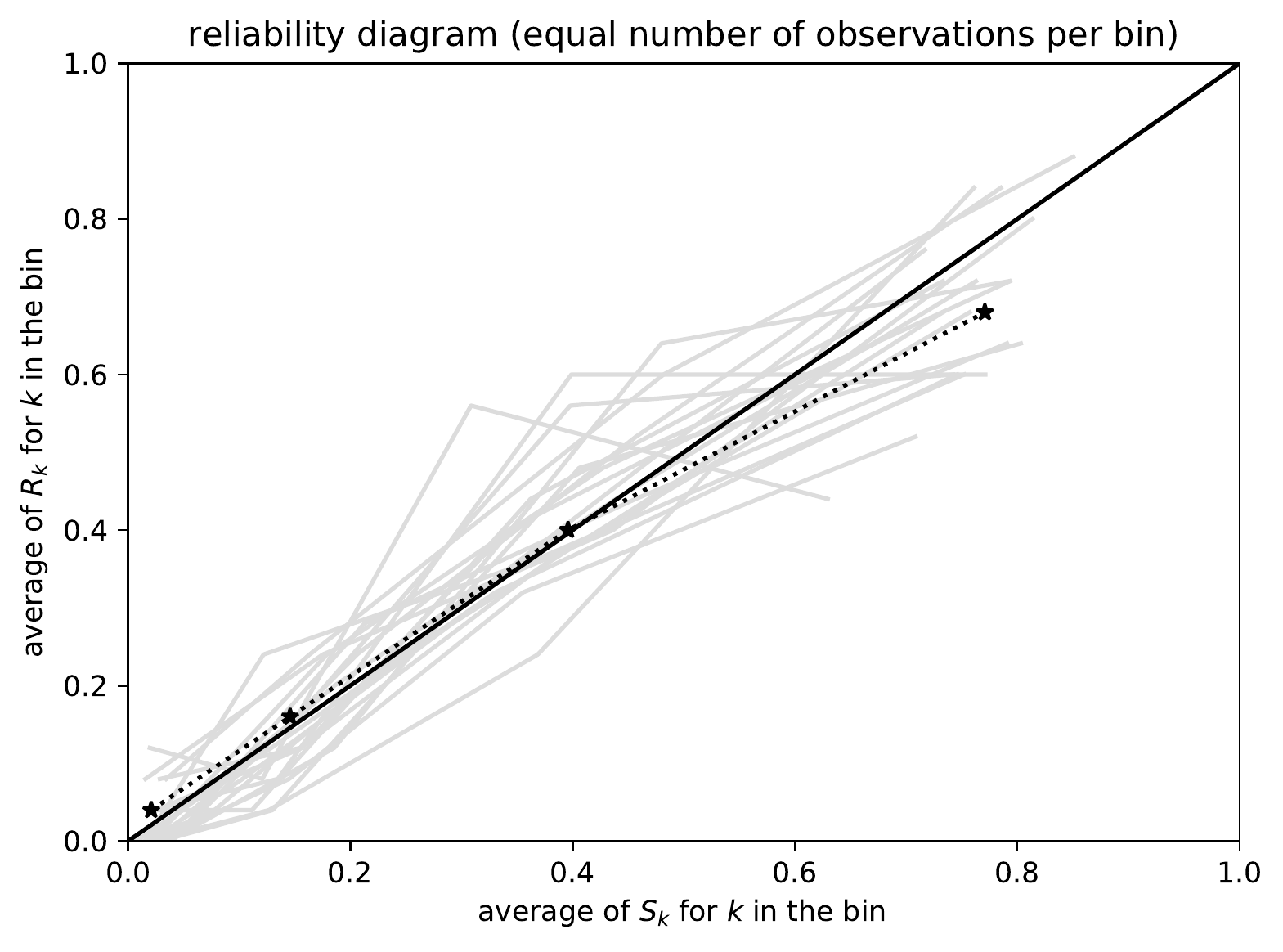}}

\end{centering}
\caption{$n =$ 100; $S_1$, $S_2$, \dots, $S_n$ are denser near 0;
         Kuiper's statistic is $0.03541 / \sigma = 0.9696$,
         Kolmogorov's and Smirnov's is $0.02050 / \sigma = 0.5615$.
Figure~\ref{100_00e} displays the ground-truth reliability diagram.
}
\label{100_00}
\end{figure}

\begin{figure}
\begin{centering}

\parbox{\imsize}{\includegraphics[width=\imsize]
                {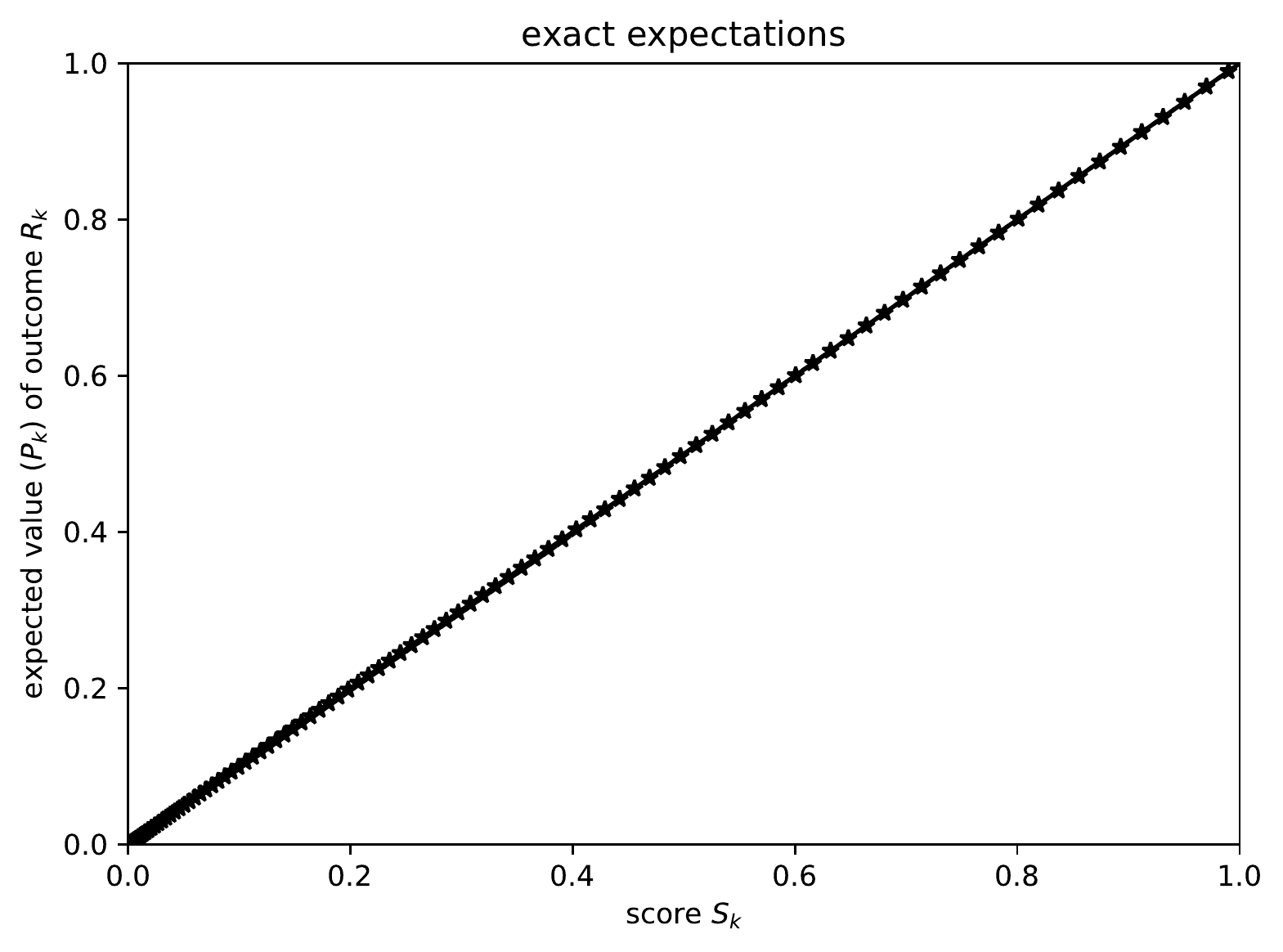}}

\end{centering}
\caption{Ground-truth reliability diagram for Figure~\ref{100_00}}
\label{100_00e}
\end{figure}

\clearpage

\bibliography{subpopulation}
\bibliographystyle{vancouver}

\end{document}